%% file: lbne-sci-opp.tex
\documentclass{lbnepaper}
\pdfoutput=1            
\input{lbnepaper/print}                   
\input{preamble}

\begin{document}

\pagestyle{empty}
\pagenumbering{roman}

\includepdf[noautoscale=true,height=11.1in,width=8.5in]{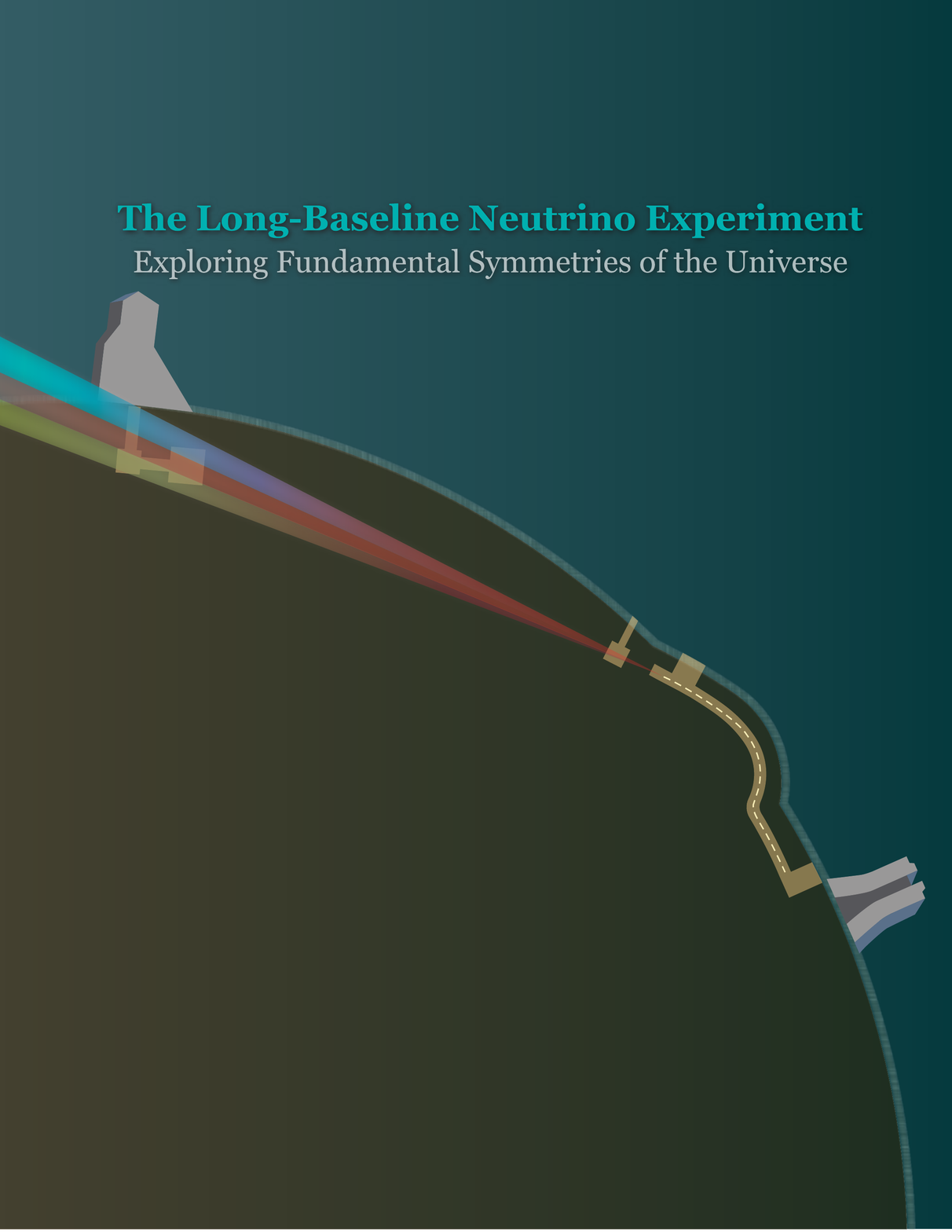}
\cleardoublepage

\input{lbne-sci-opp-main}            

\input{lbne-sci-opp-thebib}

\clearpage
\pagestyle{empty}
\cleardoublepage
\includepdf[noautoscale=true,height=11.1in,width=8.5in,pages={{},1}]{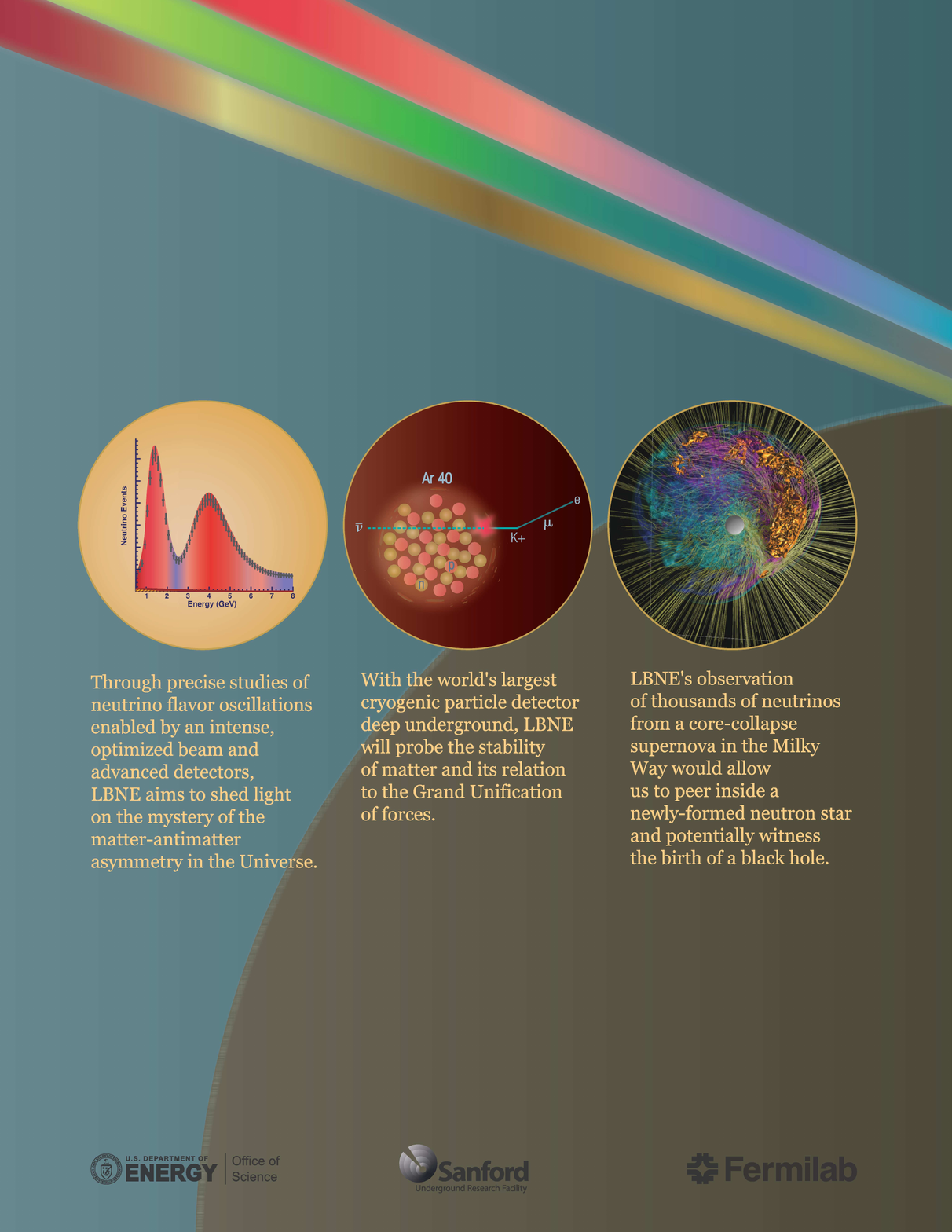}
\end{document}

%% file: lbnepaper/print.tex
\renewcommand{\lbnepagestyle}{glossy}

%% file: preamble.tex
\input{lbnepaper/color-palette}      
\input{lbnepaper/introbox}            
\input{lbnepaper/headers}

\input{lbnepaper/sections}

\input{lbnepaper/nomenclature}
\input{lbnepaper/appendix}

\input{lbnepaper/feynman}

\input{lbnepaper/defs}
\input{lbnepaper/moredefs}

\input{lbnepaper/units}                 
\input{lbnepaper/tables}                
\input{lbnepaper/hyperref}            

\usepackage{mhchem}                  
\usepackage{afterpage}
\usepackage{pagecolor}

\title{}

%% file: lbnepaper/color-palette.tex
\definecolor{sciopp-gold}{cmyk}            {0.03, 0.40, 1.00, 0.01}
\definecolor{sciopp-turquoise}{cmyk}       {0.89, 0.00, 0.38, 0.00}
\definecolor{sciopp-deep-red}{cmyk}        {0.03, 1.00, 1.00, 0.24}
\definecolor{sciopp-dark-lavender}{cmyk}   {0.56, 0.55, 0.00, 0.01}
\definecolor{sciopp-peachy-pink}{cmyk}     {0.04, 0.55, 0.45, 0.00}
\definecolor{sciopp-light-moss-green}{cmyk}{0.16, 0.00, 0.51, 0.36}
\definecolor{sciopp-pale-yellow}{cmyk}     {0.00, 0.20, 0.55, 0.00}
\definecolor{sciopp-dark-turquoise}{cmyk}  {0.56, 0.00, 0.20, 0.87}
\definecolor{sciopp-grass-green}{cmyk}     {0.75, 0.00, 0.98, 0.00}
\definecolor{sciopp-light-fakegrey}{cmyk}  {0.43, 0.35, 0.35, 0.01}
\definecolor{sciopp-light-grey}{cmyk}      {0.00, 0.00, 0.00, 0.50}
\definecolor{sciopp-brown}{cmyk}           {0.45, 0.54, 0.94, 0.22}
\definecolor{sciopp-sienna}{cmyk}          {0.03, 0.69, 0.70, 0.56}
\definecolor{sciopp-cherry-red}{cmyk}      {0.08, 1.00, 0.71, 0.00}
\definecolor{sciopp-medium-fakegrey}{cmyk} {0.65, 0.57, 0.52, 0.29}
\definecolor{sciopp-medium-grey}{cmyk}     {0.00, 0.00, 0.00, 0.80}
\definecolor{sciopp-bright-yellow}{cmyk}   {0.00, 0.01, 0.72, 0.00}
\definecolor{sciopp-baby-blue}{cmyk}       {0.49, 0.00, 0.05, 0.00}

\definecolor{CHAP1COL}{named}{sciopp-gold}
\definecolor{CHAP2COL}{named}{sciopp-deep-red}     
\definecolor{CHAP3COL}{named}{sciopp-brown}  
\definecolor{CHAP4COL}{named}{sciopp-dark-turquoise} 
\definecolor{CHAP5COL}{named}{sciopp-grass-green}     
\definecolor{CHAP6COL}{named}{sciopp-dark-lavender}     
\definecolor{CHAP7COL}{named}{sciopp-turquoise}   
\definecolor{CHAP8COL}{named}{sciopp-sienna}   
\definecolor{CHAP9COL}{named}{sciopp-light-moss-green}
\definecolor{CHAPACOL}{named}{sciopp-medium-grey}

%% file: lbnepaper/introbox.tex
\usepackage[xcolor,framemethod=tikz,nobreak=true]{mdframed}
\usetikzlibrary{shadows}

\definecolor{introcolor}{rgb}{255,255,0}
\newcommand{\IntroBackgroundColor}{introcolor}
\newcommand{\IntroLineColor}{introcolor}

\mdfdefinestyle{introstyle}{frametitle=}
\mdfapptodefinestyle{introstyle}{linecolor=\IntroLineColor}
\mdfapptodefinestyle{introstyle}{backgroundcolor=\IntroBackgroundColor}
\mdfapptodefinestyle{introstyle}{roundcorner=7pt}
\mdfapptodefinestyle{introstyle}{outerlinewidth=1pt}
\mdfapptodefinestyle{introstyle}{frametitlefont=\normalfont\ttfamily}
\mdfapptodefinestyle{introstyle}{frametitlealignment=\centering}
\mdfapptodefinestyle{introstyle}{skipabove=0pt}
\mdfapptodefinestyle{introstyle}{skipbelow=0pt}
\mdfapptodefinestyle{introstyle}{leftmargin=0pt}
\mdfapptodefinestyle{introstyle}{rightmargin=0pt}
\mdfapptodefinestyle{introstyle}{innerleftmargin=15pt}
\mdfapptodefinestyle{introstyle}{innerrightmargin=15pt}
\mdfapptodefinestyle{introstyle}{innertopmargin=15pt}
\mdfapptodefinestyle{introstyle}{innerbottommargin=15pt}
\mdfapptodefinestyle{introstyle}{outermargin=0pt}
\mdfapptodefinestyle{introstyle}{innermargin=0pt}
\mdfapptodefinestyle{introstyle}{extratopheight=0pt}
\mdfapptodefinestyle{introstyle}{needspace=0.5\baselineskip}

\newmdenv[style=introstyle]{introbox}

%% file: lbnepaper/headers.tex
% ...

\usepackage{tikz}   
\usetikzlibrary{snakes}
\usetikzlibrary{matrix}
\usetikzlibrary{trees}
\usetikzlibrary{positioning,arrows}
\usetikzlibrary{decorations.pathmorphing}
\usetikzlibrary{decorations.markings}

\usepackage{fancyhdr} 

\fancypagestyle{empty}{
\fancyhf{} 

}

\newcommand{\titlepagecornerblock}{}  
\fancypagestyle{titlepage}{
\fancyhf{} 

\fancyhead[RO]{\titlepagecornerblock}
}

\fancypagestyle{plain}{%
\fancyhf{} 
\fancyfoot[C]{\bfseries \thepage} 

}

\fancypagestyle{simple}{%
\fancyhf{} 
\fancyhead[RO,LE]{\textsf{\footnotesize \thechapter--\thepage}}
\fancyhead[LO,RE]{\textsf{\footnotesize \leftmark}}
\fancyfoot[CO,CE]{\textsf{\footnotesize The Long-Baseline Neutrino Experiment}}}

\newcommand{\toccolor}{black}
\newcommand{\headrulecolor}{black}
\newcommand{\ChapterTabColor}{gray}

\fancypagestyle{mary}{
\fancyhf{} 
\fancyhead[RO,LE]{\textsf{\footnotesize \thechapter--\thepage}}
\fancyhead[LO,RE]{\textsf{\footnotesize \leftmark}}
\fancyfoot[CO,CE]{\textsf{\footnotesize The Long-Baseline Neutrino Experiment}}
\fancyfoot[LE]{%
\setlength{\unitlength}{1mm}
  \begin{tikzpicture}[remember picture,overlay] 
    \draw[fill=\ChapterTabColor,\ChapterTabColor] (18,20) rectangle (22,21);
    \draw[fill=\ChapterTabColor,\ChapterTabColor] (18,18.5) rectangle (22,19.5);
    \draw[fill=\ChapterTabColor,\ChapterTabColor] (18,17) rectangle (22,18);
    \node at (17,0) {\includegraphics[scale=0.3]{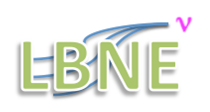}};
   \end{tikzpicture}
}
\fancyfoot[RO]{%
\setlength{\unitlength}{1mm}
  \begin{tikzpicture}[remember picture,overlay]
    \draw[fill=\ChapterTabColor,\ChapterTabColor] (-18,20) rectangle (-22,21);
    \draw[fill=\ChapterTabColor,\ChapterTabColor] (-18,18.5) rectangle (-22,19.5);
    \draw[fill=\ChapterTabColor,\ChapterTabColor] (-18,17) rectangle (-22,18);
    \node at (-17,0) {\includegraphics[scale=0.3]{lbnelogo.png}};
   \end{tikzpicture}
}

}

\newcounter{margintabsbump}
\newcommand{\margintabs}[7]{%
  \setlength{\unitlength}{1mm}
  \begin{tikzpicture}[remember picture, overlay] 
    \pgfmathsetmacro\x{#3}
    \pgfmathsetmacro\xx{#3 + #5}
    \pgfmathsetmacro\bump{#7 * \value{margintabsbump}}
    \foreach \itab in {1,...,#1} {
      \pgfmathsetmacro\y{#4 + \itab * #2 - \bump}
      \pgfmathsetmacro\yy{\y + #6}
      \draw[fill=\ChapterTabColor,\ChapterTabColor,rounded corners=7pt] (\x,\y) rectangle (\xx,\yy);
    }
  \end{tikzpicture}
}

\fancypagestyle{glossy}{%
\fancyhf{} 

\fancyhead[RO,LE]{\textsf{\footnotesize \bfseries \thepage}}
\fancyhead[CE]{\textcolor{\headrulecolor}{\textsf{\footnotesize \bfseries \thechapter~\leftmark}}}
\fancyhead[CO]{\textcolor{\headrulecolor}{\textsf{\footnotesize \bfseries \thechapter\ifnum\value{section}>0 .\arabic{section} \fi~\rightmark}}}
\fancyfoot[CO,CE]{\textsf{\footnotesize \bfseries The Long-Baseline Neutrino Experiment}}
\fancyfoot[LE]{\margintabs{3}{1.5}{-0.7}{18}{-4}{1}{1.5}}
\fancyfoot[RO]{\margintabs{3}{1.5}{0.8}{18}{4}{1}{1.5}}

}

\fancypagestyle{references}{%
\fancyhf{} 
\fancyhead[RO,LE]{\textsf{\footnotesize \bfseries \thepage}}
\fancyhead[LO,RE]{\textsf{\footnotesize References}}
\fancyfoot[CO,CE]{\textsf{\footnotesize \bfseries The Long-Baseline Neutrino Experiment}}
\fancyfoot[LE]{\margintabs{3}{1.5}{-0.8}{18}{-4}{1}{1.5}}
\fancyfoot[RO]{\margintabs{3}{1.5}{0.8}{18}{4}{1}{1.5}}

}
\fancypagestyle{acknowledgement}{%
\fancyhf{} 
\fancyhead[RO,LE]{\textsf{\footnotesize \bfseries \thepage}}
\fancyhead[LO,RE]{\textsf{\footnotesize Acknowledgements}}
\fancyfoot[CO,CE]{\textsf{\footnotesize \bfseries The Long-Baseline Neutrino Experiment}}
\fancyfoot[LE]{\margintabs{3}{1.5}{-0.8}{18}{-4}{1}{1.5}}
\fancyfoot[RO]{\margintabs{3}{1.5}{0.8}{18}{4}{1}{1.5}}

}

\fancypagestyle{wiggles}{%
\fancyhf{} 
\fancyhead[EC,OC]{
\leftmark\\
\begin{tikzpicture}
  \node (a) at (0,0) {};
  \node (b) at (12,0) {};
  \draw[thin,green,snake=coil,segment aspect=0,segment amplitude=1mm,segment length=1mm] (a) -- (b);
  \draw[thick,red,snake=coil,segment aspect=0,segment amplitude=1mm,segment length=4.42mm]  (a) -- (b);
  \draw[thick,blue,snake=coil,segment aspect=0,segment amplitude=1mm,segment length=5mm] (a) -- (b);
\end{tikzpicture}}
\fancyfoot[C]{\bfseries \thepage}

}

%% file: lbnepaper/sections.tex
\usepackage{tocloft}
\cftsetpnumwidth{2em}
\tocloftpagestyle{plain}

\usepackage{tikz}
\usepackage[explicit]{titlesec}

\newcommand\ChapterTitleColor{gray!10}
\newcommand\ChapterBubbleColor{gray!20}
\newcommand\ChapterTextColor{black}
\newcommand\ChapterTableColor{black}
\newcommand\lbnechapterlabel{\chaptername \\  \thechapter }

\newlength{\ChapterBubbleWidth}
\newlength{\ChapterBubbleHeight}
\newcommand*\chapterlabel{}
\titleformat{\chapter}
{\gdef\chapterlabel{}
  \normalfont\sffamily\Huge\bfseries}
{\gdef\chapterlabel{\lbnechapterlabel}}{0pt}
{\begin{tikzpicture}[remember picture,overlay]
    \node[yshift=-3cm] at (current page.north west)
    {\begin{tikzpicture}[remember picture, overlay]
        \draw[fill=\ChapterTitleColor,\ChapterTitleColor] (0,0) rectangle (\paperwidth,3cm);
        \draw[anchor=north west, 
              xshift=1in+\hoffset+\oddsidemargin+0.5\textwidth-0.5\ChapterBubbleWidth,
              yshift=-0.5\ChapterBubbleHeight,
              rounded corners=7pt,
              inner sep=15pt, 
              line width=1pt,
              fill=\ChapterBubbleColor,
              draw=\ChapterTitleColor]
              (0,0) rectangle (\ChapterBubbleWidth, \ChapterBubbleHeight);
        \node[anchor=north west, 
              xshift=1in+\hoffset+\oddsidemargin+0.5\textwidth-0.5\ChapterBubbleWidth,
              yshift=0.65\ChapterBubbleHeight,
              text width=0.95\ChapterBubbleWidth]
              at (0,0)
              {\parbox[b]{0.23\textwidth}{\centering\color{\ChapterTextColor}\chapterlabel}\parbox[b]{0.77\textwidth}{\flushright #1}};
       \end{tikzpicture}
      };
   \end{tikzpicture}
  }
\titlespacing*{\chapter}{0pt}{50pt}{-60pt}



%% file: lbnepaper/nomenclature.tex
\usepackage[intoc]{nomencl}

\makenomenclature

%% file: lbnepaper/appendix.tex
\usepackage[toc,page]{appendix}

%% file: lbnepaper/feynman.tex
\usepackage{tikz}
\usetikzlibrary{trees}
\usetikzlibrary{positioning,arrows}
\usetikzlibrary{decorations.pathmorphing}
\usetikzlibrary{decorations.markings}
\usetikzlibrary{decorations.pathreplacing}

\newcommand{\feynmantikzset}{%
\tikzset{
  weakz/.style={draw=orange, dashed},
  weakw/.style={draw=orange, dashed},
  lepton/.style={draw=yellow, solid},
  neutrino/.style={draw=red, solid},
  quark/.style={draw=blue, solid},
  hadron/.style={draw=black, solid},
}
}

\newcommand{\feynmanNC}[5]{%
  \feynmantikzset
  \begin{minipage}[b]{#5}
    \begin{center}
\scalebox{0.8}{
\begin{tikzpicture}[ultra thick,node distance=1cm and 1.5cm]
\coordinate[label=left:{#1}] (nu1);
\coordinate[below right=of nu1] (z1);
\coordinate[above right=of z1,label=right:{#1}] (nu2);
\coordinate[below=1.25cm of z1] (z2);
\coordinate[below left=of z2,label=left:{#3}] (o1);
\coordinate[below right=of z2,label=right:{#3}] (o2);

\draw[#2] (nu1) -- (z1);
\draw[#2] (z1) -- (nu2);
\draw[#4] (o1) -- (z2);
\draw[#4] (z2) -- (o2);
\draw[weakz] (z1) -- node[label=right:{$Z^0$}] {} (z2);
\end{tikzpicture}
}
    \end{center}
  \end{minipage}
}

%% file: lbnepaper/defs.tex
\newcommand{\mdeltacp}{\delta_{\rm CP}}
\newcommand{\deltacp}{$\mdeltacp$\xspace}

\newcommand{\numu}{$\nu_\mu$\xspace}
\newcommand{\nue}{$\nu_e$\xspace}

\newcommand{\superk}{Super--Kamiokande\xspace}

\newcommand{\SURF}{Sanford Underground Research Facility\xspace}

%% file: lbnepaper/moredefs.tex
\providecommand{\mybold}{}
\providecommand{\nm}{\mbox{\mybold $\nu_\mu$}}
\providecommand{\anm}{ \mbox{\mybold $\overline{\nu}_\mu$}} 
\providecommand{\ne}{\mbox{\mybold $\nu_e$}} 
\providecommand{\ane}{\mbox{\mybold $\overline{\nu}_e$}}

\newcommand{\degs}{\mbox{$^{\circ}$}}

\newcommand{\thetatwothree}{\mbox{$\theta_{23}$}}

\newcommand{\dchisq}{\mbox{$\Delta\chi^2$}}

%% file: lbnepaper/units.tex
\usepackage[detect-all=true,group-digits=true,group-separator={,},group-minimum-digits=4]{siunitx}

\DeclareSIUnit \kton {\kilo\tonne}
\DeclareSIUnit \kt {\kilo\tonne}
\DeclareSIUnit \Mt {\mega\tonne}
\DeclareSIUnit \eV {\electronvolt}
\DeclareSIUnit \keV {\kilo\electronvolt}
\DeclareSIUnit \MeV {\mega\electronvolt}
\DeclareSIUnit \GeV {\giga\electronvolt}
\DeclareSIUnit \km {\kilo\meter}
\DeclareSIUnit \kW {\kilo\watt}
\DeclareSIUnit \MW {\mega\watt}
\DeclareSIUnit \MHz {\mega\hertz}
\DeclareSIUnit \mrad {\milli\radian}
\DeclareSIUnit \year {year}
\DeclareSIUnit \POT {POT}
\DeclareSIUnit \sig {$\sigma$}
\DeclareSIUnit\parsec{pc}
\DeclareSIUnit\lightyear{ly}
\DeclareSIUnit\foot{ft}
\DeclareSIUnit\ft{ft}
\def\ktyr{\si[inter-unit-product=\ensuremath{{}\cdot{}}]{\kt\year}\xspace}
\def\Mtyr{\si[inter-unit-product=\ensuremath{{}\cdot{}}]{\Mt\year}\xspace}
\def\msr{\si[inter-unit-product=\ensuremath{{}\cdot{}}]{\meter\steradian}\xspace}

\newcommand{\SIadj}[2]{\SI[number-unit-product = -]{#1}{#2}}

\newcommand{\ktadj}[1]{\SIadj{#1}{\kt}}
\newcommand{\kmadj}[1]{\SIadj{#1}{\km}}
\newcommand{\keVadj}[1]{\SIadj{#1}{\keV}}
\newcommand{\MeVadj}[1]{\SIadj{#1}{\MeV}}
\newcommand{\GeVadj}[1]{\SIadj{#1}{\GeV}}
\newcommand{\MWadj}[1]{\SIadj{#1}{\MW}}
\newcommand{\kWadj}[1]{\SIadj{#1}{\kW}}
\newcommand{\tonneadj}[1]{\SIadj{#1}{\tonne}}
\newcommand{\ftadj}[1]{\SIadj{#1}{\ft}}
\newcommand{\ktmwyr}[1]{\SI[inter-unit-product=\ensuremath{{}\cdot{}}]{#1}{\kt\MW\year}}

%% file: lbnepaper/tables.tex
\usepackage{amsbsy} 

\usepackage{array} 

\usepackage{tabu}

\newcolumntype{$}{>{\color{\ChapterTableColor}\global\let\currentrowstyle\relax}}
\newcolumntype{^}{>{\color{\ChapterTableColor}\currentrowstyle}}
\newcommand{\rowstyle}[1]{\gdef\currentrowstyle{#1}%
  #1\ignorespaces
}

\newcolumntype{L}{%
  >{\color{\ChapterTableColor}\raggedright
    \bfseries
    \boldmath
  }l}%

\let\oldmc\multicolumn
\makeatletter
\newcommand{\mcinherit}{
  \renewcommand{\multicolumn}[3]{%
    \oldmc{##1}{##2}{\@rowcolor ##3}%
  }}
\makeatother

\newcommand{\rowtitlestyle}{
  \rowcolor{\ChapterBubbleColor}
  \rowstyle{\bfseries \boldmath} 
  \mcinherit
}

\newcommand{\colhline}{
  \arrayrulecolor{\ChapterBubbleColor}
  \specialrule{0.5pt}{0pt}{1pt}
  \arrayrulecolor{black}
}

\renewcommand{\toprule}{
  \arrayrulecolor{\ChapterTabColor}
  \specialrule{1.5pt}{2pt}{0pt}
  \arrayrulecolor{black}
}

\newcommand{\toprowrule}{
  \arrayrulecolor{\ChapterTabColor}
  \specialrule{1.2pt}{0pt}{1pt}
  \arrayrulecolor{black}
}

\renewcommand{\midrule}{
  \arrayrulecolor{\ChapterTabColor}
  \specialrule{1pt}{1pt}{1pt}
  \arrayrulecolor{black}
}

\renewcommand{\bottomrule}{
  \arrayrulecolor{\ChapterTabColor}
  \specialrule{1.5pt}{0pt}{2pt}
  \arrayrulecolor{black}
}

%% file: lbnepaper/hyperref.tex
\usepackage[pdftex,bookmarks]{hyperref}

\AfterPreamble{\hypersetup{
    pdftitle={The Long-Baseline Neutrino Experiment. Exploring
    Fundamental Symmetries of the Universe.},
    pdfauthor={LBNE Collaboration},
    final=true,
    colorlinks=false,
    linktocpage=true,
    linkbordercolor=blue,
    citebordercolor=green,
    urlbordercolor=magenta,
    filecolor=black,
    pdfpagemode=UseOutlines,
    pdfborderstyle={/S/U},  
}}
\usepackage[hyperref]{lbnebackref}

\renewcommand{\backrefxxx}[3]{~\hyperlink{page.#1}{#2~(pg.#1)}}

\renewcommand*{\backref}[1]{}
\renewcommand*{\backrefalt}[4]{%
  \ifcase #1 %
    \relax  
  \or
    Cited in Section~#2.%
  \else
    Cited in Sections~#2.%
  \fi%
}
\AtBeginEnvironment{figure}{\backrefsetup{disable}}
\AtBeginEnvironment{table}{\backrefsetup{disable}}
\AtBeginEnvironment{footnote}{\backrefsetup{disable}}

%% file: lbne-sci-opp-main.tex
\input{chapter-frontmatter}

\pagestyle{\lbnepagestyle}

\renewcommand{\chaptermark}[1]{ \markboth{#1}{#1} } 
\renewcommand{\sectionmark}[1]{ \markright{#1} }

\cleardoublepage
\renewcommand{\headrulecolor}{CHAP1COL!80}
\renewcommand{\toccolor}{CHAP1COL!80}
\renewcommand{\ChapterTableColor}{CHAP1COL!100}
\renewcommand{\ChapterTitleColor}{CHAP1COL!30}
\renewcommand{\ChapterBubbleColor}{CHAP1COL!15}
\renewcommand{\ChapterTabColor}{CHAP1COL!30}
\renewcommand{\IntroBackgroundColor}{CHAP1COL!15}
\renewcommand{\IntroLineColor}{CHAP1COL!30}
\chapter[\textcolor{\toccolor}{Introduction and Executive Summary}]{Introduction and \\ Executive Summary}
\addtocounter{margintabsbump}{1}
\label{exec-sum-chap}
\input{chapter-execsum}
\cleardoublepage
\renewcommand{\headrulecolor}{CHAP2COL!70}
\renewcommand{\toccolor}{CHAP2COL!70}
\renewcommand{\ChapterTableColor}{CHAP2COL!100}
\renewcommand{\ChapterTitleColor}{CHAP2COL!30}
\renewcommand{\ChapterBubbleColor}{CHAP2COL!15}
\renewcommand{\ChapterTabColor}{CHAP2COL!30}
\renewcommand{\IntroBackgroundColor}{CHAP2COL!15}
\renewcommand{\IntroLineColor}{CHAP2COL!30}
\chapter[\textcolor{\toccolor}{The Science of LBNE}]{The Science \\ of LBNE}
\addtocounter{margintabsbump}{1}
\label{intro-chap}
\input{chapter-intro}

\cleardoublepage
\renewcommand{\headrulecolor}{CHAP3COL!70}
\renewcommand{\toccolor}{CHAP3COL!70}
\renewcommand{\ChapterTableColor}{CHAP3COL!100}
\renewcommand\ChapterTitleColor{CHAP3COL!30}
\renewcommand\ChapterBubbleColor{CHAP3COL!15}
\renewcommand\ChapterTabColor{CHAP3COL!30}
\renewcommand{\IntroBackgroundColor}{CHAP3COL!15}
\renewcommand{\IntroLineColor}{CHAP3COL!30}
\chapter[\textcolor{\toccolor}{Project and Design}]{Project and  \\ Design}
\addtocounter{margintabsbump}{1}
\label{project-chap}
\input{chapter-project}


\cleardoublepage
\renewcommand{\headrulecolor}{CHAP4COL!70}
\renewcommand{\toccolor}{CHAP4COL!70}
\renewcommand{\ChapterTableColor}{CHAP4COL!100}
\renewcommand\ChapterTitleColor{CHAP4COL!30}
\renewcommand\ChapterBubbleColor{CHAP4COL!15}
\renewcommand\ChapterTabColor{CHAP4COL!30}
\renewcommand{\IntroBackgroundColor}{CHAP4COL!15}
\renewcommand{\IntroLineColor}{CHAP4COL!30}
\chapter[\textcolor{\toccolor}{Neutrino Mixing, Mass Hierarchy, and CP Violation}]{Neutrino Mixing,  Mass \\ Hierarchy and CP Violation}
\addtocounter{margintabsbump}{1}
\label{nu-oscil-chap}
\input{chapter-nu-oscil}


\cleardoublepage
\renewcommand{\headrulecolor}{CHAP5COL!70}
\renewcommand{\toccolor}{CHAP5COL!70}
\renewcommand{\ChapterTableColor}{CHAP5COL!100}
\renewcommand\ChapterTitleColor{CHAP5COL!30}
\renewcommand\ChapterBubbleColor{CHAP5COL!15}
\renewcommand\ChapterTabColor{CHAP5COL!30}
\renewcommand{\IntroBackgroundColor}{CHAP5COL!15}
\renewcommand{\IntroLineColor}{CHAP5COL!30}
\chapter[\textcolor{\toccolor}{Nucleon Decay Motivated by Grand Unified Theories}]{Nucleon Decay  Motivated \\ by Grand Unified Theories}
\addtocounter{margintabsbump}{1}
\label{pdk-chap}
\input{chapter-pdk}


\cleardoublepage
\renewcommand{\headrulecolor}{CHAP6COL!70}
\renewcommand{\toccolor}{CHAP6COL!70}
\renewcommand{\ChapterTableColor}{CHAP6COL!70!black}
\renewcommand\ChapterTitleColor{CHAP6COL!30}
\renewcommand\ChapterBubbleColor{CHAP6COL!15}
\renewcommand\ChapterTabColor{CHAP6COL!30}
\renewcommand{\IntroBackgroundColor}{CHAP6COL!15}
\renewcommand{\IntroLineColor}{CHAP6COL!30}
\chapter[\textcolor{\toccolor}{Core-Collapse Supernova Neutrinos}]{Core-Collapse \\ Supernova Neutrinos}
\addtocounter{margintabsbump}{1}
\label{sn-chap}
\input{chapter-sn}

\cleardoublepage
\renewcommand{\headrulecolor}{CHAP7COL!70}
\renewcommand{\toccolor}{CHAP7COL!70}
\renewcommand{\ChapterTableColor}{CHAP7COL!70!black}
\renewcommand\ChapterTitleColor{CHAP7COL!30}
\renewcommand\ChapterBubbleColor{CHAP7COL!15}
\renewcommand\ChapterTabColor{CHAP7COL!30}
\renewcommand{\IntroBackgroundColor}{CHAP7COL!15}
\renewcommand{\IntroLineColor}{CHAP7COL!30}
\chapter[\textcolor{\toccolor}{Precision Measurements with a High-Intensity Neutrino Beam}]{Precision Measurements with  a \\ High-Intensity Neutrino Beam}
\addtocounter{margintabsbump}{1}
\label{nd-physics-chap}
\input{chapter-nd-physics} 

\cleardoublepage
\renewcommand{\headrulecolor}{CHAP8COL!70}
\renewcommand{\toccolor}{CHAP8COL!70}
\renewcommand{\ChapterTableColor}{CHAP8COL!70!black}
\renewcommand\ChapterTitleColor{CHAP8COL!30}
\renewcommand\ChapterBubbleColor{CHAP8COL!15}
\renewcommand\ChapterTabColor{CHAP8COL!30}
\renewcommand{\IntroBackgroundColor}{CHAP8COL!15}
\renewcommand{\IntroLineColor}{CHAP8COL!30}
\chapter[\textcolor{\toccolor}{\bf Additional Far Detector Physics Opportunities}]{Additional Far  Detector \\ Physics Opportunities}
\addtocounter{margintabsbump}{1}
\label{chap-other-goals}
\input{chapter-other-goals}

\cleardoublepage
\renewcommand{\headrulecolor}{CHAP9COL!80}
\renewcommand{\toccolor}{CHAP9COL!80}
\renewcommand{\ChapterTableColor}{CHAP9COL!100}
\renewcommand\ChapterTitleColor{CHAP9COL!30}
\renewcommand\ChapterBubbleColor{CHAP9COL!15}
\renewcommand\ChapterTabColor{CHAP9COL!30}
\renewcommand{\IntroBackgroundColor}{CHAP9COL!15}
\renewcommand{\IntroLineColor}{CHAP9COL!30}
\chapter[\textcolor{\toccolor}{Summary and Conclusion}]{Summary and \\ Conclusion}
\addtocounter{margintabsbump}{1}
\label{conclusion-chap}
\input{chapter-conclusion}



\appendix

\cleardoublepage
\renewcommand{\headrulecolor}{CHAPACOL!70}
\renewcommand{\toccolor}{CHAPACOL!70}
\renewcommand{\ChapterTableColor}{CHAPACOL!100}
\renewcommand\ChapterTitleColor{CHAPACOL!30}
\renewcommand\ChapterBubbleColor{CHAPACOL!15}
\renewcommand\ChapterTabColor{CHAPACOL!30}
\renewcommand{\IntroBackgroundColor}{CHAPACOL!15}
\renewcommand{\IntroLineColor}{CHAPACOL!30}
\renewcommand\chaptername{Appendix}
\chapter[\textcolor{\toccolor}{LBNE Detector Simulation and Reconstruction}]{LBNE Detector Simulation \\ and Reconstruction}
\addtocounter{margintabsbump}{1}
\label{app-sim}
\input{appendix-fd-simulation}


\cleardoublepage
\renewcommand{\headrulecolor}{CHAPACOL!70}
\renewcommand{\toccolor}{CHAPACOL!70}
\renewcommand{\ChapterTableColor}{CHAPACOL!100}
\renewcommand\ChapterTitleColor{CHAPACOL!30}
\renewcommand\ChapterBubbleColor{CHAPACOL!15}
\renewcommand\ChapterTabColor{CHAPACOL!30}
\renewcommand{\IntroBackgroundColor}{CHAPACOL!15}
\renewcommand{\IntroLineColor}{CHAPACOL!30}
\renewcommand\chaptername{Appendix}
\chapter[\textcolor{\toccolor}{Neutrino-Nucleon Scattering Kinematics}]{Neutrino-Nucleon \\ Scattering Kinematics}
\addtocounter{margintabsbump}{1}
\label{app-dis}
\input{appendix-dis-kinematics}


\cleardoublepage
\renewcommand{\headrulecolor}{CHAPACOL!70}
\renewcommand{\toccolor}{CHAPACOL!70}
\renewcommand{\ChapterTableColor}{CHAPACOL!100}
\renewcommand\ChapterTitleColor{CHAPACOL!30}
\renewcommand\ChapterBubbleColor{CHAPACOL!15}
\renewcommand\ChapterTabColor{CHAPACOL!30}
\renewcommand{\IntroBackgroundColor}{CHAPACOL!15}
\renewcommand{\IntroLineColor}{CHAPACOL!30}
\renewcommand\lbnechapterlabel{\ }
\clearpage
\pagestyle{\lbneackstyle}
\setcounter{secnumdepth}{-1}
\chapter[\textcolor{\toccolor}{Acknowledgments}]{Acknowledgments}
\addtocounter{margintabsbump}{100}
\label{acknowl}
\input{acknowl}


%% file: chapter-frontmatter.tex
\renewcommand{\titlepagecornerblock}{ 
  \begin{flushright}
    BNL-101354-2014-JA\\
    FERMILAB-PUB-14-022\\
    LA-UR-14-20881\\
    arXiv:1307.7335
  \end{flushright}}

\pagestyle{titlepage}

\vspace*{5cm}

\begin{center}
\centerline{\normalfont\sffamily\Huge\bfseries\scshape The Long-Baseline Neutrino Experiment}  
\vskip 0.2in
 {\normalfont\sffamily\Large\bfseries\scshape Exploring Fundamental Symmetries of the Universe} 
  
April 15, 2014

\end{center}

\clearpage

\pagestyle{empty}
\vspace*{0.8\textheight}
Cover design by Diana Brandonisio, Fermilab Visual Media Services
\vskip 0.2in
Typeset in \LaTeX

\clearpage

\cleardoublepage

\includepdf[pages=-]{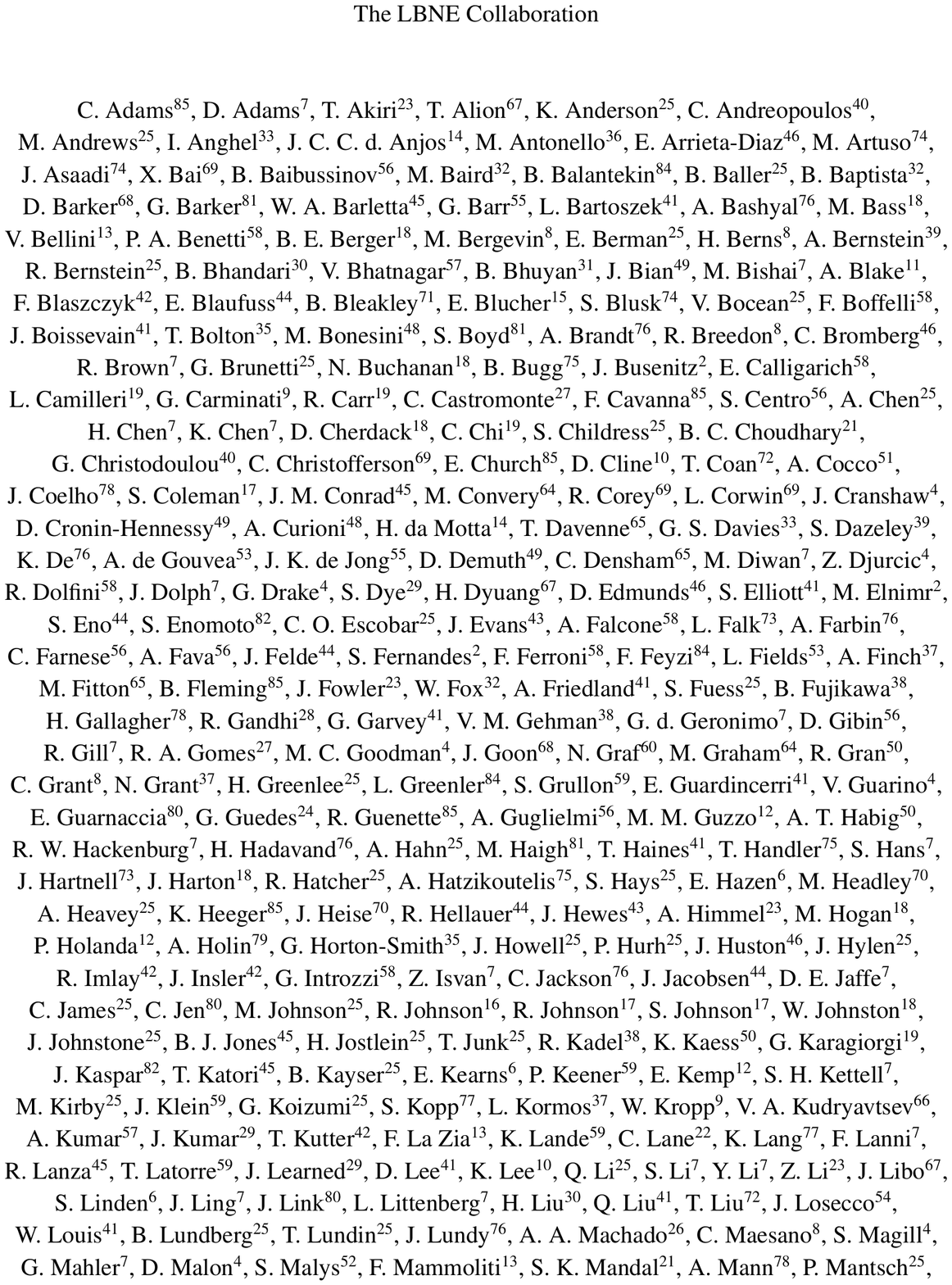}

\cleardoublepage
\pagestyle{empty}
\input{abstract}

\cleardoublepage
\pagenumbering{roman}
\pagestyle{plain}
\setcounter{tocdepth}{3}

\phantomsection
\addtocontents{toc}{\protect{\pdfbookmark[0]{Table of Contents}{toc}}}
\label{sec:toc}
\textsf{\tableofcontents}
\cleardoublepage

\printnomenclature
\cleardoublepage

\phantomsection
\addcontentsline{toc}{chapter}{\listfigurename}
\label{sec:lof}
\textsf{\listoffigures}
\cleardoublepage

\phantomsection
\addcontentsline{toc}{chapter}{\listtablename}
\label{sec:lot}
\textsf{\listoftables}

\cleardoublepage
\renewcommand{\headrulecolor}{CHAP2COL!70}
\renewcommand{\ChapterTitleColor}{CHAP2COL!30}
\renewcommand{\ChapterBubbleColor}{CHAP2COL!15}
\renewcommand{\ChapterTabColor}{CHAP2COL!30}
\renewcommand{\IntroBackgroundColor}{CHAP2COL!15}
\renewcommand{\IntroLineColor}{CHAP2COL!30}
\input{reader-guidelines}

\cleardoublepage
\pagenumbering{arabic}

%% file: abstract.tex
\begin{abstract}

  The preponderance of matter over antimatter in the early Universe, the
  dynamics of the supernova bursts that produced the heavy elements
  necessary for life and whether protons eventually decay ---
  these mysteries at the forefront of particle physics and
  astrophysics are key to understanding the early evolution of our
  Universe, its current state and its eventual fate. The Long-Baseline
  Neutrino Experiment (LBNE) represents an extensively developed plan
  for a world-class experiment dedicated to addressing these questions.

  Experiments carried out over the past half century have revealed
  that neutrinos are found in three states, or \emph{flavors}, and can
  transform from one flavor into another. These
  results indicate that each neutrino flavor state is a mixture of
  three different nonzero mass states, and to date offer the most
  compelling evidence for physics beyond the Standard Model. In a
  single experiment, LBNE will enable a broad exploration of the
  three-flavor model of neutrino physics with unprecedented
  detail. Chief among its potential discoveries is that of
  matter-antimatter asymmetries (through the mechanism of
  charge-parity violation) in neutrino flavor mixing --- a step toward
  unraveling the mystery of matter generation in the early Universe.
  Independently, determination of the unknown neutrino mass ordering
  and precise measurement of neutrino mixing parameters by LBNE may
  reveal new fundamental symmetries of Nature.

  Grand Unified Theories, which attempt to describe the unification of
  the known forces, predict rates for proton decay that cover a range
  directly accessible with the next generation of large underground
  detectors such as LBNE's. The experiment's sensitivity to key proton
  decay channels will offer unique opportunities for the
  ground-breaking discovery of this phenomenon.

  Neutrinos emitted in the first few seconds of a core-collapse
  supernova carry with them the potential for great insight into the
  evolution of the Universe. LBNE's capability to collect and analyze
  this high-statistics neutrino signal from a supernova within our
  galaxy would provide a rare opportunity to peer inside a
  newly-formed neutron star and potentially witness the birth of a
  black hole.

  To achieve its goals, LBNE is conceived around three central
  components: (1) a new, high-intensity neutrino source generated from
  a megawatt-class proton accelerator at Fermi National Accelerator
  Laboratory, (2) a fine-grained near neutrino detector installed just
  downstream of the source, and (3) a massive liquid argon
  time-projection chamber deployed as a far detector deep underground
  at the Sanford Underground Research Facility. This facility, located
  at the site of the former Homestake Mine in Lead, South Dakota, is
  $\sim$1,300~km from the neutrino source at Fermilab --- a distance
  (baseline) that delivers optimal sensitivity to neutrino
  charge-parity symmetry violation and mass ordering effects.  This
  ambitious yet cost-effective design incorporates scalability and
  flexibility and can accommodate a variety of upgrades and
  contributions.

  With its exceptional combination of experimental configuration,
    technical capabilities, and potential for transformative
    discoveries, LBNE promises to be a vital facility for the
    field of particle physics worldwide, providing physicists
    from institutions around the globe with opportunities to collaborate
    in a twenty to thirty year program of exciting science.

\end{abstract}

%% file: reader-guidelines.tex
\chapter*{How to Read this Document}
\addcontentsline{toc}{chapter}{How to Read this Document}

The LBNE science document is intended to inform a diverse readership about the goals and capabilities 
of the LBNE experiment. Your approach to reading this document will depend upon your purpose
 as well as your level of knowledge about high energy and neutrino physics.  

\begin{figure}[H]
\includegraphics[width=.05\textwidth]{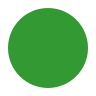}
\end{figure}

\begin{introbox}
The colored boxes distributed throughout the document highlight the important take-away points.
They are integral to the document, but to the extent possible, are written in language accessible to the nonscientist.  
\end{introbox}

\begin{figure}[H]
\includegraphics[width=.05\textwidth]{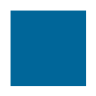}
\end{figure}

The three chapters Chapter~\ref{exec-sum-chap} \emph{Introduction
  and Executive Summary}, Chapter~\ref{project-chap} \emph{Project and
  Design} and Chapter~\ref{conclusion-chap} \emph{Summary and
  Conclusion} together provide a comprehensive overview of LBNE's
scientific objectives, its place in the landscape of neutrino physics
experiments worldwide, the technologies it will incorporate and the
capabilities it will possess.  Much of the information in these
chapters is accessible to the lay reader, but of course, the
scientific concepts, goals and methods around which LBNE is designed
are by their nature highly specialized, and the text in certain
sections is correspondingly technical.

\begin{figure}[H]
\includegraphics[width=.05\textwidth]{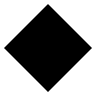}
\end{figure}

In Chapter~\ref{intro-chap} \emph{The Science of LBNE}, the initial
paragraphs in each section provide some introductory information, but 
in general this chapter assumes a working knowledge of high energy 
physics and, ideally, familiarity with neutrino physics.

The three chapters that delve into the areas corresponding to the
scientific objectives of LBNE: Chapter~\ref{nu-oscil-chap} \emph{Neutrino
  Mixing, Mass Hierarchy and CP Violation}, Chapter~\ref{pdk-chap}
\emph{Nucleon Decay Motivated by Grand Unified Theories} and
Chapter~\ref{sn-chap} \emph{Core-Collapse Supernova Neutrinos},
assume a working knowledge of high energy physics and particle astrophysics.  
This is
also true of Chapter~\ref{nd-physics-chap} \emph{Precision
  Measurements with a High-Intensity Neutrino Beam} and
Chapter~\ref{chap-other-goals} \emph{Additional Far Detector
  Physics Opportunities}, as well as the appendices.

%% file: chapter-execsum.tex

\begin{introbox}

  The Long-Baseline Neutrino Experiment (LBNE) will provide a unique, 
  world-leading program for the exploration of key questions at the
  forefront of particle physics and astrophysics.

  Chief among its potential discoveries is that of matter-antimatter
  symmetry violation in neutrino flavor mixing --- a step toward
  unraveling the mystery of matter generation in the early
  Universe. Independently, determination of the neutrino mass ordering
  and precise measurement of neutrino mixing parameters by LBNE may
  reveal new fundamental symmetries of Nature.

To achieve its ambitious physics objectives as a world-class facility,
LBNE has been conceived around three central components: 
\begin{enumerate}
\item an intense, wide-band neutrino beam 
\item a fine-grained \emph{near} neutrino detector just downstream of the neutrino
source
\item a massive liquid argon time-projection chamber (LArTPC) deployed
as a \emph{far} neutrino detector deep underground, \SI{1300}{\km} downstream; 
this distance between the neutrino source and far detector --- the \emph{baseline} --- is measured along the line of travel through the Earth
\end{enumerate}

The neutrino beam and near detector will be installed at the Fermi
National Accelerator Laboratory (Fermilab), in Batavia, Illinois. The
far detector will be installed at the Sanford Underground Research
Facility in Lead, South Dakota. 

  The location of its massive high-resolution far 
detector
  deep underground will enable LBNE to significantly expand the search
  for proton decay as predicted by Grand Unified Theories, as well as
  study the dynamics of core-collapse supernovae through observation
  of their neutrino bursts, should any occur in our galaxy during
  LBNE's operating lifetime.

  The near neutrino detector will enable high-precision measurements
  of neutrino oscillations, thereby enhancing the sensitivity to
  matter-antimatter symmetry violations and will exploit the
  potential of high-intensity neutrino beams as probes of new physics.

  With its extensively developed design and flexible configuration,
  LBNE provides a blueprint for an experimental program made even more
  relevant by recent neutrino mixing parameter measurements.
\end{introbox}

\clearpage
\section{Overview}
\label{exec-sum-overview}

Although neutrinos are the most abundant of known matter particles
(fermions) in the Universe, their properties are the least well
understood. The very existence of neutrino mass constitutes evidence of
physics beyond the Standard Model. Understanding the nature of neutrinos
has consequently become an
essential goal for particle physics.

Observations of oscillations of neutrinos from one type (flavor) to
another in numerous recent experiments have provided evidence for
neutrino flavor mixing and for small, but nonzero, neutrino masses.
The framework characterizing these observations is similar to that
describing corresponding phenomena in the quark sector, but with a
very different pattern of mixing angle values.  As in the quark case,
this framework involves a phase parameter, \deltacp, that changes
sign under combined charge conjugation and parity (CP) reversal
operations and thus would lead to CP symmetry-violating asymmetries
between the pattern of oscillations for neutrinos and antineutrinos.
While groundbreaking on its own, the observation of such asymmetries
would also provide an experimental underpinning for the basic idea of
leptogenesis\footnote{Leptogenesis refers to the mechanisms that generated an
asymmetry between leptons and antileptons in the early Universe, described in 
Section~\ref{sec:cpv-quark-lepton}.} as an explanation for the Baryon Asymmetry of the
Universe (BAU).

Neutrino oscillation data so far tell us about differences in the
squared masses of the neutrino mass states, and about the sign of the
mass-squared difference between two of the states, but not 
about the difference
of those with
respect to the third, which may be heavier (\emph{normal} ordering) 
or lighter (\emph{inverted} ordering) than the other
two.  
Resolving this neutrino mass hierarchy ambiguity, along with precise
measurements of neutrino mixing angles, would have significant
theoretical, cosmological and experimental
implications. 
One important consequence of mass hierarchy determination, in particular, 
would be the impact on future experiments designed to determine 
whether --- uniquely among the fundamental fermions --- neutrinos 
are their own antiparticles, so-called \emph{Majorana} particles.  
Though long suspected, 
this hypothesis that neutrinos are  
Majorana particles has yet to be either established or ruled out.  
Strong evidence for the \emph{inverted} hierarchy would establish 
conditions required by the next generation of neutrinoless double-beta decay 
searches to settle this question even with a null result (no observation). 
Because the
forward scattering of neutrinos in matter alters the oscillation
pattern in a hierarchy-dependent way, the long baseline of LBNE 
--- with the neutrinos traveling through the Earth's mantle ---
enables a decisive determination of the hierarchy, independent of the
value of \deltacp.  

Additionally, the high-precision determination of
oscillation parameters such as mixing angles and squared-mass
differences will provide insight into the differences between the
quark and lepton mixing patterns, which is necessary for deciphering
the flavor structure of physics in the Standard Model.  Taken
together, the above suite of measurements will thoroughly test the standard
three-neutrino flavor paradigm that guides our current understanding,
and will provide greatly extended sensitivity to signatures for
nonstandard neutrino interactions in matter.

The arena of non-accelerator physics using massive underground
detectors such as the LBNE far detector is also ripe with discovery
potential.  The observation of nucleon decay would be a watershed
event for the understanding of physics at high energy scales.
Neutrinos from supernovae are expected to provide key insights into
the physics of gravitational collapse, and may also reveal
fundamental properties of the neutrino.

Among massive detectors designed for neutrino and nucleon decay physics, 
the LArTPC technology offers 
unmatched capabilities for 
position and energy resolution 
and for
high-precision reconstruction of complex interaction topologies
over a broad energy range.  It also provides a compact, scalable approach for 
achieving the required sensitivity to the 
primary physics signatures to be explored by LBNE.  
As these capabilities are also important for non-accelerator neutrino physics, 
 LBNE will complement the large, underground water Cherenkov
and/or scintillator-based detectors that may be operating in parallel.
LArTPC detectors are especially well-suited to proton decay modes 
such as the supersymmetry-favored $p \to K^+\overline{\nu}$ mode, 
uniquely providing 
detection efficiency and background rejection 
sufficient to enable a discovery with a single well-reconstructed event.  
With regard to supernova-neutrino detection, 
liquid argon detectors are primarily sensitive to the $\nu_e$ 
component of the flux, while  $\overline{\nu}_e$ interactions
dominate for water and scintillator-based detectors.  Thus, LBNE will be 
sensitive to different features of the supernova-neutrino production 
process.  Finally, 
the LArTPC technology opens up an avenue for precision studies of oscillation
physics with atmospheric neutrinos, thereby augmenting the results of the 
beam-based measurements at the core of the experiment.

The highly capable near detector will measure the absolute flux and
energy scales of all four neutrino species in the LBNE beam, as well
as neutrino cross sections on argon, water, and other nuclear targets in the 
beam's energy range.  These measurements are needed to attain the
ultimately desired precision of the oscillation parameter
measurements.  Additionally, the near detector will enable a broad
range of precision neutrino-interaction measurements, thereby adding a
compelling scientific program of its own.

The unique combination in LBNE of a \kmadj{1300} baseline, exceptional
resolution, large target mass and deep underground location 
offers opportunity for discovery of entirely unanticipated phenomena.
History shows that ambitious scientific endeavors with leading-edge
instruments have often been rewarded with unexpected signatures of new
physics.  

LBNE is an extensively developed  
experiment whose execution will have substantial
impact on the overall direction of 
high energy physics (HEP) in the
U.S.  The U.S. Department of Energy (DOE) has endorsed the science objectives
of LBNE, 
envisioning the experiment as a phased program, and 
has
given first stage (CD-1) approval with a budget of \$867M toward the
initial phase.  The science scope of this and subsequent phases will
depend on the level of investment by additional national and
international partners.  



This document outlines the LBNE physics
program and how it may evolve in the context
of long-term planning studies~\cite{snowmass2013}.  The physics reach
of this program is summarized under scenarios that are consistent with
short-, medium- and long-term considerations.
%
The general conclusions regarding the scientific capabilities of LBNE
in a phased program are twofold: 
\begin{enumerate}

\item A full-scope LBNE will provide an exciting
broad-based physics program with exceptional capabilities for all of
the identified core physics objectives, and many additional ones.

\item A first phase with a LArTPC far detector of fiducial\footnote{In neutrino 
experiments, not all neutrino interactions in
    the instrumented (active) volume of a detector are used for
    physics studies. Only interactions that are well contained within
    the instrumented volume are used. The smaller volume of detector
    that encompasses the neutrino interactions is known as the
    \emph{fiducial volume} and the target mass contained within it is
    known as the \emph{fiducial mass}. Unless otherwise noted, this
    document will use fiducial mass to characterize the far detector
    size.} 
   mass \SI{10}{\kt}\footnote{The \emph{kt} refers to a
    metric kiloton, equivalent to \SI{1000}{kg}.}  or greater will substantially advance the field of
  neutrino oscillation physics while laying the foundations for 
a broader physics program in a later phase.
\end{enumerate}

Section~\ref{exec-sum-develop} provides the context for development 
of LBNE as a phased program that maintains flexibility for enhancements in
each of its stages through the contributions of additional
partners.  The physics reach of LBNE at various stages is summarized 
in Section~\ref{exec-sum-physics}.

\clearpage
\section{Development of a World-Class Experiment}
\label{exec-sum-develop}

\begin{introbox}

To 
achieve the transformative physics goals of LBNE in an era of 
highly constrained funding for basic research in the U.S., 
the conceptual design has evolved so as to provide a scalable, 
phased and global approach, 
while maintaining a U.S. leadership role as the host for a 
global facility.  International partnerships 
are being actively pursued to both enhance and accelerate the LBNE Project. 



LBNE's primary beamline is designed to operate initially with a beam
power of \SI{1.2}{\MW}, upgradable to \SI{2.3}{\MW}.  This beamline
extracts protons with energies from \num{60} to \SI{120}{\GeV} from
the Fermilab Main Injector. The protons collide with a target to
generate a secondary beam of charged particles, which in turn decay to
generate the neutrino beam.

The liquid argon TPC far detector technology  
combines fine-grained
tracking with total absorption calorimetry. Installed \SI{4850}{ft}
underground to minimize backgrounds, this detector will be a powerful
tool for long-baseline neutrino oscillation physics and underground
physics such as proton decay, supernova neutrinos and atmospheric
neutrinos.  The far detector design is scalable and flexible, allowing
for a phased approach, with an initial fiducial mass of at least
\SI{10}{\kt} and a final configuration of at least \SI{34}{\kt}.

A high-precision near detector is planned as a separate facility 
allowing maximal flexibility in phasing and deployment. 
\end{introbox}

The concept of a high-intensity neutrino beam directed toward a
distant, massive underground detector to simultaneously investigate the
nature of the neutrino, proton decay and astrophysical sources of neutrinos 
has been under serious investigation since the late 1990s~\cite{Diwan:2000yb,Marciano:2001tz,Shrock:2003pb,Diwan:2002xc,Diwan:2003bp,Weng:2004zz,Diwan:2006qf,Barger:2007yw}.
Since that time both the science goals and concepts for implementation
have been the subject of intense study and review by distinguished
panels. These panels include the National Academies Neutrino
Facilities Assessment Committee in 2003~\cite{nusandbeyond}, the
National Science and Technology Council Committee on Science
in 2004~\cite{potu},  
the
National Academies EPP2010 panel in 2006~\cite{epp2010}, the
HEPAP/NSAC Neutrino Scientific Assessment Group in 2007~\cite{nusag},
the HEPAP Particle Physics Project Prioritization Panel (P5) in
2008~\cite{p5report}, the National Academies ad hoc Committee to
Assess the Science Proposed for DUSEL in 2011~\cite{nasdusel}, and
most recently the HEPAP Facilities Subpanel in
2013~\cite{facilitiesreport}.  High-level studies performed in Europe
and Asia have come to similar conclusions (e.g.,~\cite{europeanstrategy}) 
about the merits and feasibility of such a program. 

\clearpage
\subsection{Long-Term Vision}

LBNE as described in this document has been developed by a collaboration
formally established in 2009, which currently comprises over 475
collaborators from over 80 institutions in six countries.  In
January 2010 the 
DOE formally recognized the
LBNE science objectives with approval of the mission need statement
(CD-0)~\cite{cdzero}. This action established LBNE as a DOE project.
 Fermilab has
recognized LBNE as a central component of its long-term future 
program.

The central role of LBNE within the U.S. particle physics program has
been acknowledged in other documents prepared for the 2013 particle
physics community planning exercise~\cite{snowmass2013}, including the
Project X Physics Book~\cite{Kronfeld:2013uoa} and the reports from Intensity
Frontier working groups on neutrino physics~\cite{deGouvea:2013onf}
and baryon number violation~\cite{Babu:2013jba}.

The LBNE conceptual design 
reflects a flexible and cost-effective approach to next-generation
neutrino physics experiments that maintains a world-leadership role
for the U.S. over the long term.  
The full-scope LBNE includes a \ktadj{34} fiducial mass 
(\ktadj{50} total) far detector located
in a new experimental area to
be excavated at the 4,850-ft level of the \SURF\footnote{Much larger
  detectors could also be accommodated at this facility.} in the
former Homestake Mine, and a fine-grained near neutrino detector
located on the Fermilab site.  
Simultaneous construction of a new
neutrino beamline at Fermilab would permit operation with 
 an initial beam power of \SI{1.2}{\MW}, enabled
by upgrades to the front end of the accelerator complex carried
out within the Proton Improvement Plan-II (PIP-II)
program~\cite{PIPII}.  
 In anticipation of potential enhancements
beyond PIP-II~\cite{Holmes:2013vpa},
the beamline is designed to
support upgrades to accommodate a beam power of \SI{2.3}{\MW}.  The \kmadj{1300} baseline
is in the optimal range for the neutrino oscillation program.  The 
cosmic ray shielding provided by the deep underground site for the far detector
enables the non-accelerator portion of the physics program, including
proton decay searches, detailed studies of neutrino bursts from
galactic supernovae, and precision analyses of atmospheric-neutrino
samples.
 
The overall physics reach of LBNE is predominantly limited by detector
mass.  From the outset, a guiding principle of the far detector design
has been scalability.  The conceptual design for the full-scope detector,
consisting of two identical \ktadj{17} (\ktadj{25} total) TPC 
modules housed within separate vessels (cryostats), employs technology
developed by the liquefied natural gas (LNG) storage and transport
industry.  
The TPC modules themselves consist of arrays of
modular anode and cathode plane assemblies (APAs and CPAs) that are
suspended from rails affixed to the top of the cryostats.  The APA/CPA
dimensions are chosen for ease of transportation and installation.
The modularity of the detectors allows flexibility in the geometry and
phased construction of the LBNE far detector complex.  Cost-effective designs 
for larger detector masses are readily obtained by increasing the vessel size 
and simply adding APA/CPA units, thereby also exploiting economies 
of scale and benefiting from an increased ratio of volume to surface area.
Detector mass may also be increased through the
addition of distinct detectors of the same or a different technology,
either during initial construction or in a later phase.

\subsection{Present Status of the LBNE Project}

Since DOE CD-0 approval, a compete conceptual design for the
full-scope LBNE has been developed, consisting of a \ktadj{34} LArTPC
far detector located \num{4850} feet underground, a \kmadj{1300}
baseline, a highly capable near neutrino detector, and a
multi-megawatt-capable neutrino beamline. This design has been
thoroughly reviewed, and found to be sound, most recently at a
Fermilab Director's CD-1 Readiness Review in March
2012~\cite{mar2012review}. Since then, considerable effort has been
devoted to understanding how the LBNE Project can be staged so as to
accommodate anticipated budget conditions while maintaining compelling
physics output at each stage~\cite{LBNEreconfig}. This process led to
a first-phase configuration that was reviewed by the DOE in
October~\cite{oct2012review} and November 2012 ~\cite{nov2012review},
and that received CD-1 approval~\cite{dec2012approv} in December 2012. This
configuration~\cite{CDRv1,CDRv2,CDRv3,CDRv4,CDRv5,CDRv6} maintained
the most important aspects of LBNE: the \kmadj{1300} baseline to the
\SURF, a large --- of order tens of kilotons in fiducial mass ---  LArTPC
far detector design, and a multi-megawatt-capable, wide-band
neutrino/antineutrino beam. However, the far detector size was limited at CD-1
to \SI{10}{\kt} 
and placed at the surface under minimal
overburden, and the near detector was deferred to a later phase.

The DOE CD-1 approval document~\cite{dec2012approv} explicitly allows
adjustment of the scope of the first phase of LBNE in advance of CD-2 if additional
partners bring significant contributions to LBNE. \emph{Using the CD-1
  DOE funding as the foundation, the goal for the first phase of LBNE
  is a deep underground far detector of at least \SI{10}{\kt}, placed in a 
cavern that will accommodate up to a \ktadj{34} detector, coupled
  with a \MWadj{1.2} neutrino beamline, and a highly capable near 
  detector.}  This goal has been endorsed by the LBNE Collaboration, the
LBNE Project, the Fermilab directorate, and the DOE Office of High Energy
Physics.  Since a large portion of the LBNE Project cost is in civil
infrastructure, 
funding contributions from new partners could have considerable impact
on 
the experimental facilities, and therefore the physics scope, in the first phase.

\subsection{Global Partnerships} 
\label{sec:global-partner}

Global conditions are favorable for significant international
partnerships in developing and building LBNE.  As an example, the 2013
update~\cite{europeanstrategy} of the European Strategy for Particle
Physics document places long-baseline neutrino physics among the
highest-priority large-scale activities for Europe, 
recognizing that it requires ``significant resources, sizeable collaborations and sustained
commitment.'' It includes the primary recommendation of exploring ``the
possibility of major participation in leading long-baseline neutrino
projects in the U.S. and Japan.''  As of March 2014 the LBNE Collaboration
includes institutions from the U.S., Brazil, India, Italy, Japan and the United
Kingdom.  Discussions with a number of potential international
partners are underway --- some already at an advanced stage.  A
summary of recent progress in these discussions can be found in the
presentation of LBNE status to the U.S. Particle Physics Projects Prioritization Panel in 
November 2013~\cite{wilson-nov2013}.

\subsection{Context for Discussion of Physics Sensitivities} 

To reflect the physics reach of various phasing scenarios, this
document presents many of the parameter sensitivities for the
accelerator-based neutrino topics as functions of exposure, defined as
the product of detector fiducial mass, beam power and run time.  As
needed, the capabilities of both a \ktadj{10} first-phase configuration
and the full \ktadj{34} configuration are explicitly highlighted, each
benchmarked for six to ten years of operations with a \MWadj{1.2}
beam power from the PIP-II accelerator upgrades at Fermilab.  Since the
U.S. program planning exercises currently under way look beyond the
present decade, this document also presents the long-term physics
impact of the full-scope LBNE operating with the \MWadj{2.3} beam
power available with further anticipated upgrades to the Fermilab
accelerator complex.

\clearpage

\section{The LBNE Physics Program}
\label{exec-sum-physics}

\begin{introbox}
  The technologies and configuration of the planned LBNE facilities
  offer excellent sensitivity to a range of physics processes:
 \begin{itemize}
 \item The muon-neutrino ($\nu_\mu$) beam produced at Fermilab with a peak flux at
   \SI{2.5}{\GeV}, coupled to the 
baseline of
   \SI{1300}{\km}, will present near-optimal sensitivity to neutrino/antineutrino    
   charge-parity (CP) symmetry violation effects. 
 \item The long baseline of LBNE will ensure a large matter-induced
   asymmetry in the oscillations of neutrinos and antineutrinos, thus
   providing a clear, unambiguous determination of the mass ordering of the
   neutrino states.
 \item The near detector located just downstream of the neutrino
   beamline at Fermilab will enable high-precision long-baseline
   oscillation measurements as well as precise measurements and searches for new phenomena
on its own using the high-intensity neutrino beam.
 \item The deep-underground LArTPC far detector will provide superior
   sensitivities to proton decay modes with kaons in the final states,
   modes that are favored by many Grand Unified and supersymmetric
   theoretical models. 
 \item Liquid argon as a target material will provide unique
   sensitivity to the electron-neutrino ($\nu_e$) component of the initial burst
   of neutrinos from a core-collapse supernova.
 \item The excellent energy and directional resolution of the 
   LArTPC will allow novel physics studies with atmospheric neutrinos.
 \end{itemize}
\end{introbox}

This section summarizes LBNE's potential for achieving its core
physics objectives based on the current 
experimental landscape, scenarios for staging LBNE, and 
the technical capabilities of LBNE at each stage.

LBNE's capability to achieve the physics objectives described in
this document has been subject to extensive review over a number of
years.  In addition to the various reviews of the LBNE Project
described in Section~\ref{exec-sum-develop}, reviews 
that focused strongly on LBNE's science program include the DOE Office
of Science Independent Review of Options for Underground Science in
the spring of 2011~\cite{marxcommittee}, the LBNE Science Capabilities
Review (by an external panel commissioned by LBNE)~\cite{scicapreview}
in the fall of 2011, and the LBNE Reconfiguration
Review~\cite{LBNEreconfig} in the summer of 2012.

\subsection{Neutrino Mixing, Mass Hierarchy and CP Violation}

\textbf{Neutrino Mass Hierarchy:} The \SIadj{1300}{\km} baseline
establishes one of LBNE's key strengths: sensitivity to the matter
effect. This effect leads to a large discrete asymmetry in the
$\nu_\mu\to \nu_e$ versus $\overline{\nu}_\mu \to \overline{\nu}_e$
oscillation probabilities, the sign of which depends on the mass
hierarchy (MH).  At \SI{1300}{\km} this asymmetry is approximately
$\pm 40\%$ in the region of the peak flux; this is larger than the
maximal possible CP-violating asymmetry associated with \deltacp,
meaning that both the MH and \deltacp can be
determined unambiguously with high confidence within the same
experiment using the beam neutrinos.

\begin{figure}[!htb]
\centerline{
\includegraphics[width=0.5\textwidth]{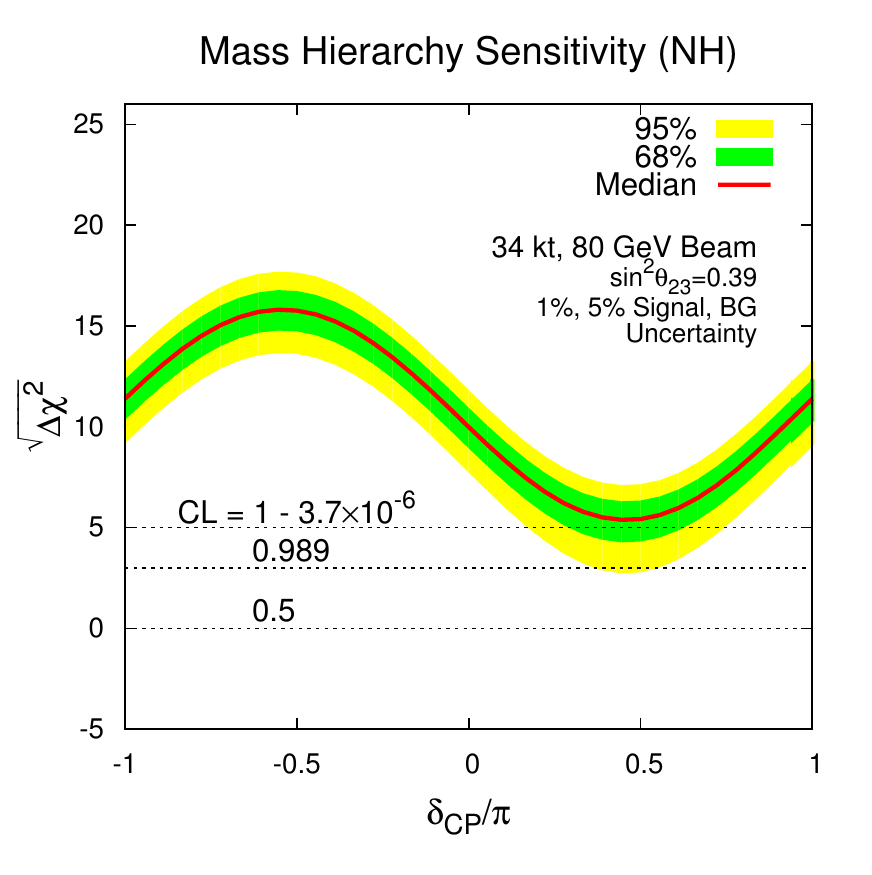}%
\includegraphics[width=0.5\textwidth]{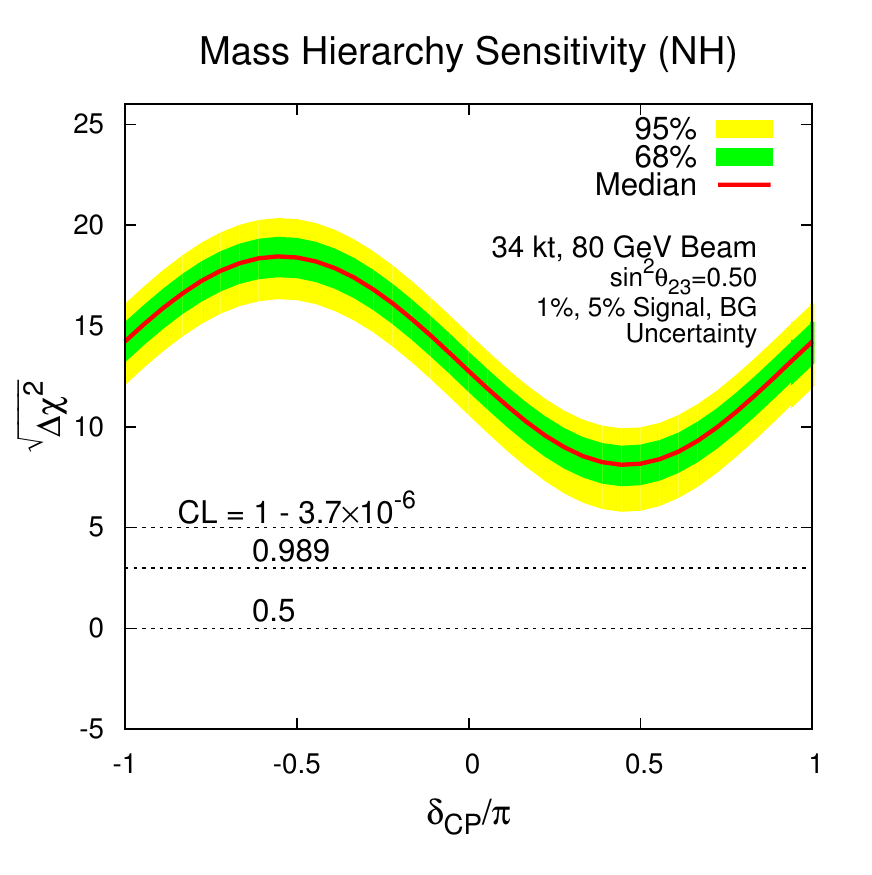}
}
\caption[Summary of mass hierarchy sensitivities]{The square root of
  the mass hierarchy discrimination metric $\Delta \chi^2$
  is plotted as a function of the unknown value of \deltacp for
  the full-scope LBNE with \SI{34}{\kt},  3+3 ($\nu+\overline{\nu}$) years of 
  running in a \SIadj{1.2}{\MW} beam, assuming normal hierarchy.  
  The plot on the left is for an assumed value of $\sin^ 2 \theta_{23}
  = 0.39$ (based on global fits and assuming worst-case $\theta_{23}$ 
  octant), while that on the right is for $\sin^ 2 \theta_{23}
  = 0.5$ (maximal mixing).  In each plot, 
  the red curve represents the median experimental value expected 
  ($\sqrt{\overline{\Delta \chi^2}}$),
  estimated using a data set absent statistical fluctuations, while 
  the green and yellow bands represent the
  range of $\Delta \chi^2$ values expected in 68\% and 95\% of all
  possible experimental instances, respectively. 
  For certain values of $\sqrt{\Delta \chi^2}$,  
  horizontal lines are shown, indicating the corresponding 
  confidence levels ($1-\alpha$ in the language of hypothesis 
  testing) with which a typical experiment ($\beta = 0.5$)
  correctly determines the MH, 
  computed according to a Bayesian 
  statistical formulation   
  (Section~\ref{sec-mh-statistics} for further discussion).
}
\label{fig:mhexec}
\end{figure}

In detail, the sensitivity of LBNE depends on the actual values of
poorly known mixing parameters (mainly \deltacp 
and $\sin^2{\theta_{23}}$), as well as the true value of the MH itself. 
The discrimination between the two MH hypotheses is characterized as a
function of the \emph{a priori} unknown true value of \deltacp by
considering the difference, denoted $\dchisq$, between the
$-2\log{\cal L}$ values calculated for a data set with respect to these
hypotheses, considering all possible values of
\deltacp\footnote{For the case of the MH
  determination, the usual association of this test statistic 
  with a $\chi^2$ distribution for one degree of freedom is incorrect; 
  additionally the assumption of a Gaussian probability density 
  implicit in this notation is not exact.  The discussion in
  Chapter~\ref{nu-oscil-chap} provides a brief description of the
  statistical considerations.}.  
In terms of this test statistic, the MH sensitivity of LBNE with \SI{34}{\kt}, 
and 
running three years each in $\nu$ and $\overline{\nu}$ modes in a \SIadj{1.2}{\MW} beam 
is illustrated in Figure~\ref{fig:mhexec} for the case of normal 
hierarchy for two different values of $\sin^2 \theta_{23}$. 
Across the overwhelming majority of the parameter space for the mixing parameters that
are not well known 
(mainly \deltacp and $\sin^2{\theta_{23}}$), 
LBNE's determination of the MH will be definitive, but even for unfavorable 
combinations of the parameter values, a statistically ambiguous outcome is highly unlikely.

The least favorable scenario corresponds to a true value of 
\deltacp in which the MH asymmetry
is maximally offset by the leptonic CP asymmetry, and where, independently, 
$\sin^2{\theta_{23}}$ takes on a value at the low end of its 
experimentally allowed range.  For this 
scenario, studies indicate that with a \SIadj{34}{\kt} 
LArTPC operating for 
six years in a \SIadj{1.2}{\MW} beam, LBNE on its own can (in a 
typical data set) distinguish between normal and
inverted hierarchy with 
$|\Delta \chi^2| = \overline{|\Delta \chi^2|} = 25$. This 
corresponds to a $\geq 99.9996\%$ probability of determining the 
correct hierarchy.  In $>97.5\%$ of data sets, LBNE will measure 
$|\Delta \chi^2| > 9$ in this scenario, where measuring 
$|\Delta\chi^2| = 9$ 
with an expected value of \num{25} corresponds to a significance 
in excess of three Gaussian standard deviations.

Concurrent analysis of the corresponding atmospheric-neutrino samples
in an underground detector will improve the precision with which the
MH is resolved.  It is important to note that for the
initial stages of LBNE, a greatly improved level of precision in the
determination of the MH can be achieved by incorporating
constraints from NO$\nu$A and T2K data.  
With an initial \ktadj{10} detector, for half
the range of possible \deltacp values, the expected significance 
exceeds $\overline{\Delta \chi^2} = 25$; again this corresponds to a $\geq
99.9996\%$ probability of determining the correct hierarchy.  To put
this in context, it is notable that even an extended NO$\nu$A
program~\cite{Messier:2013sfa} at four times its nominal exposure (of six
years of operation at \SI{700}{\kW}), would have coverage at the
$\overline{\Delta \chi^2} = 9$ level or better for only $40\%$ of the
\deltacp range.

\textbf{CP Violation and the Measurement of {\boldmath \deltacp}:}
The LBNE program has two somewhat distinct objectives with regard to CP
symmetry violation in the $\nu_\mu \to \nu_e$ oscillation channel.
First, LBNE aims to make a precise determination of the value of
\deltacp within the context of the standard three-flavor mixing
scenario described by the PMNS matrix (discussed in Section~\ref{ss:oscphysics}).  Second, and perhaps more
significantly, LBNE aims to observe a signal for leptonic CP
violation, independent of the underlying nature of neutrino
oscillation phenomenology.  Within the standard three-flavor mixing
scenario, such a signal will be observable, provided \deltacp is
not too close to either of the values for which there is no
CP violation  (zero and $\pi$).  
Together, the pursuit of these two goals provides a thorough test of the 
standard three-flavor scenario.

\begin{figure}[!htb]
\centering\includegraphics[width=0.6\textwidth]{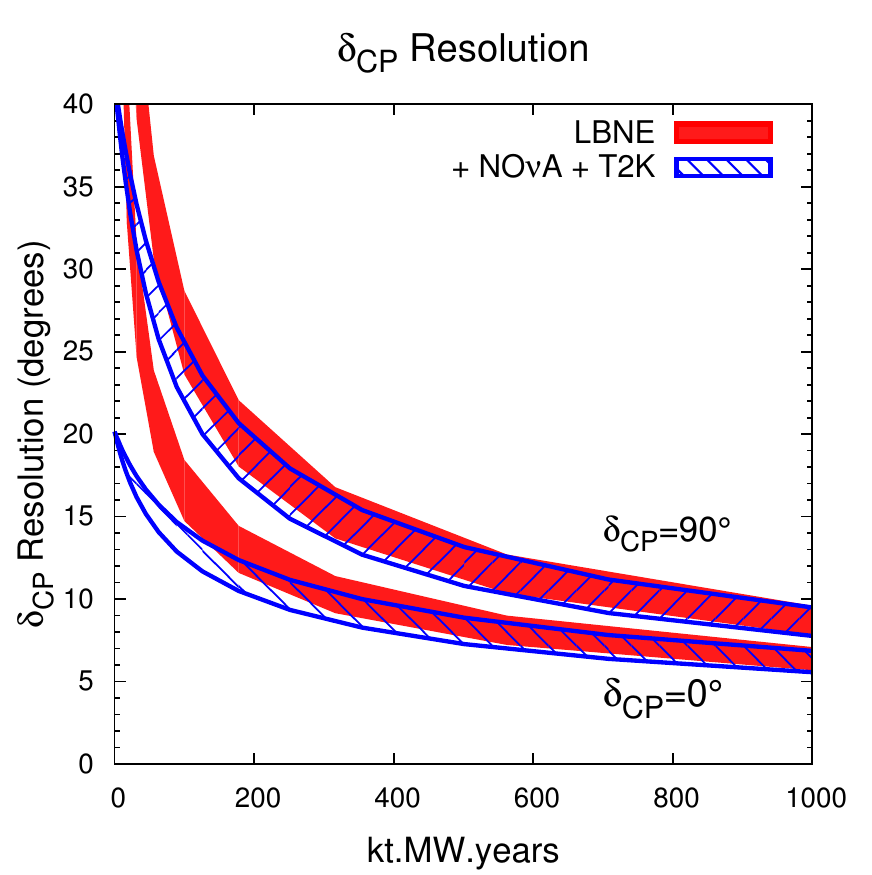}
\caption[Expected 1$\sigma$ resolution for \deltacp as a
  function of exposure]  {The
  expected 1$\sigma$ resolution for \deltacp as a function of
  exposure in detector mass (kiloton) $\times$ beam power (MW) $\times$ time (years).  The red curve is the
  precision that could be obtained from LBNE alone, while the blue curve
  represents the combined precision from LBNE plus the T2K and NO$\nu$A
  experiments.  The width of the bands represents variation with 
  the range of beamline design parameters and proton energy values 
  being considered.} 
\label{fig:execsummaryCP}
\end{figure}

Figure~\ref{fig:execsummaryCP} shows the expected 1$\sigma$ resolution
for \deltacp as a function of exposure for a proton beam power of
\SI{1.2}{\MW}. 
At this beam power, in a six-year run, a \ktadj{10} far detector will be able to
measure \deltacp to $\pm$ 20$^\circ$ $-$ 30$^\circ$ (depending on
its value), independent of other experiments. 
A full-scope 
LBNE operating with multi-megawatt 
beam power in a later phase, will achieve a precision better than $\pm
10^\circ$, comparable to the current precision on the CP phase in the
CKM matrix in the quark sector.

LBNE with a \ktadj{10} detector, in combination with T2K and NO$\nu$A, will
determine leptonic CP violation with a precision of $3\sigma$ or
greater for $\approx 40\%$ of \deltacp values in a 
six-year run with \SIadj{1.2}{\MW} beam power. It is important to note
that LBNE alone dominates the combined sensitivity and that T2K and
NO$\nu$A have very limited 
sensitivity to CP violation on their
own. To reach $5\sigma$ for an appreciable fraction of the range of
\deltacp, the full-scope 
LBNE 
will be needed to control systematic errors while
accumulating large enough samples in the far detector to reach this
level of sensitivity.  No experiment can provide coverage at 100\%,
since CP violation effects vanish as $\mdeltacp\to 0$ or $\pi$.

\textbf{\boldmath Determination of $\sin^2{2\theta_{23}}$ and Octant
  Resolution:}  In long-baseline experiments with $\nu_\mu$ beams, the
magnitude of $\nu_\mu$ disappearance and $\nu_e$ appearance signals is
proportional to $\sin^2{2\theta_{23}}$ and $\sin^2{\theta_{23}}$,
respectively, in the standard three-flavor mixing scenario.  Current
$\nu_\mu$ disappearance data are consistent with maximal mixing,
$\theta_{23} = 45^\circ$.  To obtain the best sensitivity to both the
magnitude of its deviation from $45^\circ$ as well as
its sign ($\theta_{23}$ octant), a combined analysis of the two
channels is needed~\cite{Huber:2010dx}.  As demonstrated in
Chapter~\ref{nu-oscil-chap}, a \ktadj{10} LBNE detector will be able to resolve the
$\theta_{23}$ octant at the $3\sigma$ level or better for
$\theta_{23}$ values less than $40^\circ$ or greater than $50^\circ$,
provided \deltacp is not too close to zero or $\pi$.  A full-scope 
LBNE will measure $\theta_{23}$ with a precision of 
$1^\circ$ or less, even for values within a few degrees of $45^\circ$.

\subsection{Nucleon Decay Physics Motivated by Grand Unified Theories}

\begin{figure}[!htp]
\centering\includegraphics[width=0.7\textwidth]{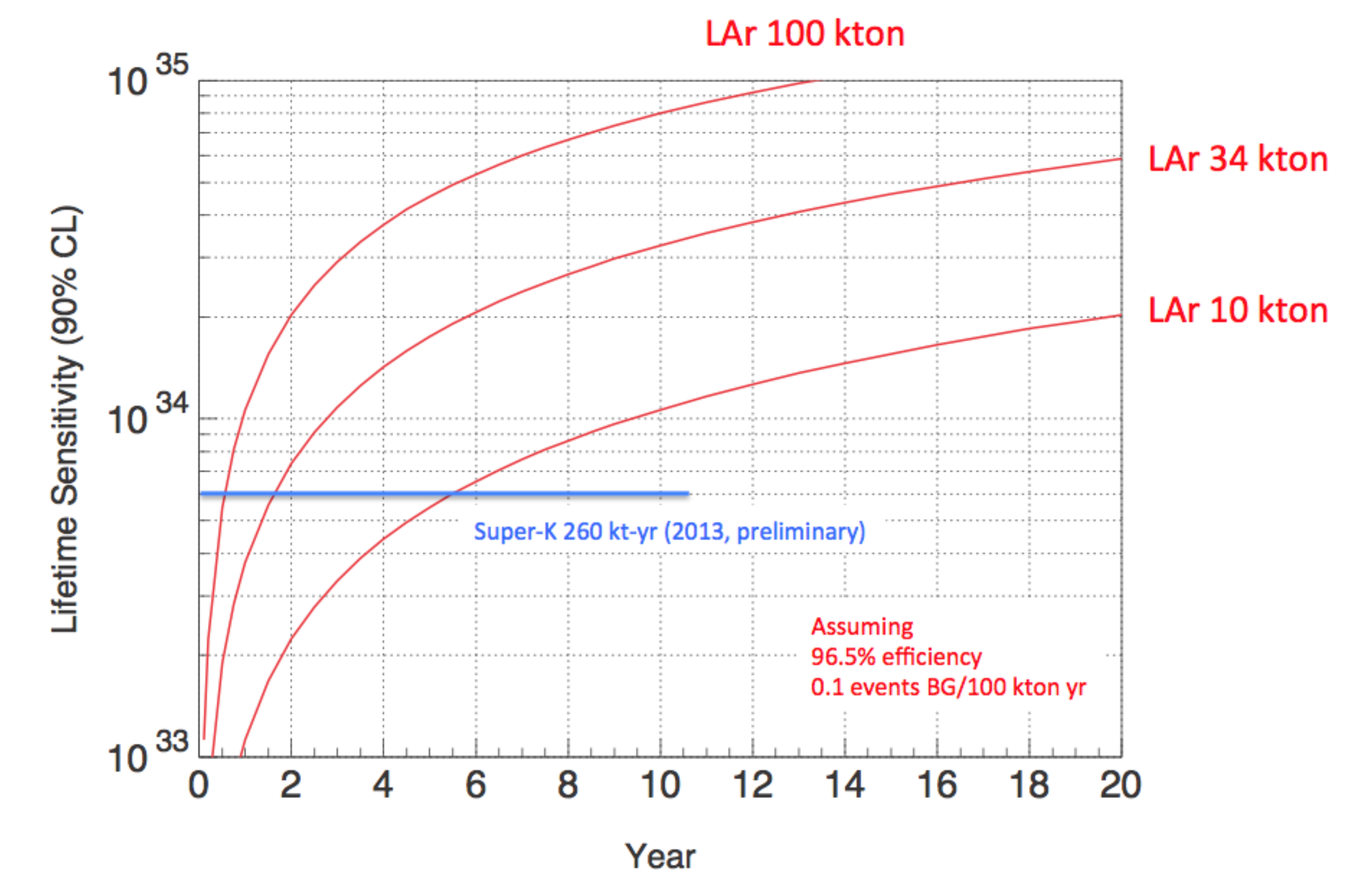}
\caption[Sensitivity to the decay $p\to K^+ \overline{\nu}$ with liquid argon detectors]
  {Sensitivity to the decay $p\to K^+ \overline{\nu}$ 
  as a function of time for underground liquid argon detectors with different masses.}
\label{fig:execsummarypdk}
\end{figure}

The LBNE far detector will significantly extend lifetime sensitivity for specific nucleon decay
modes by virtue of its high detection efficiency relative to water Cherenkov detectors and its 
low background rates.  As an example, LBNE has
enhanced capability for detecting the $p\to K^+\overline{\nu}$
channel, where lifetime predictions from supersymmetric models extend
beyond, but remain close to, the current (preliminary)
Super-Kamiokande limit of  $\tau/B > \SI{5.9e33}{year}$ (90\% CL)
from a \SI[number-unit-product = -, inter-unit-product=\ensuremath{{}\cdot{}}]{260}{\kt\year}
exposure~\cite{kearns_isoups}\footnote{The lifetime shown here is divided by the branching
fraction for this decay mode, $\tau/B$, and as such is a \emph{partial lifetime}.}.  
The signature for an
isolated semi-monochromatic charged kaon in a LArTPC is distinctive,
with multiple levels of redundancy.  A \ktadj{34} LBNE far detector deep
underground will reach a limit of \SI{3e34}{\year} after ten years
of operation (Figure~\ref{fig:execsummarypdk}), and would see nine
events with a background of 0.3 should $\tau/B$ be \SI{1e34}{\year}, just beyond the
current limit.
Even a \ktadj{10} detector 
(placed underground) 
would yield an intriguing signal of a few events after a ten-year exposure 
in this scenario.

\subsection{Supernova-Neutrino Physics and Astrophysics}

The neutrinos from a core-collapse supernova are emitted in a burst of
a few tens of seconds duration, with about half in the first
second. Energies are in the range of a few tens of MeV, and the
luminosity is divided roughly equally between the three known neutrino
flavors.  Currently, experiments worldwide are sensitive primarily to
electron antineutrinos ($\ane$), with detection through the inverse-beta decay
process on free protons\footnote{This refers to neutrino interactions with the nucleus of a
hydrogen atom in H$_2$O in water detectors or in hydrocarbon chains in 
liquid scintillator detectors.},
 which dominates the interaction rate in water
and liquid-scintillator detectors.  Liquid argon has a unique sensitivity to
the electron-neutrino ($\nu_e$) component of the flux, via the absorption
interaction on $^{40}$Ar as follows:
\begin{eqnarray*}
\nu_e +{}^{40}{\rm Ar} & \rightarrow & e^-+{}^{40}{\rm K^*}
\end{eqnarray*} 
This interaction can be tagged via the coincidence of the emitted
electron and the accompanying photon cascade from the $^{40}{\rm K^*}$
de-excitation.  About 900 events would be expected in a \ktadj{10}
fiducial mass liquid argon detector for a supernova at a distance of
\SI{10}{\kilo\parsec}.  In the neutrino channel the oscillation
features are in general more pronounced, since the $\nu_e$ spectrum is
always significantly different from the $\nu_\mu$ ($\nu_\tau$) spectra
in the initial core-collapse stages, to a larger degree than is the
case for the corresponding \ane\ spectrum.  Detection of a large
neutrino signal in LBNE would help provide critical information on key
astrophysical phenomena such as
\begin{enumerate}
\item the neutronization burst
\item formation of a black hole
\item shock wave effects
\item shock instability oscillations
\item turbulence effects
\end{enumerate}

%
%
\subsection{Precision Measurements with a High-Intensity Neutrino Source and High-Resolution Near Detector}

The near neutrino detector will provide precision measurements of
neutrino interactions, which in the medium to long term are essential
for controlling the systematic uncertainties in the long-baseline
oscillation physics program.  The near detector, which will include argon
targets, will measure the absolute flux and energy-dependent shape of
all four neutrino species, \nm, \anm, $\nu_e$ and \ane\, to accurately
predict for each species the far/near flux ratio as a function of
energy.  It will also measure the four-momenta of secondary hadrons,
such as charged and neutral mesons, produced in the neutral and
charged current interactions that constitute the dominant backgrounds
to the oscillation signals.

With  \num{240000} (\num{85000}) $\nu_\mu$ ($\overline{\nu}_\mu$) charged current 
and \num{90000} (\num{35000})  neutral current interactions per ton per $1 \times 10^{20}$
protons-on-target at \SI{120}{GeV}  in the $\nu$ ($\overline\nu$) beam, the near detector
will also be the source of data for a rich program of neutrino-interaction 
physics in its own right.  These numbers correspond to
\num{e7}  neutrino interactions per year for the range of beam
configurations and near detector designs under consideration.
Measurement of fluxes, cross sections and particle production over a
large energy range of \SIrange{0.5}{50}{\GeV} (which can also help constrain
backgrounds to proton decay signals from atmospheric neutrinos) are the key
elements of this program.  
Furthermore, since the near detector data will feature very large
samples of events that are amenable to precision reconstruction and
analysis, they can be exploited for sensitive studies of electroweak
physics and nucleon structure, as well as for searches for new physics
in unexplored regions (heavy sterile neutrinos, high-$\Delta m^2$
oscillations, light Dark Matter particles, and so on). 


%
\section{Summary}

\begin{introbox} 
The LBNE physics program has been identified as a priority of the 
global HEP community for the coming decades. The facilities available 
in the U.S. are the best suited internationally to carry out this program 
 and the substantially developed LBNE design 
is at the forefront of technical innovations in the field. 
Timely implementation of LBNE will significantly advance the 
global HEP program and assure continued intellectual leadership 
for the U.S. within this community.
\end{introbox}

This chapter has touched only briefly on the most prominent portion of
the full suite of physics opportunities enabled by LBNE.  
The following chapters cover these in detail, as well as
topics that were omitted here in the interest of brevity and focus.
In Chapter~\ref{conclusion-chap} progress toward
LBNE physics milestones is addressed, based on one potential scenario
for the operation of successive stages of LBNE detector and PIP-II
implementations,
and the broad role of LBNE is discussed 
in the context of such scenarios.
The present chapter concludes with a summary of its key points.

The primary science goals of LBNE are drivers for the advancement of
particle physics. The questions being addressed are of wide-ranging
consequence: the origin of flavor and the generation structure of the
fermions (i.e., the existence of three families of quark and lepton flavors), 
the physical mechanism that provides the CP violation needed
to generate the Baryon Asymmetry of the Universe, 
and the high energy
physics that would lead to the instability of matter.
Achieving these goals requires a dedicated, ambitious and long-term
program.  No other proposed long-baseline neutrino oscillation program
with the scientific scope and sensitivity of LBNE is as advanced in terms of
engineering development and project planning.  A phased program with a
far detector of even modest size in the initial stage (e.g., 10 kt) will
enable exciting physics in the intermediate term, including a
definitive mass hierarchy determination and a measurement of the CP
phase without ambiguities, while providing the fastest route toward
achieving the full range of LBNE's science objectives.  Should LBNE find
that the CP phase is not zero or $\pi$, it will have found strong
indications ($>3\sigma$) of leptonic CP violation.  Global interest is
favorable for contributions from international partners to accelerate
and enhance this program, including the LBNE first-phase scope.

Implementing the vision that has brought LBNE to this point will allow
the U.S. to host this world-leading program, bringing together the
world's neutrino community to explore key questions at the forefront of
particle physics and astrophysics. 
Moreover, the excitement generated
by both the technical challenges of mounting LBNE and the potential
physics payoffs are widely shared --- among the 
generation of 
scientists who have been paving the way for these innovations, as well as 
the young scientists for whom LBNE will provide numerous research opportunities
over the next two decades.

%% file: chapter-intro.tex

\begin{introbox}
  The Standard Model of particle physics describes all of the known
  fundamental particles and the electroweak and strong forces that, in
  combination with gravity, 
  govern today's Universe. The observation
  that neutrinos have mass is one demonstration that the Standard Model is
  incomplete. By exploring physics beyond the Standard Model, LBNE
  will address fundamental questions about the Universe:
\begin{description}
\item [What is the origin of the matter-antimatter asymmetry in the
  Universe?] 
Immediately after the Big Bang, matter and antimatter
  were created equally, yet matter now dominates. By studying the
  properties of neutrino and antineutrino oscillations, LBNE is pursuing the
  most promising avenue for understanding this asymmetry.
\item[What are the fundamental underlying symmetries of the Universe?]
  Resolution by LBNE of the detailed mixing patterns and ordering of neutrino mass
  states, and comparisons to the corresponding phenomena in the quark
  sector, could reveal underlying symmetries that are as
  yet unknown.
\item[Is there a Grand Unified Theory of the Universe?] Experimental
  evidence hints 
that the physical forces observed today were unified
  into one force at the birth of the Universe. Grand Unified Theories
  (GUTs), which attempt to describe the unification of forces, predict
  that protons should decay, a process that has never been observed. 
  LBNE will probe proton lifetimes predicted by
  a wide range of GUT models.
\item[How do supernovae explode?] The heavy elements that are the key
  components of life --- such as carbon --- were created in the
  super-hot cores of collapsing stars.  LBNE's design will enable it 
 to detect the
  neutrino burst from core-collapse supernovae. By measuring
  the time structure and energy spectrum of a neutrino burst, LBNE
  will be able to elucidate critical information about the dynamics of
  this special astrophysical phenomenon.
\item[What more can LBNE discover about the Standard Model?] The high
  intensity of the 
LBNE neutrino beam  will provide a unique probe for
  precision tests of Standard Model processes as well as searches for new physics in unexplored regions. 
\end{description} 
\end{introbox}

LBNE 
has been designed to address a
wide range of scientific topics using well-characterized,
high-intensity, accelerator-based neutrino beams, a long baseline for
neutrino oscillations, and a very large, deep-underground detector
with excellent particle identification capabilities over a large range
of energies.
While maximizing the 
reach for 
a core set of scientific objectives, its design --- described in
Chapter~\ref{project-chap} --- accommodates the
flexibility to extend the scope of measurements as additional
resources become available.  

\section{Scientific Objectives of LBNE}

The scientific objectives of LBNE have been categorized into primary,
secondary, and additional secondary objectives according to priorities
developed and agreed upon by the LBNE community and accepted as part
of the CD-0 (Mission Need) approval by the U.S. Department of Energy~\cite{DOCDB3056}.  

{\bf Primary 
objectives} of LBNE,  in priority order, are
the following measurements:

\begin{enumerate}
\item precision measurements of the parameters that govern $\nu_{\mu}
  \rightarrow \nu_e$ oscillations; this includes precision measurement
  of the third mixing angle $\theta_{13}$, measurement of the
  charge-parity (CP) violating phase \deltacp, and determination
  of the neutrino mass ordering (the sign of $\Delta m^{2}_{31} = m_3^2 - m_1^2$), 
 the so-called \emph{mass hierarchy}

\item precision measurements of the mixing angle $\theta_{23}$, including the
determination of the octant in which this angle lies, and the value of the mass difference, |$\Delta
m^{2}_{32}$|, in $\nu_{\mu} \rightarrow \nu_{e,\mu}$ oscillations

\item {search for proton decay, yielding significant improvement
  in the current limits on the partial
  lifetime of the proton ($\tau$/BR) in one or more important
  candidate decay modes, e.g., $p \rightarrow K^+\overline\nu$}

\item {detection and measurement of the neutrino flux from a
  core-collapse supernova within our galaxy, should one occur during
  the lifetime of LBNE}
\end{enumerate}

In a phased approach to LBNE, the goal of the first phase is to
maximize the effectiveness of the facility to achieve the first
two objectives, above.  The mass hierarchy determination and the
precision determination of $\theta_{23}$ will most likely be complete
in the first phase of LBNE; while the precision determination of CP
violation will require the full-scope LBNE, an initial measurement of
the CP phase parameter \deltacp will be performed in earlier phases.

{\bf Secondary objectives}, which may also be enabled by the facility designed 
to achieve the primary objectives, include:
\begin{enumerate}
\item other accelerator-based, neutrino oscillation measurements; these could 
include further sensitivity to Beyond Standard Model (BSM) physics 
such as nonstandard interactions
\item {measurements of neutrino oscillation phenomena using atmospheric
  neutrinos}
\item {measurement of other astrophysical phenomena using medium-energy
  neutrinos}
\end{enumerate}


{\bf Additional secondary objectives}, the achievement of which may
require upgrades to the facility that is designed to achieve the
primary physics objectives (e.g., deployment of 
additional detector mass or alternate detector technologies), include:
\begin{enumerate}
\item {detection and measurement of the diffuse supernova-neutrino
  flux}
\item {measurements of neutrino oscillation phenomena and of solar
  physics using solar neutrinos}
\item {measurements of astrophysical and geophysical neutrinos of
  low energy}
\end{enumerate}

In addition, a rich set of
science objectives enabled by a sophisticated near neutrino detector
have been identified. A primary and a secondary objective, respectively, are:
\begin{enumerate}
\item measurements
necessary to achieve the primary physics research objectives listed above

\item studies of neutrino interactions that may be enabled either by the facility 
designed to achieve the primary
objectives or by future upgrades to the
facility and detectors; these include precision studies of 
the weak interaction, studies of nuclear and nucleon structure, and
searches for new physics
\end{enumerate}

\clearpage
\section{Neutrino Three-Flavor Mixing, CP Violation and the Mass Hierarchy}
\label{ss:oscphysics}

\begin{figure}[!htb]
\centerline{\includegraphics[width=0.9\textwidth]{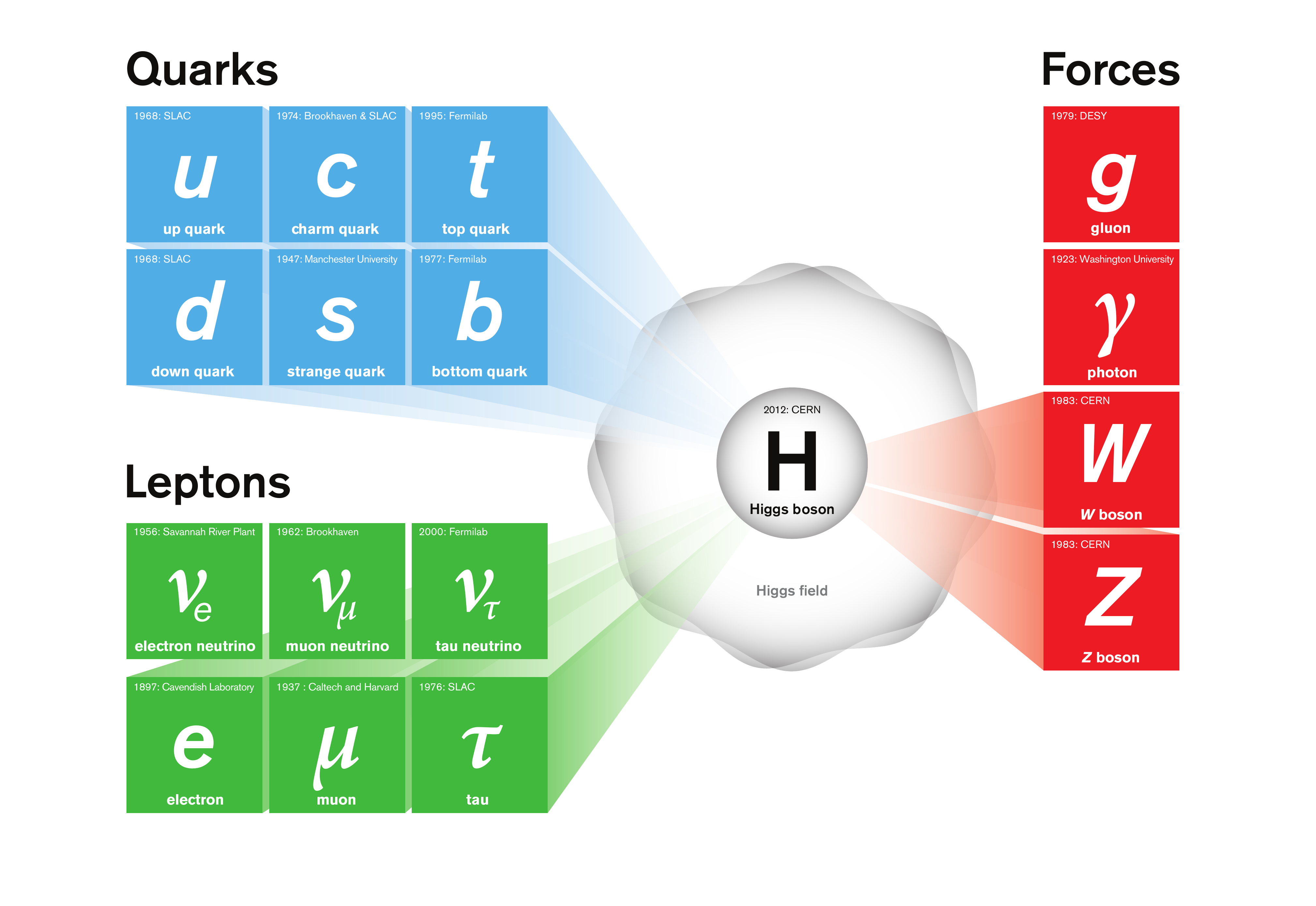}}
\caption[Standard Model particles]{Known particles and forces in the
Standard Model of particle physics.
The quarks and leptons are arranged in
pairs into three generations: $(u,d), (c,s), (t,b)$ and $(\nu_e,e),
(\nu_\mu,\mu), (\nu_\tau,\tau)$, respectively. There are three known
neutrino mass states $\nu_1, \nu_2, \nu_3$ which are mixtures of the
three neutrino flavors $\nu_e, \nu_\mu, \nu_\tau$ shown in this figure.
The Standard Model includes the gluon ($g$), photon ($\gamma$) and
$(W^{\pm},Z^0)$ bosons that are the mediators of the strong,
electromagnetic and weak interactions, respectively. The Higgs boson is a
manifestation of the Higgs field that endows all the known particles with
mass.}
\label{fig:standardmodel}
\end{figure}

The Standard Model of particle physics 
(Figure~\ref{fig:standardmodel}) presents a remarkably accurate
description of the elementary particles and their
interactions. However, its limitations beg deeper questions about
Nature. The unexplained patterns of quarks, leptons, flavors and
generations imply that a more fundamental underlying theory must
exist.  
LBNE plans to pursue a detailed
study of neutrino mixing, resolve the neutrino mass ordering,
and search for CP violation in the lepton sector by
studying the oscillation patterns of high-intensity
$\nu_\mu$ and $\overline{\nu}_\mu$ beams measured over a long baseline. 

\begin{figure}[!htb]
\centerline{\includegraphics[width=0.5\textwidth]{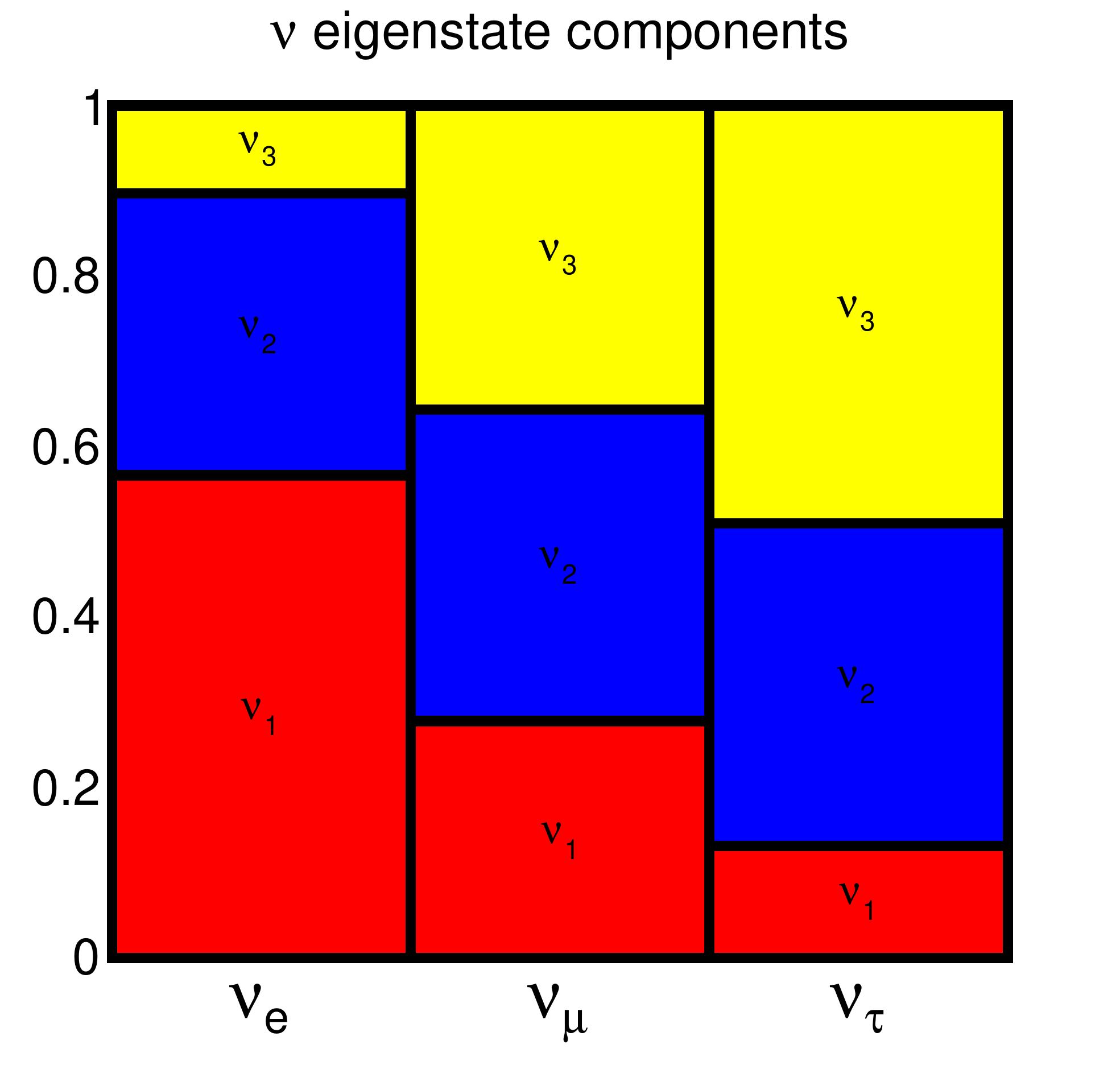}}
\caption[Neutrino mass eigenstate components of flavor eigenstates ]{The neutrino mass eigenstate components of the known flavor eigenstates.}
\label{fig:pmns}
\end{figure}

\begin{introbox}
Results from the last decade, indicating that the three known types of
neutrinos  have nonzero mass, mix
with one another and oscillate between generations, imply physics beyond
the Standard Model~\cite{Mohapatra:2005wg}. 
Each of the three flavors 
of neutrinos, $\nu_e, \nu_\mu$ and $\nu_\tau$ (Figure~\ref{fig:standardmodel}),  is 
known to be 
a different mix of three mass eigenstates 
$\nu_1,\nu_2$ and $\nu_3$ (Figure~\ref{fig:pmns}).
In the Standard Model, the
simple Higgs mechanism, which has now been confirmed by the observation of
the Higgs boson~\cite{Aad:2012tfa,Chatrchyan:2012ufa}, is responsible for
both quark and lepton masses, mixing and charge-parity (CP) violation (the
mechanism responsible for matter-antimatter asymmetries). However, the
small size of neutrino masses and their relatively large mixing bears
little resemblance to quark masses and mixing, suggesting that different
physics --- and possibly different mass scales --- in the two sectors may
be present, and motivating precision study of mixing and CP violation in
the lepton sector.
\end{introbox}
Neutrino oscillation arises from mixing between the flavor and mass eigenstates of neutrinos,
corresponding to the weak and gravitational interactions, respectively. 
This three-flavor-mixing
scenario can be described by a rotation between the weak-interaction
eigenstate basis $(\nu_e,\, \nu_\mu,\, \nu_\tau)$ and the basis of
states of definite mass $(\nu_1,\, \nu_2,\, \nu_3)$.  In direct
correspondence with mixing in the quark sector, the transformations
between basis states is expressed in the form of a complex unitary
matrix, known as the PMNS matrix :
\begin{equation}
\left(\begin{array}{ccc} \nu_e \\ \nu_\mu \\ \nu_\tau \end{array} \right)= 
\underbrace{
  \left(\begin{array}{ccc}
      U_{e 1} &  U_{e 2} & U_{e 3} \\ 
      U_{\mu1} &  U_{\mu2} & U_{\mu 3} \\ 
      U_{\tau 1} &  U_{\tau 2} & U_{\tau 3} 
    \end{array} \right)
}_{U_{\rm PMNS}} \left(\begin{array}{ccc} \nu_1 \\ \nu_2 \\ \nu_3 \end{array} \right).
\label{eqn:pmns0}
\end{equation}
The PMNS matrix in full generality depends on just three mixing angles
and a CP-violating phase.  The mixing angles and phase are designated
as $(\theta_{12},\, \theta_{23},\, \theta_{13})$ and
\deltacp.  
This matrix can be parameterized as the product of three
two-flavor mixing matrices as follows, where $c_{\alpha \beta}=\cos \theta_{\alpha \beta}$ and $s_{\alpha
  \beta}=\sin \theta_{\alpha \beta}$:
\begin{equation}
U_{\rm PMNS} = 
  \underbrace{
    \left( \begin{array}{ccc}
        1 & 0 & 0 \\ 
        0 & c_{23} & s_{23} \\ 
        0 & -s_{23} & c_{23}
    \end{array} \right)
  }_{\rm I}
\underbrace{
  \left( \begin{array}{ccc}
      c_{13} & 0 & e^{i\mdeltacp}s_{13} \\ 
      0 & 1 & 0 \\ 
      -e^{i\mdeltacp}s_{13} & 0 & c_{13}
  \end{array} \right) 
}_{\rm II}
\underbrace{
 \left( \begin{array}{ccc}
      c_{12} & s_{12} & 0 \\ 
      -s_{12} & c_{12} & 0 \\ 
      0 & 0 & 1
  \end{array} \right) 
}_{\rm III}
\label{eqn:pmns}
\end{equation}
The parameters of the PMNS
matrix determine the probability amplitudes of the neutrino
oscillation phenomena that arise from mixing. 
\begin{introbox}
The relationship between the three mixing angles $\theta_{12}, \theta_{23}$, and $\theta_{13}$ and the mixing between the neutrino flavor and mass states can 
be described as follows~\cite{neutrinomatrix}:
\begin{eqnarray}
\tan^2 \theta_{12} & : & \frac{\rm amount \ of \ \nu_e \ in \ \nu_2}{\rm amount \ of \ \nu_e \ in \ \nu_1} \\
\tan^2 \theta_{23} & : & {\rm ratio \ of \ } \nu_\mu {\rm \ to \ } \nu_{\tau}{\rm \ in \ \nu_3} \\
\sin^2 \theta_{13} & : & {\rm amount \ of \ \nu_e \ in \ \nu_3}
\end{eqnarray}

The frequency of neutrino oscillation among the weak-interaction (flavor)
eigenstates depends on the difference in the squares of the neutrino
masses, $\Delta m^{2}_{ij} \equiv m^{2}_{i} - m^{2}_{j}$; a set of three
neutrino mass states implies two independent mass-squared differences
($\Delta m^{2}_{21}$ and $\Delta m^{2}_{32}$). The ordering of the
mass states is known as the \emph{neutrino mass hierarchy}. An ordering of
$m_1 < m_2 < m_3$ is known as the \emph{normal hierarchy} since it matches
the ordering of the quarks in the Standard Model, whereas an ordering of $m_3 < m_1 < m_2$
is referred to as the \emph{inverted hierarchy}.

Since each flavor eigenstate is a mixture of three mass eigenstates,
there can be an overall phase difference between the quantum states, 
referred to as
\deltacp. A nonzero value of this phase implies that
neutrinos and antineutrinos oscillate differently --- a phenomenon
known as charge-parity (CP) violation. \deltacp is therefore 
often referred to as the \emph{CP phase} or the \emph{CP-violating phase}.
 
\end{introbox}

The entire complement of neutrino experiments to date has measured
five of the mixing parameters: the three angles $\theta_{12}$,
$\theta_{23}$ and (recently) $\theta_{13}$, and the two mass differences
$\Delta m^{2}_{21}$ and $\Delta m^{2}_{32}$. The sign of $\Delta
m^{2}_{21}$ is known, but not that of $\Delta m^{2}_{32}$, which 
is the crux of the 
mass hierarchy ambiguity.
The values of $\theta_{12}$ and $\theta_{23}$ are large, while 
$\theta_{13}$ is smaller~\cite{An:2012bu}. The value of \deltacp is unknown.
The real values of the entries of the PMNS mixing matrix, which
contains information on the strength of flavor-changing weak decays in
the lepton sector, can be expressed in approximate form as

\begin{equation}
|U_{\rm PMNS}|\sim \left(\begin{array}{ccc} 0.8 & 0.5 & 0.2 \\ 0.5 & 0.6 & 0.6 \\ 0.2 & 0.6 & 0.8\end{array} \right).
\label{eq:pmnsmatrix}
\end{equation}


The three-flavor-mixing scenario for neutrinos is now well
established. However, the mixing parameters are not known to the same precision 
as are those in the
corresponding quark sector, and several important quantities, including
the value of \deltacp and the sign of the large mass splitting, are
still undetermined. In addition, several recent
anomalous experimental results count among their possible
interpretations phenomena that do not fit this 
model~\cite{Aguilar:2001ty,AguilarArevalo:2007it,Aguilar-Arevalo:2013pmq,Mention:2011rk}.

The relationships between the values of the parameters in the neutrino
and quark sectors suggest that mixing in the two sectors is
qualitatively different. Illustrating this difference, the value of
the entries of the CKM quark-mixing matrix (analogous to the PMNS matrix for
neutrinos, and thus indicative of the strength of flavor-changing weak
decays in the quark sector) can be expressed in approximate form as
\begin{equation}
|V_{\rm CKM}|\sim \left(\begin{array}{ccc} 1 & 0.2 & 0.004\\ 0.2 & 1 & 0.04 \\ 0.008 & 0.04 & 1\end{array} \right)
\label{eq:ckmmatrix}
\end{equation}
and compared to the entries of the PMNS matrix given in Equation~\ref{eq:pmnsmatrix}.
As discussed in \cite{King:2014nza}, the question of why the quark mixing angles are
smaller than the lepton mixing angles is an important part of the ``flavor problem.''

Quoting the discussion in~\cite{deGouvea:2013onf}, ``while the CKM
matrix is almost proportional to the identity matrix plus
hierarchically ordered off-diagonal elements, the PMNS matrix is far
from diagonal and, with the possible exception of the $U_{e3}$
element, all elements are ${\cal O}(1)$.''
One theoretical method often used to address this question involves the use of non-Abelian discrete
subgroups of $SU(3)$ as flavor symmetries; the popularity of this method comes partially from
the fact that these symmetries can give rise to the nearly \emph{tri-bi-maximal}\footnote{Tri-bi-maximal mixing refers to a form of the neutrino mixing matrix with effective bimaximal mixing of $\nu_\mu$ and $\nu_\tau$
at the atmospheric scale ($L/E \sim$ \SI{500}{\km / \GeV}) and effective trimaximal
mixing for $\nu_e$ with $\nu_\mu$ and $\nu_\tau$ 
at the solar scale ($L/E \sim$ \SI{15000}{\km / \GeV})~\cite{Harrison:2002er}.} 
structure of the PMNS matrix.
Whether employing these flavor symmetries or other methods,
any theoretical principle that attempts to describe the fundamental
symmetries implied by the observed organization of quark and neutrino
mixing --- such as those proposed in unification models --- leads to
testable predictions such as sum rules between CKM and PMNS
parameters~\cite{deGouvea:2013onf,Mohapatra:2005wg,King:2014nza,Albright:2006cw}.
%
Data on the patterns of neutrino mixing 
are already proving crucial in the quest for a 
relationship between quarks and leptons and their seemingly arbitrary generation
structure.  
Table~\ref{tab:params} displays the comparison between quark and lepton mixing
in terms of the fundamental parameters 
and the precision to which they are known\footnote{A global fit~\cite{Fogli:2012ua} to 
 existing results from current
 experiments sensitive to neutrino oscillation effects is the source for the PMNS matrix values.}, 
highlighting the limited precision  of the neutrino-mixing parameter measurements.

\begin{table}[!htb]
\caption[Best-fit values of the neutrino mixing parameters in the PMNS
  matrix]{Best-fit values of the neutrino mixing parameters in the PMNS
  matrix (assumes normal hierarchy) from~\cite{Fogli:2012ua}, their $1\sigma$ uncertainties and comparison to 
the analogous values in the CKM
 matrix~\cite{Beringer:1900zz}. $\Delta M^2$ is defined as $m_3^2
- (m_1^2+m_2^2)/2$.}
\begin{center}
\begin{tabular}[!htb]{$L^c^c}
\toprule
\rowtitlestyle
Parameter  & Value (neutrino PMNS matrix) & Value (quark CKM matrix)\\
\toprowrule
\rowstyle{\normalfont}

$\theta_{12}$     & $34 \pm 1^{\circ}$ & $13.04 \pm 0.05^{\circ}$\\ 
\colhline
$\theta_{23}$     & $38 \pm 1^{\circ}$ & $2.38 \pm 0.06^{\circ}$\\ 
\colhline
$\theta_{13}$     & $8.9 \pm 0.5^{\circ}$  & $0.201 \pm 0.011^{\circ}$\\ 
\colhline
$\Delta m^2_{21}$  & $+(7.54 \pm 0.22) \times 10^{-5}$ eV$^2$ & \\ 
\colhline
$|\Delta M^2|$    & $(2.43^{+0.10}_{-0.06}) \times 10^{-3}$ eV$^2$ & $m_3 >> m_2$\\ 
\colhline
$\mdeltacp$  & $-170 \pm 54^{\circ}$ &  $67 \pm 5^{\circ}$\\
\bottomrule
\end{tabular}
\end{center}
\label{tab:params}
\end{table}

Clearly much work remains in order to complete the standard three-flavor 
mixing picture, particularly 
with regard to $\theta_{23}$ (is it less than, greater than, or equal
to $45^\circ$?), mass hierarchy (normal or inverted?) 
and \deltacp. 
Additionally, there is 
great value in obtaining a set of measurements for multiple parameters 
\emph{from a single experiment}, so that correlations and systematic 
uncertainties can be handled properly.  Such an experiment would also be 
well positioned to extensively test the standard picture of three-flavor mixing.  
LBNE is designed to be this experiment.

\subsection{CP Violation in the Quark and Lepton Sectors}
\label{sec:cpv-quark-lepton}

In the particular parameterization of the PMNS matrix shown in
Equation~\ref{eqn:pmns}, the middle factor, labeled `II', describes
the mixing between the $\nu_1$ and $\nu_3$ mass states, and depends on
the CP-violating phase \deltacp.  
In the three-flavor model, leptonic CP violation in an oscillation mode
occurs due to the interference of contributions from terms in this factor ---
some of which contain \deltacp (i.e., involve the 
$\nu_1$-$\nu_3$ mixing directly) and some of which do not.  
The presence of nonzero CP-odd terms, e.g., Equation~\ref{eqn:papprox3}, (which requires $\mdeltacp \neq
0$ or $\pi$) in the interference patterns would result in an
asymmetry in neutrino versus antineutrino oscillations. The magnitude of the
CP-violating terms in the oscillation depends most directly on the size of 
the Jarlskog Invariant~\cite{Jarlskog:1985cw}, a function that was
introduced to provide a measure of CP violation independent of
mixing-matrix parameterization. In terms
of the three mixing angles and the (as yet unmeasured) 
CP-violating phase, the Jarlskog Invariant is:
\begin{equation}
J_{CP}^{\rm PMNS} \equiv \frac{1}{8} \sin 2 \theta_{12} \sin 2 \theta_{13}
\sin 2 \theta_{23} \cos \theta_{13} \sin \mdeltacp.
\end{equation}

The relatively large values of the mixing angles in the lepton sector imply that
leptonic CP-violation effects may be quite large ---  
depending on the value of the phase \deltacp, which is currently unknown. 
Experimentally, it is unconstrained
at the 2$\sigma$ level by the global fit~\cite{Fogli:2012ua}. Many theoretical 
models, 
examples of which include~\cite{Meroni:2012ty,Ding:2013bpa,Luhn:2013lkn,Ding:2013nsa,Antusch:2013wn,King:2013hoa}, 
provide predictions for \deltacp, but these predictions range over all
possible values so do not yet provide any guidance.

Given the current best-fit values of the mixing angles~\cite{Fogli:2012ua} 
and assuming normal hierarchy,
\begin{equation}
J_{CP}^{\rm PMNS} \approx 0.03 \sin \mdeltacp.
\end{equation}
This is in sharp contrast to the very small mixing in the quark sector,  
which leads to a very small value of the corresponding quark-sector
Jarlskog Invariant~\cite{Beringer:1900zz},
\begin{equation}
J_{CP}^{\rm CKM} \approx 3 \times 10^{-5},
\end{equation}
despite the large value of $\delta^{\rm CKM}_{CP}\approx70^{\circ}$.

To date, all observed CP-violating effects have occurred in
experiments involving systems of quarks, in particular strange and
$b$-mesons \cite{Beringer:1900zz}.  Furthermore, in spite of several
decades of experimental searches for other sources of CP violation,
all of these effects are explained by the CKM quark-mixing paradigm,
and all are functions of the quark-sector CP phase parameter,
$\mdeltacp^{\rm CKM}$.  In cosmology, successful synthesis of the
light elements after the Big Bang~\cite{kolb94,weinberg08} (Big Bang
Nucleosynthesis) requires that there be an imbalance in the number of
baryons and antibaryons to one part in a billion when the Universe is
a few minutes old~\cite{Steigman:2007xt}.  CP violation in the quark
sector has not, however, been able to explain the observed Baryon
Asymmetry of the Universe (BAU), due to the small value of
$J_{CP}^{CKM}$.

Baryogenesis~\cite{Fukugita:1986hr} is a likely mechanism for
generating the observed matter-antimatter asymmetry of our
Universe. One way that it is elegantly achieved is by first having
\emph{leptogenesis} in the very early Universe. That mechanism can
come about from the production and decay of very heavy right-handed
neutrinos, if they are Majorana states (i.e. do not conserve lepton
number\footnote{In the Standard Model, lepton number ($L$) and
  baryon number ($B$) are conserved quantum numbers. Leptons
  have $B=0$ and $L = 1$ and antileptons have $L=-1$. A
  quark has $L=0$ and $B = 1/3$ and an antiquark has
  $B=-1/3$. \label{bnumber}}), CP symmetry is violated in their decays
(thus distinguishing particles and antiparticles) and the Universe is
in non-equilibrium.  Leptogenesis will lead to an early dominance of
antileptons over leptons. When the cooling Universe reaches the
electroweak phase transition, $T \sim$ \SI{250}{\GeV}, a baryon number
excess is generated from the lepton asymmetry by a $B-L^{\ddagger}$
conserving mechanism (analogous to proton decay in that it violates
$B$ and $L$ separately but conserves $B-L$) already present in the
Standard Model.

The heavy Majorana right-handed neutrino states that could give rise
to leptogenesis in the very early Universe are also a natural
consequence of the GUT-based \emph{seesaw}
mechanism~\cite{Yanagida:1980xy} --- the simplest and most natural
explanation of the observed super-light neutrino mass scales. The
seesaw mechanism is a theoretical attempt to reconcile the very small
masses of neutrinos to the much larger masses of the other elementary
particles in the Standard Model. The seesaw mechanism achieves this
unification by assuming an unknown new physics scale that connects the
observed low-energy neutrino masses with a higher mass scale that
involves very heavy sterile neutrino states. The seesaw mechanism as
generator of neutrino mass is in addition to the Higgs mechanism that
is now known to be responsible for the generation of the quark,
charged lepton, and vector boson masses.

The no-equilibrium leptogenesis ingredient is expected in a hot Big
Bang scenario, but the Majorana nature of the heavy neutrinos and
needed CP violation can only be indirectly inferred from light
neutrino experiments by finding lepton number violation (validating
their Majorana nature via neutrinoless double-beta decay) and
observing CP violation in ordinary neutrino oscillations.
\begin{introbox}
  Recent theoretical advances have demonstrated that CP violation,
  necessary for the generation of the Baryon Asymmetry of the Universe
  at the GUT scale (baryogenesis), can be directly related to the
  low-energy CP violation in the lepton sector that could manifest in
  neutrino oscillations. As an example, the theoretical model
  described in~\cite{Pascoli:2006ci} predicts that
  leptogenesis, the generation of the analogous lepton asymmetry, can
  be achieved if
  \begin{equation}
    | \sin \theta_{13} \sin \mdeltacp| \gtrsim  0.11
    \label{eqn:leptogenesis}
  \end{equation}
   This implies $|\sin \mdeltacp| \gtrsim 0.7$ given the latest global fit
  value of $|\sin \theta_{13}|$~\cite{Capozzi:2013csa}.
\end{introbox}

The goal of
establishing an experimental basis for assessing this possibility
should rank very high on the list of programmatic priorities within
particle physics, and can be effectively addressed by LBNE. 

\subsection{Observation of CP-Violating Effects in Neutrino Oscillation Experiments}
\label{sec:oscil-cpv}
 
Whereas the Standard Model allows for violation of charge-parity (CP) symmetries in weak interactions, 
CP transformations followed by time-reversal transformations (CPT) are invariant. 
Under CPT invariance, the probabilities of neutrino oscillation and antineutrino oscillation
are equivalent, i.e., $P(\nu_l \rightarrow \nu_l) =
P(\overline{\nu}_l\rightarrow \overline{\nu}_l)$
where $l = e,\, \mu,\, \tau$. Measurements of $\nu_l \rightarrow
\nu_l$ oscillations in which the flavor of the neutrino before and after
oscillations remains the same are referred to as \emph{disappearance} or \emph{survival}
measurements.  CPT invariance in neutrino oscillations was recently
 tested by measurements of $\nu_\mu
\rightarrow \nu_\mu$ and $\overline{\nu}_\mu \rightarrow \overline{\nu}_\mu$
oscillations~\cite{Adamson:2011ch}; no evidence for CPT
violation was found.  Therefore, asymmetries in neutrino versus antineutrino
oscillations arising from CP violation effects can only be accessed in
\emph{appearance} experiments, defined as oscillations of $\nu_l \rightarrow
\nu_{l'}$, in which the flavor of the neutrino 
after
oscillations 
has changed.  Because of the intrinsic challenges of
producing and detecting $\nu_\tau$'s, the oscillation modes
$\nu_{\mu,e} \rightarrow \nu_{e,\mu}$ provide the most promising
experimental signatures of leptonic CP violation.

For $\nu_{\mu,e} \rightarrow \nu_{e,\mu}$ 
oscillations that occur as the neutrinos propagate through matter,  
as in terrestrial long-baseline experiments, 
the coherent forward scattering of $\nu_e$'s on electrons in matter 
modifies the energy and path-length dependence of the vacuum oscillation 
probability in a way that depends on the magnitude \emph{and} sign of $\Delta m^2_{32}$. 
This is  the Mikheyev-Smirnov-Wolfenstein (MSW) effect~\cite{Mikheev:1986gs,Wolfenstein:1977ue}
that has already been observed in solar-neutrino oscillation (disappearance) 
experiments~\cite{Bellini:2008mr,Bellini:2011rx,Aharmim:2011vm,Renshaw:2013}. 
The oscillation probability of $\nu_{\mu,e}
\rightarrow \nu_{e,\mu}$ through matter, in a constant density
approximation, keeping terms up to second order in 
$\alpha \equiv
|\Delta m_{21}^2|/|\Delta m_{31}^2|$ and $\sin ^2 \theta_{13}$, 
is~\cite{Freund:2001pn,Beringer:1900zz}:
\begin{equation}
P(\nu_\mu \rightarrow \nu_e) \cong  P(\nu_e \rightarrow \nu_\mu) \cong 
P_0 + \underbrace{P_{\sin \delta}}_{\rm CP \ violating} + P_{\cos \delta} + P_3
\label{eqn:papprox0}
\end{equation}
where
\begin{eqnarray}
P_0 &=& \sin^2 \theta_{23} \frac{\sin^2 2 \theta_{13}}{(A-1)^2}\sin^2[(A-1)\Delta], \label{eqn:papprox1}\\
P_3 &=& \alpha^2 \cos^2 \theta_{23} \frac{\sin^2 2 \theta_{12}}{A^2}\sin^2 (A \Delta), \label{eqn:papprox2}\\
P_{\sin \delta} &=& \alpha \frac{8 J_{cp}}{A(1-A)} \sin \Delta \sin(A \Delta) \sin [(1-A)\Delta],  \label{eqn:papprox3}\\
P_{\cos \delta} &=& \alpha \frac{8 J_{cp} \cot \mdeltacp}{A(1-A)} \cos
\Delta \sin(A \Delta) \sin [(1-A)\Delta], \label{eqn:papprox4}
\end{eqnarray}
and where 
  \[ \Delta = \Delta m^2_{31} L/4E,\ and \ A=\sqrt{3} G_F N_e 2E/\Delta m^2_{31}. \]
In the above, the CP phase \deltacp appears (via $J_{cp}$)
in the expressions for $P_{\sin\delta}$ (the CP-odd term) which 
switches sign in going from 
$\nu_\mu \to \nu_e$ to the $\overline{\nu}_\mu \to \overline{\nu}_e$ channel, 
and $P_{\cos\delta}$ (the CP-conserving term) which does not.  
The matter effect 
also introduces a neutrino-antineutrino asymmetry, the origin of which 
is simply the presence of electrons and absence of positrons in the Earth.  

Recall that in Equation~\ref{eqn:pmns}, the CP phase 
appears in the PMNS matrix through the mixing of the $\nu_{1}$ and
$\nu_3$ mass states. 
The physical characteristics of an appearance experiment
are therefore determined by the baseline and neutrino energy at which the mixing
between the $\nu_{1}$ and $\nu_3$ states is maximal, as follows:
\begin{eqnarray}
\frac{L ({\rm km})}{E_\nu ({\rm GeV})} &=& 
(2n-1) \frac{\pi}{2} \frac{1}{1.27 \times \Delta m^2_{31} {\rm (eV^2)}}  \\
&\approx& (2n-1) \times 510~{\rm km/GeV} 
\label{eqn:nodes}
\end{eqnarray}
where $ n = 1,2,3...$ denotes the oscillation nodes at which the
appearance probability is maximal. 

\begin{figure}[!tb]
\centerline{
\includegraphics[width=0.5\textwidth]{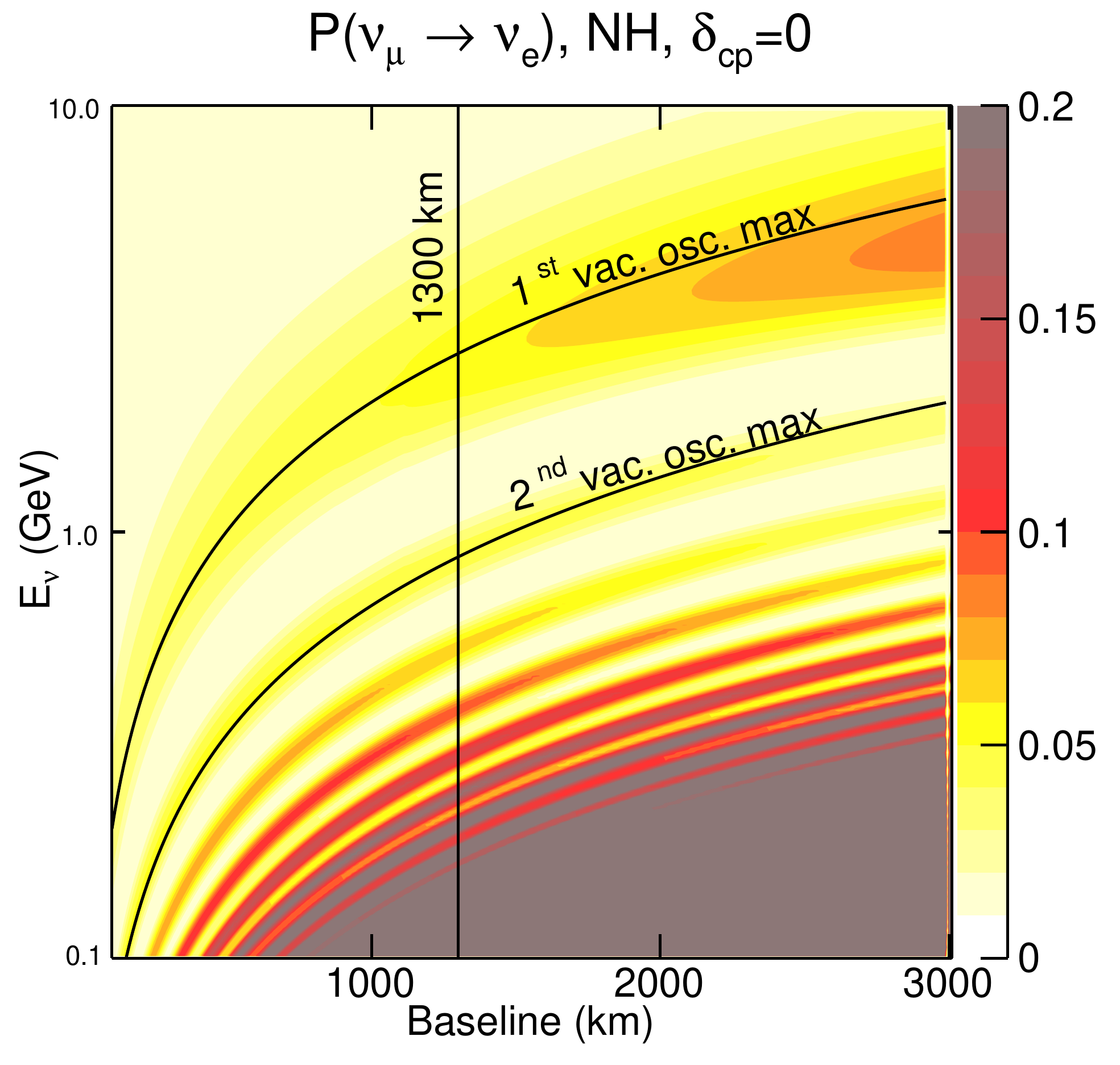}
\includegraphics[width=0.5\textwidth]{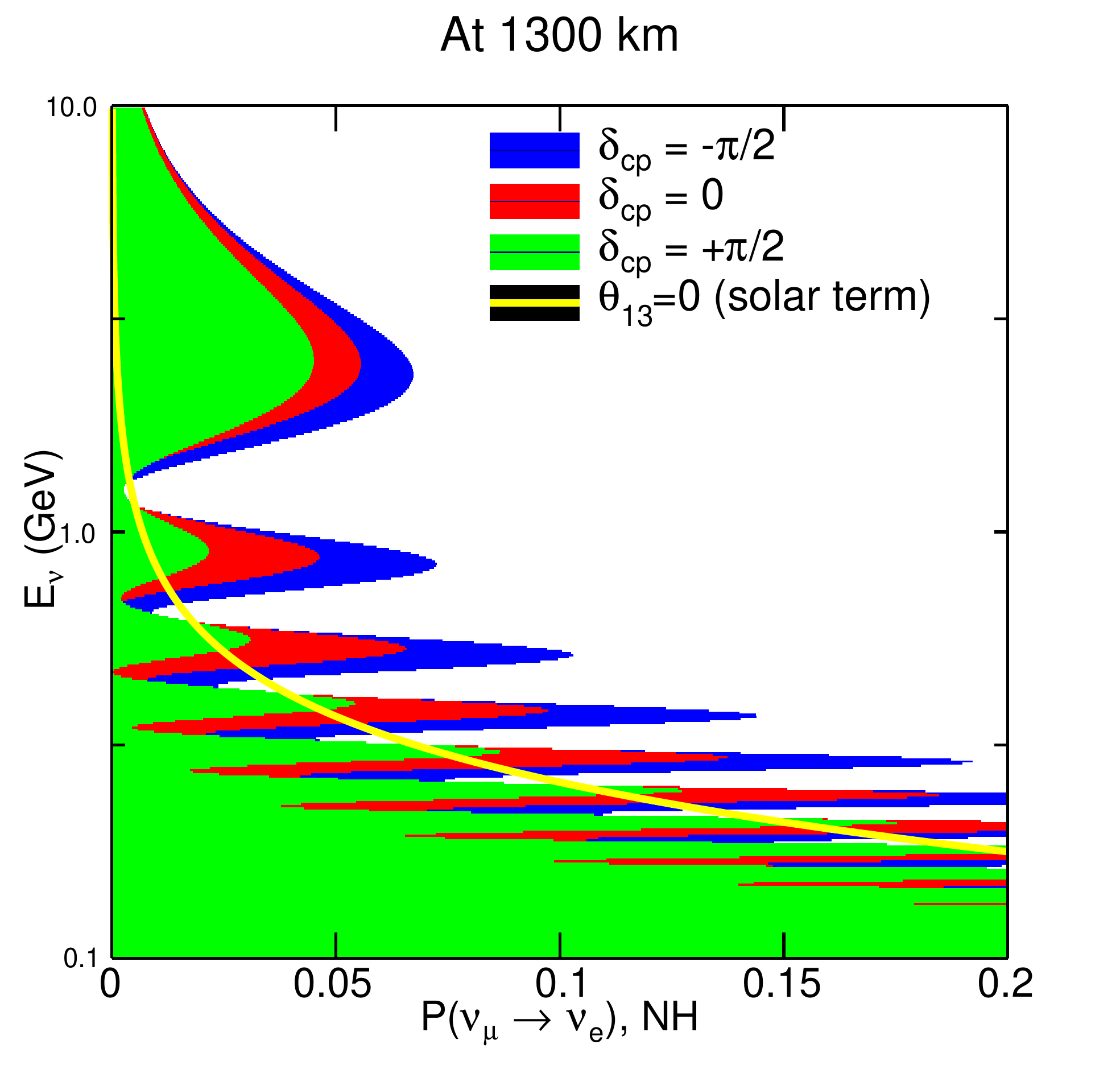}
}
\centerline{
\includegraphics[width=0.5\textwidth]{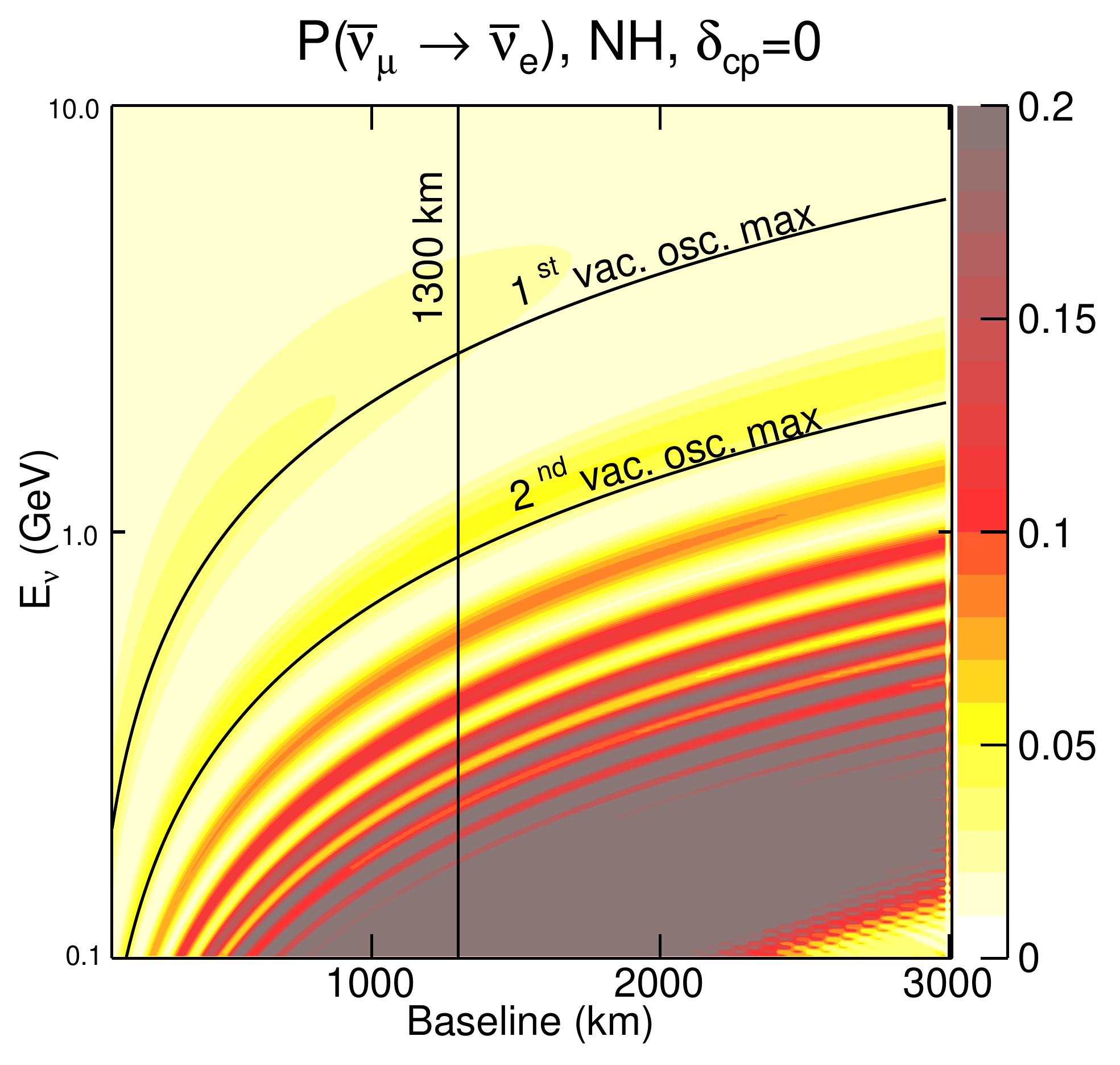}
\includegraphics[width=0.5\textwidth]{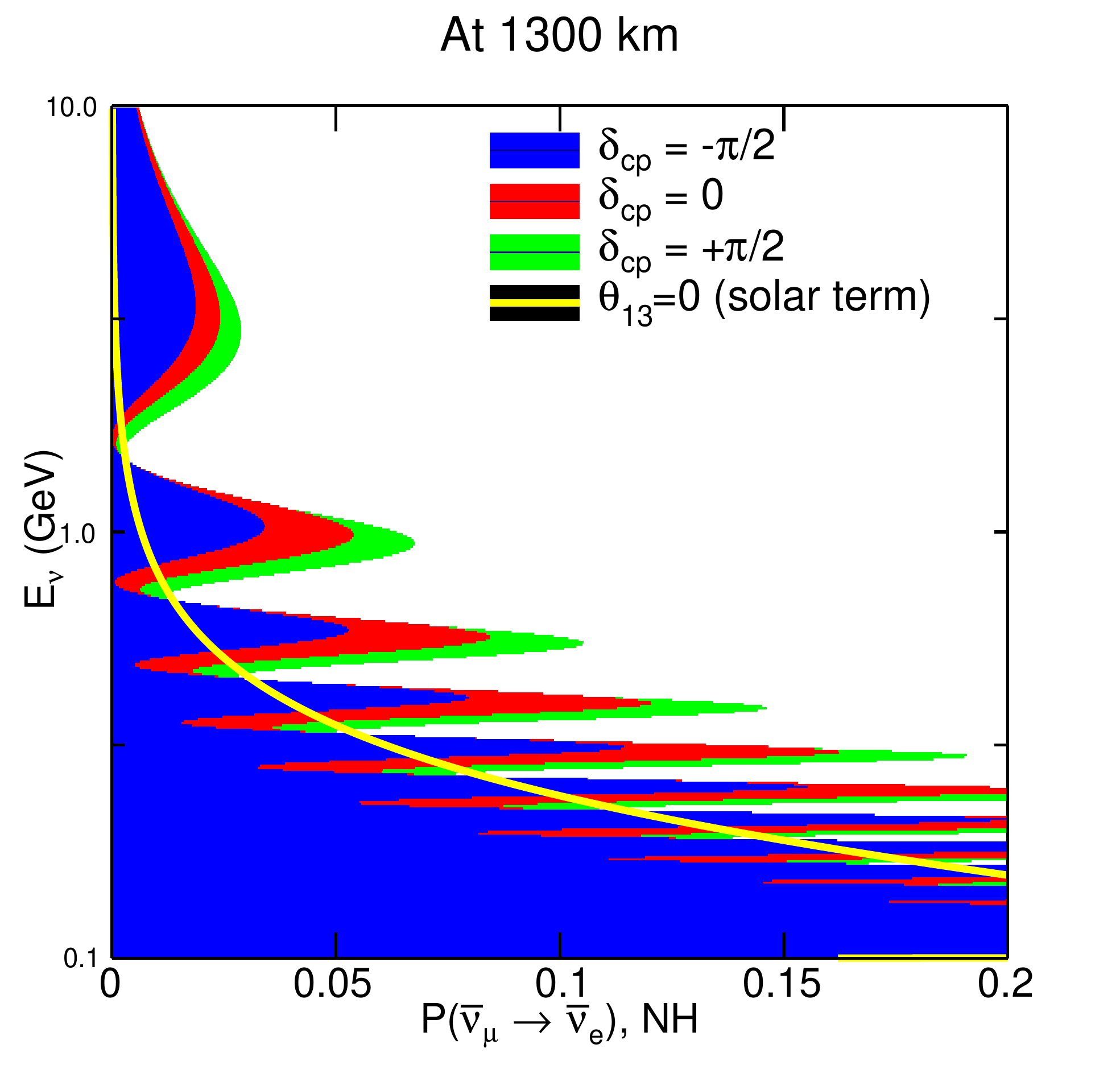}
}
\caption[$\nu$ oscillation probabilities versus $E$ 
  and baseline for various  \deltacp, normal MH]{Neutrino oscillation
  probabilities as a function of energy and baseline, for
  different values of \deltacp, \emph{normal hierarchy}. The oscillograms on the left
  show the $\nu_\mu \rightarrow \nu_e$ oscillation probabilities as a
  function of baseline and energy for \emph{neutrinos} (top left) and
  \emph{antineutrinos} (bottom left) with $\mdeltacp =0$. The figures on the right show the projection of the
  oscillation probability on the neutrino energy axis at a baseline of
  \SI{1300}{km} for $\mdeltacp =0$ (red),  $\mdeltacp =+\pi/2$ (green),
  and  $\mdeltacp =-\pi/2$ (blue) for neutrinos (top right) and
  antineutrinos (bottom right). The yellow curve is the $\nu_e$
  appearance solely from the ``solar term'' due to $\nu_{1}$ to 
$\nu_2$ mixing as given
  by Equation~\ref{eqn:papprox2}. }
\label{fig:oscnodes1a}
\end{figure}

\begin{figure}[!tb]
\centerline{
\includegraphics[width=0.5\textwidth]{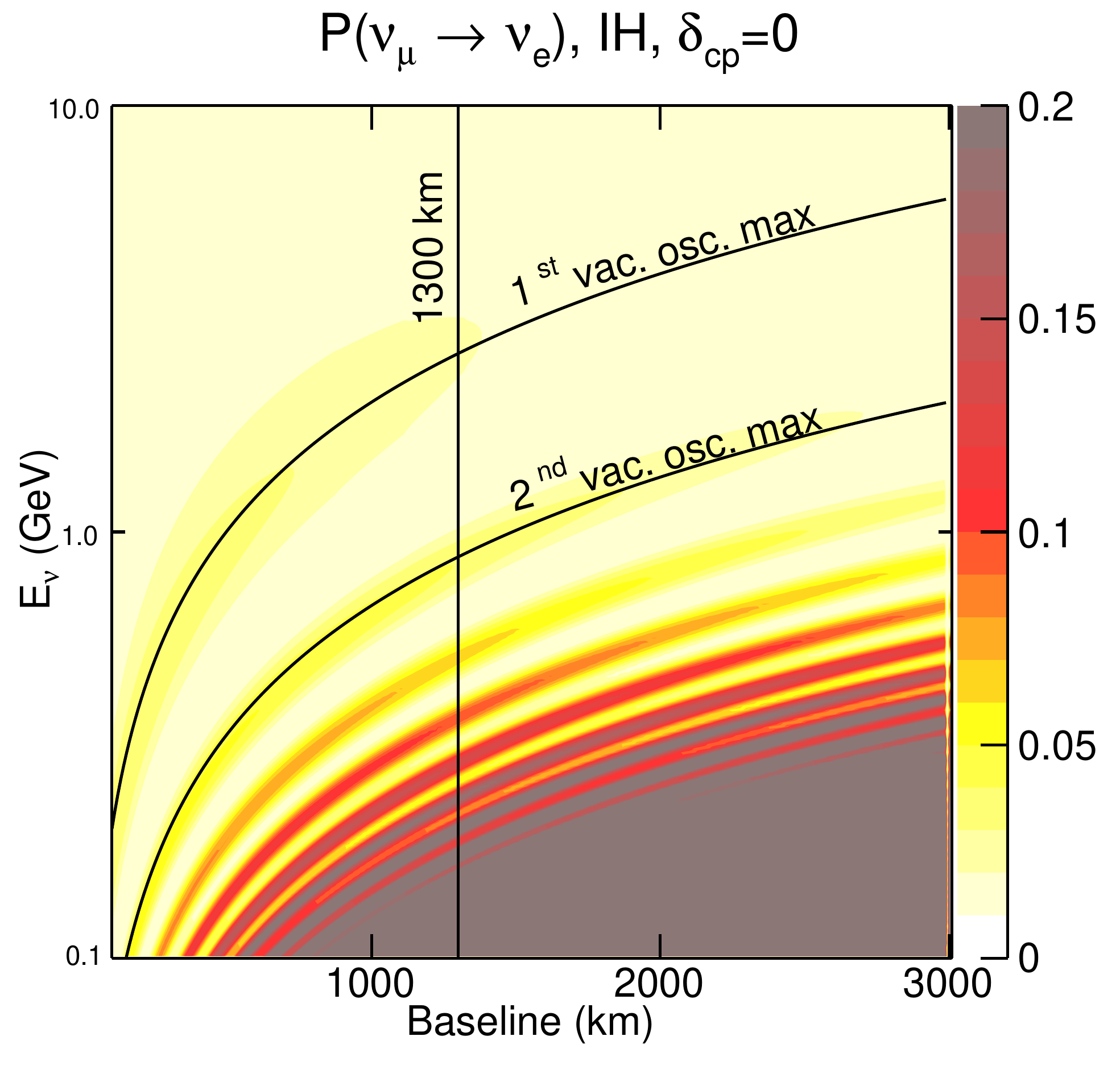}
\includegraphics[width=0.5\textwidth]{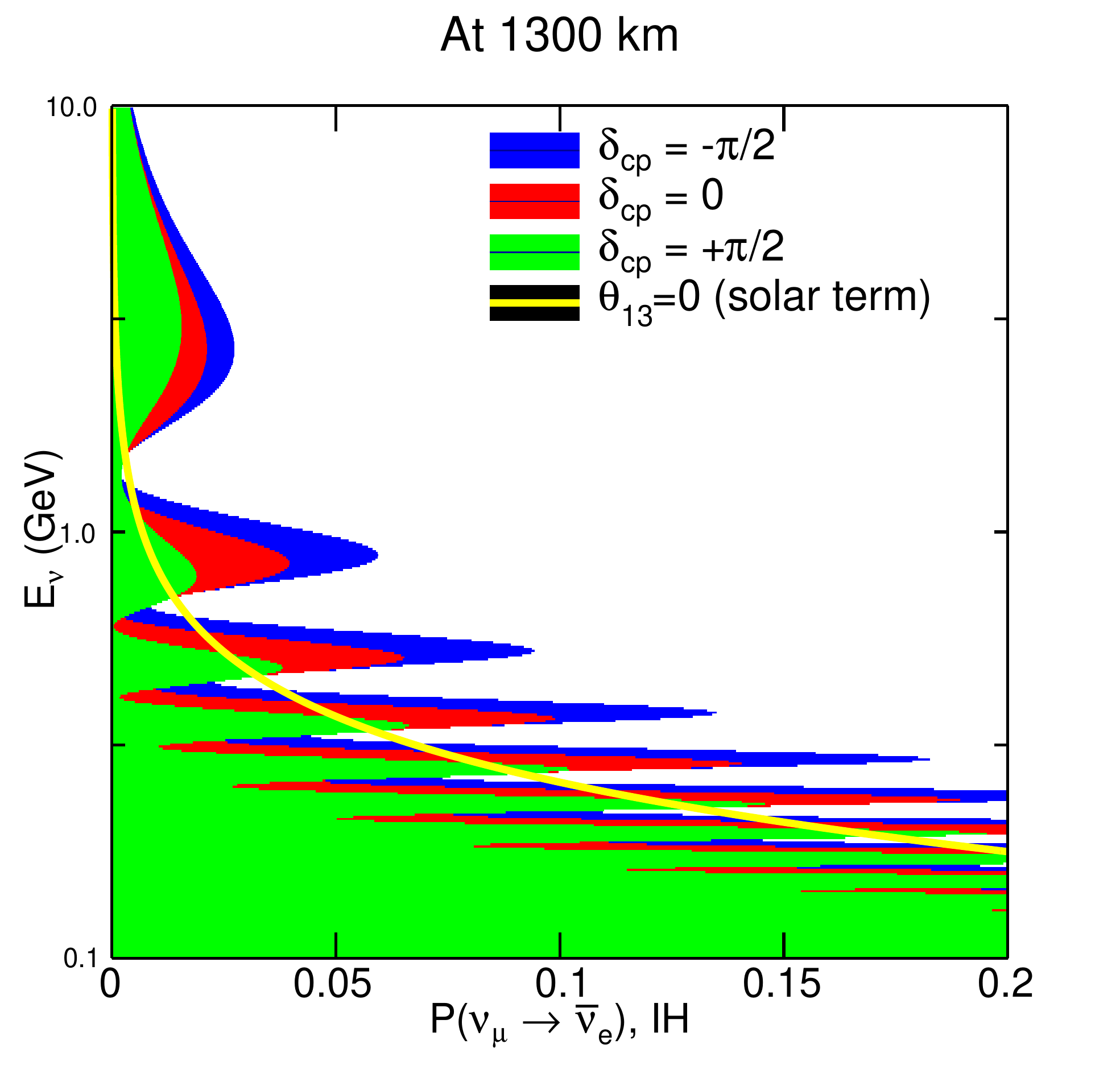}
}
\centerline{
\includegraphics[width=0.5\textwidth]{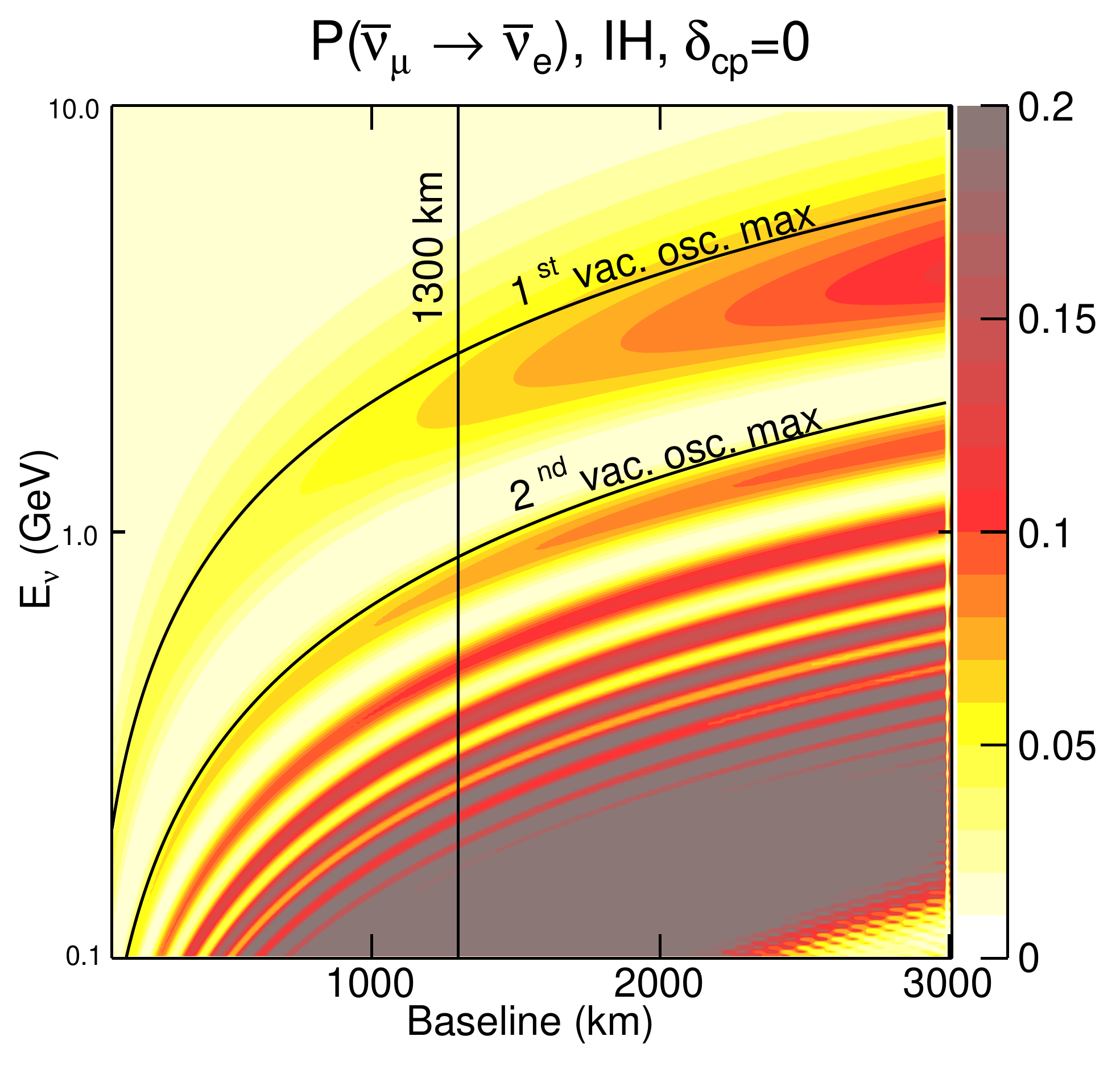}
\includegraphics[width=0.5\textwidth]{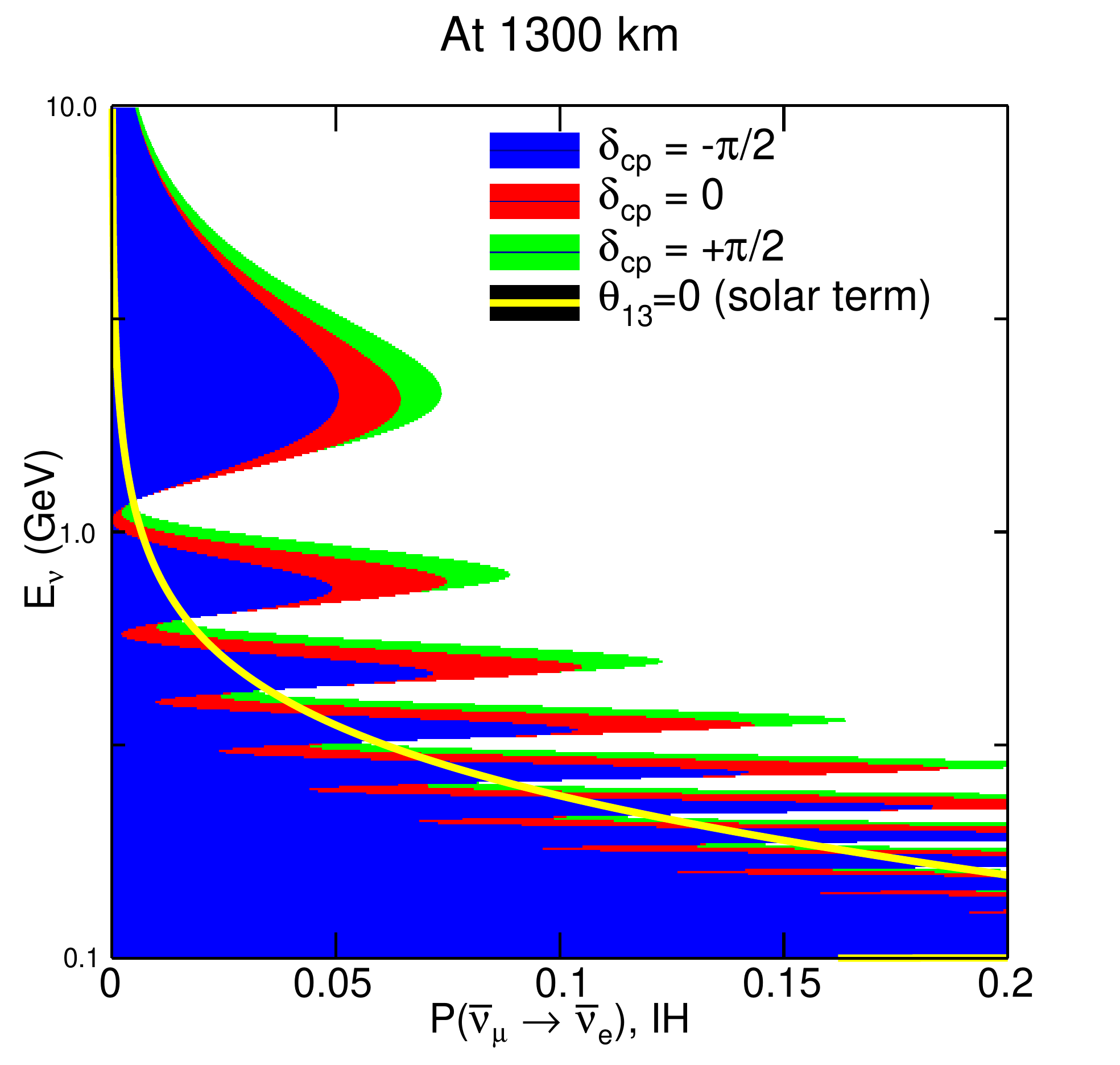}
}
\caption[$\nu$ oscillation probabilities versus $E$ 
  and baseline for various  \deltacp, inverted MH]{Neutrino oscillation
  probabilities as a function of energy and baseline, for
  different values of \deltacp, \emph{inverted
  hierarchy}. The oscillograms on the left
  show the $\nu_\mu \rightarrow \nu_e$ oscillation probabilities as a
  function of baseline and energy for \emph{neutrinos} (top left) and
  \emph{antineutrinos} (bottom left) with $\mdeltacp =0$. The figures on the right show the projection of the
  oscillation probability on the neutrino energy axis at a baseline of
  \SI{1300}{km}  for  $\mdeltacp =0$ (red),  $\mdeltacp =+\pi/2$ (green),
  and  $\mdeltacp =-\pi/2$ (blue) for neutrinos (top right) and
  antineutrinos (bottom right).The yellow curve is the $\nu_e$
  appearance solely from the ``solar term'' due to $\nu_{1}$ to 
$\nu_2$ mixing as given
  by Equation~\ref{eqn:papprox2}. }
\label{fig:oscnodes1b}
\end{figure}

The dependences on $E_\nu$  of the oscillation probability for the LBNE baseline of 
$L=$1,300~km are plotted on the right in Figures~\ref{fig:oscnodes1a} 
and~\ref{fig:oscnodes1b}. The colored curves demonstrate the
variation in the $\nu_e$ appearance probability as a function of $E_\nu$, 
for three different values of \deltacp. 

The variation in the $\nu_\mu \rightarrow
\nu_e$ oscillation probabilities with the value of \deltacp
indicates that it is experimentally possible to measure the value of
\deltacp at a fixed baseline using only the observed shape of the
$\nu_\mu \rightarrow \nu_e$ {\em or} the 
$\overline{\nu}_\mu \rightarrow \overline{\nu}_e$
appearance signal measured over an energy range that encompasses at
least one full oscillation interval. A measurement of the value of
$\mdeltacp \neq 0 \ {\rm or} \ \pi$, assuming that neutrino mixing follows the three-flavor model, would imply CP violation.  
The CP asymmetry,
$\mathcal{A}_{CP}$, is defined as 
\begin{equation}
\label{eqn:cp-asymm}
 \mathcal{A}_{CP} = \frac{P(\nu_\mu \rightarrow \nu_e) -
  P(\overline{\nu}_\mu \rightarrow \overline{\nu}_e)}{P(\nu_\mu \rightarrow
  \nu_e) + P(\overline{\nu}_\mu \rightarrow \overline{\nu}_e)}.
\end{equation}
In the three-flavor model the asymmetry can be approximated to leading
order in $\Delta m_{21}^2$ as~\cite{Marciano:2006uc}:
\begin{equation}
\mathcal{A}_{CP} \sim \frac{\cos \theta_{23} \sin 2 \theta_{12}
  {\sin \mdeltacp}}{\sin \theta_{23} \sin \theta_{13}}
\left(\frac{\Delta m^2_{21} L}{ 4 E_{\nu}}\right) + {\rm matter
  \ effects}
\label{eqn:cpasym}
\end{equation}
Regardless of the 
value obtained for \deltacp, it is clear that the explicit observation
of an asymmetry between $P(\nu_l \rightarrow \nu_{l'})$ and
$P(\overline{\nu}_l \rightarrow \overline{\nu}_{l'})$ 
is sought to directly demonstrate the
leptonic CP violation effect that a value of \deltacp different 
from zero or $\pi$ implies.
For long-baseline experiments such as LBNE, where the neutrino beam propagates through 
the Earth's mantle, the leptonic CP-violation effects must be disentangled from 
the matter effects.



\subsection{Probing the Neutrino Mass Hierarchy via the Matter Effect}
\label{ssec:matter:effects:mass:hier}

The asymmetry induced by matter effects as neutrinos pass through  
the Earth arises from the change in sign of the
factors proportional to $\Delta m^2_{31}$ (namely $A$, $\Delta$ and
$\alpha$; Equations~\ref{eqn:papprox0} to~\ref{eqn:papprox4}) in going from the normal to the inverted
neutrino mass hierarchy.  This sign change provides a means for determining the
currently unknown mass hierarchy.  
The oscillation probabilities given
in these approximate equations 
for $\nu_\mu
\rightarrow \nu_e$ as a function of baseline in kilometers and energy in GeV
are calculated numerically with an exact formalism~\cite{nuosc} and
shown in the oscillograms of Figure~\ref{fig:oscnodes1a} and
\ref{fig:oscnodes1b} for $\mdeltacp =0$, for normal and inverted
hierarchies, respectively. The oscillograms include the matter effect,
assuming an Earth density and electron fraction described
by~\cite{PREM}.  These values are taken as a constant average over paths
through regions of the Earth with continuous density change.
Any baseline long enough to pass through a discontinuity is split into
three or more segments each of constant average density and electron fraction.
The solid
black curves in the oscillograms indicate the location of the first
and second oscillation maxima as given by Equation~\ref{eqn:nodes},
assuming oscillations in a vacuum; matter effects will change the
neutrino energy values at which the mixing between the $\nu_{1}$ and
$\nu_{3}$ mass states
is maximal.  
\begin{introbox}{ The 
    significant impact of the matter effect on the $\nu_\mu
    \rightarrow \nu_e$ and $\overline{\nu}_\mu \rightarrow
    \overline{\nu}_e$ oscillation probabilities at longer baselines (Figures~\ref{fig:oscnodes1a} and~\ref{fig:oscnodes1b}) 
    implies that $\nu_e$ appearance measurements over long distances
    through the Earth provide a powerful probe into the neutrino
    mass hierarchy question: is $m_1 > m_3$ or vice-versa?
}
\end{introbox}
The dependence of the matter effect on the mass hierarchy is
illustrated in the oscillograms plotted on the left hand side of
Figures~\ref{fig:oscnodes1a} and~\ref{fig:oscnodes1b}, and can be
characterized as follows:
%

%
\begin{itemize}
\item For normal hierarchy, $P(\nu_\mu \rightarrow \nu_e)$ is enhanced
  and $P(\overline{\nu}_\mu \rightarrow \overline{\nu}_e)$ is suppressed. The
  effect increases with baseline at a fixed $L/E$.
\item For inverted hierarchy, $P(\nu_\mu \rightarrow \nu_e)$ is
  suppressed and $P(\overline{\nu}_\mu \rightarrow \overline{\nu}_e)$ is
  enhanced. The effect increases with baseline at a fixed $L/E$.
\item The matter effect has the largest impact on the probability
  amplitude at the first oscillation maximum.
\item The matter effect introduces a phase shift in the oscillation
  pattern, shifting it to a lower energy for a
  given baseline when the hierarchy changes from normal to
  inverted. The shift is approximately $-100$ MeV.
\end{itemize}

\subsection{Disentangling CP-Violating and Matter Effects}

\begin{figure}[!htb]
\centerline{
\includegraphics[width=0.49\textwidth]{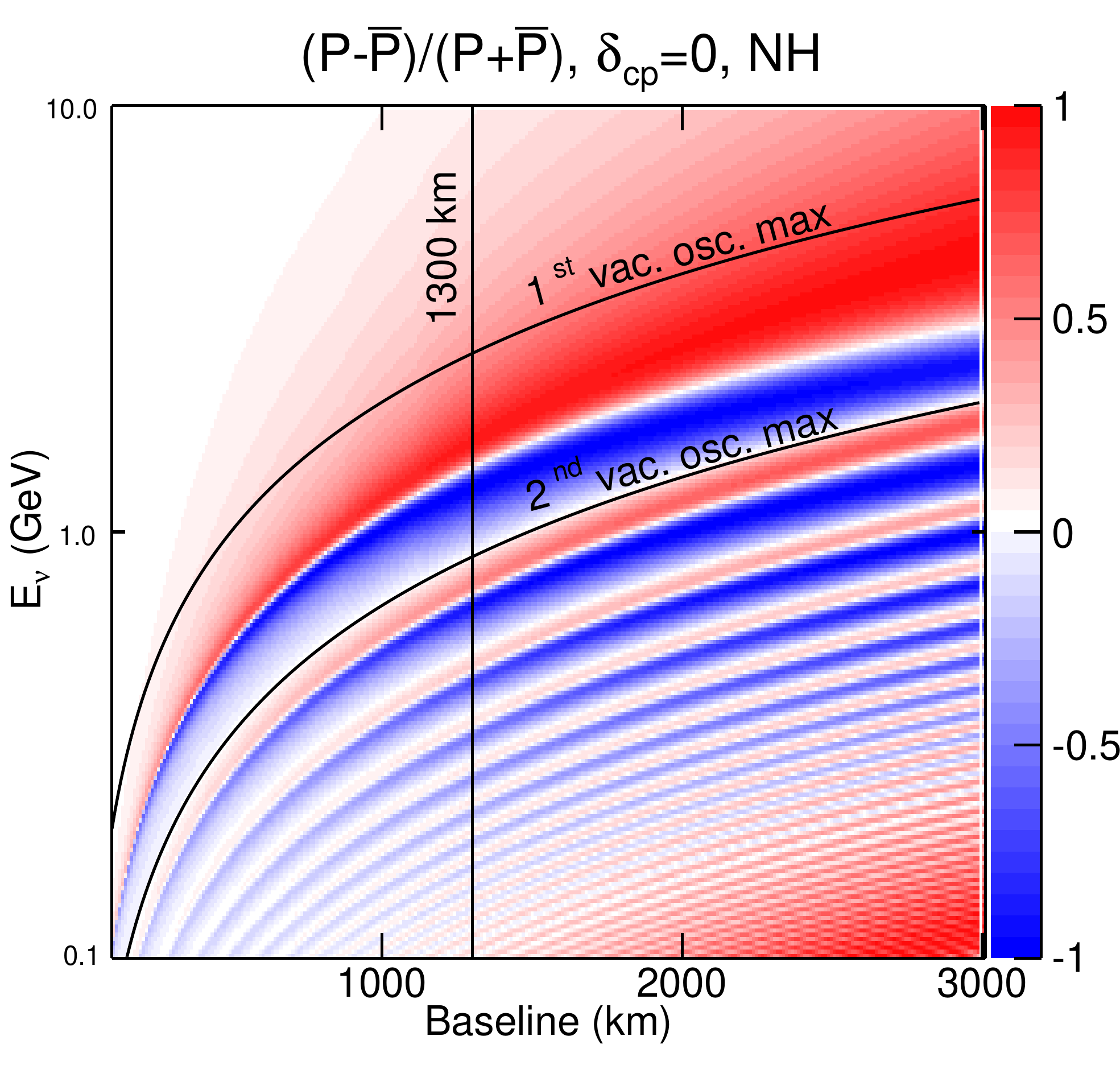}
\includegraphics[width=0.49\textwidth]{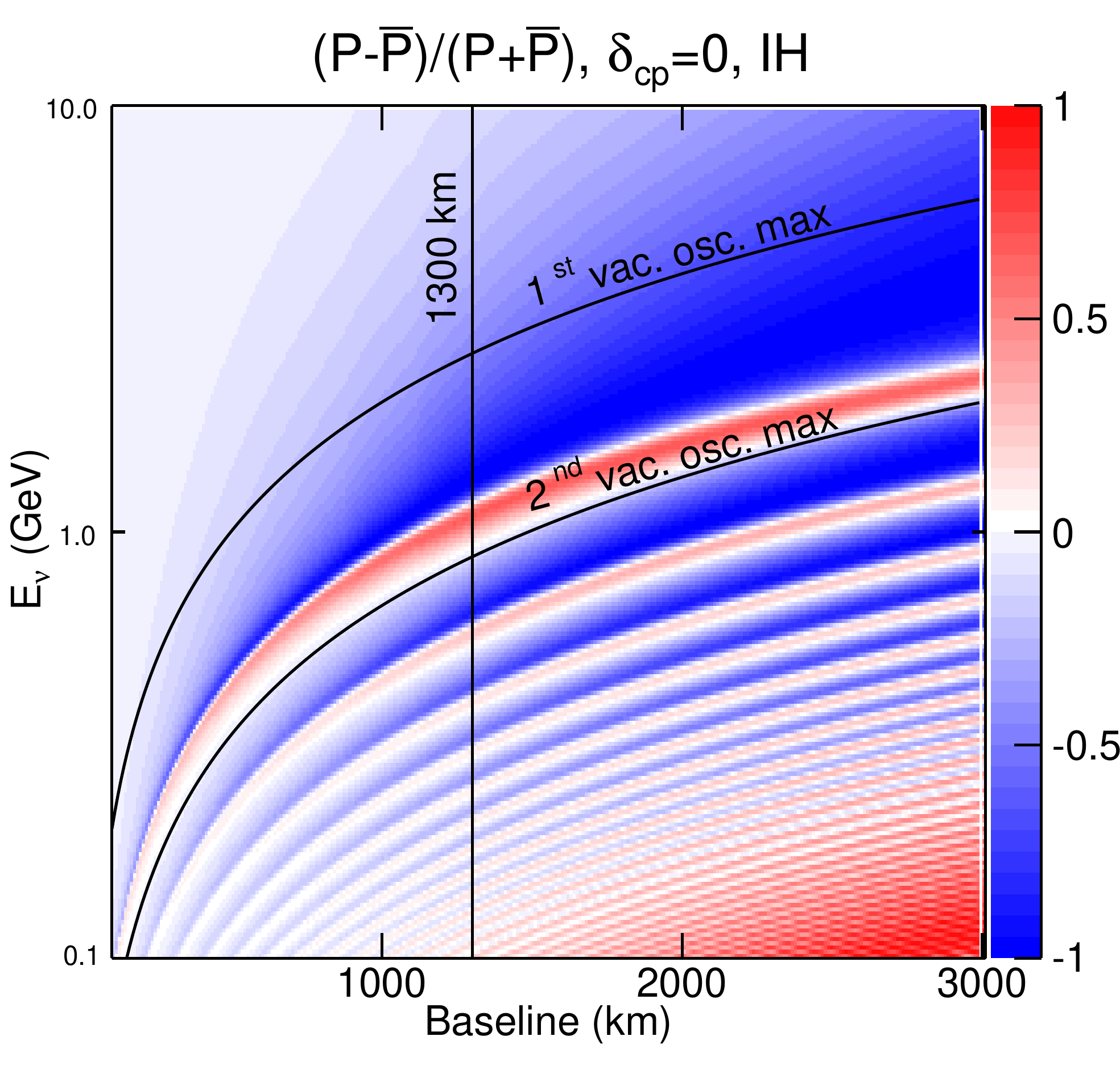}
}

\centerline{
\includegraphics[width=0.49\textwidth]{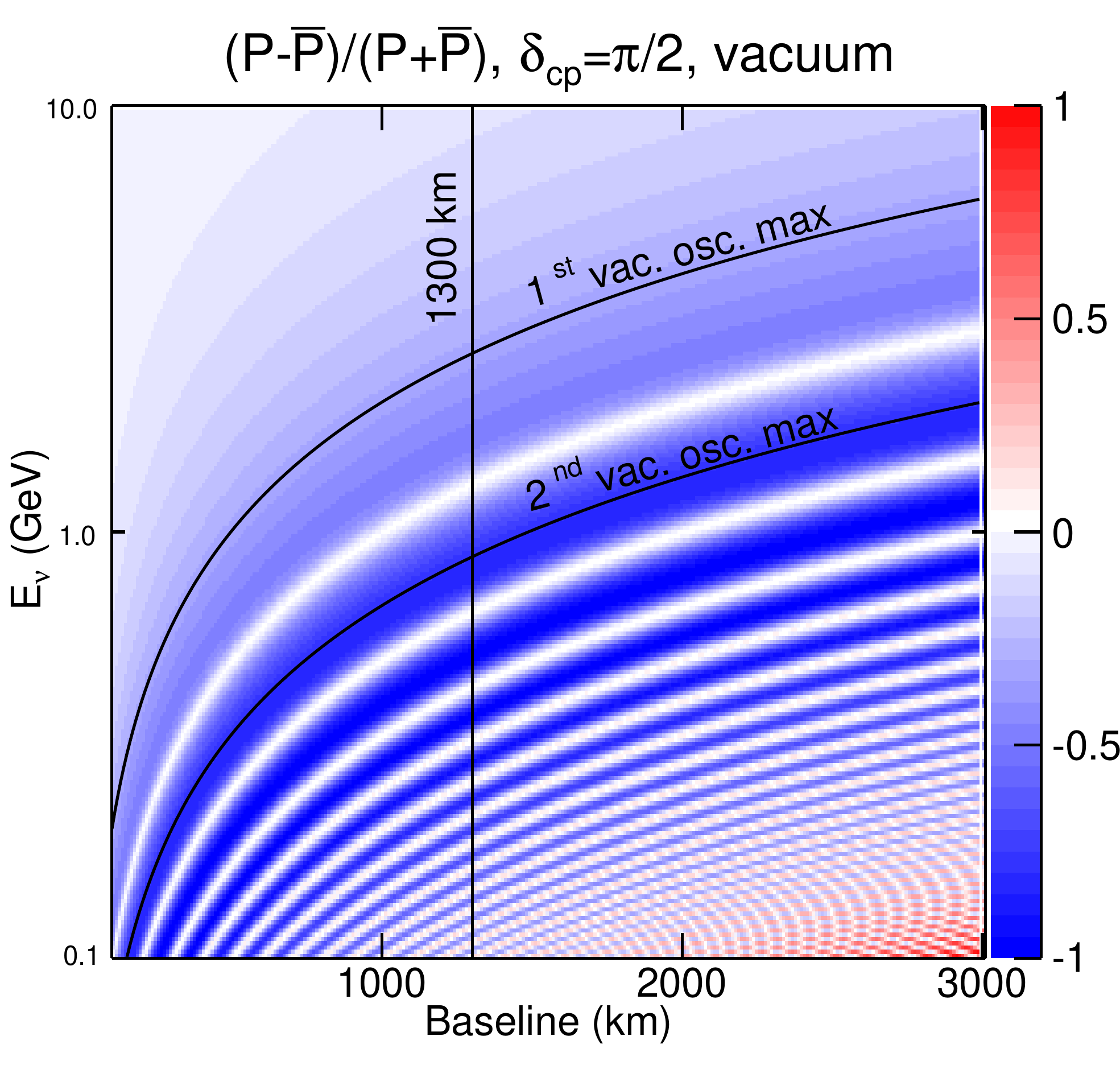}
\includegraphics[width=0.49\textwidth]{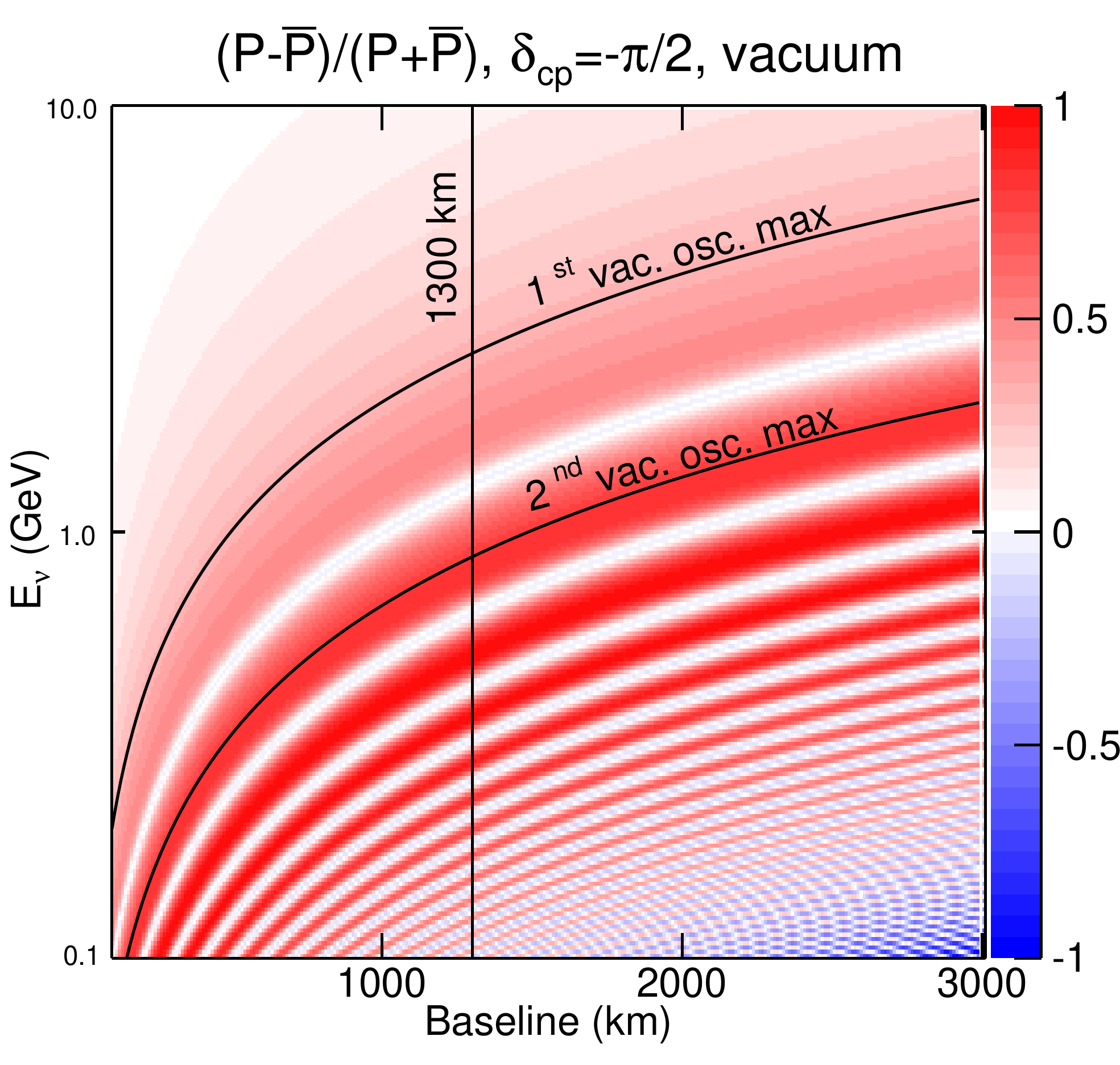}
}

\caption[CP asymmetry versus baseline]{The $\nu$/$\overline\nu$ oscillation probability asymmetries as a function of baseline. The top two
  figures 
show the asymmetry induced by the matter effect only for
  normal (top left) and inverted (top right) hierarchies. 
The bottom
  figures 
show the asymmetry induced through the CP-violating phase
\deltacp in vacuum, for $\mdeltacp = +\pi/2$ (bottom left) and 
$\mdeltacp = -\pi/2$ (bottom right) }
\label{fig:oscnodes2}
\end{figure}

In Figure~\ref{fig:oscnodes2}, the asymmetries induced by matter and
maximal CP violation (at $\mdeltacp =\pm \pi/2$) are shown separately
as 2D oscillograms in baseline and neutrino energy.  The matter effect induces an asymmetry in $P(\nu_l \rightarrow
\nu_{l'})$ and $P(\overline{\nu}_l \rightarrow \overline{\nu}_{l'})$ that
adds to the CP asymmetry. 
At longer baselines ($>1000\,$km), the
matter asymmetry in the energy region of the first oscillation node is
driven primarily by the change in the $\nu_e$ appearance amplitude. At
shorter baselines ($\mathcal{O}(100\,\rm{km})$) the asymmetry is driven by the
phase shift. The dependence of the asymmetry on baseline and energy, where the
oscillation probabilities peak and the appearance signals are largest, can
be approximated as follows:
\begin{eqnarray}
\mathcal{A}_{cp} &\propto& L/E ,\\ 
\mathcal{A}_{matter} &\propto& L \times E .
\label{eqn:asyms}
\end{eqnarray}
The phenomenology of $\nu_\mu \rightarrow \nu_e$ oscillations
described in Section~\ref{sec:oscil-cpv}
implies that the experimental sensitivity to CP
violation and the mass hierarchy from measurements of the total
asymmetry between $P(\nu_l \rightarrow \nu_{l'})$ and $P(\overline{\nu}_l
\rightarrow \overline{\nu}_{l'})$ requires the disambiguation of the
asymmetry induced by the matter effect and that induced by CP
violation. This is particularly true for experiments designed to
access mixing between the $\nu_{1}$ and $\nu_{3}$ mass states 
using neutrino beams of
$\mathcal{O}(1\,\rm{GeV})$. Such beams require baselines of
at least several hundred kilometers, at which the matter asymmetries are
significant.  The currently known values of the oscillation parameters
permit calculation of the magnitude of the matter asymmetry within an
uncertainty of $< 10\%$; only the sign of the asymmetry, which
depends on the sign of $\Delta m^2_{31}$, is unknown. Since the
magnitude of the matter asymmetry is known, baselines at which the
size of the matter asymmetry exceeds that of the maximal possible CP
asymmetry are required in order to separate the two effects. 

\begin{figure}[!htb]
\centerline{
\includegraphics[width=0.5\textwidth]{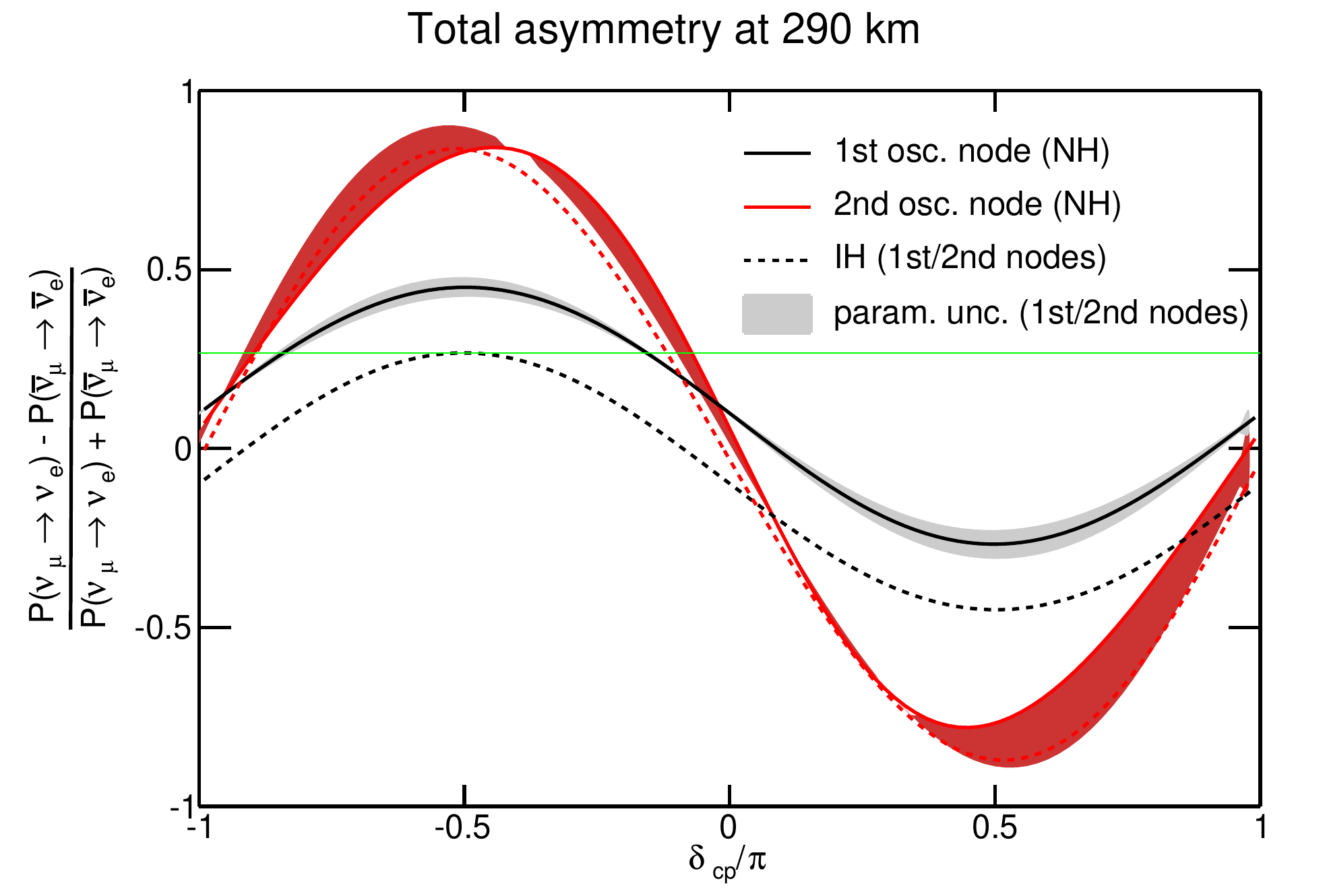}
\includegraphics[width=0.5\textwidth]{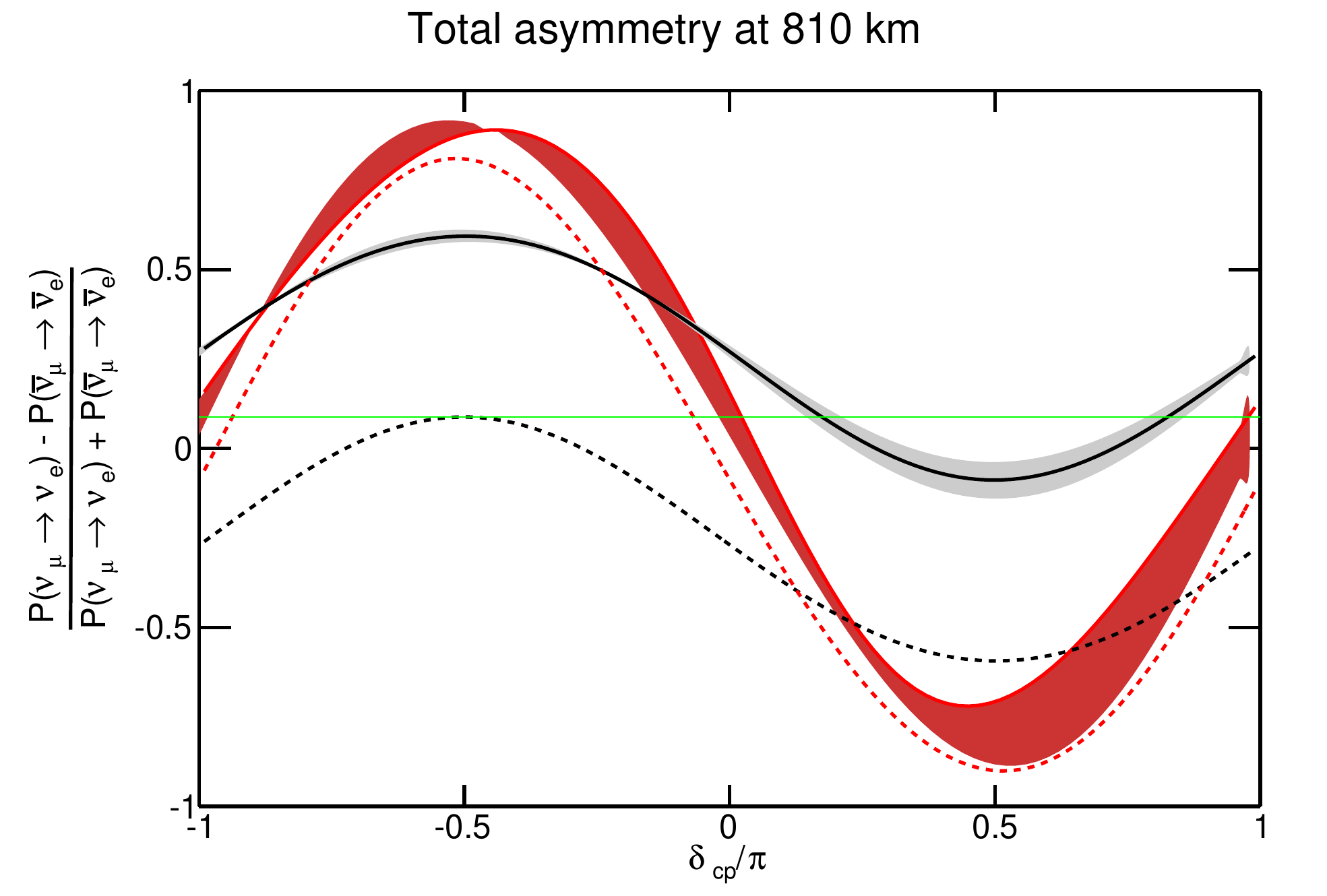}
}
\centerline{
\includegraphics[width=0.5\textwidth]{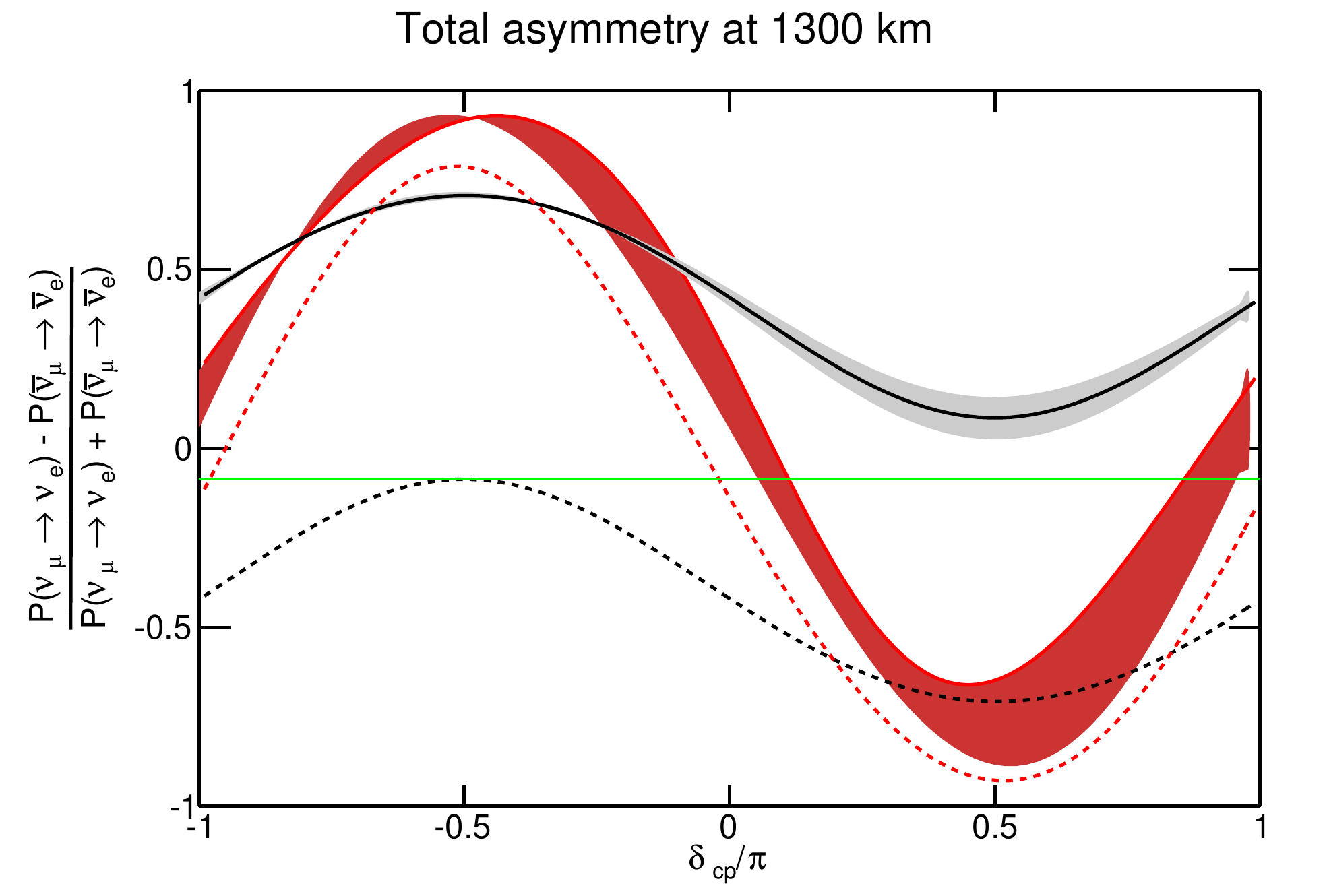}
\includegraphics[width=0.5\textwidth]{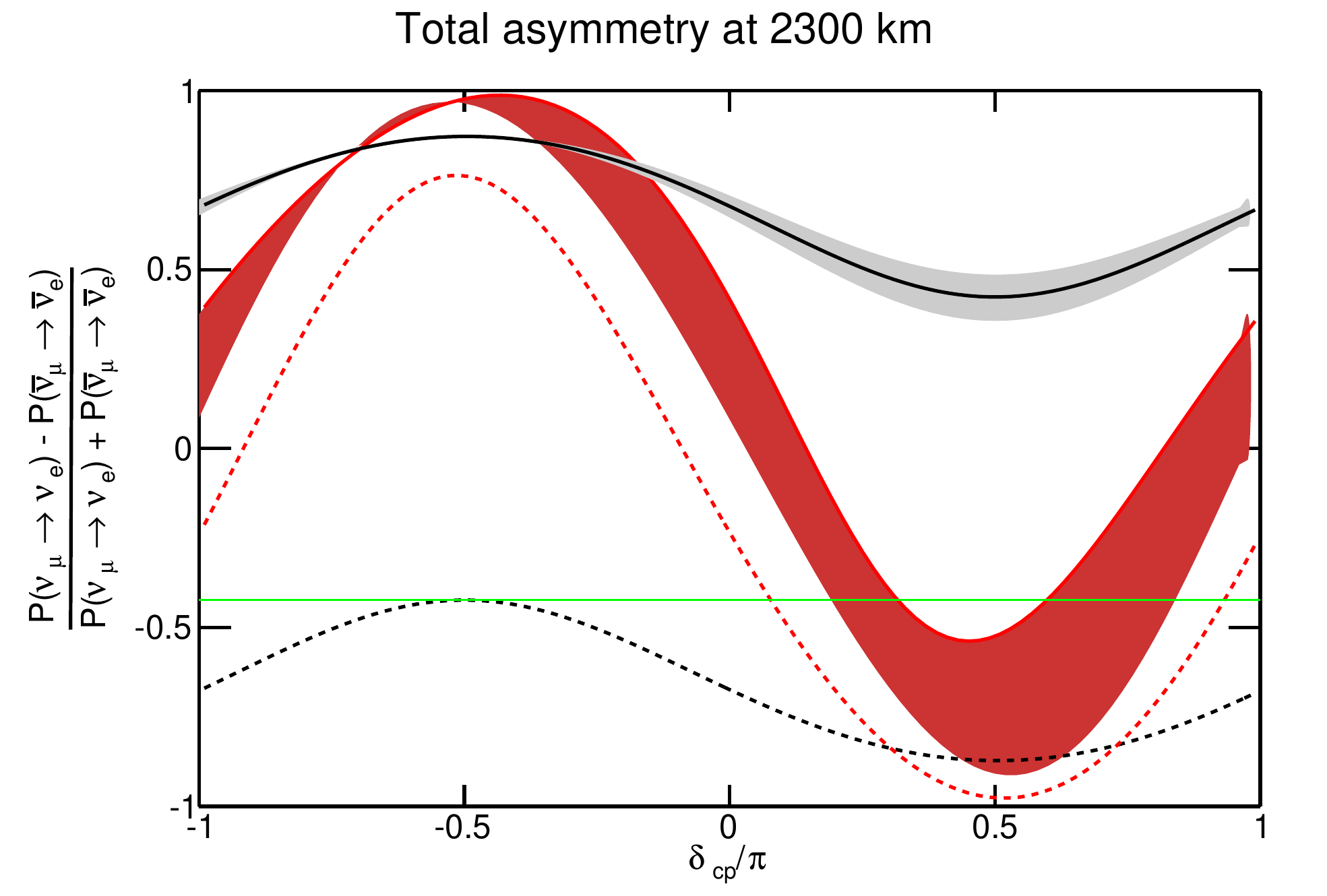}
}
\caption[$\nu/\overline{\nu}$ oscillation asymmetries versus \deltacp
  at the first two oscillation nodes]{The $\nu$/$\overline\nu$ oscillation probability 
asymmetries versus \deltacp at the first two oscillation
  nodes. Clockwise from top left: 290$\,$km, 810$\,$km, 2,300$\,$km
  and 1,300$\,$km. The solid/dashed black line is the total asymmetry at the
  first oscillation node for normal/inverted hierarchy. The red lines
  indicate the asymmetries at the second node. }
\label{fig:oscnodes4}
\end{figure}

Figure~\ref{fig:oscnodes4} illustrates the ambiguities that can arise
from the interference of the matter and CP asymmetries.  The plots
show the total asymmetry as a function of \deltacp at four
baseline values (clockwise from top left): 290$\,$km, 810$\,$km,
2,300$\,$km and 1,300$\,$km. The curves in black and red illustrate
the asymmetries at the first and second oscillation nodes,
respectively. The solid lines represent normal hierarchy, and the
dashed lines represent inverted hierarchy. 
The plots demonstrate that experimental
measurements of the asymmetry (Equation~\ref{eqn:cp-asymm}) at the
first oscillation node could yield ambiguous results for short
baselines if the hierarchy is unknown.  This occurs in regions of the
($L,E,\mdeltacp$) phase space where the matter and CP asymmetries
cancel partially or totally.  For example, the green lines in
Figure~\ref{fig:oscnodes4} indicate the asymmetry at the first node
for maximal CP violation ($\mdeltacp = \pi/2$) with an inverted
hierarchy. At a baseline of 290$\,$km, the measured asymmetry at
$\mdeltacp = \pi/2$ (inverted hierarchy) is degenerate with that
at $\mdeltacp \sim 0$ (normal hierarchy) at the first
node. Measurements of the asymmetry at different $L/E$ or at different
baselines can break the degeneracies (Equation~\ref{eqn:asyms}).  At
very long baselines, for which the matter asymmetry exceeds the
maximal CP asymmetry at the first oscillation node, there are no
degeneracies and the mass hierarchy and CP asymmetries can be resolved
within the same experiment. For the current best-fit values of the
oscillation parameters, the matter asymmetry exceeds the maximal
possible CP asymmetry at baselines of $\geq$ 1,200~km.
%

\subsection{Optimization of the Oscillation Baseline for CPV and Mass Hierarchy}
\begin{figure}[!htb]
\centerline{
\includegraphics[width=0.7\textwidth]{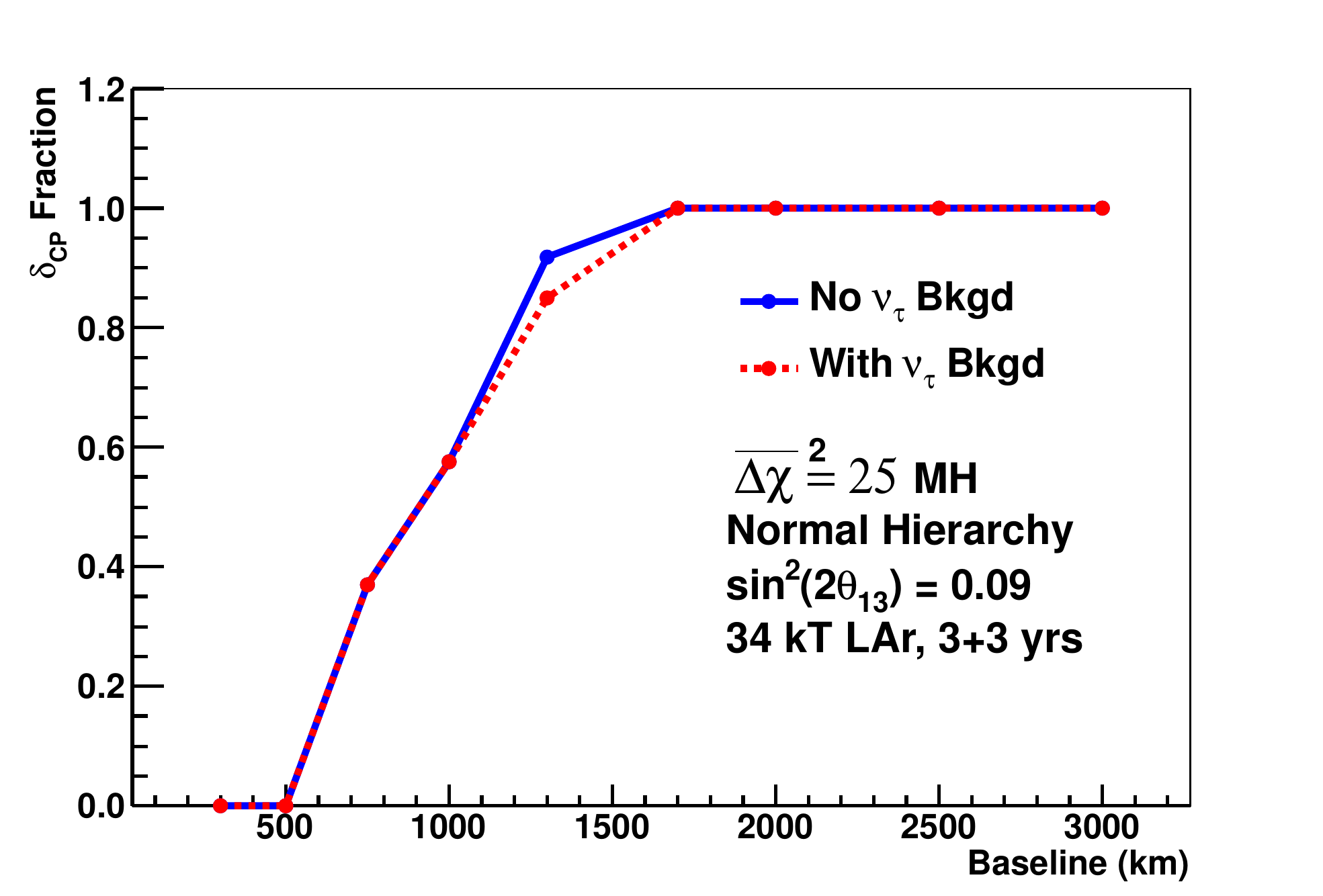}
}
\centerline{
\includegraphics[width=0.7\textwidth]{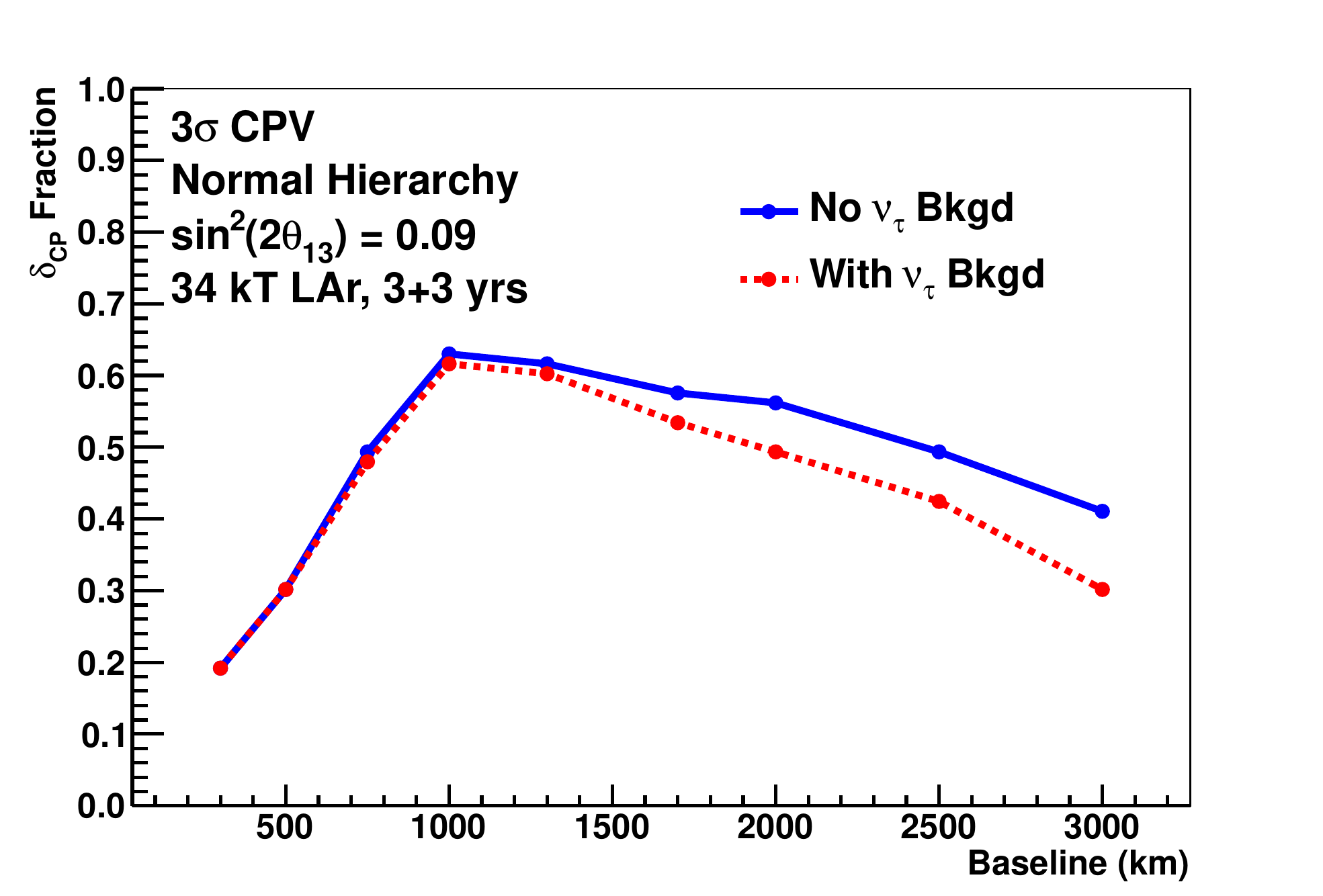}
}
\caption[Fraction of \deltacp values covered at a given
  significance for CPV
and MH vs baseline]{The fraction of \deltacp values 
 for which the mass hierarchy can be determined with an average
$|\overline{\Delta \chi^2}| = 25$  or greater as a function of baseline (top) and
the fraction of \deltacp values which CP violation can be
determined at the 3$\sigma$ level or greater as a function of
baseline (bottom). A NuMI based beam design with a \GeVadj{120} 
beam was optimized for each baseline. Projections assume
$\sin^22\theta_{13} = 0.09$ and a \ktadj{34} LArTPC as the far
detector~\cite{LBNEreconfigPWG}. An exposure of 3yrs+3yrs of
neutrino+antineutrino running with \MWadj{1.2} beam power is assumed.}
\label{fig:BLcpfrac}
\end{figure}


The simple arguments above suggest that a baseline $\geq$ 1,200~km is required to
search for CP violation and determine the mass hierarchy
simultaneously in a single long-baseline neutrino oscillation experiment. 
To understand the performance of a long-baseline experiment as a
function of baseline using realistic 
neutrino beamline designs,  a
study of the sensitivities to CP violation and the mass hierarchy as a
function of baseline was carried out using a neutrino
beamline design optimized individually
for each baseline. A \ktadj{34} LArTPC 
neutrino detector at the far site was assumed since it has a high 
$\nu_e$-identification efficiency that is flat over a large range of energies
(Chapter~\ref{nu-oscil-chap}). 
The beamline design was based on the NuMI beamline utilizing the
\GeVadj{120}, \MWadj{1.2}  proton beam from the Fermilab Main Injector and was fully simulated
using GEANT3~\cite{Brun:1987ma}. 
Varying the distance between the target
and the first horn allowed selection of a beam spectrum that covered the
first oscillation node and part of the second. The design incorporated
an evacuated decay pipe of 4-m diameter and a length that varied from
280 to \SI{580}{\meter}.  For baselines less than \SI{1000}{\meter}, the oscillation
occurs at neutrino energies where on-axis beams produce too little
flux. Therefore, off-axis beams --- which produce narrow-band,
low-energy neutrino fluxes --- were simulated for these baselines, with
the off-axis angle chosen to provide the most coverage of the first
oscillation node.  The results of this study~\cite{LBNEreconfigPWG} are summarized in
Figure~\ref{fig:BLcpfrac}.  The sensitivity to CP violation (bottom
plot) assumes that the mass hierarchy is unknown. An updated study with more detail is 
available~\cite{Bass:2013vcg}.
The baseline study indicates that with realistic experimental
conditions, 
baselines between 1,000 and 1,300~km are near optimal for
determination of CP violation. With baselines $>\,$1,500$\,$km, the
correct mass hierarchy could be determined with a probability greater
than 99\% for all values of \deltacp with a large LArTPC far
detector. However, at very long baselines,
in one of the neutrino beam polarities
($\nu$/$\overline{\nu}$ for inverted/normal hierarchy) the event rate
suppression due to 
the matter effect
becomes very
large,
making it difficult to observe an explicit CP-violation asymmetry.

\subsection{Physics from Precision Measurements of Neutrino Mixing}

Precision measurements of the neutrino mixing parameters in long-baseline oscillations not only reveal the neutrino mixing patterns in
greater detail, but also serve as probes of new physics that manifests as
perturbations in the oscillation patterns driven by three-flavor mixing.

The determination of whether there is maximal mixing between $\nu_{\mu}$
and $\nu_{\tau}$
 --- or a measurement of the deviation from maximal --- is of great interest 
theoretically~\cite{Luhn:2013lkn,Raidal:2004iw,Minakata:2004xt,Smirnov:2013uba,Harada:2013aja,Hu:2012eb}. Models of quark-lepton universality propose
that the quark and lepton mixing matrices (Equations \ref{eq:ckmmatrix} and 
\ref{eq:pmnsmatrix}, respectively) are given by
\begin{eqnarray}
U^{\rm CKM} & = & 1 + \epsilon_{\rm Cabbibo} {\rm~and} \\
U^{\rm PMNS} & = & T + \epsilon_{\rm Cabbibo}, \\ \nonumber
\end{eqnarray}
where $T$ is determined by Majorana
physics~\cite{ramond_isoups} and $\epsilon_{\rm Cabbibo}$ refers to
small terms driven by the Cabbibo weak mixing angle
($\theta_C = \theta_{12}^{\rm CKM}$).  
In such models $\theta_{23} \sim \pi/4 +
\Delta \theta$, where $\Delta \theta$ is of order the Cabbibo angle,
$\theta_C$, and $\theta_{13} \sim \theta_C/\sqrt{2}$.  It is therefore
important to determine  experimentally both the value of $\sin ^2
\theta_{23}$ and 
the octant of $\theta_{23}$ if $\theta_{23}\ne45^{\circ}$.   

\begin{introbox}
Studying $\nu_{\mu}$ disappearance probes $\sin^ 2 2
\theta_{23}$ and |$\Delta m^{2}_{32}$| with very high
precision. Disappearance measurements can therefore determine whether
$\nu_{\mu}$-$\nu_{\tau}$ 
mixing is maximal or near maximal such that $\sin^ 2 2 \theta_{23} =
1$, but they cannot resolve the octant of $\theta_{23}$ if $\nu_\mu$-$\nu_\tau$
mixing is less
than maximal. Combining the $\nu_{\mu}$ disappearance signal with the
$\nu_e$ appearance signal can help determine the $\theta_{23}$ octant
and constrain some of the theoretical models of quark-lepton
universality.
\end{introbox}

Direct unitarity tests, in which the individual components of the PMNS
matrix are measured separately, are challenging due to limited
experimentally available oscillation
channels~\cite{Antusch:2006vwa,Qian:2013ora}.
Application of the ``proof by contradiction'' principle offers another way to perform the
unitarity tests. 
In these tests, the mixing angles are extracted from the
data by assuming unitarity in the standard three-flavor framework.  
If measurements of the
same mixing angle by two different processes are
inconsistent, 
then the standard three-flavor framework is insufficient and new physics beyond this framework is
required.  Observation of unitarity violation will constrain the phase
space of possible new physics.  In particular, the precision
measurement of $\sin^22\theta_{13}$ provides the most promising
unitarity test~\cite{Qian:2013ora} for the PMNS matrix.  It is
important to note that several theoretical models of new physics, such
as the existence of sterile neutrinos or nonstandard interactions, 
could lead to apparent deviations of the $\sin^22\theta_{13}$ value
measured in $\nu_e$ appearance experiments from that measured in
reactor ($\overline{\nu}_e$ disappearance) experiments. 

Precision measurements of  $\nu_\mu$ and $\overline{\nu}_\mu$
survival over long baselines could reveal nonstandard physics 
driven by new interactions in matter. Examples of some of 
these effects and the experimental signatures in long-baseline oscillations 
are discussed in Chapter~\ref{nu-oscil-chap}.

In addition, experiments with long enough baselines and sufficient
neutrino flux at $E_{\nu}> 3$ GeV, coupled with high-resolution tracking
detectors, as in the LBNE design, can also probe $\nu_\mu \rightarrow
\nu_\tau$ appearance with higher precision than is currently possible
using $\nu_\tau$ charged-current interactions. The 
combination of $\nu_\mu \rightarrow \nu_\mu$, $\nu_\mu
\rightarrow \nu_e$, and $\nu_\mu \rightarrow \nu_\tau$ can
ultimately over-constrain the three-flavor model of neutrino 
oscillations both in
neutrino and antineutrino modes. 



\subsection{Oscillation Physics with Atmospheric Neutrinos}

\begin{figure}[!htb]
\centerline{
\includegraphics[width=0.5\textwidth]{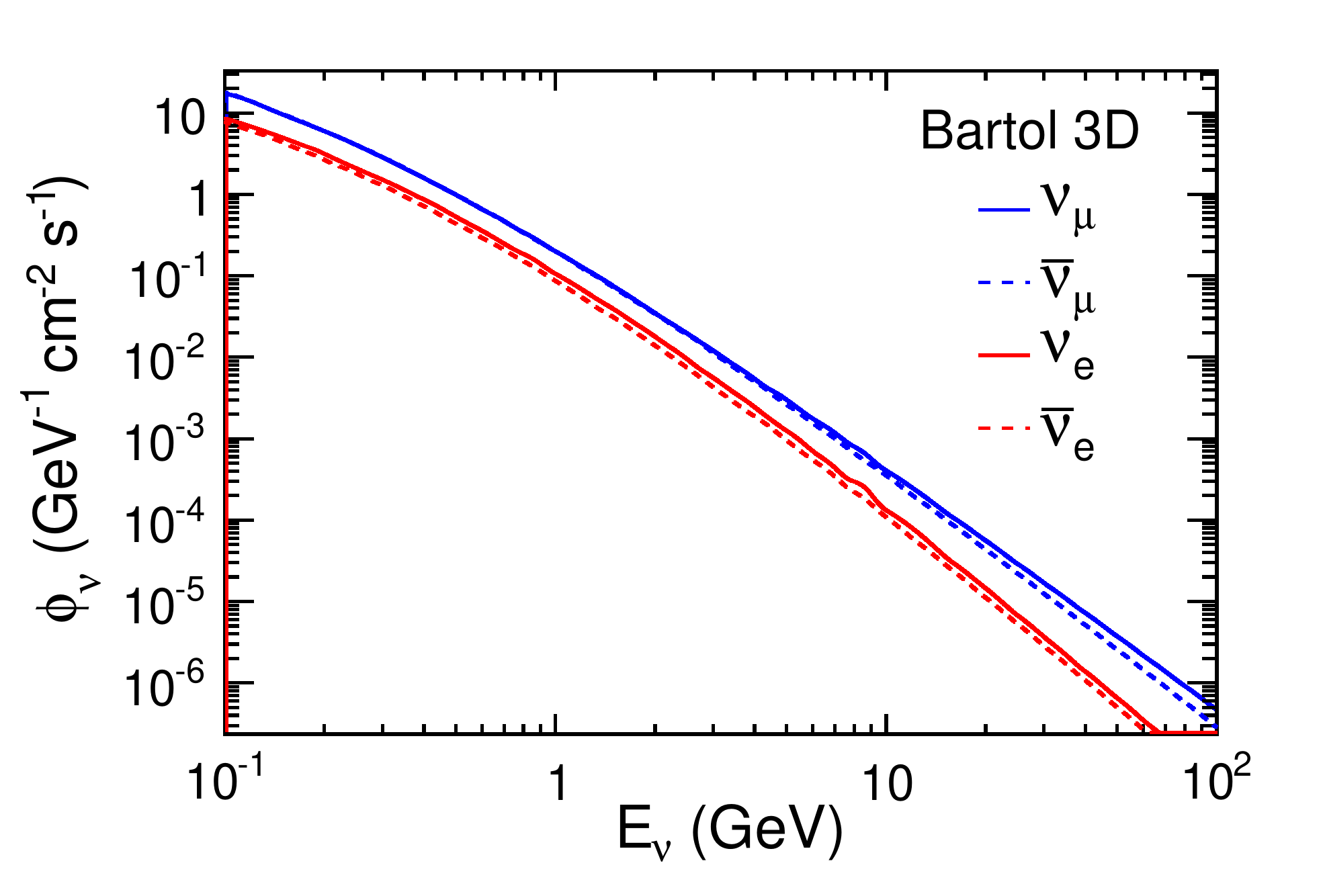}
\includegraphics[width=0.5\textwidth]{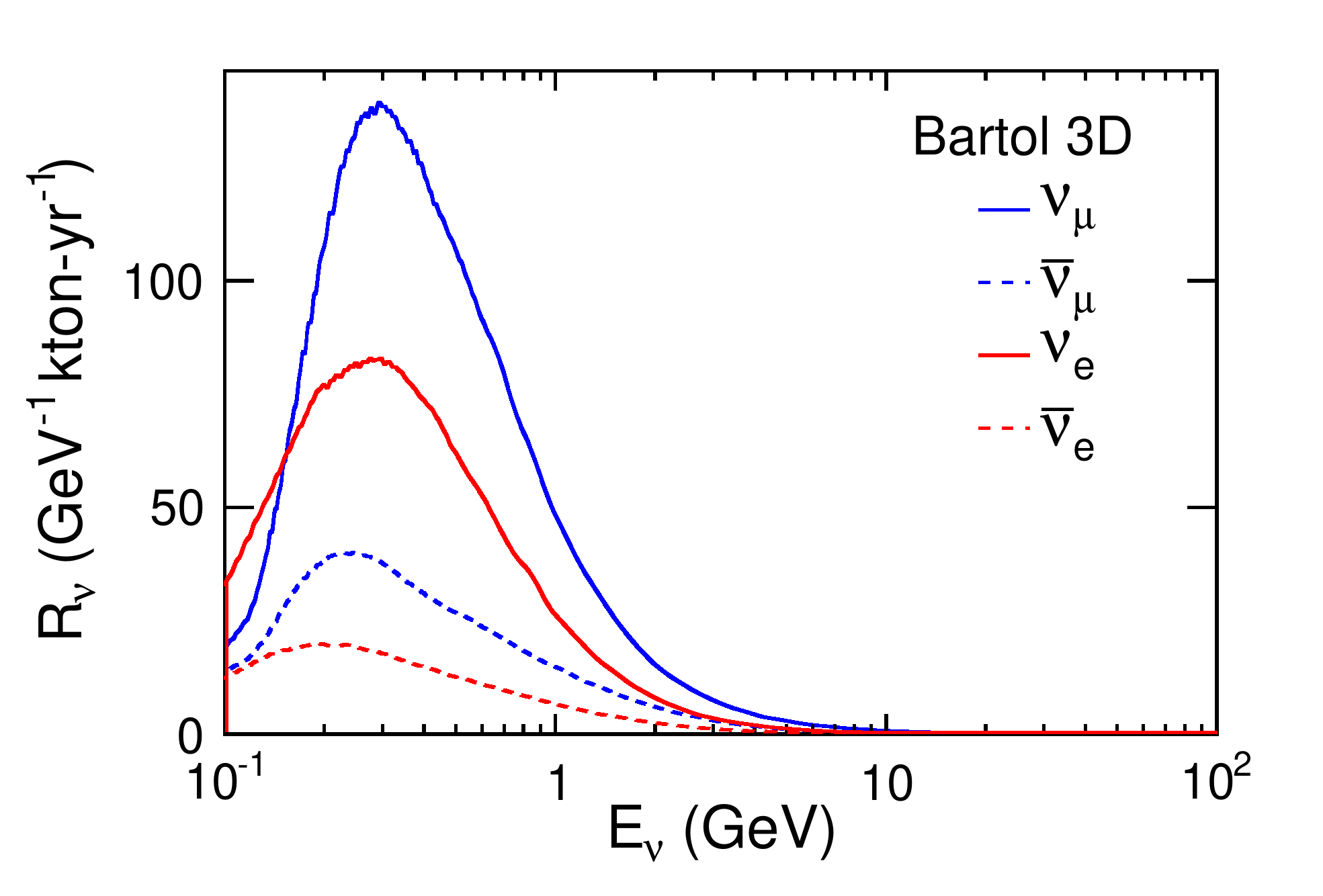}
}
\caption[Atmospheric neutrino flux and spectrum]{The atmospheric
  neutrino flux in neutrinos per second per square centimeter as a function of
  neutrino energy for different flavors (left). The
  atmospheric neutrino spectrum per GeV per kt per year for the
  different species (right). }
\label{fig:atmflux}
\end{figure}

\begin{figure}[!htb]
\centering\includegraphics[width=\textwidth]{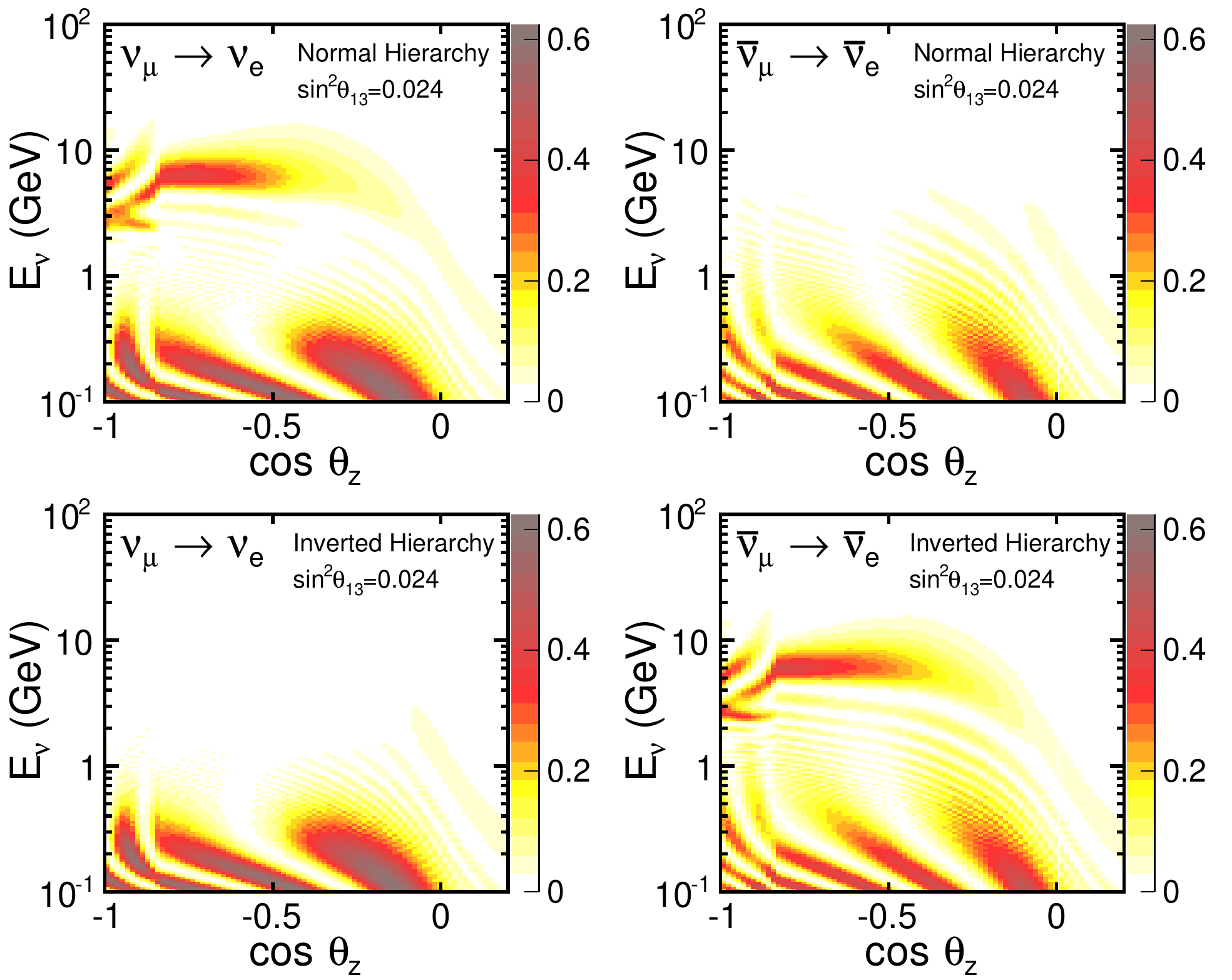}
\caption[Probabilities of atmospheric $\nu_\mu \rightarrow \nu_e$ oscillations versus zenith angle]{The probabilities of atmospheric $\nu_\mu \rightarrow \nu_e$ (left) and
$\overline{\nu}_\mu \rightarrow \overline{\nu}_e$ (right) oscillations for normal (top) and
inverted (bottom) hierarchies as a function of zenith angle.}
\label{fig:oscatm}
\end{figure}

Atmospheric neutrinos are unique among sources used to study
oscillations; the flux contains neutrinos and antineutrinos of all
flavors, matter effects play a significant role, both $\Delta m^2$
values contribute to the oscillation patterns, and the oscillation
phenomenology occurs over several orders of magnitude in both energy
(Figure~\ref{fig:atmflux}) and path length.  These characteristics
make atmospheric neutrinos ideal for the study of oscillations and
provide a laboratory suitable to search for exotic phenomena for which
the dependence of the flavor-transition and survival probabilities on
energy and path length can be defined. The probabilities of atmospheric
$\nu_\mu \rightarrow \nu_e$ and $\overline{\nu}_\mu \rightarrow
\overline{\nu}_e$ oscillations for normal and inverted hierarchies are
shown as a function of zenith angle in Figure~\ref{fig:oscatm}.

Even with dedicated long-baseline experiments exploring the large mass
splitting ($\Delta m^2_{32}$) for nearly a decade, atmospheric data
continue to contribute substantially to our understanding of the
neutrino sector. Broadly speaking:
\begin{itemize}
\item The data demonstrate {\em complementarity} with beam results via
  two- and three-flavor fits and the measurement of a 
  $\nu_\tau$ appearance
  signal consistent with expectations.
\item The data serve to increase measurement {\em precision} through
  global fits, given that the sensitivity of atmospheric neutrinos to
  the mass hierarchy is largely independent of \deltacp and the
  octant of $\theta_{23}$.
\item {\em New physics} searches with atmospheric neutrinos have
  placed limits on CPT violation, nonstandard interactions,
  mass-varying neutrinos and Lorentz-invariance violation.
\end{itemize}

Atmospheric neutrinos can continue to play these roles in the LBNE era
given LBNE's deep-underground far detector. In particular,
complementarity will be vital in a future where, worldwide, the number
of high-precision, long-baseline beam/detector facilities is
small. The physics potential of a large underground liquid argon
detector for measuring atmospheric neutrinos is discussed in
Section~\ref{atmnu}.

\section{Nucleon Decay Physics Motivated by Grand Unified Theories}
\label{ss:bnonconservation}
\begin{introbox}{
Searches for proton decay, bound-neutron decay and similar processes such as
di-nucleon decay and neutron-antineutron oscillations test the apparent
but unexplained conservation law of baryon number. These decays are
already known to be rare based on decades of prior searches, all of
which have produced negative results. If measurable event rates or even a
single-candidate event were to be found, it would be sensible to presume that 
they occurred via unknown virtual processes based on physics beyond
the Standard Model. The impact of demonstrating the existence of a
baryon-number-violating process would be profound.
}
\end{introbox}

\subsection{Theoretical Motivation from GUTs} 

The class of theories known as Grand Unified Theories (GUTs) make
predictions about both baryon number violation and  proton lifetime 
that may be within reach of the full-scope LBNE experiment. 
The theoretical motivation for the study of proton decay
has a long and distinguished
history~\cite{Pati:1973rp,Georgi:1974sy,Dimopoulos:1981dw} and has
been reviewed many
times~\cite{Langacker:1980js,deBoer:1994dg,Nath:2006ut}.
Early GUTs provided the original motivation for proton decay searches
in kiloton-scale detectors placed deep underground to limit backgrounds.  The
\ktadj{22.5} \superk\ experiment extended the search for
proton decay by more than an order of magnitude relative to the previous generation of experiments.
Contemporary reviews~\cite{Raby:2008pd,Senjanovic:2009kr,Li:2010dp}
discuss the strict limits already set by \superk\ and the
context of the proposed next generation of larger underground
experiments such as Hyper-Kamiokande and LBNE.

Although no evidence for proton decay has been detected, the lifetime
limits from the current generation of experiments already constrain
the construction of many contemporary GUT 
models. In some cases, these lifetime limits are approaching the upper limits allowed
by GUT models.
This situation points naturally toward continuing the search with new, larger detectors.
These searches are motivated by a range of scientific issues:
\begin{itemize}
\item Conservation laws arise from underlying symmetries in Nature~\cite{Noether:1918zz}.  
  Conservation of baryon number is therefore
  unexplained since it corresponds to no known long-range force or
  symmetry.
\item Baryon number non-conservation has cosmological consequences,
  such as a role in inflation and the matter-antimatter asymmetry of the
  Universe.
\item Proton decay is predicted at some level by almost all GUTs. 
\item Some GUTs can accommodate
  neutrinos with nonzero mass and characteristics consistent with
  experimental observations.
\item GUTs incorporate other previously unexplained features of
  the Standard Model such as the relationship between quark and lepton
  electric charges. 
\item The unification scale is suggested both experimentally and
  theoretically by the apparent convergence of the running coupling
  constants of the Standard Model. The unification scale is in excess of \SI{e15}{GeV}.
\item The unification scale is not accessible by any accelerator
  experiment; it can only be probed by virtual processes such as with
  proton decay.
\item GUTs usually predict the relative branching
  fractions of different nucleon decay modes. Testing these predictions
  would, however, require a sizeable sample of proton decay events.
\item The dominant proton decay mode of a GUT is often sufficient to roughly
  identify the likely characteristics of the GUT,
  such as gauge mediation or the involvement of supersymmetry.
\end{itemize}

\begin{introbox}
  The observation of even a single unambiguous proton decay event
  would corroborate the idea of unification and the signature
  of the decay would give strong guidance as to the nature of the
  underlying theory. 
\end{introbox}

\subsection{Proton Decay Modes} 

\tikzset{
particle/.style={draw=blue, postaction={decorate},
    decoration={markings,mark=at position .6 with {\arrow[blue]{triangle 45}}}},
noarrow/.style={draw=blue},
boson/.style={draw=blue,dashed},
}

\begin{figure}[!htb]
  \begin{center}
\begin{tikzpicture}
  \coordinate (cross);
  \coordinate[above left=of cross, label={[yshift=-6mm]$d$}](d);
  \coordinate[below left=of cross, label={[yshift=1mm]$u$}](u);  
  \coordinate[above right=of cross](tau);
  \coordinate[below right=of cross](tee);
  \coordinate[right=of tau, label=below:$\overline{\nu}_\tau$] (vtau);
  \coordinate[right=of tee, label=above:$\overline{s}$] (sbar);
  \coordinate[below=of u, label=above:$u$](u2);
  \coordinate[below=of sbar, label=above:$u$](u3);

  \draw[particle] (d) -- (cross);
  \draw[particle] (u) -- (cross);
  \draw[noarrow] (cross) -- node[label=above left:$\widetilde{\tau}$] {} (tau);
  \draw[noarrow] (cross) -- node[label=below left:$\widetilde{t}$] {} (tee);
  \draw[boson] (tau) -- node[label=right:$\widetilde{W}$] {} (tee);
  \draw[particle] (vtau) -- (tau);
  \draw[particle] (sbar) -- (tee);
  \draw[particle] (u2) -- (u3);

  \draw [decorate,decoration={brace,mirror,raise=3mm}] (d) -- (u2) node [black,midway,xshift=-8mm] {$P$};
  \draw [decorate,decoration={brace,raise=3mm}] (sbar) -- (u3) node [black,midway,xshift=8mm] {$K^+$};

\end{tikzpicture}%
\begin{tikzpicture}
  \coordinate (x);
  \coordinate[above left=of x, label={[yshift=-5mm]$u$}](u);
  \coordinate[below left=of x, label={[yshift=1mm]$u$}](u2);  
  \coordinate[right=of x](x2);  
  \coordinate[above right=of x2, label={[yshift=-6mm]$e^+$}](eplus);  
  \coordinate[below right=of x2, label={[yshift=1mm]$\overline{d}$}](dbar);
  \coordinate[below=of u2,label=above:$d$](d);  
  \coordinate[below=of dbar, label=above:$d$](d2);  

  \draw[particle] (u) -- (x);
  \draw[particle] (u2) -- (x);
  \draw[boson] (x) -- node[label=above:$X$] {} (x2);
  \draw[particle] (eplus) -- (x2);
  \draw[particle] (dbar) -- (x2);
  \draw[particle] (d) -- (d2);

  \draw [decorate,decoration={brace,mirror,raise=3mm}] (u) -- (d) node [black,midway,xshift=-8mm] {$P$};
  \draw [decorate,decoration={brace,raise=3mm}] (dbar) -- (d2) node [black,midway,xshift=8mm] {$\pi^0$};

\end{tikzpicture}
  \end{center}
\caption[Proton decay modes from SUSY and gauge-mediation models]{Feynman diagrams for proton decay modes from
supersymmetric GUT, $p^+ \rightarrow K^+ \overline{\nu}$  (left) and gauge-mediation GUT models, $p^+ \rightarrow e^+ \pi^0$ (right).}
\label{pdk_feyn}
\end{figure}
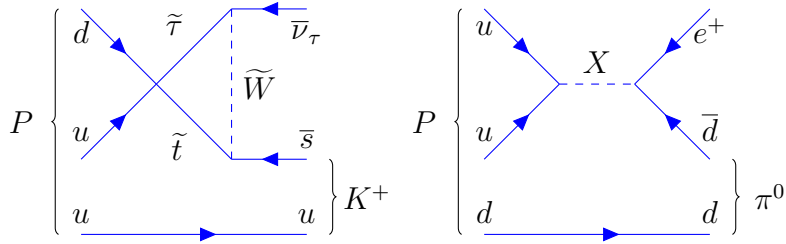
From the body of literature, two decay modes (shown in Figure~\ref{pdk_feyn})
emerge that dominate the LBNE experimental design. The more
well-known of the two, the decay mode of $p \rightarrow e^+ \pi^0$,
arises from gauge mediation.  It is often predicted to have the higher
branching fraction and is also demonstrably the more straightforward
experimental signature for a water Cherenkov detector. In this mode,
the total mass of the proton is converted into the electromagnetic
shower energy of the positron and two photons from $\pi^0$ decay,
with a net momentum vector near zero. 

The second key mode is $p \rightarrow K^+ \overline{\nu}$. 
This mode is dominant
in most supersymmetric GUTs, many of which also favor additional
modes involving kaons in the final state. This decay mode with a
charged kaon is uniquely interesting; since stopping kaons have a higher ionization density than other particles, a LArTPC could detect it with extremely high efficiency, as described in Chapter~\ref{pdk-chap}. 
In addition, many final states of $K^+$ decay would be fully
reconstructable in a LArTPC.

There are many other allowed modes of proton or bound neutron into
antilepton plus meson decay that conserve $B-L$\footnote{In these
  models, the quantum number $B-L$ is expected to be conserved even
  though $B$ and $L$ are not individually conserved.}, but none of
these will influence the design of a next-generation experiment. The
most stringent limits, besides those on $p \rightarrow e^+ \pi^0$,
include the lifetime limits on $p \rightarrow \mu^+ \pi^0$ and $p
\rightarrow e^+ \eta$, both of which are greater than \num{4e33} 
years~\cite{Nishino:2012ipa}. Any experiment that will do well for $p \rightarrow e^+ \pi^0$
will also do well for these decay modes.  The decays $p \rightarrow
\overline\nu \pi^+$ or $n \rightarrow \overline\nu \pi^0$ may have
large theoretically predicted branching fractions, but they are
experimentally difficult due to the sizeable backgrounds from
atmospheric-neutrino interactions. The decay $p \rightarrow \mu^+ K^0$
can be detected relatively efficiently by either water Cherenkov or
LArTPC detectors.

A number of other possible modes exist, such as those that conserve
$B+L$, that violate only baryon number, or that decay into only
leptons. These possibilities are less well-motivated theoretically, as
they do not appear in a wide range of models, and are therefore not
considered here.

\begin{figure}[!htb]
\centering
\includegraphics[width=0.9\textwidth]{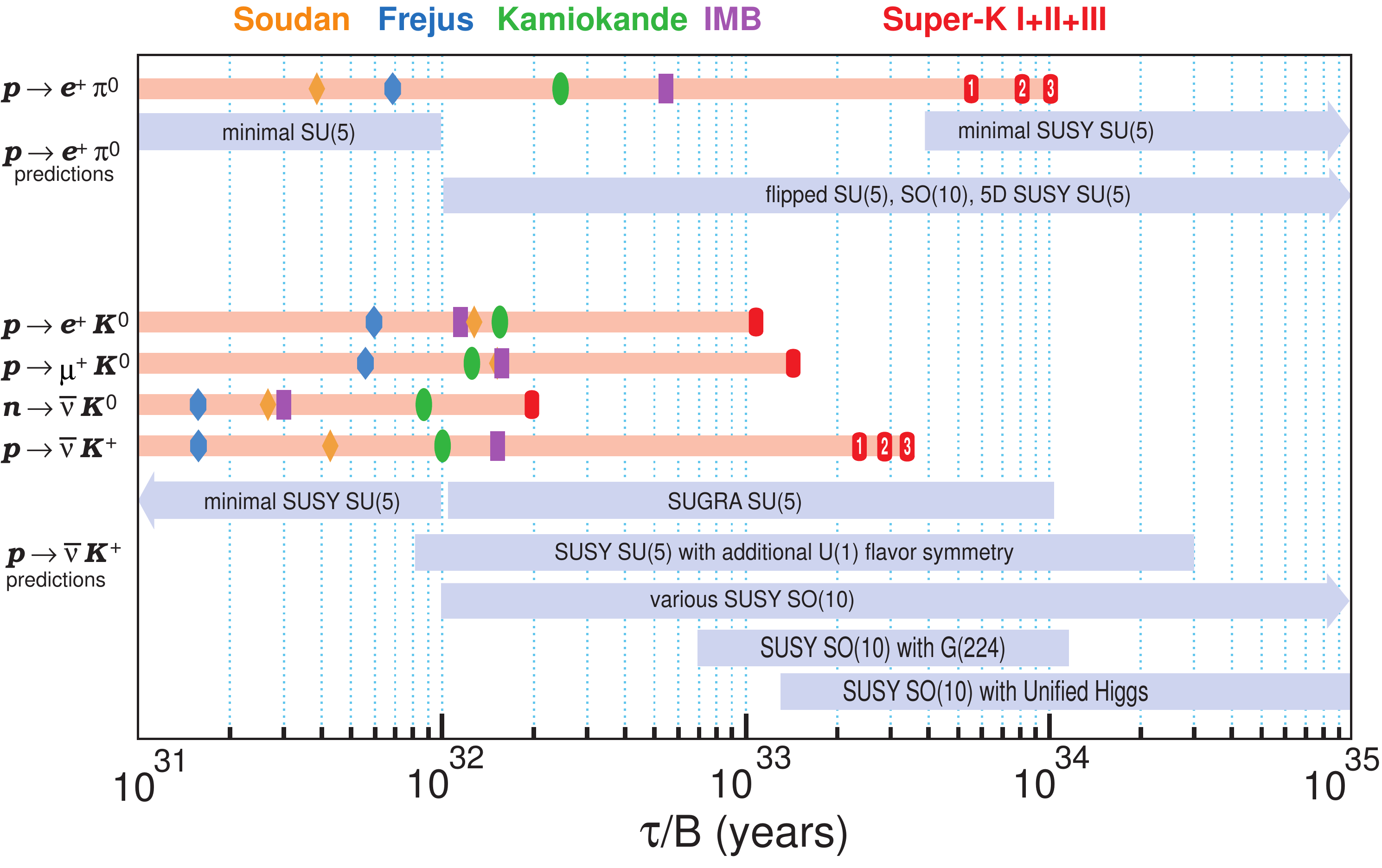}
\caption[Proton decay lifetime limits  compared to lifetime ranges
  predicted by GUTs]{Proton decay lifetime limits~\cite{Beringer:1900zz,Nishino:2012ipa} compared to lifetime ranges
  predicted by Grand Unified Theories. The upper section is for
  $p \rightarrow e^+ \pi^0$, most commonly caused by gauge mediation.
  The lower section is for SUSY-motivated models, which commonly
  predict decay modes with kaons in the final state. The
  marker symbols indicate published experimental limits, as indicated
  by the sequence and colors on top of the figure.}
\label{PDK-limits-theory}
\end{figure}

Figure~\ref{PDK-limits-theory} shows a comparison of experimental
limits, dominated by recent results from \superk\ to the
ranges of lifetimes predicted by an assortment of GUTs. At this time,
the theory literature does not attempt to precisely predict lifetimes,
concentrating instead on suggesting the dominant decay modes and
relative branching ratios. The uncertainty in the lifetime
predictions comes from details of the theory, such as masses and
coupling constants of unknown heavy particles, as well as poorly known
details of matrix elements for quarks within the nucleon.

It is apparent from Figure~\ref{PDK-limits-theory} that a continued
search for proton decay is by no means assured of obtaining a positive
result.  With that caveat, an experiment with sensitivity to proton lifetimes
between $10^{33}$ and $10^{35}$ years is searching in the right territory over
virtually all GUTs; even if no proton decay is detected,
stringent lifetime limits will provide strong constraints on such
models.  Minimal SU(5) was ruled out by the early work of IMB and
Kamiokande and minimal SUSY~SU(5) is considered to be ruled out by \superk.
In most cases, another order of magnitude in improved limits will not rule out
specific models but will constrain their allowed parameters;
this could allow identification of models which must be fine-tuned
in order to accommodate the data, and are thus less favored.

As Chapter~\ref{pdk-chap} will show, the
performance and scalability of the LArTPC technology opens up nucleon
decay channels that are not as readily accessible in existing and
proposed water Cherenkov detectors, providing LBNE with a unique and
compelling opportunity for discovery.

\clearpage
\section{Supernova-Neutrino Physics and Astrophysics}
\label{ss:snphysics}

For over half a century, researchers have been grappling to understand
the physics of the neutrino-driven core-collapse supernova.
The interest in observing the core-collapse supernova explosion
mechanism comes from the key role supernovae of this
type have played in the history of the Universe. Without taking supernova
feedback into account, for example, modern simulations of galaxy
formation cannot reproduce the structure of our galactic disk. More poetically,
the heavy elements that are the basis of life on Earth were synthesized
inside stars and ejected by supernova explosions. 
 
Neutrinos from a core-collapse supernova are emitted in a burst of a
few tens of seconds duration, with about half emitted in the first
second. They record  the information about the physical processes 
in the center of the explosion during the first
several seconds --- as it is happening. Energies are in the few-tens-of-MeV 
range and luminosity is
divided roughly equally between flavors.   
The basic model of core collapse was confirmed by the
observation of neutrino events from
SN1987A, a supernova in the Large Magellanic Cloud --- outside the 
Milky Way --- 50~kpc (kiloparsecs)
away. Nineteen events were detected in two water Cherenkov 
detectors~\cite{Bionta:1987qt,Hirata:1987hu} and additional events were reported in a scintillator 
detector~\cite{Alekseev:1987ej}.  The neutrino
signal from a core-collapse supernova in the Milky Way is expected to 
generate a high-statistics
signal from which LBNE could extract a wealth of 
information~\cite{Scholberg:2007nu,Dighe:2008dq}.
\begin{introbox}
The expected rate of core-collapse supernovae is two to three per
century in the Milky Way~\cite{Tammann:1994ev,Cappellaro:1999qy}.
 In a 20-year experimental
run, LBNE's probability of observing neutrinos from a core-collapse
supernova in the Milky Way is about 40\%.  The detection of thousands
of supernova-burst neutrinos from this event would dramatically expand
the science reach of the experiment, allowing observation of the
development of the explosion in the star's core and probing the
equation-of-state of matter at nuclear densities. In addition,
independent measurements of the neutrino mass hierarchy and the
$\theta_{13}$ mixing angle are possible, as well as additional
constraints on physics beyond the Standard Model.

Each of the topics that can be addressed by studying supernova-burst
neutrinos represent important outstanding
problems in modern physics, each worthy of a separate, dedicated
experiment, and the neutrino physics and astrophysics communities 
would receive payback simultaneously. 
The opportunity of targeting these topics in a single experiment is very
attractive, especially since it may come only at incremental cost to
the LBNE Project. 
\end{introbox}

The explosion mechanism is thought to have three distinct stages: the
collapse of the iron core, with the formation of the shock and its
breakout through the neutrinosphere; the accretion phase, in which the
shock temporarily stalls at a radius of about 200~km while the
material keeps raining in; and the cooling stage, in which the hot
proto-neutron star loses its energy and trapped lepton number, while
the re-energized shock expands to push out the rest of the star. Each
of these three stages is predicted to have a distinct signature in the
neutrino signal. Thus, it should be possible to directly observe, for
example, how long the shock is stalled.  More exotic features of the
collapse may be observable in the neutrino flux as well, such as
possible transitions to quark matter or to a black hole.  (An
observation in conjunction with a gravitational wave detection would
be especially interesting; e.g.~\cite{Pagliaroli:2009qy,Ott:2012jq}.)


Over the last two decades, neutrino flavor oscillations have been
firmly established in solar neutrinos and a variety of terrestrial
sources. 
The physics of
the oscillations in the supernova environment promises to be much
richer than in any of the cases measured to date, for a variety of
reasons:
\begin{itemize}
\item Neutrinos travel
through the changing profile of the explosion with stochastic density
fluctuations behind the expanding shock and, due to their coherent scattering off of 
each other, their flavor states are coupled. 
\item The oscillation patterns come out very differently for the normal and
inverted mass hierarchies. 
\item The expanding shock and turbulence leave a unique imprint in the neutrino signal. 
\item Additional information on oscillation parameters, free of
supernova model-dependence, will be available if matter effects due to the Earth 
can be observed in detectors at different locations around the
world~\cite{Mirizzi:2006xx,Choubey:2010up}.  
\item The observation of this
potentially copious source of neutrinos will also allow limits on
coupling to axions, large extra dimensions, and other exotic physics
(e.g.,~\cite{Raffelt:1997ac,Hannestad:2001jv}).
\item The oscillations of neutrinos and antineutrinos from a core-collapse supernova manifest very differently. In the neutrino channel, the oscillation features are in
general more pronounced, since the initial spectra of $\nu_e$ and
$\nu_\mu$ ($\nu_\tau$) are always significantly different.  It would be
extremely valuable to detect both neutrino and antineutrino
channels with high statistics.
\end{itemize}

Only about two dozen neutrinos were observed from SN1987A, which occurred in a nearby galaxy;
in contrast, the currently proposed next-generation detectors would register thousands
or tens of thousands of interactions from a core-collapse
supernova in our galaxy.
The type of observed interactions will depend on the detector
technology: a water-Cherenkov detector is primarily sensitive to $\overline{\nu}_e$'s,
whereas a LArTPC detector has excellent
sensitivity to $\nu_e$'s. 
In each case, the high event rate
implies that it should be possible to measure not only the
time-integrated spectra, but also their second-by-second
evolution. This is a key feature of the supernova-burst physics potential of the
planned LBNE experiment.

\begin{figure}[!htb]
\centering
\includegraphics[width=0.8\textwidth]{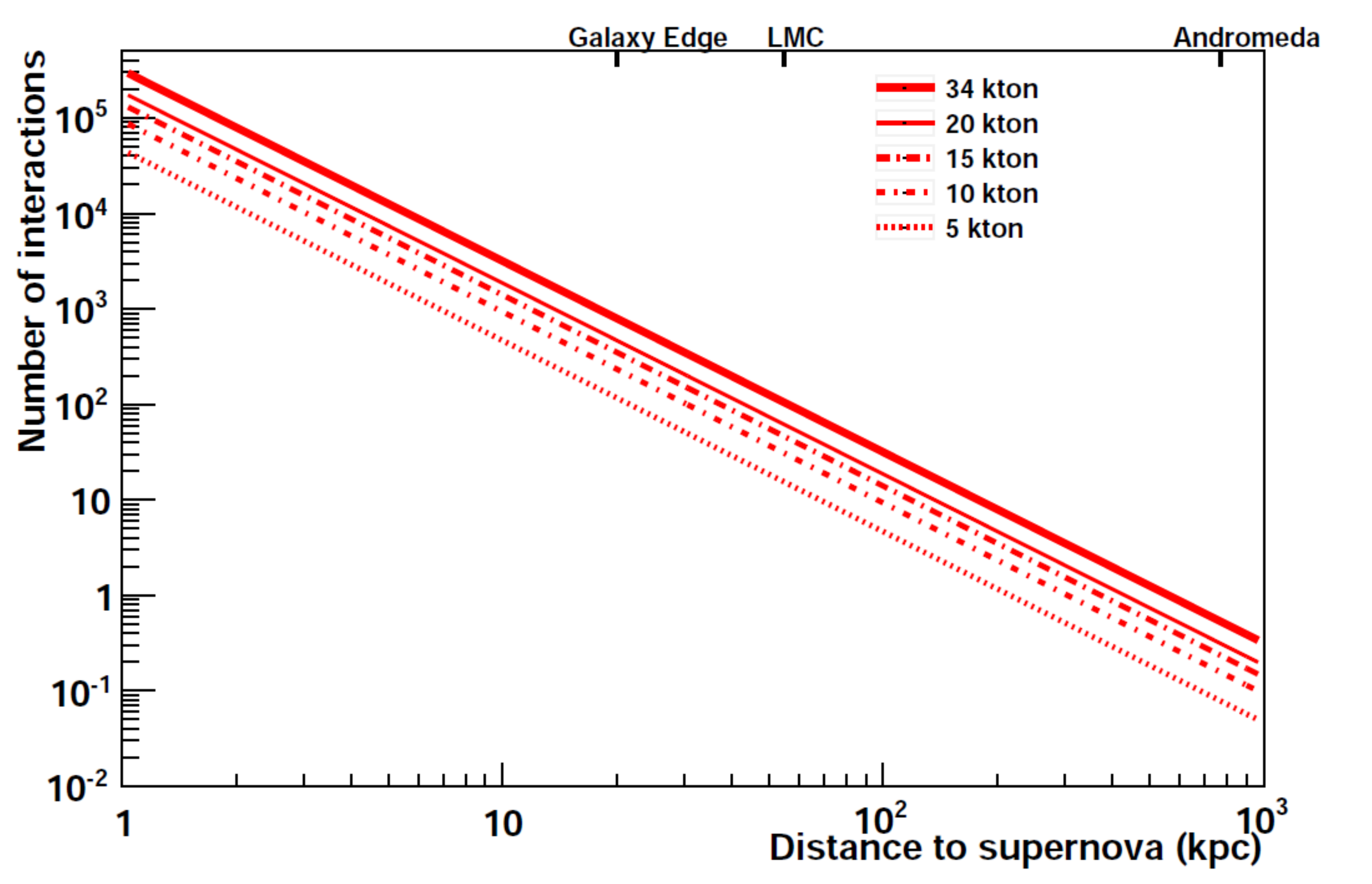}
\caption[Number of supernova-neutrino interactions in a LAr detector vs distance]{Number of supernova neutrino interactions in a liquid argon detector
as a function of distance to the supernova for different detector
masses.  Core collapses are expected to occur
a few times per century, at a most-likely distance from \SIrange{10}{15}{\kilo\parsec}.}
\label{fig:snevents}
\end{figure}

Currently, experiments worldwide are sensitive primarily to  $\overline{\nu}_e$'s, 
via inverse-beta decay on free protons, which dominates
the interaction rate in water and liquid-scintillator detectors.
Liquid argon exhibits a unique sensitivity to the $\nu_e$ component of the flux, via the absorption interaction on
$^{40}$Ar, $\nu_e +{}^{40}{\rm Ar} \rightarrow e^- +{}^{40}{\rm
  K^*}$. In principle, this interaction can be tagged via the
coincidence of the electron and the $^{40}{\rm K^*}$ de-excitation
gamma cascade.  About 900 events would be expected in a \ktadj{10} fiducial
liquid argon detector for a core-collapse supernova at 10~kpc.
The number of signal events
scales with mass and the inverse square of distance, as shown in
Figure~\ref{fig:snevents}.
For a collapse in the Andromeda galaxy, massive detectors of hundreds of kilotons
would be required to observe a handful of events. However, for supernovae
within the Milky Way, even a relatively small
\ktadj{10} detector would gather a significant $\nu_e$ signal.

Because the neutrinos emerge promptly after core
collapse, in contrast to the electromagnetic radiation which must beat
its way out of the stellar envelope, an observation could
provide a prompt supernova alert~\cite{Antonioli:2004zb,Scholberg:2008fa}, allowing 
astronomers to find the supernova in early light turn-on stages, which
could yield information about the progenitor (in turn, important for
understanding oscillations). 
 Further, observations and measurements by
multiple, geographically separated detectors during a core collapse ---
of which several are expected to be online over the next few
decades~\cite{Scholberg:2007nu,Scholberg:2010zz} --- will enhance the
potential science yield from such a rare and spectacular event~\cite{Mirizzi:2006xx}.

%% file: chapter-project.tex

\begin{introbox}
  The LBNE Project was formed to design and construct
  the  Long-Baseline Neutrino Experiment.  The experiment will
  comprise a new, high-intensity neutrino source generated from a
megawatt-class proton accelerator at Fermi
National Accelerator Laboratory  (Fermilab) directed at a large far detector at the Sanford Underground Research Facility in Lead, SD. A near detector will be located about
\SI{500}{\meter} downstream of the neutrino production target.  LBNE is currently planned as a phased program, with increased scientific  capabilities at each phase.

\begin{itemize}
\item 
The experimental facilities are designed to meet the primary scientific objectives of
  the experiment: (1) fully characterize neutrino oscillations,
  including measuring the value of the unknown CP-violating phase, $\delta_{\rm
    CP}$, and determining the ordering of the neutrino mass states, (2)
  significantly improve proton decay lifetime limits, and (3) measure
  the neutrino flux from potential core-collapse supernovae in our galaxy.
\item The LBNE beamline, based on the existing \emph{Neutrinos at the
    Main Injector} (NuMI) beamline design, is designed to deliver a
  wide-band, high-purity $\nu_\mu$ beam with a peak flux at \SI{
    2.5}{GeV}, which 
  optimizes the oscillation physics potential at the \kmadj{1300}
  baseline.  The beamline will operate initially at \SI{1.2}{\MW} and
  will be upgradable to \SI{2.3}{\MW} utilizing a proton beam with
  energy tunable from 60 to 120 GeV. 
\item The full-scope LBNE far detector is a liquid argon time-projection chamber (LArTPC) of fiducial mass \SI{34}{kt}.

The TPC design is modular, allowing flexibility in the choice of initial detector size.
\item The LBNE far detector will be located 4,850~feet underground, a
  depth favorable for LBNE's search for proton decay and detection of
  the neutrino flux from a core-collapse supernova.
\item The high-precision near detector and its conventional facilities can be built
  as an independent project, at the same time as the far detector and
  beamline, or later. 
\end{itemize}
\end{introbox}

\clearpage

\section{LBNE and the U.S. Neutrino Physics Program}
\label{sec:lbneproject}

\begin{figure}[!htb]
\centering\includegraphics[width=0.8\textwidth]{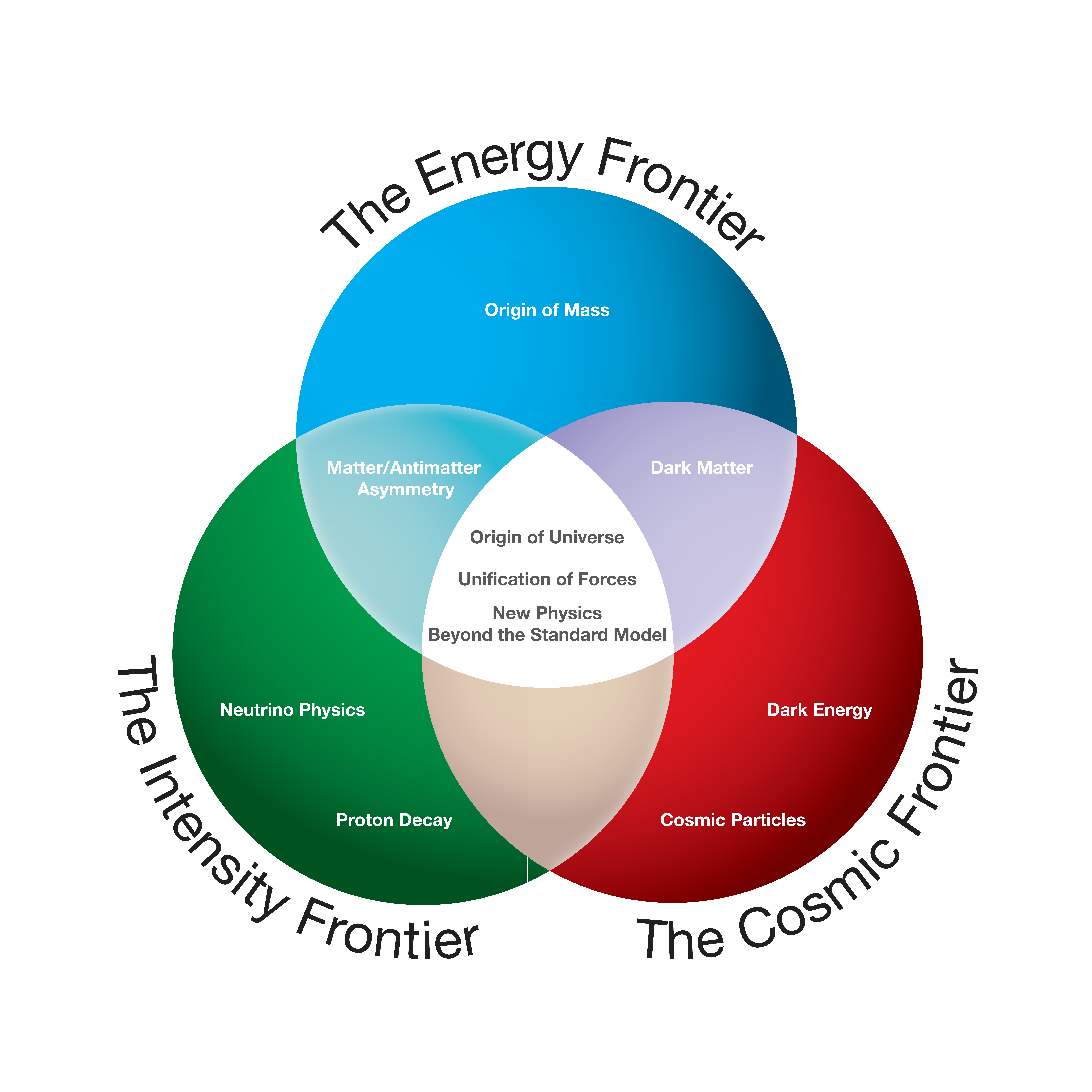}
\caption[Three frontiers of particle physics in the U.S.]{
Three frontiers of research in particle physics form
an interlocking framework that addresses
fundamental questions about the laws of Nature
and the cosmos. Each frontier, essential to the whole, has a unique approach to making discoveries~\cite{p5report}.
}
\label{fig:frontiers}
\end{figure}
In its 2008 report, the U.S. Particle Physics Project Prioritization
Panel (P5)\footnote{P5 is an advisory panel to the two main funding
  bodies for particle physics in the United States, the Department of
  Energy (DOE) and the National Science Foundation (NSF).} recommended
a world-class neutrino physics program as a core component of a
U.S. particle physics program~\cite{p5report} that revolves around
three research frontiers as shown in Figure~\ref{fig:frontiers}.  Included
in the report is the long-term vision of a large far detector at the
site of the former Homestake Mine in Lead, SD, and a high-intensity,
wide-band neutrino source at Fermilab.  At the time, the proposed
Deep Underground Science and Engineering Laboratory (DUSEL) was
planned to occupy the site of the former mine; it is now the \SURF.

On January 8, 2010 the DOE
approved the Mission Need~\cite{cdzero} statement\footnote{A \emph{Mission
  Need} statement initiates the process and provides initial funding
  toward developing the conceptual design of a DOE scientific
  project.} for a new long-baseline neutrino experiment that would
enable this world-class program and firmly establish the U.S. as the
leader in neutrino science. The LBNE experiment is designed to meet
this Mission Need.

With the facilities provided by the LBNE Project and the unique
features of the experiment --- in particular the long baseline of
1,300$\,$km, the wide-band beam and the high-resolution, underground far
detector --- LBNE will conduct a broad scientific program addressing
key physics questions concerning the nature of our Universe as
described in Chapter~\ref{intro-chap}.  The focus of the long-baseline
neutrino program will be the explicit demonstration of leptonic CP
violation, if it exists, and the determination of the neutrino mass
hierarchy. 

\begin{introbox}{ The \kmadj{1300} baseline has been determined to
    provide optimal sensitivity to CP violation and the measurement of
    \deltacp, and is long enough to enable an unambiguous
    determination of the neutrino mass hierarchy~\cite{Bass:2013vcg}.
  }
\end{introbox} 
The focus of the non-beam scientific program will be to search for
proton decay, to enable detailed studies of atmospheric neutrinos, and
to detect and measure the neutrino flux from a supernova, should one
occur within our galaxy.

It is currently planned to implement LBNE as a phased program, with
increased scientific capabilities at each phase. The initial
phase 
of LBNE will achieve significant advances 
with respect to its primary scientific objectives as compared to
current experiments.  The \emph{goal} for the initial phase of LBNE  is:
\begin{enumerate}
\item A new neutrino beamline at Fermilab driven by a 60 to \SI{120}{\GeV} proton beam with power of up to \SI{1.2}{\MW}. 
\item A liquid argon time-projection chamber (LArTPC) detector of fiducial mass at least 10 kt located at the \SURF at a depth of 4,850 feet. 
\item A high-precision near neutrino detector on the Fermilab site.
\end{enumerate}

The cost for this initial phase (with a \ktadj{10} far detector) is
estimated to be 1.2B U.S.\$ according to DOE standard project
accounting.

In December of 2012, the DOE issued CD-1 (Conceptual Design phase) approval for a budget of 867M\$ 
U.S. based on a reduced scope that excluded the near neutrino detector and the underground 
placement of the far detector. Domestic and international partners are being sought to enable 
construction of the full first-phase scope outlined above. Subsequent phases of LBNE are expected to 
include additional  far detector mass and upgrades of the beam to $\geq$\MWadj{2.3} capability.

\clearpage
\section{Near Site: Fermi National Accelerator Laboratory}
\label{intro-fnal}

\begin{introbox}

Fermilab, located 40 miles west of Chicago, Illinois, is a DOE-funded
laboratory dedicated to high energy physics.  The laboratory builds
and operates accelerators, detectors and other facilities that
physicists from all over the world use to carry out forefront
research.  

Dramatic discoveries in high energy physics have revolutionized our
understanding of the interactions of the particles and forces that
determine the nature of matter in the Universe.  Two major components
of the Standard Model of Fundamental Particles and Forces were
discovered at Fermilab: the bottom quark (May-June 1977) and the top
quark (February 1995). In July 2000, Fermilab experimenters announced
the first direct observation of the tau neutrino, thus filling the
final slot in the lepton sector of the Standard Model.  Run II of the
Fermilab Tevatron Collider 
was inaugurated in March 2001. The Tevatron was the world's
highest-energy particle accelerator and collider until the Large
Hadron Collider at CERN came online in 2011.

While CERN now hosts the world's highest-energy particle collider, the
Fermilab accelerator complex is being retooled to produce the world's
highest-intensity beams of protons, muons and
neutrinos. Scientists from around the world can exploit this capability to
pursue cutting-edge research in the lepton sector of the
Standard Model where strong hints of new physics have surfaced.

The beamline and near detector for LBNE will be constructed
at Fermilab, referred to as the \emph{Near Site}.
\end{introbox}

Fermi National Accelerator Laboratory, originally named the National
Accelerator Laboratory, was commissioned by the U.S. Atomic Energy
Commission, under a bill signed by President Lyndon B. Johnson on
November 21, 1967.  On May 11, 1974, the laboratory was renamed in
honor of 1938 Nobel Prize winner Enrico Fermi, one of the preeminent
physicists of the atomic age.

Today, the DOE operates national
laboratories throughout the United States, including Fermilab. The DOE
awarded to Fermi Research Alliance (FRA) the management and operating
contract for Fermilab, effective January 1, 2007. The FRA is a
tax-exempt, limited liability company (LLC) organized and operated for
charitable, scientific and educational purposes under Section
501(c)(3) of the Internal Revenue Code.  The two members of FRA are
the University of Chicago and the Universities Research Association
(URA). FRA has earned extensions to the Fermilab contract through
Dec. 31, 2015.

At Fermilab, a robust scientific program pushes forward on the three interrelated scientific 
frontiers specified by the P5 panel in 2008~\cite{p5report} and illustrated in Figure~\ref{fig:frontiers}:
\begin{enumerate}
\item At the Energy Frontier,
  Fermilab scientists are significant contributors to the LHC and to
  the CMS experiment.
\item At the Intensity Frontier, Fermilab operates two neutrino beams that support a number of 
experiments.  In the next few years several new neutrino and muon experiments will be coming 
online, of which LBNE will be the largest.
\item At the Cosmic Frontier, 
  Fermilab runs and/or participates in several experiments, with
  instruments installed in North America, South America and Europe.
\end{enumerate}

\begin{figure}[!htb]
\centering
\includegraphics[width=\textwidth]{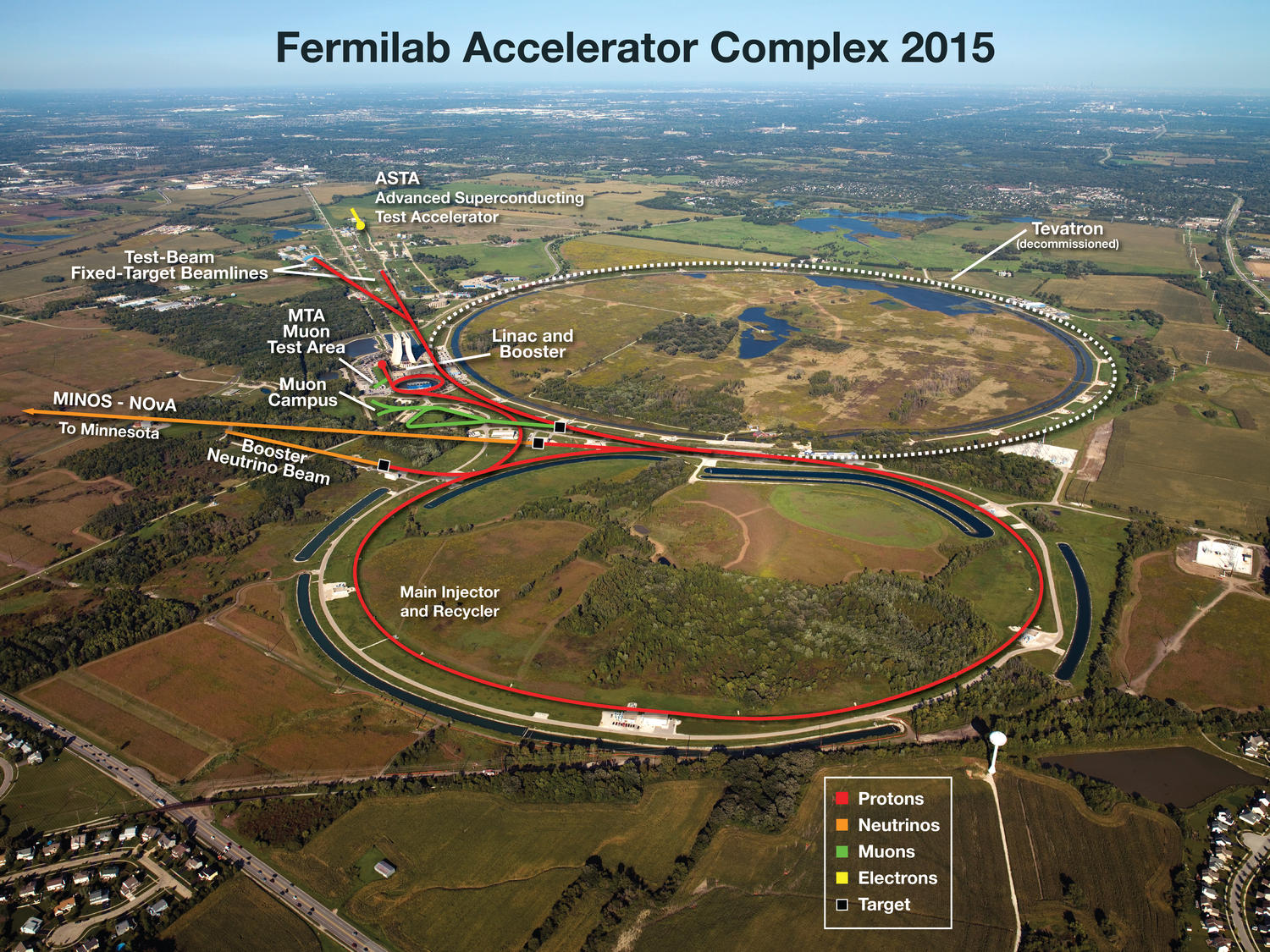}
\caption[Fermilab's accelerator chain]{The accelerator chain at Fermi
  National Accelerator Laboratory. A \MeVadj{400} linear accelerator (linac) feeds into the 15-Hz
  Booster, which produces an \GeVadj{8} beam. The Booster beam is used for
  the Booster Neutrino Beamline experiments. The Booster feeds into
  the \GeVadj{120} Main Injector. The Main Injector is the source for the NuMI beamline, which
  supplies a high-power, high-energy neutrino beam to the MINOS/MINOS+
  and NO$\nu$A experiments.}
\label{fig:accels}
\end{figure}
The neutrino beams at Fermilab come from two of the lab's proton accelerators (Figure~\ref{fig:accels}), the \GeVadj{8} Booster, which feeds the \emph{Booster
Neutrino Beamline} (BNB), and the \GeVadj{120} Main Injector (MI), which feeds the
NuMI beamline.
The LBNE beamline, described in Section~\ref{beamline-chap}, will
utilize the MI beam. 

\begin{figure}[!htb]
\centering
\includegraphics[width=0.8\textwidth]{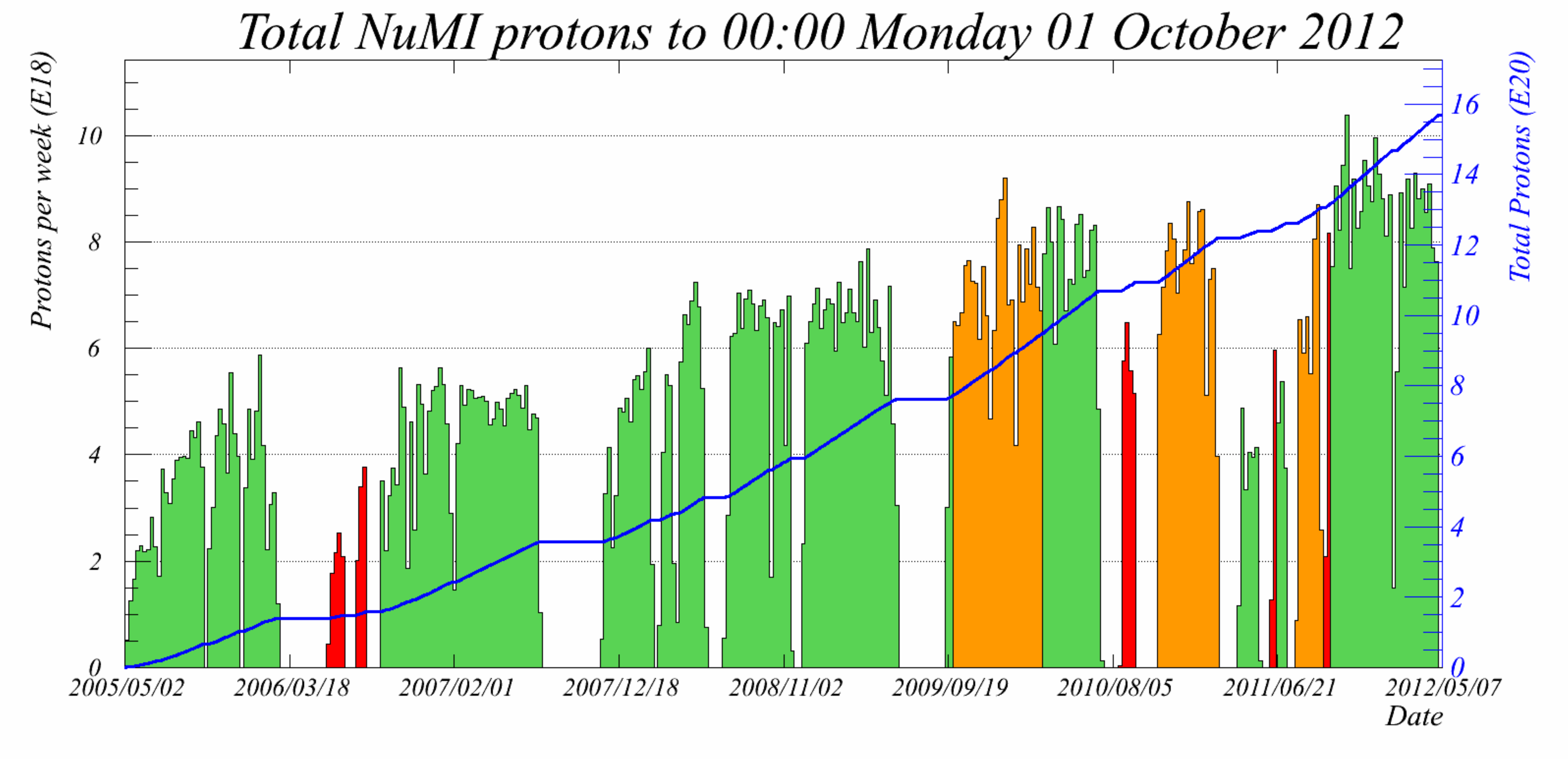}
\caption[NuMI beamline performance]{The NuMI beamline performance}
\label{fig:numipot}
\end{figure}

NuMI, on which LBNE's beamline design is based, is a high-energy neutrino beam that has been operating since
2004. It was designed for steady \kWadj{400} operation and achieved that
goal by the end of the MINOS experimental run in 2012. As shown in
Figure~\ref{fig:numipot}, the NuMI beamline was running with an
average of $9\times 10^{18}$ protons per week ($\approx 2.7 \times
10^{20}$ protons-on-target per year) in mid 2012.
\begin{figure}[!htb]
\centering
\includegraphics[width=0.8\textwidth]{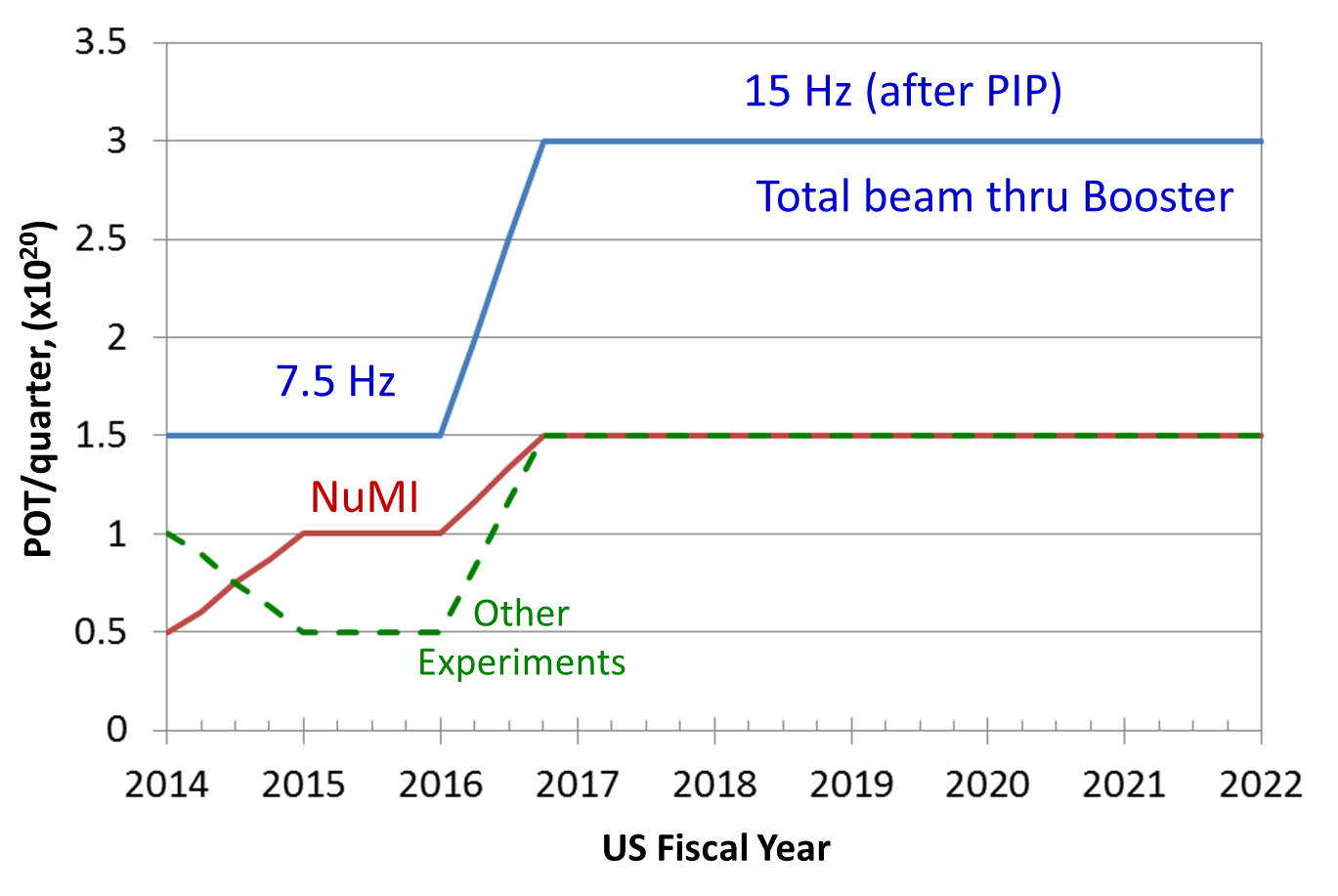}
\caption[Possible flux ramp-up scenario for Intensity Frontier experiments]{A possible ramp-up
scenario for proton flux from Fermilab's proton source for the Intensity Frontier experiments.}
\label{fig:pip}
\end{figure}

Upgrades to the Recycler\footnote{The Recycler, a fixed 8-GeV kinetic
  energy storage ring located directly
  above the MI beamline, stores protons from the 8-GeV Booster 
during MI ramp up. }  and MI as part of the NO$\nu$A
Project, as well as the Proton Improvement Plan (PIP) that is
currently underway, comprise a set of improvements to the existing
Linac, Booster and MI aimed
at supporting 15-Hz beam operations from the Booster (Figure~\ref{fig:pip}). 

\begin{figure}[!htb]
\centering
\includegraphics[width=0.7\textwidth]{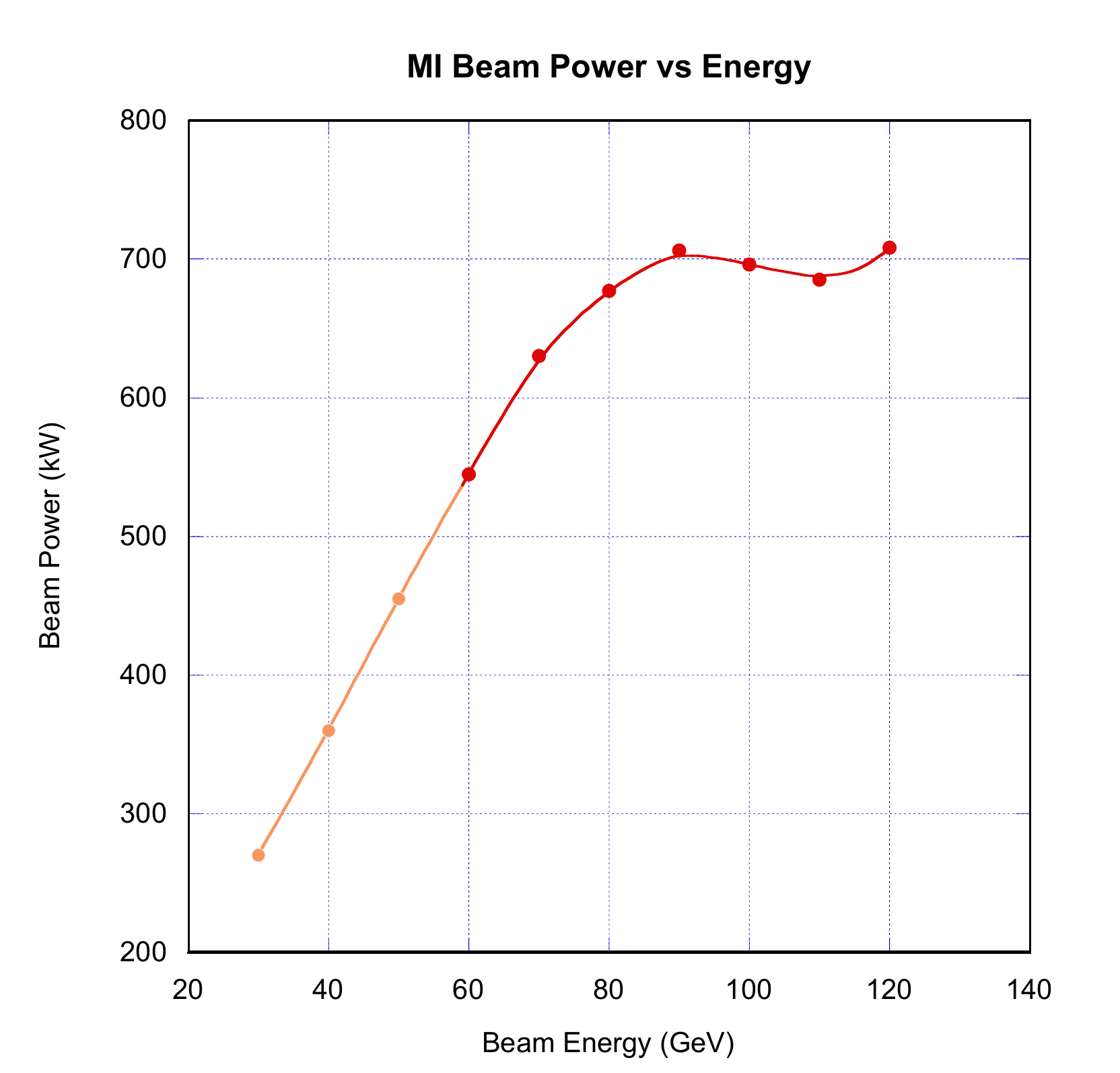}
\caption[Proton beam power versus beam energy]{Proton beam power expected to be available 
as a function of MI beam energy after proton-improvement-plan (PIP) upgrades.
}
\label{fig:pippower}
\end{figure}

In combination, the NO$\nu$A
upgrades and the PIP create a capability of delivering \SI{700}{\kW}
from the MI at 120~GeV ($\approx 6 \times 10^{20}$
proton-on-target per year) by 2016. The proton beam power expected to
be available as a function of MI beam energy after
completion of the PIP upgrades is shown in Figure~\ref{fig:pippower}.

A conceptual plan for further upgrades to the Fermilab accelerator
complex has been completed. Called the \emph{Proton Improvement
  Plan-II} (PIP-II)~\cite{PIPII}, its goal is to increase the
capabilities of the existing accelerator complex to support delivery
of 1.2~MW of beam power to the LBNE production target at the
initiation of operations, while simultaneously providing a platform
for subsequent upgrades of the complex to multi-MW capability. The
starting point of this plan is the \emph{Project X Reference Design Report}~\cite{Holmes:2013vpa}.  

\begin{figure}[!htb]
\centering
\includegraphics[width=0.9\textwidth]{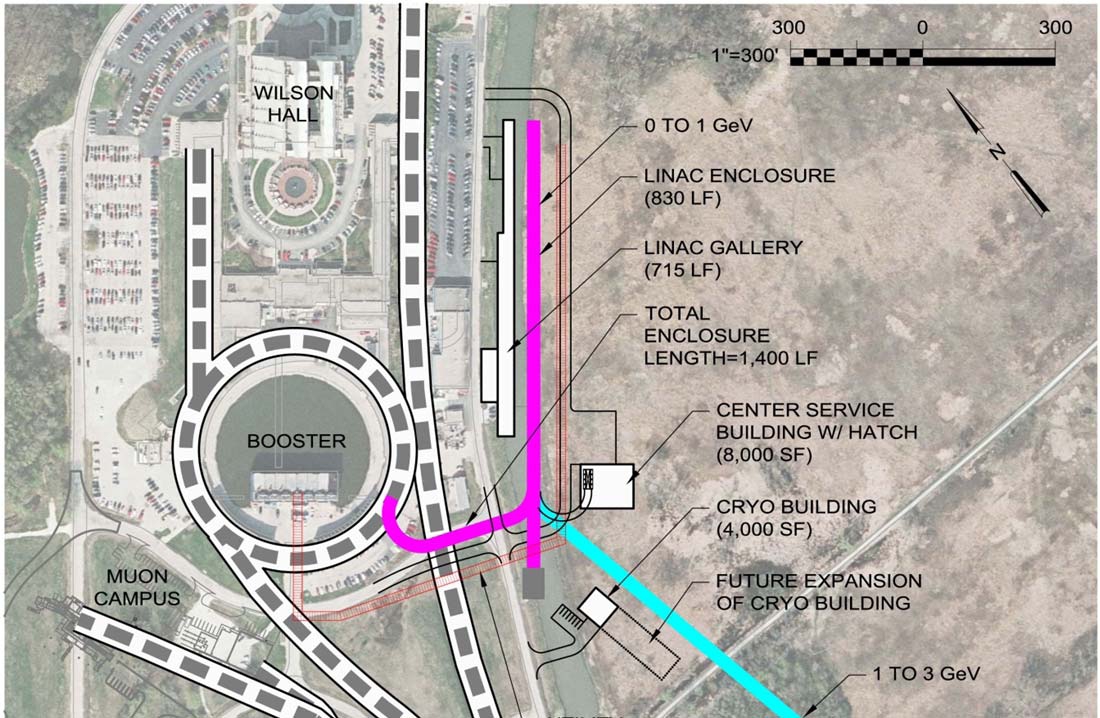}
\vskip -0.02in
\includegraphics[width=0.9\textwidth]{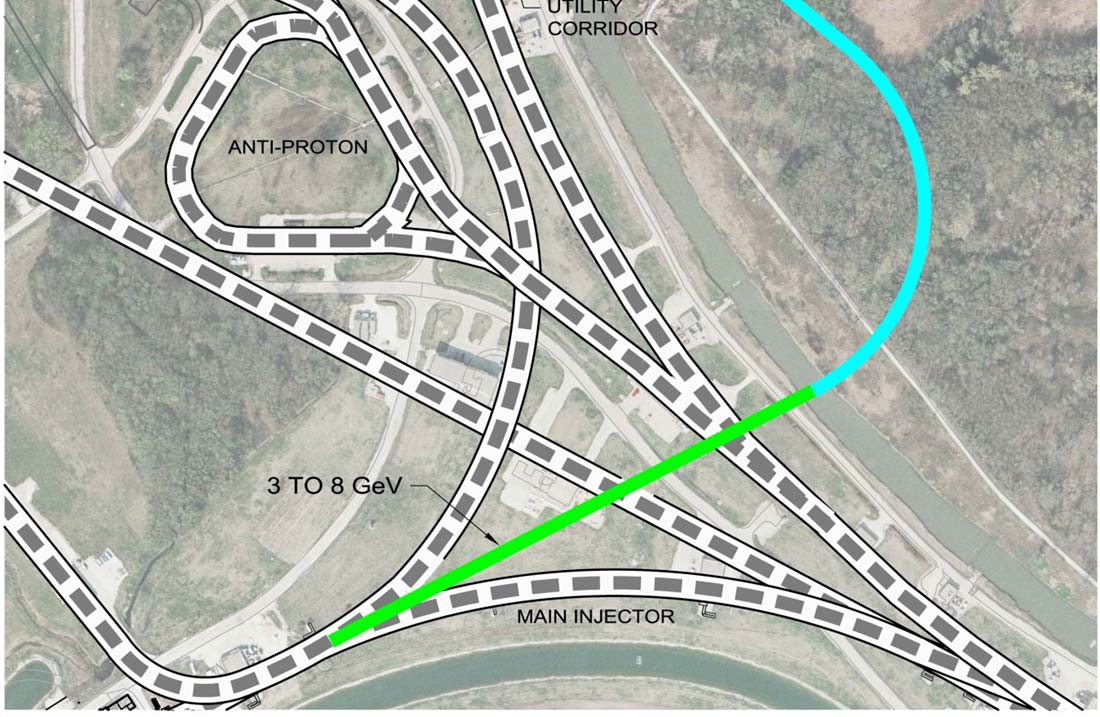}
\caption[Site layout of PIP- II]{Site layout of PIP- II is shown as
  the magenta line which is the 800 MeV linac enclosure and transfer
  line. New construction includes the linac enclosure, transfer line
  enclosure, linac gallery, center service building, utility corridor,
  and cryo building. Dashed areas represent existing or planned
  underground enclosures. Further possible upgrades to the Fermilab
  complex beyond PIP- II are shown in the bottom half of the figure:
  cyan is a 1-3 GeV CW linac and transfer line, and green is a 3-8 GeV
  pulsed linac~\cite{PIPII}. }
\label{fig:pip2-layout}
\end{figure}

The primary bottleneck to providing increased beam power at Fermilab
is the Fermilab Booster, limited by space-charge forces at
injection. In the intermediate term the most cost-effective approach
to removing this bottleneck is to increase the injection energy into
the Booster. The PIP-II meets this goal via an \MeVadj{800} superconducting
linear accelerator (linac), operated at low duty factor, but
constructed of accelerating modules that are capable of
continuous-wave (CW) operations if provided with sufficient cryogenic
cooling and appropriate RF power. This is expected to increase the
beam intensity delivered from the Booster by 50\% relative to current
operations. Shortening the MI cycle time to \num{1.2}~s yields
a beam power of \SI{1.2}{\MW} at 120~GeV.  The conceptual site layout
of PIP-II is shown in Figure~\ref{fig:pip2-layout}. Further possible
upgrades beyond PIP-II would require replacing the \GeVadj{8} Booster with
a superconducting linac injecting into the MI at energies
between 6 and 8 GeV as shown in Figure~\ref{fig:pip2-layout}, eventually
increasing the power from the MI to 2.0--2.3 MW at 60--120
GeV.

\clearpage

\section{Far Site: Sanford Underground Research Facility}
\label{intro-surf}

\begin{introbox}{ The Sanford Underground Research
    Facility~\cite{SURF} is a laboratory located on the site of the
    former Homestake gold mine in Lead, SD that is dedicated to underground
    science.  This laboratory has been selected as the location of the far
    detector for LBNE, and is referred to as the \emph{Far
      Site}.  

Underground neutrino experiments in the former mine date
    back to 1967 when nuclear chemist Ray Davis installed a solar
    neutrino experiment 4,850 feet below the
    surface~\cite{Cleveland:1998nv}. Ray Davis earned a share of the
    Nobel Prize for physics in 2002 for his experiment, which ran
    until 1993. 

LBNE is envisioned as the next-generation, multi-decade neutrino
    experiment at this site seeking groundbreaking discoveries.}
\end{introbox}
\begin{figure}[!htb]
\centering
\includegraphics[height=0.9\textheight]{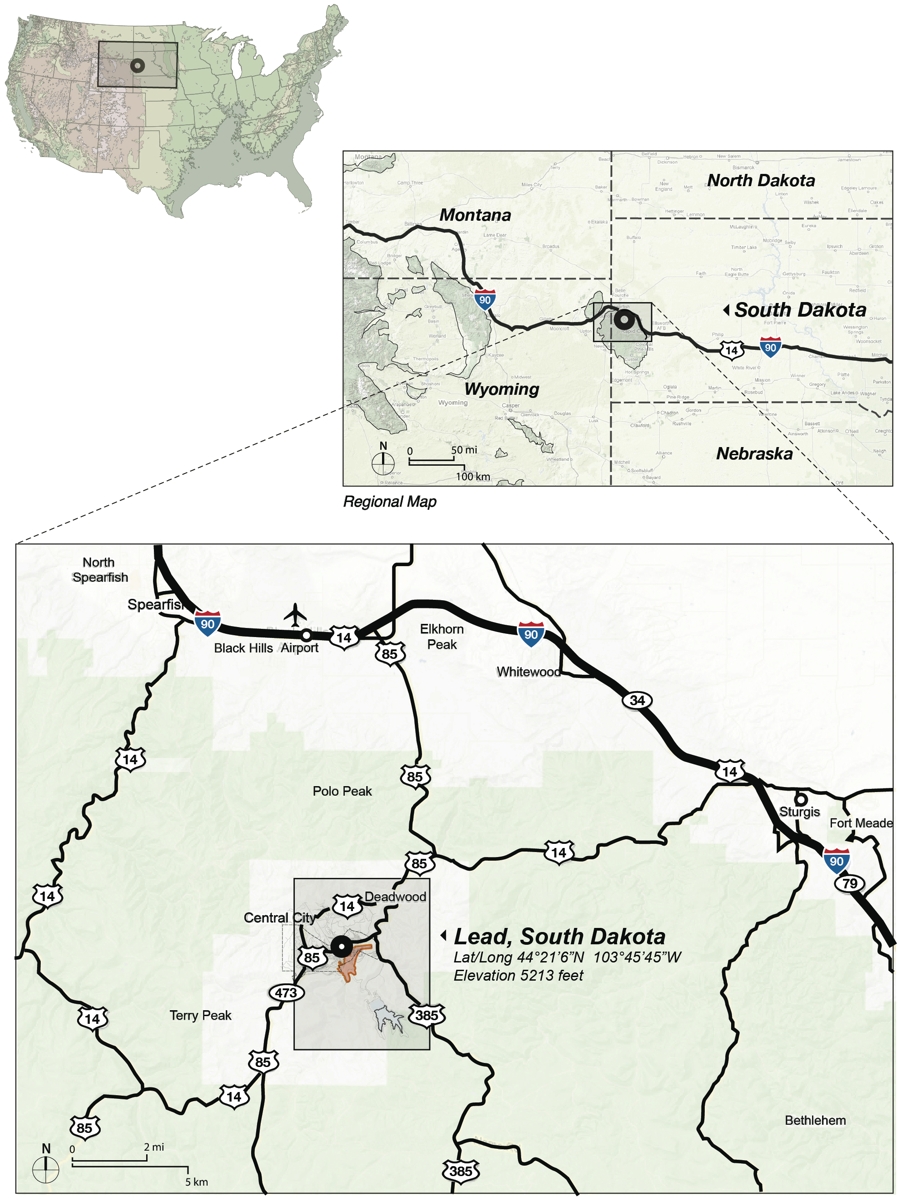}
\caption[Location of Lead, South Dakota]{Location of the town of Lead,
  South Dakota - the site of the former Homestake Gold Mine.}
\label{fig:lead}
\end{figure}
\begin{figure}[!htbp]
\centering
\includegraphics[width=0.9\textheight,angle=90]{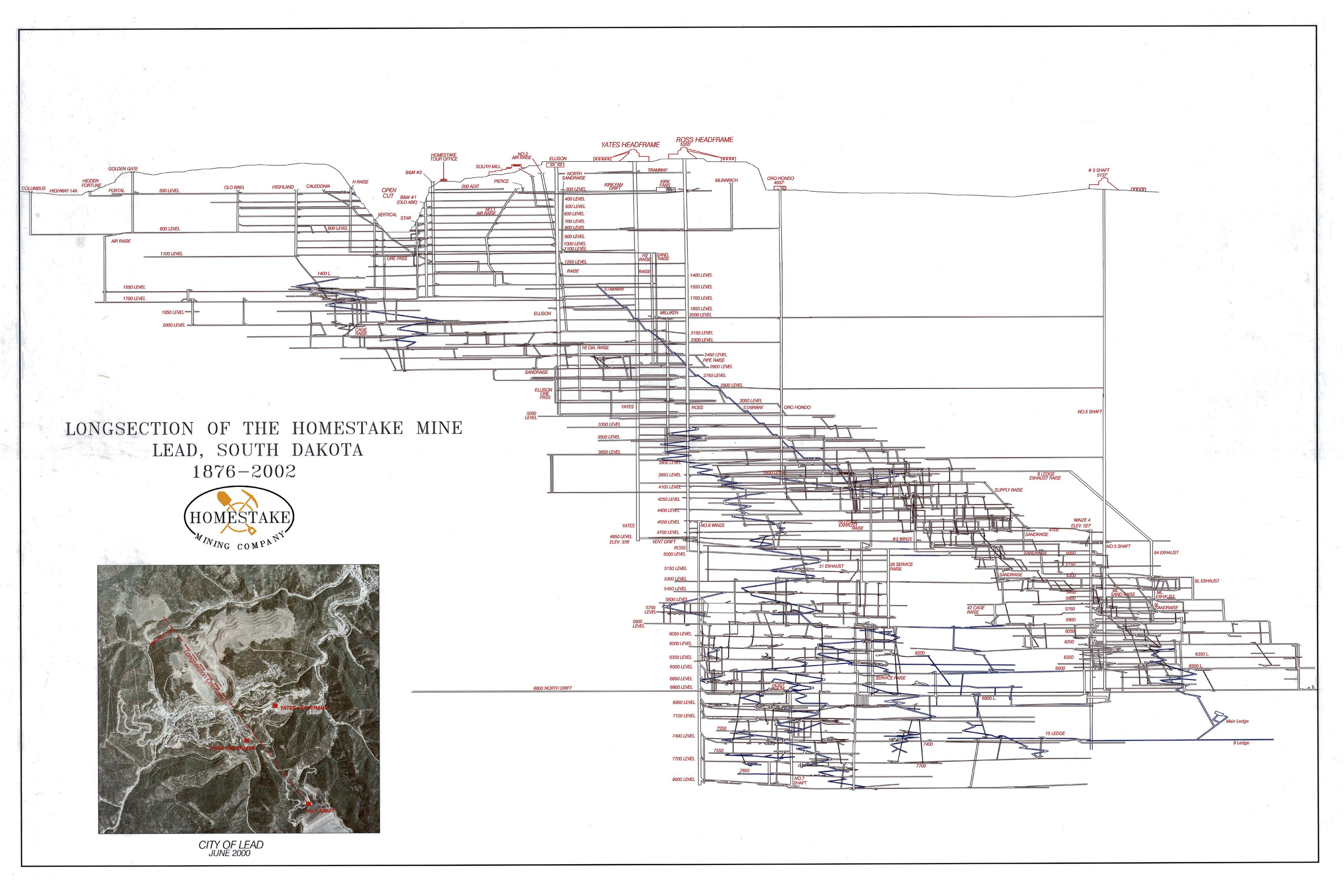}
\caption[Former Homestake Mine vertical
long section]{The long section of the former Homestake Gold
  Mine. This figure illustrates the 60 underground levels extending to depths 
  greater than 8,000 feet. The location of cross section
  is indicated in the inset along a NW to SE plane. The projection
  extends for 5.2 km along this plane }
\label{fig:hstake-xsec1}
\end{figure}
\begin{figure}[!htbp]
\centering
\includegraphics[height=0.9\textheight,clip,trim=2.2cm 3.5cm 2.2cm 1cm]{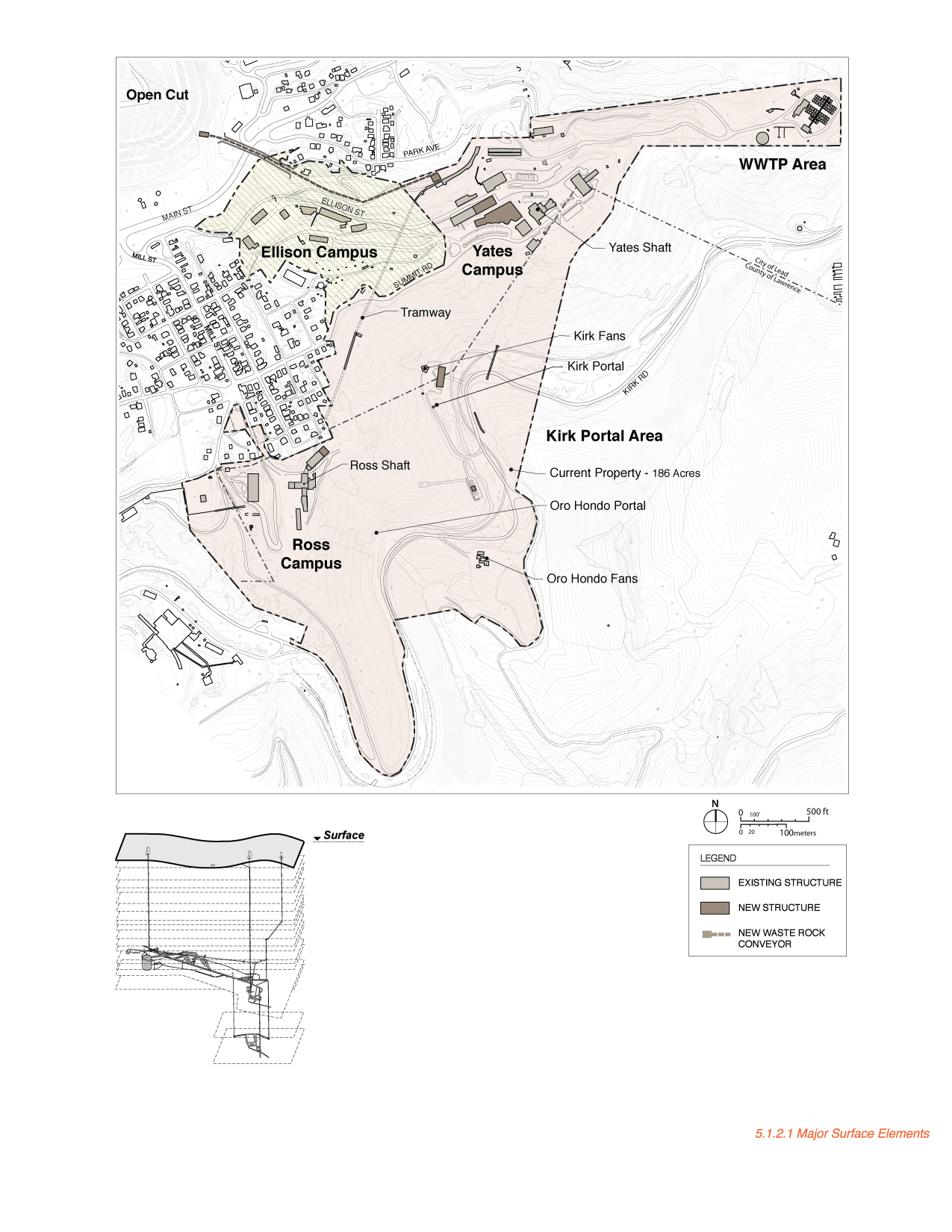}
\caption[\SURF campuses]{The surface and
  underground campuses of the \SURF. The 3D inset image illustrates the
  plans to develop the 4850L and 7400L. 
Most current experiments are at the 4850L.}
\label{fig:surf_larloc}
\end{figure}
In 2006, Barrick Gold Corporation donated the Homestake Gold Mine
site, located in Lead, South Dakota (Figure~\ref{fig:lead}) 
to the State of South Dakota,
following over 125 years of mining. 
Mining operations created over 600
km of tunnels and shafts in the facility, extending from the surface
to over 8,000 feet below ground. The mining levels are distributed $\sim$150
feet apart and are referenced by their depth below the facility entrance, 
e.g., the level 4,850 feet below ground is referred to as the 
\emph{4850L}. 
This former mine encompasses the deepest
caverns in the western hemisphere, offering extensive drifts both
vertically and laterally.  A detailed vertical cross section of the 60
underground levels developed for mining is shown in
Figure~\ref{fig:hstake-xsec1}. 
%

In 2004, the South Dakota state legislature created the South Dakota
Science and Technology Authority (SDSTA) to foster scientific and
technological investigations, experimentation and development in South
Dakota. A six-member board of directors appointed by the governor of
South Dakota directs the SDSTA. The SDSTA's first task was to reopen
the former Homestake 
site to the 4,850-foot level for scientific research.  At this site,
the SDSTA now operates and maintains the Sanford Underground Research
Facility through a contract managed and overseen by a dedicated
operations office at Lawrence Berkeley National Laboratory 
as a deep-underground research
laboratory. 
The \SURF property comprises 186 acres on the surface and 7,700 acres
underground. The surface campus includes approximately 253,000 gross
square feet of existing structures.  A surface schematic of the 
campus is shown in Figure~\ref{fig:surf_larloc}.

The state legislature has since committed more than \$40 million in
state funds to the development of the \SURF, and the state
has also obtained a \$10 million Community Development Block Grant to
help rehabilitate the site. In addition, a \$70 million donation from
philanthropist T. Denny Sanford has been used to reopen the site for
science and to establish the Sanford Center for Science Education.
The initial concepts for the facility were developed with the support of the
U.S. National Science Foundation (NSF) as the primary site for the NSF's
Deep Underground Science and Engineering Laboratory (DUSEL). With the
National Science Board's decision to halt development of the 
NSF-supported underground laboratory, the 
DOE now supports the operation of the facility in addition to state
and private funding. Both the NSF and the DOE support experiments at the site.

\begin{figure}[!htb]
\centering
\includegraphics[width=\textwidth]{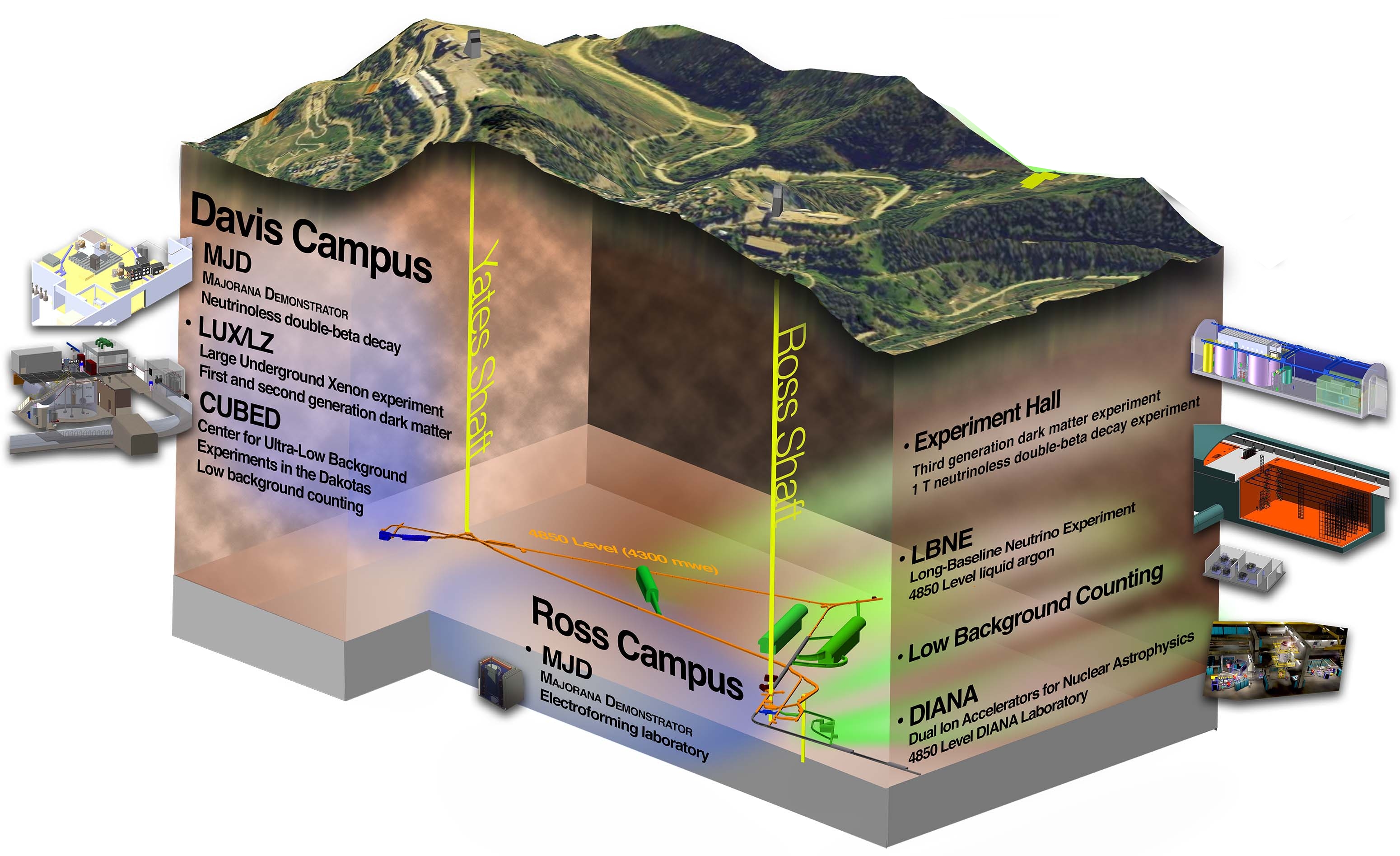}
\caption[Layout at the \ftadj{4850} level in the
\SURF]{Layout of experiments at the \ftadj{4850} level in the \SURF}
\label{fig:4850l}
\end{figure}

\begin{figure}[!htb]
\centering
\includegraphics[width=0.8\textwidth]{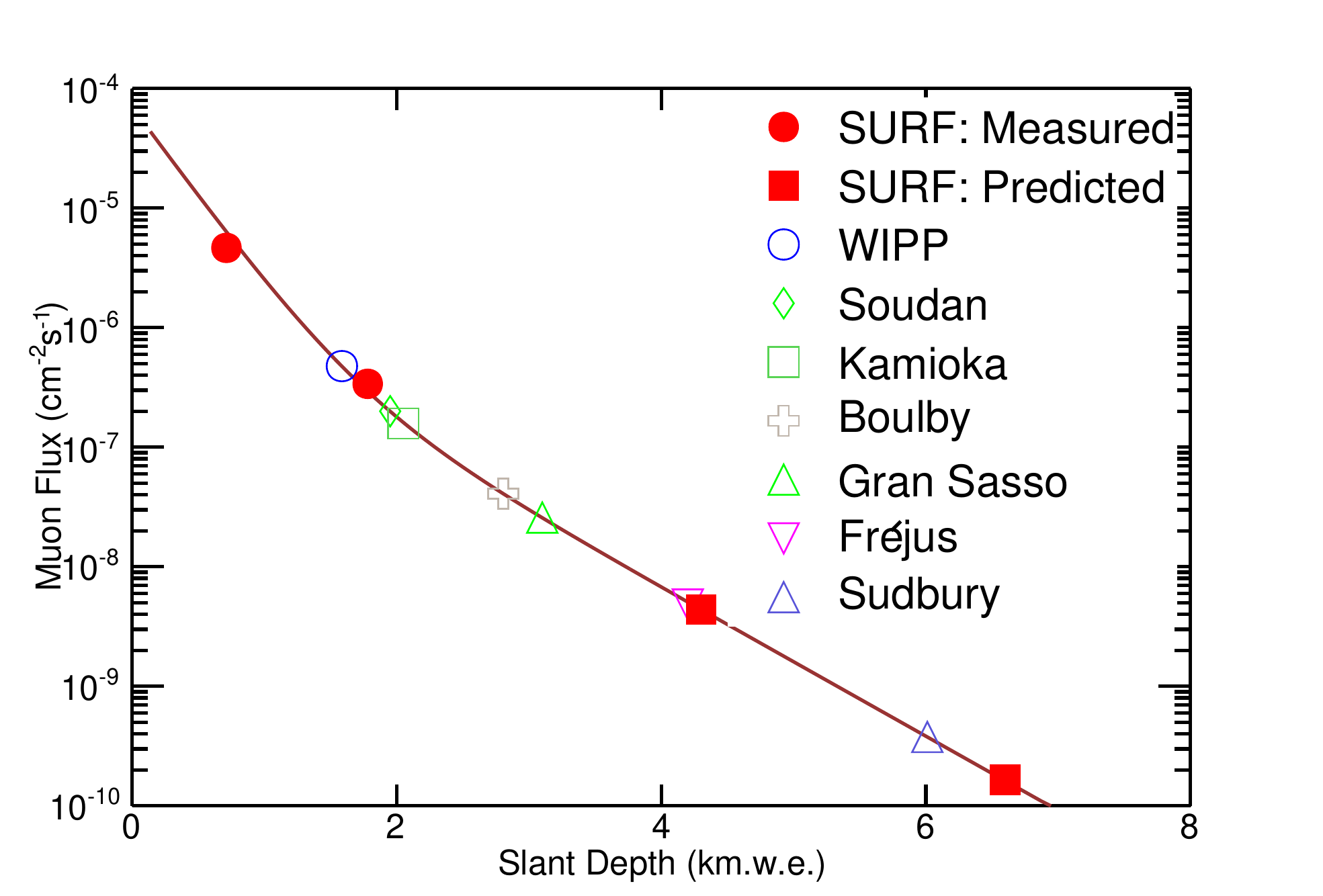}
\caption[Predicted cosmic-ray flux as a function of depth]{Predicted cosmic-ray flux as a function of depth. 
The predicted muon flux at the
\SI{4850}{ft} and \SI{8000}{ft} levels of the Sanford Underground
Research Facility (SURF) are show as red squares.  Two measured depths in the
facility are shown as red circles.  Values for other underground
laboratories are also shown~\cite{Gray:2010nc}.  The line shows a parameterized model
of the muon flux.}
\label{fig:hstake_cosmic}
\end{figure}

Access to the underground areas has been reestablished and the primary
access rehabilitated and improved. The facility has been stabilized
and the accumulated underground water has been pumped out below
5,680~ft. The area around the Davis cavern at the 4850L, named for the
late Ray Davis, has been enlarged and adapted primarily for current
and next-generation dark matter and neutrinoless double-beta decay
experiments. This upgraded area of the 4850L is now called the Davis
Campus.
Additional science efforts are located throughout the facility, including
an ultrapure detector development laboratory, geophysics and
geological efforts, and a public outreach program.  A 3D schematic
highlighting the planned development of the 4850L 
is shown in Figure~\ref{fig:4850l}.
The LBNE far detector 
will be located 
in new excavated spaces near the bottom of the Ross Shaft, about 1~km
from the Davis Campus.  The \ftadj{4850} depth
makes it an extremely
competitive location in terms of cosmic-ray background suppression
for 
undertaking the nucleon decay and supernova neutrino studies that LBNE
plans to address.  Figure~\ref{fig:hstake_cosmic} shows the predicted
cosmic-ray flux at this site~\cite{Gray:2010nc} as compared to other
underground laboratories worldwide.
\begin{figure}[!htb]
\centering
\includegraphics[height=0.8\textheight]{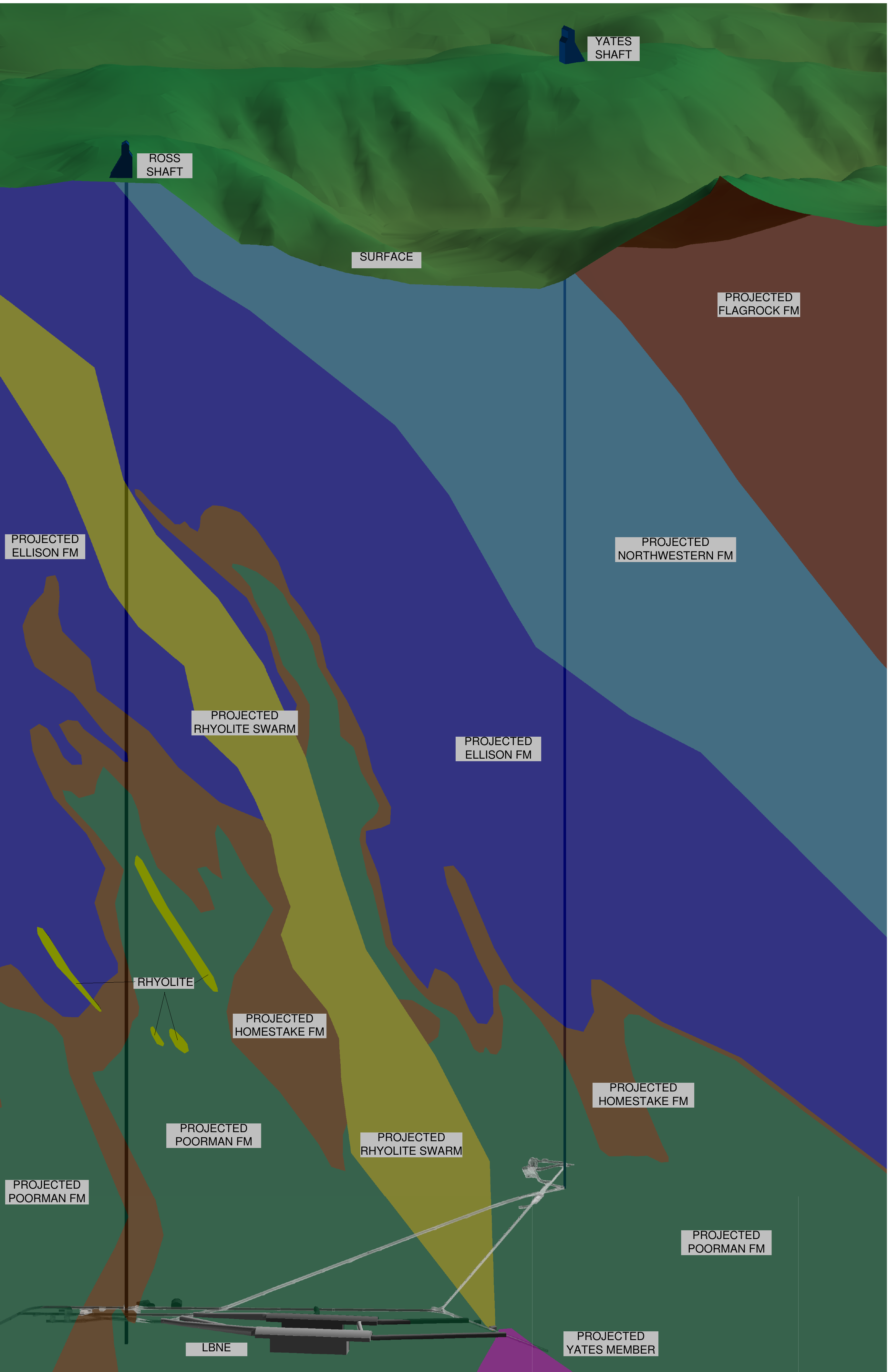}
\caption[Geologic long section of Sanford Underground Research
  Facility]{Geologic long section of Sanford Underground Research
  Facility showing the main rock formations. The dark green rock is
  the Poorman formation, and the yellow areas indicate a projected
  rhyolite swarm. The proposed location of two LBNE detector caverns
  are shown in the foreground.}
\label{fig:surfrock}
\end{figure}

\begin{figure}[!htb]
\centering
\includegraphics[width=\textwidth]{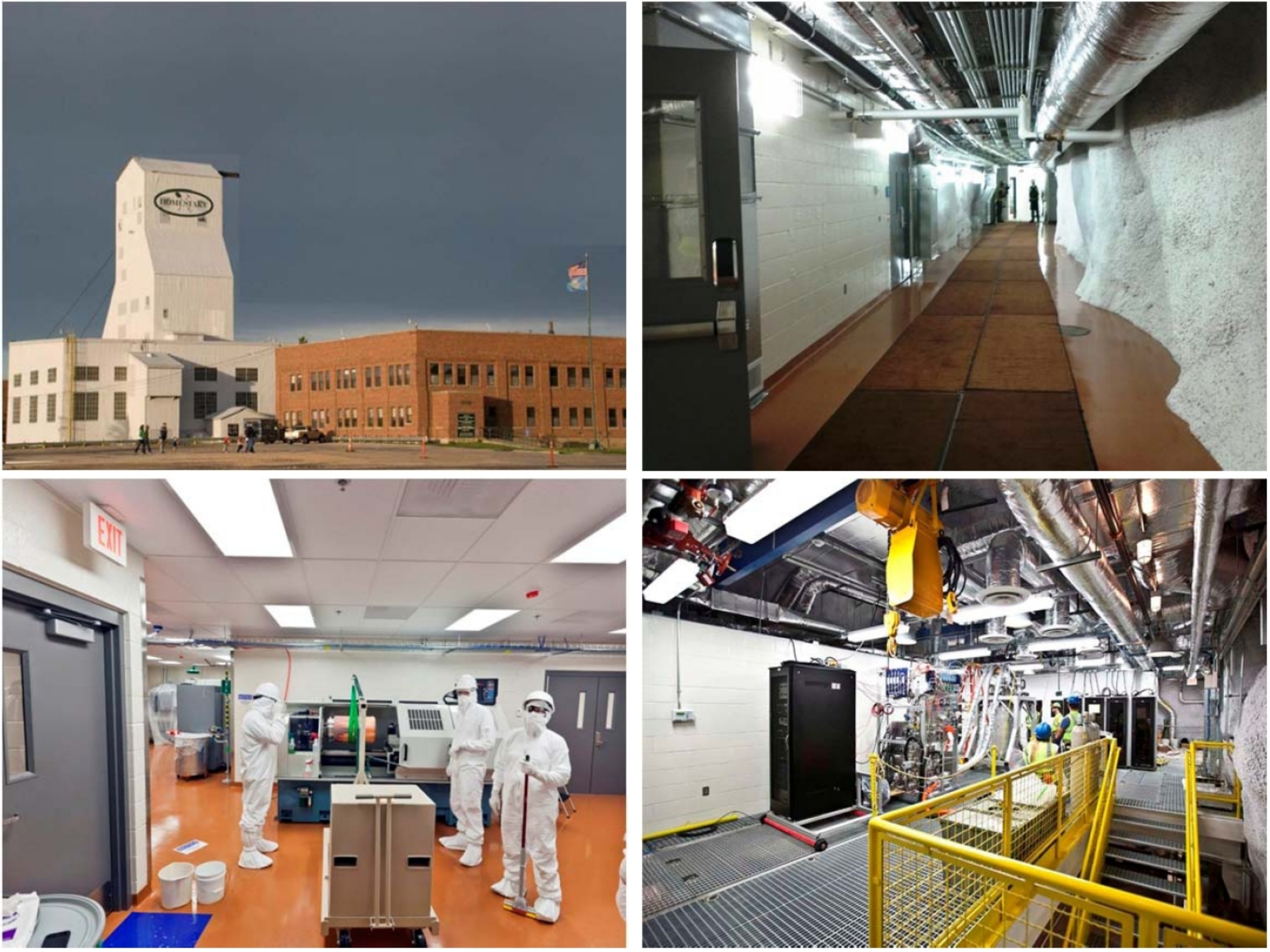}
\caption[Photos from the \SURF]{Sanford Underground Research
  Facility: Administration building and Yates shaft headframe (top
  left); corridor at \SI{4850}{\ft} (\SI{1480}{\meter}) depth leading to clean rooms and
  experimental halls (top right); billet of radiopure electroformed
  copper for the MJD experiment being placed on a
  lathe in a clean room at \SI{4850}{\ft} depth (bottom left); LUX experiment
at \SI{4850}{\ft} depth (bottom right).}
\label{fig:lux}
\end{figure}

Another advantage of the 4850L \SURF site for LBNE is the low level of
rock radioactivity that could contribute backgrounds to the supernova
burst neutrino signal and other low-energy physics searches. It was
found that the U/Th/K radioactivity for the underground bedrocks at
Homestake is in general very low when compared to common construction
materials such as concrete and shotcrete; some samples are in the
sub-ppm levels. However, samples from rhyolite intrusions, {\em a very
small fraction of the total}, show a relatively high content of U, Th,
and K more typical of the levels found in other laboratories, in
particular those in granitic formations. Regions of potential rhyolite
intrusions have been identified and documented as shown in
Figure~\ref{fig:surfrock}. 
In some cases local
shielding significantly mitigates the impact of the rhyolite
intrusions. Table ~\ref{tab:rockrad} presents some of the assay
results, obtained by direct gamma counting for rock samples from the
mine, including those collected close to the 4850L~\cite{surfrock}.
\begin{table}[!htb]
  \caption[Partial U/Th/K assay results for far site rock samples]{Partial U/Th/K assay results for \SURF rock samples. Overall errors estimated to be $\sim$10-20\%. Also shown are results for various construction materials (shotcrete/concrete).}
\label{tab:rockrad}
\begin{center}
\begin{tabular}{$L^l^l^l^l} 
\toprule
\rowtitlestyle
& Uranium (ppm) & Thorium (ppm) & Potassium (\%) \\
\rowtitlestyle
& Ave. [Range] & Ave. [Range] & Ave. [Range] \\ 
\toprowrule
U/G Country Rock
& 0.22 [0.06-0.77]
& 0.33 [0.24-1.59]
& 0.96 [0.10-1.94] \\ \colhline
Shotcrete 
& 1.89 [1.74-2.23]
& 2.85 [2.00-3.46]
& 0.88 [0.41-1.27] \\ \colhline
Concrete Blocks 
&  2.16 [2.14-2.18]
& 3.20 [3.08-3.32]
& 1.23 [1.27-1.19] \\ \colhline
Rhyolite Dike
& 8.75 [8.00-10.90]
& 10.86 [8.60-12.20]
& 4.17 [1.69-6.86] \\ 
\bottomrule
\end{tabular}
\end{center}
\end{table}
%
The Large Underground Xenon (LUX) experiment is now operating in the
cavern first excavated for Davis in the 1960s. LUX is the most
sensitive detector yet to search for dark
matter~\cite{Akerib:2013owa}.  The Majorana Demonstrator experiment (MJD),
also being installed in a newly excavated space adjacent to the
original Davis cavern, will search for neutrinoless double-beta
decay. Figure~\ref{fig:lux} shows four photographs of facilities and
activities at the \SURF related to the LUX and MJD 
at the 4850L. 
%
The LBNE far detector will benefit from the common infrastructure
being developed to house large experiments underground. The layout of
the different proposed experiments at the 4850L, including the LBNE
detector, is shown in Figure~\ref{fig:4850l}.

\begin{figure}[!htb]
\centering
\includegraphics[width=0.9\textheight,trim=1cm 5cm 2cm 2cm, clip,angle=90]{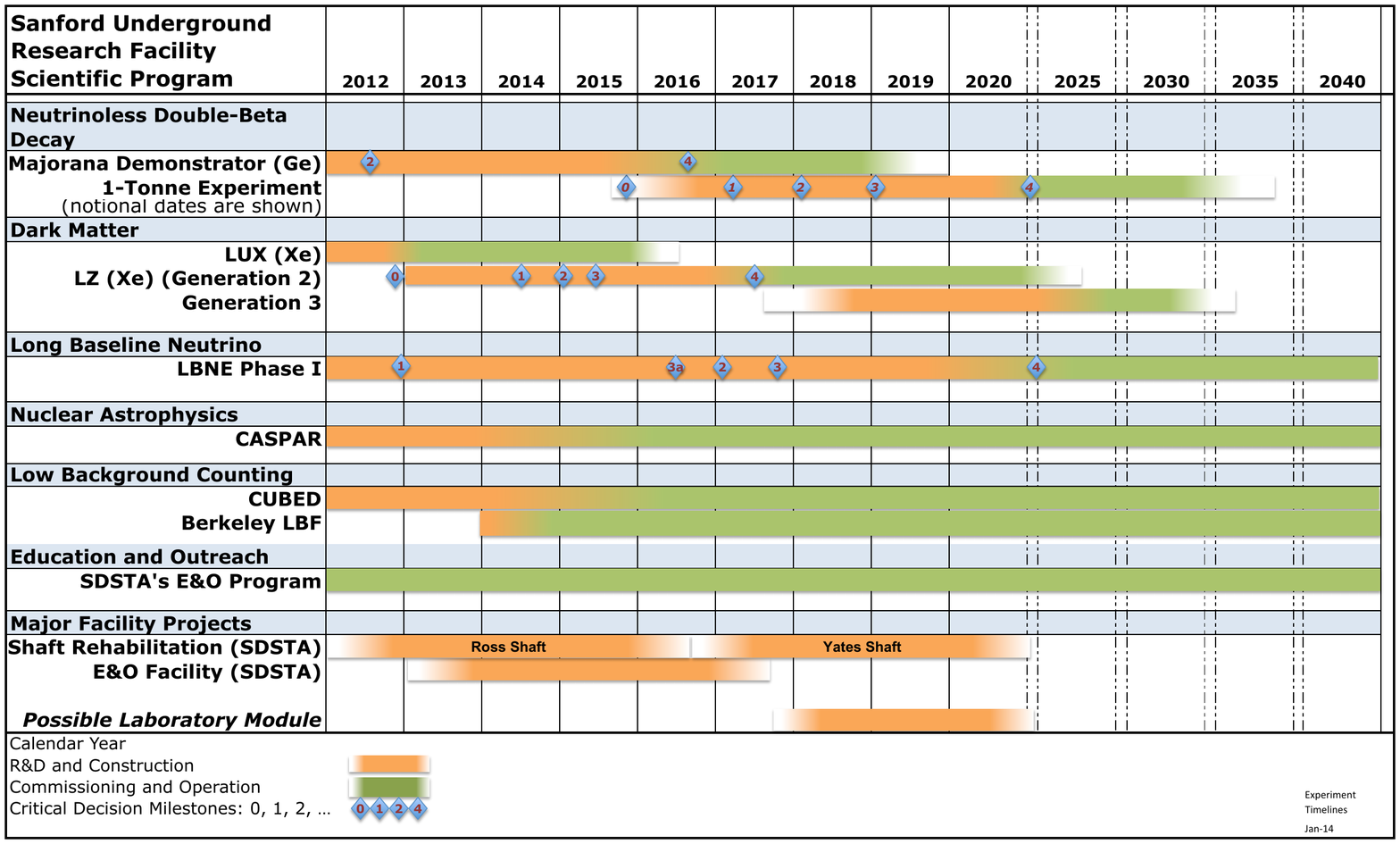}
\caption[Experiment timeline for the \SURF]{Timeline exploring the
  long-term potential of deep science experiments at the \SURF. 
  Figure courtesy of Mike Headley, the \SURF.
}
\label{fig:hstake_timeline}
\end{figure}

In addition to LBNE, LUX and MJD, the \SURF science program for the
coming five to ten years (Figure ~\ref{fig:hstake_timeline}) consists of the expansion of the LUX dark matter search, the 
Center for Ultralow Background Experiments at
Dakota (CUBED), and the geoscience installations. 
Long-term plans are
being developed to host a nuclear astrophysics program \linebreak
involving underground particle accelerators (CASPAR and DIANA), and second- and
third-generation dark matter experiments.

\clearpage
\section{Beamline} 

\label{beamline-chap}

\begin{introbox}
  The LBNE neutrino beamline, located at Fermilab, utilizes a conventional horn-focused
  neutrino beam produced from pion decay-in-flight, based largely on
  the highly successful NuMI beamline design:
\begin{itemize} 
\item The primary beam utilizes 60- to \GeVadj{120} protons from the
  Main Injector accelerator. The primary beamline is
  embedded in an engineered earthen embankment 
  --- a novel construction concept to reduce costs and improve
  radiological controls.
\item The beamline is designed to operate at \SI{1.2}{\MW} and to
  support an upgrade to \MWadj{2.3} operation.
\item The beamline will generate a wide-band, high-purity beam,
  selectable for muon neutrinos or muon antineutrinos. Its tunable
  energies from 
  \num{60} to \SI{120}{GeV} will be well matched to the \kmadj{1300} neutrino
  oscillation baseline. 
\end{itemize}
\end{introbox}

The LBNE beamline facility 
will aim a beam of neutrinos 
toward the LBNE far detector located 1,300~km away at the \SURF.  The
beamline facility, which will be fully contained within Fermilab
property, will consist of a primary (proton) beamline, a neutrino
beamline, and conventional facilities to support the technical
components of the primary and neutrino beamlines~\cite{CDRv2}.  
The LBNE beamline reference design parameters approved at CD-1 are 
summarized in Table~\ref{tab:beam_reference_params}.  
Improvements to this design that have been made or are being considered 
are described in this section, including the important change to an initial beam 
power of \SI{1.2}{\MW}, enabled by the planned PIP-II.  The beamline needed 
for the full-scope LBNE will be 
realized in the first phase of LBNE and will be  
upgradable to \SI{2.3}{\MW}.

\begin{table}[!htb]
  \caption[Parameters for LBNE beamline reference design at CD-1]{Partial set of 
    parameters for the elements of the LBNE Beamline reference design at CD-1 from Volume 2 of the CDR~\cite{CDRv2}. 
    The reference design 
    described a \SI{700}{\kW} beam; 
    it has since been changed to \SI{1.2}{\MW}.
    For each parameter the third column lists the range that had been studied prior to  CD-1. Distances between beam elements are given from the upstream face (the end facing the proton beam) 
    with respect to the upstream (front) face of Horn 1.} 
\label{tab:beam_reference_params}
\begin{tabular}{$L^l^l^l}
\toprule
\rowtitlestyle
Element  & Parameter &  Range studied & Reference design  \\ 
\rowtitlestyle
& &  & value  (\SI{700}{\kW})\\ \toprowrule
Proton Beam & energy & \SIrange{60}{120}{\GeV} & \SI{120}{\GeV} \\
& protons per pulse  & & 4.9$\times10^{13}$ \\
& cycle time between pulses & & \SI{1.33}{\second} \\
& size at target $\sigma_{x,y}$ & \SIrange{1}{2}{\mm} & \SI{1.3}{\mm} \\
& duration & & 1.0$\times10^{-5}$ sec \\
& POT per year & & 6.5$\times10^{20}$ \\ \colhline
Target  
 &  material  & graphite, beryllium &  graphite \\
 &            & hybrid~\cite{DOCDB3151} &  \\
 &  length & $\geq$ 2 interaction lengths  & \SI{966}{\mm}\\
 &  profile  & rectangular, & rectangular \\
 &           & round ($r=$ \SIrange{5}{16}{\mm}) &  \SI{7.4}{\mm} x \SI{15.4}{\mm}\\
 & dist. from Horn 1 (front) & \SIrange{0}{-250}{\cm} & \SIrange{-35}{-285}{\cm} \\
\colhline
Focusing Horn 1~\cite{DOCDB8398}
 & shape & cylindrical-parabolic,  & double-parabolic \\
 &       & double-parabolic & (NuMI) \\
 & length (focusing region) &  \SIrange{2500}{3500}{\mm} & \SI{3000}{\mm} \\
 & current  & \SIrange{180}{350}{\kA} & \SI{200}{\kA} \\ 
 & minimum inner radius  &   & \SI{9.0}{\mm} \\
 & maximum outer radius  &   & \SI{174.6}{\mm} \\
 \colhline
Focusing Horn 2 
 & shape & double-parabolic & NuMI Horn 2 \\
 & length (focusing region) & \SIrange{3000}{4000}{\mm} & \SI{3000}{\mm}  \\
 & current  & \SIrange{180}{350}{\kA}  & \SI{200}{\kA} \\
 & minimum inner radius  &   & \SI{39.0}{\mm} \\
 & maximum outer radius  &   & \SI{395.4}{\mm} \\
 & dist. from Horn 1 (front) & \SIrange{4000}{10000}{\mm} & \SI{6600}{\mm} \\ 
\colhline
Decay Pipe 
 & length & \SIrange{200}{350}{\m} & \SI{204}{\m}\\
 & radius & \SIrange{1.0}{3.0}{\m}  & \SI{2}{\m}\\
 & atmosphere  &  Air, He, vacuum & air at atm. pressure\\
 & dist. from Horn 1 (front) & \SIrange{11}{23}{\m} & \SI{17.3}{\m} \\
\bottomrule
\end{tabular} 
\end{table}

\begin{figure}[!htb]
\centering
\includegraphics[width=0.9\textwidth]{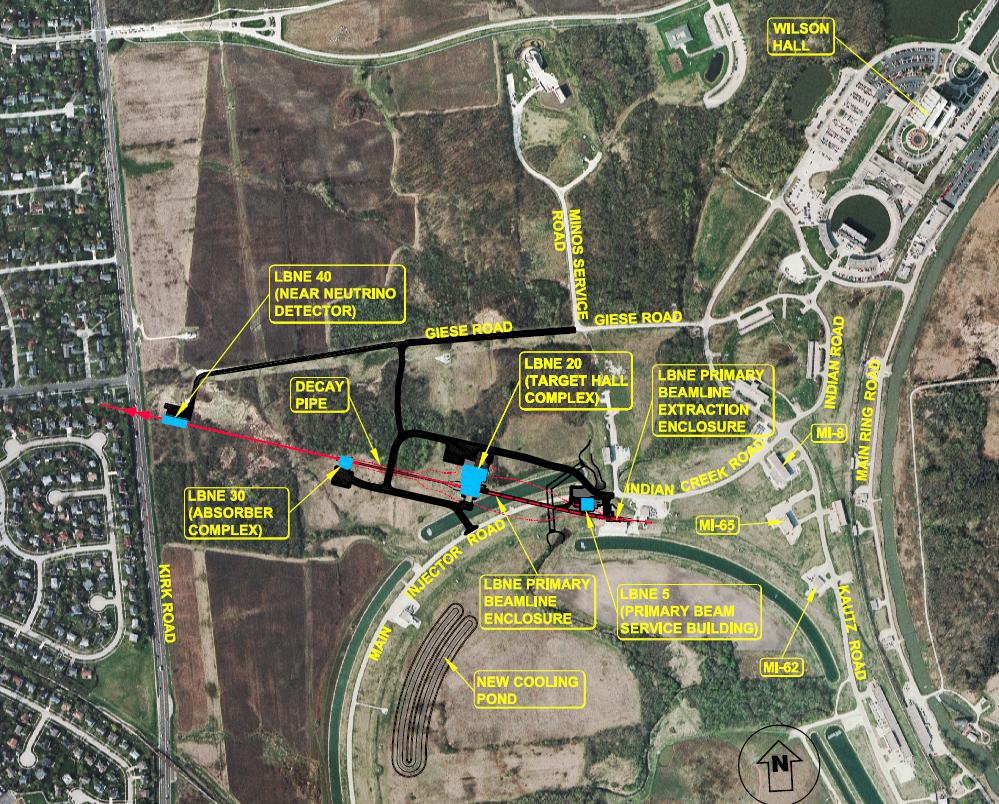}
\caption[LBNE project layout at Fermilab]{Plan view of the
  overall Near Site project layout showing locations for the LBNE Beamline
  extraction point from the MI, the primary beamline,
  target hall, decay pipe, absorber and near neutrino detector. }
\label{fig:nscf_layout}
\end{figure}

\begin{figure}[!htb]
\centering
\includegraphics[width=\textwidth]{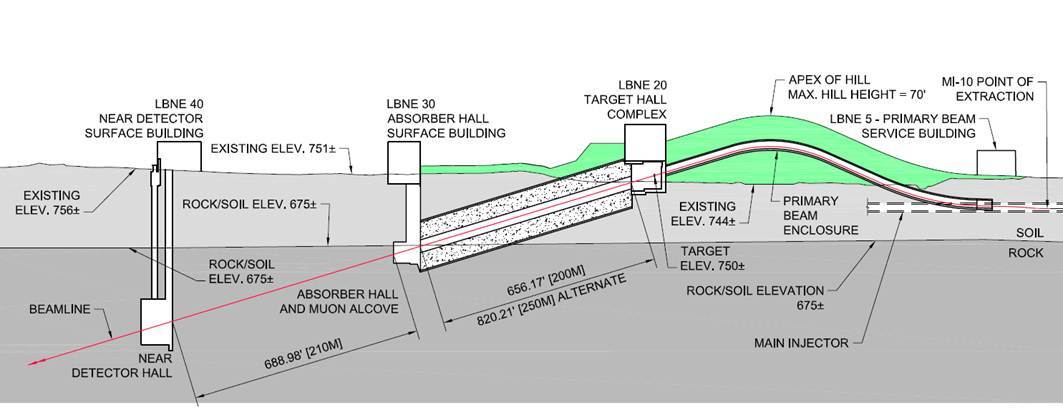}
\caption[Longitudinal section of the LBNE beamline
facility.]{Longitudinal section of the LBNE Beamline facility. The
  beam enters from the right in the figure, the protons being
  extracted from the MI-10 extraction point at the Main Injector.}
\label{v-beam-fig:intro_elev_overview}
\end{figure}

The primary beam, composed of protons in the energy range of 60-120
GeV, will be extracted from the MI-10 straight section of the
Main Injector using single-turn extraction. The beam will then be
transported to the target area within a beam enclosure embedded in an
engineered earthen embankment (hill).  The primary-beam transport
section is designed for very low losses. The embankment's dimensions
are designed to be commensurate with the bending strength of the
required dipole magnets so as to provide a net 5.8$^\circ$ downward
vertical bend to the neutrino beam (Figures~\ref{fig:nscf_layout}
and \ref{v-beam-fig:intro_elev_overview}). The beamline is then buried
by soil shielding that is placed at a stable angle of repose,
resulting in the embankment final geometry.

For \GeVadj{120} operation and with the MI upgrades
implemented for the NO$\nu$A experiment~\cite{Ayres:2007tu}, the fast,
single-turn extraction will deliver $4.9 \times 10^{13}$ protons to
the LBNE target in a \SIadj{10}{\micro\second} pulse.
With a \SIadj{1.33}{\second} cycle time, the beam power for NO$\nu$A is
\SI{700}{\kW}. Additional accelerator upgrades planned as
PIP-II~\cite{PIPII} will increase the protons per cycle to $7.5\times
10^{13}$ and reduce the cycle time to \SI{1.2}{\second}, resulting in
an initial beam power for LBNE of 1.2 MW. The LBNE beamline is
designed to support additional beam power upgrades beyond PIP-II,
discussed in Section~\ref{intro-fnal}, that can increase the beam power up to \SI{2.3}{MW}. At \MWadj{1.2}
operation the accelerator and primary beamline complex are expected
to deliver $11\times 10^{20}$ protons per year to the
target. 

\begin{figure}[!htb]
\centering
\includegraphics[width=\textwidth]{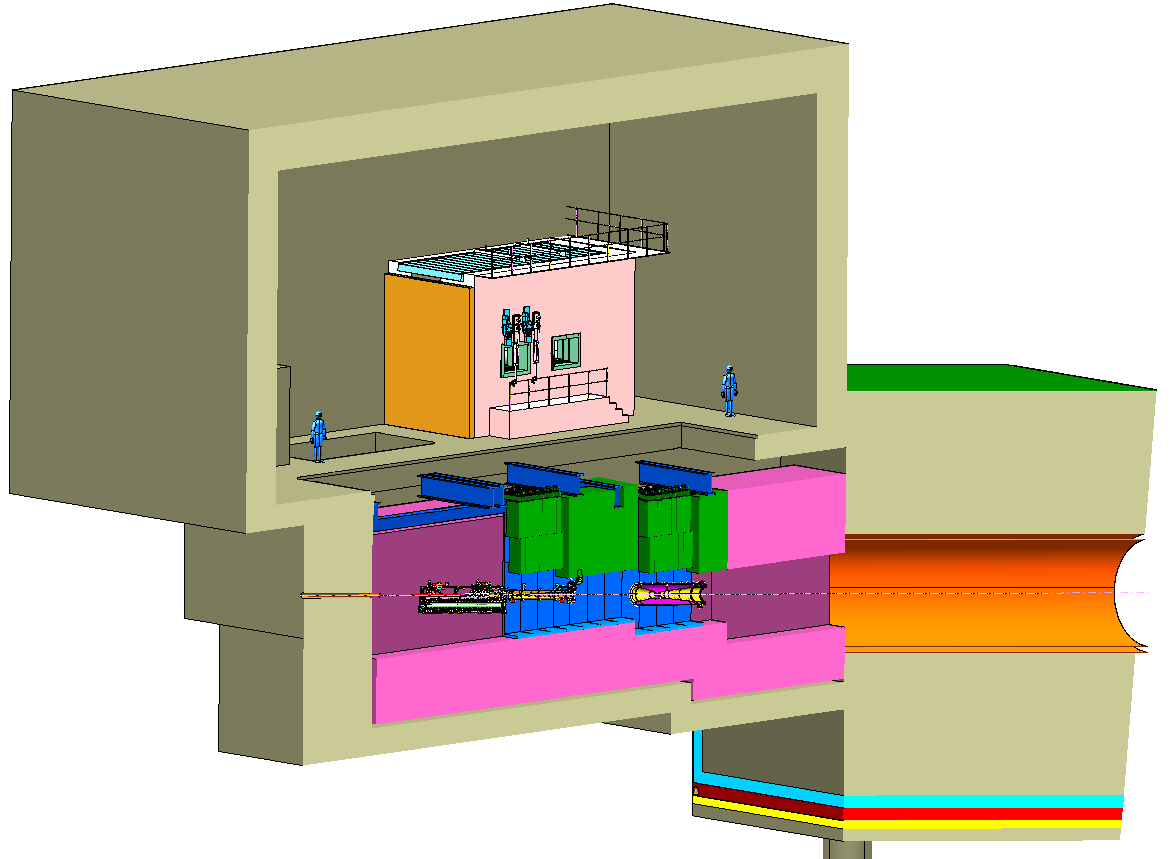}
\caption[Neutrino beamline components]{Schematic of the upstream
  portion of the LBNE neutrino beamline showing the major components
  of the neutrino beam. The target chase bulk steel shielding is shown in magenta. Inside the target chase from left to right (the direction of the
  beam) pointing downwards: the beam window, horn-protection baffle and target mounted on a carrier, the two toroidal 
  focusing horns (the green custom shielding blocks are part of the horn support modules that are not shown) and the decay pipe (orange). Above the chase and to the right is the work cell for horn and target system repairs. The beige areas indicate concrete shielding.
}
\label{fig:schematic-nu-beamline-ndc}
\end{figure}

Approximately 85\% of the protons interact with the solid
target, producing pions and kaons that subsequently get focused by a
set of magnetic horns into a decay pipe where they decay into muons
and neutrinos (Figure~\ref{fig:schematic-nu-beamline-ndc}). The
neutrinos form a wide-band, sign-selected neutrino or antineutrino
beam, designed to provide flux in the energy range of 0.5 to 5
GeV. This energy range will cover the first and second
neutrino-oscillation maxima, which for a \kmadj{1300} baseline are at
approximately 2.5 and 0.8 GeV, respectively.

\begin{figure}[!htb]
\centerline{
\includegraphics[width=0.5\textwidth]{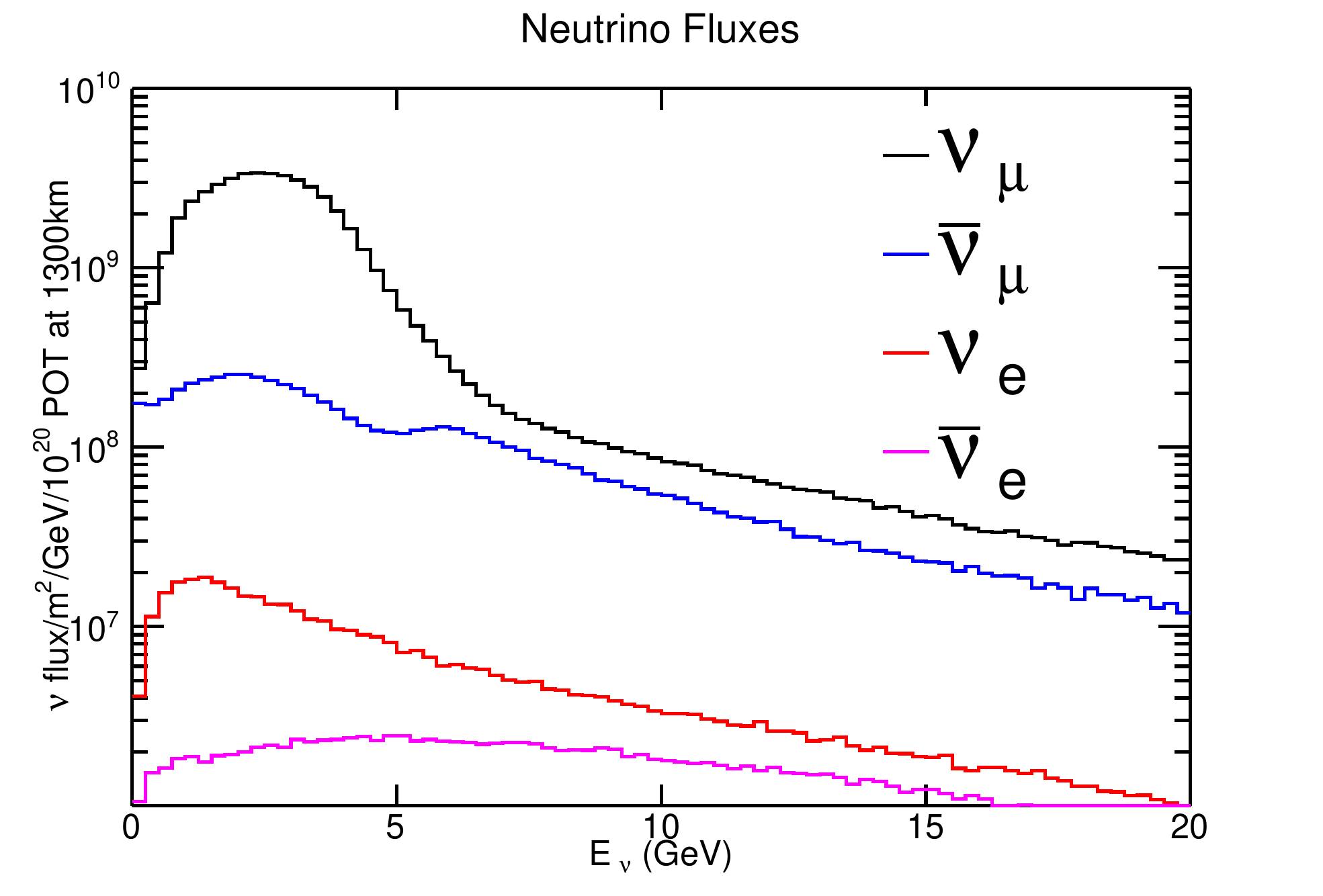}
\includegraphics[width=0.5\textwidth]{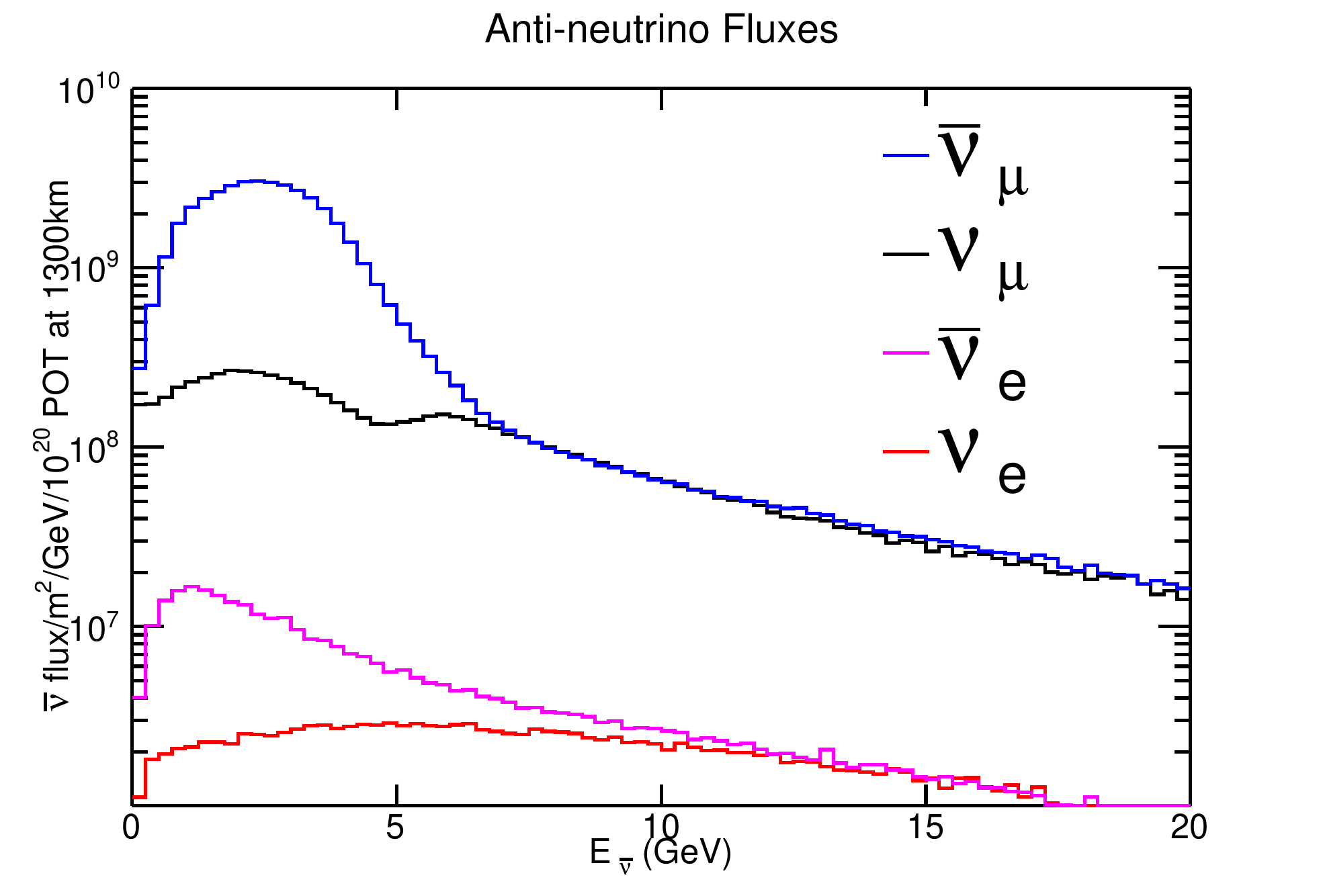}
}
\caption[Neutrino fluxes from the LBNE beam]{The neutrino beam fluxes
  (left) and antineutrino beam fluxes (right) produced by a Geant4
  simulation of the LBNE beamline. The horn current assumed is 200~kA,
  the target is located 35~cm in front of horn 1, the decay
  pipe is air-filled, \SI{4}{\m} in diameter and \SI{204}{\m} in length.}
\label{fig:cdrflux}
\end{figure}

\begin{figure}[!htb]
\centering
\includegraphics[width=0.8\textwidth]{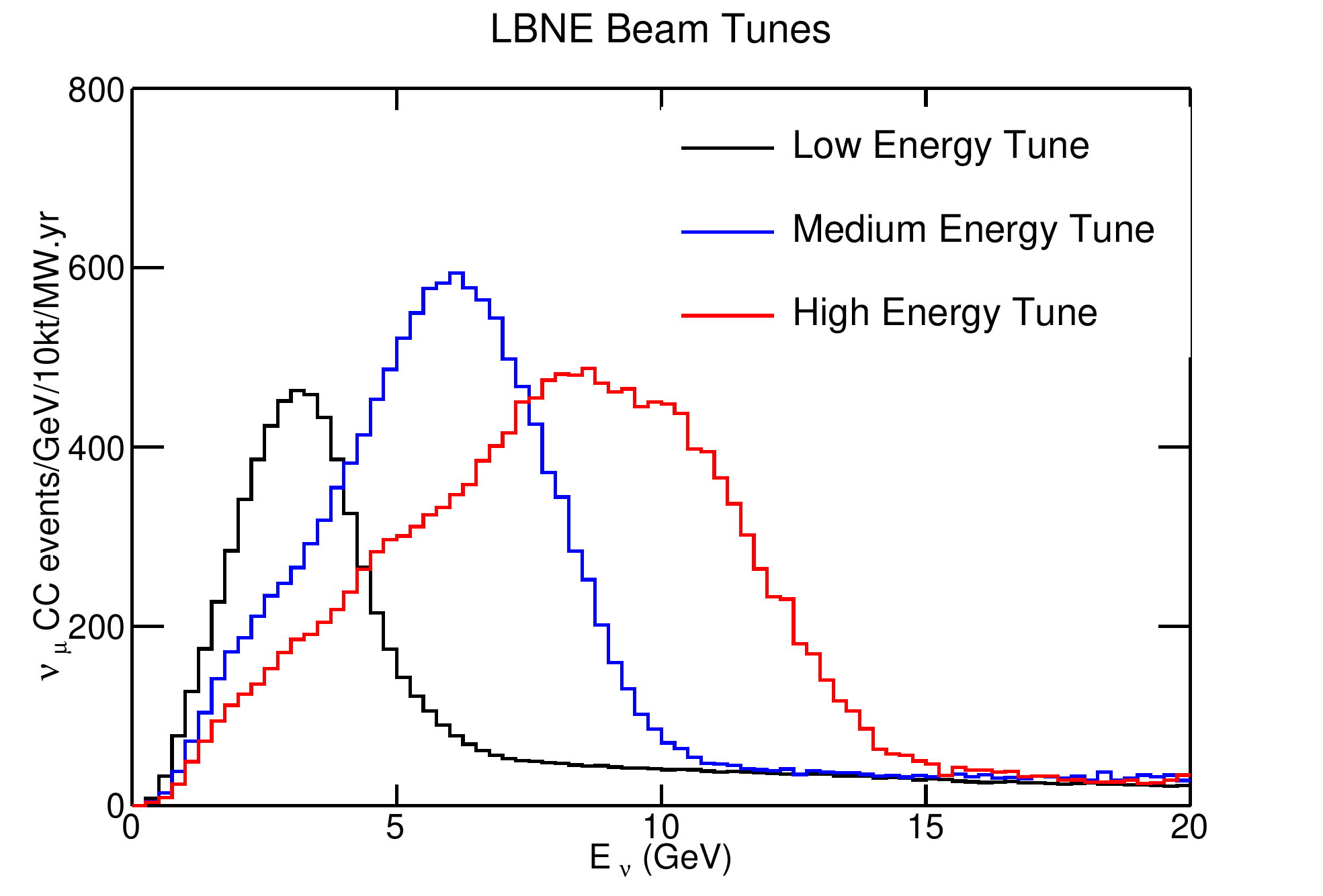}
\caption[Neutrino beam spectra from different beam tunes]{
  Event interaction rates at the LBNE far detector in the absence of oscillations
  and due to neutrinos produced by a \SI{120}{\GeV} proton beam for several target positions
  relative to Horn 1.  The black curve shows the expected interaction spectrum for the
 low-energy tune
  (LE) where the upstream face of the target is located \SI{35}{cm} upstream  
  of Horn 1, the blue curve is a sample medium-energy
  (ME) tune with the target located \SI{1.5}{m} upstream of Horn 1 and the red
  curve is the high-energy tune (HE) with the target
  located \SI{2.5}{m} upstream of Horn 1.  The horn current assumed is
  200~kA, the decay pipe is air-filled, \SI{4}{\m} in diameter and \SI{204}{\m} in
  length.}
\label{fig:beamtunes}
\end{figure}

The reference target design for LBNE is an upgraded version of the
NuMI-LE (Low Energy) target that was used for eight years to deliver
beam to the MINOS experiment. The target consists of 47 segments, each
2~cm long, of POCO graphite ZXF-5Q. Focusing of charged particles is
achieved by two magnetic horns in series, the first of which partially
surrounds the target. They are both NuMI/NO$\nu$A-design horns with
double-paraboloid inner conductor profiles. The NuMI/NO$\nu$A-design horns 
currently operate at \SIrange{185}{200}{\kA}.  The horns have been
evaluated and found to be operable with currents up to \SI{230}{\kA} but the
striplines that supply the horn currents are still under evaluation.
Additional development of the target and horns is required to adapt
the existing designs from the \kWadj{700} beam power used by NO$\nu$A
to \SI{1.2}{\MW} for LBNE.
The horn current polarity
can be changed to selectively focus positive or negative hadrons, thus
producing high purity ($> 90\%$ in oscillation region) $\nu_\mu$ or
$\overline{\nu}_\mu$ beams. Each beam polarity will have a $< 10\%$
contamination of neutrinos of the ``wrong sign'' in the oscillation
energy region ($\overline{\nu}$'s in the $\nu$ beam and vice-versa)
from decays of wrong-sign hadrons that propagate down the center of
the focusing horns --- where there is no magnetic field --- into the decay
volume. 
In addition, a $\leq 1\%$ contamination of $\nu_e$ and $\overline{\nu}_e$
in the $\nu_e$ appearance signal region is produced by the decays of
tertiary muons from pion decays, and decays of kaons.  The neutrino
flux components from the LBNE CD-1 beamline design produced using a
full Geant4 simulation of both horn polarities are shown in
Figure~\ref{fig:cdrflux}. 
The beamline design provides a wide-band neutrino beam with a peak
flux at 2.5~GeV, which matches the location of the first $\nu_\mu
\rightarrow \nu_e$ oscillation maximum. The NuMI reference target
design used for LBNE allows the target to be moved with respect to
Horn 1. The location of the upstream face~\footnote{The proton beam
  direction determines the upstream and downstream conventions. The
  upstream (front) face of Horn 1 is therefore the Horn 1 face closest to the
  proton beam window.} of the target with respect to the
upstream face of Horn 1 can be varied from $-$35~cm (default location)
to $-$2.85~m, thus the LBNE beamline can produce a wide range of beam
spectra.  Three possible far-site beam spectra, produced by moving the
target from $-$35~cm (low-energy) to $-$1.5~m (medium-energy) to
$-$2.5~m (high energy) are shown in Figure~\ref{fig:beamtunes}.

The decay volume design for LBNE is a helium-filled, 
air-cooled pipe of circular cross section with a diameter of 4~m and
length from \SI{204}{m} to \SI{250}{m} optimized such that decays of the pions and
kaons result in neutrinos in the energy range useful for the
experiment. A \SIadj{250}{\meter} decay pipe is the maximum length that
will allow the near neutrino detector complex to fit within the
Fermilab site boundaries. At the end of the decay region, the
absorber, a water-cooled structure of aluminum and steel, is designed
to remove any residual hadronic particles; it 
must absorb a large fraction of the incident beam power of up to
2.3~MW. Instrumentation immediately upstream of the absorber measures
the transverse distribution of the resultant hadronic showers to
monitor the beam on a pulse-by-pulse basis.

An array of muon detectors in a small alcove immediately downstream of
the absorber measures tertiary-beam muons and thereby indirectly
provides information on the direction, profile and flux of the
neutrino beam. This will be described in Section~\ref{sec:ndproj}.

The beamline conventional facilities include the civil
construction required to house the beamline components in their
planned layout as shown in Figures~\ref{fig:nscf_layout}
and \ref{v-beam-fig:intro_elev_overview}.  Following the beam from
southeast to northwest, or roughly from right to left in
Figure~\ref{fig:nscf_layout}, the elements include the underground
Extraction Enclosure, the Primary Beam Enclosure (inside the
embankment) and its accompanying surface-based Service Building (LBNE
5), the Target Complex (LBNE 20) located in the embankment, the Decay
Pipe, the underground Absorber Hall with the muon alcove, and its
surface-based Service Building (LBNE 30). The embankment will need to
be approximately 290~m long and 18~m above grade at its peak.  The
planned near neutrino detector facility is located as near as is
feasible to the west site boundary of Fermilab, along the line-of-sight 
indicated in red in Figure~\ref{fig:nscf_layout}. 

The parameters of the beamline facility were determined taking into
account several factors including the physics goals, the Monte Carlo
modeling of the facility, spatial and radiological constraints and the
experience gained by operating the NuMI facility at Fermilab. The
relevant radiological concerns, prompt dose, residual dose, air
activation and tritium production have been extensively modeled and
the results implemented in the system design. The beamline facility
design described above minimizes expensive underground construction
and significantly enhances capability for ground-water radiological
protection.  In general, components of the LBNE beamline system that
cannot be replaced or easily modified after substantial irradiation are 
being designed for \MWadj{2.3} operation. Examples of such
components are the shielding of the target chase and decay pipe, and
the absorber with its associated shielding.

The following LBNE beamline design improvements beyond the CD-1
conceptual design are being assessed:
\begin{itemize}
\item An increase in the length of the decay pipe up to 250~m (the maximum
  length allowed by the existing Fermilab site boundaries), and also
  possibly an increase in its diameter up to 6~m. Increases to the decay
  pipe size would require additional cost of the order several tens of
  millions of dollars. Increasing the length of the decay pipe from
  200 to 250~m increases the overall event rate in the oscillation
  region by 12\%. Increases in the decay pipe diameter produce a 6\%
  increase in the low-energy neutrino event rate as shown in Table~\ref{tab:heairtable}.

\item It has recently been decided to fill the decay pipe with helium instead of air. The total
  $\nu_\mu$ event rate increases by about 11\%, with a decrease in
  $\overline\nu$ contamination in the neutrino beam.  Introducing helium
  in the decay pipe requires the design and construction of a
  decay pipe window.

\item An increase in the horn current of the horns by a modest
  amount (from 200$\,$kA to 230$\,$kA); this is expected to increase
  the neutrino event rates by about 10-12\% at the first oscillation
  maximum~\cite{docdb-6599}. A Finite Element Analysis simulation and
  a cooling test of the horns are underway to evaluate this option.

\item Use of an alternate material to the POCO graphite for the target to
  increase the target longevity. This would involve additional R\&D
  effort and design work. A beryllium target, for example, could be made
  shorter, potentially improving the horn focusing.

\item Development of more advanced horn designs that could boost the 
 low-energy flux in the region of the second oscillation maximum.
 It should be noted that the target and horn systems can be modified 
 or replaced even after operations have begun if improved designs 
 enable higher 
 beam flux.

\end{itemize}

Table~\ref{tab:heairtable} summarizes the impact of the beam design
improvements after CD-1 and the additional costs required.  Together,
the changes are anticipated to result in an increase of $\sim 50\%$ in
the $\nu_e$ appearance signal rate at the far detector. A 30\%
increase in signal event rate at the far detector can be achieved for $<$ 10 M\$ 
without changing the CD-1 decay pipe size (\SI{4}{\meter} diameter $\times$ \SI{204}{\meter}
length) 
by changing from an air-filled to a helium-filled decay pipe. Increasing the decay pipe
size to \SI{6}{\meter} diameter $\times$ \SI{250}{\meter} length would result in an
additional 15\% increase in flux but would cost an additional $\sim
47$~M\$ --- this includes the cost of a redesigned absorber.
\begin{table}[!htb]
  \caption[Impact of the beam improvements on the $\nu_\mu \rightarrow 
  \nu_e$ CC appearance rates]{Impact of the beam improvements under study on the neutrino $\nu_\mu \rightarrow 
    \nu_e$ CC appearance rates at the far detector in the range of the first and second oscillation maxima,
shown as the ratio of appearance rates: the \emph{improved} rate divided by the rate from the beam design described in the Conceptual Design Report. }
\label{tab:heairtable}

\begin{center}
\begin{tabular}{$L^l^l^l^l} 
\toprule
\rowtitlestyle
Changes     &            0.5 to 2~GeV &         2 to 5~GeV & Extra Cost\\
\toprowrule

Horn current 200$\,$kA $\rightarrow$ 230$\,$kA &   1.00   &          1.12 & none \\ \colhline
Proton beam 120 $\rightarrow$ 80$\,$GeV at constant power & 1.14  &          1.05 & none \\ \colhline
Target NuMI-style graphite $\rightarrow$ Be cylinder 
&        1.10    &        1.00 & $<$ 1~M\$\\ \colhline
Decay pipe Air $\rightarrow$ He        &   1.07     &  1.11 & $\sim$ 8~M\$\\ \colhline
Decay pipe diameter 4$\,$m $\rightarrow$ 6$\,$m             & 1.06           &  1.02 & $\sim$ 17~M\$ \\  \colhline
Decay pipe length 200$\,$m $\rightarrow$ 250$\,$m      &   1.04   &         1.12 & $\sim$ 30~M\$\\ 
\toprule
\rowtitlestyle
Total       &            1.48    &        1.50 & \\ 
\bottomrule
\end{tabular}
\end{center}
\end{table}

\section{Near Detector}
\label{sec:ndproj}

\begin{introbox}{
A high-resolution near neutrino detector located approximately \SI{500}{\m} downstream of the LBNE neutrino production target,
as shown in Figure~\ref{v-beam-fig:intro_elev_overview}, is a key component of the full LBNE scientific program: 
\begin{itemize}
\item 
The near neutrino detector will enable the LBNE experiment to achieve its primary
scientific goals  --- in particular discovery-level sensitivity
to CP violation and high-precision measurements of the neutrino
oscillation parameters, including the unknown CP-violating phase, \deltacp. 
\item A rich program of LBNE physics measurements at the near detector
  will exploit the potential of high-intensity neutrino beams as
  probes of new physics.  
\end{itemize}
}
\end{introbox}

To achieve the precision required to make a significant advancement in the measurement of neutrino
oscillation parameters  over current
experiments  and to reach the desired
$5\sigma$ sensitivity to CP violation (discussed in
Chapters~\ref{nu-oscil-chap} and \ref{nd-physics-chap}), LBNE will need to
measure the unoscillated flux spectrum, to a few percent, for all neutrino species in
the beam: $\nu_\mu,\, \nu_e,\, \overline{\nu}_\mu$ and
$\overline{\nu}_e$. 
This requires a high-resolution, magnetized near neutrino detector
with high efficiency for identifying and measuring electrons and
muons. To measure the small $\nu_e$ contamination in the beam with
greater precision, the detector would need to be able to distinguish
$e^+$ from $e^-$; this would require a low-density detector with a
commensurately long physical radiation length.  In addition, 
use of an argon target nucleus --- similar to the far detector ---
would allow cancellation of systematic errors.
A reference design has been developed for a near neutrino detector that 
will meet these requirements; in particular it will measure the
neutrino event rates and cross sections on argon, water and other
nuclear targets for both $\nu_e$ and $\nu_\mu$ charged current (CC)
and neutral current (NC) scattering events. 

\begin{figure}[!htb]
\centering
\includegraphics[width=0.9\textwidth]{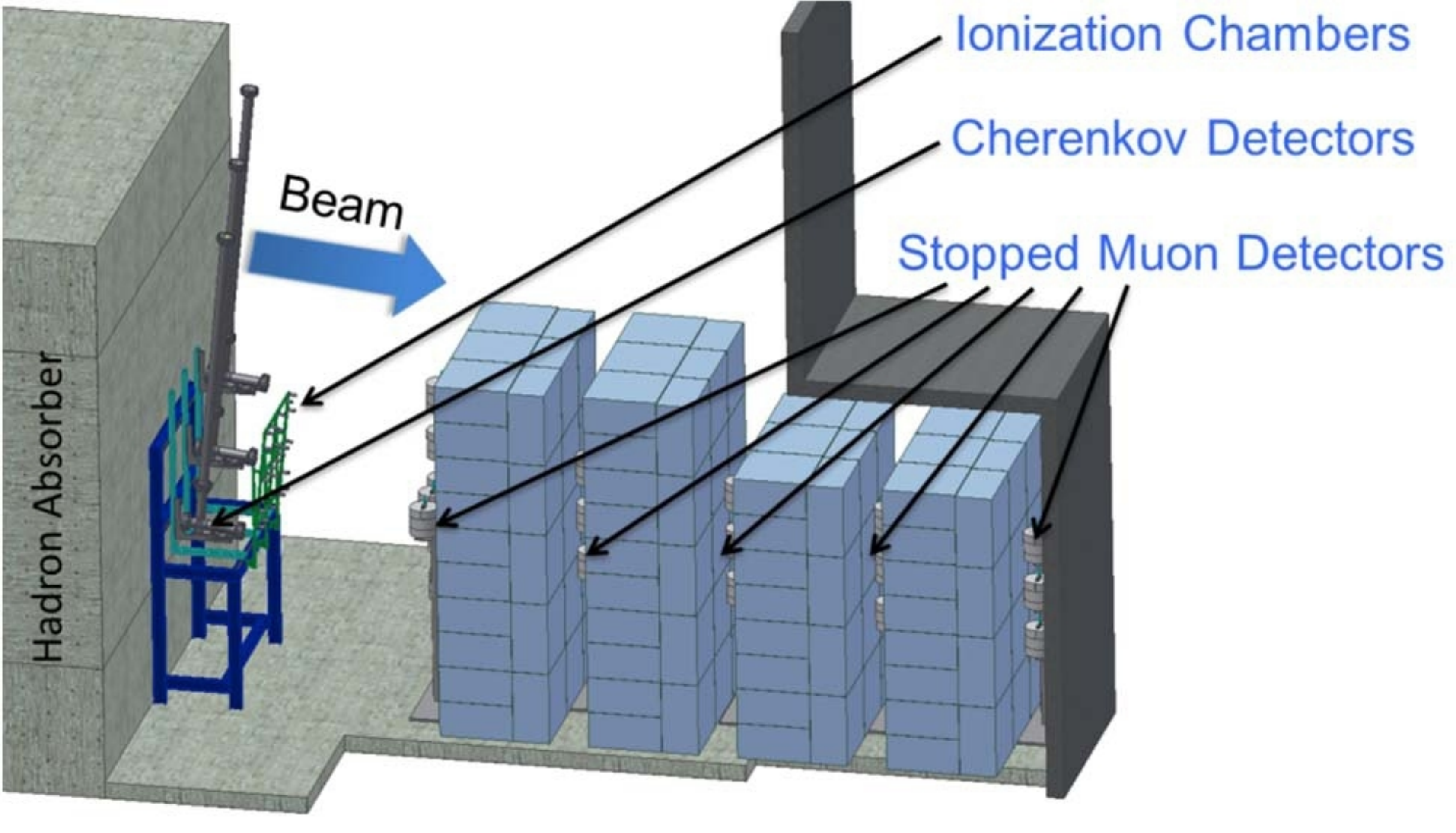}
\caption[System of tertiary-beam muon detectors]{System of tertiary-beam muon detectors, 
located downstream of the LBNE beamline absorber, for monitoring the muon flux from the LBNE beamline.} 
\label{fig:muons}
\end{figure}
In addition to the near neutrino detector, a sophisticated array of
muon detectors will be placed just downstream of the absorber.
The muon detectors, shown in Figure~\ref{fig:muons}, detect mostly
muons from the two-body decays of $\pi^{+(-)} \rightarrow \mu^{+(-)}
\nu_\mu (\overline{\nu}_\mu)$ in the beamline, thus the measured muon
and $\nu_\mu$ flux distributions are highly correlated.  The
ionization chamber array will provide pulse-by-pulse monitoring of the
beam profile and direction.  The variable-threshold gas Cherenkov
detectors will map the energy spectrum of the muons exiting the
absorber on an on-going basis.  The stopped muon detectors will sample
the lowest-energy muons, which are known to correlate with the
neutrino flux above 3~GeV --- equivalent to about half the neutrino
flux near the first oscillation maximum --- and a decreasing fraction
of it at lower energy.
This system, together with the existing level of understanding of the
similar NuMI beam and experience in previous neutrino oscillation
experiments, will provide additional constraints on the understanding
of the neutrino beam, and will thus support and complement the near
neutrino detector measurements.

\begin{figure}[!htb]
\centering\includegraphics[height=0.53\textheight]{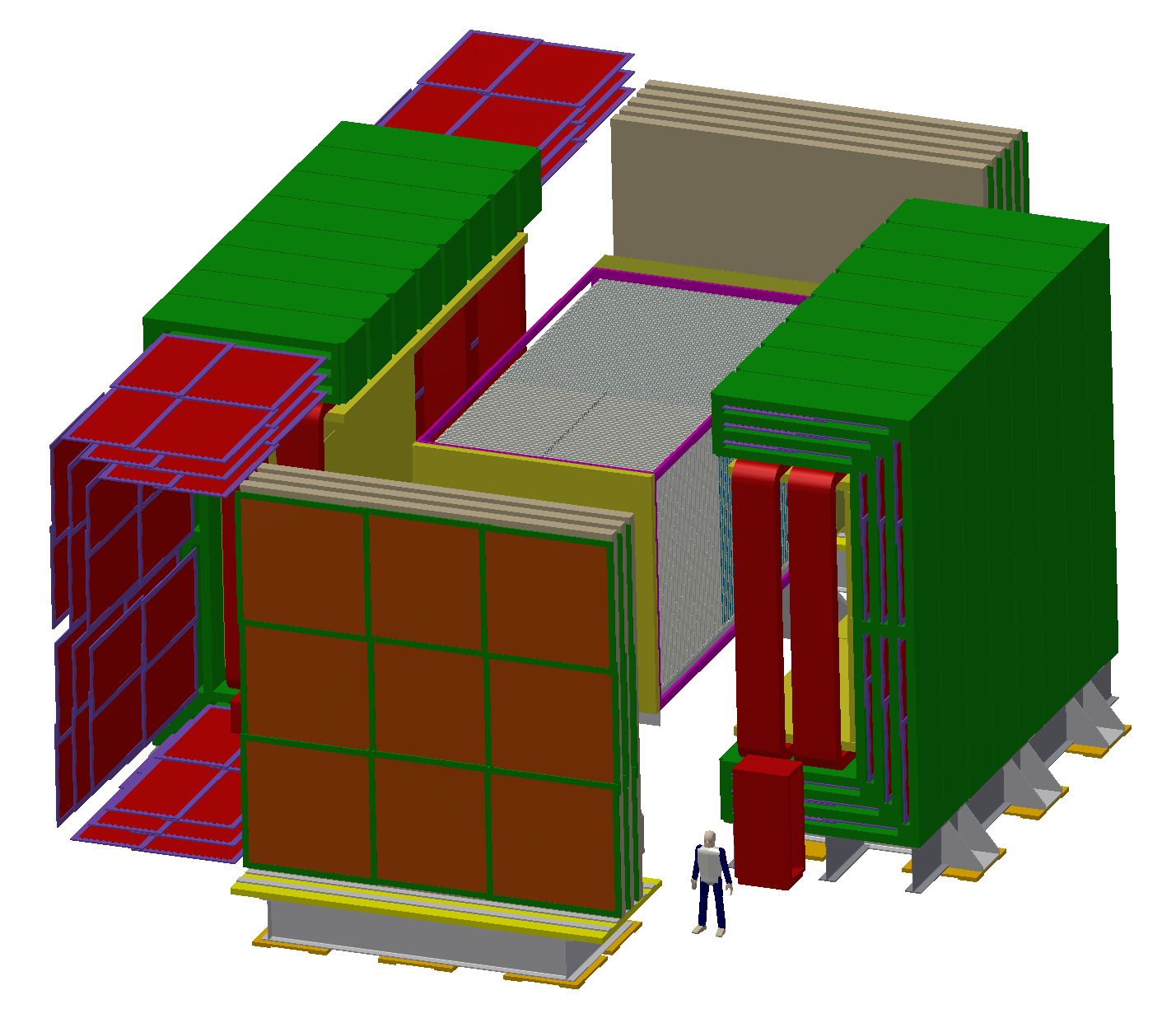}
\caption[Magnetized straw-tube tracker near detector design]{The LBNE
  near neutrino detector reference design with the dipole magnet open
  to show the straw-tube tracker (grey) and electromagnetic
  calorimeter (yellow).  RPCs for muon identification (red squares)
  are embedded in the yoke steel and up- and downstream steel walls.
}
\label{fig:nd_designs}
\end{figure}
\begin{figure}[!hbt]
\centering\includegraphics[height=0.4\textheight]{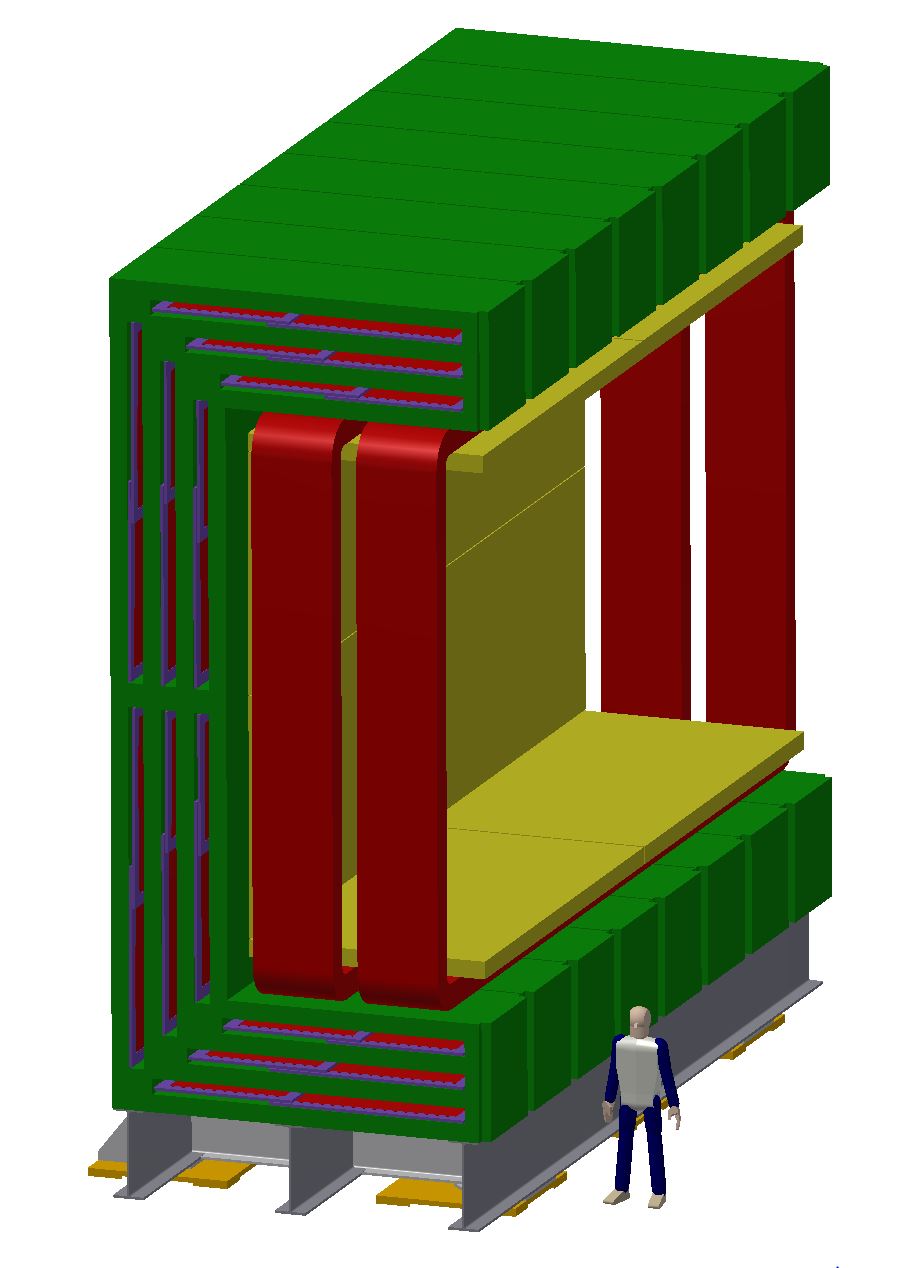}
\caption[Schematic drawing of near detector ECAL]{A schematic drawing of the ECAL (yellow modules) next to the magnet coils (red) and
MuID (blue modules) interspersed in the magnet steel (green).
}
\label{fig:nd-fgt-ecal}
\end{figure}
The reference design for the
near neutrino detector is a fine-grained tracker~\cite{Mishra:2008nx},
illustrated in Figure~\ref{fig:nd_designs}. 
It consists of a $3 \times 3 \times 7.04$ \si{\metre^{3}} straw-tube
tracking detector (STT) and electromagnetic calorimeter inside of a
0.4-T dipole magnet, illustrated in Figure~\ref{fig:nd-fgt-ecal}, and
resistive plate chambers for muon identification (MuID) located in the
steel of the magnet and also upstream and downstream of the tracker.
High-pressure argon gas targets, as well as water and other nuclear
targets, are embedded in the upstream part of the tracking volume.
The nominal active volume of the STT corresponds to eight tons of
mass.  The STT is required to contain sufficient mass of argon gas in tubes (Al
or composite material) to provide at least a factor of ten more statistics than
expected in the far detector.
Table~\ref{table:fgt-perf} summarizes the performance for the
fine-grained tracker's configuration, and Table~\ref{table:fgt-params}
lists its parameters.
\begin{table}[!htb]
\caption{Summary of the performance for the fine-grained tracker configuration}
\label{table:fgt-perf}
\begin{tabular}[!htb]{$L^l^l}
\toprule
\rowtitlestyle
Performance Metric & Value  \\
\toprowrule
Vertex resolution &  \SI{0.1}{\milli\meter} \\ \colhline
Angular resolution &  \SI{2}{\milli\radian} \\ \colhline
$E_e$ resolution & 5\% \\ \colhline
$E_\mu$ resolution & 5\% \\  \colhline
$\nu_\mu$/$\overline{\nu}_\mu$ ID &  Yes \\ \colhline
$\nu_e$/$\overline{\nu}_e$ ID &  Yes \\ \colhline
NC$\pi^0$/CC$e$ rejection &  0.1\% \\ \colhline
NC$\gamma$/CC$e$ rejection &  0.2\% \\ \colhline
NC$\mu$/CC$e$ rejection &  0.01\% \\ 
\bottomrule
\end{tabular} 
\end{table}

\begin{table}[!htb]
  \caption[Parameters for the fine-grained tracker]{Parameters for the fine-grained tracker. 
}
\label{table:fgt-params}
\begin{tabular}[!htb]{$L^l^l}
\toprule
\rowtitlestyle
Parameter & Value  \\
\toprowrule
STT detector volume &  $3 \times 3 \times 7.04$ \si{\metre^{3}}\\ \colhline
STT detector mass &  8 tons\\ \colhline
Number of straws in STT & 123,904 \\ \colhline
Inner magnetic volume &   $4.5 \times 4.5 \times 8.0$ \si{\metre^{3}} \\ \colhline
Targets & 1.27-cm thick argon ($\sim$ \SI{50}{\kg}), water and others \\ \colhline
Transition radiation radiators & 2.5 cm thick \\  \colhline
ECAL $X_0$&   10 barrel, 10 backward, 18 forward \\ \colhline
Number of scintillator bars in ECAL &  32,320 \\ \colhline
Dipole magnet &   2.4-MW power; 60-cm steel thickness \\ \colhline
Magnetic field and uniformity &  0.4 T; < 2\% variation over inner volume \\ \colhline
MuID configuration &  32 RPC planes interspersed between 20-cm thick \\
& layers of steel \\  
\bottomrule
\end{tabular} 
\end{table}

Figure~\ref{fig:nd-fgt-ecal} 
shows the locations of the electromagnetic calorimeter and MuID next to the
magnet steel and magnet coils. 
The fine-grained tracker has excellent position and angular
resolutions due to its low-density ($\sim$ \SI{0.1}{g/cm^3}), high-precision STT. The low density
and magnetic field allow it to distinguish $e^+$ from $e^-$ on an event-by-event basis.
The high resolution is important for determining the neutrino vertex and determining whether the neutrino
interaction occurs in a water or argon target.
Electrons are distinguished from hadrons using transition radiation. 

The design of the near neutrino detector is the subject of study by
the LBNE Collaboration, and alternatives such as a magnetized liquid argon
TPC will be investigated further. A detailed description of the
fine-grained tracker can be found in~\cite{docdb-6704}, and
descriptions of it and the alternative LArTPC design are
presented in the March 2012 LBNE CDR (Volume 3 of~\cite{CDRv3}).

High-intensity neutrino beams can be used as probes of new
physics and given the broad energy range of the
LBNE beam, a diverse range of physics measurements 
is possible in the high-resolution near neutrino
detector. These potentially wide-ranging physics measurements would
complement other physics programs, such as those at Jefferson Laboratory,
that are using proton, electron or ion beams from colliders and
fixed-target facilities. A detailed discussion of the physics
capabilities of a high-resolution near detector is presented in
Chapter~\ref{nd-physics-chap} and in~\cite{docdb-6704}.  

\section{Far Detector}
\label{sec:fdproj}

\begin{introbox}{ The full-scope LBNE far detector is a liquid argon time-projection
chamber of  fiducial mass \SI{34}{kt}
    located at the \SIadj{4850}{\ft} level of the Sanford Underground
    Research Facility.  The LArTPC technology allows for
    high-precision identification of neutrino flavors, offers
    excellent sensitivity to proton decay modes with kaons in the
    final state and provides unique sensitivity to electron neutrinos
    from a core-collapse supernova.  The full detector size and its
    location at a depth of 4,850 feet will enable LBNE to meet the
    primary scientific goals --- in particular, to find evidence for
    CP violation over a large range of \deltacp values, and to
    significantly advance proton-decay lifetime limits.  Conceptual
    designs of the \ktadj{34} underground detector are well developed.
  }
\end{introbox}

The liquid argon TPC technology chosen for LBNE combines fine-grained
tracking with total absorption calorimetry to provide a detailed view
of particle interactions, making it a powerful tool for neutrino
physics and underground physics such as proton decay and
supernova-neutrino observation.  It provides millimeter-scale
resolution in 3D for all charged particles.  Particle types can be
identified both by their $dE/dx$ and by track patterns, e.g., the
decays of stopping particles.  The modest radiation length (14~cm) is
sufficiently short to identify and contain electromagnetic showers
from electrons and photons, but long enough to provide good $e/\gamma$
separation by $dE/dx$ (one versus two minimum ionizing particles) at
the beginning of the shower.  In addition, photons can be
distinguished from electrons emanating from an event vertex by the
flight path before their first interaction. These characteristics
allow the LArTPC to identify and reconstruct signal events with high
efficiency while rejecting backgrounds to provide a high-purity data
sample. The principal design parameters of the full-scope LBNE LArTPC
far detector are given in Table~\ref{tab:param-summ-larfd}.
\begin{table}[!htb]
\caption[Parameters for LArTPC far detector]{Principal design parameters of the full-scope LBNE LArTPC far detector from~\cite{CDRv4}.}
\label{tab:param-summ-larfd}
\centering
 \begin{tabular}{$L^l}
\toprule
\rowtitlestyle
Parameter & Value  \\
\toprowrule
Total/Active/Fiducial Mass &   50/40/34~kt \\
\colhline
Number of Detector Modules (Cryostats) &  2 \\
\colhline
Drift Cell Configuration within Module &   3~wide $\times$ 2~high $\times$ 18~long drift cells\\
\colhline
Drift Cell Dimensions  &  2 $\times$ 3.7~m wide (drift) $\times$ 7~m high $\times$ 2.5~m long \\
\colhline
Detector Module Dimensions &  22.4~m wide $\times$ 14~m high $\times$  45.6~m long \\
\colhline
Anode Wire Spacing &  $\sim$5~mm \\
\colhline
Wire Planes (Orientation from vertical) & Grid (0$^\circ$), Induction 1 (45$^\circ$), Induction 2 (-45$^\circ$) \\
&  Collection (0$^\circ$) \\
\colhline
Drift Electric Field &  500~V/cm \\ 
\colhline
Maximum Drift Time & 2.3~ms \\
\bottomrule
\end{tabular} 
\end{table}

Scalability has been a design consideration of critical importance for
the LBNE Project, and for the far detector in particular, since the
Project's inception in 2009. 
A \ktadj{10} LArTPC far detector
configuration has been identified as the minimal
initial configuration of LBNE 
that can make significant advances toward the
primary scientific goals of LBNE.
Because of the scalability built into the LBNE design, other, more capable, 
configurations could be accomplished either in
the initial phase with the identification of additional resources, or
at a later stage. 

Other important considerations for the construction 
of LBNE's large LArTPC far detector include: 
\begin{enumerate}
\item cryogenic safety and the elimination of hazards associated with large cryogenic liquid volumes 
\item  attainment of stringent argon purity requirements 
 with respect to electronegative contaminants (e.g., $<0.2\,$ppb O$_2$ concentration)
\item ease of transport and assembly of TPC mechanical systems
\item efficient deployment of high-sensitivity/low-noise electronics for readout of the ionization 
signal
\end{enumerate}

The far detector complex for both the first-phase ($\geq$ \ktadj{10})
and full \ktadj{34} options will be outfitted with two separately
instrumented detector vessels instead of a single, larger vessel ---
an approach which has several benefits.  First, this design enables
each cryostat and TPC to be filled and commissioned while the other
remains available for liquid storage, allowing for repairs to be made
after the start of commissioning, should that be necessary.  Secondly,
it allows deployment of TPCs of different designs.  This would enable,
for example, international partners to contribute a detector of an
alternate design, based on their own experience, or one that would
emphasize a particular research interest.

The detector vessels will be constructed using technology standards
from the liquefied natural gas (LNG) industry.  With similar
requirements and geometries, adaptation of industrial LNG cryostat
design provides a high-performance, extensively tested approach to the
challenge of liquid argon containment for LBNE.  The cryostats in
large LNG tanker ships are constructed using a thin (1--2 mm),
polished, stainless steel inner membrane surrounded by thick foam
passive insulation.  With stainless steel as the only wetted surface,
this is an inherently clean design, ideal for liquid argon detectors
where high purity is essential.

The underground detector placement at the 4850L of the \SURF was
studied in detail during the Conceptual Design Phase of LBNE and
presented at the Fermilab Director's Independent Conceptual Design
Review in March of 2012~\cite{mar2012review}.  Significant effort has
been invested to minimize the (dominant) cost of the far site
conventional facilities.

\subsection{The \ktadj{10} Detector Design} 
\label{proj:10kt-design}

\begin{introbox}  
\begin{itemize} 
\item The far detector for the initial phase of LBNE will have fiducial mass of \emph{at
    least} \SI{10}{\kt}.  This mass allows for high probability
  determination of the neutrino mass hierarchy
  and can provide evidence for CP violation, if this effect is
  large.  \item The detector needs to be located deep underground to
  provide sensitivity for proton decay searches in the kaon modes and
  for measuring neutrinos from potential supernovae in the galaxy.
\item A conceptual design for a \ktadj{10} LArTPC has been developed,
  thoroughly reviewed and found to be sound.
\item LBNE is working with international partners in an effort to deploy a more
  massive far detector in the initial phase. 
\end{itemize}
\end{introbox}

\begin{figure}[!htb]
\centering
\includegraphics[width=\linewidth]{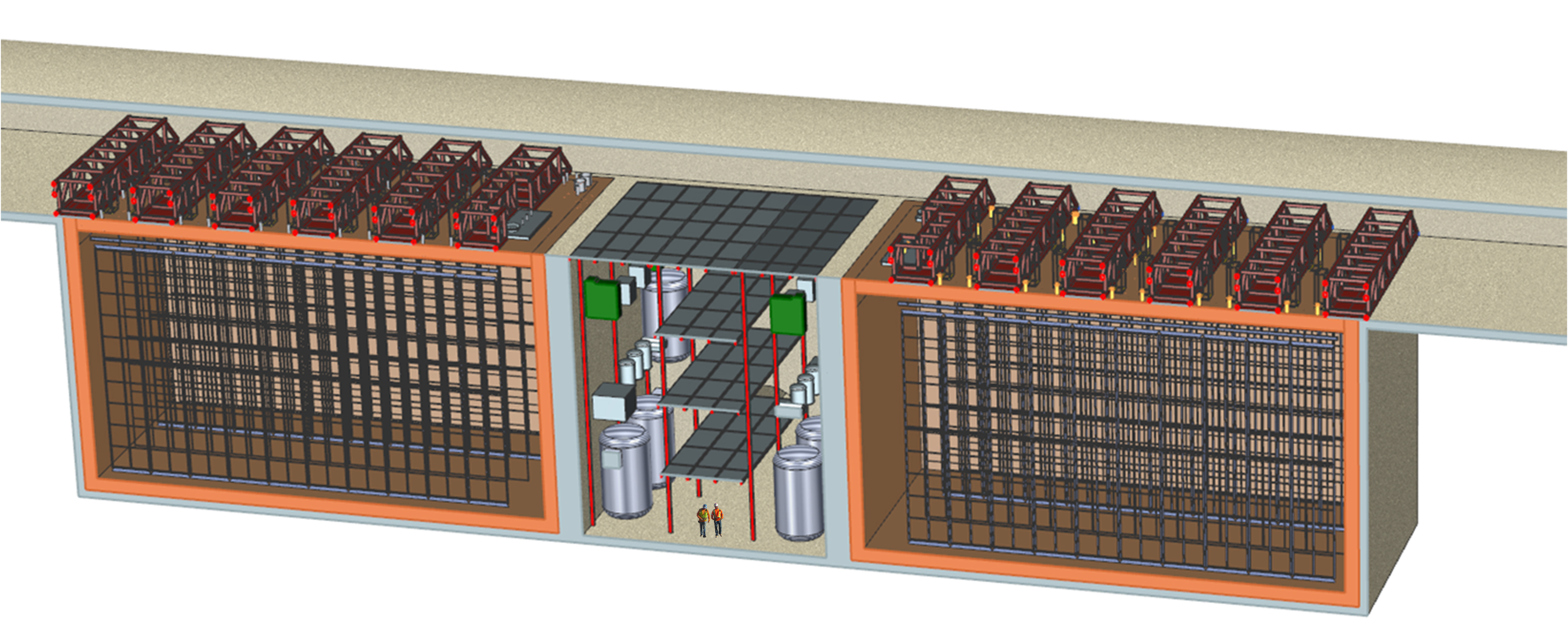}
\caption[View of a \ktadj{10} far detector showing the two vessels]{3D view
  of the \ktadj{10} far detector showing a lateral cross section of the two
  \ktadj{5} fiducial-mass LArTPC vessels }
\label{fig:10kton}
\end{figure}
A conceptual design for the initial \ktadj{10} far detector for the
first-phase LBNE Project has been developed that is easily scalable to
larger detectors.  Many of the detector elements, in particular the
modular TPC design and readout electronics, utilize full-scale modules
and designs that can easily be replicated in larger numbers to
instrument a larger detector.
This design consists of two \ktadj{9.4} liquid argon
vessels~\cite{CDRv4}, each designed to hold a \ktadj{5} fiducial-mass
LArTPC as shown in Figure~\ref{fig:10kton}.

The cryogenics systems for the \ktadj{10} detector will consist of two \kWadj{85} liquid nitrogen
liquefaction plants, a liquid argon receiving station, a liquid argon
circulation system with liquid purifiers, and a liquid argon
re-condensing system.  All the cryogenics systems
are similar to large-scale systems found in industrial applications.

\begin{figure}[!htb]
\centering
\includegraphics[width=\linewidth]{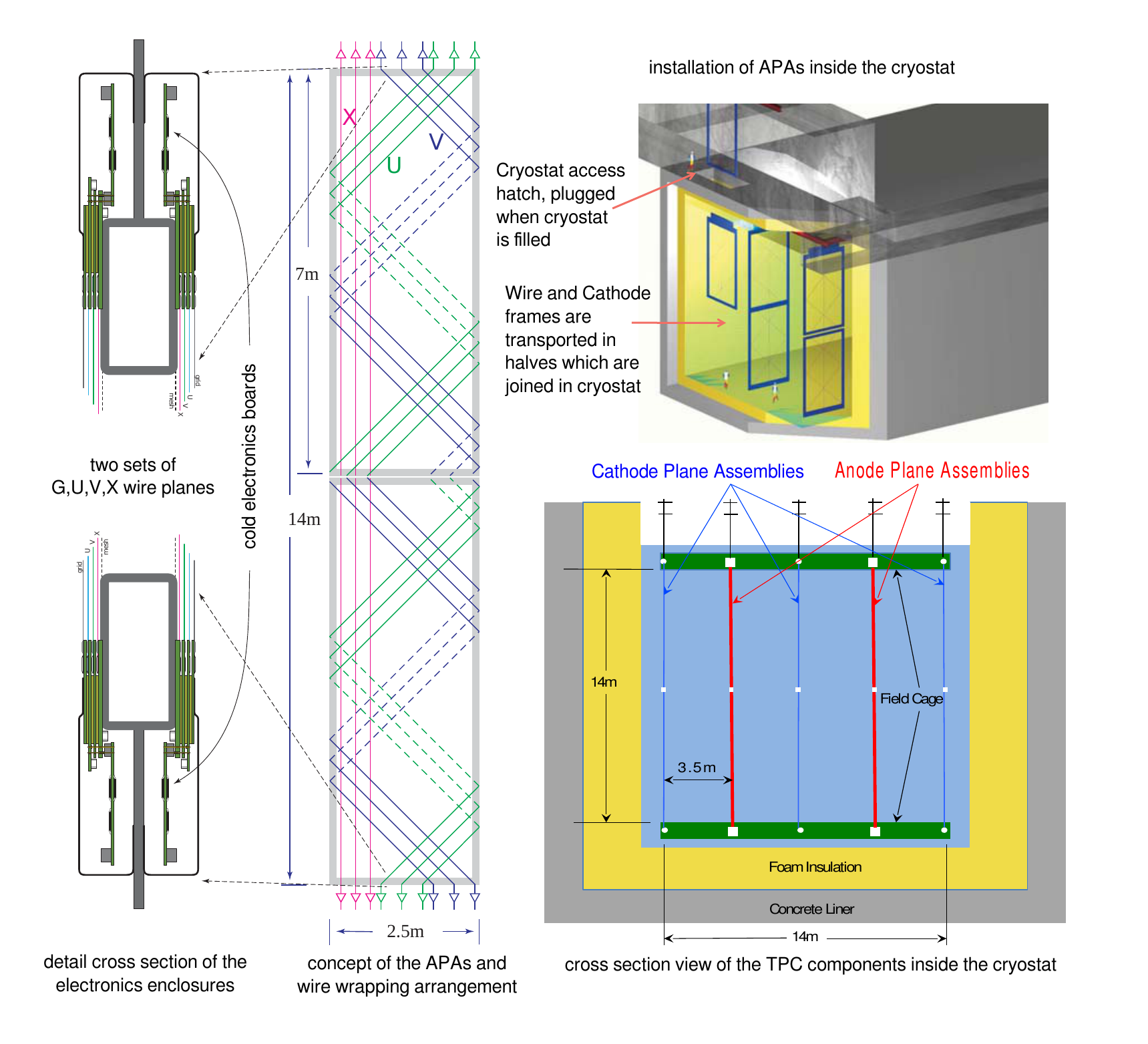}
\caption{The LBNE TPC modular construction concept}
\label{fig:tpc-concept}
\end{figure}

The LBNE TPC design for the \SIadj{10}{\kt} detector consists of three
rows of cathode plane assemblies (CPAs) interspersed with two rows of
anode plane assemblies (APAs), similar to the layout concept shown in
Figure~\ref{fig:tpc-concept} bottom right, with readout electronics
mounted directly on the APA frames (Figure~\ref{fig:tpc-concept},
left).  These elements run the length of a cryostat module, save for
space at one end allocated to the cryogenics systems.  A field cage
for shaping the electric field covers the top, bottom, and ends of the
detector.  The spacing between the CPA and APA rows is 3.48~m and the
cathode planes will be operated at 173~kV, establishing a drift field
of 500~V/cm and a corresponding maximum drift time of~2.16 ms.

The APAs and CPAs are designed in a modular fashion as illustrated in
Figure~\ref{fig:tpc-concept}, top right.  Each APA/CPA is constructed
with a support frame \SI{2.5}{\meter} long and \SI{7}{\meter} high;
these dimensions are chosen for ease of transportation to the detector
site and installation within the cryostat.  During installation, two
APAs are connected end-to-end to form a \SI{14}{\meter} tall,
\SI{2.5}{\meter} long unit, which is transported to its final position
in the detector and suspended there using a rail system at the top of
the detector.  Pairs of CPAs are installed in a similar fashion.  This
system of \SI{2.5}{\meter} long detector elements is easily scalable
to any desired detector size. A total of 40 APAs and 60 CPAs per
cryostat are needed for the \ktadj{10} detector design, configured as
two rows of APAs, ten APA pairs long.

Three sense wire planes (two \emph{induction} planes and one
\emph{collection} plane) with wire pitches of 4.8 mm are mounted on
each side of an APA frame, for sensitivity to ionization signals
originating within the TPC cell on either side.  The wires on these
planes are oriented vertically (collection) and at $\pm 45^\circ$
(induction)\footnote{The current design uses a $36^\circ$ orientation to
remove hit assignment ambiguities.}.  The induction plane wires are wrapped around the APA
frame, and are therefore sensitive to charge arriving from either side
of the APA, depending on where the charge arrives along the length of
the wires.  This configuration allows placement of
readout electronics at the top and bottom of each two-APA unit.
(Cables from the bottom APA are routed up through the support frame,
thereby eliminating any obstruction they would otherwise cause.)  In
this way, adjacent APA-pairs can be abutted so as to minimize the
uninstrumented region in the gaps between them along the length of the
detector.

Low-noise, low-power CMOS (Complementary Metal Oxide Semiconductor) 
preamplifier and ADC ASICS (Application Specific Integrated Circuit) have been
developed for deployment on circuit boards mounted directly on the APA
frames. This scheme ensures good signal-to-noise performance, even
allowing for some attenuation of long-drift ionization signals due to
residual impurities in the argon.  It also offers the possibility of
digital signal processing, including multiplexing and zero suppression
at the front end, thereby limiting the cable plant  within the cryostat
and the number of penetrations required, while also easing
requirements on the downstream readout/DAQ systems located outside the
cryostat.  The ASICS have been laid out following design rules
developed explicitly for long-term operation at cryogenic
temperatures.

In order to separate neutrino beam events from other interactions
--- particularly for proton decay and supernova neutrino signals ---
it is necessary to accurately determine the event time relative to the
neutrino beam time window or an incoming cosmic muon.  If the event
time is known at the microsecond level then out-of-time cosmic-ray
backgrounds for beam neutrinos can be rejected to the level of
$10^{-5}$ (the beam spill duty factor).  The slow ionization-electron
drift velocity gives the TPC its 3D imaging capability, but an
independent fast signal is required to localize events in time and in
space along the drift direction.  The excellent scintillation
properties of liquid argon (${\cal O}(10^4)$ photons per MeV of energy
deposition) are exploited to address this issue.  A photon detection
system is planned for detection of the 128-nm scintillation light
that, in turn, allows determination of the event timing.  Several
photon detector designs are under study.  The most advanced
design uses cast acrylic bars coated with wavelength shifter, and
SiPMs (silicon photomultipliers) at the ends for read-out.  These bars
will be assembled into paddles of dimensions 10 cm by 2 m, and mounted
on the APA frames, fitting within the 5-cm gap between the sets of
wire planes located on both sides of the frames.  Initial studies
indicate a light yield of 0.1 to 0.5 photoelectrons per MeV.

\subsection{The \ktadj{34} Detector Design}

\begin{figure}[!htb]
\centering
\includegraphics[width=\linewidth]{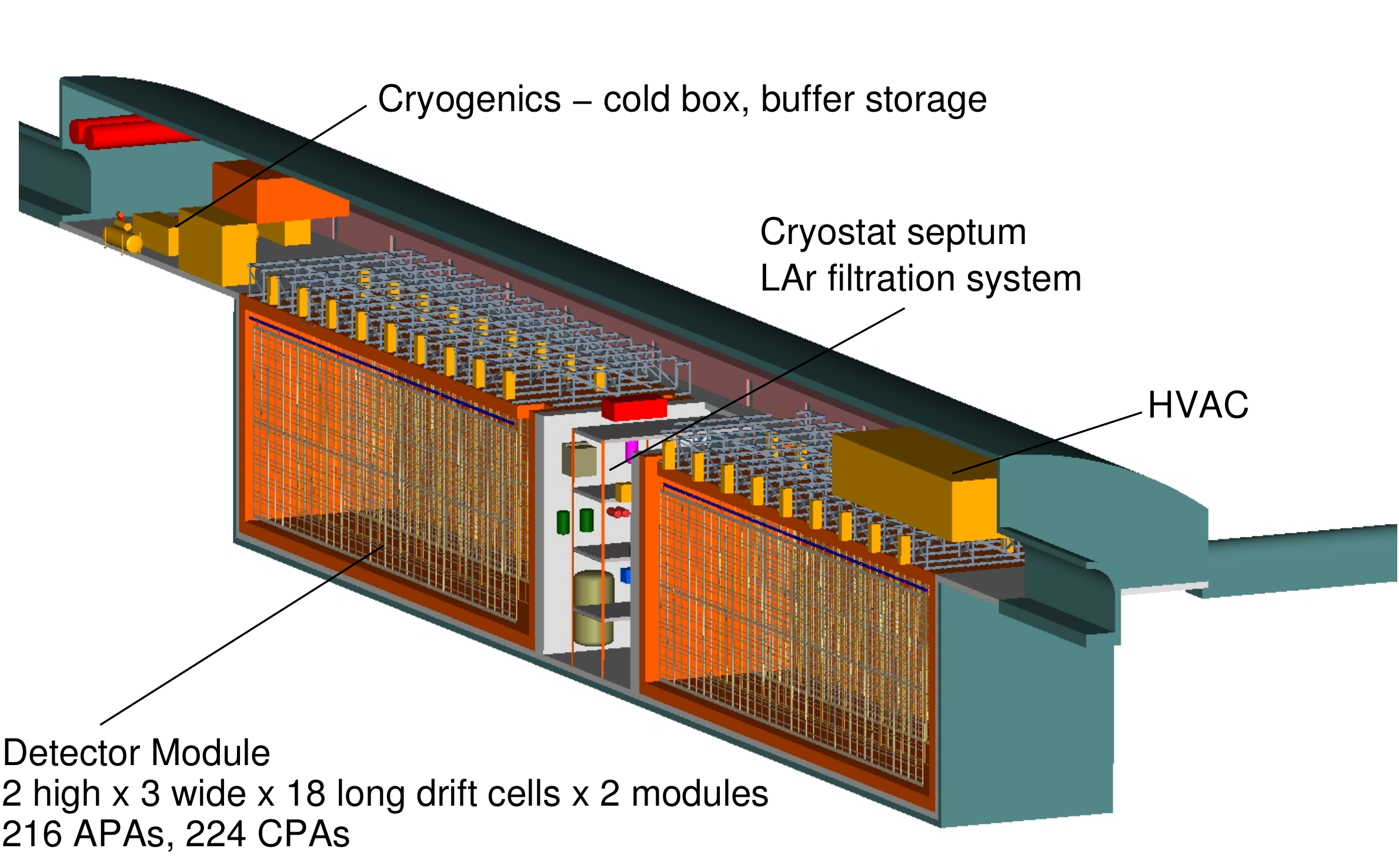}
\caption[Schematic of a \ktadj{34} LArTPC design]{Schematic of a \ktadj{34}
  LArTPC design. The detector comprises two \ktadj{17} 
LArTPC
  vessels.}
\label{fig:fd_34kt}
\end{figure}

One possible design of a \ktadj{34} detector is two \ktadj{17} modules placed 
end-to-end in a common cavern at the \ftadj{4850} level of the \SURF,
as shown in Figure~\ref{fig:fd_34kt}. This design was
reviewed at the Fermilab Director's Independent Conceptual Design
Review in March of 2012~\cite{mar2012review}.  

\begin{figure}[!htb]
\centering\includegraphics[width=\linewidth]{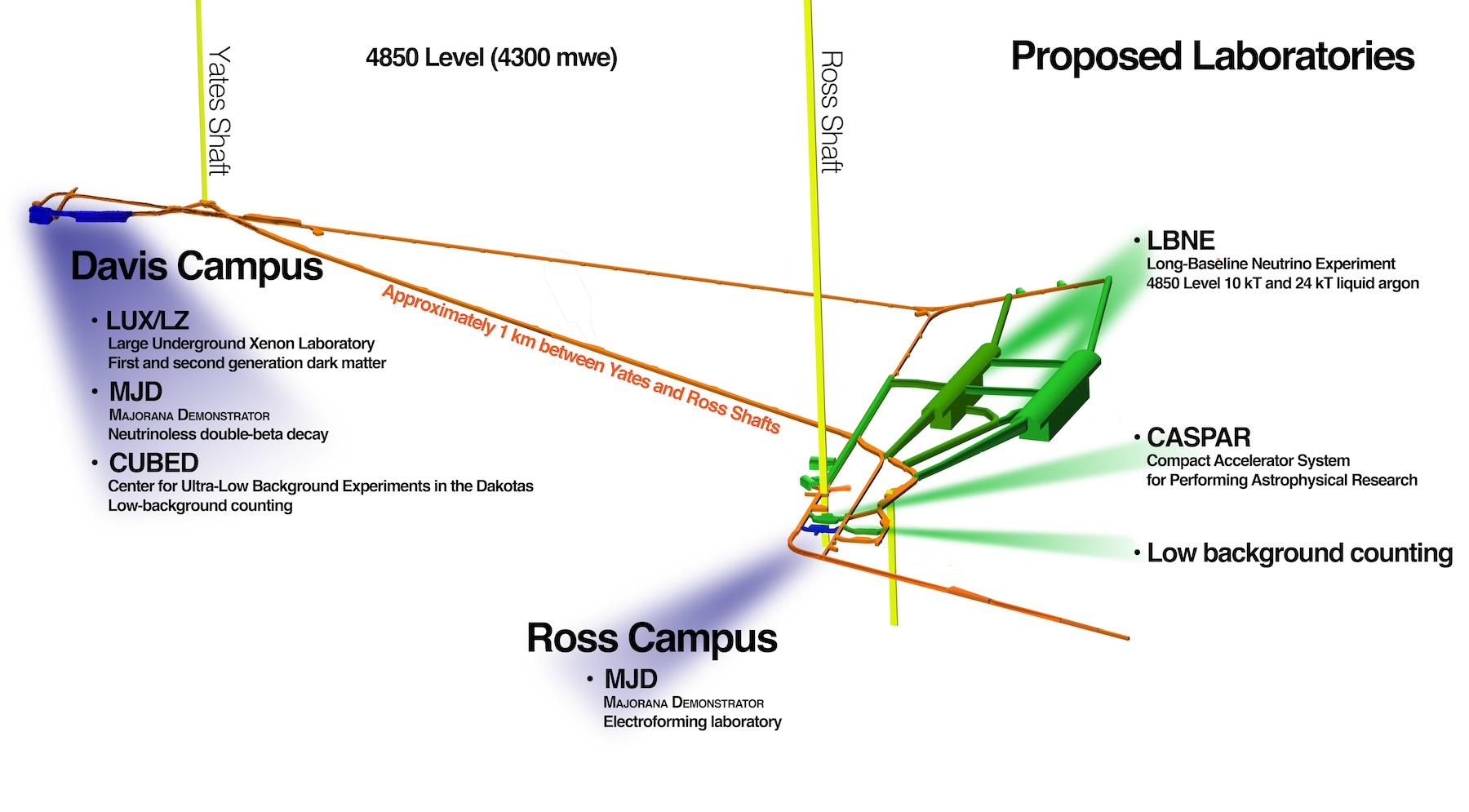}
\caption[Layout of the \ktadj{10} + \ktadj{24} LArTPC detector halls at
the \ftadj{4850} level]{Layout of the \ktadj{10} + \ktadj{24} LArTPC
  detector halls at the \ftadj{4850} level of the \SURF.}
\label{fig:fd_4850}
\end{figure}

Alternatively, the \ktadj{34} detector can be realized by adding a
roughly \ktadj{24} detector of essentially the same design as the
\ktadj{10} detector, housed in a set of two cryostats, each holding
\SI{12}{kt} (\SI{20}{kt} total) of liquid argon.  In this
configuration the additional cryostats each have three APA rows (total
84 APAs) and four CPA rows (total 112 CPAs), making them wider than
the \ktadj{10} design described in Section~\ref{proj:10kt-design}.
The APA-to-CPA row spacing is expanded to 3.77~m and the length of
each is increased to 14 APA units long.The cryogenics system installed
for the \ktadj{10} design will simply be expanded from two to four
\kWadj{85} refrigerators to service both the \ktadj{10} and the
\ktadj{24} detector. The \ktadj{24} detector hall will be excavated
parallel to the \ktadj{10} detector hall as shown in
Figure~\ref{fig:fd_4850}.

Given the modular design of the detector and the use of industrial
technologies in the cryogenics system, there is a great deal of
flexibility in possible contributions from new partners to expand the
size of the detector. The details of any scope change would depend on the interests,
capabilities and resources of the new partners.

\begin{figure}[!htb]
\centering
\includegraphics[width=\textwidth]{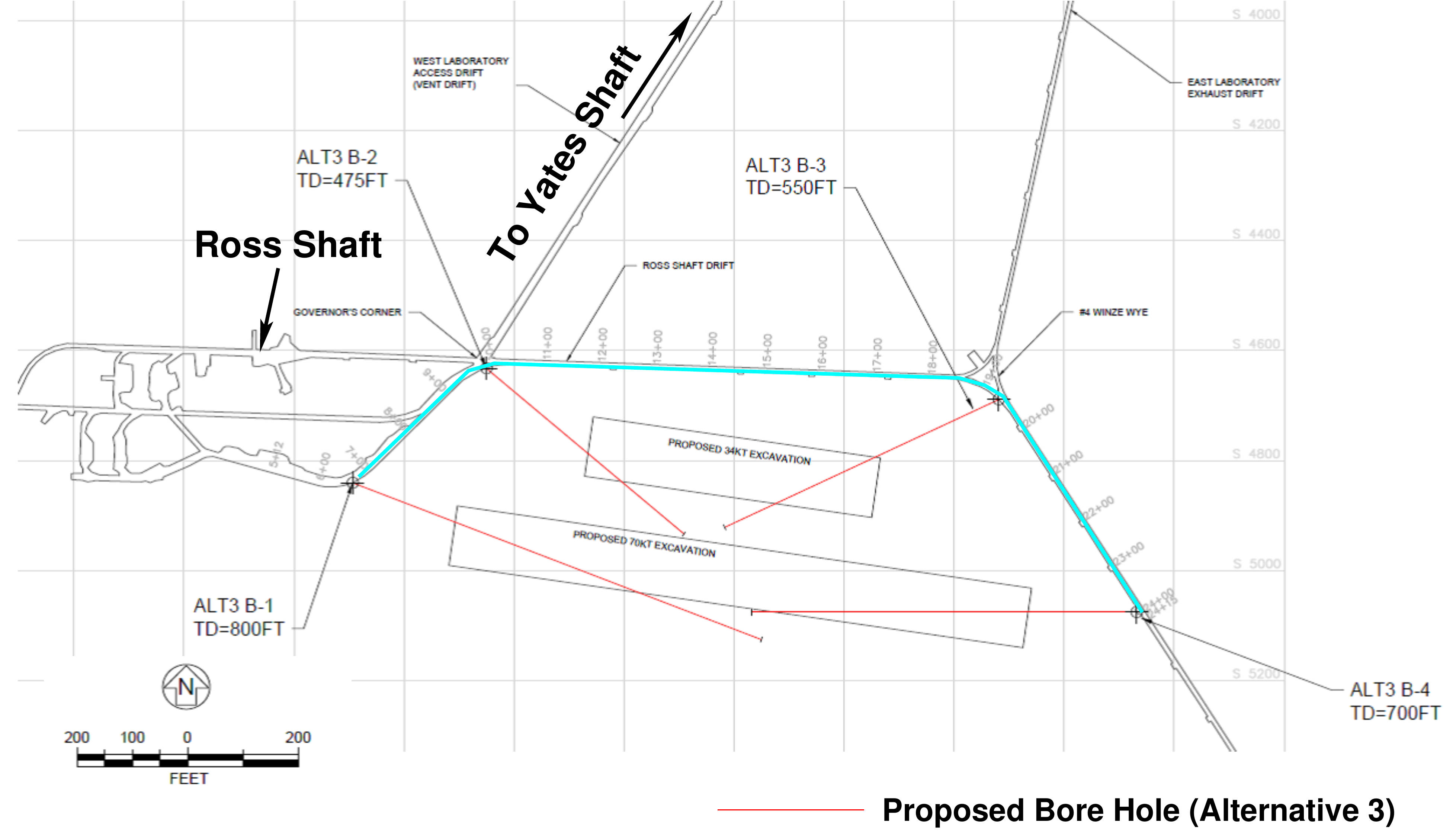}
\caption[Far site geotechnical investigation plan] {Geotechnical
  site investigation plan, showing the drifts that have been mapped
  (blue) and the planned core borings (red) overlaid on possible
  locations of caverns that would accommodate the \ktadj{34} or
  larger (\SIadj{70}{\kt} shown as an example) LArTPC detectors.}
\label{fig:fd_4850_100kt}
\end{figure}

A full geotechnical site investigation is underway to characterize the
rock mass in which it is planned to site the LBNE far detector.
Mapping of existing drifts in the vicinity of the proposed detector
location has been completed and a core boring program was launched
in early 2014. This investigation will explore the area with enough
breadth to allow flexibility in siting and sizing detector modules in
the future before design work begins. The proposed boring layouts are
shown in Figure~\ref{fig:fd_4850_100kt} overlaid with possible
\ktadj{34} and \ktadj{70} modules to demonstrate the large
capacity of this location.

%% file: chapter-nu-oscil.tex
\begin{introbox}
  LBNE is designed to address the science of neutrino oscillations with
  superior sensitivity to many mixing parameters in a single
  experiment, in particular,

\begin{enumerate}
\item precision measurements of the parameters that
  govern $\nu_{\mu} \rightarrow \nu_e$ and $\overline\nu_{\mu} \rightarrow \overline\nu_e$ oscillations; this includes
  precision measurement of the third mixing angle $\theta_{13}$,
  measurement of the
  CP-violating phase \deltacp, and determination of the mass
  ordering (the sign of $\Delta m^{2}_{32}$)

\item precision measurements of $\sin ^2 2 \theta_{23}$ and |$\Delta
  m^{2}_{32}$| in the $\nu_{\mu}/\overline\nu_{\mu}$ disappearance channel

\item determination of the $\theta_{23}$ octant using combined precision measurements of the
$\nu_{e}/\overline\nu_e$ appearance and $\nu_{\mu}/\overline\nu_{\mu}$ disappearance channels

\item search for nonstandard physics that can manifest itself as
  differences in higher-precision measurements of $\nu_\mu$
  and $\overline{\nu}_\mu$ oscillations over long baselines
\end{enumerate}

\end{introbox}

\section{Experimental Requirements Based on Oscillation Phenomenology} 

The experimental requirements for designing a neutrino oscillation
experiment to simultaneously address neutrino CP violation and the
mass hierarchy (MH) can be extrapolated as follows from the phenomenology
summarized in Chapter~\ref{intro-chap}:

\begin{enumerate}
\item \emph{Phenomenology: An appearance experiment is necessary to extract the CP-violating effects. } 

{Experimental requirements:}
\begin{itemize}
\item The experiment will probe oscillations of $\nu_{\mu,e} \rightarrow \nu_{e,\mu}$.

\item The experiment will identify $\nu_e$ and $\nu_\mu$ 
  with high efficiency and
  purity in order to tag (or otherwise know) the flavor of the neutrino 
  before and after flavor transformations. 

\item The experiment requires $E_\nu >$\SI{100}{\MeV} so that
it will be possible to perform flavor-tagging of muon 
neutrinos using the lepton flavor produced in a 
charged current (CC) interaction ($\nu_\mu + N \rightarrow \mu N' X$).

 \end{itemize}

\item \emph{Phenomenology: In the three-flavor mixing model, the CP-violating 
Jarlskog invariant arises in the interference term
    $P_{\sin \delta}$ as given by Equation~\ref{eqn:papprox3}; the
    oscillation scale where the interference term is maximal is that
    determined by the mixing between the $\nu_{1}$ and $\nu_{3}$ states.} 

{Experimental  requirements:}
\begin{itemize}
\item The experimental baseline and corresponding neutrino energy are
  chosen according to Equation~\ref{eqn:nodes} such that $L/E$ equals \SI{510}{km/GeV}
  to maximize sensitivity to the CP-violating term in the
  neutrino flavor mixing.

\item Flavor-tagging of muon neutrinos that can be produced either at
  the source or after flavor-mixing requires $E_\nu >$ \SI{100}{\MeV};
  therefore, the experimental baselines over which to measure neutrino
  oscillations are $L >$ \SI{50}{\km}\footnote{Neutrino experiments using
    beams from pion decay-at-rest experiments such as DAE$\delta$ALUS
    are exceptions since the $\overline{\nu}_\mu$ production spectrum is
    well known and only the $\overline{\nu}_e$ flavor after oscillations is
    tagged through inverse-beta decay. The neutrino energies are 
    $\sim$\SI{50}{\MeV} below the CC muon-production threshold.}.
\end{itemize}

\item \emph{
Phenomenology: In the three-flavor model $\nu_{\mu,e}
    \rightarrow \nu_{e,\mu}$ oscillations depend on all parameters in
    the neutrino mixing matrix as well as on the mass differences, 
    as shown in Equations~\ref{eqn:papprox0} to~\ref{eqn:papprox3}.
}

{Experimental requirements:}
\begin{itemize}
\item The precision with which
  \deltacp can be determined --- and the sensitivity to small CP-violating effects or CP violation outside the three-flavor model --- requires precision
  determination of all the other mixing parameters, preferably in the
  same experiment. The experiment will be designed so as to minimize
  dependence on external measurements of the oscillation parameters.
\end{itemize}

\item \emph{Phenomenology: Observation of CP violation requires the explicit observation
  of an asymmetry between 
$P(\nu \rightarrow \nu)$ and $P(\overline{\nu} \rightarrow \overline{\nu})$.}

{Experimental requirements:}
\begin{itemize}
\item The experiment will
  probe the oscillations of both neutrinos and antineutrinos in an
  unambiguous way. 

\item The experiment will be capable of charge tagging in addition to flavor tagging. 
Charge tagging can be achieved at detection using the lepton charge and/or at production by selecting 
 beams purely of neutrinos or antineutrinos.

\item The experiment will
  be capable of resolving degeneracies between matter and 
  CP asymmetries in order to determine the MH. This can 
  be achieved by using a baseline greater than \SI{1000}{\km} or
  with measurements probing oscillations over a range of $L/E$ values. 
\end{itemize}
\item {\it Phenomenology: CP asymmetries are maximal at the secondary
    oscillation nodes.} 

{Experimental requirements:}
\begin{itemize}
\item Coverage of the $L/E$ scale of the secondary oscillation nodes
  improves experimental sensitivity to small values of \deltacp
  by enabling measurements of the asymmetry at the secondary nodes
  where the CP asymmetries are much larger and where there is no
  degeneracy with the matter asymmetries. 
  The experiment will be performed with a wide-band
  beam to provide sensitivity to the $L/E$ scale of both the first and
  second oscillation nodes.

\item The experimental
  baseline will be $>$\SI{150}{\km}, given that muon
  flavor tagging is required at either production or detection.
The secondary oscillation nodes are located at scales set by
  Equation~\ref{eqn:nodes} where $n>1$. The second oscillation
  maximum is located at scales given by $L/E \sim$\SI{1500}{\km/\GeV}. 
\end{itemize}
\end{enumerate}

Based on the experimental requirements prescribed by the neutrino
oscillation phenomenology detailed above, pursuit of the primary
science objectives for LBNE dictates the need for a very large mass
(\SIrange{10}{100}{\kt}) neutrino detector located at a distance greater than
\SI{1000}{\km} from the neutrino source. This large mass coupled with a powerful
wide-band beam and long exposures is required to accumulate enough
neutrino interactions --- $\mathcal{O}$(\num{1000}) events --- to make
precision measurements of the parameters that govern the subdominant
$\nu_\mu \rightarrow \nu_e$ oscillations. At \SI{1300}{\km}, the baseline
chosen for LBNE, both the first and second oscillation nodes are at
neutrino energies $>$ \SI{0.5}{\GeV}, as shown in Figure~\ref{fig:beamspectrum}. 
This places both neutrino oscillation nodes
in a region that is well matched to the energy spectrum of the high-power conventional neutrino beams that can be obtained using the
\SIrange{60}{120}{\GeV} Main Injector (MI) proton accelerator at Fermilab.

\section{Simulation of Neutrino Oscillation Experiments}
\label{sec:expt-sim}

To evaluate the sensitivity of LBNE and to optimize the experiment design,
it is important to accurately predict the neutrino flux produced by the
neutrino beamline, the neutrino interaction rate at the far detector, 
and the far detector performance. This is achieved using Monte Carlo (MC) simulations and
the GLoBES~\cite{Huber:2004ka,Huber:2007ji} package. The simulations and experimental
assumptions that are used to evaluate the sensitivity of LBNE to neutrino
mixing parameters, to the neutrino mass hierarchy (MH) and to CP violation are 
described in this section.

\subsection{Expected Signal}  

\begin{figure}[!htb]
\centering 
\includegraphics[width=0.8\textwidth]{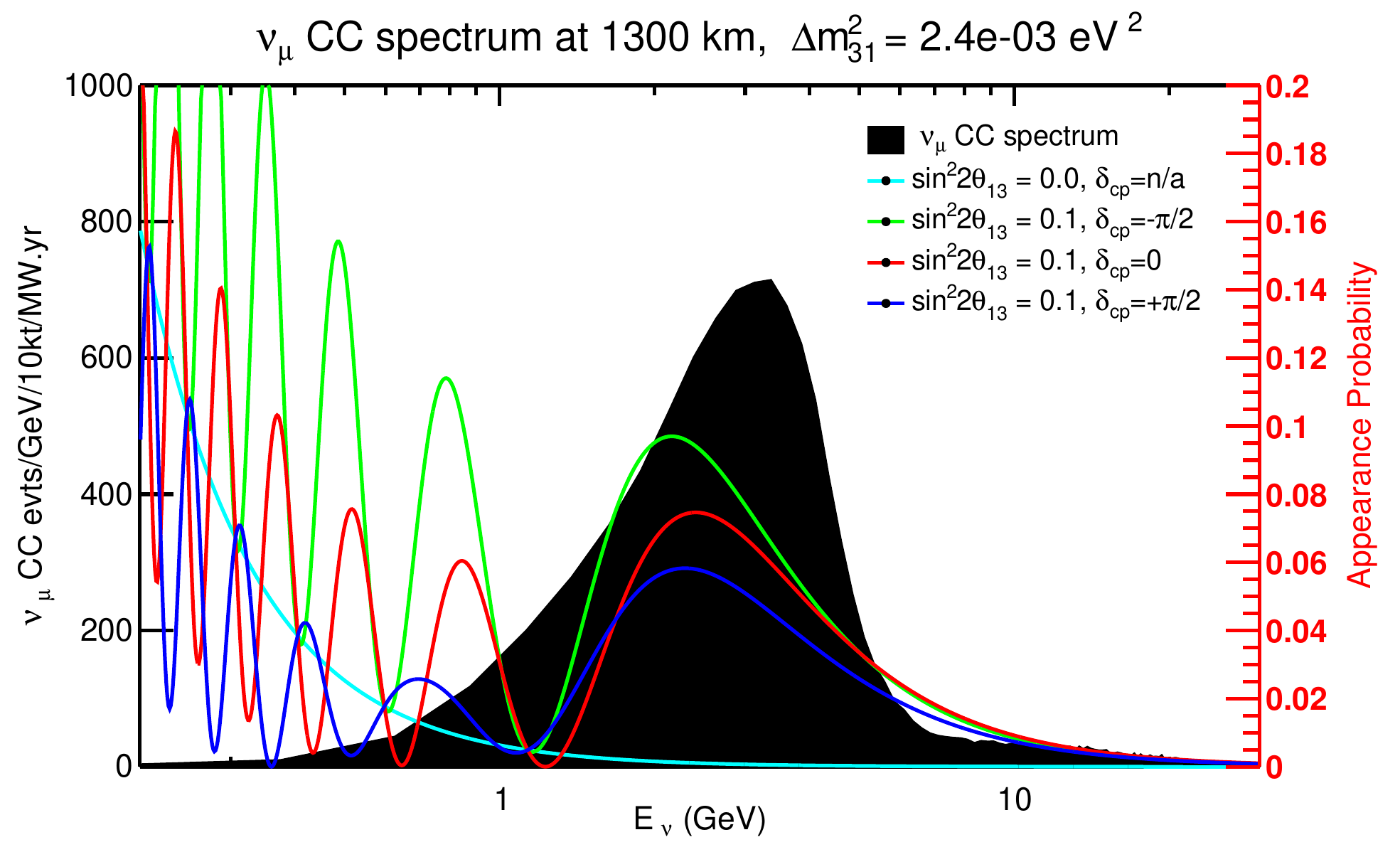}
\caption[Unoscillated spectrum of $\nu_\mu$ events and $\nu_\mu \rightarrow \nu_e$
  oscillation probabilities]{The simulated unoscillated spectrum 
of $\nu_\mu$ events from the LBNE
  beam (black histogram) overlaid with the $\nu_\mu \rightarrow \nu_e$
  oscillation probabilities (colored curves) for different values of \deltacp and
  normal hierarchy.}
\label{fig:beamspectrum}
\end{figure}

The LBNE beamline design, described in Section~\ref{beamline-chap}, 
is simulated using Geant4~\cite{Agostinelli:2002hh}.
The simulated $\nu_\mu$ spectrum (unoscillated flux $\times$ cross section) 
at \SI{1300}{\km} obtained from the LBNE beamline using \GeVadj{80} protons from the
MI is shown as the black histogram in Figure~\ref{fig:beamspectrum}.
At this baseline,
there is no degeneracy between matter and CP asymmetries
at the first oscillation node where the LBNE neutrino beam spectrum
peaks.  The wide coverage of the oscillation patterns
enables the search for physics beyond the three-flavor model because
new physics effects may interfere with the standard oscillations and induce 
a distortion in
the oscillation patterns. As a next-generation neutrino oscillation
experiment, LBNE aims to study in detail the spectral shape
of neutrino mixing over the range of energies where the mixing effects
are largest. This is
crucial for advancing the science beyond the current generation of
experiments, which depend primarily on rate asymmetries.

The LBNE reconfiguration
study~\cite{LBNEreconfig} determined that the far detector location at
the \SURF provides an optimal baseline for precision measurement
of neutrino oscillations using a conventional neutrino beam from
Fermilab. The \kmadj{1300} baseline optimizes sensitivity to CP
violation and is long enough to resolve the MH with a high
level of confidence, as shown in Figure~\ref{fig:BLcpfrac}.

\begin{table}[!htb] 
  \caption[Raw $\nu$ oscillation event rates at the LBNE far site with $E_{\nu} < 10$ GeV]{
    Raw $\nu$ oscillation event rates at the LBNE far site with $E_{\nu} < 10$ GeV. Assumes
    \num{1.8e7} seconds/year (Fermilab). \emph{POT} is \emph{protons-on-target}. 
    Oscillation parameters used are: $\theta_{12} = 0.587$, $\theta_{13} =
    0.156$, $\theta_{23} = 0.670$, $\Delta m^2_{21} =~$\SI{7.54e-5}{eV^2},
    and $\Delta m^2_{31} = +$\SI{2.47e-3}{eV^2} (normal hierarchy). The NC event rate
    is for events with visible energy $> 0.5$ GeV. For comparison, the rates at other neutrino oscillation experiments 
    (current and proposed) are shown for similar exposure in mass and time. 
     No detector effects are included.}
\label{tab:lbl_event_rates}
{\footnotesize
\begin{tabular}{$l^c^c^c^c^c^c^c^c^c} 
\toprule
\rowtitlestyle
Experiment & Baseline & $\nu_{\mu}$ unosc. & $\nu_{\mu}$ osc. & $\nu_e$ beam & $\nu_{\mu}$  
           & $\nu_{\mu}\rightarrow\nu_{\tau}$ & \multicolumn{3}{^>{\columncolor{\ChapterBubbleColor}}c}{$\nu_{\mu}\rightarrow\nu_e$ CC}  \\
\rowtitlestyle
{\it details} & km  & CC & CC & CC &  NC  & CC 
& $\mdeltacp = -\frac{\displaystyle \pi}{\displaystyle 2}$, & $0$,& $\frac{\displaystyle \pi}{\displaystyle 2}$ \\ 
\toprowrule
{\bf LBNE LE} & \num{1300}  & & & & & & \multicolumn{3}{c}{}\\ 
\multicolumn{2}{^l}{\it 80 GeV, \SI{1.2}{\MW}}  & & & & & \multicolumn{3}{c}{}\\
\multicolumn{2}{^l}{\it \num{1.5e21}  POT/year} & & & & & & \multicolumn{3}{c}{}\\
50 \ktyr $\nu$ & & 12721 & 4339 & 108 & 3348 & 156 & 605 & 480 & 350 \\
50 \ktyr $\overline{\nu}$ & & 4248 & 1392 & 34 & 1502 & 48 & 51 & 86 & 106 \\ \colhline
{\bf LBNE ME} & \num{1300} & & & & & & \multicolumn{3}{c}{}\\
\multicolumn{2}{^l}{\it 120 GeV, \SI{1.2}{\MW}} & & & & & & \multicolumn{3}{c}{}\\ 
\multicolumn{2}{^l}{\it \num{1e21} POT/year} & & & & & & \multicolumn{3}{c}{}\\ 
50 \ktyr $\nu$ & & 19613 & 12317 & 72 & 5808 & 686 & 435 & 399 & 293 \\ \colhline
{\bf T2K} & 295  & & & & & & \multicolumn{3}{c}{}\\
\multicolumn{2}{^l}{\it 30 GeV, \SI{750}{\kW}}  & & & & & & \multicolumn{3}{c}{}\\
\multicolumn{2}{^l}{\it \num{9e20}  POT/year} & & & &  & & \multicolumn{3}{c}{}\\ 
50 \ktyr $\nu$ & & 2100 & 898 & 41 & 360 & $<1$ & 73 & 58 & 39 \\ \colhline
{\bf MINOS LE} & 735  & & & & & & \multicolumn{3}{c}{}\\
\multicolumn{2}{^l}{\it 120 GeV, \SI{700}{\kW}}& & & & & & \multicolumn{3}{c}{}\\ 
\multicolumn{2}{^l}{\it \num{6e20}  POT/year} & & & & & & \multicolumn{3}{c}{}\\
50 \ktyr $\nu$ & & 17574 & 11223 & 178  & 4806 & 115 & 345 & 326 & 232 \\
50 \ktyr $\overline{\nu}$ & & 5607 & 3350 & 56 & 2017 & 32 & 58 & 85 & 88 \\ \colhline
{\bf NOvA ME} &  810  & & & & & & \multicolumn{3}{c}{}\\
\multicolumn{2}{^l}{\it 120 GeV, \SI{700}{\kW}} & & & & & & \multicolumn{3}{c}{}\\ 
\multicolumn{2}{^l}{\it \num{6e20}  POT/year} & & & & & & \multicolumn{3}{c}{}\\
50 \ktyr $\nu$ & & 4676 & 1460 & 74 & 1188 & 10 & 196 & 168 & 116 \\
50 \ktyr $\overline{\nu}$ & & 1388 &  428 & 19 & 485 & 2 & 22 & 35 & 41 \\ \colhline
{\bf LBNO} & \num{2300}   & & & & & & \multicolumn{3}{c}{}\\ 
\multicolumn{2}{^l}{\it 50 GeV  $\sim$ \SI{2}{\MW}}   & & & & & & \multicolumn{3}{c}{}\\ 
\multicolumn{2}{^l}{\it \num{3e21}  POT/year} & & & & & & \multicolumn{3}{c}{}\\
50 \ktyr $\nu$ & & 8553 & 2472 & 48 & 2454 & 570 & 534 & 426 & 336 \\
50 \ktyr $\overline{\nu}$ & & 3066 & 828 & 15 & 1140 & 255 & 24 & 45 & 54  \\ 
\toprule
\rowtitlestyle
$\nu$-Factory & & $\nu_{\mu}$ unosc. & $\nu_{\mu}$ osc. &  & $\nu_{\mu}$  & $\nu_{\mu}\rightarrow\nu_{\tau}$ & 
\multicolumn{3}{^>{\columncolor{\ChapterBubbleColor}}c} {$ \nu_e \rightarrow \nu_\mu$ CC}  \\ 
\rowtitlestyle
{\it details} &  & CC & CC & &  NC  & CC 
& $\mdeltacp = -\frac{\displaystyle \pi}{\displaystyle 2}$, & $0$,& $\frac{\displaystyle \pi}{\displaystyle 2}$ \\ 
\toprowrule
{\bf NuMAX I} &  \num{1300}   & & & & & & \multicolumn{3}{c}{}\\ 
\multicolumn{2}{^l}{\it 3 GeV, \SI{1}{\MW}} & & & & & & \multicolumn{3}{^c}{}\\ 
\multicolumn{2}{^l}{\it \num{.94e20}  $\mu$/year} & & & & & & \multicolumn{3}{c}{}\\ 
50 \ktyr $\mu^+$ & & 1039 & 339 & & 484 & 28 & 71 & 97 & 117  \\ 
50 \ktyr $\mu^-$ & & 2743 & 904 &  & 945 & 89 & 24 & 19 & 12 \\ \colhline
{\bf NuMAX II} & \num{1300}  & & & & & & \multicolumn{3}{c}{}\\ 
\multicolumn{2}{^l}{\it 3 GeV,  \SI{3}{\MW}} & & & & & & \multicolumn{3}{c}{}\\ 
\multicolumn{2}{^l}{\it \num{5.6e20}  $\mu$/year} & & & & & & \multicolumn{3}{c}{}\\ 
50 \ktyr $\mu^+$ & & 6197 & 2018 & & 2787 & 300 & 420 & 580 & 700  \\ 
50 \ktyr $\mu^-$ & & 16349 & 5390 &  & 5635 & 534 & 139 & 115 & 85 \\ 
\bottomrule 
\end{tabular}
}
\end{table}
Table~\ref{tab:lbl_event_rates} lists the beam neutrino interaction
rates for all three known species of neutrinos as expected at the LBNE
far detector. This table shows only the raw interaction rates using
the neutrino flux from the Geant4 simulations of the LBNE beamline and
the default interaction cross sections included in the GLoBeS 
package~\cite{Huber:2004ka} with {\em no detector effects included}.  A
tunable LBNE beam spectrum, obtained by varying the distance between
the target and the first focusing horn (Horn~1), is assumed. The
higher-energy tunes are chosen to enhance the $\nu_\tau$ appearance 
signal and improve the oscillation fits to the three-flavor paradigm.
To estimate the NC event rates based on visible energies above \SI{0.5}{GeV},
a true-to-visible energy smearing function based on output from the
GENIE neutrino MC generator~\cite{Andreopoulos:2009zz} is
used.  For comparison, the rates at current neutrino oscillation
experiments such as T2K~\cite{Abe:2011ks}, MINOS~\cite{minos-numi-url} 
and NO$\nu$A~\cite{Ayres:2007tu} are shown for
similar exposure in mass and time and using the same interaction
cross sections.  The raw interaction rates from other proposed
neutrino oscillation experiments such as LBNO~\cite{Rubbia:2013zqa} and the NuMAX neutrino factory designs~\cite{Delahaye:2013jla} 
are also shown\footnote{T2K uses a JPARC neutrino beam, MINOS and NO$\nu$A use the Fermilab NuMI neutrino beam and LBNO uses a CERN neutrino beam.}. 
 It is important to note that
the duty factors for the JPARC and CERN beams are $\sim1/3$ and $\sim1/2$ 
of NuMI/LBNE respectively.  
For LBNO, the event rates are
obtained using the optimized beam from the HP-PS2 \GeVadj{50}
synchrotron~\cite{Longhin:2012ae} with an exposure of
\SI{3e21}{\POT/\year}.  The LBNO duty cycle is assumed to be
$\sim$\num{e7} seconds/year, which corresponds to a beam power of
\SI{2}{\MW}. Note that for Stage 1 and Stage 2 of the NuMAX neutrino
factory proposal~\cite{Delahaye:2013jla}, Project X beams~\cite{Holmes:2013vpa} 
at 3~GeV with 1 and 3~MW, respectively, are
needed\footnote{Project X has been superseded by PIP-II as of late 2013; PIP-II is briefly described in Section~\ref{beamline-chap}. }
 It is clear that the LBNE beam design and baseline produce
high rates of $\nu_e$ appearance coupled with large rate asymmetries
when CP-violating effects are included. For example, LBNE has
significantly higher appearance rates with a Main Injector
\MWadj{1.2} beam when compared to Stage 1 of the NuMAX neutrino
factory with a \MWadj{1} beam from a \GeVadj{3} linac. The $\nu_e$ appearance 
rates are very similar in LBNE and LBNO with normal
hierarchy (NH), but the $\overline{\nu}_e$ appearance rates (NH) in LBNO are $\approx1/2$ that of LBNE due to the
suppression from the larger matter effect (longer baseline) in LBNO.

\subsection{Detector Simulation using the  GLoBES Package}

For the sensitivity studies presented here, 
the GLoBES
package~\cite{Huber:2004ka,Huber:2007ji} was used to simulate the detector
response using simple smearing and using detector efficiency values
based on results from ICARUS and earlier simulation efforts as
documented in~\cite{CDRv1}.  The values used in GLoBES are shown in
Table~\ref{tab:lar-nuosc-totaltable}.

\begin{table}[!htb]
\caption[LArTPC detector performance parameters]{Estimated range of the LArTPC detector
  performance parameters for the primary oscillation physics. Signal
  efficiencies, background levels, and resolutions are obtained from
ICARUS and earlier simulation efforts
(middle column) and the value
  chosen for the baseline LBNE neutrino oscillation sensitivity
  calculations (right column). }
\label{tab:lar-nuosc-totaltable}
\begin{center}
\begin{tabular}{$L^c^c}
\toprule
\rowtitlestyle
Parameter & Range of Values &    Value Used for  LBNE Sensitivities\\ 
\rowtitlestyle
& \multicolumn{2}{^>{\columncolor{\ChapterBubbleColor}}c}{For $\nu_e$-CC appearance studies} \\ 
\toprowrule
$\nu_e$-CC efficiency          & 70-95\%               & 80\%   \\ \colhline
$\nu_\mu$-NC misidentification rate  & 0.4-2.0\%   & 1\%   \\ \colhline
$\nu_\mu$-CC misidentification rate  & 0.5-2.0\%     & 1\%   \\ \colhline
Other background                 & 0\%                   & 0\% \\ \colhline
Signal normalization error       & 1-5\%                 & 1-5\% \\ \colhline 
Background normalization error   & 2-15\%                 & 5-15\% \\  
\toprule
\rowtitlestyle
& \multicolumn{2}{^>{\columncolor{\ChapterBubbleColor}}c}{For $\nu_\mu$-CC disappearance studies} \\ 
\toprowrule
$\nu_\mu$-CC efficiency          & 80-95\%               & 85\%   \\ \colhline
$\nu_\mu$-NC misidentification rate  & 0.5--10\%   & 1\%   \\ \colhline
Other background                 & 0\%                   & 0\% \\ \colhline
Signal normalization error       & 1-10\%                 & 5--10\% \\ \colhline 
Background normalization error   & 2-20\%                 & 10-20\% \\ 
\toprule
\rowtitlestyle
& \multicolumn{2}{^>{\columncolor{\ChapterBubbleColor}}c}{For $\nu$-NC disappearance studies} \\ 
\toprowrule
$\nu$-NC efficiency          & 70-95\%               & 90\%   \\ \colhline
$\nu_\mu$-CC misidentification rate  & 2-10\%   & 10\%    \\ \colhline
$\nu_e$-CC misidentification rate  & 1-10\%   & 10\%   \\ \colhline
Other background                 & 0\%                   & 0\% \\ \colhline
Signal normalization error       & 1-5\%                 &  under study\\ \colhline 
Background normalization error   & 2-10\%                &  under study \\ 
\toprule
\rowtitlestyle
& \multicolumn{2}{^>{\columncolor{\ChapterBubbleColor}}c}{ Neutrino energy resolutions} \\ 
\toprowrule
$\nu_e$-CC energy resolution & $15\%/\sqrt{E(GeV)}$ & $15\%/\sqrt{E(GeV)}$ \\ \colhline
$\nu_\mu$-CC energy resolution & $20\%/\sqrt{E(GeV)}$ & $20\%/\sqrt{E(GeV)}$ \\ \colhline
$E_{\nu_e}$ scale uncertainty & under study & under study\\ \colhline
$E_{\nu_\mu}$  scale uncertainty & 1-5\% & 2\%\\ \bottomrule
\end{tabular}
\end{center} 
\end{table}

Studies from ICARUS have estimated and measured single-particle energy
resolutions in liquid argon.  Below \SI{50}{\MeV}, the energy resolution of electrons
is $11\%/\sqrt{E[\mathrm{MeV}]} + 2\%$. The energy resolution of an
electromagnetic shower with energy in the range (50--5000)~MeV is
$33\%/\sqrt{E(\mathrm{MeV})} + 1\%$~\cite{Amoruso:2003sw} and that of
hadronic showers  
is $\approx 30\%/\sqrt{E(\mathrm{GeV})}$.  A
significant fraction of the $\nu_e$-CC signal in LBNE in the range of
\SIrange{1}{6}{\GeV} comes from non-quasi-elastic CC interactions with a large component
of the visible energy in the hadronic system. From recent simulations
of neutrino interactions in this region it has been determined
that $< E_{\rm lepton}/E_{\nu}>\approx 0.6$. For this reason, the
total $\nu_e$ energy resolution for the neutrino oscillation
sensitivity calculation is chosen to be $15\%/\sqrt{E(\mathrm{GeV})}$. In a
non-magnetized LArTPC, the muon momentum can be obtained from 
measurements of range and multiple scattering.
The muon momentum resolution for partially contained muons
is found to be in
the range $10-15\%$~\cite{T2K2kmprop,Ankowski:2006ts} for muons
in the \SIrange{0.5}{3}{\GeV} range. The $\nu_\mu$ total energy
resolution in LBNE is, therefore, assumed to be 
$20\%/\sqrt{E(\mathrm{GeV})}$; the resolution will be significantly better
than this for the small subsample of events in which muons are fully contained 
by the detector.

\begin{figure}[!htp]
\centerline{
\includegraphics[width=0.5\textwidth]{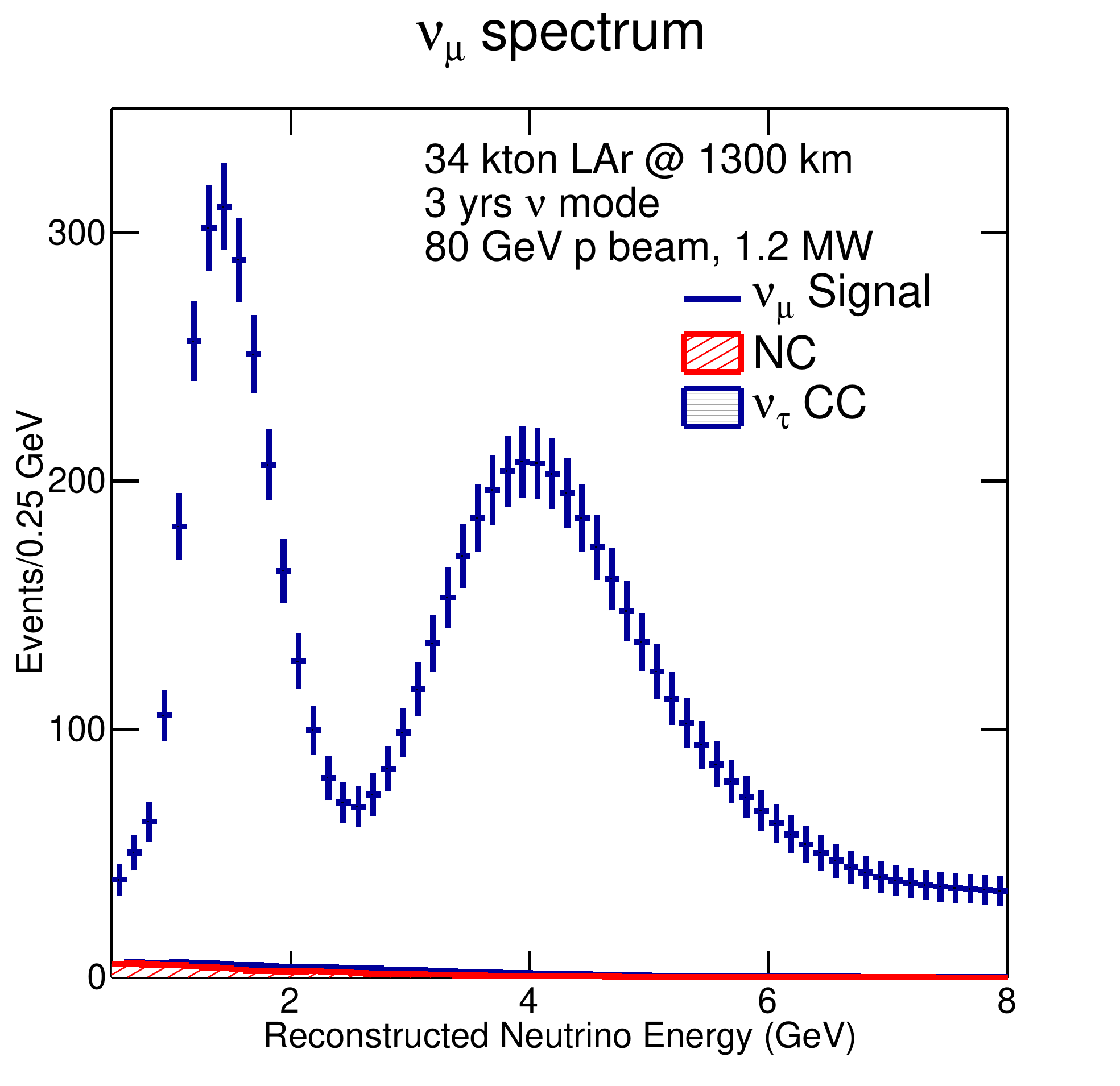}
\includegraphics[width=0.5\textwidth]{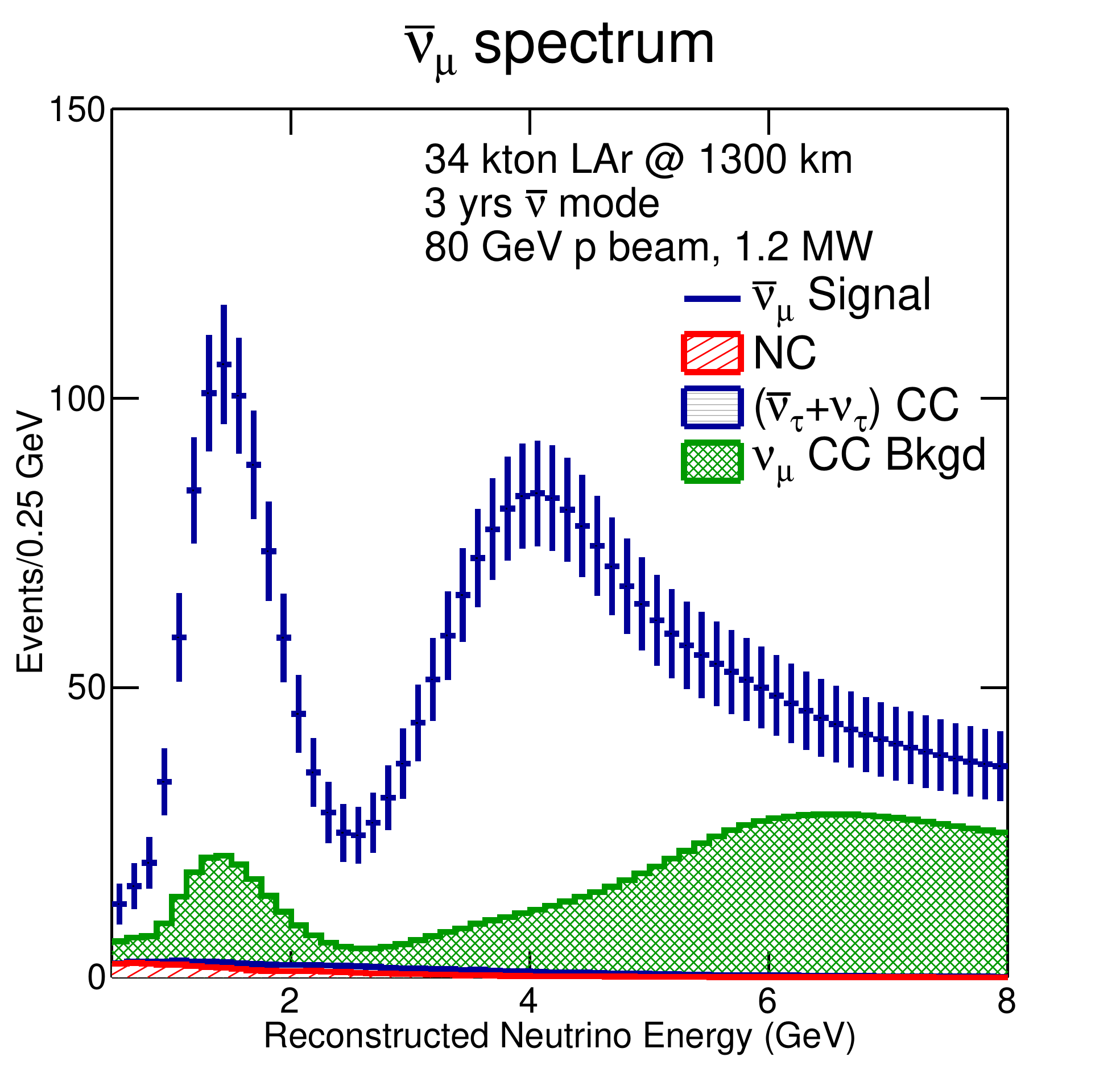}}
\caption[Disappearance spectra in a LArTPC]{The expected reconstructed 
neutrino energy spectrum of
  $\nu_{\mu}$ or $\overline{\nu}_{\mu}$ events in a \SIadj{34}{\kt} LArTPC for
  three  years of neutrino (left) and antineutrino (right) running with
  a \SIadj{1.2}{\MW} beam.}
\label{fig:lar-disapp-spectrum}
\end{figure}
\begin{figure}[!tp]
\centerline{
\includegraphics[width=0.5\textwidth]{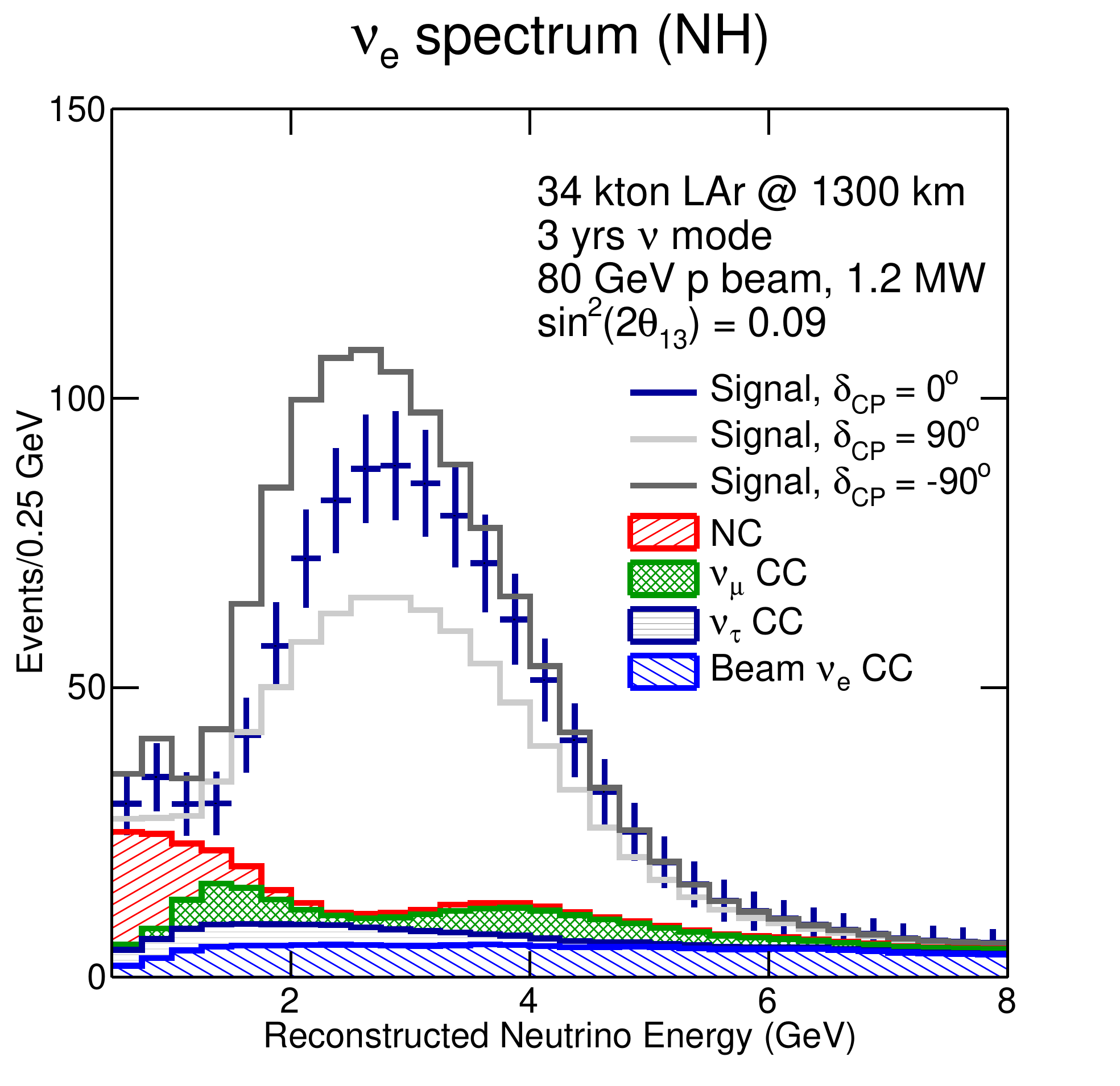}
\includegraphics[width=0.5\textwidth]{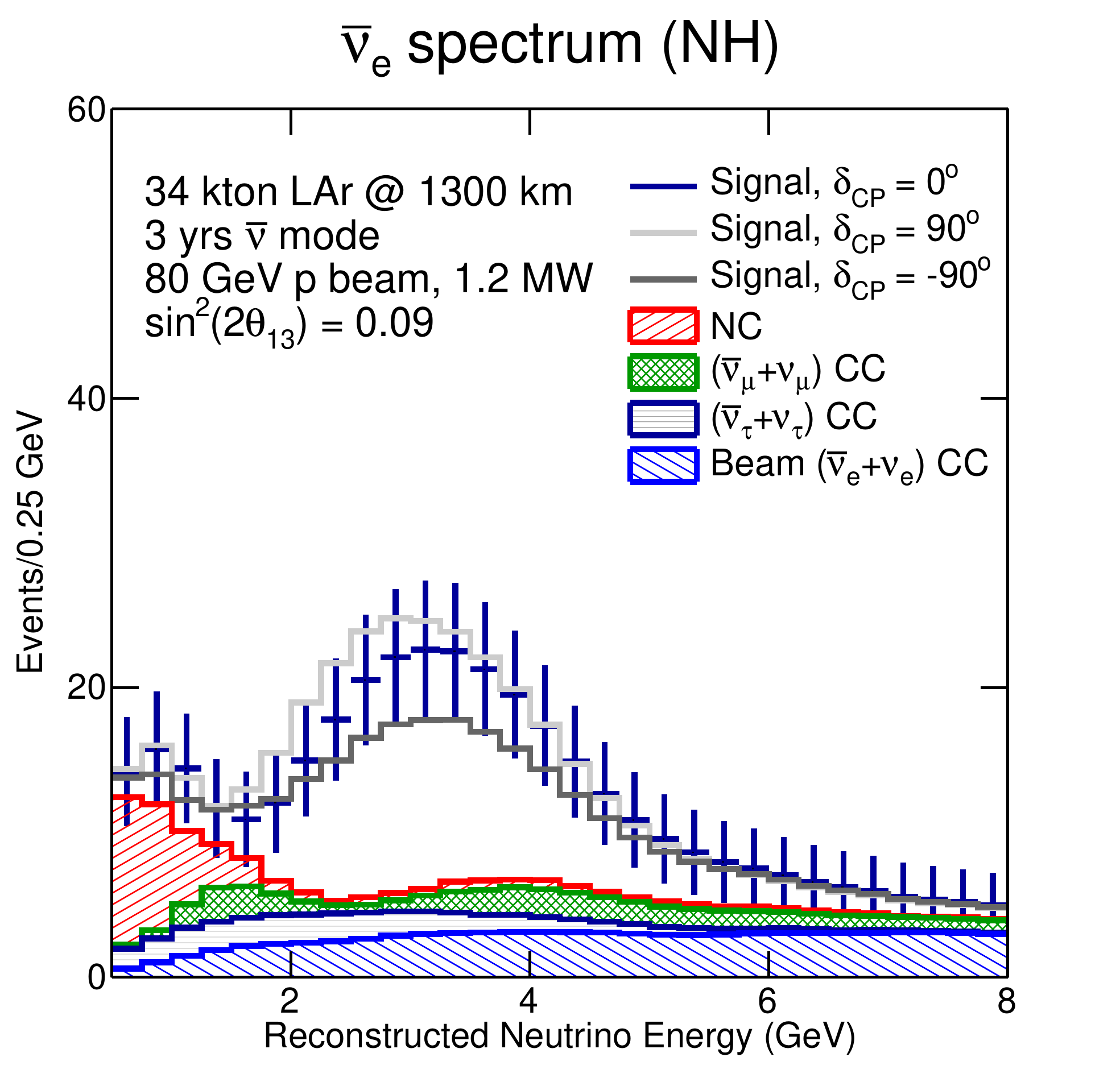}
}
\centerline{
\includegraphics[width=0.5\textwidth]{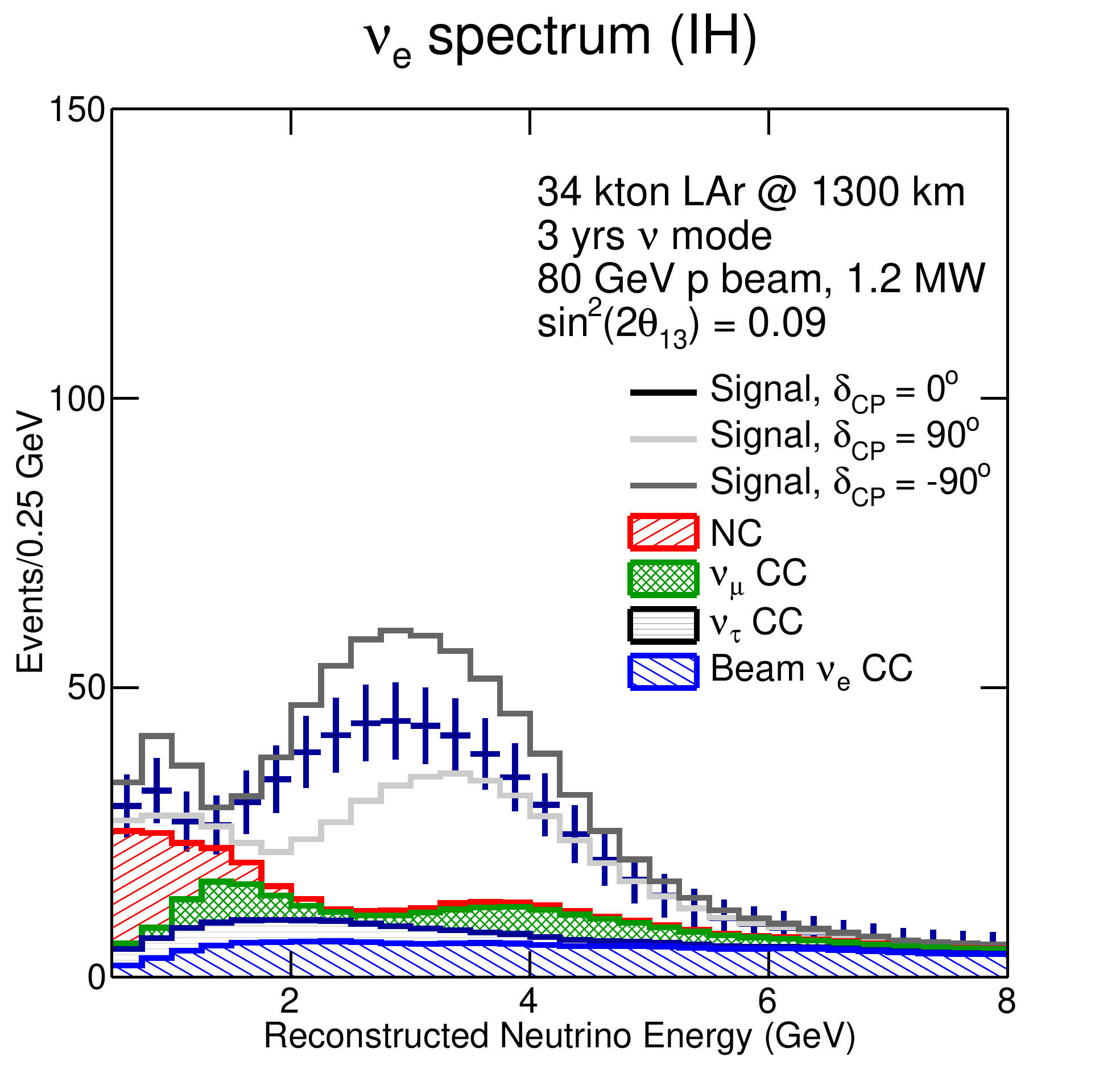}
\includegraphics[width=0.5\textwidth]{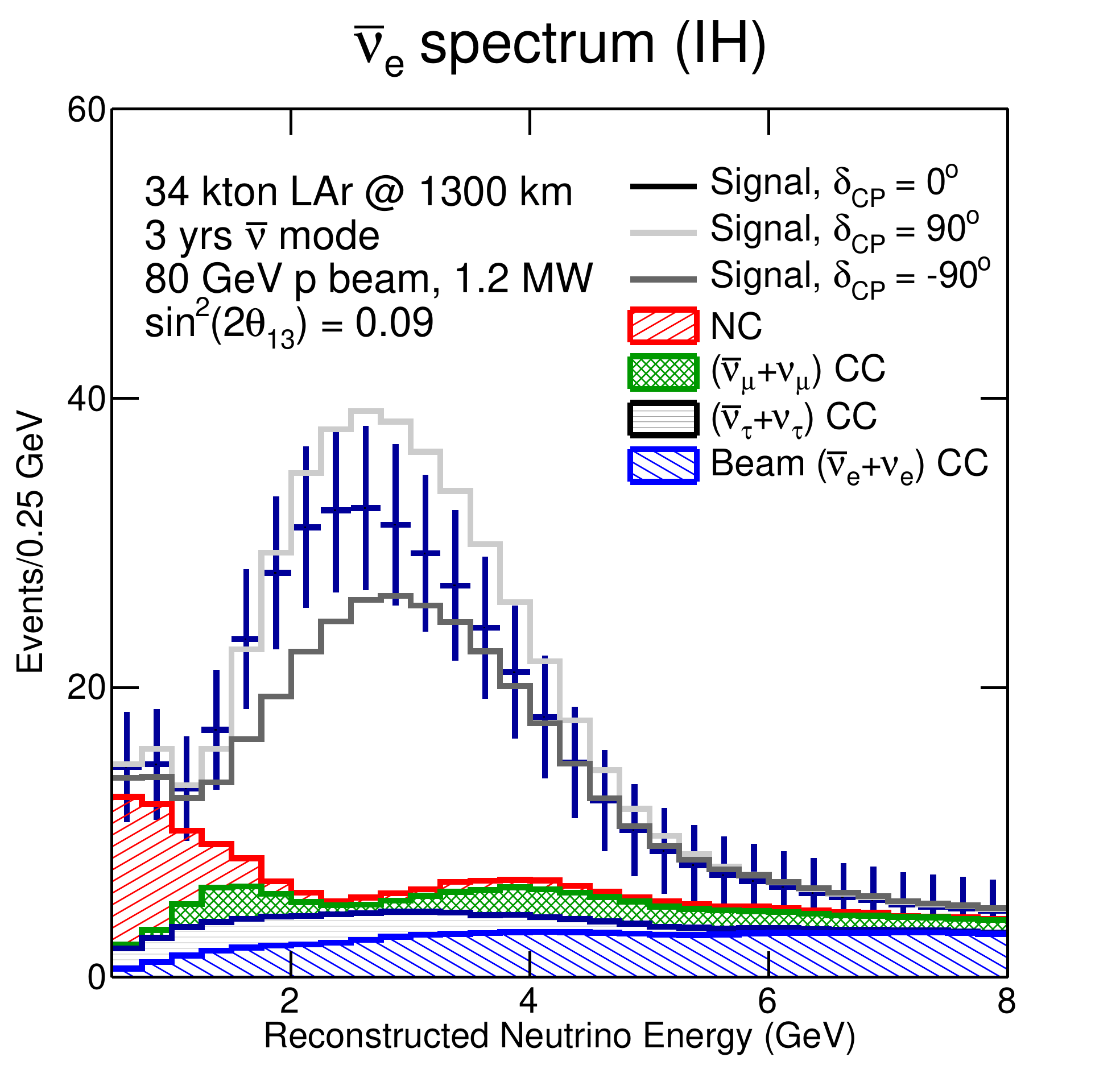}}
\caption[Event spectra of neutrino interactions in a LArTPC]{The
  expected reconstructed neutrino energy spectrum of $\nu_e$ or $\overline{\nu}_e$ oscillation
  events in a \ktadj{34} LArTPC for three years of neutrino (left) and
  antineutrino (right) running with a 
\MWadj{1.2}, \GeVadj{80} beam assuming
  $\sin^2(2\theta_{13}) = 0.09$. The plots on the
top are for normal hierarchy and the plots on the bottom are for
inverted hierarchy. }
\label{fig:lar-event-spectrum}
\end{figure}

\begin{table}[!htb]
\caption[Expected number of $\nu$ oscillation signal and beam
  background events]{Expected number of neutrino oscillation
  signal and background events in the energy range \SIrange{0.5}{8.0}{\GeV}  at
  the far detector after detector smearing and event selection. The
  calculation assumes $\sin^2(2\theta_{13})=0.09$ and
  $\mdeltacp=0$. The event rates are given per \ktadj{10} LArTPC 
and three years of running with the improved \GeVadj{80} LBNE beam at \SI{1.2}{\MW}. For signal, the number of $\nu$ and $\overline{\nu}$ events are shown separately, while for the background estimates $\nu$ and $\overline{\nu}$ events are combined. The MH has negligible impact on $\nu_\mu$ disappearance signals. }
\begin{tabular}{$L^l^c^c^c^c^c^c}
\toprule
\rowtitlestyle
Beam & Hierarchy  &  Signal Events & \multicolumn{5}{^>{\columncolor{\ChapterBubbleColor}}c}{ Background Events} \\
\rowtitlestyle
 &   & $\nu_x/\overline{\nu}_x$ CC &  $\nu_{\mu}$ NC &  $\nu_{\mu}$ CC & $\nu_e$ Beam & $\nu_\tau$ CC & Total \\ 
\toprowrule
\rowtitlestyle
&  \multicolumn{7}{^>{\columncolor{\ChapterBubbleColor}}c}{$\nu_\mu \rightarrow \nu_{x=\mu}$ (disappearance)} \\ \toprowrule 
Neutrino &    -      & 2056/96 & 23  & N/A & - & 18 &  41\\ \colhline
Antineutrino & -     & 280/655 & 10  & N/A & - & 10 & 20 \\ 
\rowtitlestyle
&  \multicolumn{7}{^>{\columncolor{\ChapterBubbleColor}}c}{$\nu_\mu \rightarrow \nu_{x=e}$ (appearance)} \\ \toprowrule 
Neutrino & Normal         & 229/3 & 21  & 25 & 47 & 14 & 107 \\ \colhline
Neutrino & Inverted       & 101/5  & 21  & 25 & 49 & 17 & 112 \\ \colhline
Antineutrino & Normal    & 15/41  & 11  & 11 & 24 & 9 & 55 \\ \colhline
Antineutrino & Inverted  & 7/75  & 11  & 11 & 24 & 9 & 55 \\ \bottomrule
\end{tabular}
\label{tab:eventcounts}
\end{table}
Figures~\ref{fig:lar-disapp-spectrum} and~\ref{fig:lar-event-spectrum}
show the predicted spectra of observed 
signal and background events
in LBNE  produced from the GLoBES implementation, including the
effects of neutrino oscillation.
Figure~\ref{fig:lar-disapp-spectrum} shows the 
$\nu_\mu$ and $\overline{\nu}_\mu$-CC
sample and Figure~\ref{fig:lar-event-spectrum} shows
the $\nu_e$ and $\overline{\nu}_e$-CC appearance sample. Table~\ref{tab:eventcounts}
shows the expected LBNE signal and background event rates in 
$\nu_\mu$ disappearance and $\nu_e$ appearance modes for neutrinos and
antineutrinos, for normal (NH) and inverted (IH) hierarchy. The rates are given per \SI{10}{\kt} of fiducial LArTPC mass.

The GLoBES implementation used in the sensitivity studies presented here
appears to be in good agreement with more recent results 
from the Fast MC, described in Section~\ref{appxsec:fastmc}. 
Updated sensitivity and systematics studies are currently underway using the
Fast MC for detector simulation, and customized GLoBES-based software 
for the oscillation fits
and propagation of systematics. A full MC simulation of the
far detector and automated event reconstruction is being developed;
this is also described in Appendix~\ref{app-sim}. 

\clearpage
\section{Measurements of Mass Hierarchy and the CP-Violating Phase}

The neutrino mass hierarchy (MH) and the value of the CP-violating phase, 
\deltacp, are currently unknown. Knowledge of the MH
has significant theoretical, cosmological and experimental implications.
A determination of the \deltacp value to be neither zero ($0$) nor $\pi$ would
constitute the first observation of CP violation in the lepton sector.

The expected performance of 
a \ktadj{10} LArTPC far detector \SI{1300}{\km} downstream from 
a 
neutrino source is detailed in the LBNE Conceptual Design Report
Volume~1~\cite{CDRv1}.  Estimated sensitivities to the determination of the MH
 and discovery of CP violation, presented both here and in the CDR, are
calculated using the GLoBES
package. The detector response assumed in these calculations
is summarized in 
Table~\ref{tab:lar-nuosc-totaltable}. The sensitivities are obtained by
simultaneously fitting the $\nu_\mu \rightarrow \nu_\mu$,
$\overline{\nu}_\mu \rightarrow \overline{\nu}_\mu$, $\nu_\mu \rightarrow \nu_e$, 
and  $\overline{\nu}_\mu \rightarrow \overline{\nu}_e$ oscillated spectra, examples of
which are shown in Figures~\ref{fig:lar-disapp-spectrum} and~\ref{fig:lar-event-spectrum}.
The $\nu_\tau$ background is not used in the sensitivity calculations
since it is expected that further analysis will reduce this
background to negligible levels. 

In these calculations, experimental sensitivity is
quantified using $\Delta\chi^2$ parameters, which are determined
by comparing the predicted spectra for various scenarios. 
These quantities are defined, differently for neutrino MH
and CP-violation sensitivity, to be:
\begin{eqnarray}
\Delta\chi^2_{MH} & = & |\chi^2_{MH^{test}=IH} - \chi^2_{MH^{test}=NH}|, \\
\Delta\chi^2_{CPV} & = & \min\left(\Delta\chi^2_{CP}(\mdeltacp^{test}=0),\Delta\chi^2_{CP}(\mdeltacp^{test}=\pi)\right)\mbox{, where} \\
\Delta\chi^2_{CP} & = & \chi^2_{\mdeltacp^{test}} - \chi^2_{\mdeltacp^{true}}. \\ \nonumber
\end{eqnarray}
These sensitivities are evaluated separately for true NH and IH. 
Since the true value of \deltacp is unknown, a scan is  performed over
all possible values of $\mdeltacp^{true}$.
The individual $\chi^2$ values are calculated using
\begin{equation}
\chi^2({\bf n}^{true},{\bf n}^{test},f) = 2 \sum_{i}^{N_{reco}} (n_i^{true} ln\frac{n_i^{true}}{n_i^{test}(f)} + n_i^{test}(f) - n_i^{true}) + f^2,
\label{eq:globes_chisq}
\end{equation}
where {\bf n} are event rate vectors in $N_{reco}$ bins of reconstructed energy and $f$ represents
a nuisance parameter 
to be profiled. Nuisance parameters include the values of mixing angles, mass splittings, and signal and
background normalization. The nuisance parameters are constrained by Gaussian priors; in the
case of the oscillation parameters, the Gaussian prior has standard deviation
determined by taking 1/6 of the 3$\sigma$ range allowed by the global fit~\cite{Fogli:2012ua}. 

With the exception of results reported in Section~\ref{sec-mh-statistics},
where more information on the statistical interpretation of MH 
sensitivity is provided, the sensitivities presented here are for the
\emph{typical experiment} with no statistical fluctuations considered.
In the absence of statistical
fluctuations, the $\chi^2$ value for the \emph{true} spectra is identically 
zero. Statistical fluctuations are incorporated by repeatedly varying the
contents of each energy bin in each sample by drawing from a Poisson
distribution with the expected number of events in that bin as the
mean. 

This section presents the sensitivities 
of various LBNE configurations to determination of the MH
and CP violation.  In particular, a \SIadj{10}{\kt} far detector and the
full-scope \ktadj{34} far detector 
 are considered. In each case, the
performance of LBNE with both the \GeVadj{120} beamline design
presented in the CDR~\cite{CDRv2} as well as the upgraded
\GeVadj{80} beam described in Section~\ref{beamline-chap} is
studied. In addition, the sensitivities at different possible stages
of LBNE with increases to far detector mass and Main Injector beam
upgrades are estimated.

Figure~\ref{fig:10ktonsens} summarizes the sensitivities for
determining the MH and CP violation ($\mdeltacp \neq 0
\ {\rm or} \ \pi$) as a function of the true value of \deltacp
with a \ktadj{10} LArTPC.  The red band shows the sensitivity that is
achieved with an 
exposure of six years with equal exposures in $\nu$ and
$\overline{\nu}$ mode in a \MWadj{1.2} beam.  The cyan band shows the
sensitivity obtained by combining the \SIadj{10}{\kt} LBNE with T2K and
NO$\nu$A, where the T2K exposure is \num{7.8e21} POT in
$\nu$ mode only and the NO$\nu$A exposure is six years
(assuming {\num{6e20} POT per year) with equal
exposures in $\nu$ and $\overline{\nu}$ mode. The bands indicate the
sensitivity range corresponding to different levels of signal and
background normalization uncertainties and different possible beam
designs. The gray curves are the expected sensitivities for the
combination of NO$\nu$A and T2K.  The known mixing parameters are
allowed to float in the fit, but are constrained (using a Gaussian prior)
by the uncertainties from the
2012 global best fit~\cite{Fogli:2012ua}.
The reactor mixing angle, $\sin^2 2\theta_{13}$, is constrained to be 
$0.094 \pm 0.005$. The uncertainty is equal to the size of the current systematic
uncertainty from the Daya Bay Experiment~\cite{An:2013zwz} and is used as
a conservative estimate of the precision
that will be achieved by the current generation of reactor experiments.
Figure~\ref{fig:35kton} shows the sensitivities for determining the MH and CP violation as a function of the true value of \deltacp after six years of running in the LBNE \ktadj{34} configuration under the same assumptions.

\begin{figure}[!htb]
\centerline{
\includegraphics[width=0.5\textwidth,trim=0cm 0.3cm 0cm 0.3cm,clip]{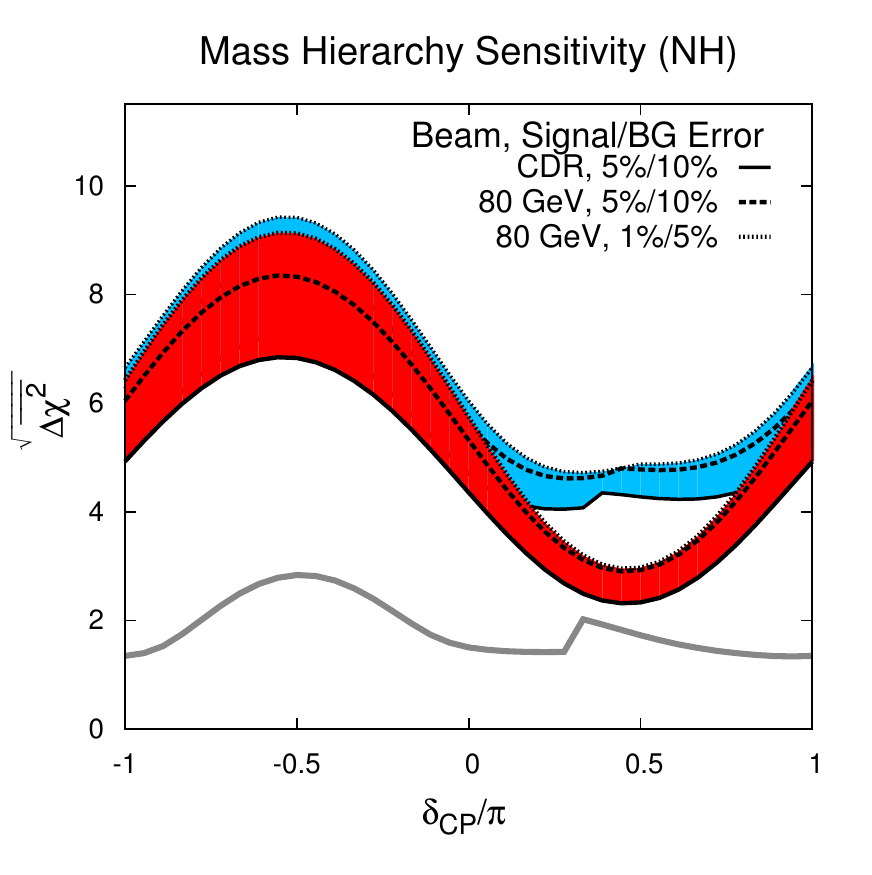}
\includegraphics[width=0.5\textwidth,trim=0cm 0.3cm 0cm 0.3cm,clip]{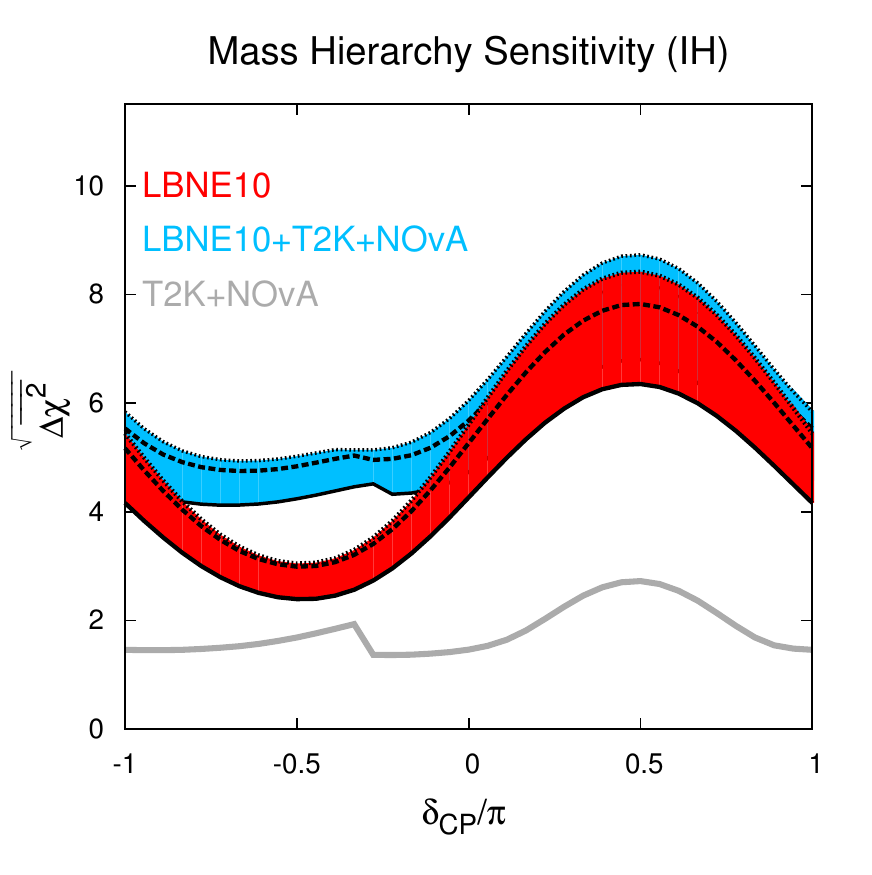}
}
\centerline{
\includegraphics[width=0.5\textwidth,trim=0cm 0.3cm 0cm 0.3cm,clip]{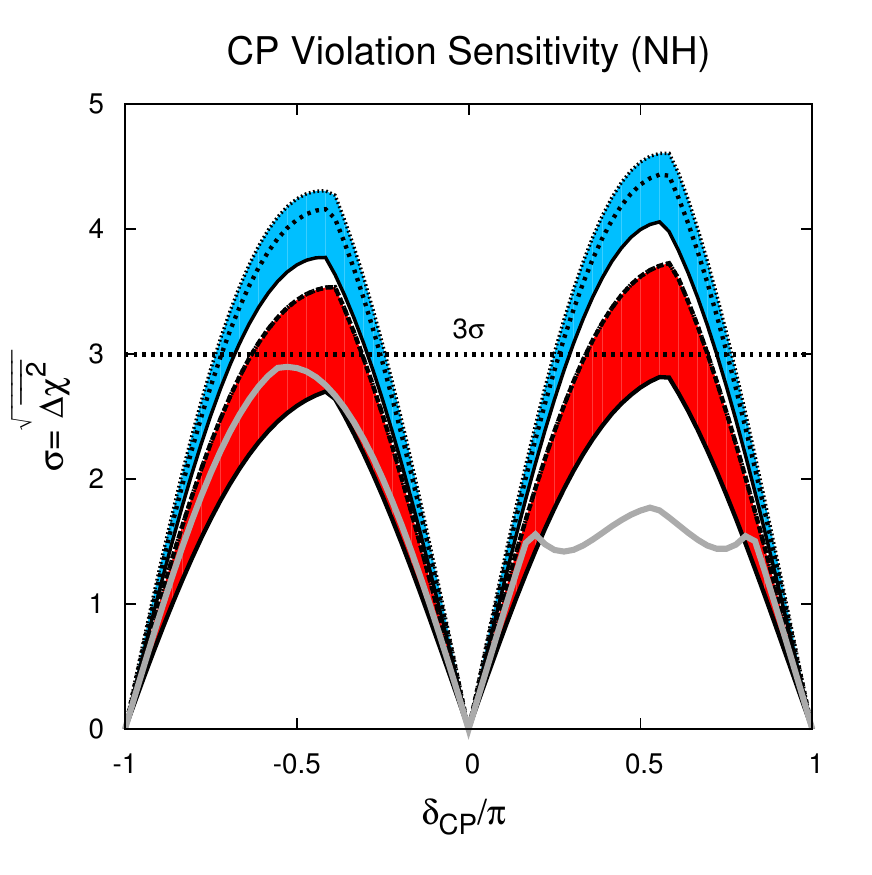}
\includegraphics[width=0.5\textwidth,trim=0cm 0.3cm 0cm 0.3cm,clip]{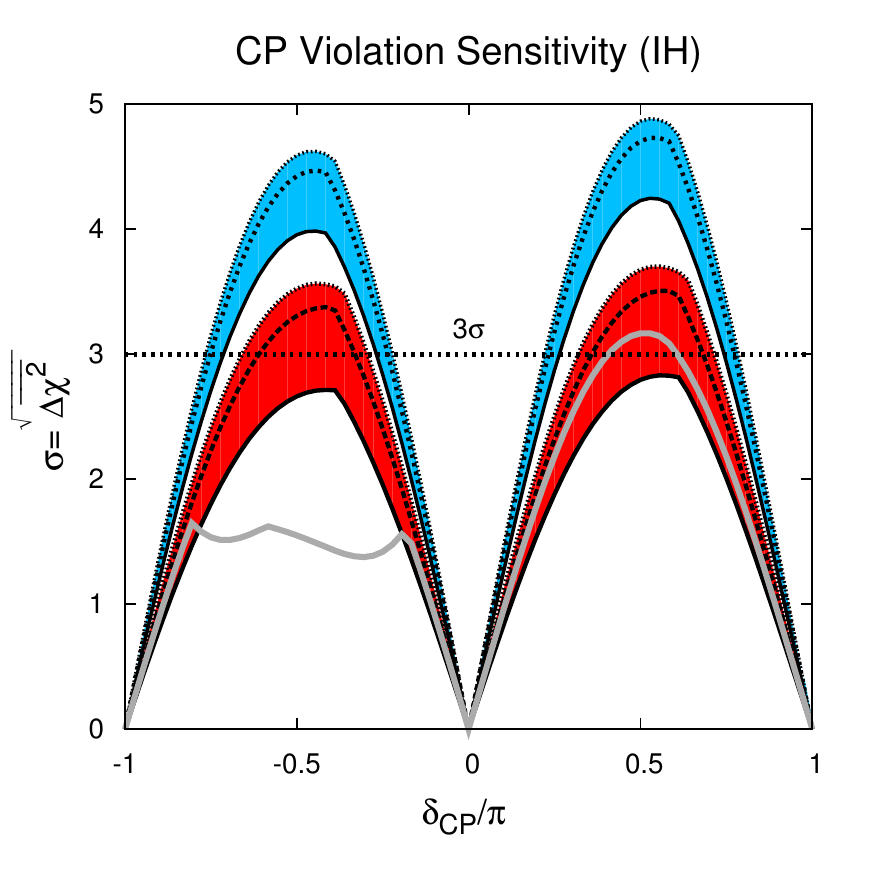}
}
\caption[Sensitivity to MH and CP violation in a \ktadj{10}
LArTPC]{The significance with which the mass hierarchy (top) and
  CP violation ($\mdeltacp \neq 0 \ {\rm or} \ \pi$, bottom) can
  be determined as a function of the value of \deltacp.  
  The plots on the left are for
  normal hierarchy and the plots on the right are for inverted
  hierarchy. The red band shows the sensitivity that is achieved
  by a typical experiment with
  the LBNE \ktadj{10} configuration alone, where the width of the band shows the range of
  sensitivities obtained by varying the beam design and the signal and background
  uncertainties as described in the text. The cyan band shows the sensitivity
  obtained by combining the \SIadj{10}{\kt} LBNE with T2K and NO$\nu$A, and the 
  gray curves are the expected
  sensitivities for the combination of NO$\nu$A and T2K; the assumed
  exposures for each experiment are described in the text. For the CP-violation 
sensitivities, the MH is assumed to be
  unknown.} 
\label{fig:10ktonsens}
\end{figure}
\begin{figure}[!htb]
\centerline{
\includegraphics[width=0.5\textwidth]{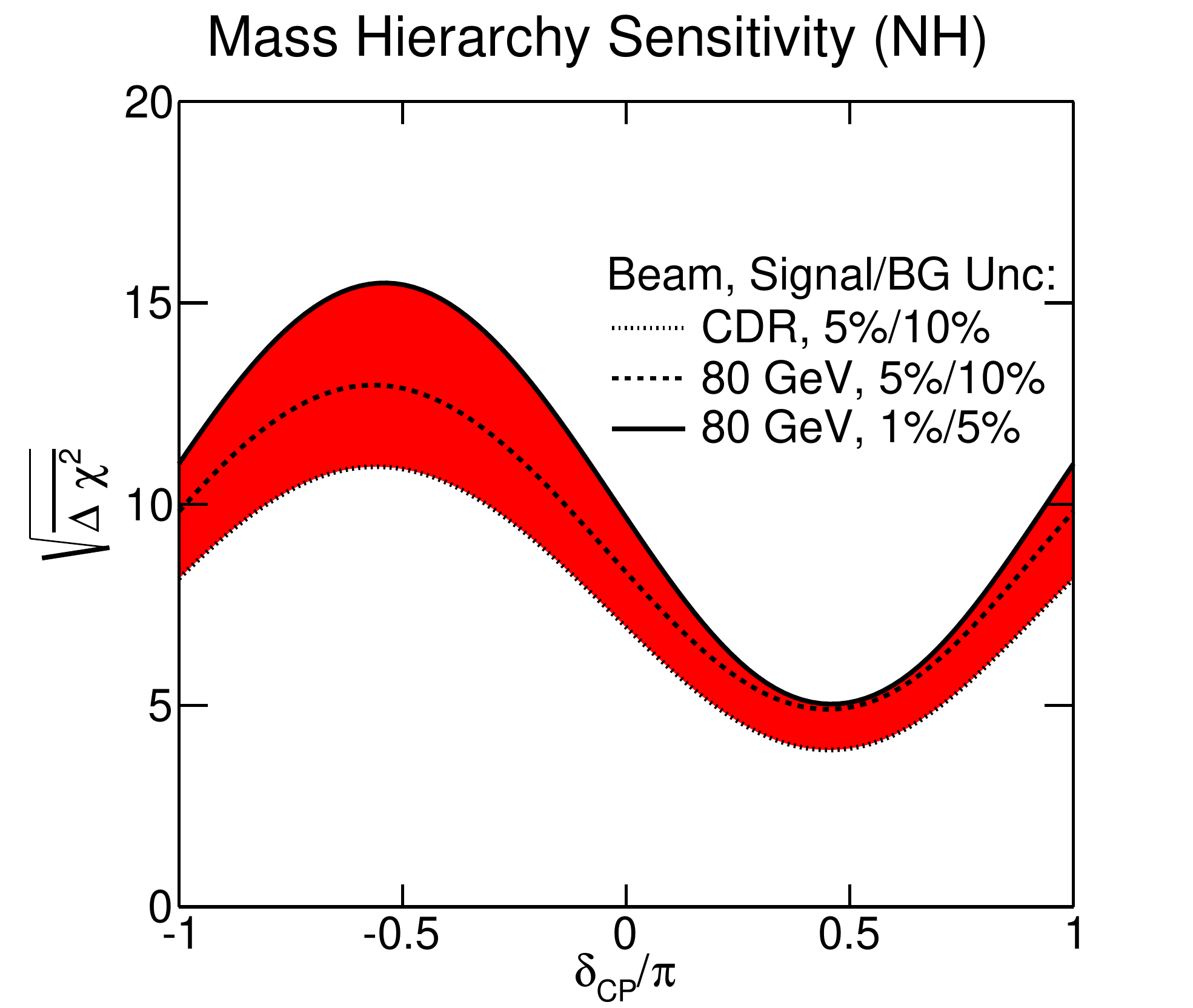}
\includegraphics[width=0.5\textwidth]{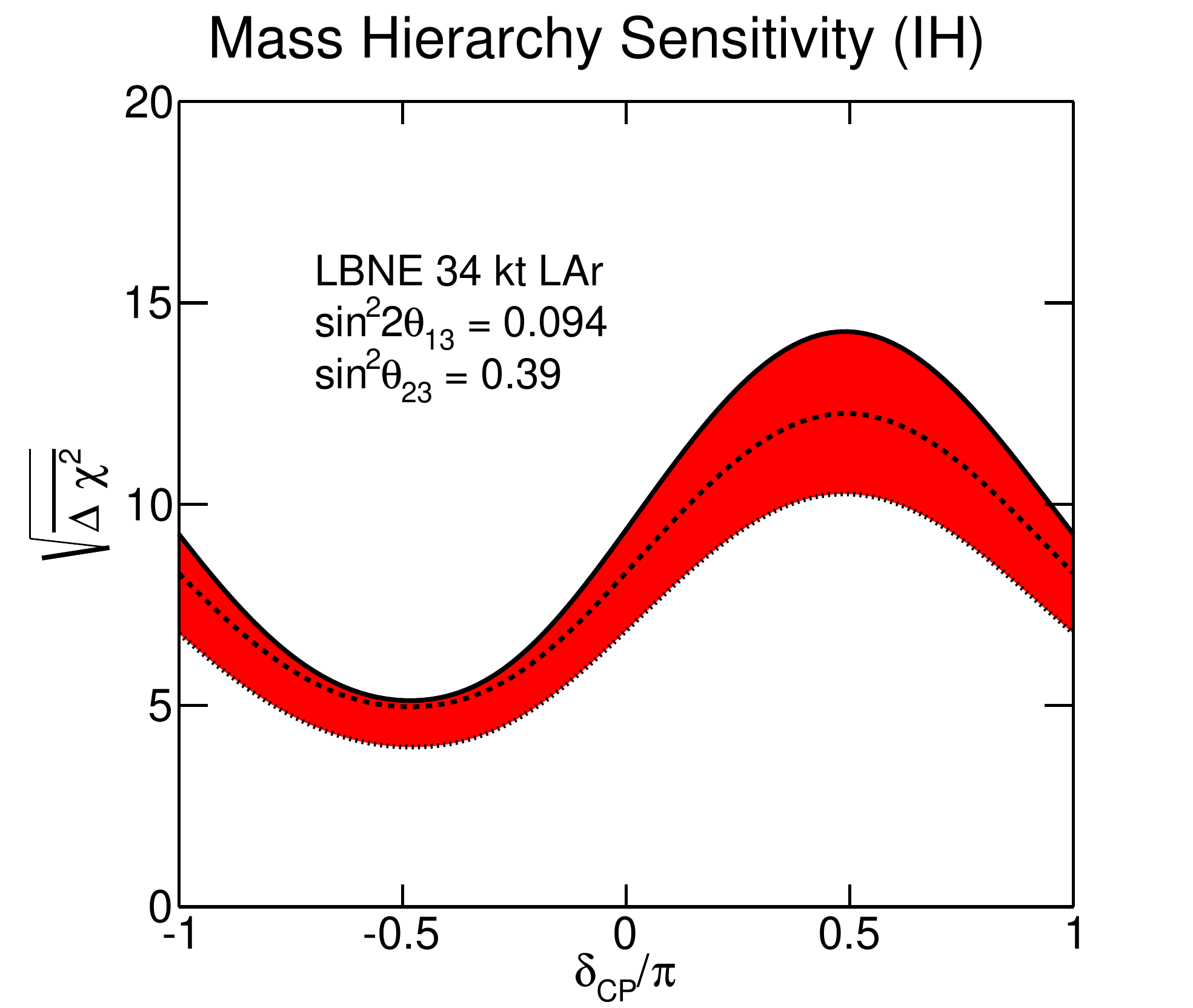}
}
\vskip 0.27in
\centerline{
\includegraphics[width=0.5\textwidth]{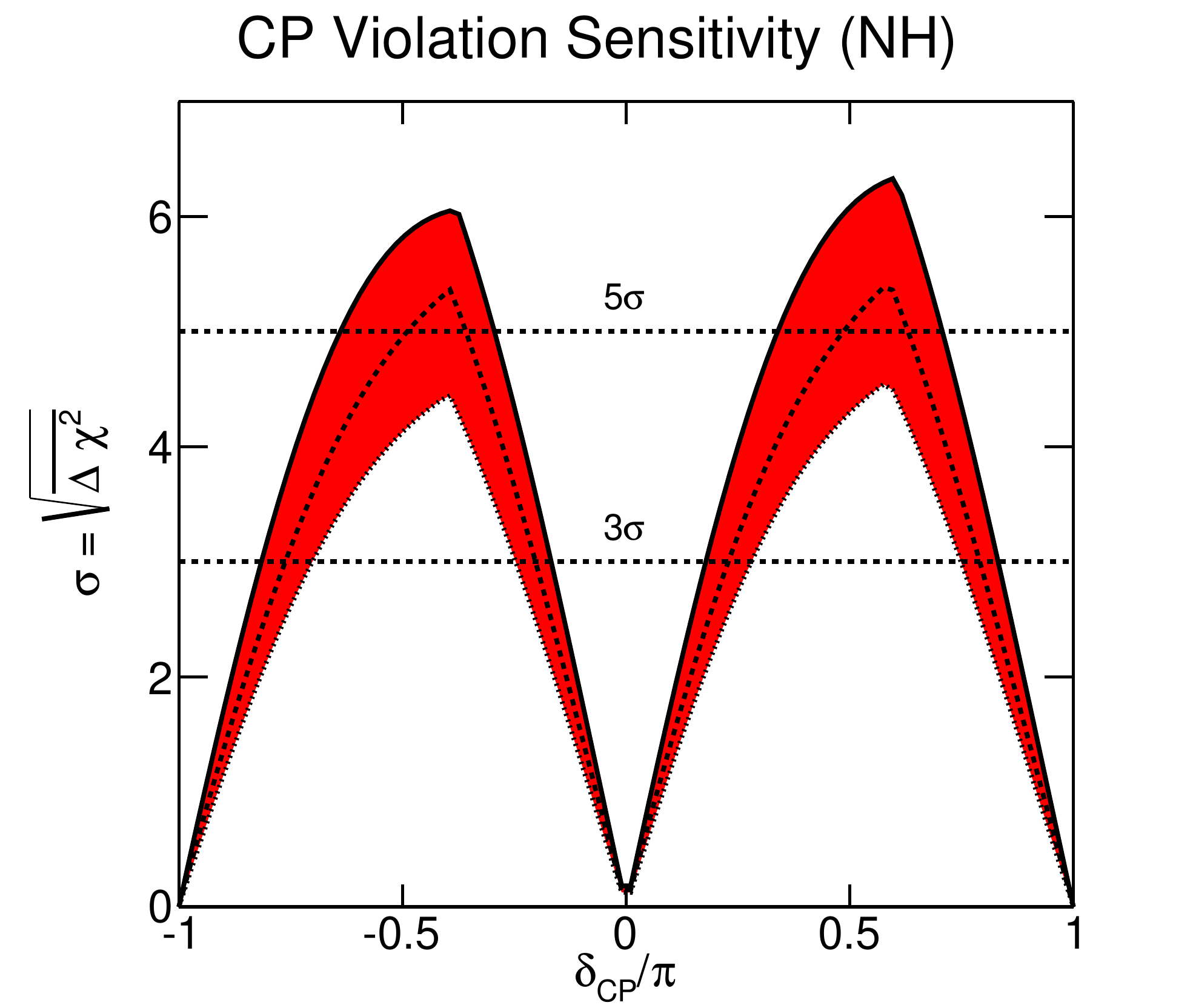}
\includegraphics[width=0.5\textwidth]{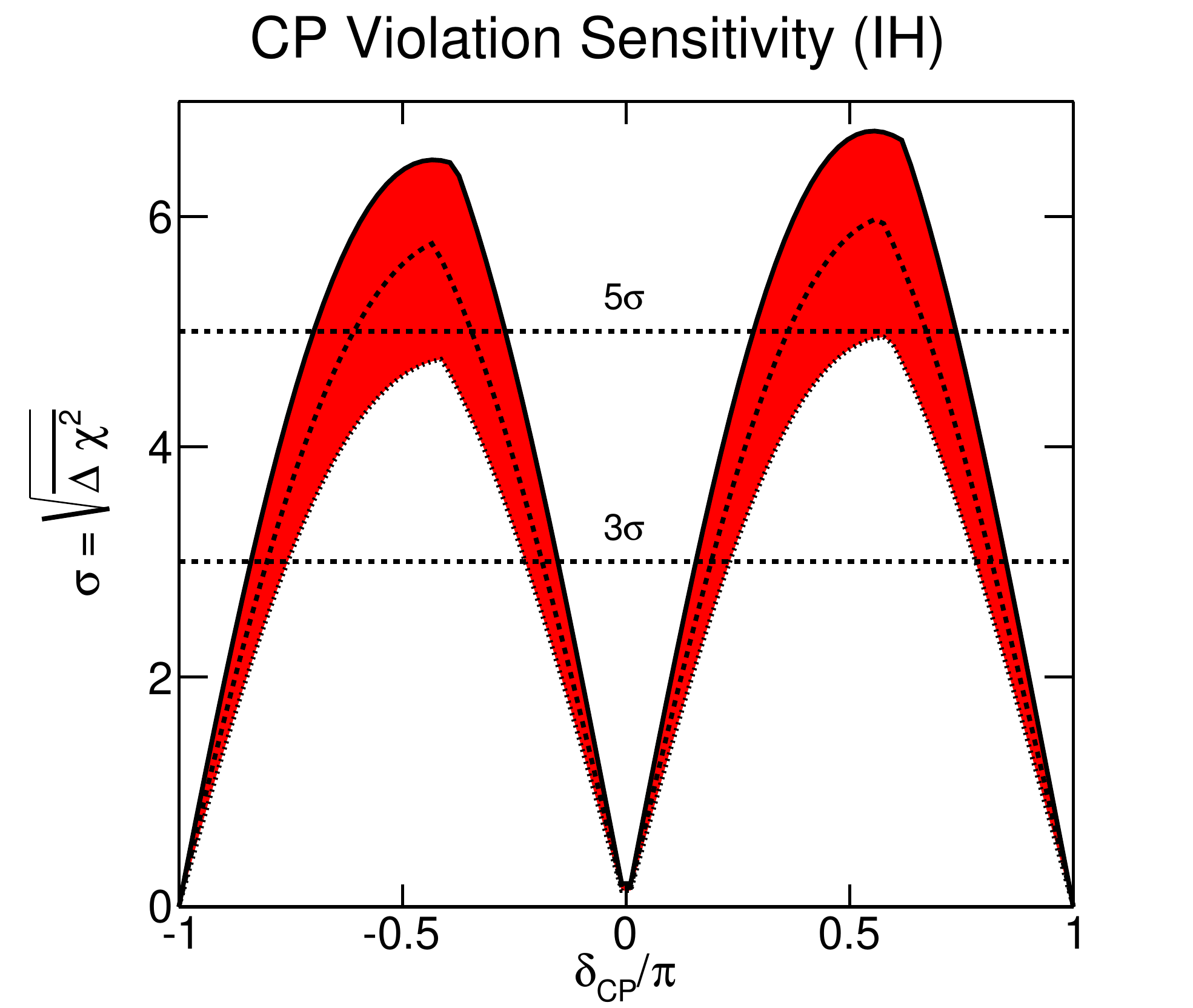}
}
\vskip 0.27in
\caption[Sensitivity to MH and CP violation in a \ktadj{34}
  LArTPC]{The significance with which the mass hierarchy (top) and
  CP violation ($\mdeltacp \neq 0 \ {\rm or} \ \pi$, bottom) can
  be determined by a typical LBNE experiment with a \ktadj{34} far
  detector as a function of the value of \deltacp.  The plots on
  the left are for normal hierarchy and the plots on the right are for
  inverted hierarchy. The width of the red band shows the range of
  sensitivities that can be achieved by LBNE when varying the beam
  design and the signal and background uncertainties as described in
  the text.}
\label{fig:35kton}
\end{figure}

The sensitivity bands in Figures~\ref{fig:10ktonsens} and \ref{fig:35kton}
represent the variation in sensitivity as a function of the beam
design and normalization uncertainties 
on the signal and
background. The solid curve at the lower end of the red band
represents the beamline design described in the LBNE CDR Volume
2~\cite{CDRv2} for which there is no near detector.  The
dashed line above the solid curve represents the sensitivity with the
beam design improvements currently under study as described in
Section~\ref{beamline-chap}, still without a near detector. The dashed
line at the upper end of the red band represents the case in which
both the beam design improvements and a high-resolution, highly capable
near detector are implemented. The key design goal of the LBNE near
detector and beamline simulation software is to enable a prediction of the
far detector unoscillated flux with a precision of $\leq 2\%$.
Therefore, the total signal and background normalization uncertainties
on the $\nu_\mu$ disappearance signal are assumed to be 5\% and 10\%,
respectively.  The default $\nu_e$ appearance signal {\em
  uncorrelated} normalization uncertainties for the full-scope 
LBNE presented in this chapter are assumed to be 1\%. The $\nu_e$ appearance 
background uncertainty is expected to be at least as good
as the $\sim5\%$~\cite{Adamson:2013ue} achieved by the $\nu_e$ appearance
 search in the MINOS experiment.

A detailed discussion of the systematics assumptions for
LBNE is presented in Section~\ref{sec:systs}.  
In the case that LBNE has no near neutrino detector, the uncertainties on signal
and background are expected to be $5\%$ and $10\%$, respectively,
extrapolating from the performance and detailed knowledge of the NuMI beam on which the
LBNE beamline is modeled, in situ measurements of the muon flux at the
near site as described in~\cite{CDRv1}, the expectation of
improved hadron production measurements with the NA61 and MIPP 
experiments, and the experience of previous $\nu_e$ appearance
experiments as summarized in Table~\ref{tab:pastexpts}.

\clearpage
 \begin{table}[!htb]
  \begin{center}
    \begin{tabular}{$L^c^c^c^c^l} \toprule 
      \rowtitlestyle
     Experiment &  Year & \numu-NC/CC  & \nue-CC &  Background  &  Comment \\ 
      \rowtitlestyle
      & &  Events  &  Events           &         Syst.Error     &         \\ \toprowrule
     BNL E734~\cite{Murtagh:1987xu}   &  1985 &  235 & 418 & 20\% & No ND \\ \colhline
     BNL E776(NBB)~\cite{Seto:1988vg} & 1989  & 10     &  9   & 20\%    & No ND \\  \colhline
     BNL E776 (WBB)~\cite{Borodovsky:1992pn}    &  1992 & 95     &  40  & 14\%    & No ND \\ \colhline
     NOMAD~\cite{Astier:2003gs}       & 2003 & $<$300  & 5500 & $<5$\%  & No ND \\ \colhline
     MiniBooNE~\cite{AguilarArevalo:2008rc} & 2008 & 460   & 380  & 9\%     & No ND \\ \colhline
     MiniBooNE~\cite{Aguilar-Arevalo:2013pmq} & 2013 & 536   & 782  & 5\%     & SciBooNE \\ \colhline
     MINOS~\cite{Adamson:2013ue}     & 2013   &  111     &   36  & 4\%   & ND--FD \\ \colhline
     T2K~\cite{Abe:2013hdq} & 2013 & 1.1 & 26 & 9\%* & ND--FD \\ \bottomrule
    \end{tabular}
    \caption[Achieved systematic error performance in prior
      \numu\ $\rightarrow$ \nue\ oscillation experiments]{Summary of
      achieved systematic error performance in several select prior
      \numu\ $\rightarrow$ \nue\ oscillation experiments.  These
      numbers were extracted from publications and may not correspond
      exactly to the description in the text. NBB/WBB indicates a
      narrow/wide band beam. \emph{No ND} indicates there was no near
      detector, and \emph{ND-FD} indicates a two (near-far) detector
      experiment with extrapolation of the expected background and
      signal from the near to the far detector. In the case of T2K, the
      quoted systematic (*) is actually the total uncertainty on the observed events,
      which are predominately signal.}
    \label{tab:pastexpts}
    \end{center}
\end{table}

\subsection{Interpretation of Mass Hierarchy Sensitivities}
\label{sec-mh-statistics}

\begin{introbox}
  LBNE will be definitive in its ability to discriminate between 
  normal and inverted mass hierarchy for the allowed range 
  of unknown parameters such as \deltacp and 
  $\sin^2{\theta_{23}}$.  To assess the sensitivity of  
  LBNE to this physics, particularly for the case of less 
  favorable parameter values, detailed understanding of 
  statistical significance is essential.

  At the true values of \deltacp for which the mass hierarchy
  asymmetry is maximally offset by the leptonic CP asymmetry,
  LBNE's sensitivity to the mass hierarchy is at its minimum.  Even in
  this case, with a \ktadj{34} LArTPC operating for six years in a
  \MWadj{1.2} beam, the $|\Delta\chi^2|$ value obtained in a typical
  data set will exceed \num{25}, allowing LBNE on its own to rule out
  the incorrect mass ordering at a confidence level above $1-3.7\times
  10^{-6}$.  Considering fluctuations, LBNE will measure, in
  $\geq97.5\%$ of all possible data sets for this 
  least favorable scenario, a value of $|\Delta \chi^2|$ equal to
  \num{9} or higher, which corresponds to a $\geq 99\%$ probability of
ruling out the incorrect hierarchy hypothesis.
\end{introbox}

In the mass hierarchy (MH) determination, only two possible results are
considered, as the true MH is either normal (NH)  or inverted (IH).
Reference~\cite{Qian:2012zn} presents the statistical considerations
of determining the sensitivity of an experiment to the MH,
framed partly in the context of two separate but related questions:
\clearpage
\begin{enumerate}
\item \emph {Given real experimental data, with what significance can
the MH be determined?}
\item \emph {When evaluating future experimental sensitivities, 
what is the probability that
a particular experimental design will be able to determine the MH
with a given significance?}
\end{enumerate}

Once data are in hand, a number of techniques based either within Bayesian or frequentist statistics 
make it possible to determine the level of confidence at which one MH hypothesis or 
the other can be ruled out.  
In assessing the sensitivity of future experiments, it is common 
practice to generate a simulated data set (for an assumed true 
MH) that does not include statistical fluctuations.  
The expected sensitivity can be reported 
as $\overline{\Delta\chi^2}$,
representative of the mean or the most likely value of $\Delta\chi^2$ that 
would be obtained in an ensemble of experiments for a particular true MH.
With the exception of Figure~\ref{fig:mh_statbands}, the sensitivity plots
in this document have been generated using this method.

However, addressing the expected sensitivity of an experiment per the
second question above requires consideration of the effect of
statistical fluctuations and variations in systematics.  If the
experiment is repeated many times, a distribution of $\Delta\chi^2$
values will appear.  Studies in~\cite{Qian:2012zn} and elsewhere
(e.g.,~\cite{Blennow:2013oma}) show that the $\Delta \chi^2$ metric
employed here {\em does not} follow the commonly expected $\chi^2$
function for one degree of freedom, which has a mean of
$\overline{\Delta\chi^2}$ and can be interpreted using a Gaussian
distribution with a standard deviation of
$\sqrt{|\overline{\Delta\chi^2}|}$. Rather, these studies show that
when the observed counts in the experiment are large enough,
the distribution of $\Delta\chi^2$ used here approximately follows
a Gaussian distribution with a
mean and standard deviation of $\overline{\Delta\chi^2}$ and
$2\sqrt{|\overline{\Delta\chi^2}|}$, respectively~\cite{Qian:2012zn}.

Figure~\ref{fig:mh_toy} shows the expected distribution of $\Delta\chi^2$ 
values in LBNE from toy Monte Carlo studies. 
\begin{figure}[!htb]
\centerline{
\includegraphics[width=0.5\textwidth,trim=0.5cm 0.5cm 0.5cm 0.5cm,clip]{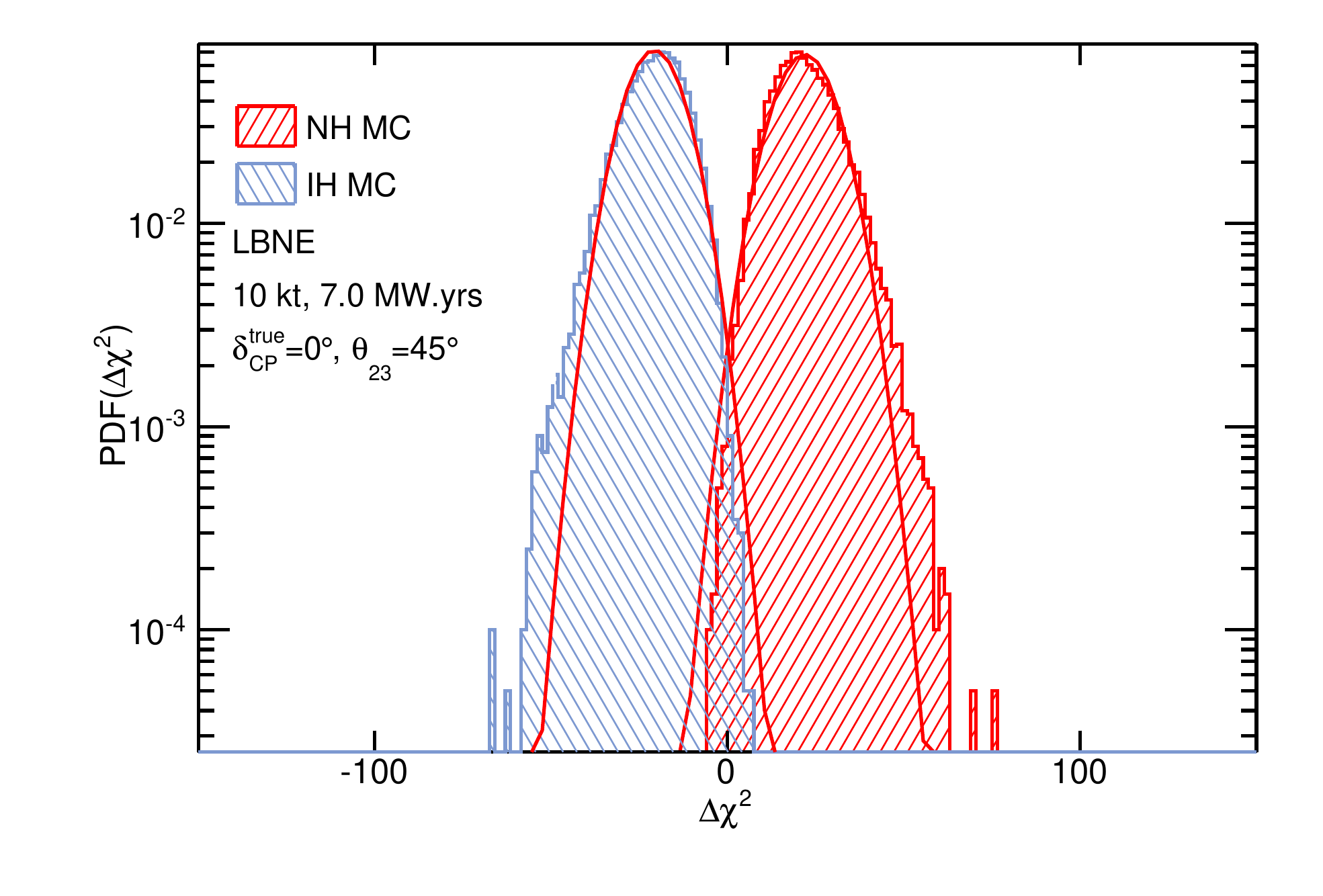}
\includegraphics[width=0.5\textwidth,trim=0.5cm 0.5cm 0.5cm 0.5cm,clip]{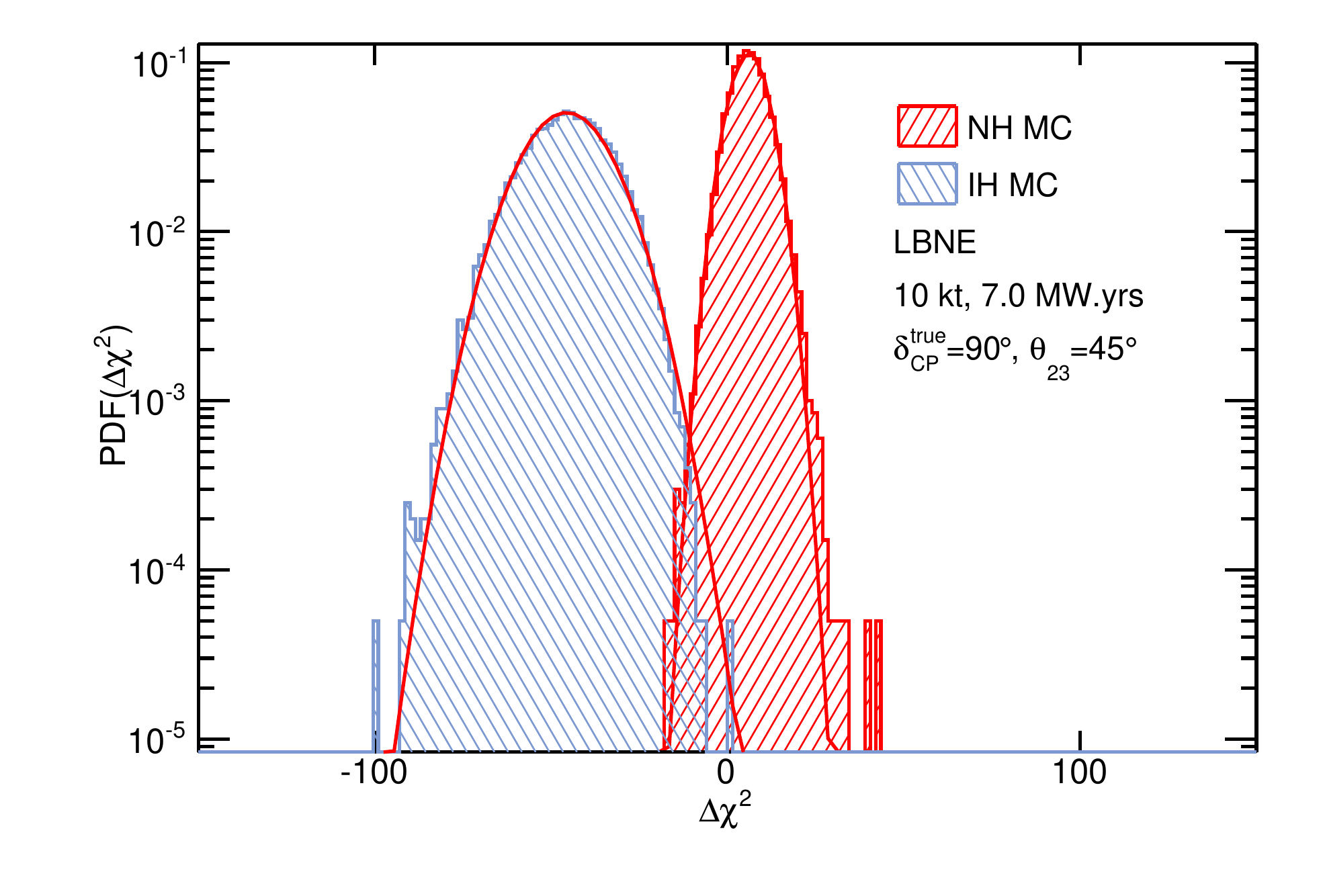}
}
\centerline{
\includegraphics[width=0.5\textwidth,trim=0.5cm 0.5cm 0.5cm 0.5cm,clip]{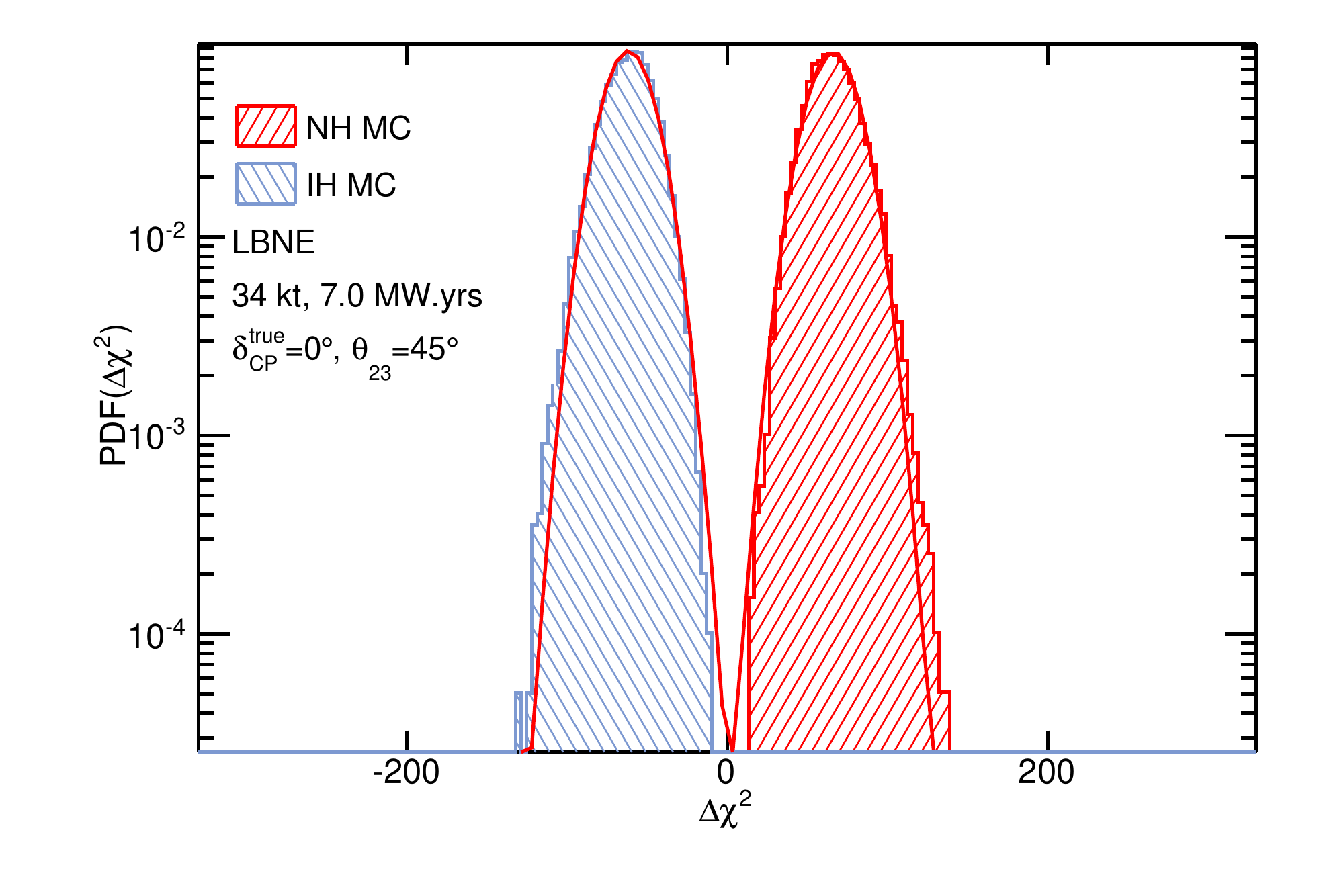}
\includegraphics[width=0.5\textwidth,trim=0.5cm 0.5cm 0.5cm 0.5cm,clip]{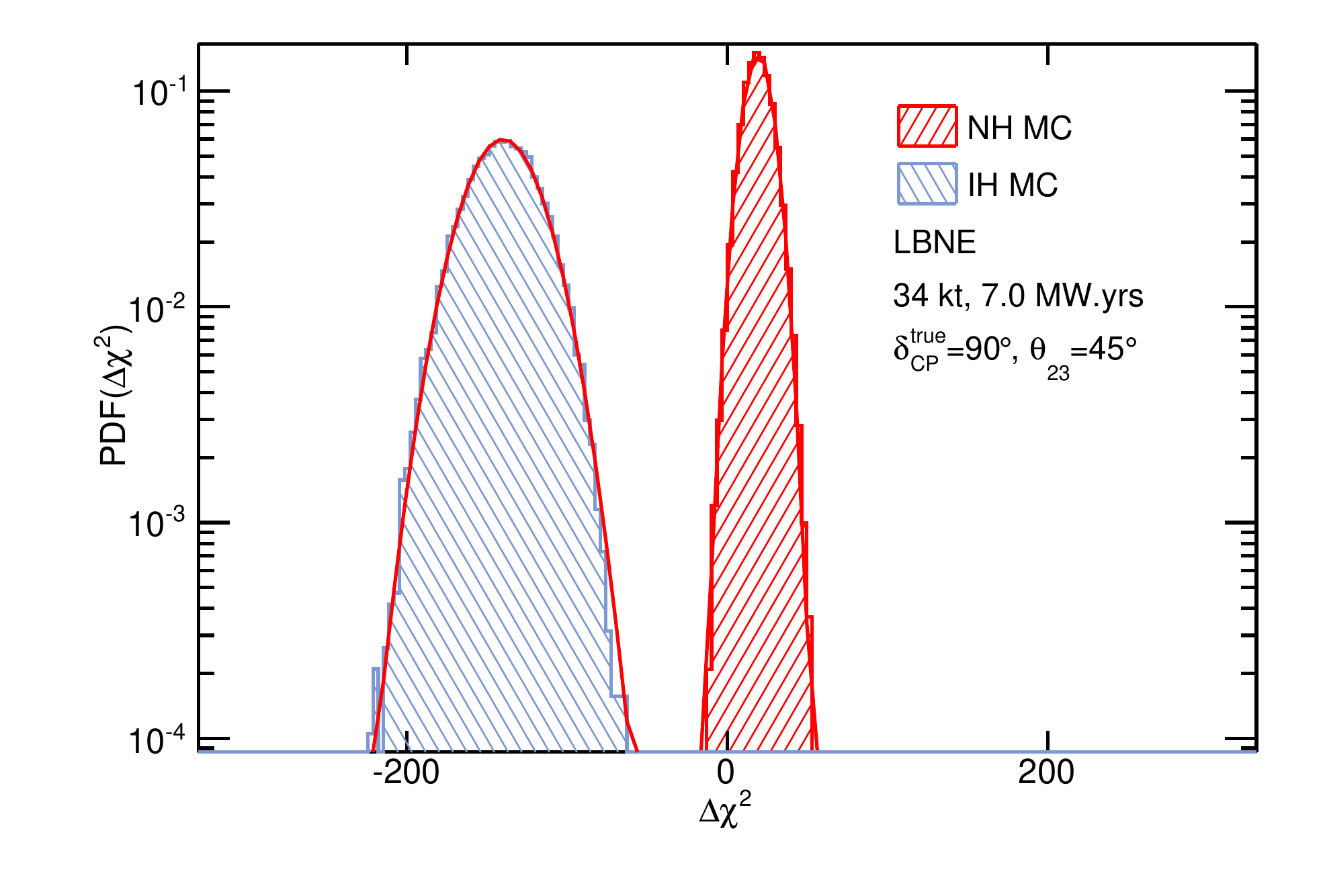}
}
\caption[$\Delta\chi^2_{\rm MH=NH}$ and $\Delta\chi^2_{\rm MH=IH}$
distributions for LBNE from toy MC studies]{ $\Delta\chi^2_{\rm
    MH=NH}$ (red) and $\Delta\chi^2_{\rm MH=IH}$ (blue) distributions
  for LBNE from Toy MC studies.  The top set of figures are for a
  \SIadj{10}{\kt} detector operating six years in a \MWadj{1.2} beam. The
  bottom set is for a \SIadj{34}{\kt} detector operating six
  years in a \MWadj{1.2} beam. The figures on the left are for 
  $\delta^{true}_{cp} = 0$ and the figures on the right are 
  for $\delta^{true}_{cp} = 90^\circ$. The value of \deltacp 
  is unconstrained in the fit.} 
\label{fig:mh_toy}
\end{figure}
The interpretation of pairs of distributions, such as those in the
various panels of this figure, depends on the information being sought. For
example, one is not necessarily interested simply in the fraction of
experiments where $\Delta\chi^2$ has the ``right'' sign.  (An
experiment that obtains a small value of $\Delta\chi^2$, even with the
``right'' sign, would not be particularly constraining since there is
no way {\em a priori} to know which is the right sign --- this is what
the experiment is attempting to measure.)  It should also be noted
that in general $|\overline{\Delta\chi^2_{\rm MH=NH}}|$ , i.e., true NH, is not necessarily equal to
$|\overline{\Delta\chi^2_{\rm MH=IH}}|$, i.e., true IH, nor do the corresponding distributions necessarily have the
same shape.  For some ranges in \deltacp, for example, the event
rate in LBNE is sufficiently 
different for the two MH hypotheses
that the corresponding distributions in $\Delta\chi^2$ are quite
distinct.

The plots shown on the left in Figure~\ref{fig:mh_toy} illustrate the 
case for a true value of $\mdeltacp=0^\circ$, where the 
$\Delta \chi^2$ distributions for NH and IH scenarios are similar.
Shown on the right are the corresponding distributions for the case of 
 $\mdeltacp = 90^\circ$, where for NH 
the matter asymmetry is maximally offset by the CP asymmetry, leading 
to poorer MH discrimination.  For the IH case, 
these effects go in the same direction, leading to better MH discrimination.
The converse is the case for $\mdeltacp=-90^\circ$.
Since the true value of \deltacp is unknown (although a best-fit value
and confidence interval will emerge from the analysis of the data 
collected), comparison of a given value of $\Delta \chi^2$ with 
expected distributions for NH and IH cases for the \emph{same} 
value of \deltacp does not in general provide the appropriate test.  
For simplicity, following~\cite{Blennow:2013oma}, 
the discussion below focuses on the respective values 
of \deltacp for which the experiment will have poorest sensitivity 
for NH ($+90^\circ$) and IH ($-90^\circ$) scenarios.

Given the above introduction to the statistical fluctuation issues,
it is natural to employ the statistical language of hypothesis testing
in projecting LBNE's MH sensitivity.  Specifically, 
 $\alpha$ is defined as the desired Type-I error rate --- that is,
the probability of rejecting a particular hypothesis, e.g., NH, in the case where 
this is the true hypothesis.  
One can then ask what the corresponding Type-II error rate
$\beta$ would be, defined as the probability of accepting
the hypothesis being tested (NH in this example), when in fact the alternate
hypothesis (IH) is true.
The pair of $\alpha$ and $\beta$ would correspond to a particular
value of $\overline{\Delta\chi^2}$
chosen (in advance of the experiment)
as a criterion for deciding whether to 
rule out the NH (or IH).
Historically, many experiments have characterized their 
anticipated sensitivity by
reporting $\alpha$ for the case of $\beta = 0.5$,
which is nothing more than that given by the median value of the test
statistic (in this case, $\Delta \chi^2 = \overline{\Delta \chi^2}$) as 
described above. Sometimes, the sensitivity is also reported
as the square root of $\overline{\Delta \chi^2}$. 

Due to the approximate symmetry of the MH 
ambiguity as a function of \deltacp for the two MH scenarios
and the desire to be able to reject 
exactly one of the two possible mass orderings~\cite{Blennow:2013oma},
it is also natural to report a value of $\alpha$ for an experiment such 
that $\alpha = \beta$~\cite{Cousins:priv2013,Cousins:2005pq,Blennow:2013oma}.
In this way, it is possible to express just how \emph{unlucky} an experiment can be while 
maintaining a corresponding sensitivity $\alpha$. 
In the case of LBNE, a reasonable benchmark for comparison 
corresponds to 
$\overline{\Delta\chi^2} = 36$.  For this case, specifying $\alpha = \beta$ 
yields $\alpha=0.0013$, which means that the experiment will have a 0.13\% 
probability of ruling out 
the true MH hypothesis and of accepting the wrong MH
hypothesis.

As described above, and as is evident in the plots presented, such as those 
in Figures~\ref{fig:10ktonsens} and \ref{fig:35kton},
the sensitivity of LBNE
is strongly dependent on the true value of \deltacp; 
Figure~\ref{fig:mh_statbands}   
shows that it also depends on the true value of $\sin^2{\theta_{23}}$.  
While plotting the value of
$\alpha$ (for some choice of $\beta$, such as $\beta = 0.5$ or
$\beta = \alpha$) as a
function of these parameters encapsulates the sensitivity, 
a visually helpful presentation is obtained by plotting the
expected mean value, $\overline{\Delta\chi^2}$, as well as ranges of
possible values corresponding to the expected distribution in
$\Delta\chi^2$.  Thus,
Figure~\ref{fig:mh_statbands} shows the dependence of
$\sqrt{|\overline{\Delta\chi^2}|}$ on the true value of \deltacp
for the typical LBNE data set, for two possible values of
$\sin^2\theta_{23}$, as well as the corresponding 
expectation bands within which 68\% (green) and 95\%
(yellow) of LBNE sensitivities will fall.  
These expectation bands give a
semi-quantitative picture of the likely range of outcomes for the
experiment.

The horizontal dashed lines on
Figure~\ref{fig:mh_statbands} specify the confidence level of
an experiment with a particular value of $\Delta \chi^2$ such that:
\begin{equation}
{\rm CL} = P({\rm favored \ MH} | {\rm data} \,  x) / (P({\rm favored
  \ MH} | {\rm data} \,  x) + P({\rm unfavored \ MH} | {\rm data} \,  x) ),
\end{equation}
following the convention in \cite{Qian:2012zn}, where 
the notation $P({\rm A}|{\rm B})$ represents the probability of 
A given condition B, and these probabilities
are inferred from the corresponding likelihoods via Bayes' Theorem.
Alternatively, the $\Delta\chi^2$ values shown in these plots can be 
approximately translated to sensitivities in terms of $\alpha$, for whatever 
choice of $\beta$ is desired,
following, for example,  the prescription described in~\cite{Blennow:2013oma}.   
\begin{figure}[!htb]
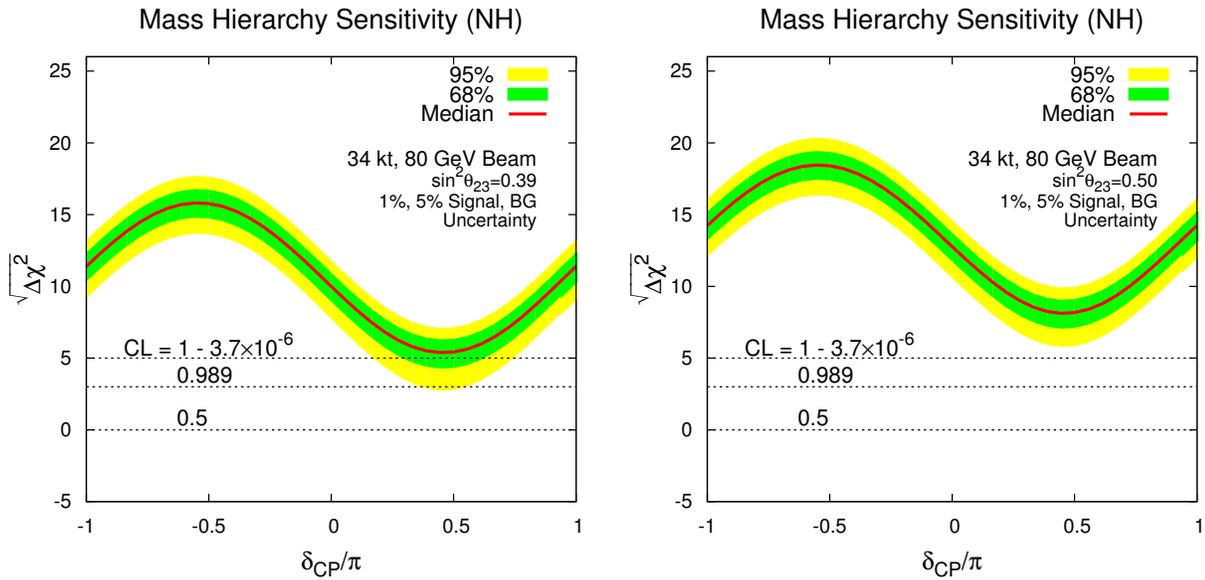

\centerline{
\includegraphics[width=0.5\textwidth,trim=0cm 0.3cm 0cm 0.3cm,clip]{bassfig/mh_lbne35BU.pdf}%
\includegraphics[width=0.5\textwidth,trim=0cm 0.3cm 0cm 0.3cm,clip]{bassfig/mh_lbne35BU_o2.pdf}
}
\caption[Expected $\sqrt{|\Delta\chi^2|}$ values for the typical LBNE
  experiment]
  {The square root of
  the mass hierarchy discrimination metric $\Delta \chi^2$
  is plotted as a function of the unknown value of \deltacp for
  the full-scope LBNE with \SI{34}{\kt}, 3+3 ($\nu+\overline{\nu}$) years of
  running in a \MWadj{1.2} beam, for true NH.  
  The red curve represents the most likely experimental value obtained,  
  estimated using a data set absent statistical fluctuations, while 
  the green and yellow bands represent the
  range of $\Delta \chi^2$ values expected in 68\% and 95\% of all
  possible experimental cases, respectively. 
  The horizontal lines indicate the probability that an
  experiment with that value of $\Delta \chi^2$ correctly
  determines the MH, computed according to a Bayesian 
  statistical formulation. 
  The plot on the left assumes a value of $\sin^ 2 \theta_{23}
  = 0.39$~\cite{Fogli:2012ua},  while that on the right 
  assumes $\sin^ 2 \theta_{23}
  = 0.5$ (maximal $\nu_\mu$-$\nu_\tau$ mixing).}
\label{fig:mh_statbands}
\end{figure}

As seen in Figure~\ref{fig:mh_statbands}, a typical LBNE data set with a \SIadj{34}{\kt}
detector can determine the MH with
$|\overline{\Delta\chi^2}| \geq 25$ for all values of \deltacp
(for the left plot, where $\sin^2 \theta_{23} = 0.39$).  
From a Bayesian analysis,  
the probability that an experiment measuring $|\Delta\chi^2| = 25$ 
has ruled out the true MH hypothesis is $3.7\times10^{-6}$, 
as indicated for the corresponding horizontal dashed line 
in the plots in this figure.  
When considering the effect of statistical fluctuations, 
for the same value of $\theta_{23}$,
about 97.5\% of experiments will determine the MH 
with $|\Delta\chi^2| > 9$ for the least favorable value
of \deltacp, 
where $|\Delta\chi^2| = 9$ corresponds to a CL of 98.9\%.

For the bulk of the range of \deltacp, the sensitivity 
of LBNE is vastly better than for the least favorable value 
described above.  Furthermore, newer data
prefer values of $\theta_{23}$ closer to 
maximal~\cite{Capozzi:2013csa}, which results in significantly  
enhanced LBNE MH sensitivity.  As shown in the right-hand plot of 
Figure~\ref{fig:mh_statbands}, if $\sin^2 \theta_{23} = 0.5$,
the expected MH
sensitivity for the typical LBNE experiment at the least favorable \deltacp 
point is $|\overline{\Delta\chi^2}| \approx 64$,
which is significantly larger than the sensitivity of $|\overline{\Delta\chi^2}| \approx 25$ 
expected for the same value of \deltacp if $\sin^2
\theta_{23} = 0.39$.
 This suggests that a typical LBNE 
data set will determine the MH with
$|\Delta\chi^2|$ well above the benchmark value of $36$ 
mentioned above for even the least favorable values of \deltacp.  

In addition to detailed LBNE-specific frequentist studies 
reported in~\cite{Blennow:2013oma},
an LBNE-specific update (using both Bayesian and frequentist approaches)
to the general statistical studies reported in~\cite{Qian:2012zn} 
is in preparation.

\subsection{Sensitivities and Systematics}
\label{sec:systs}

The main systematic uncertainties in any experiment are determined by the 
analysis strategy employed and the performance of the detector.
Figure~\ref{fig:nueanalysis}
outlines the analysis strategy commonly employed to extract oscillation
parameters in two-detector
long-baseline neutrino oscillation experiments.
\begin{figure}
\tikzstyle{NDDATA} = [rectangle, draw=black, rounded corners, fill=blue!40, text centered, text width=0.2\textwidth, minimum size=0.05\textheight]
\tikzstyle{FDPREDICT} = [rectangle, draw=black, rounded corners, fill=green!40, text centered, text width=0.2\textwidth, minimum size=0.05\textheight]
\tikzstyle{FDDATA} = [rectangle, draw=black, rounded corners, fill=magenta!40, text centered, text width=0.2\textwidth, minimum size=0.05\textheight]
\tikzstyle{ANALYSIS} = [rectangle, draw=black, rounded corners, fill=yellow!40, text centered, text width=0.2\textwidth, minimum size=0.05\textheight]
\tikzstyle{EMPTY} = [rectangle, draw=white, rounded corners, fill=white, text centered, text width=0.2\textwidth, minimum size=0.05\textheight]
\tikzstyle{ratioedge} = [blue, thick, double]
\tikzstyle{oscedge} = [red, thick, double]
\tikzstyle{addedge} = [green, thick, double]
\tikzstyle{anaedge} = [orange, thick, double]
  \begin{tikzpicture}
    \matrix [matrix of nodes, column sep=-0.1\textwidth, row sep=13mm] {
      \node (NDnueCC) [NDDATA] {ND $\nu_e$-CC candidates (all Bkgd) \\ $N_{\rm ND}^{data}(\nu_e)$}; &      &  &     & &
      \node (NDnumuCC) [NDDATA] {
        ND $\nu_\mu$-CC candidates \\ 
        $ N^{data}_{\rm ND} ( \nu_\mu) = \Phi_{\rm ND} (\nu_\mu) \cdot\varepsilon_{\rm ND}  (\nu_\mu)  \cdot \sigma_{\rm ND} (\nu_\mu)$}; &            \\
      &    &  & & &
      \node (FDnumuCC) [FDPREDICT] {FD $\nu_\mu$-CC Unoscillated Prediction};  &        \\ 
      & & & & & & \\
      \node (FDnueCCvarBG) [FDPREDICT] {FD $\nu_e$-CC Bkgd prediction (NC+ Beam $\nu_e$-CC)};  & & 
      \node (FDnueCCnutauBG)[FDPREDICT] {FD $\nu_e$-CC Bkgd Prediction ($\nu_\tau$-CC)}; & &
      \node (FDnueCCsig)[FDPREDICT] {FD $\nu_e$-CC Signal Prediction };  & &
      \node (FDnumuCCsig)[FDPREDICT] {FD $\nu_\mu$-CC Signal Prediction \\ $N^{expected}_{\rm FD} ( \nu_\mu)$};  \\
      &  
       \node (FDnueCCpred) [FDPREDICT] {FD $\nu_e$-CC predicted candidates \\ $N^{expected}_{\rm FD} ( \nu_e)$};  & & & & &  \\
      &    &    & 
      \node (Analysis) [ANALYSIS]{ 3-flavor Analysis }; &   & & \\
    & & 
     \node (FDnueCCcand) [FDDATA] {FD $\nu_e$-CC Candidates \\ $N^{data}_{\rm FD} ( \nu_e)$}; & & 
      \node (FDnumuCCcand) [FDDATA] {FD $\nu_\mu$-CC Candidates \\ $N^{data}_{\rm FD} ( \nu_\mu)$}; & & \\
    }; 
    \path[->]
    (NDnueCC) edge[ratioedge] 
    node [left,text width=20mm, outer sep=1mm] {Beam MC \\ $\Phi_{\rm FD}/\Phi_{\rm ND}$} 
    (FDnueCCvarBG)
    (NDnumuCC) edge[ratioedge] 
    node[left,text width=20mm, outer sep=1mm] {Beam MC \\ $\Phi_{\rm FD}/\Phi_{\rm ND}$} 
    (FDnumuCC)
    (FDnumuCC) edge[oscedge] 
    (FDnueCCnutauBG) 
    (FDnumuCC) edge[oscedge] 
    (FDnueCCsig) 
    (FDnumuCC) edge[oscedge] node [right, text width=30mm, outer sep=5mm] {
      Oscillations \\ 
      $\varepsilon_{\rm FD}/\varepsilon_{\rm ND}$   \\ 
      $\sigma_{\rm FD}/\sigma_{\rm ND}$ } 
    (FDnumuCCsig)
    (FDnueCCvarBG) edge[addedge] node[left] {Add}  (FDnueCCpred)
    (FDnueCCnutauBG) edge[addedge] (FDnueCCpred)
    (FDnueCCsig) edge[addedge] (FDnueCCpred)
    (FDnueCCpred) edge[anaedge] (Analysis)
    (FDnueCCcand) edge[anaedge] (Analysis)
    (FDnumuCCcand) edge[anaedge] (Analysis)
    (FDnumuCCsig) edge[anaedge] (Analysis)
    ;
  \end{tikzpicture}
  \caption[Flow chart of $\nu_e$ appearance analysis method]{Flow
    chart of the $\nu_e$ appearance analysis method in a two-detector
    long-baseline experiment. $\Phi$ refers to the beam flux,
    $\varepsilon$ refers to detector efficiencies and smearing, and
    $\sigma$ refers to neutrino interaction modeling. The terms ND and FD refer to
    the near and far detector, respectively.}
\label{fig:nueanalysis}
\end{figure}
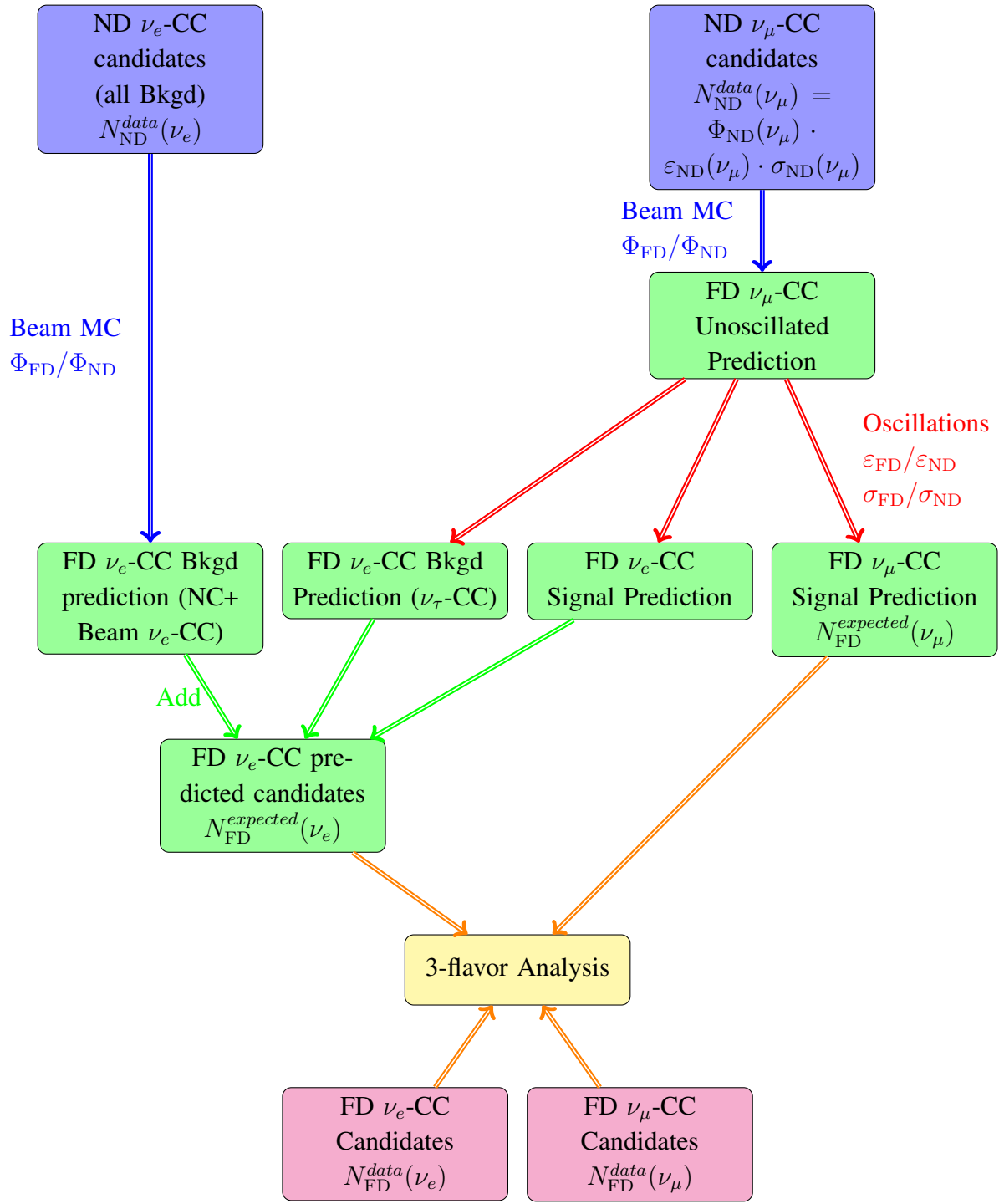
The measured spectrum of $\nu_\mu$ events in the near detector,
$N^{data}_{\rm ND} (\nu_\mu)$ is extrapolated to the far detector and
is used to predict both the $\nu_\mu$ and $\nu_e$ appearance
signals in the far detector, $N^{expected}_{\rm FD}(\nu_\mu)$ and
$N^{expected}_{\rm FD}(\nu_e)$ respectively.  The measured spectrum of
$\nu_e$ candidates in the near detector, $N^{data}_{\rm ND}
(\nu_e)$, which comprises mostly 
the beam $\nu_e$ events and NC
$\pi^0$ misidentified events, is used to predict the background to the
$\nu_e$ appearance signal in the far detector. In LBNE, neutrino
oscillation parameters will be extracted using a fit to four far detector data
samples: $\nu_e$, $\overline\nu_e$, $\nu_{\mu}$, and $\overline\nu_{\mu}$,
which will allow for partial cancellation of uncertainties.

In the current
generation of experiments, the measured spectrum of neutrino events in
the near detector is a product of beam flux ($\Phi$), detector
efficiency and smearing ($\varepsilon$), and neutrino interaction
dynamics ($\sigma$). 
To extrapolate the observed spectra in the near
detector to the far detector, corrections have to be made for: 
\begin{enumerate}
\item Differences in the beam flux in the near and far detectors, $\Phi_{\rm
    FD}/\Phi_{\rm ND}$: The near detector is much closer to the
  neutrino beamline and sees an extended source of neutrinos from the
  decay pipe as compared to the far detector, which observes a point
  source. A beam MC is used to correct for these
  differences. Uncertainties arise from inaccuracies in the simulation
  of the hadron production from the target, the focusing of the horns,
  the material in the beamline (which absorbs hadrons before they can
  decay), and the decay channel geometry.
\item Differences in near and far detector smearing and efficiencies,
  $\varepsilon_{\rm FD}/\varepsilon_{\rm ND}$: The largest
  uncertainties arise from the different event selection efficiencies
  in the near and far detectors and, in particular, the imperfect modeling of the
  energy scales of the near and far detectors. Identical near and far
  detectors allow most of these uncertainties to cancel in the
  extrapolation in the case of the $\nu_\mu$ signal prediction. The
  $\nu_e$ signal prediction is extrapolated from $N^{data}_{\rm ND}
  (\nu_\mu)$; thus there are irreducible residual uncertainties
  arising from different criteria used to select $\nu_e$ and $\nu_\mu$
  candidate events and different detector response functions.

\item Differences in the interactions of neutrinos in the near and far
  detector, $\sigma_{\rm FD}/\sigma_{\rm ND}$: In the case in which both
  near and far detectors use the same target nucleus, the
  differences cancel for extrapolation of the $\nu_\mu$ signal from
  the near to the far detector. When using the $\nu_\mu$ signal in the near
  detector to predict the $\nu_e$ (and $\nu_\tau$) signals in the far
  detector, uncertainties arising from differences in $\nu_e$
  ($\nu_\tau$) and $\nu_\mu$ interactions, $\sigma_{\rm
    FD}(\nu_e)/\sigma_{\rm ND}(\nu_\mu)$, dominate. These
  uncertainties are limited by theoretical uncertainties 
  and are typically smaller at higher energies.
\end{enumerate}
The estimation of the expected signals at the far detector can be summarized thus:
\begin{eqnarray}
N^{data}_{\rm ND} ( \nu_\mu) & = & \Phi_{\rm ND} (\nu_\mu) \otimes\varepsilon_{\rm ND}  (\nu_\mu)  \otimes \sigma_{\rm ND} (\nu_\mu) \\
N^{expected}_{\rm FD} ( \nu_\mu) & = & N^{data}_{\rm ND} ( \nu_\mu) \otimes 
\frac{\Phi_{\rm FD}(\nu_\mu)}{\Phi_{\rm ND}(\nu_\mu)} \otimes 
P(\nu_\mu \rightarrow \nu_\mu) \otimes 
\frac{\varepsilon_{\rm FD}(\nu_\mu)}{\varepsilon_{\rm ND}(\nu_\mu)} \otimes 
\frac{\sigma_{\rm FD}(\nu_\mu)}{\sigma_{\rm ND}(\nu_\mu)} 
\end{eqnarray}
\begin{eqnarray}
\nonumber
N^{expected}_{\rm FD} ( \nu_e) & = & \underbrace{
N^{data}_{\rm ND} ( \nu_\mu) \otimes 
\frac{\Phi_{\rm FD}(\nu_\mu)}{\Phi_{\rm ND}(\nu_\mu)} \otimes 
P(\nu_\mu \rightarrow \nu_e) \otimes 
\frac{\varepsilon_{\rm FD}(\nu_e)}{\varepsilon_{\rm ND}(\nu_\mu)} \otimes 
\frac{\sigma_{\rm FD}(\nu_e)}{\sigma_{\rm ND}(\nu_\mu)}
}_{\rm Expected \  signal \  events}
 \\ \nonumber
&  & + \underbrace{N^{data}_{\rm ND} ( \nu_e) \otimes 
\frac{\Phi_{\rm FD}(\nu_e)}{\Phi_{\rm ND}(\nu_e)} \otimes 
P(\nu_e \rightarrow \nu_e) \otimes 
\frac{\varepsilon_{\rm FD}(\nu_e)}{\varepsilon_{\rm ND}(\nu_e)} \otimes 
\frac{\sigma_{\rm FD}(\nu_e)}{\sigma_{\rm ND}(\nu_e)}
}_{\rm Beam \ \nu_e \ events} \\ \nonumber
&  & + {\rm NC \ background \ extrapolated \ from \ } N^{data}_{\rm
   ND} ( \nu_e) \\ 
&  & + {\rm \nu_\tau \ background \ extrapolated \ from \ } N^{data}_{\rm ND} ( \nu_\mu)
\end{eqnarray}
Expected systematic uncertainties on the LBNE $\nu_e$ appearance and
$\nu_\mu$ signal samples in the three-flavor fit for LBNE
(Table~\ref{tab:lar-nuosc-totaltable}) are extrapolated from the
current performance of the \linebreak MINOS~\cite{Adamson:2013ue,Adamson:2011qu}
and T2K~\cite{Abe:2013hdq} experiments.
The dominant uncertainties on the
current $\nu_e$ appearance analysis from MINOS and T2K and the
expected corresponding uncertainties in LBNE are shown in
Table~\ref{tab:nuesysts}. 
\begin{table}[!hbt]
  \caption[$\nu_e$ appearance signal dominant systematic uncertainties MINOS/T2K/LBNE]
{ The dominant systematic uncertainties on the $\nu_e$ appearance signal prediction in MINOS and T2K and a projection of the expected uncertainties in LBNE. For the MINOS uncertainties \emph{absolute} refers to the total uncertainty and $\nu_e$ is the effect on the $\nu_e$ appearance signal only. The LBNE uncertainties are the total \emph{expected} uncertainties on the $\nu_e$ appearance signal which include both correlated and uncorrelated uncertainties in the three-flavor fit.}
\label{tab:nuesysts}
\begin{tabular}{$L^c^c^c^l}
\toprule
\rowtitlestyle
Source of & MINOS & T2K & LBNE & Comments \\ 
\rowtitlestyle
Uncertainty & Absolute/$\nu_e$ & $\nu_e$ & $\nu_e$ & \\ \toprowrule
Beam Flux & 3\%/0.3\% & 2.9\% & 2\% & MINOS is normalization only.\\ 
after N/F & & & & LBNE normalization  and shape   \\
extrapolation & & & & highly correlated between $\nu_\mu/\nu_e$. \\ \toprule
\multicolumn{5}{^c}{Detector effects}  \\ \toprowrule
Energy scale  & 7\%/3.5\% & included& (2\%) & Included in LBNE $\nu_\mu$ sample  \\ 
($\nu_\mu$) & & above & &  uncertainty only in three-flavor fit. \\
& & & & MINOS dominated by hadronic scale. \\ \colhline
Absolute energy  & 5.7\%/2.7\% & 3.4\% & 2\% & Totally active LArTPC with calibration \\
scale ($\nu_e$) & & includes & & and test beam data lowers uncertainty. \\
 & & all FD & & \\
 & & effects & & \\ \colhline 
Fiducial & 2.4\%/2.4\% & 1\% & 1\% & Larger detectors = smaller uncertainty. \\ 
volume & & & & \\ \toprule
\multicolumn{5}{^c}{Neutrino interaction modeling}  \\ \toprowrule
Simulation & 2.7\%/2.7\% & 7.5\% & $\sim 2\%$ & Hadronization models are better  \\
includes: & & & & constrained in the LBNE LArTPC. \\
hadronization & & & &  N/F cancellation larger in MINOS/LBNE. \\ 
cross sections & & & & X-section uncertainties larger at T2K energies. \\ 
nuclear models & & & & Spectral analysis in LBNE provides \\ 
& & & & extra constraint. \\ \toprule
\rowtitlestyle
Total  & 5.7\% & 8.8\% & 3.6 \% & Uncorrelated $\nu_e$ uncertainty in  \\ 
\rowtitlestyle
& & & & full LBNE three-flavor fit = 1-2\%. \\ \bottomrule 
\end{tabular}
\end{table}
The categorization of the dominant
experimental uncertainties in Table~\ref{tab:nuesysts} are not
always in exact correspondence since T2K and MINOS are very different
experiments and deploy different analysis techniques.  A detailed
description of the expected LBNE performance on each of the dominant
uncertainties follows.

\textbf{Beam flux uncertainties:} The LBNE high-resolution near detector is
  being designed with the goal of accurately measuring the
  unoscillated beam flux at the near site with a precision $\leq
  2\%$ for both shape and absolute normalization.
  Table~\ref{tab:fluxes} summarizes the precision that can be achieved
  using different near detector analysis techniques, described in
  detail in Section~\ref{sec-fluxosc}, to measure the absolute
  normalization and shape of the different components of this flux. 
\begin{table}[!htb]
    \caption[Near detector precisions for in situ $\nu_\mu$ and $\nu_e$ flux measurements]
{ Precisions achievable from in situ $\nu_\mu$ and $\nu_e$ flux measurements
      in the fine-grained, high-resolution ND with different techniques. }
  \label{tab:fluxes}
  \begin{tabular}{$L^c^c^c^l}
\toprule
\rowtitlestyle
    Technique            & Flavor                     & Absolute      & Relative           & Near Detector 	 \\
\rowtitlestyle
                         &                               & normalization & flux $\Phi(E_\nu)$ &   requirements                    	 \\ 
\toprowrule
     NC Scattering                & $\nu_\mu$            & 2.5\%         & $\sim5\%$  	  & $e$ ID 		  	 \\
    $\nu_\mu e^- \to \nu_\mu e^-$  &                      &  	&	  & $\theta_e$ Resolution 	 \\ 
                          &                              &  	&	  & $e^-/e^+$ Separation   	 \\  \colhline
    Inverse muon  & $\nu_\mu$            & 3\%  	     &  		  & $\mu$ ID 		  	 \\ 
    decay   &  		 &  	     &  		  & $\theta_\mu$ Resolution 	 \\ 
    $\nu_\mu e^- \to \mu^- \nu_e$    &  	 &  	     &  		  & 2-Track ($\mu+$X) Resolution \\       
                         &  				&  &  		  & $\mu$ energy scale 	         \\ \colhline
   CC QE                 & $\nu_\mu$        & $3-5\%$       & $5-10\%$ 	  & $D$ target  		 \\
   $\nu_\mu n \to \mu^- p$       &                       &               &            & $p$ Angular resolution \\ 
   $Q^2 \to 0$          &   			        &  	     &  		  & $p$ energy resolution \\ 
                        &                         &	 &          	  & Back-Subtraction  		     \\  \colhline
   CC QE    & $\overline\nu_\mu$      & $5\%$  	     & $10\%$ 		  & $H$ target   \\
  $\overline\nu_\mu p \to \mu^+ n$ &                        &   	     &  		  & Back-Subtraction  \\  
   $Q^2 \to 0$                          &                               &               &                    & \\
                             &                               &               &                    & \\ \colhline
  Low-$\nu_0$   	 & $\nu_\mu$                     & 	       & 2.0\% 		  & $\mu^-$ vs $\mu^+$ 		     \\ 
                         &                         	 &   	     &          	  & $E_\mu$-Scale 		     \\ 
                         &                         	 &   	     &          	  & Low-$E_{Had}$ Resolution         \\ \colhline
    Low-$\nu_0$   	 & \anm\                         &    	     & 2.0\% 		  & $\mu^-$ vs $\mu^+$  	     \\ 
                         &                         	 &   	     &          	  & $E_\mu$-Scale 		     \\ 
                         &                         	 &   	     &          	  & Low-$E_{Had}$ Resolution         \\ \colhline
   Low-$\nu_0$           & $\nu_e$/$\overline{\nu}_e$    & 1-3\%  	     & 2.0\%  		  & $e^-/e^+$ Separation ($K^0_L$)   \\ \hline
   CC                    & $\nu_e$/$\nu_{\mu}$           & <1\%         & $\sim$2\%        & $e^-$ ID \& $\mu^-$ ID \\
                         &                               &              &                  & $p_e$/$p_{\mu}$ Resolution \\ \hline
   CC                    & $\overline\nu_e$/$\overline\nu_{\mu}$      & <1\%         & $\sim$2\%        & $e^+$ ID \& $\mu^+$ ID \\
                         &                               &            &                  & $p_e$/$p_{\mu}$ Resolution \\ \hline
   Low-$\nu_0$/CohPi     & $\overline\nu_{\mu}$/$\nu_{\mu}$ & $\sim$2\% & $\sim$2\% & $\mu^+$ ID \& $\mu^-$ ID \\
      &                  &                               &                 &  $p_{\mu}$ Resolution\\
      &                  &                               &                 &   $E_{Had}$ Resolution \\
\bottomrule
  \end{tabular}
\end{table}
It is important to note that
several of these techniques have already been used and {\em proven to
  work} in neutrino experiments such as MINOS~\cite{Adamson:2009ju}
and NOMAD~\cite{Wu:2007ab,Lyubushkin:2008pe}. In particular, the inclusive neutrino charged current (CC) 
cross-section measurement in the MINOS near detector reported in~\cite{Adamson:2009ju} 
has already achieved a normalization
uncertainty of $\sim2\%$ in the range of $3<E_\nu<9$ GeV using the
low-$\nu_0$ method described in Section~\ref{sec-fluxosc}.  The total
systematic uncertainty on the NuMI neutrino flux determination by the
MINOS near detector reported in~\cite{Adamson:2009ju} was $\sim6\%$ 
and was limited by the detector performance. Recent
independent studies on extraction of the neutrino flux using the
low-$\nu_0$ method~\cite{Bodek:2012cm} indicate that the technique
can be reliably extended down to \SI{1}{\GeV}.  

The LBNE near detector is being designed to significantly 
improve performance relative to the 
current generation of high-intensity 
neutrino detectors. A detailed beamline simulation will enable 
the extrapolation of the
LBNE near detector flux measurements to the unoscillated far detector
spectrum with high precision using techniques similar to those used by
MINOS~\cite{Adamson:2007gu}. The near-to-far $\nu_\mu$
unoscillated-spectrum extrapolation uncertainties already achieved by
MINOS are $< 3\%$ in the MINOS (and also in the LBNE) appearance
signal range of $1<E_\nu<8$\SI{}{\GeV}~\cite{Bishai:2012kta,Adamson:2007gu}.  
The MINOS extrapolation does
not include any independent constraints on the hadron production
spectrum from the proton target or information on the horn focusing
performance from the muon flux measurements at the near site. 
The NuMI beamline
--- the design of which is very similar to LBNE's --- is expected to operate
for more than a decade with improved flux measurements using the much
more 
capable MINER$\nu$A detector~\cite{Osmanov:2011ig} in
both the low-energy and high-energy tunes. MINER$\nu$A is designed to
measure the absolute NuMI flux with a precision of $\sim 5\%$ or better;
data from
MINER$\nu$A will be used to further improve the accuracy of the
LBNE beamline simulation, reducing the uncertainties on
the extrapolation of the flux. 
A new
program of hadron production measurements at the 
NA61/SHINE~\cite{Korzenev:2013gia} experiment will also reduce the near-to-far
extrapolation uncertainties from the LBNE beamline simulation.    
The combination of LBNE
near detector flux measurements and improved beamline simulation is
expected to enable a prediction of the far detector 
$\nu_e$ appearance signal with a precision of $<2$\% 
total normalization and shape uncertainty. Since this 
uncertainty is highly correlated among the four data samples in the three-flavor fit, 
 the final uncorrelated uncertainty on the $\nu_e$ signal sample will be significantly smaller.

 \textbf{$\boldmath{\nu_\mu}$ energy-scale uncertainty:} Both T2K and MINOS use
 the reconstructed $\nu_\mu$ event spectrum in the near detector to
 predict the $\nu_e$ appearance signal at the far detector.  Therefore
 the $\nu_\mu$ energy-scale uncertainty in the near detector is
 propagated as an uncertainty on the $\nu_e$ appearance signal at the
 far detector. In MINOS --- which has a high proportion of non-QE
 events --- the $\nu_\mu$ energy-scale uncertainty
 is dominated by uncertainty in the hadronic
 energy scale (7\% for $E_\nu< 3$ GeV)~\cite{Adamson:2011ig} and the
 muon energy scale (2.5\%). 
 Utilization of the low-$\nu_0$ method
 for energies less than 3~GeV in LBNE reduces the hadronic energy-scale
 contribution to the uncertainty in the
 $\nu_\mu$ energy scale in the near detector. 
 As discussed in Chapter~\ref{nd-physics-chap},
 it is expected that both the muon and hadronic energy-scale
 uncertainties in the near detector will be $<$1\%,
 so far detector energy-scale uncertainties will dominate
 the uncertainty in the $\nu_\mu$ signal prediction.
 The high-resolution
 LArTPC far detector and an active program of hadron test-beam 
 experiments planned
 for LBNE will reduce far detector hadronic energy-scale
 uncertainties, which also contribute to uncertainty in the energy scale of
 the far detector $\nu_\mu$ 
 signal used in the three-flavor analysis.
 Extrapolating from MINOS,
 the LBNE $\nu_\mu$ energy-scale uncertainty is thus estimated to be
 $\sim 2\%$. 

 In MINOS, the 7\% $\nu_\mu$ energy-scale uncertainty
 resulted in a residual uncertainty of 3.5\% on the $\nu_e$ signal
 prediction. In the LBNE full three-flavor analysis, this uncertainty
 is 100\% correlated between the predicted $\nu_\mu$ and $\nu_e$
 signal samples; therefore a $E_{\nu_\mu}$ energy-scale uncertainty of
 2\% is assigned to the $\nu_\mu$ signal prediction in LBNE. The
 residual uncorrelated uncertainty on the $\nu_e$ signal prediction
 is considered to be negligible.

\textbf{Absolute} $\boldmath{\nu_e}$ \textbf{energy-scale uncertainties:} In 
  Figure~\ref{fig:systs2},
  the MH and CP-violation sensitivity
  obtained using a rate-only, a shape-only and a rate+shape analysis
  of $\nu_e$ appearance is shown.  This study demonstrates that a
  critical component of LBNE's oscillation sensitivity is an
  accurate measurement of the shape of the $\nu_e$ appearance
  signal. 
\begin{figure}[!htb]
\centerline{
\includegraphics[width=0.5\linewidth,trim=0cm 0.3cm 0cm 0.0cm,clip]{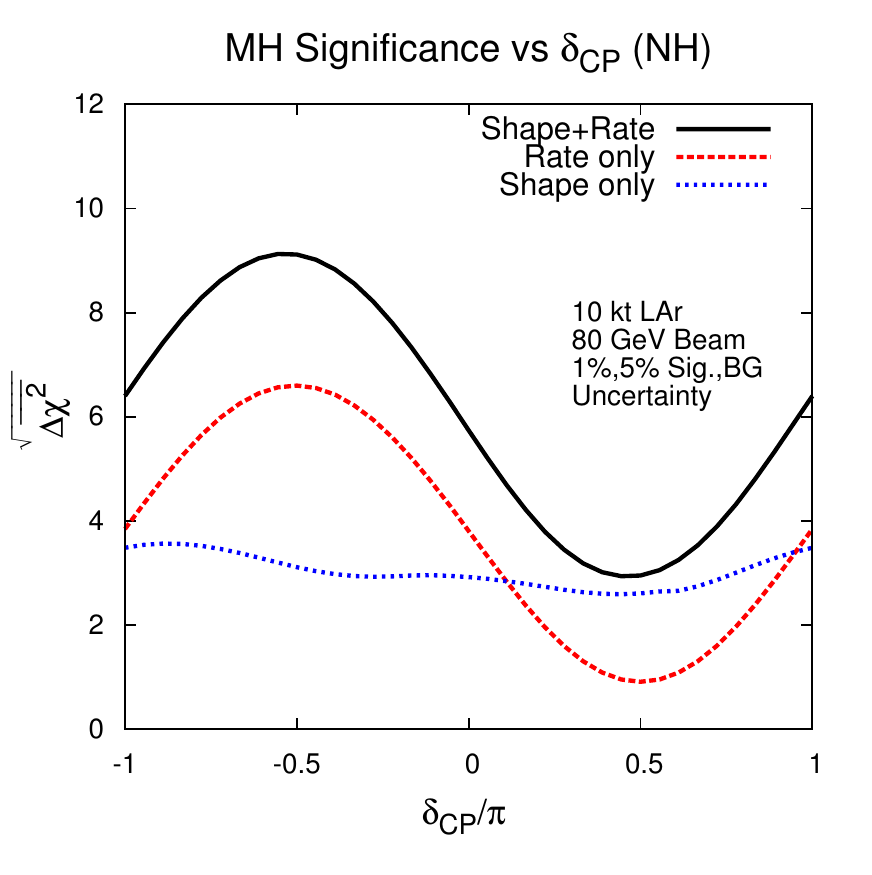}
\includegraphics[width=0.5\linewidth,trim=0cm 0.3cm 0cm 0.0cm,clip]{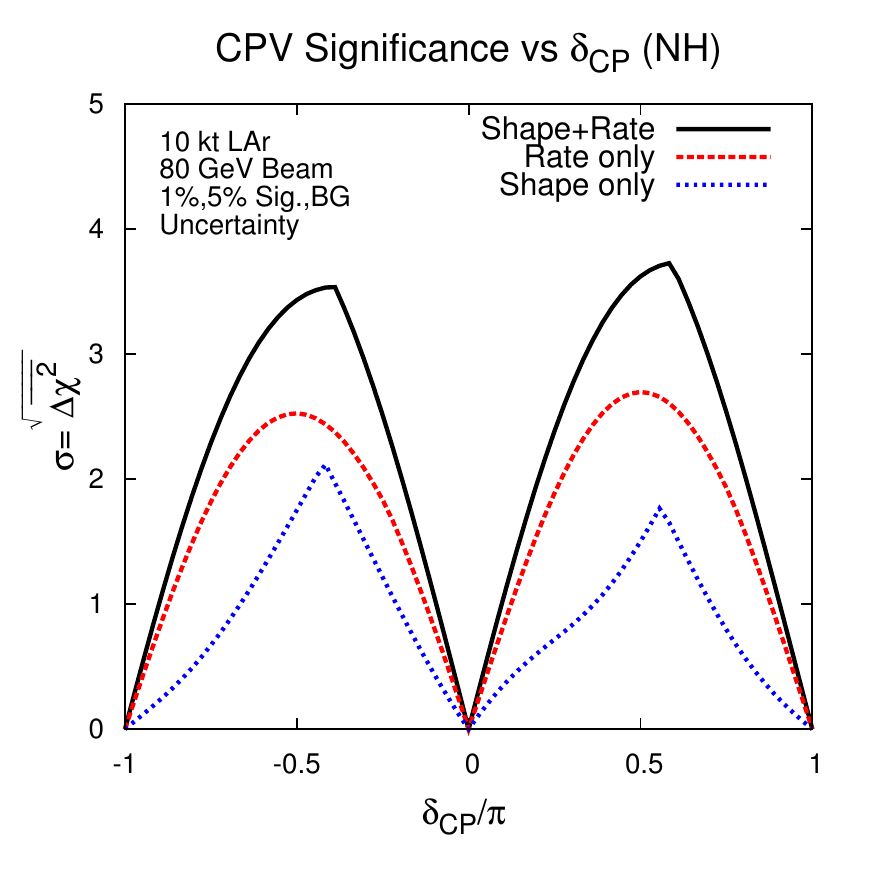}
}
\caption[MH and CP violation
  sensitivities from shape, rate, and shape+rate]{The mass hierarchy (left) and CP violation (right)
  sensitivities from shape, rate, and shape+rate. The sensitivity is
  for a \SIadj{10}{\kt} detector, \SIadj{1.2}{\MW} beam, 3+3 ($\nu$ + $\overline{\nu}$) years, for true normal hierarchy.}
\label{fig:systs2}
\end{figure}
  This measurement depends on the precision with which the
  detector response to $\nu_e$ interactions is understood. The $\nu_e$
  energy-scale uncertainty, which is not yet included in 
  the current
  sensitivity calculation with the GLoBES framework,
  is therefore expected to be an important
  systematic uncertainty in the LBNE oscillation analysis. 

  The effect of $\nu_e$ energy-scale uncertainty on the $\nu_e$ signal
  normalization, determined by the precision of
  detector calibration, was 2.7\% in MINOS and 3.4\% in T2K, where the T2K
  uncertainty actually includes most far detector effects.
  LBNE's LArTPC detector
  technology is expected to outperform both the MINOS sampling calorimeter
  and the T2K water Cherenkov
  detector in reconstruction of the $\nu_e$ interaction. For example,
  the proton produced from the $\nu_e$-QE interaction --- the interaction 
  with potentially 
  the best $\nu_e$ energy resolution --- is clearly visible in a
  LArTPC~\cite{TingjunYang:2013vva}, whereas it is often below
  Cherenkov threshold in T2K. An active program of test beam
  experiments with LArTPCs is currently being planned to address the
  detector response to electrons and hadrons. Results from the test
  beam experiments and the projected performance of the in situ
  calibration will enable LBNE to limit the detector energy-scale 
  uncertainties below the level achieved by the current
  generation of experiments.

  Hadronic energy is expected to contribute more than half
  of the total energy 
  deposit for many $\nu_e$ and $\nu_{\mu}$ interactions in LBNE.
  The hadronic energy scale does not depend on neutrino flavor; since it
  should be identical for $\nu_e$ and $\nu_{\mu}$ interactions, this portion
  of the absolute energy-scale uncertainty is expected to largely cancel
  in the LBNE three-flavor analysis. This cancellation may be reduced to
  the extent that event-selection criteria vary the hadronic
  energy fraction among the samples.

\textbf{Simulation uncertainties:} The simulation uncertainties listed in
  Table~\ref{tab:nuesysts} refer primarily to uncertainties in
  modeling neutrino interactions with the target nucleus in the
  near and far detectors. These uncertainties include $\nu_e$ and
  $\nu_\mu$ cross-section uncertainties, uncertainties arising from
  the modeling of the structure of the target nucleus, modeling of final-state
  interactions within the nucleus, and hadronization model
  uncertainties arising from the break up of the target nucleus in
  higher-energy inelastic interactions. 
The deployment of identical
  nuclear targets in the MINOS (iron) and LBNE (argon) near and far
  detectors allows for a larger cancellation of the simulation
  uncertainties as compared to T2K, which used dissimilar target nuclei
  in its near detector (carbon) and far detector (oxygen). A high-resolution near detector such as
  that being designed for LBNE will enable further constraints on the
  hadronization models by resolving many of the individual particles
  produced in resonance and deep inelastic interactions, which
  represent $\sim$75\% of LBNE neutrino interactions.  

  The MINOS $\nu_e$ appearance
  analysis achieved a 2.7\% residual uncertainty from simulation after
  the near-to-far extrapolation.  The MINOS simulation uncertainty is
  dominated by hadronization uncertainties, because cross-section 
  uncertainties largely cancel between the identical nuclei in the near
  and far detectors. The T2K residual
  uncertainty after near-to-far extrapolation is 7\%. 
  Additionally, the T2K analysis includes more sources of cross-section
  uncertainties than MINOS and, at the lower T2K energies,
  larger differences in $\nu_\mu$/$\nu_e$ cross sections (2.9 \%)
  persist after extrapolating the $\nu_\mu$ spectrum in the near
  detector to the $\nu_e$ signal prediction in the far. 

  The LBNE near
  detector design is required to achieve a cancellation of near-to-far
  cross-section and hadronization-model uncertainties at the same
  level as MINOS or better. The $\nu_e$ appearance 
  signal in LBNE peaks at \SI{2.5}{\GeV}; these higher energies will result
  in lower uncertainties from the cross-section effects considered by
  T2K. In addition, since cross-section variations impact the
  observed $\nu_e$ and $\nu_\mu$ spectra differently when compared to
  oscillation effects, the fit to the wide-band spectrum in LBNE
  could constrain some of these uncertainties further.
  Therefore, it is expected that LBNE could reduce the total
  $\nu_e$ appearance simulation uncertainties to a level of 2\%.
  Preliminary results from the LBNE Fast MC simulation (described in
  Section~\ref{appxsec:fastmc}) indicate that many cross-section
  uncertainties cancel out when combining the $\nu_\mu$ disappearance
  and $\nu_e$ appearance signal samples in a three-flavor fit, 
  resulting in a 
  much smaller uncorrelated uncertainty on the $\nu_e$ signal sample.

  It is important to note that some $\nu/\overline{\nu}$ simulation
  uncertainties may not cancel out in the near-to-far extrapolation or
  in the combined fit; in particular, uncertainties due to nuclear
  models and intra-nuclear effects are different for
  $\nu/\overline{\nu}$ interactions. New models of intra-nuclear
  effects are being evaluated to determine the size of these
  irreducible residual uncertainties. Additionally, there are 
  uncertainties at the level of 1-2\%
  in the cross sections that will not cancel between $\nu_e$ and $\nu_{\mu}$~\cite{Day:2012gb}.
  In the absence of theoretical progress, these should also be considered
  irreducible.

\textbf{Fiducial volume uncertainties:} One of the dominant uncertainties
  in the MINOS $\nu_\mu$ disappearance analysis ---  a high-precision 
  oscillation analysis based on a detailed spectral shape
  --- was the fiducial-volume uncertainty, which included near and far
  detector reconstruction uncertainties. The uncertainty on the fiducial
  volume of the MINOS far detector alone was 2.4\%.
  T2K, with a much larger far
  detector (\SI{22.5}{\kt} fiducial), was able to reduce this
  uncertainty to the 1\% level. It is expected that LBNE will be able
  to achieve this level of uncertainty on the $\nu_e$ appearance
  signal. With the combination of all four signal samples ($\nu_\mu,
  \overline{\nu}_\mu, \nu_e, \overline{\nu}_e$) in a three-flavor fit, the
  $\nu_e$ uncorrelated portion of this uncertainty is expected to be
  smaller than 1\%.

$\boldmath{\nu_e}$ \textbf{appearance background systematic uncertainties:} %
The $\nu_e$ appearance normalization uncertainty is expected to
be at least as good as the $\sim5\%$~\cite{Adamson:2013ue} achieved
by the $\nu_e$ appearance search in the MINOS experiment, using the technique
of predicting intrinsic-beam and neutral current (NC) background levels from near detector
measurements. The LBNE far detector should be able to provide
additional constraints on the background level by independently measuring
NC and $\nu_\tau$ background.

In Figure~\ref{fig:systs}, the MH and CP-violation
sensitivities as a function of exposure are evaluated using three
different sets of assumptions regarding the uncorrelated $\nu_e$
signal/background 
normalization uncertainties: 1\%/5\% (the goal of the LBNE
scientific program), 2\%/5\% and 5\%/10\%. The last is a conservative
estimate of the uncertainties that can be achieved in LBNE without
unoscillated neutrino beam measurements at the near site. The impact
of signal and background normalization uncertainties on the
MH sensitivity is small even at high exposures given
the large $\nu/\overline{\nu}$ asymmetry at \SI{1300}{\km} and the fact
that much of the sensitivity to the MH comes from analysis of
the spectral shapes (Figure~\ref{fig:systs2}). 
For CP
violation, however, the impact of normalization uncertainties
is significant at exposures $\geq$
\SI[inter-unit-product=\ensuremath{{}\cdot{}}]{100}{\kt\MW\year}s.
\begin{figure}[!htb]
\centerline{
\includegraphics[width=0.5\linewidth]{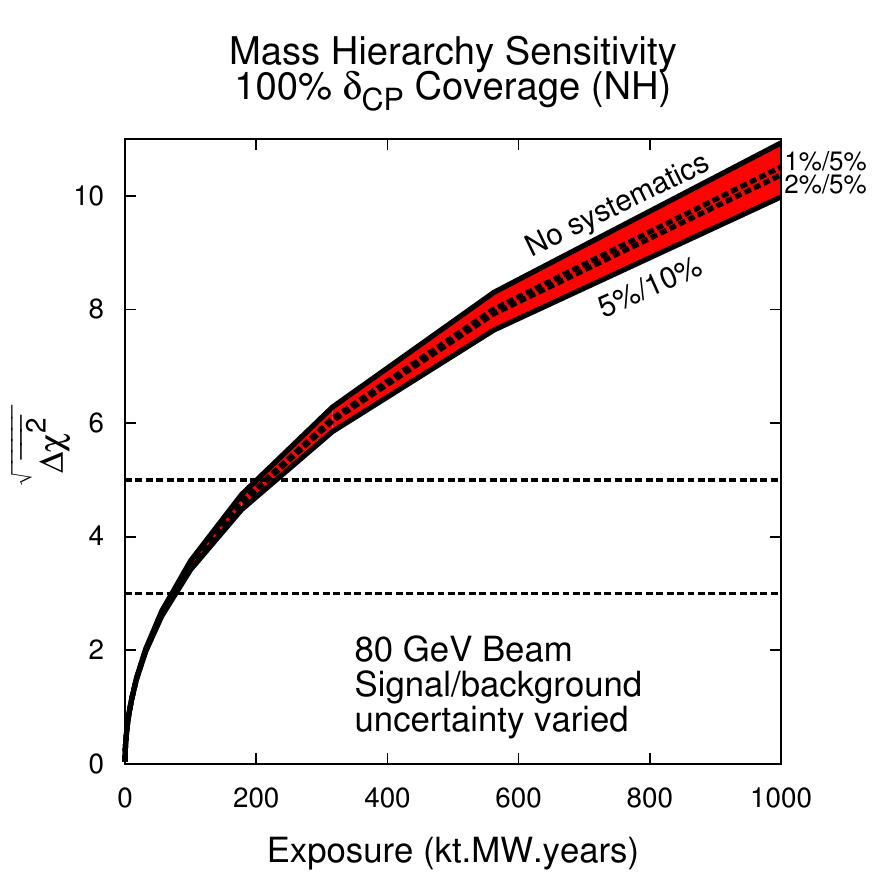}
\includegraphics[width=0.5\linewidth]{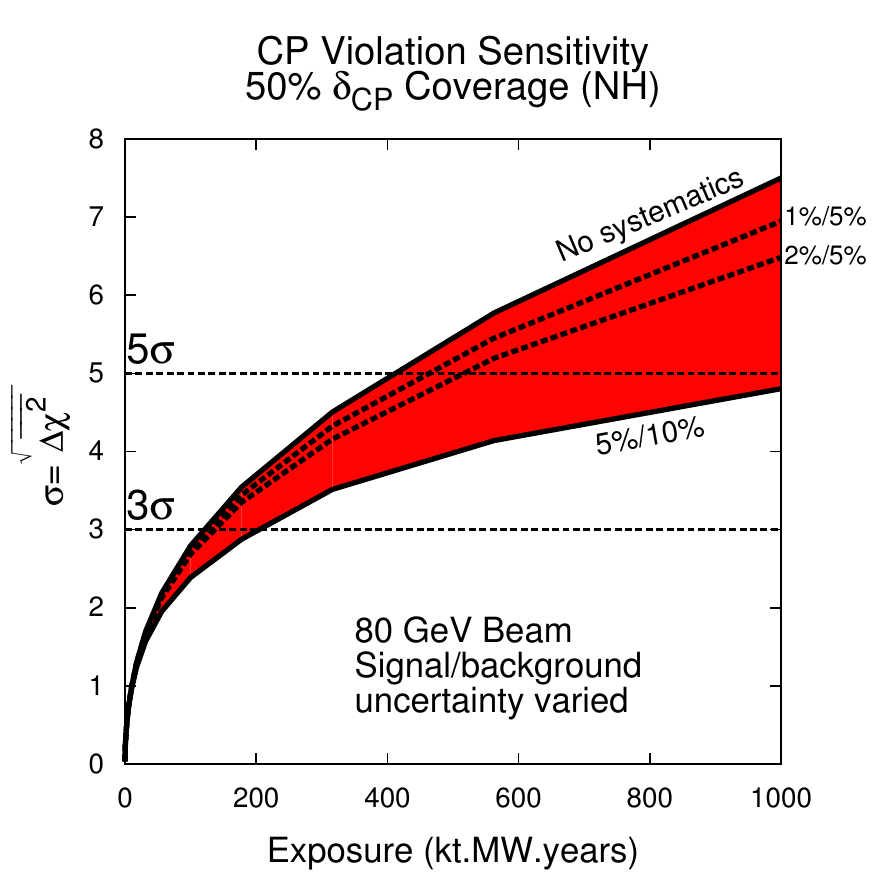}
}
\caption[MH and CP violation 
  sensitivities as a function of exposure in \ktyr]{The mass hierarchy (left) and CP violation (right)
  sensitivities as a function of exposure in \ktyr, for true normal hierarchy. 
 The band
  represents the range of signal and background normalization errors.}
\label{fig:systs}
\end{figure}
\begin{table}[!htb]
  \caption[Exposures to reach
  $3\sigma$ and $5\sigma$ sensitivity to CPV for $\geq$50\% of \deltacp values]
  { The exposures required to reach
    $3\sigma$ and $5\sigma$ sensitivity to CP violation for at least 50\% of all
    possible values of \deltacp as a function of systematic uncertainties assumed on the $\nu_e$ appearance signal. 
    The uncertainties varied are the uncorrelated signal normalization uncertainty (Sig) and the background normalization uncertainty (Bkgd).}
\label{tab:senssysts}
\begin{tabular}{$L^c^c^c}
\toprule
\rowtitlestyle
Systematic uncertainty & \multicolumn{2}{^>{\columncolor{\ChapterBubbleColor}}c}{CPV Sensitivity} & Required Exposure \\ 
\rowtitlestyle
 & \deltacp Fraction & ($\sqrt{\overline{\Delta\chi^2}}$) &  \\ 
\toprowrule
0 (statistical only) & 50\% \deltacp & 3 $\sigma$ & \ktmwyr{100} \\ 
                     & 50\% \deltacp & 5 $\sigma$ & \ktmwyr{400}\\ \colhline
1\%/5\% (Sig/bkgd)   & 50\% \deltacp & 3 $\sigma$ & \ktmwyr{100} \\ 
                     & 50\% \deltacp & 5 $\sigma$ & \ktmwyr{450} \\ \colhline 
2\%/5\% (Sig/bkgd)   & 50\% \deltacp & 3 $\sigma$ & \ktmwyr{120} \\ 
                     & 50\% \deltacp & 5 $\sigma$ & \ktmwyr{500} \\ \colhline
5\%/10\% (no near $\nu$ det.)  & 50\% \deltacp & 3 $\sigma$ & \ktmwyr{200} \\
\bottomrule
\end{tabular}
\end{table}

Table~\ref{tab:senssysts} summarizes the LBNE exposures required to
reach $3\sigma$ and $5\sigma$ sensitivity to CP violation for at least 50\% of all
possible values of \deltacp.  The exposures vary depending on the assumptions made about the
normalization uncertainties that can be achieved in LBNE. The normalization uncertainty
assumptions range from 1-2\%/5\% on signal/background to 5\%/10\%.
The uncertainties listed in Table~\ref{tab:senssysts} and shown in the
sensitivity figures pertain to the $\nu_e$ appearance signal and
background normalization. In Figure~\ref{fig:systs2} the sensitivities
obtained from the rate only, shape only and rate+shape of the
appearance spectrum are shown for a \SIadj{10}{\kt} detector with an \GeVadj{80} beam. 
For CP violation (right), the rate information
dominates the sensitivity,  
but the shape information enables the detector to exceed $3 \sigma$
sensitivity for large CP violation.  For the MH
sensitivity, Figure~\ref{fig:systs2} (left) demonstrates that the sensitivity
in the least favorable range of \deltacp values is dominated by the shape
information.  Further analysis has shown that it is the region of the
second oscillation node that is responsible for this effect. The shape
of the signal in this region will enable LBNE to determine the sign of
\deltacp, which is sufficient to break the degeneracy with
MH effects and determine the correct sign of the mass
ordering.

Figures~\ref{fig:th13_variation},~\ref{fig:th23_variation},
and~\ref{fig:dm2_variation} show the variation in sensitivity to CP
violation and MH when the true value of the oscillation
parameters $\theta_{13}$, $\theta_{23}$ and $\Delta
m^2_{31}$ 
are varied within the 3$\sigma$ range allowed by the 2012 $3 \nu$
global fit~\cite{Fogli:2012ua}.  These sensitivities are calculated
for six years with equal exposures in $\nu$ and $\overline{\nu}$
mode in a \MWadj{1.2} beam for the case in which an upgraded
\GeVadj{80} beam and a near detector have both been implemented.  
\begin{figure}[!htb]
\centerline{
\includegraphics[width=0.5\textwidth]{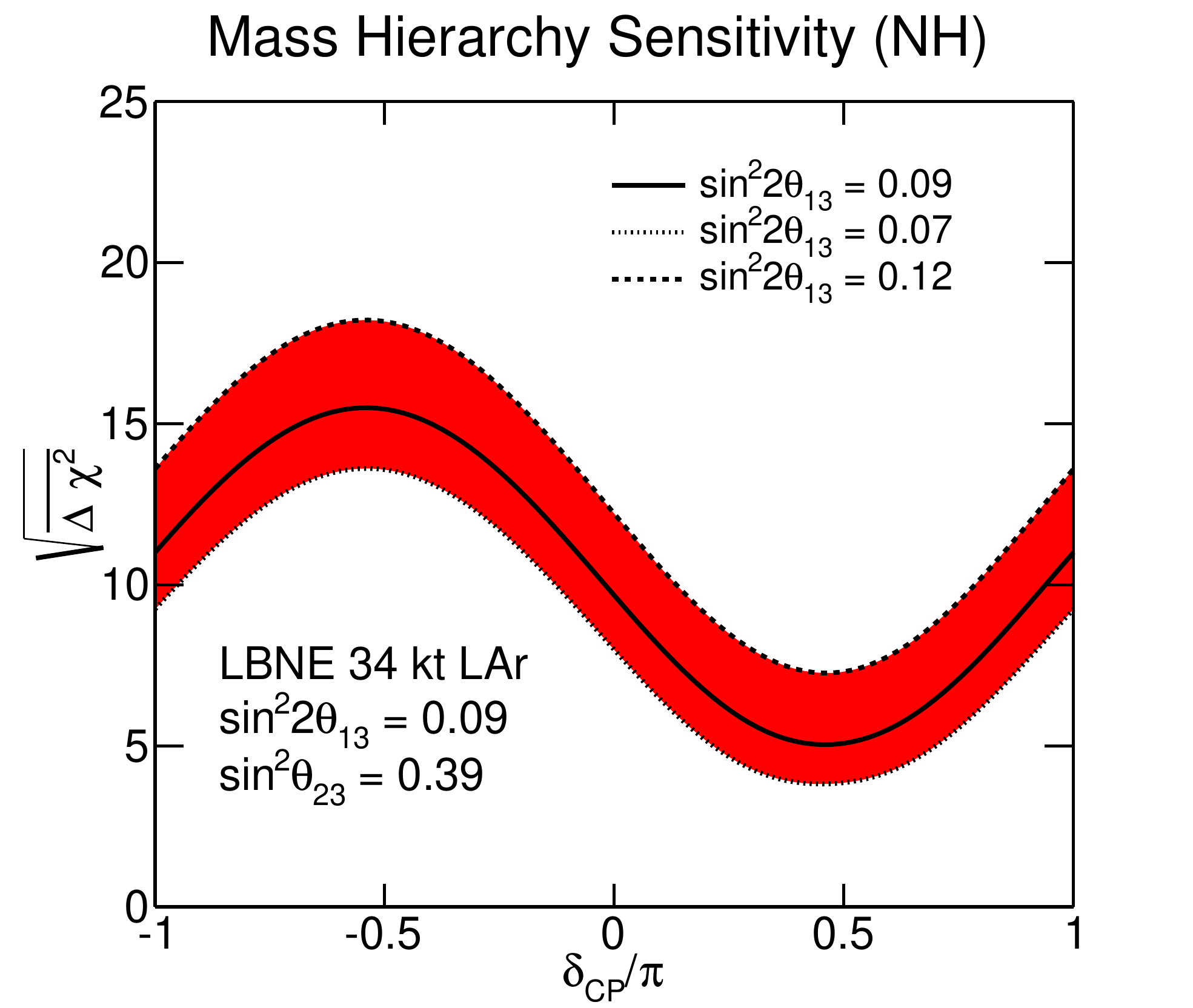}
\includegraphics[width=0.5\textwidth]{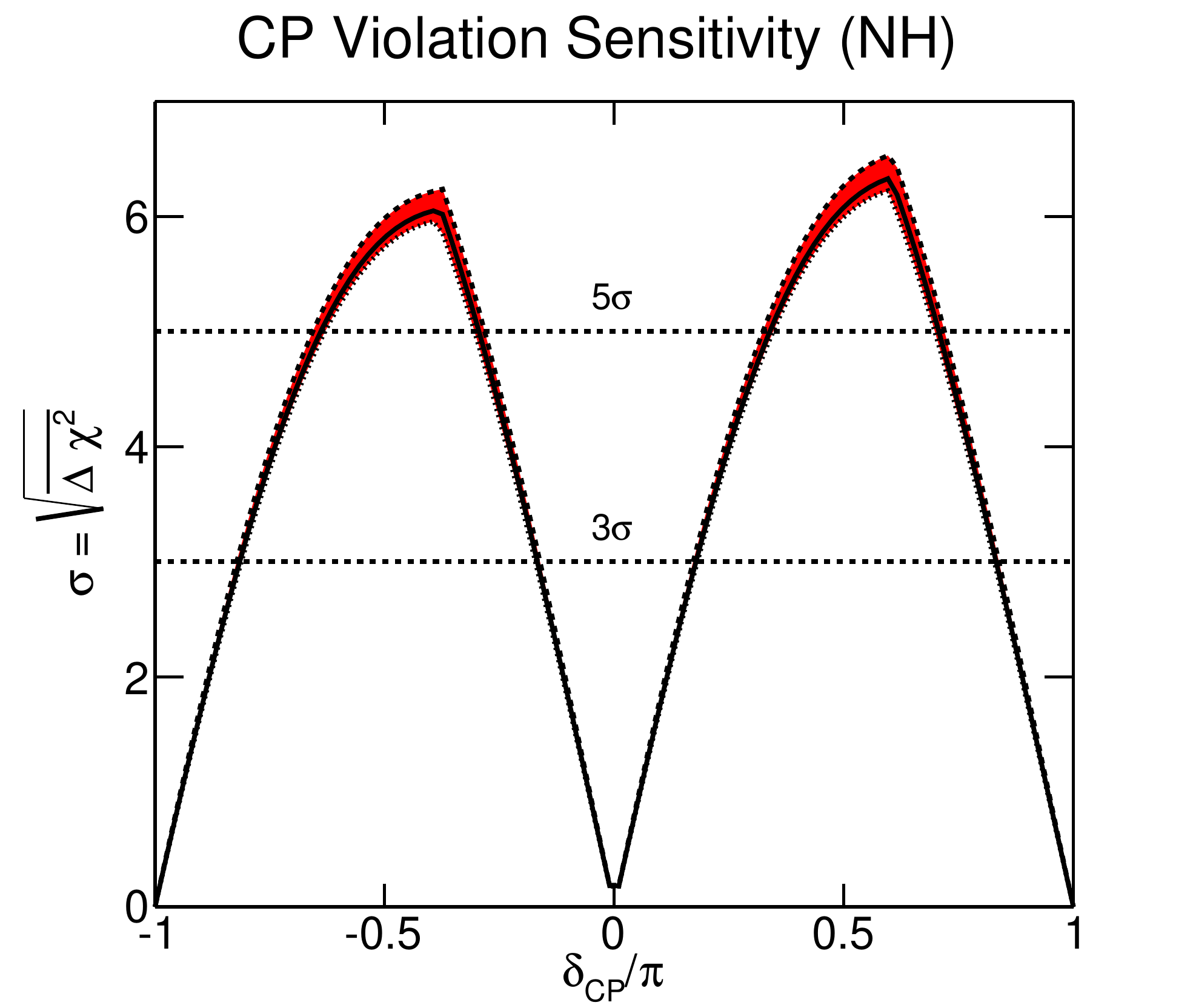}
}
\caption[Variation of sensitivity to MH and CP violation with $\theta_{13}$]{The 
  significance with which the mass hierarchy (left) and
  CP violation, i.e., $\mdeltacp \neq 0 \ {\rm or} \ \pi$, (right) can
  be determined by a typical LBNE experiment as a function of the value of \deltacp for an
  allowed range of $\theta_{13}$ values and for normal hierarchy; assumes a \SIadj{34}{\kt} far detector.}  
\label{fig:th13_variation}
\end{figure}

\begin{figure}[!htb]
\centerline{
\includegraphics[width=0.5\textwidth]{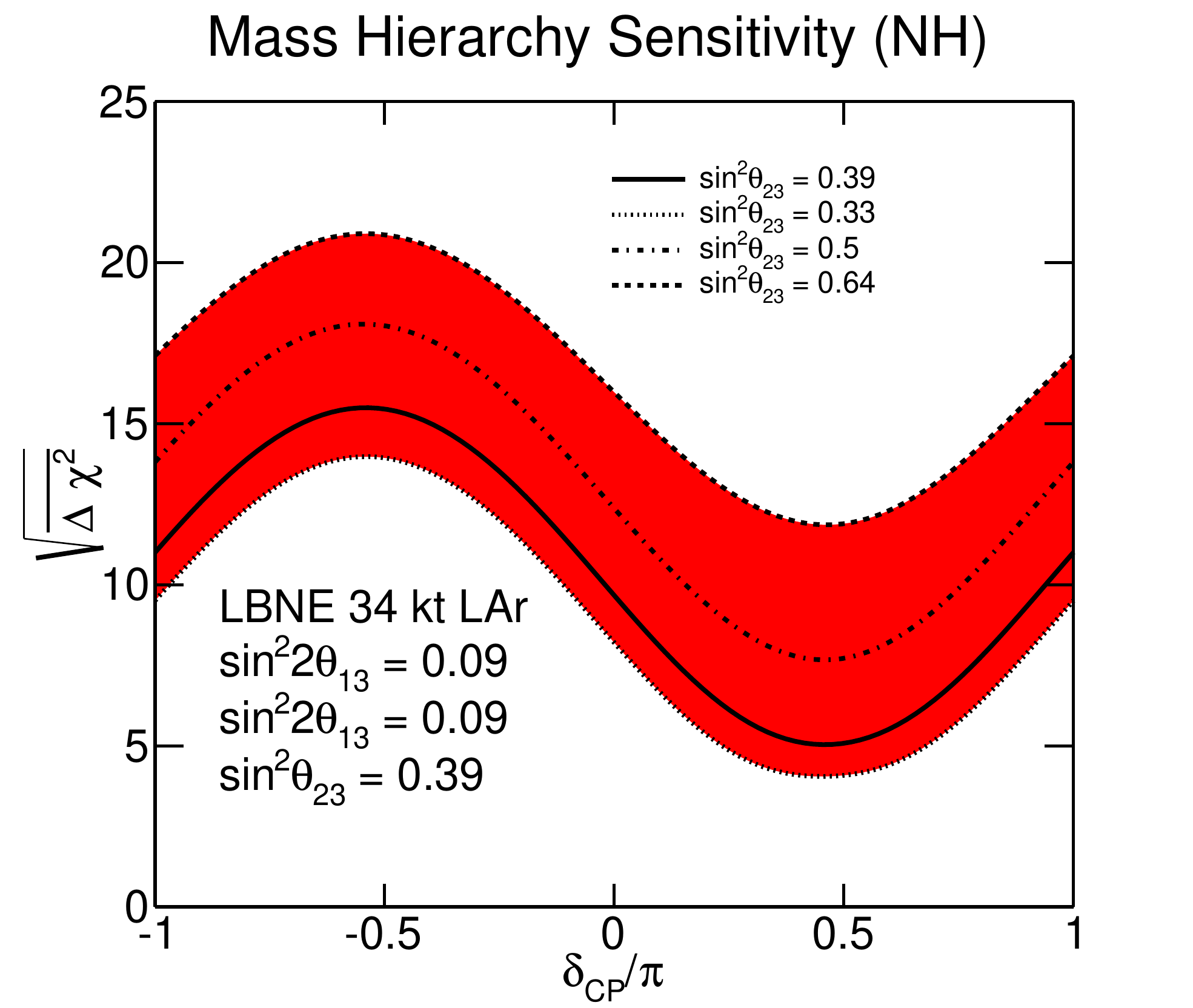}
\includegraphics[width=0.5\textwidth]{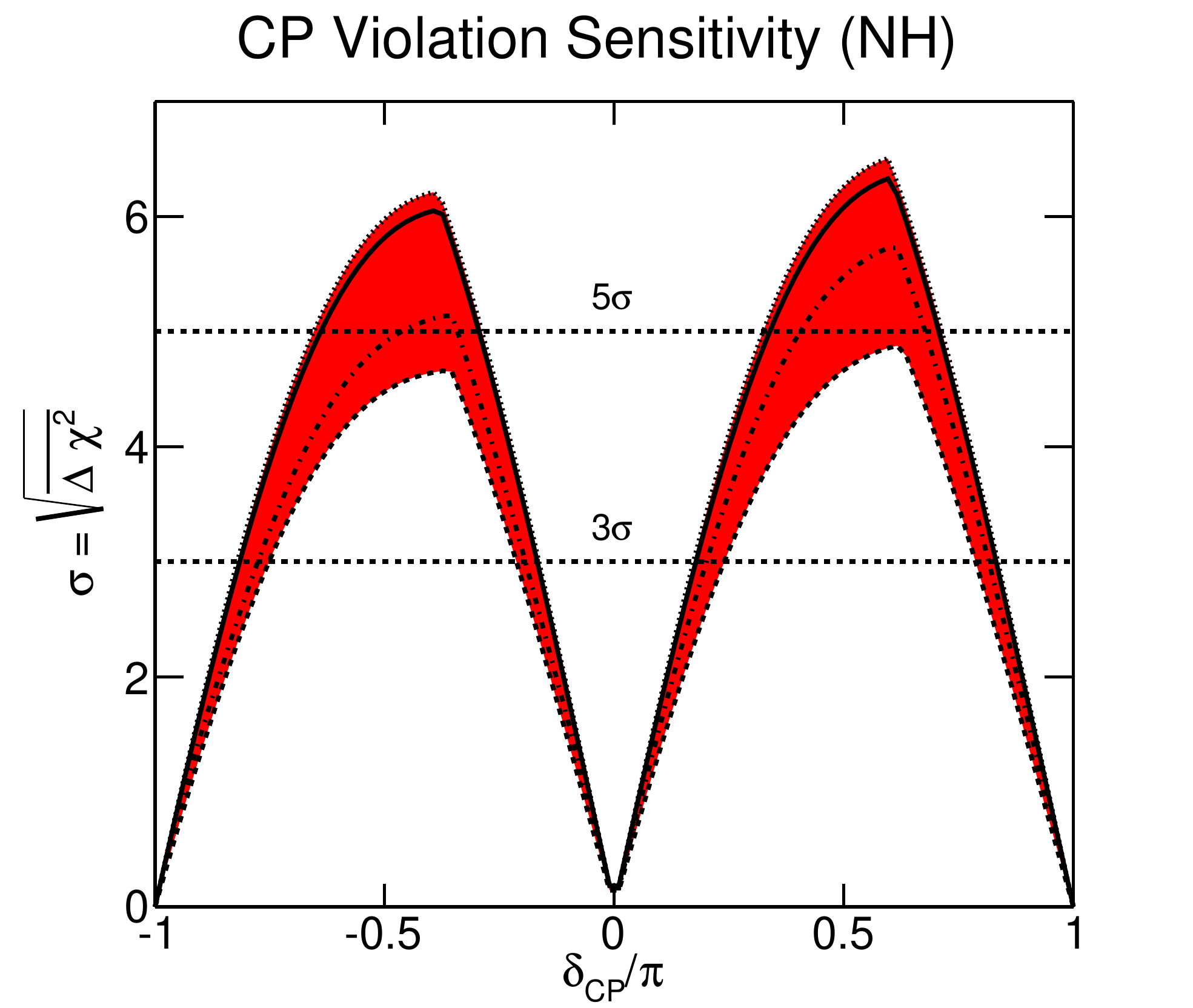}
}
\caption[Variation of sensitivity to MH and CP violation with $\theta_{23}$]{The 
  significance with which the mass hierarchy (left) and
  CP violation, i.e., $\mdeltacp \neq 0 \ {\rm or} \ \pi$, (right) can
  be determined by a typical LBNE experiment as a function of the value of \deltacp for an
  allowed range of $\theta_{23}$ values and for normal hierarchy; assumes a \SIadj{34}{\kt} far detector.}  
\label{fig:th23_variation}
\end{figure}
\begin{figure}[!htb]
\centerline{
\includegraphics[width=0.5\textwidth]{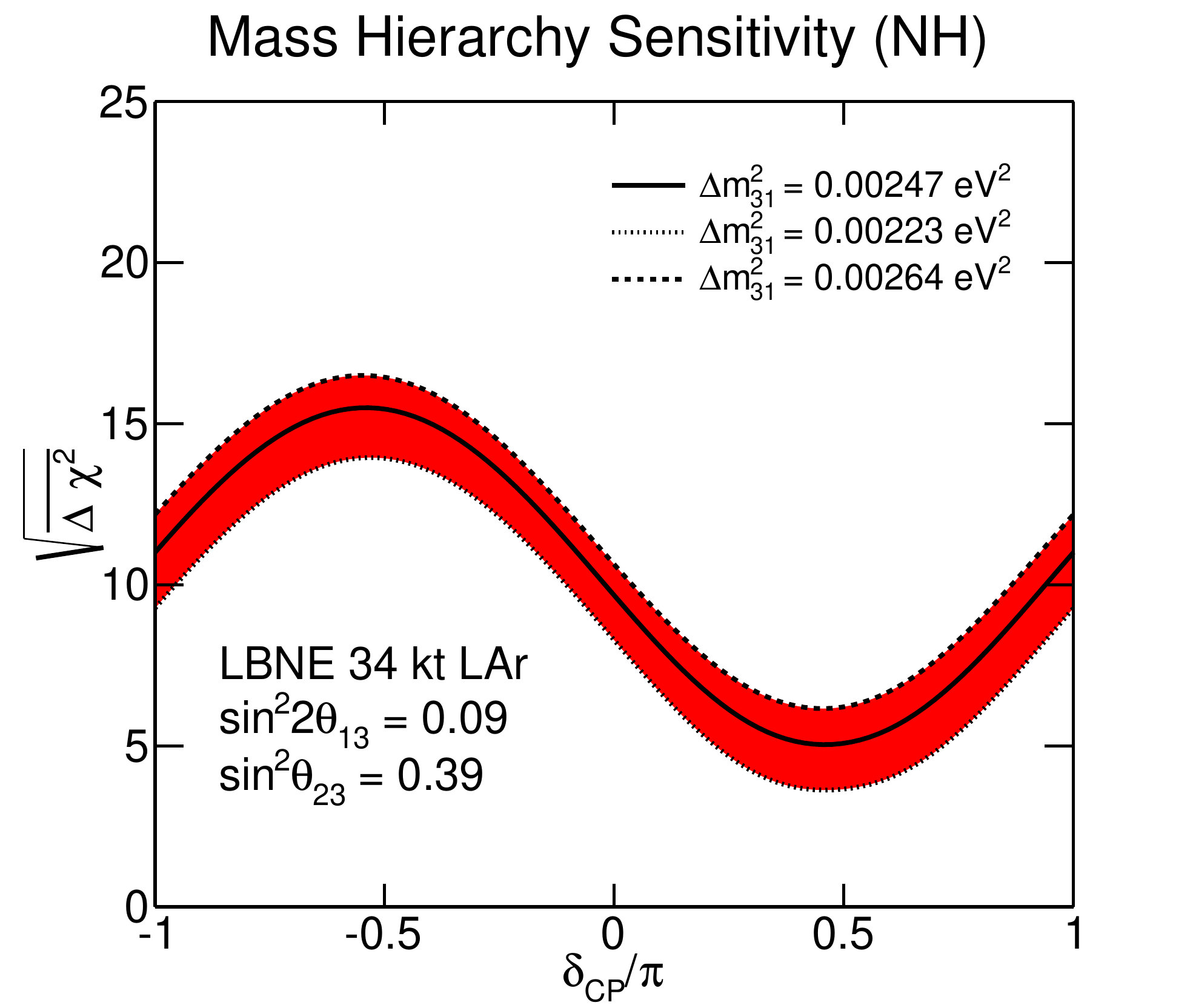}
\includegraphics[width=0.5\textwidth]{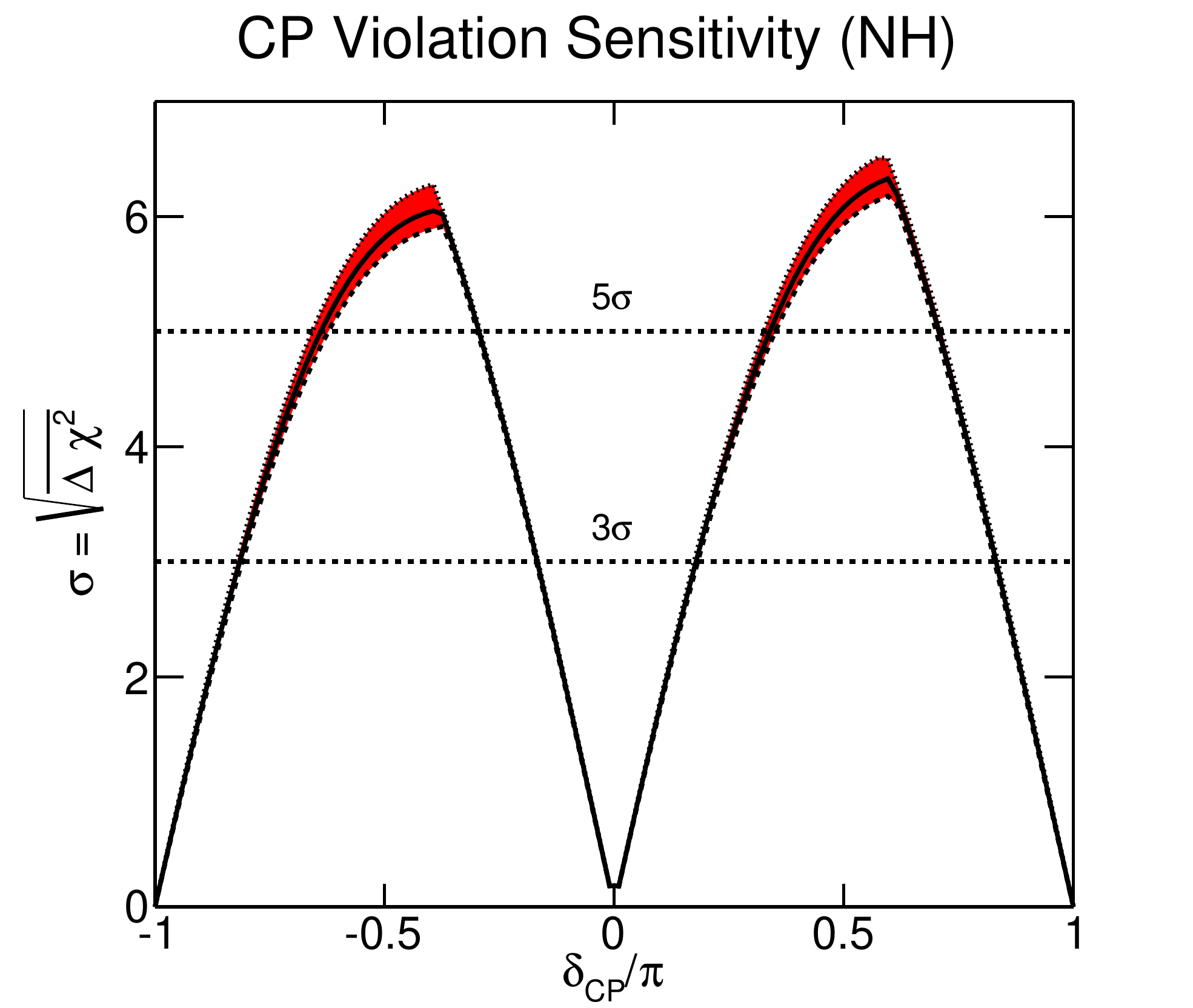}
}
\caption[Variation of sensitivity to MH and CP violation with $\Delta m^2_{31}$]{The 
  significance with which the mass hierarchy (left) and
  CP violation, i.e., $\mdeltacp \neq 0 \ {\rm or} \ \pi$, (right) can
  be determined by a typical LBNE experiment as a function of the value of \deltacp for an
  allowed range of $\Delta m^2_{31}$ values and for normal hierarchy; assumes a \SIadj{34}{\kt} far detector.}  
\label{fig:dm2_variation}
\end{figure}

In
comparing Figures~\ref{fig:th13_variation},~\ref{fig:th23_variation}
and~\ref{fig:dm2_variation}, the dependence on the true value of
$\theta_{23}$ is particularly striking. As $\sin^2 \theta_{23}$
increases, the sensitivity to CP violation decreases because the CP
asymmetry that LBNE measures is inversely proportional to $|\sin
\theta_{23}|$ as demonstrated in Equation~\ref{eqn:cpasym}. For the
same reason, as $\theta_{23}$ increases, the degeneracy between the CP
and matter asymmetries is broken, which increases the LBNE sensitivity
to neutrino MH. 
The explicit dependence of MH
sensitivity on the value of $\sin^2\theta_{23}$ is shown in Figure~\ref{fig:mhvsth23}.
\begin{figure}[!htb]
\centerline{
\includegraphics[width=0.7\textwidth]{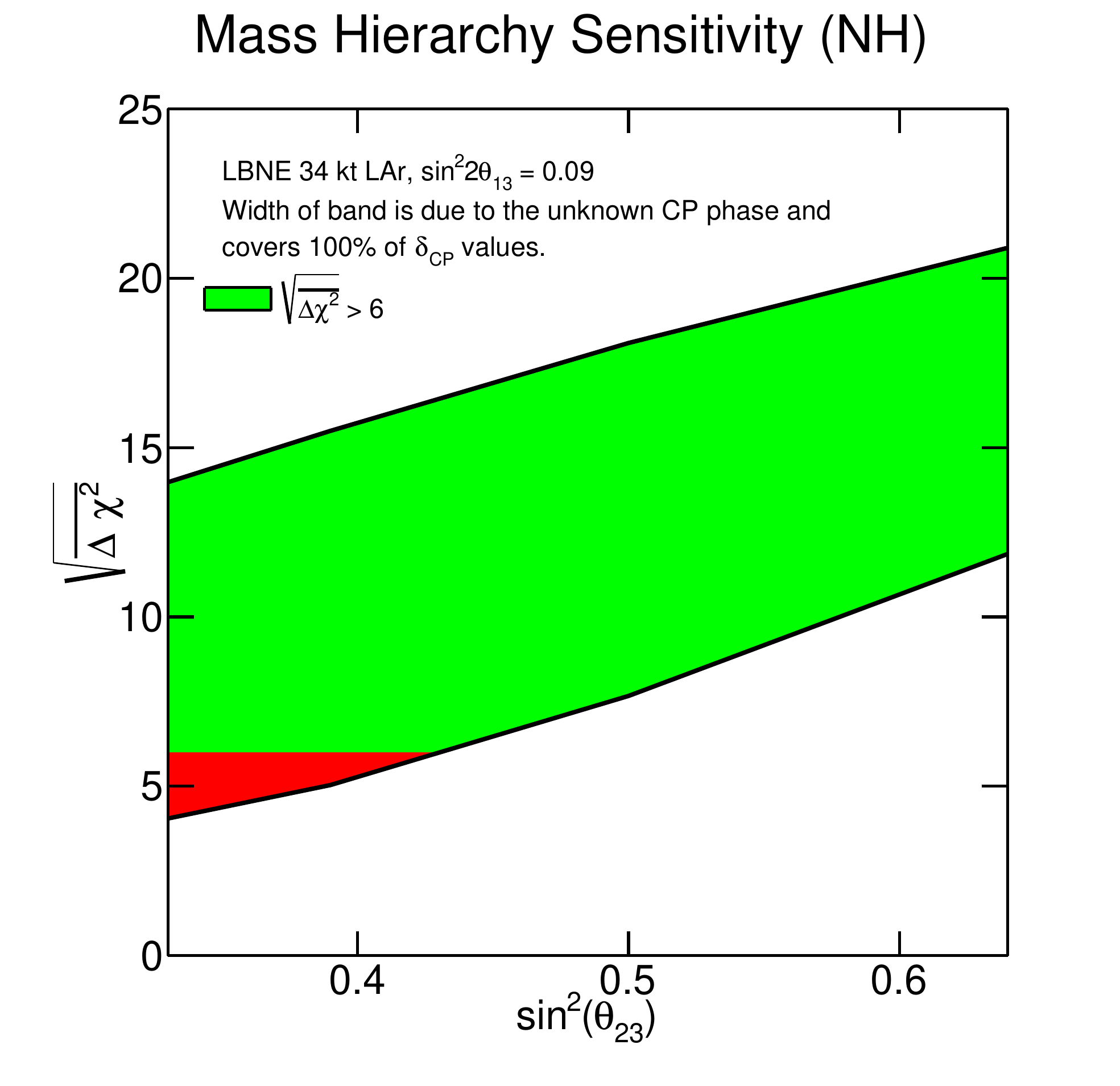}
}
\caption[Sensitivity to MH versus $\sin^2\theta_{23}$]{The
  significance with which the MH can
  be determined by a typical LBNE experiment as a function of the value of 
  $\sin^2\theta_{23}$, for the 3$\sigma$ 
  allowed range of $\sin^2\theta_{23}$, for true normal hierarchy. The width of the
  band is due to the unknown value of \deltacp and covers all possible values of
  \deltacp. The green region shows the parameter space for which  
  $\sqrt{\overline{\Delta\chi^2}}>6$. Assumes a \SIadj{34}{\kt} far detector with 6 years of running in a \SI{1.2}{\MW} beam.}
\label{fig:mhvsth23}
\end{figure}
As this plot makes clear, LBNE resolves the MH with a significance
of $\sqrt{\overline{\Delta\chi^2}}>6$ for nearly all allowed values of $sin^2\theta_{23}$
and \deltacp.

\subsection{Summary of CP-Violation and Mass Hierarchy Sensitivities}
\label{sect:cpmhsummary}

For the  \ktadj{10} LBNE, the statistical
uncertainties are much larger than the systematic uncertainties. 
Combining the sensitivity from the  \ktadj{10} LBNE with expected
knowledge from the NO$\nu$A and T2K 
experiments
would allow LBNE 
to achieve a $\geq 4\sigma$ sensitivity for detecting CP violation
for $30\%$ of the allowed values of \deltacp and a 
$\geq 3\sigma$ sensitivity for $50\%$ of these values. 
It is clear that the  \ktadj{10} LBNE sensitivity would be the
dominant contribution in the combined sensitivities and
would therefore represent a significant advance in the search for leptonic CP
violation over the current
generation of experiments, particularly in the region
where the CP and matter effects are degenerate.  

The combination with T2K and NO$\nu$A
would allow the MH to be determined with a {\em minimum}
precision of
$|\overline{\Delta\chi^2}| \geq 25$ over 60\% \deltacp values and
$|\overline{\Delta\chi^2}| \geq 16$
for all possible values of \deltacp.
Due to the low event
statistics in these experiments,
the combination with NO$\nu$A and T2K only helps the sensitivity
in the region of $\mdeltacp > 0$ (NH) or 
\deltacp < 0 (IH) where there are residual degeneracies
between matter and CP-violating effects. As will be
discussed in Section~\ref{atmnu}, the combination with atmospheric
neutrino oscillation studies can also be used to improve the MH sensitivity in this region for the LBNE \ktadj{10} configuration.

\begin{introbox}
  Assuming the normal hierarchy, the most recent global fit of
  experimental data for the three-neutrino paradigm favors a value of \deltacp close to $-\pi/2$
  with $\sin \mdeltacp < 0$ at a confidence level of $\sim 90\%$~\cite{Capozzi:2013csa} (Figure~\ref{fig:octchisq}). LBNE alone
  with a \SIadj{10}{\kt} detector and six years of running would resolve
 with $\geq3 \sigma$ precision the question of 
  whether CP is violated for the
  currently favored value of \deltacp. With a \SIadj{34}{\kt} detector running for six
  years, LBNE, alone will achieve a precision approaching
  $6\sigma$. 
\end{introbox}

\clearpage

Table~\ref{tab:10ktsens} summarizes the MH and CP sensitivities
that can be reached by a typical experiment with the 
LBNE \ktadj{10} and \ktadj{34} configurations assuming a running time 
of 3+3 ($\nu+\overline{\nu}$) years with a
\MWadj{1.2} beam under a variety of scenarios.
\begin{table}[!htb]
  \caption[Summary of \SIadj{10}{\kt} and \SIadj{34}{\kt} far detector
    sensitivities]{The mass hierarchy and CP violation sensitivities that can be
    reached with a typical data set from the LBNE \SIadj{10}{\kt} and
    \SIadj{34}{\kt} configurations with a \SI{1.2}{\MW} beam, no near
    neutrino detector (ND) unless otherwise stated, and a run time of
    3+3 $\nu +\overline{\nu}$ years under a variety of beam and
    systematic scenarios, for normal hierarchy. Note that the
    sensitivities for inverted hierarchy are similar but not
    identical. As discussed in the text, the significance of the MH determination should not be interpreted using Gaussian
    probabilities.}
\label{tab:10ktsens}
\begin{tabular}{$L^c^c^c^c} 
\toprule
\rowtitlestyle
Scenario ($\sin^2 \theta_{23} = 0.39$) & \multicolumn{2}{^>{\columncolor{\ChapterBubbleColor}}c}{MH sensitivity} & \multicolumn{2}{^>{\columncolor{\ChapterBubbleColor}}c}{CPV sensitivity} \\ 
\rowtitlestyle
         & \deltacp Fraction & ($\sqrt{\overline{\Delta\chi^2}}$) & \deltacp Fraction & ($\sqrt{\overline{\Delta\chi^2}}$)\\ 
\toprowrule
LBNE \SI{10}{\kt}, CDR beam 
& 50\% & $\geq 4$ & 
  40\% & $\geq 2 \sigma$\\
& 100\% & $\geq 2$ &
  - & - \\ \colhline
LBNE \SI{10}{\kt}, \SIadj{80}{\GeV} upgraded beam
& 50\% & $\geq 5$
& 23\% & $\geq 3 \sigma$ \\
& 100\% & $\geq 3$ &
  55\%  & $\geq 2 \sigma$ \\ \colhline
LBNE \SI{10}{\kt}, \SIadj{80}{\GeV} beam, with $\nu$ ND  
& 50\% & $\geq 5$ 
& 33\% & $\geq 3 \sigma$\\
& 100\% & $\geq 3$ 
& 60\% & $\geq 2 \sigma$\\ \colhline
+ NO$\nu$A (6 yrs), T2K (\num{7.8e21} POT) 
& 75\% & $\geq 5$ 
& 30\% & $\geq 4 \sigma$\\
& 100\% & $\geq 4$ 
& 50\% & $\geq 3 \sigma$\\ \colhline
LBNE \SI{34}{\kt} , CDR beam 
& 50\% & $\geq 7$ 
& 20\% & $\geq 4 \sigma$\\
& 100\% & $\geq 4$ 
& 50\% & $\geq 3 \sigma$\\ \colhline
LBNE \SI{34}{\kt}, \SIadj{80}{\GeV} upgraded beam
& 50\% & $\geq 8$ 
& 15\% & $\geq  5\sigma$\\
& 100\% & $\geq 5$ 
& 35\% & $\geq 4 \sigma$\\ \colhline
LBNE \SI{34}{\kt}, \SIadj{80}{\GeV} beam, with $\nu$ ND  
& 50\% & $\geq 9$ 
& 35\% & $\geq  5\sigma$\\
& 100\% & $\geq 5$ 
& 50\% & $\geq 4 \sigma$\\ \bottomrule
\end{tabular}
\end{table}
%
\begin{figure}[!htb]
  \centering\includegraphics[width=0.8\textwidth,clip,trim=2cm 50mm 2cm 4cm]{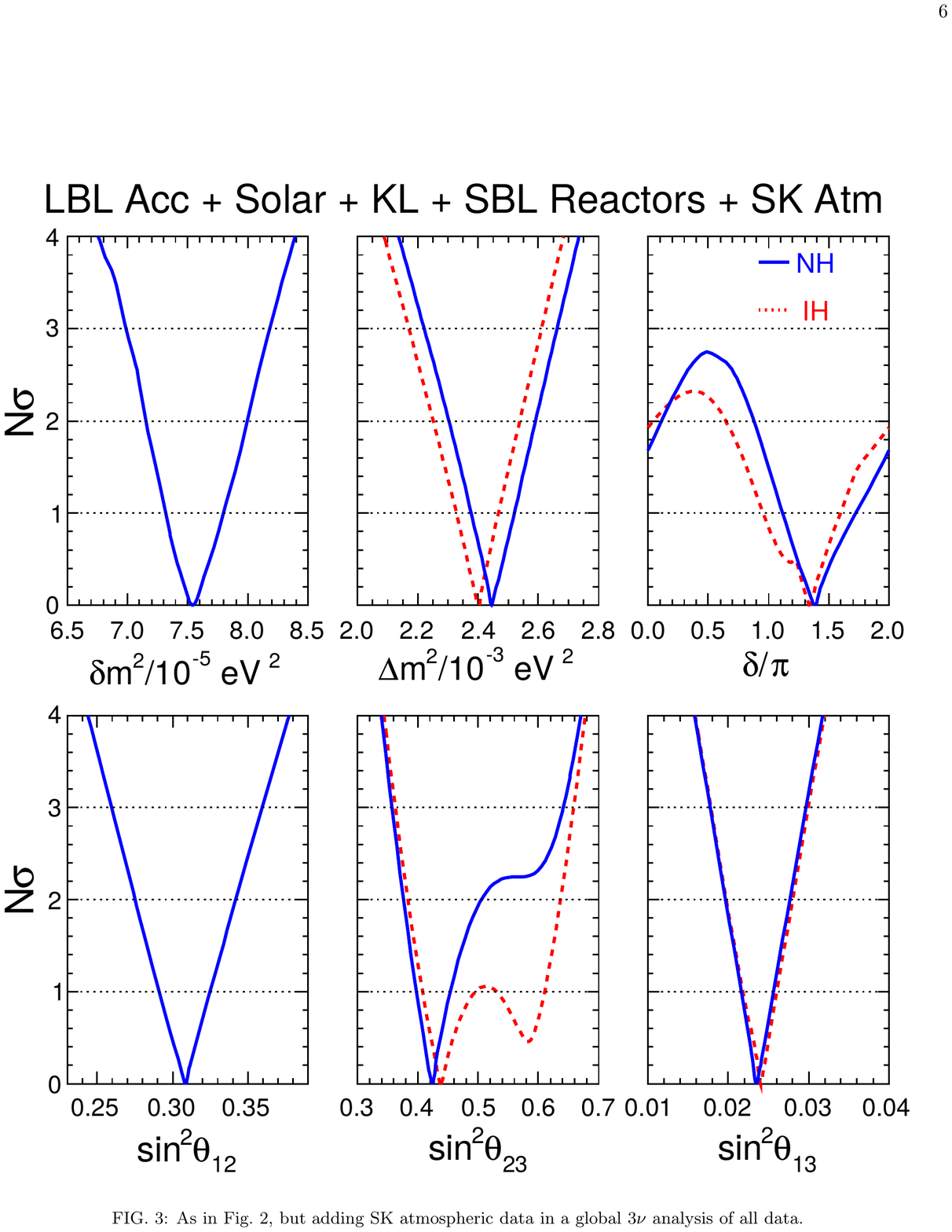}
  \caption[Measurement of the mixing parameters from Capozzi
  {\em et al.}]{Results of the 2013 global analysis from Capozzi {\em et al.} shown as
  N$\sigma$ bounds on the six parameters governing three $\nu$ flavor 
  oscillations. Blue (solid)
    and red (dashed) curves refer to NH and IH, respectively. 
    Figure is from~\cite{Capozzi:2013csa}.}
  \label{fig:octchisq}
\end{figure}

\subsection{CP-Violating and Mass Hierarchy Sensitivities with Increased Exposures}

Figure~\ref{fig:lar-cp-frac} shows the minimum significance with which the
MH can be resolved and CP violation determined by LBNE 
as a function of increased exposure in units of 
mass $\times$  beam power $\times$ time\footnote{Time is denoted in years of running at Fermilab. One year of running at Fermilab corresponds to $\approx 1.7 \times 10^7$
  seconds.}.  For this study, the LBNE beamline improvements discussed in
Section~\ref{beamline-chap} are used with $E_p =$ \SI{80}{\GeV}, and the
signal and background normalization uncertainties are assumed to be 1\% and 5\%,
respectively. Both $\nu_e$ and $\nu_\mu$ appearance signals are used
in a combined analysis.
\begin{figure}[!htb]
\centerline{
\centering\includegraphics[width=0.5\textwidth]{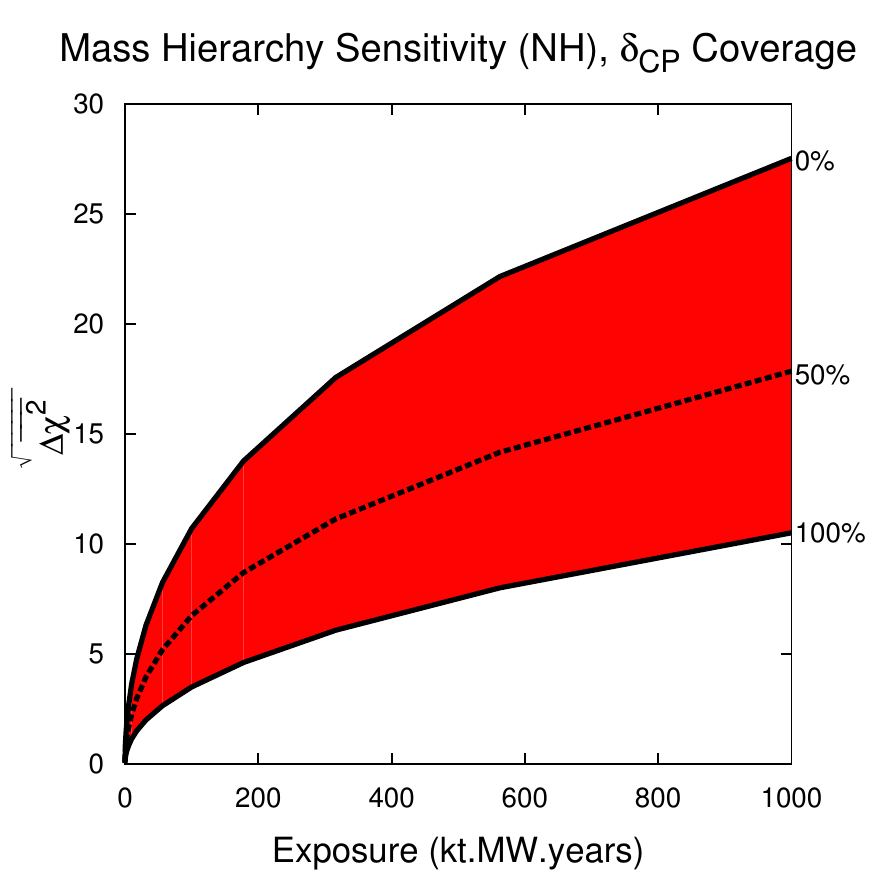}
\centering\includegraphics[width=0.5\textwidth]{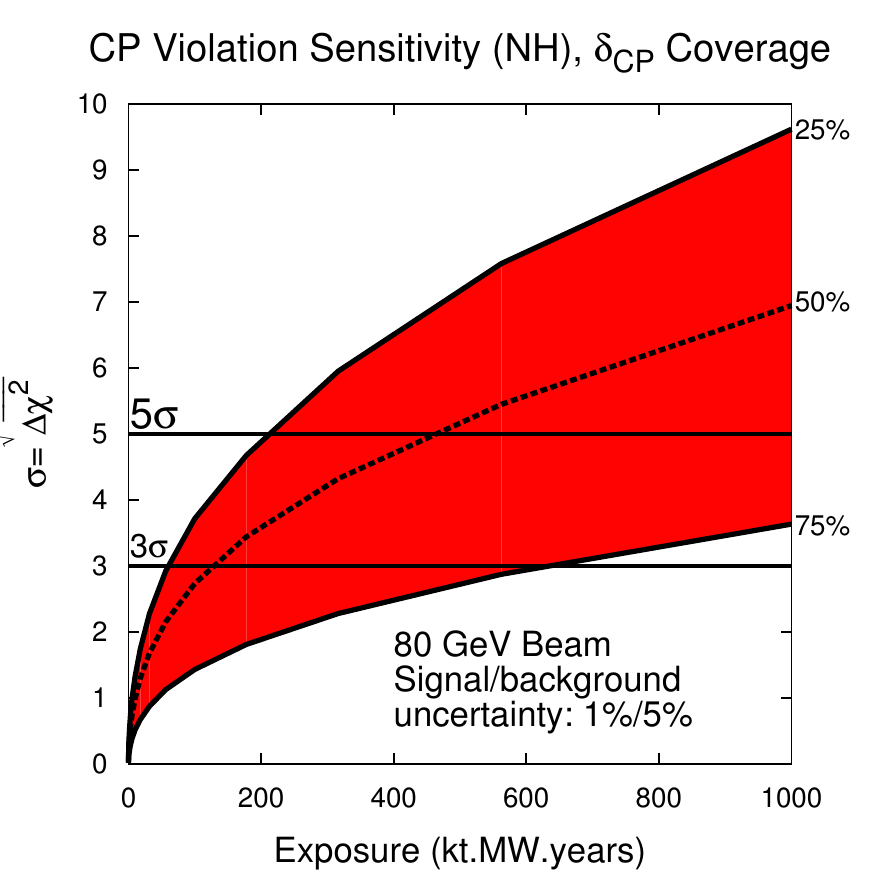}
}
\caption[Mass hierarchy and CPV versus exposure in mass, beam
  power and time]{The minimum significance with which the mass
  hierarchy (left) and CP violation (right) can be resolved as a
  function of exposure in detector mass (kiloton) $\times$ beam power
  (MW) $\times$ time (years), for true NH. 
  The red band represents the fraction of
  \deltacp values for which the sensitivity can be achieved with
  at least the minimal significance on the y-axis. }
\label{fig:lar-cp-frac}
\end{figure}
Due to the long baseline, the
determination of the MH in LBNE to high precision does not
require a large exposure; a sensitivity of
$\sqrt{\overline{\Delta\chi^2}} = 5$ for the worst case (NH,
$\mdeltacp = \pi/2$ or IH, $\mdeltacp = -\pi/2$) requires an
exposure of $\sim$
\SI[inter-unit-product=\ensuremath{{}\cdot{}}]{200}{\kt\MW\year}s, but
$\sqrt{\overline{\Delta\chi^2}} = 5$ sensitivity can be reached for
$50\%$ of the allowed values of \deltacp with an exposure of less
than \SI[inter-unit-product=\ensuremath{{}\cdot{}}]{100}{\kt\MW\year}s.
On the other hand, reaching discovery-level sensitivity ($\geq 5
\sigma$) to leptonic CP violation for at least 50\% of the possible
values of \deltacp will require large exposures of $\approx$ 
\SI[inter-unit-product=\ensuremath{{}\cdot{}}]{450}{\kt\MW\year}s. Figure~\ref{fig:lar-cp-frac2} demonstrates the
sensitivity to CP violation as a function of \deltacp and
exposure that can be achieved with various stages of the Fermilab
Proton-Improvement-Plan (PIP-II and upgrades to PIP-II). In this
study, the PIP-II upgrades are assumed to provide LBNE with
\SI{1.2}{\MW}\footnote{The assumed exposures are only accurate to
  the level of 15\% due to incomplete knowledge of the PIP-II final design
  parameters and running conditions.} at 80~GeV, followed by further
upgrades in which the booster is replaced with a linac that will
provide \SI{2.3}{\MW} from the Main Injector (MI), also at 80~GeV.  The
study demonstrates that it is possible to reach 5$\sigma$ sensitivity
to CP violation over at least 40\% of \deltacp values running for
a little over 10 years, starting with the PIP-II MI power
and a LArTPC greater than \SI{10}{\kt}, and phasing in more detector mass.
Other possible staging scenarios of detector mass and beam power are
discussed in Chapter~\ref{conclusion-chap}.

\begin{table}[!htb]
\centering
  \caption[Summary of CPV sensitivities with PIP II, mass upgrades and beyond]
  {The CP violation sensitivities
    that can be reached by LBNE alone starting with the LBNE \SIadj{10}{\kt} 
    configuration 
    with a \MWadj{1.2} beam and a run time of 3+3 ($\nu +\overline{\nu}$) years and phasing in
    additional far detector mass and beam power upgrades beyond the
    current PIP-II.
    In all cases, the sensitivities are calculated using
    the \SI{80}{\GeV} upgraded beam and 1\%/5\% signal/background normalization uncertainties, for true normal hierarchy.
    The sensitivity for each stage includes exposure from the 
    previous stage(s) of the experiment.}
\label{tab:px}
\begin{tabu}{$L^l^c^c} 
\toprule
\rowtitlestyle
Exposure & Possible Scenario & \multicolumn{2}{^>{\columncolor{\ChapterBubbleColor}}c}{CPV sensitivity} \\ 
\rowtitlestyle
 &        &  \deltacp Fraction & ($\sqrt{\overline{\Delta\chi^2}}$)\\ 
\toprowrule 
\SI[inter-unit-product=\ensuremath{{}\cdot{}}]{60}{\kt\year}s \SI{1.2}{\MW} beam & PIP-II, \SI{10}{\kt}, \SI{6}{\year}s & 
60\% \deltacp & $\geq 2 \sigma$ \\ 
& & 
33\% \deltacp & $\geq 3 \sigma$ \\ \colhline
+ \SI[inter-unit-product=\ensuremath{{}\cdot{}}]{200}{\kt\year}s \SI{1.2}{\MW} beam & PIP-II, \SI{34}{\kt}, \SI{6}{\year}s & 
40\% \deltacp & $\geq 5\sigma$ \\ \colhline
+ \SI[inter-unit-product=\ensuremath{{}\cdot{}}]{200}{\kt\year}s \SI{2.3}{\MW} beam & Booster replaced,  \SI{34}{\kt}, \SI{6}{\year}s
& 60\% \deltacp & $\geq 5\sigma$\\ \bottomrule
\end{tabu}
\end{table}

\begin{figure}[!htb]
  \centering\includegraphics[width=0.7\textwidth]{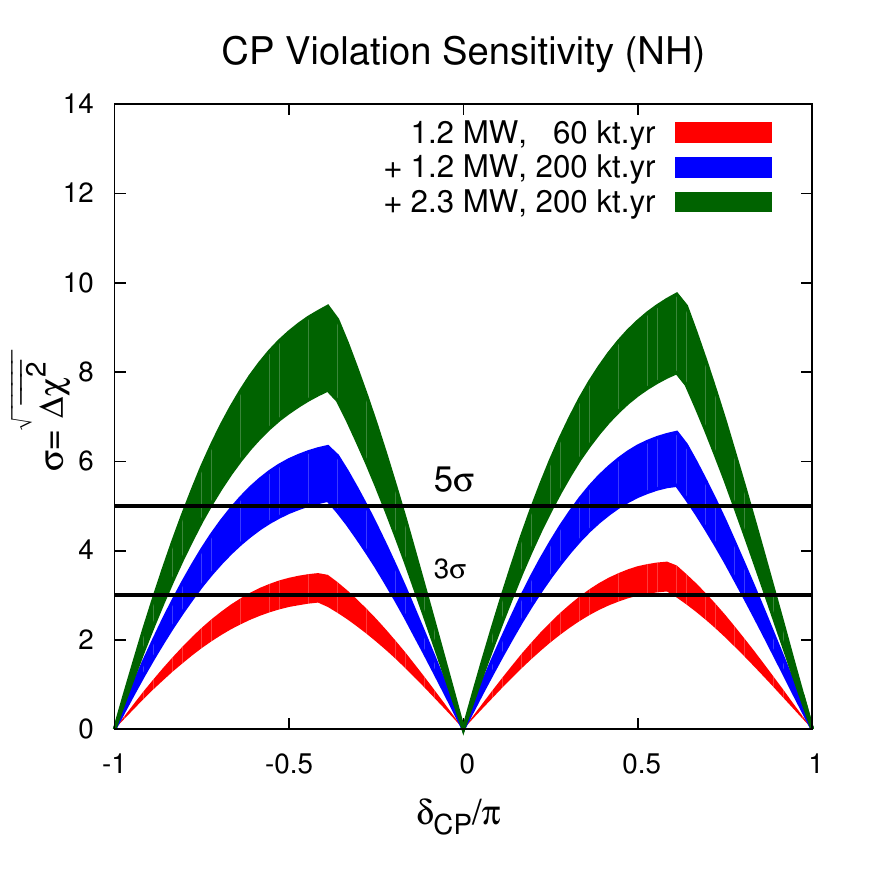}
\caption[CPV determination versus \deltacp]{The significance with which
  CP violation --- $\mdeltacp \neq 0 \ {\rm or} \ \pi$ --- can be
  determined as a function of \deltacp.
  The different color curves represent possible exposures
  from different stages of PIP and detector mass upgrades as follows: \SI{1.2}{\MW}, \SI{60}{\kt$\cdot$\year}s
  (red) +  \SI{1.2}{\MW}, \SI{200}{\kt$\cdot$\year}s (blue) +  \SI{2.3}{\MW}, \SI{200}{\kt$\cdot$\year}s (green). The sensitivity
  for each higher exposure is in addition to that from all lower 
  exposures. The bands represent the range of sensitivities obtained from the
  improvements to the CDR beamline design.}
\label{fig:lar-cp-frac2}
\end{figure}


\section{Measurement of $\theta_{23}$ and Determination of the Octant}
\label{sec:octant}

The value of $\textrm{sin}^{2}2\theta_{23}$ is measured to be $>$~0.95 at 90\% CL 
using atmospheric neutrino oscillations~\cite{Abe:2011ph}. 
This corresponds to a value of 
$\thetatwothree$ near 45$\degs$, but leaves an ambiguity
as to whether the value of $\thetatwothree$ is in the lower octant 
(less than 45$\degs$), the upper octant (greater than 45$\degs$)
or exactly 45$\degs$. 
The value of $\sin^2 \theta_{23}$ from
the 2013 global fit reported by~\cite{Capozzi:2013csa} is $\sin ^2 \theta_{23} = 0.425
^{+0.029} _{-0.027} (1 \sigma)$ for normal hierarchy (NH), but as shown
in Figure~\ref{fig:octchisq}, the distribution of the $\chi^2$ from
the global fit has another local minimum --- particularly if the MH 
is inverted --- at $\sin^2 \theta_{23}\approx 0.59$. A
\emph{maximal} mixing value of $\sin^2 \theta_{23} =0.5$ is therefore still allowed
by the data and the octant is still largely undetermined.
As discussed in Chapter~\ref{intro-chap},
a value of $\thetatwothree$ exactly equal to 45$\degs$ would indicate that 
$\nu_{\mu}$ and $\nu_{\tau}$ have equal contributions from $\nu_3$,
which could be evidence for a previously unknown symmetry. 
It is
therefore important experimentally to determine the value of
$\sin ^2 \theta_{23}$ 
with sufficient precision to determine 
the octant of $\theta_{23}$. 

The measurement of $\nu_\mu \rightarrow \nu_\mu$ oscillations is
sensitive to $\sin ^2 2 \theta_{23}$, whereas the measurement of
$\nu_\mu \rightarrow \nu_e$ oscillations is sensitive to $\sin^2
\theta_{23}$. 
A combination of both $\nu_e$ appearance and $\nu_\mu$ disappearance
measurements can probe both maximal mixing and the $\theta_{23}$
octant. With the large statistics and rich spectral structure in a
wide-band, long-baseline experiment such as LBNE
(Figure~\ref{fig:lar-disapp-spectrum}), precision measurements of $\sin
^2 \theta_{23}$ can be significantly improved compared to existing
experiments, particularly for values of $\theta_{23}$ near
$45^\circ$. 
\begin{figure}[!htb]
  \centering\includegraphics[width=0.8\textwidth]{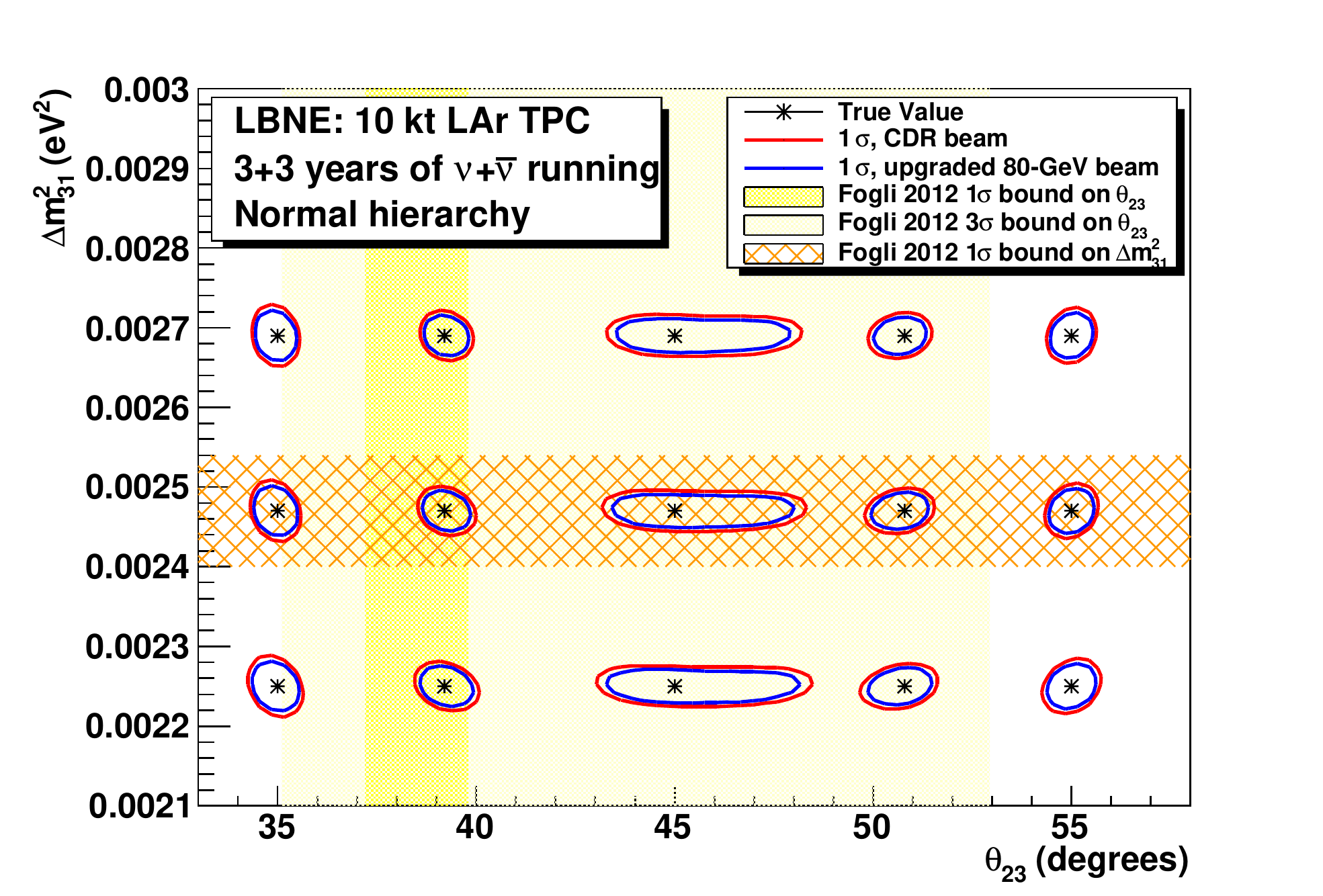}
  \caption[Measurement of $\theta_{23}$ and $\Delta m^2_{31}$ with the \SIadj{10}{\kt}
    LBNE]{The precision with which a simultaneous measurement of
    $\theta_{23}$ and $\Delta m^2_{31}$ can be determined with \SI{10}{\kt} and 3+3 years  of $\nu+\overline{\nu}$
    running in a \MWadj{1.2} beam. The yellow bands represent the
    $1\sigma$ and $3\sigma$ allowed ranges of $\theta_{23}$ and the orange hatched region represents the
    $1\sigma$ allowed range of $\Delta m^2_{31}$ from~\cite{Fogli:2012ua}.}
  \label{fig:lbl_theta23_octant0}
\end{figure}
Figure~\ref{fig:lbl_theta23_octant0} demonstrates the
measurement precision of $\theta_{23}$ and $\Delta m^2_{31}$ that can
be achieved for different true values of these parameters by a
\SIadj{10}{\kt} LBNE detector. The subdominant 
$\nu_\mu \rightarrow \nu_e$ appearance signal in a
\SIadj{10}{\kt} detector is limited by statistical uncertainties.

The significance with which  a \SIadj{10}{\kt} LBNE detector can determine 
the $\theta_{23}$ octant is shown in the top plot of 
Figure~\ref{fig:lbl_theta23_octant1}. The $\Delta\chi^2$ metric is defined as:
\begin{eqnarray}
\Delta\chi^2_{octant} & = & |\chi^2_{\theta_{23}^{test}>45^\circ} - \chi^2_{\theta_{23}^{test}<45^\circ}|, \\ \nonumber
\end{eqnarray}
where the value of $\theta_{23}$ in the \emph{wrong} octant is constrained 
only to have a value within the \emph{wrong} octant (i.e., it is not required
to have the same value of $\sin^22\theta_{23}$ as the true value).
The individual $\chi^2$ values are given by 
Equation~\ref{eq:globes_chisq}. As in the  $\Delta\chi^2$ metrics for 
MH and CP violation, the $\chi^2$ value for the \emph{true} octant 
is identically zero in the absence of statistical fluctuations.
If $\theta_{23}$ is 
within the $1\sigma$ bound of the global 
fit~\cite{Fogli:2012ua}, an LBNE \SIadj{10}{\kt} detector alone will 
determine the octant with $> 3\sigma$ significance for all values 
of \deltacp. Figure~\ref{fig:lbl_theta23_octant1} (bottom) demonstrates the increasing
sensitivity to the $\theta_{23}$ octant for values closer to maximal
$\nu_\mu$-$\nu_\tau$ mixing that can be achieved with subsequent phases of LBNE coupled
with upgrades in beam power from the Main Injector. 
\begin{introbox}
With sufficient
exposure, LBNE can resolve the $\theta_{23}$ octant with $> 3 \sigma$
significance even if $\theta_{23}$ is within a few degrees of
$45^\circ$, the value at which the mixing between the $\nu_\mu$ and $\nu_\tau$ neutrino states is maximal.
\end{introbox}
\begin{figure}[!htb]
  \centering\includegraphics[width=0.6\textwidth]{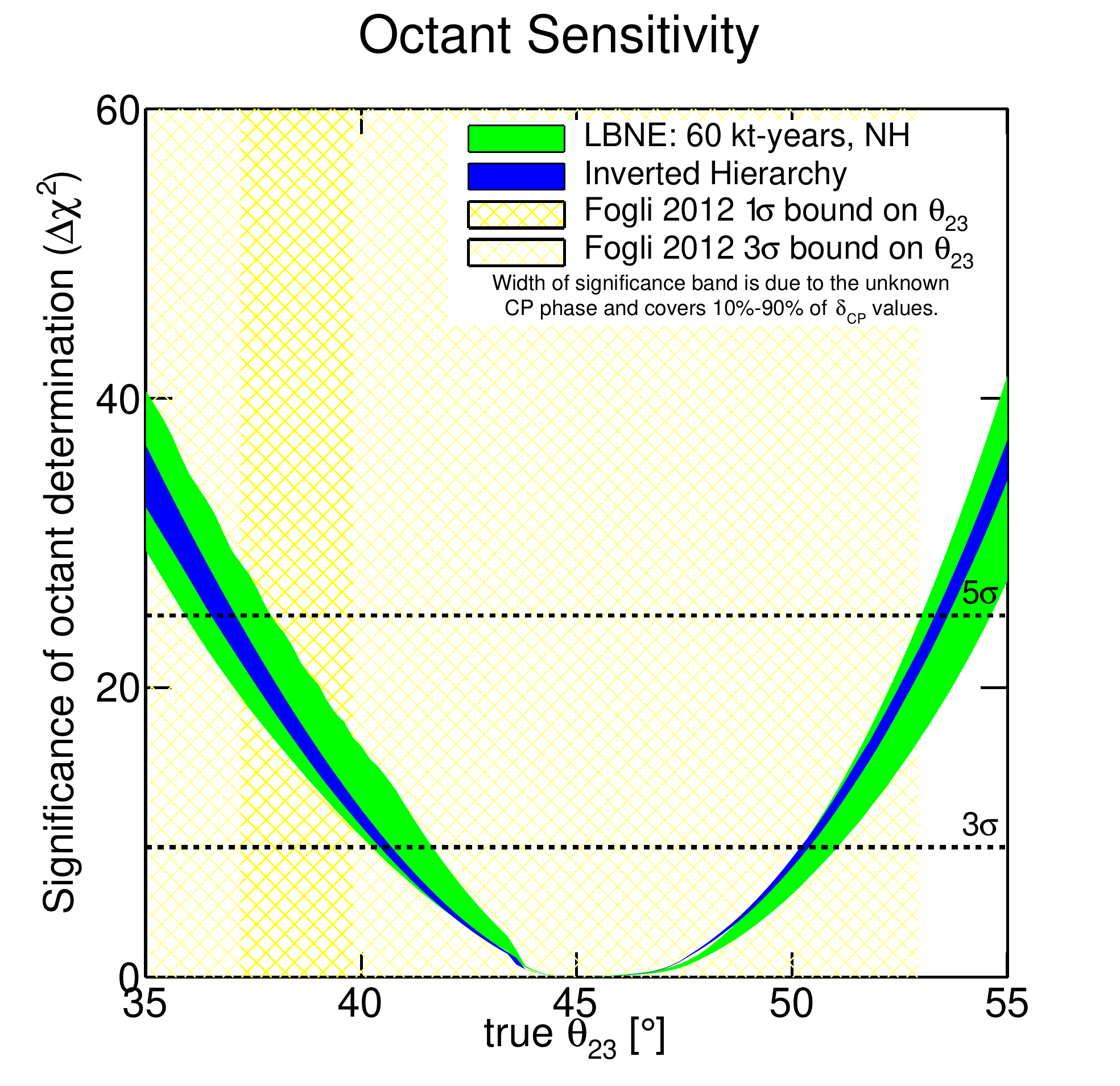}
  \centering\includegraphics[width=0.6\textwidth]{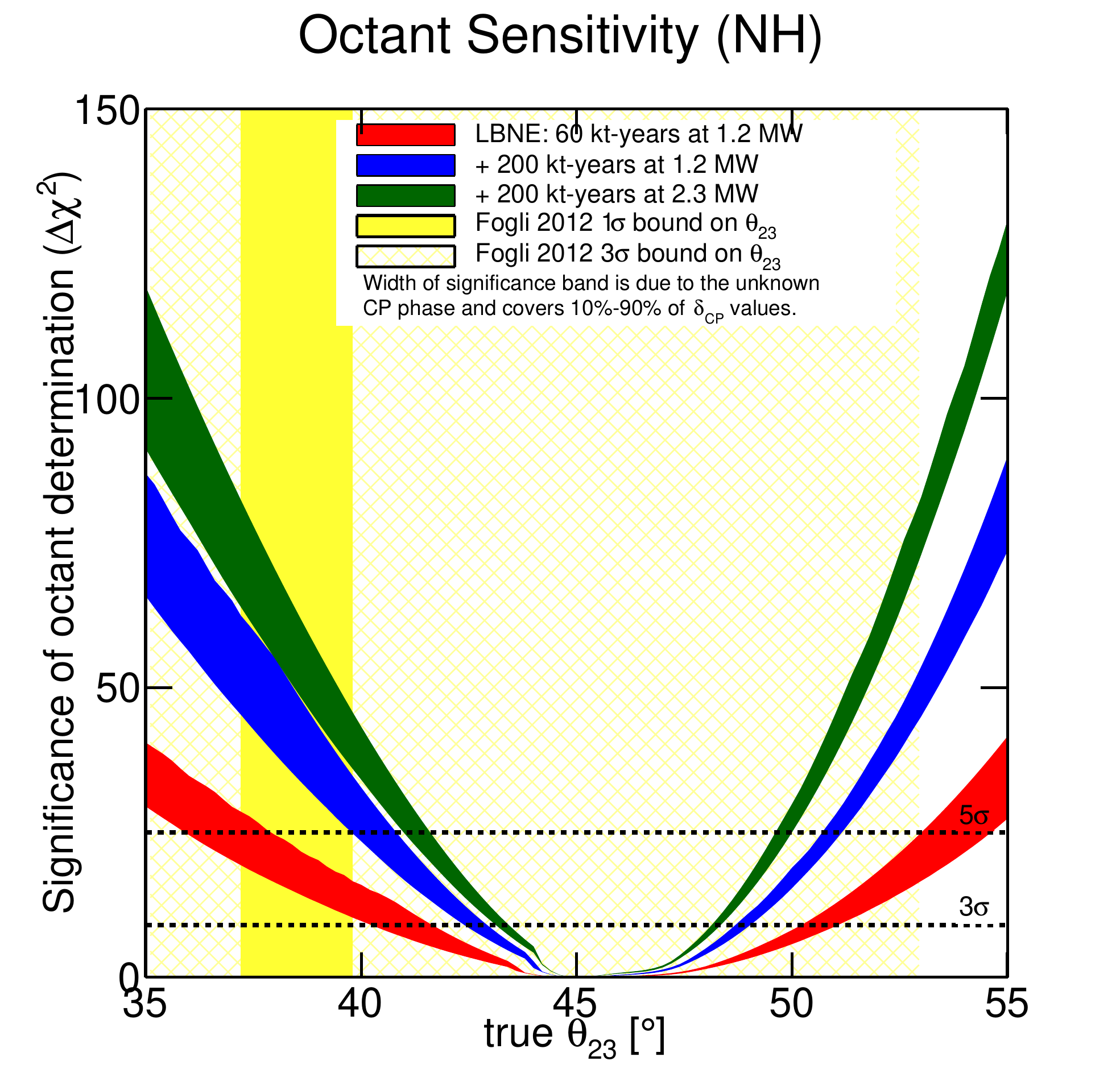}
  \caption[Sensitivity to the determination of the
  $\theta_{23}$ octant]{Top: significance with which LBNE can resolve the
    $\theta_{23}$ octant degeneracy for 3+3 years of
    $\nu$+$\overline{\nu}$ running at \SI{1.2}{\MW} with a
    \SIadj{10}{\kt} detector. The bands are for normal (green) and
    inverted (blue) hierarchy.  The widths of the bands correspond to
    the fraction of $\mdeltacp$ values covered at this
    significance or higher, ranging from $10\%$ to $90\%$.  The yellow
    bands represent the $1\sigma$ and $3\sigma$ allowed ranges of
    $\theta_{23}$ from ~\cite{Fogli:2012ua}. 
Bottom:
significance with which LBNE can
    resolve the $\theta_{23}$ octant degeneracy (normal hierarchy) for equal
    $\nu$+$\overline{\nu}$ running  with increased exposure. The colored bands
  represent increasing exposures as follows: \SI{1.2}{\MW}, \SI{60}{\kt$\cdot$\year}
  (red) +  \SI{1.2}{\MW}, \SI{200}{\kt$\cdot$\year}s (blue) +  \SI{2.3}{\MW}, \SI{200}{\kt$\cdot$\year}s (green). The sensitivity
  for each higher exposure is in addition to that from all lower 
  exposures.
}
  \label{fig:lbl_theta23_octant1}
\end{figure}
%

\clearpage
\section{Precision Measurements of the Oscillation Parameters in
  the Three-Flavor Model}
\label{sec:precision}
The rich oscillation structure that can be observed by LBNE and the
excellent particle identification capability of the detector
will enable precision measurement  in a single experiment of all the mixing parameters
governing $\nu_1$-$\nu_3$ and $\nu_2$-$\nu_3$ mixing. As discussed
in Chapter~\ref{intro-chap}, theoretical models probing quark-lepton
universality predict specific values of the mixing angles and the
relations between them. The 
mixing angle $\theta_{13}$ is
expected to be measured accurately in reactor experiments by the end
of the decade with a precision that will be limited by
systematics. The systematic uncertainty on the value of $\sin ^ 2 2
\theta_{13}$ from the Daya Bay reactor neutrino experiment, which has
the lowest systematics, is currently $\sim4$\%~\cite{An:2013zwz}.

\begin{figure}[!htb]
  \centering\includegraphics[width=0.7\textwidth]{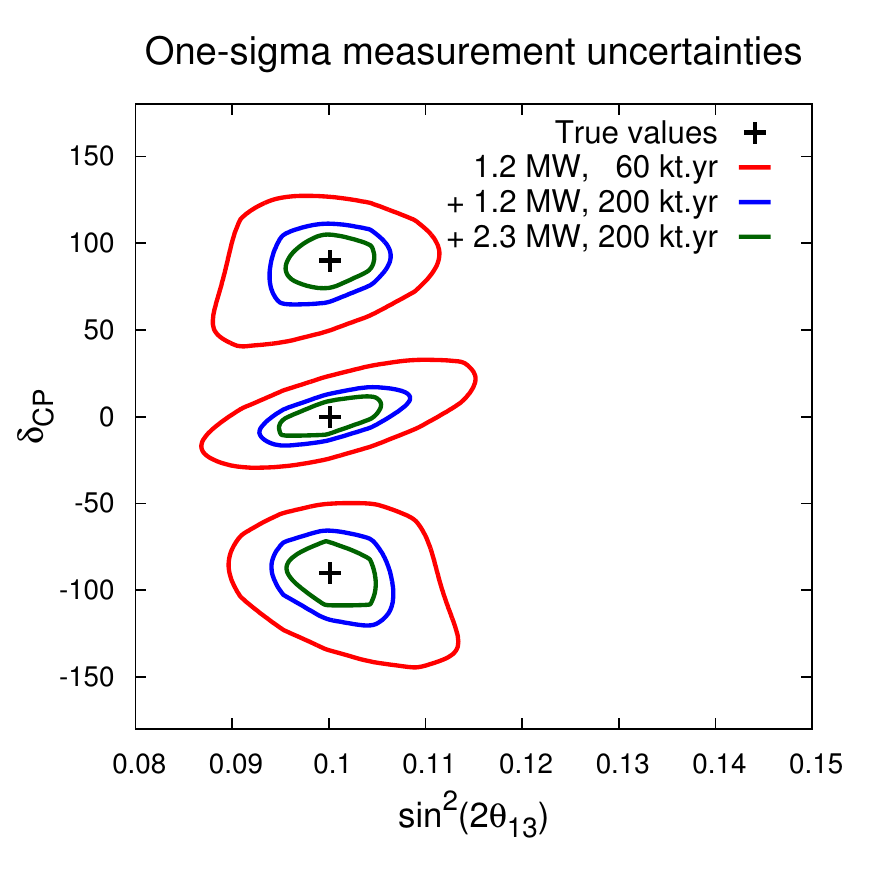}
\caption[Measurement of \deltacp and $\theta_{13}$ in LBNE with
  increased exposures]{Measurement of \deltacp and $\theta_{13}$ in LBNE with
  different exposures, for true normal hierarchy (NH). 
  The different color curves represent 
  one-sigma contours for three
possible exposures
  from different stages of PIP and detector mass upgrades as follows: \SI{1.2}{\MW}, \SI{60}{\kt$\cdot$\year}
  (red),  \SI{1.2}{\MW}, \SI{200}{\kt$\cdot$\year}s (blue) +  \SI{2.3}{\MW}, \SI{200}{\kt$\cdot$\year}s (green). The sensitivity
  for each higher exposure is in addition to that from all lower 
  exposures.}
\label{fig:t13dcp}
\end{figure}
While the constraint on $\theta_{13}$ from the reactor experiments will be
important in the
early stages of LBNE for determining CP violation, measuring
\deltacp and determining the $\theta_{23}$ octant, 
LBNE itself will eventually be able to measure
$\theta_{13}$ independently with a precision on par with the final
precision expected from the reactor experiments. 
Whereas the reactor experiments measure $\theta_{13}$ using $\overline{\nu}_e$
disappearance, LBNE will measure it through $\nu_e$ and
$\overline{\nu}_e$ appearance, thus providing an independent constraint on
the three-flavor mixing matrix. 
Figure~\ref{fig:t13dcp} demonstrates the
precision with which LBNE can measure \deltacp and $\theta_{13}$
simultaneously, with no external constraints on $\theta_{13}$, as a
function of increased exposure, for three different exposures. 
Both appearance and
disappearance modes are included in the fit using the upgraded \GeVadj{80}
beam. Signal/background normalization uncertainties of 1\%/5\% are assumed.

Figure~\ref{fig:precision} shows the expected 1$\sigma$ resolution
on different three-flavor oscillation parameters as a function of exposure
in \ktyr in a \MWadj{1.2} beam with LBNE alone and LBNE in combination with the
expected performance from T2K and NO$\nu$A. 
\begin{figure}[!htb]
\centerline{
  \includegraphics[width=0.5\textwidth]{bassfig/svl_oscvars_dcp.pdf}
  \includegraphics[width=0.5\textwidth]{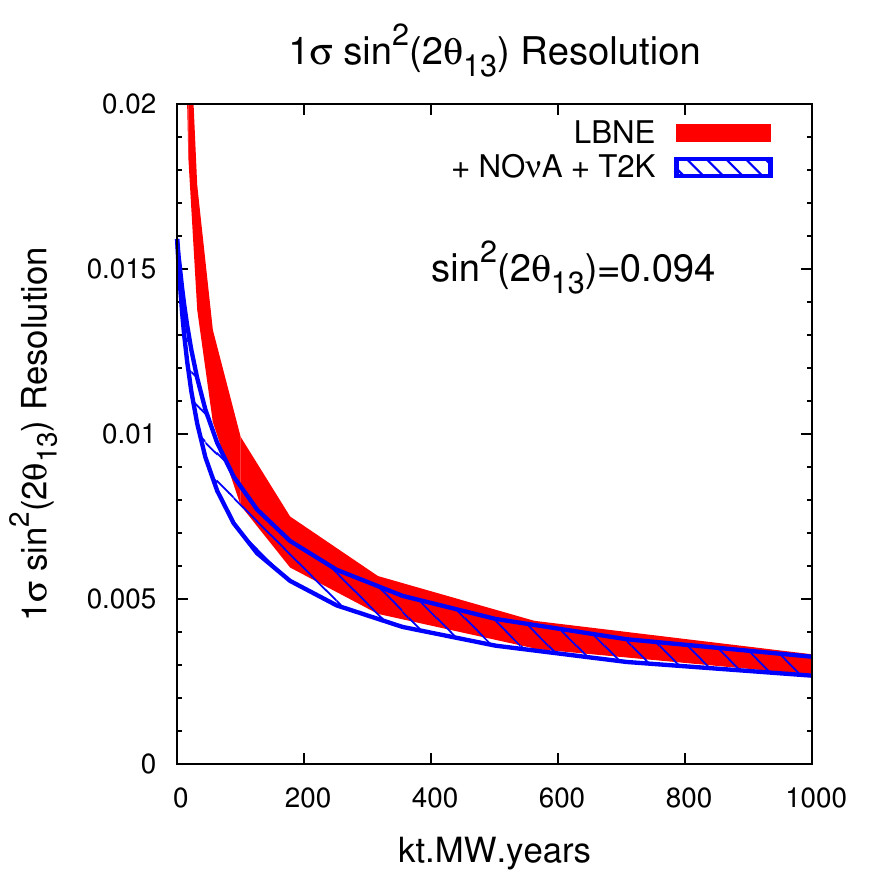}}
\centerline{
  \includegraphics[width=0.5\textwidth]{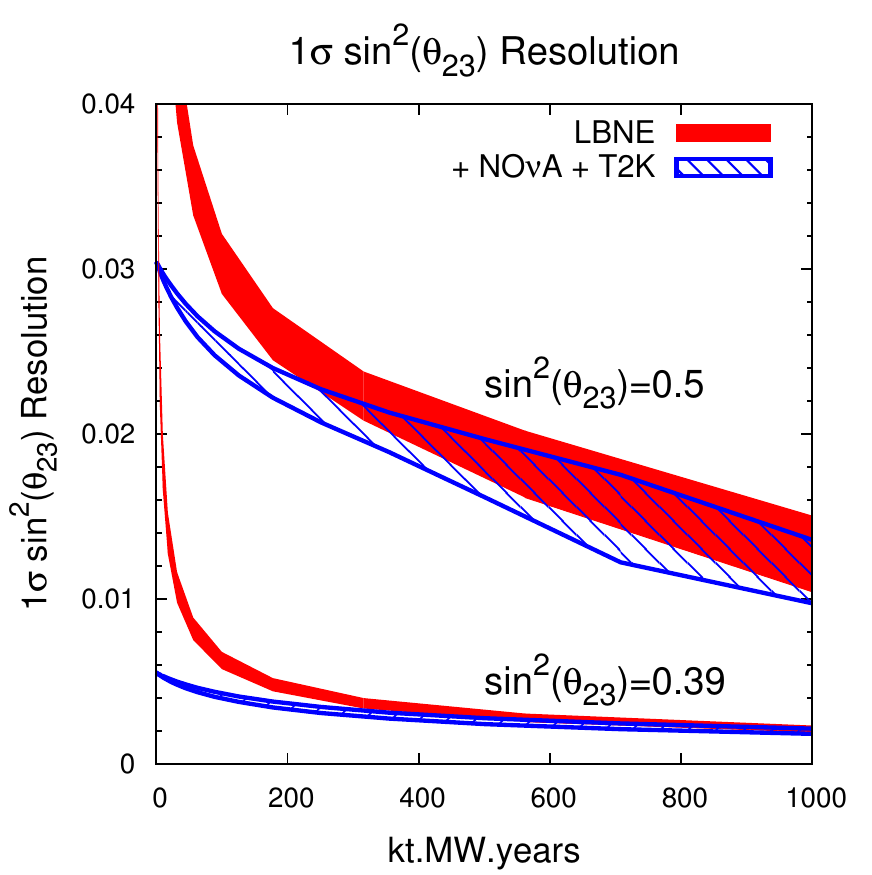}
  \includegraphics[width=0.5\textwidth]{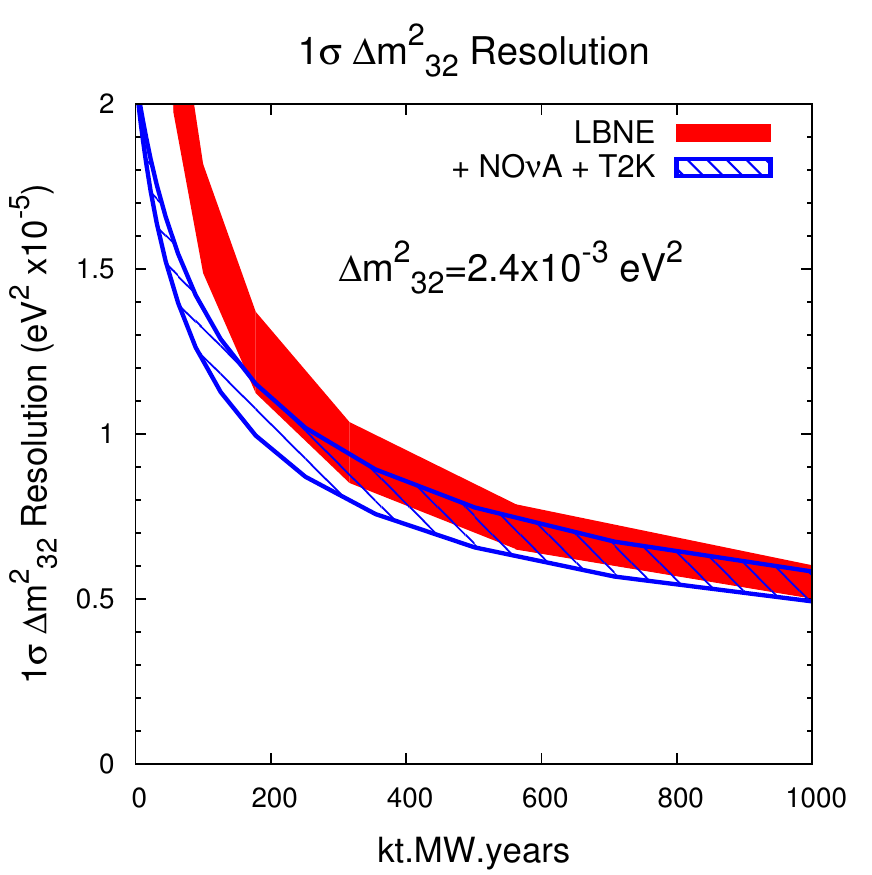}}
\caption[Expected 1$\sigma$ resolution on three-flavor
  oscillation parameters, \SIadj{1.2}{\MW} beam]{The expected 1$\sigma$
  resolution on different three-flavor oscillation parameters as a
  function of exposure in \ktmwyr{}s, for true NH. 
  The red curve indicates the precision that could
  be obtained from LBNE alone, and the blue curve represents the
  combined precision from LBNE and the T2K and NO$\nu$A experiments.
  The width of the bands represents the range of performance with the
  beam improvements under consideration. 
}
\label{fig:precision}
\end{figure}
It should be noted that LBNE alone
could reach a precision on $\sin ^ 2 2\theta_{13}$ of 0.005 with an exposure of
$\sim$\SI[inter-unit-product=\ensuremath{{}\cdot{}}]{300}{\kt\MW\year}s. LBNE can also significantly improve the
resolution on $\Delta m^2_{32}$ beyond what the combination of
NO$\nu$A and T2K can achieve, reaching a precision of \SI{1e-5}{eV^2}
with an exposure of $\sim$\SI[inter-unit-product=\ensuremath{{}\cdot{}}]{300}{\kt\MW\year}s. 
The
precision on $\Delta m^2_{32}$ will ultimately depend on tight control
of energy-scale systematics. Initial
studies of the systematics reveal that the measurement of $\nu_\mu$
disappearance in LBNE over a full oscillation interval, with two
oscillation peaks and two valleys (Figure~\ref{fig:lar-disapp-spectrum}), 
reduces the dependency of the $\Delta
m^2_{23}$ measurement on the energy-scale systematics, which limited
the measurement precision in MINOS~\cite{Adamson:2011ig}.

\clearpage
\section{Oscillation Studies Using Atmospheric Neutrinos}
\label{atmnu}

\begin{introbox}
  Atmospheric neutrinos are unique among sources used to study
  oscillations: the flux contains neutrinos and antineutrinos of all
  flavors, matter effects play a significant role, both $\Delta m^2$
  values contribute, and the oscillation phenomenology occurs over
  several orders of magnitude each in energy (Figure~\ref{fig:atmflux}) and
  path length. These characteristics make atmospheric neutrinos ideal
  for the study of oscillations (in principle sensitive to all of the
  remaining unmeasured quantities in the PMNS matrix) and provide a
  laboratory in which to search for exotic phenomena for which the
  dependence of the flavor-transition and survival probabilities on
  energy and path length can be defined. The large LBNE LArTPC far
  detector, placed at sufficient depth to shield against cosmic-ray
  background, provides a unique opportunity to study atmospheric
  neutrino interactions with excellent energy and path-length
  resolutions.
\end{introbox}

LBNE has obtained far detector physics sensitivities based on
information from atmospheric neutrinos by using a Fast MC and a
three-flavor analysis framework developed for the MINOS experiment
\cite{Adamson:2013whj}.  Four-vector-level events are generated using
the GENIE neutrino event generator~\cite{Andreopoulos:2009zz}.  For
atmospheric neutrinos the Bartol \cite{Agrawal:1995gk} flux
calculation for the Soudan, MN site was used, and for beam neutrinos
the \GeVadj{80}, \MWadj{1.2} 
beamline design described in
Section~\ref{beamline-chap} was used. In
this section, unless otherwise specified, the oscillation parameters are as specified in Table~\ref{tab:atmparams}.
\begin{table}[!htb]
\caption[Oscillation parameters used in the atmospheric-neutrino analysis]{Oscillation parameters used in the atmospheric-neutrino analysis.}
\label{tab:atmparams}
\begin{tabular}{^l^c}
\toprule
\rowtitlestyle
Parameter &   Value \\ \toprowrule
 $\Delta m^2=1/2(\Delta m_{32}^2+\Delta m_{31}^2)$ (NH)  &  
$+2.40 \times 10^{-3}$ eV$^2$ \\ \colhline 
$\sin^2 \theta_{23}$ & 0.40 \\ \colhline
$\Delta m_{21}^2$ &  $7.54 \times10^{-5}$ eV$^2$ \\ \colhline
$\sin^2 \theta_{12}$ & $0.307$ \\ \colhline 
$\sin^2 \theta_{13}$ & $0.0242$ \\ \colhline
\deltacp & 0 \\ \bottomrule
\end{tabular}
\end{table}

The expected interaction rates in 100 \ktyr are shown in 
Table~\ref{tab:atmos_event_rates}.  All interactions occur on argon and are
distributed uniformly throughout a toy detector geometry consisting of
two modules, each 14.0~m high, 23.3~m wide, and 45.4~m long.  For this
study, events with interaction vertices outside the detector volume
(e.g., events that produce upward-going stopping or through-going
muons) have not been considered.  Cosmogenic
background has not been studied in detail, but since atmospheric neutrinos
are somewhat more tolerant of background than proton decay, a depth
that is sufficient for a proton decay search is expected to also be suitable
for studies of atmospheric neutrinos.  Given the detector's \ftadj{4850} depth, a veto should
not be necessary and the full fiducial mass of the detector should be usable.

\begin{table}[!htb]
\caption[Expected atmospheric $\nu$ interaction rates in 100 \ktyr
  with a LArTPC]{Expected atmospheric $\nu$ interaction rates in a
  LArTPC with an exposure of
  \SI[inter-unit-product=\ensuremath{{}\cdot{}}]{100}{\kt\year}s for
  the Bartol flux and GENIE argon cross sections (no oscillations).}
\label{tab:atmos_event_rates}
\begin{tabular}{$L^r^r^r} 
\toprule
\rowtitlestyle
Flavor                   &      CC  &     NC  &  Total \\ \toprowrule
$\nu_\mu$                &   \num{10069}  &    \num{4240} & \num{14309}  \\ \colhline
$\overline{\nu}_{\mu}$   &    \num{2701}  &    \num{1895} &  \num{4596}  \\ \colhline 
$\nu_e$                  &    \num{5754}  &    \num{2098} &  \num{7852}  \\ \colhline
$\overline{\nu}_e$       &    \num{1230}  &     \num{782} & \num{2012}  \\ 
\toprule
\rowtitlestyle
Total:                   &   \num{19754}  &    \num{9015} & \num{28769}  \\ \bottomrule
\end{tabular}
\end{table}

A Fast MC runs on the produced four-vectors, placing events into
containment and flavor categories.  Containment is evaluated by
tracking leptons through the liquid argon detector box geometry and classifying
events as either fully contained (FC) or partially contained (PC).  A detection threshold
of 50~MeV is assumed for all particles.  Flavor determination,
in which events are placed into electron-like or muon-like categories, is
based on properties of the primary and secondary particles above detection
threshold.
Electrons are assumed to be correctly identified with 90\% probability
and other electromagnetic particles (e.g., $\pi^0$, $\gamma$) are
misidentified as electrons 5\% of the time. Muons are identified with
100\% probability and charged pions are misidentified as muons 1\% 
of the time.
Events in which neither of the two leading particles is 
identified as a muon or electron are 
placed into an \emph{NC-like} category.  With these assumptions, the
purities of the flavor-tagged samples are 97.8\% for the FC electron-like
sample, 99.7\% for the FC muon-like sample, and 99.6\% for the PC
muon-like sample.  The NC-like category is not used in this analysis,
but would be useful for $\nu_\tau$ appearance studies.
\begin{table}[!tb]
\caption[Performance assumptions for atmospheric and combined
atmospheric+beam $\nu$]{Detector performance assumptions for the atmospheric neutrino and the combined
atmospheric+beam neutrino analyses.}
\label{tab:atmosdpa}
\begin{tabular}{$L^c} 
\toprule
\rowtitlestyle
Particle & Resolution \\ \toprowrule
\multicolumn{2}{^c}{Angular Resolutions } \\ \toprowrule
  Electron        & $1^\circ$ \\ \colhline
  Muon            & $1^\circ$ \\ \colhline
  Hadronic System & $10^\circ$ \\ \toprule
\multicolumn{2}{^c}{\textbf{Energy Resolutions}}\\ \toprowrule
 Stopping Muon   & $3\%$    \\ \colhline
 Exiting Muon    & $15\%$   \\ \colhline
 Electron       & $1\%/\sqrt{E(GeV)} \oplus 1\%$ \\ \colhline
 Hadronic System & 30\%/$\sqrt{E(GeV)}$  \\ \bottomrule
\end{tabular}
\end{table}
The energy and direction of the event are then assigned by separately
smearing these quantities of the leptonic and hadronic systems,
where the width of the Gaussian resolution functions for each 
flavor/containment 
category are given in Table~\ref{tab:atmosdpa}.  Detector
performance assumptions are taken both from the LBNE CDR~\cite{CDRv1} and 
from published results from the ICARUS 
experiment~\cite{Amoruso:2003sw,ankowski2010energy,Arneodo:112001,Rubbia:2011ft}.
Including oscillations, the expected number of events in 100 \ktyr is summarized in
Table~\ref{tab:atmevts}. 
\begin{table}[!htb]
\caption[Atmospheric-neutrino event rates in 100 \ktyr with oscillations]{Atmospheric-neutrino 
event rates including oscillations in 100 \ktyr with a
LArTPC, fully or partially  
contained in the detector fiducial volume.}
\label{tab:atmevts}
\begin{tabular}{$L^c}
\toprule
\rowtitlestyle
Sample & Event rate \\ \toprowrule
fully contained electron-like sample & 4,015 \\ \colhline
fully contained muon-like sample & 5,958 \\ \colhline
partially contained muon-like sample & 1,963 \\ \bottomrule
\end{tabular}
\end{table}
\begin{figure}[!hbt]
\centering
\includegraphics[width=0.6\textwidth]{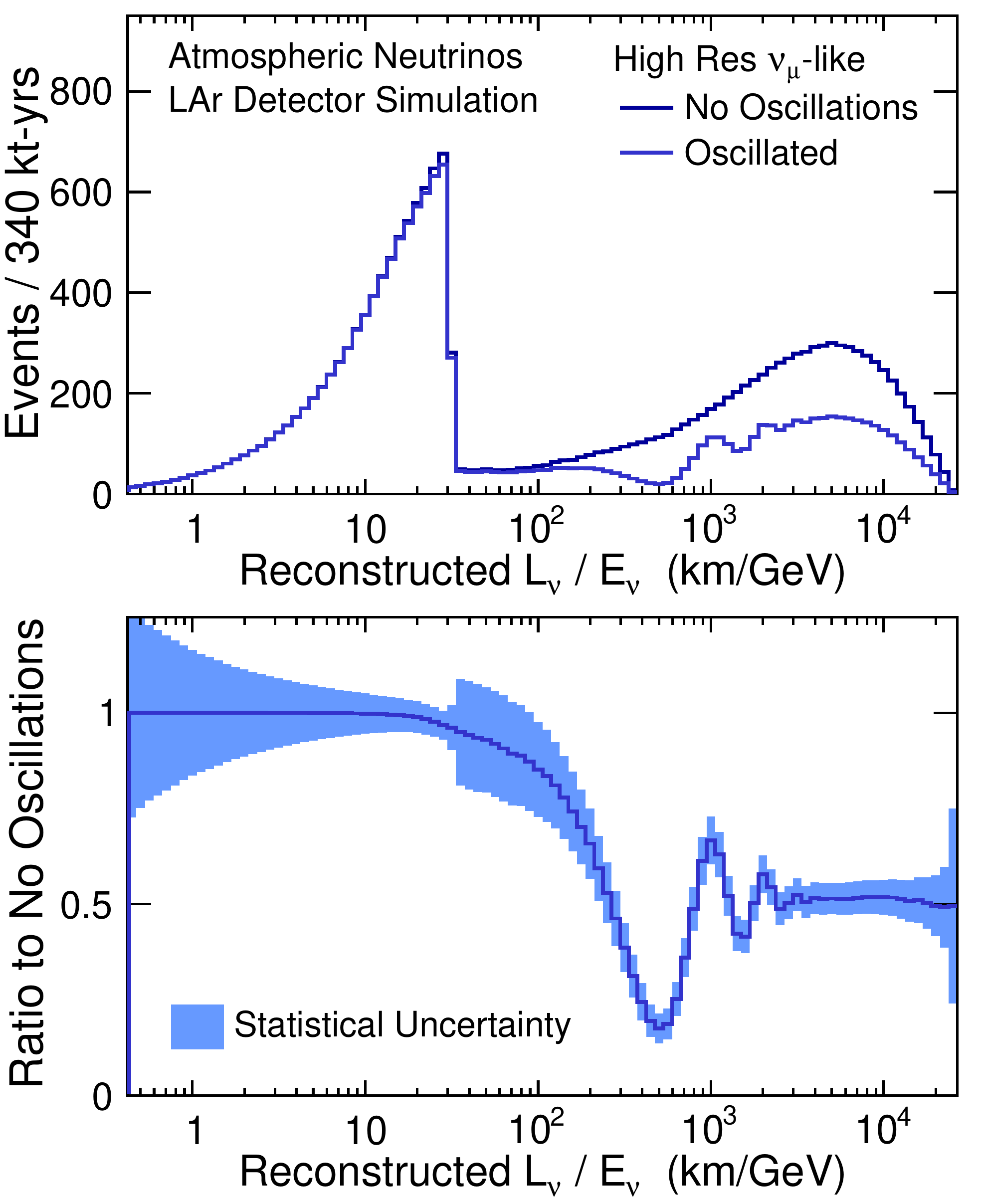}
\includegraphics[width=0.6\textwidth,height=0.3\textheight]{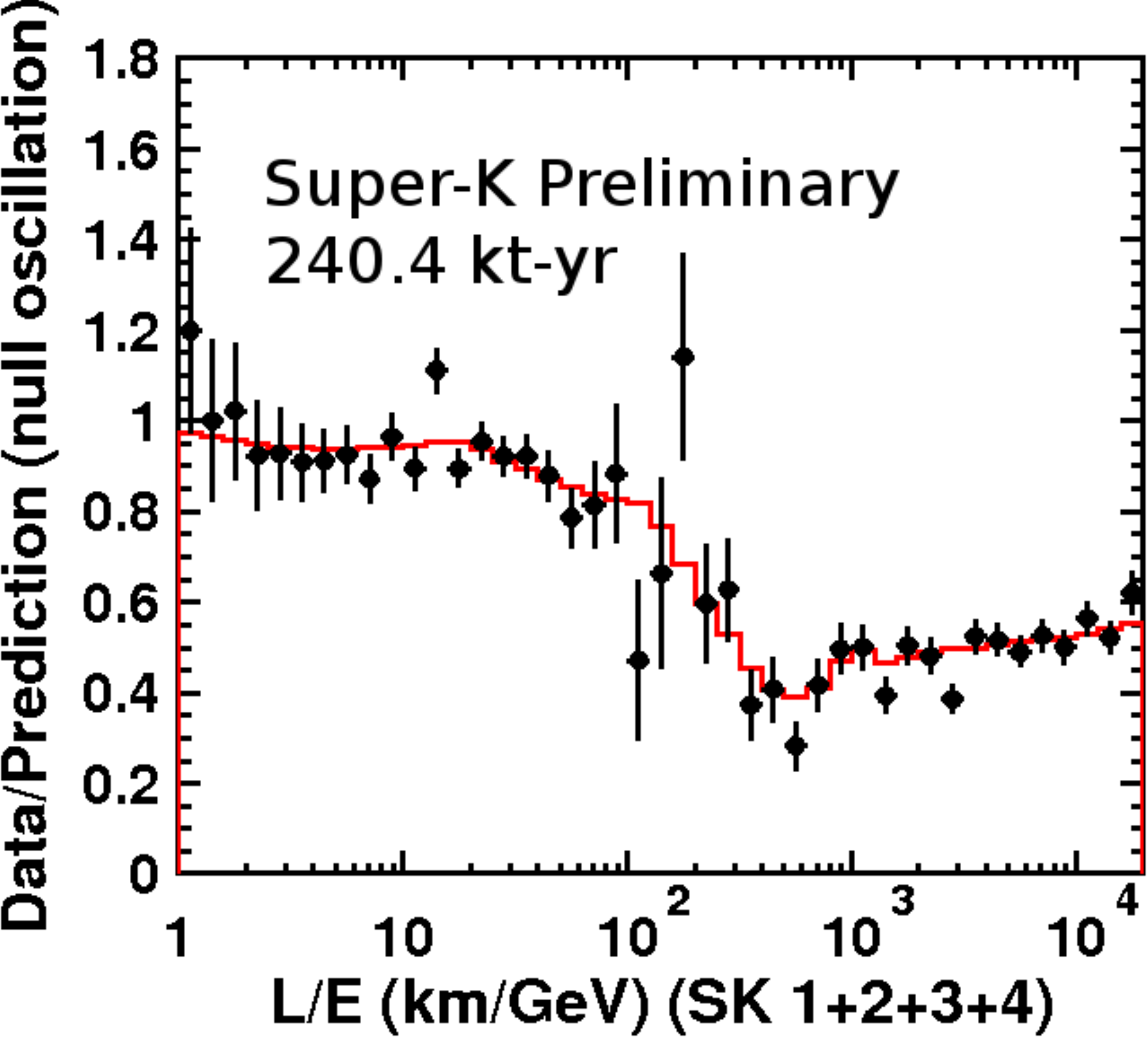}
\caption[L/E distribution of high-resolution $\mu$-like atmospheric $\nu$ events, 
\SI{340}{\kt$\cdot$\MW$\cdot$\year}
] {
Reconstructed L/E distribution of \emph{high-resolution}
  $\mu$-like atmospheric neutrino events in LBNE with a \SI{340}{\kt$\cdot$\MW$\cdot$\year}
  exposure with and without oscillations (top); the ratio of the
  two, with the shaded band indicating the size of the
  statistical uncertainty (center); the ratio of observed
  data over the null oscillation prediction from the Super-Kamiokande
  detector with \SI[inter-unit-product=\ensuremath{{}\cdot{}}]{240.4}{\kt\year}s of exposure (bottom).}
\label{fig:lovere}
\end{figure}

Figure~\ref{fig:lovere} shows the
expected L/E distribution for \emph{high-resolution} muon-like events from
a \SI[inter-unit-product=\ensuremath{{}\cdot{}}]{350}{\kt\year} exposure; 
the latest data from Super-Kamiokande are
shown for comparison.  LBNE defines high-resolution events 
similarly to Super-Kamiokande, i.e., either by excluding a region of  low-energy 
events or events pointing toward the horizon where the baseline resolution
is poor.  The data provide excellent resolution of the first two
oscillation nodes, even when taking into account the expected statistical
uncertainty. 

In performing oscillation fits, the data in each flavor/containment category
are binned in energy and zenith angle.  Figure~\ref{fig:atmnspectra} shows
the zenith angle distributions for several ranges of reconstructed energy, 
where oscillation features are clearly evident. 
\begin{figure}[!htb]
\centering
\includegraphics[width=1.0\textwidth]{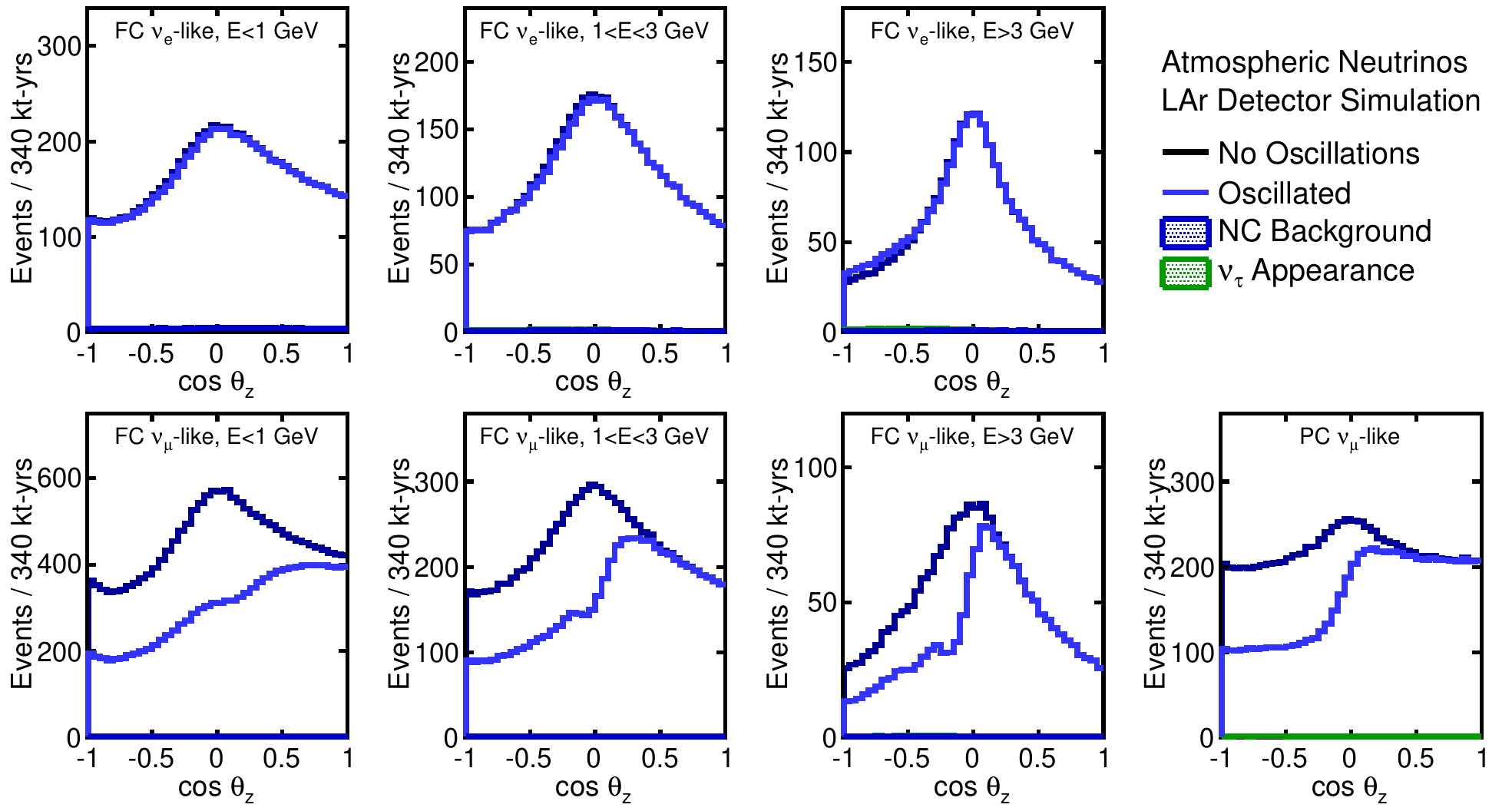}
\caption[Reconstructed zenith angle distributions in several ranges of 
energy]{Reconstructed zenith angle distributions in several ranges of 
energy for the FC $e$-like, FC $\mu$-like, and PC $\mu$-like samples.  
The small contributions from NC background and $\nu_\tau$  are also
shown. }
\label{fig:atmnspectra}
\end{figure}
\begin{figure}[!htb]
\vskip 0.125in
\centering
\includegraphics[width=0.9\textwidth]{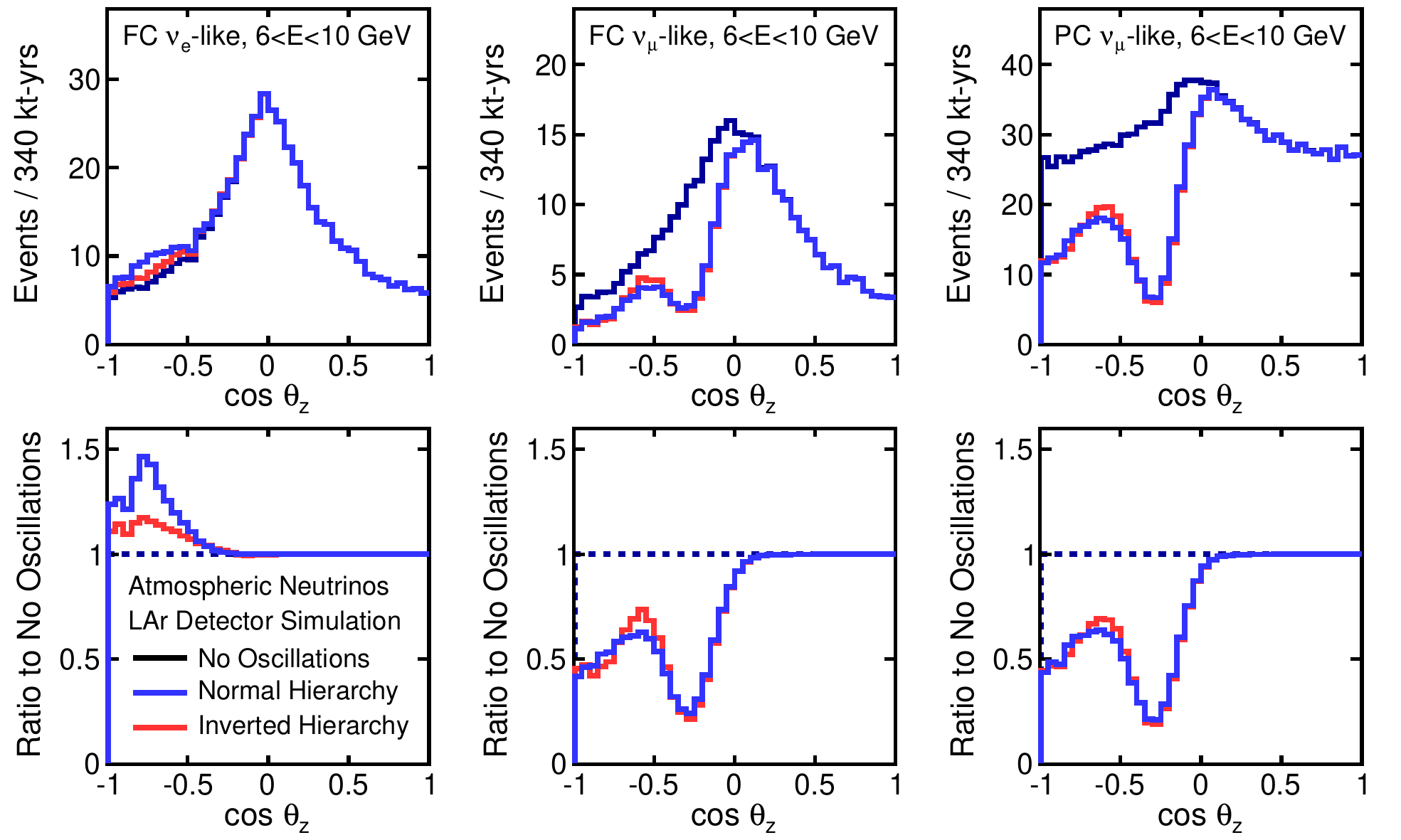}
\caption[Reconstructed zenith angle distributions for events of 6-10 GeV]{Reconstructed zenith angle distributions for 6 to \GeVadj{10}  
events in the different FC and PC samples. Top plots 
show the expected distributions for no oscillations (black), oscillations 
with normal (blue), and inverted (red) hierarchy.  
Bottom plots show the ratio of the 
normal and inverted expectations to the no-oscillation distributions for each category. 
}
\label{fig:atmnspectra2}
\end{figure}

The power to resolve the mass hierarchy (MH)  with atmospheric neutrinos comes 
primarily from the MSW enhancement of few-GeV neutrinos at large zenith
angles.  This enhancement occurs for neutrinos in the normal hierarchy and
antineutrinos in the inverted hierarchy.  Figure~\ref{fig:atmnspectra2}
shows zenith angle distributions of events in the relevant energy range
for each of the three flavor/containment categories.  Small differences
are evident in comparing the NH and IH predictions.


Since the resonance peak occurs for neutrinos in the NH
and antineutrinos in the IH, the MH sensitivity can be
greatly enhanced if neutrino and antineutrino events can be separated.
The LBNE detector will not be magnetized; however, its high-resolution
imaging offers possibilities for tagging features of events
that provide statistical discrimination between neutrinos and
antineutrinos.  For the sensitivity calculations that follow, two such
tags are included: a proton tag and a decay-electron tag.  For
low-multiplicity events, protons occur preferentially in neutrino
interactions; protons are tagged with 100\% efficiency if their
kinetic energy is greater than 50~MeV. Decay electrons are assumed to
be 100\% identifiable and are assumed to occur 100\% of the time for
$\mu^+$ and 25\% of the time for $\mu^-$, based on the $\mu^\pm$
capture probability on $^{40}$Ar.

In the oscillation analysis, 18 nuisance 
parameters are included, with
detector performance parameters correlated between beam and
atmospheric data.  In all cases, $\sin^2 \theta_{12}$, $\Delta
m^2=1/2(\Delta m_{32}^2+\Delta m_{31}^2)$, and $\Delta m_{21}^2$ are
taken to be fixed at the values given in Table~\ref{tab:atmparams}.  The fits then range
over $\theta_{23}$, $\theta_{13}$, \deltacp, and the MH. 
 A 2\% constraint is assumed on the value of $\theta_{13}$;
this value is chosen to reflect the expected ultimate precision of the
current generation of reactor-neutrino experiments.  The systematic
errors included in this analysis are given in
Table~\ref{tab:atmossyst}.
\begin{table}[!htb]
\caption[Systematic errors in the atmospheric and beam+atmospheric
$\nu$ analysis]{Systematic errors included in the atmospheric and beam+atmospheric
neutrino analysis.  The beam values assume the existence of a near detector (ND). 
Atmospheric spectrum ratios include the combined effect of flux and 
detector uncertainties (e.g., the up/down flux uncertainty as well as the 
uncertainty on the detector performance for the up/down ratio).  The 
atmospheric spectrum shape uncertainty functions are applied separately 
for $\nu_\mu, \nu_e, \overline{\nu}_{\mu}, \overline{\nu}_e$.  }
\label{tab:atmossyst}
\centering
\begin{tabular}{$L^c^c} 
\toprule
\rowtitlestyle
                 &  Atmospheric                          &   Beam (Assumes ND)    \\  \toprowrule
Normalization   &  Overall (15\%)                       &   $\mu$-like (5\%)     \\
                 &                                       &   e-like (1\%)         \\  \colhline
NC Background   &  e-like  (10\%)                       &   $\mu$-like (10\%)    \\
                 &                                       &   e-like (5\%)         \\  \colhline
Spectrum Ratios  &  up/down (2\%)                        &                        \\ 
                 &  $\nu_e/\nu_\mu$ (2\%)                &                        \\
                 &  $\overline{\nu}_{\mu}/\nu_\mu$ (5\%) &                        \\
                 &  $\overline{\nu}_e/\nu_e$ (5\%)       &                        \\  \colhline
Spectrum Shape   &  $f(E<E_0)=1+\alpha(E-E_0)/E_0$       &                        \\
                 &  $f(E>E_0)=1+\alpha \log(E/E_0)$      &                        \\
                 &  where $\sigma_\alpha$=5\%            &                        \\  \colhline
Energy Scales    &  \multicolumn{2}{^c}{ Muons (stopping 1\%, exiting 5\%) }      \\ 
(Correlated)     &  \multicolumn{2}{^c}{ Electrons (1\%) }                        \\ 
                 &  \multicolumn{2}{^c}{ Hadronic System (5\%) }                  \\  \bottomrule
\end{tabular}
\end{table}

For the determination of the MH, 
the $\overline{\Delta \chi^2}$ value is calculated
between the best-fit points in the NH and IH where, at
each, the nuisance parameters have been marginalized.  
The sensitivity in the plots that follow is given as $\sqrt{\overline{\Delta \chi^2}}$. 
Figure~\ref{fig:atmnsens1} shows the MH sensitivity from a 
340-\ktyr  
exposure of atmospheric neutrino data alone.   
\begin{figure}[!htb]
\centering
\includegraphics[width=0.7\textwidth]{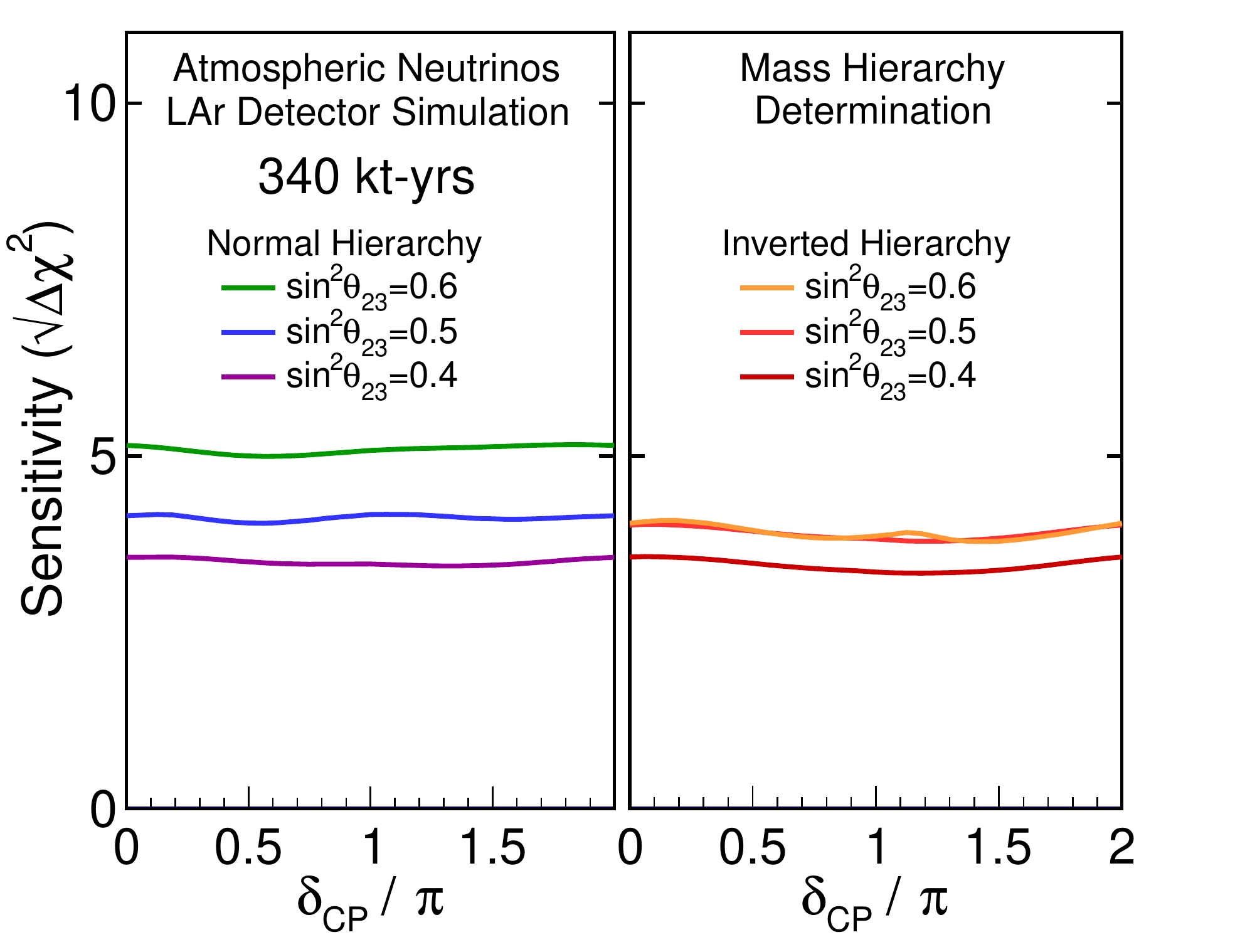}
\caption[Sensitivity for atmospheric-neutrino data to MH as a function of 
\deltacp, 340 \ktyr]{Sensitivity of \SI[inter-unit-product=\ensuremath{{}\cdot{}}]{340}{\kt\year}s of atmospheric neutrino data 
to MH as a function of 
\deltacp for true normal (left) and inverted (right) hierarchy and 
different assumed values of $\sin^2\theta_{23}$. }
\label{fig:atmnsens1}
\end{figure}
For all values of the MH and \deltacp, 
the MH can be determined at $\sqrt{\overline{\Delta \chi^2}} >3$.   
The resolution depends significantly 
on the true value of $\theta_{23}$; the sensitivity for three $\theta_{23}$ values 
is shown.  
The sensitivity depends relatively weakly on the true hierarchy and the 
true value of \deltacp.  This is in sharp contrast to the MH 
sensitivity of the beam, which 
has a strong dependence on the true value of \deltacp. 
Figure \ref{fig:atmnsens2} shows the MH sensitivity as a function of 
the fiducial exposure.  Over this range of fiducial exposures, the 
sensitivity goes essentially as 
the square root of the exposure, indicating that the measurement is not 
systematics-limited.    
\begin{figure}[!htb]
\centering
\includegraphics[width=0.6\textwidth]{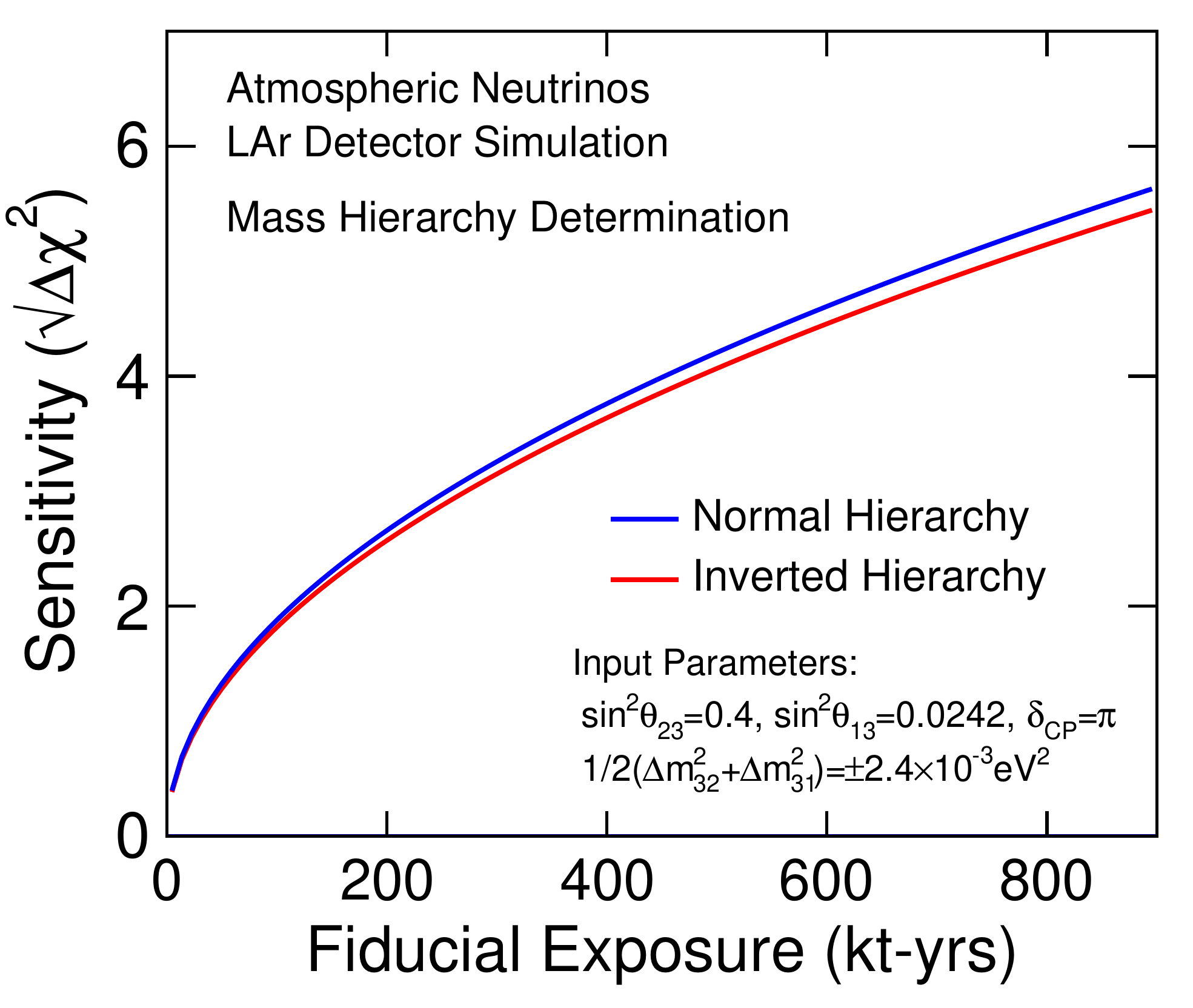}
\caption[Sensitivity to mass hierarchy using atmospheric neutrinos]{Sensitivity to mass hierarchy using atmospheric neutrinos as a function of fiducial
 exposure in a liquid argon detector.}
\label{fig:atmnsens2}
\end{figure}
Figure \ref{fig:atmnsens3} shows the octant and CPV sensitivity from a
340-\ktyr  
exposure of atmospheric neutrino data alone.  For the
determination of the octant of $\theta_{23}$, the $\overline{\Delta \chi^2}$
value is calculated between the best-fit points in the lower
($\theta_{23}<45^\circ$) and higher ($\theta_{23}>45^\circ$) octants,
where at each, the nuisance parameters have been marginalized.  The
discontinuities in the slopes of the octant sensitivity plot are real
features, indicating points at which the best fit moves from one
hierarchy to the other.  For the detection of CP violation, the
$\overline{\Delta \chi^2}$ exclusion is similarly computed for
$\mdeltacp=(0,\pi)$.
\begin{figure}[!htb]
\centerline{
\includegraphics[width=0.45\textwidth]{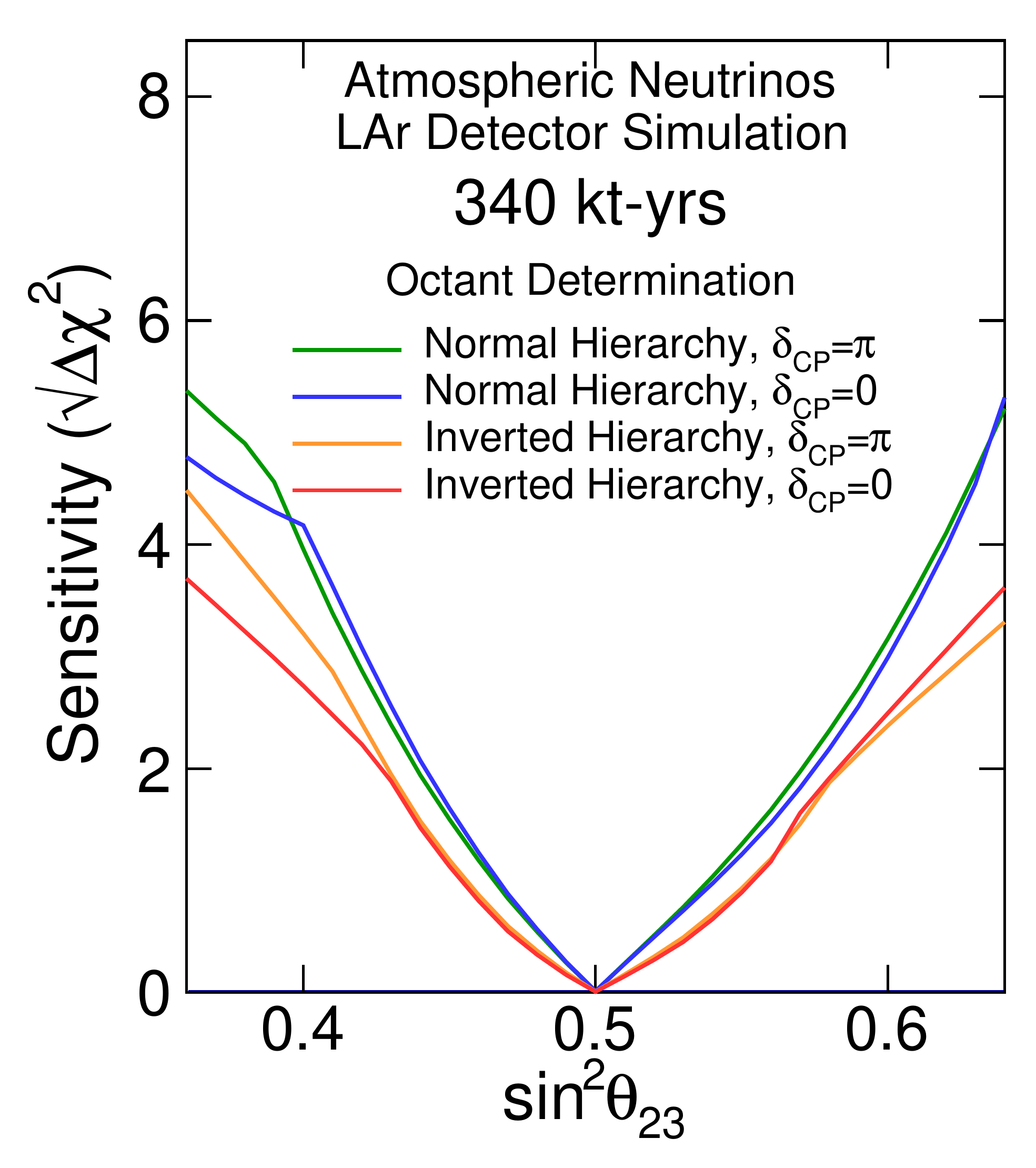}
\includegraphics[width=0.45\textwidth]{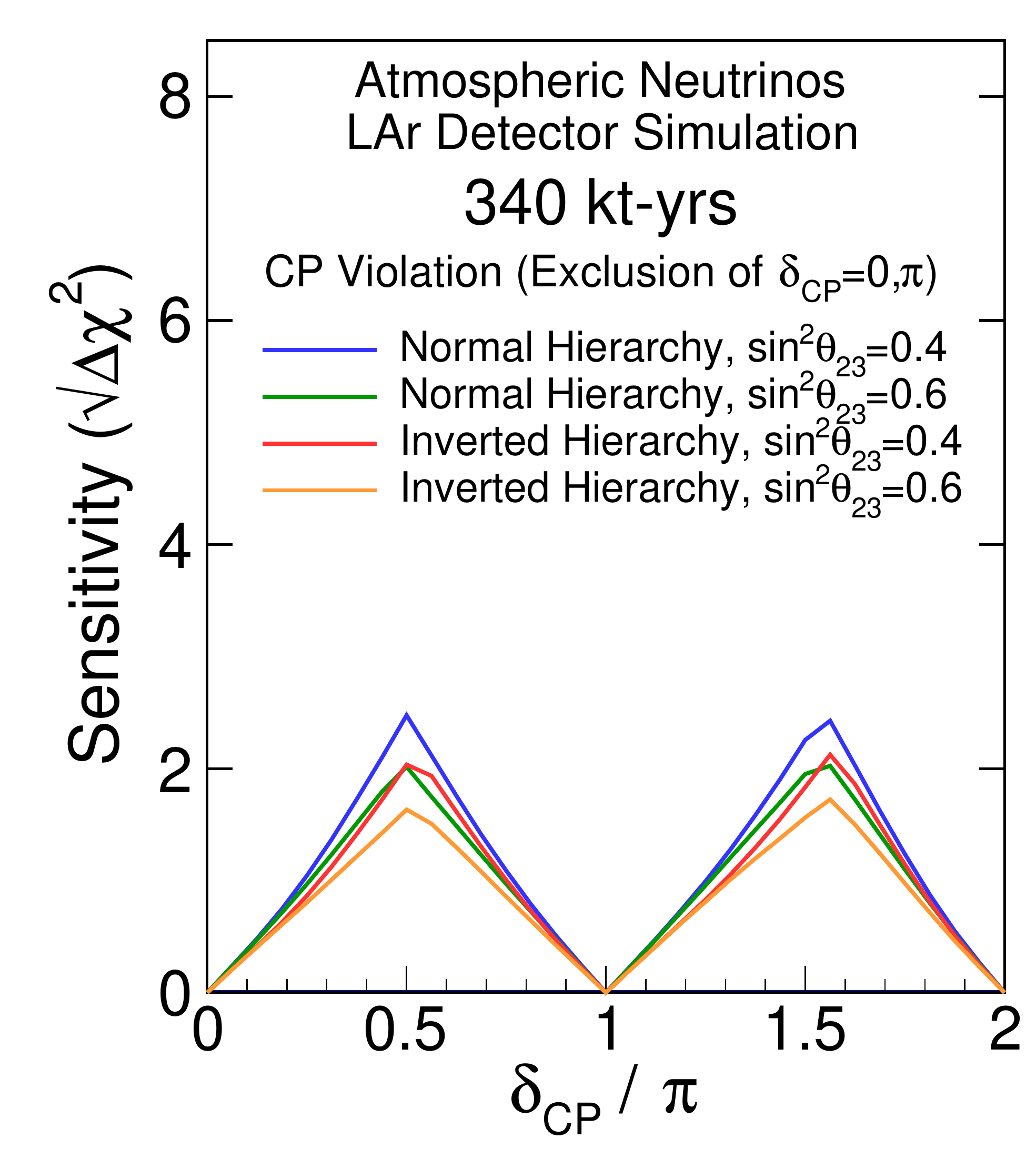}}
\caption[Sensitivity to $\theta_{23}$ octant and CPV using atmospheric neutrinos]{
Sensitivity to $\theta_{23}$ octant (left) and CPV (right) using atmospheric neutrinos.}
\label{fig:atmnsens3}
\end{figure}

Figure \ref{fig:atmnsens4} shows the combined sensitivity to beam 
and atmospheric neutrinos for determination of
the MH.  
This assumes a 10-year run with equal amounts of 
neutrino and antineutrino
running in a \MWadj{1.2} beam. 
\begin{figure}[!htb]
\centering
\includegraphics[width=0.7\textwidth]{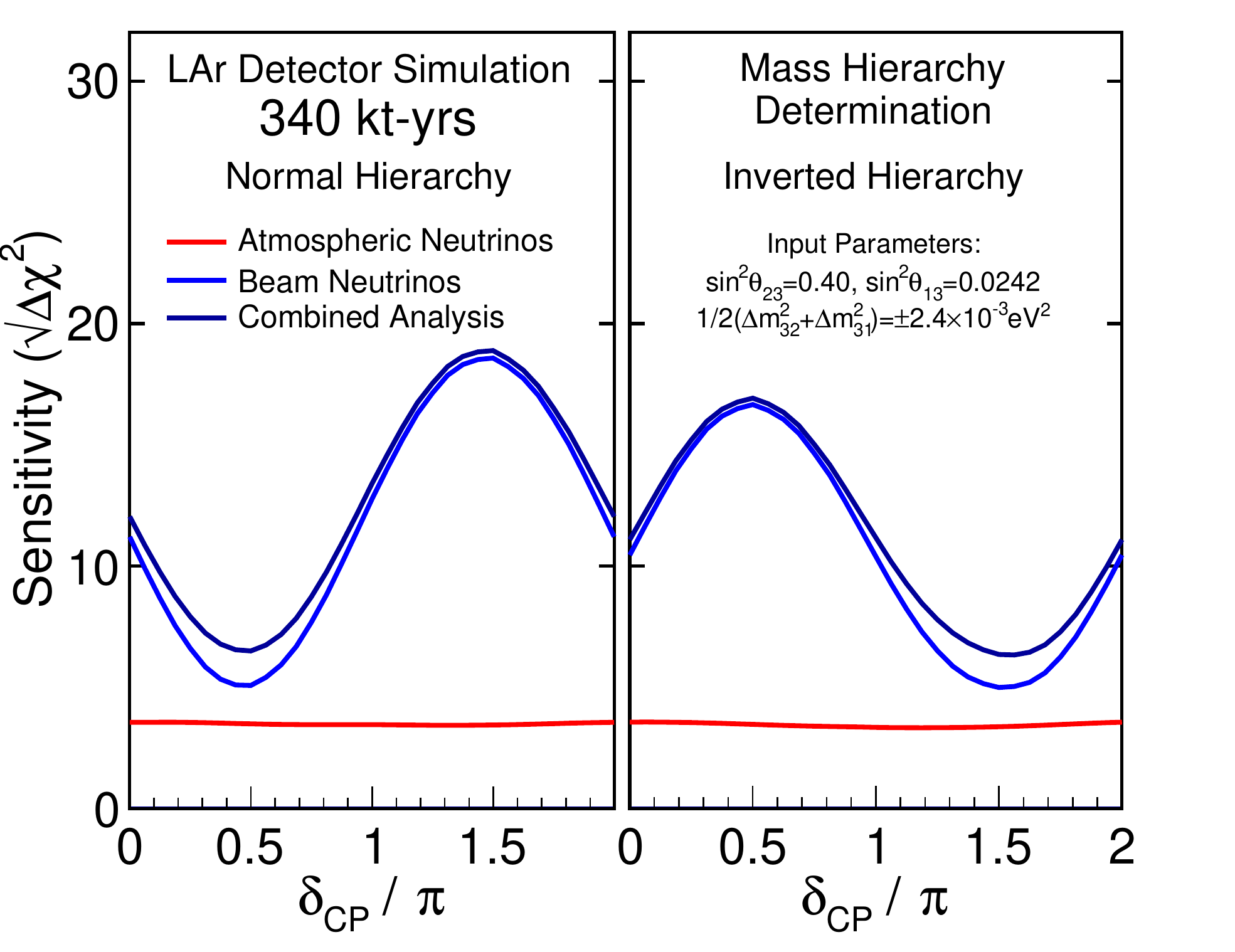}
\caption[Sensitivity to MH using atmospheric
and beam neutrinos as a function of $\mdeltacp/\pi$]{Sensitivity to mass hierarchy 
using atmospheric neutrinos combined with beam neutrinos with an
  exposure of 340 \ktyr in a \MWadj{1.2} beam for normal (left) and inverted (right) hierarchy.}
\label{fig:atmnsens4}
\end{figure}
\begin{figure}[!htb]
\centering
\includegraphics[width=0.65\textwidth]{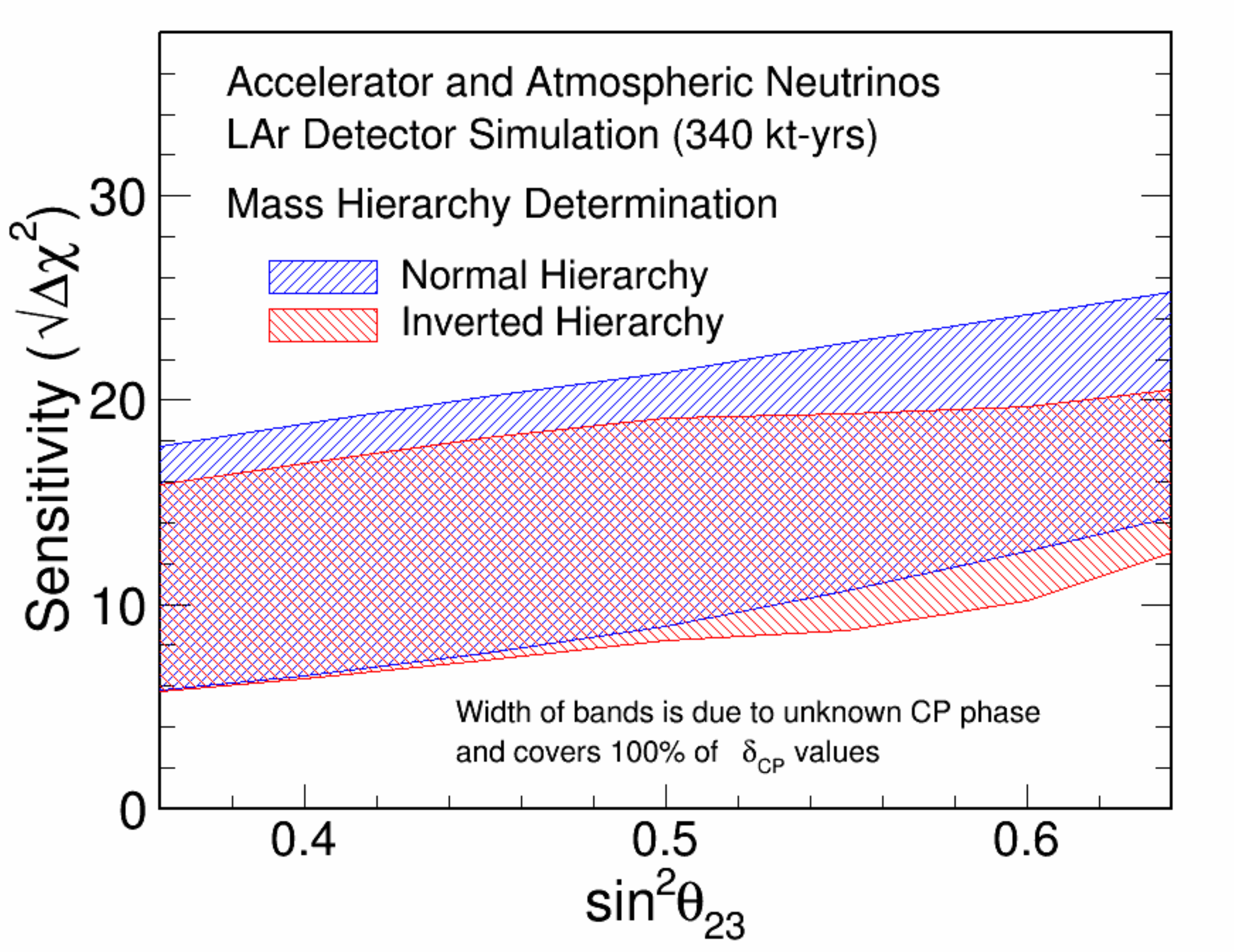}
\caption[Sensitivity to MH using atmospheric
 and beam neutrinos as a function of $\sin^2\theta_{23}$]{Sensitivity 
to mass hierarchy using atmospheric neutrinos combined with beam neutrinos 
as a function of the true value of $\sin^2\theta_{23}$, for true normal (blue)
and inverted (red) hierarchy. The width of the
  band is due to the unknown value of \deltacp and covers all possible values of
  \deltacp. Assumes an
  exposure of 340 \ktyr  in a \MWadj{1.2} beam.}
\label{fig:atm_varyth23_band}
\end{figure}
\begin{introbox}
  In the region of \deltacp where the LBNE neutrino-beam-only
  analysis is least sensitive to the mass hierarchy, atmospheric
  neutrinos measured in the same experiment offer comparable
  sensitivity. 
The combined beam and atmospheric neutrino 
 sensitivity to the mass hierarchy is $|\sqrt{\overline{\Delta \chi^2}}| > 6$ for
  all values of \deltacp ($\sin^2 \theta_{23} = 0.4$) in a
  \SIadj{34}{\kt} detector, assuming a \MWadj{1.2} beam running for ten 
  years.  It is important to note that the combined sensitivity is
  better than the sum of the separate $\Delta \chi^2$ values, as the
  atmospheric data help to remove degeneracies in the beam data.
\end{introbox}

Figure~\ref{fig:atm_varyth23_band} shows 
the combined sensitivity to beam and atmospheric neutrinos for
determination of MH as a function of the true value of 
$\sin^2\theta_{23}$, for the same 340-\ktyr exposure in a \MWadj{1.2} beam.
This can be compared to Figure~\ref{fig:mhvsth23} in 
Section~\ref{sect:cpmhsummary}, which shows the same sensitivity using
only beam neutrinos.
Figure~\ref{fig:atmnsens5} shows the combined sensitivity to beam and 
atmospheric neutrinos for 
the $\theta_{23}$ octant determination and CPV.   The role played by atmospheric 
data in resolving beam-neutrino
degeneracies is also clear from considering the combined and 
beam-only sensitivities in these plots.  
%

%
\begin{figure}[htb]
\centerline{
\includegraphics[width=0.45\textwidth]{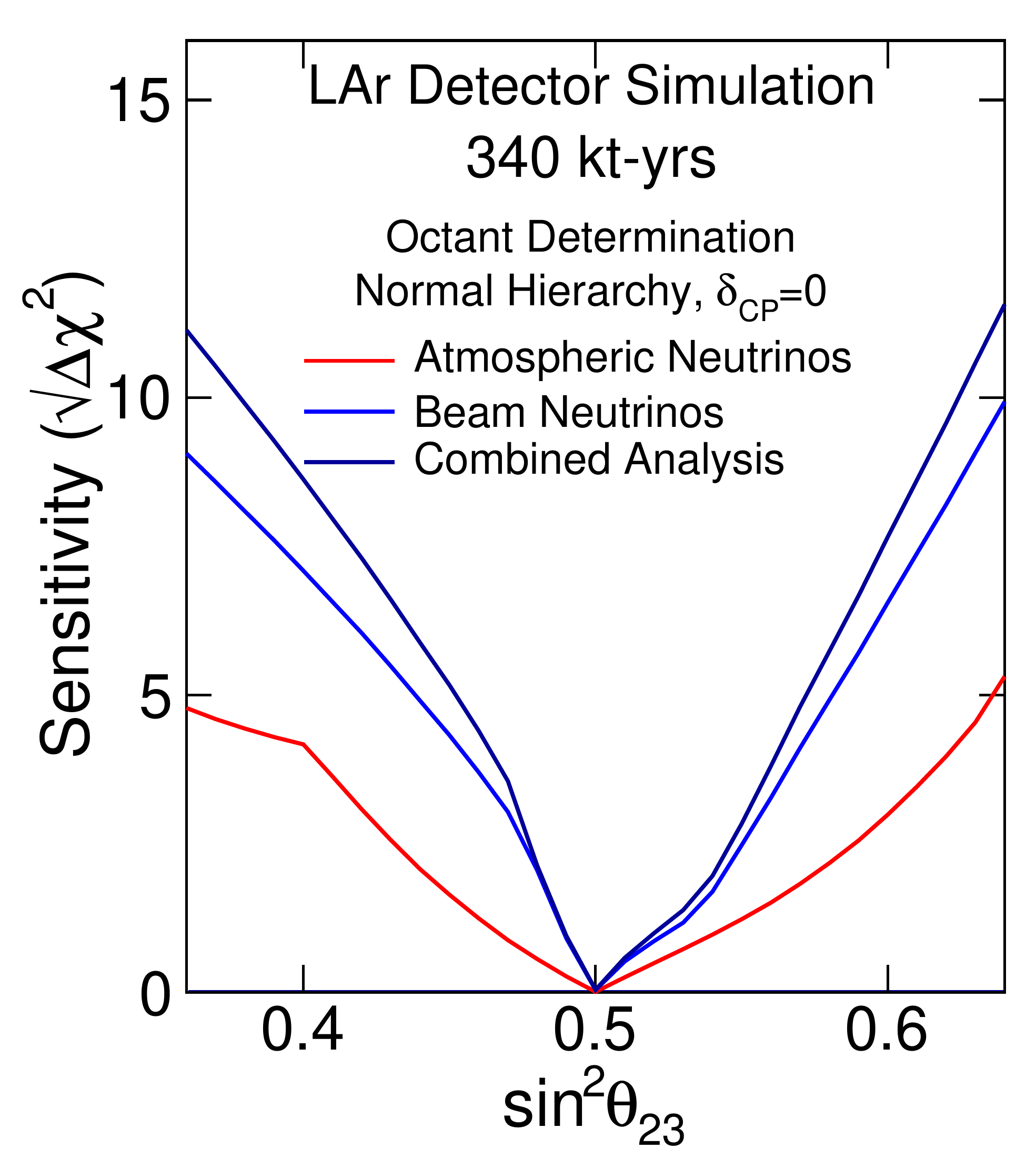}
\includegraphics[width=0.45\textwidth]{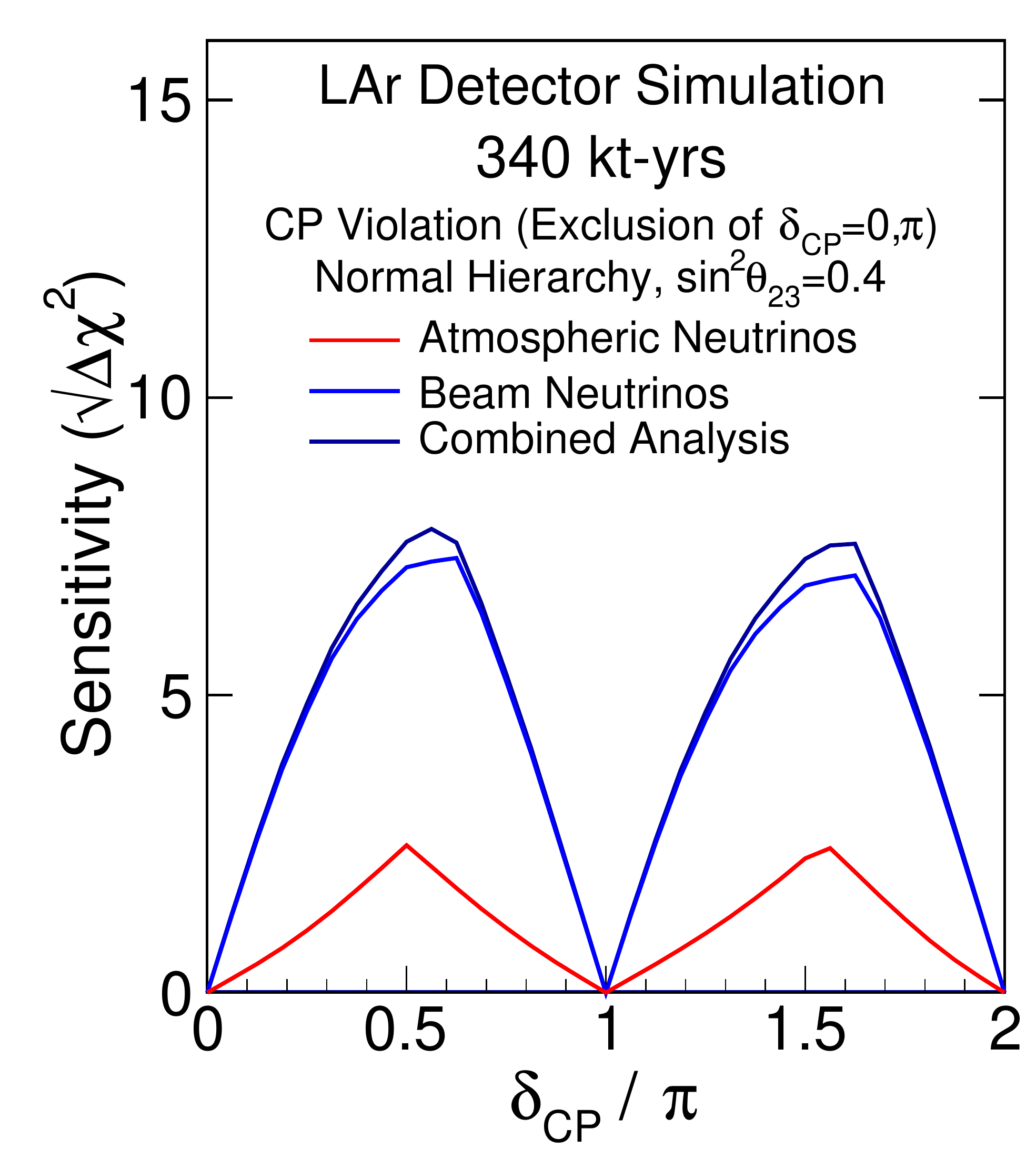} }
\caption[Sensitivity to $\theta_{23}$ octant and CPV using atmospheric and beam neutrinos]
{Sensitivity to $\theta_{23}$ octant (left) and CPV (right) using atmospheric neutrinos 
combined with beam neutrinos with an exposure of 340~\ktyr in a 
\MWadj{1.2} beam.}
\label{fig:atmnsens5}
\end{figure}
\section{Searches for Physics Beyond the Standard Three-Flavor Neutrino Oscillation Model}
\label{sec:new_physics}

\begin{introbox}
  Due to the very small masses and large mixing of neutrinos, their oscillations over a long distance
  act as an exquisitely precise interferometer with high sensitivity to very small perturbations caused by 
  new physics phenomena, such as:
  \begin{itemize}
  \item nonstandard interactions in matter that manifest in
    long-baseline oscillations as deviations from the three-flavor mixing model
  \item new long-distance potentials arising from discrete symmetries
    that manifest as small perturbations on neutrino and antineutrino
    oscillations over a long baseline
  \item sterile neutrino states that mix with the three known active neutrino states
  \item large compactified extra dimensions from String Theory models that manifest through mixing
    between the Kaluza-Klein states and the three active neutrino
    states
  \end{itemize}
  Full exploitation of LBNE's sensitivity to such new phenomena
  will require higher-precision predictions of the unoscillated
  neutrino flux at the far detector and large exposures. 
\end{introbox}


This section explores the potential of the full-scope LBNE design to
pursue physics beyond the three-flavor neutrino oscillation model.

\subsection{Search for Nonstandard Interactions}
\label{sec:nsi}

Neutral current (NC) nonstandard interactions (NSI) can be understood as nonstandard
matter effects that are visible only in a far detector at a
sufficiently long baseline. They can be parameterized as new contributions
to the MSW matrix in the neutrino-propagation Hamiltonian:

\begin{equation}
  H = U \left( \begin{array}{ccc}
           0 &                    & \\
             & \Delta m_{21}^2/2E & \\
             &                    & \Delta m_{31}^2/2E
         \end{array} \right) U^\dag + \tilde{V}_{\rm MSW} \,,
\end{equation}
with
\begin{equation}
  \tilde{V}_{\rm MSW} = \sqrt{2} G_F N_e
\left(
  \begin{array}{ccc}
    1 + \epsilon^m_{ee}       & \epsilon^m_{e\mu}       & \epsilon^m_{e\tau}  \\
        \epsilon^{m*}_{e\mu}  & \epsilon^m_{\mu\mu}     & \epsilon^m_{\mu\tau} \\
        \epsilon^{m*}_{e\tau} & \epsilon^{m*}_{\mu\tau} & \epsilon^m_{\tau\tau}
  \end{array} 
\right)
\end{equation}

Here, $U$ is the leptonic mixing matrix, and the $\epsilon$-parameters give the
magnitude of the NSI relative to standard weak interactions.  For new physics
scales of a few hundred GeV,  a value of $|\epsilon| \lesssim 0.01$ is
expected~\cite{Davidson:2003ha,GonzalezGarcia:2007ib,Biggio:2009nt}.
LBNE's \kmadj{1300} baseline provides an advantage in the detection of NSI relative
to existing beam-based experiments with shorter baselines.
Only atmospheric-neutrino experiments have longer baselines, but the sensitivity
of these experiments to NSI is limited by systematic effects. 

To assess the sensitivity of LBNE to NC NSI, the NSI discovery reach
is defined in the following way: the expected event spectra are
simulated using GLoBeS, assuming \emph{true} values for the NSI
parameters, and a fit is then attempted assuming no NSI. If the fit is
incompatible with the simulated data at a given confidence level,
the chosen \emph{true} values of the NSI parameters are considered to be
within the experimental discovery reach.  In Figure~\ref{fig:LAr-NSI},
the NSI discovery reach of LBNE is shown; only one of the
$\epsilon^m_{\alpha\beta}$ parameters at a time is taken to be
non-negligible.

\begin{figure}[!htb]
  \centering\includegraphics[width=0.8\textwidth]{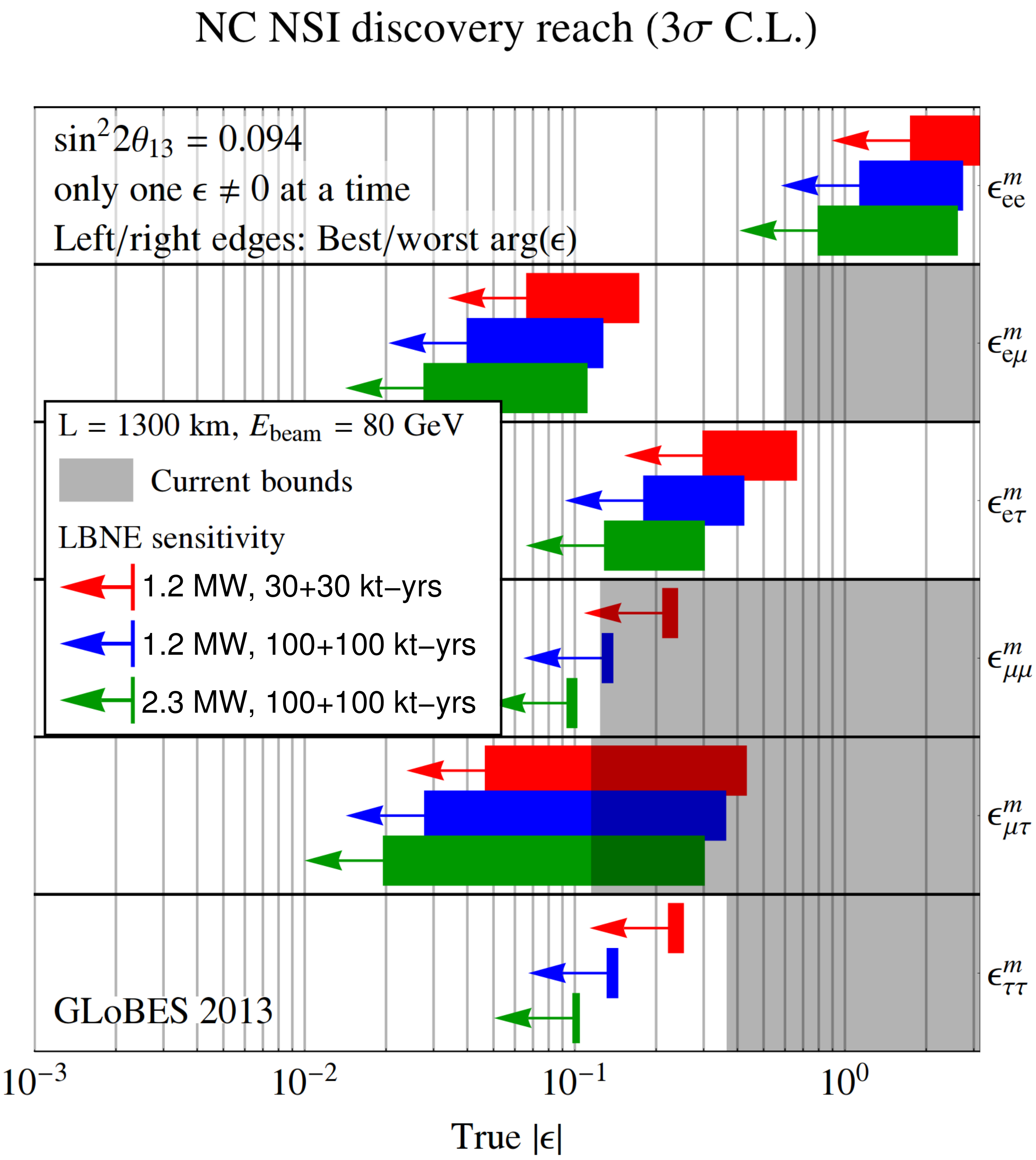}
  \caption[Sensitivity to nonstandard interactions]{Nonstandard
    interaction discovery reach in LBNE with increasing exposure: \SI{1.2}{\MW}, \SI{60}{\kt$\cdot$\year}s (red) 
+ \SI{1.2}{\MW}, 200~\ktyr (blue) + \SI{2.3}{\MW}, 200~\ktyr
    (green). The
    left and right edges of the error bars correspond to the most
    favorable and the most unfavorable values for the complex phase of
    the respective NSI parameters. The gray shaded regions indicate the current
    model-independent limits on the different parameters at 
    3$\sigma$~\cite{Davidson:2003ha,GonzalezGarcia:2007ib}. 
For this study the value of $\sin^ 2 2\theta_{13}$ was assumed to be 0.09. Figure courtesy of Joachim Kopp.}
 \label{fig:LAr-NSI}
\end{figure}

\subsection{Search for Long-Range Interactions}

The small scale of neutrino-mass differences implies that minute
differences in the interactions of neutrinos and antineutrinos with
currently unknown particles or forces may be detected through 
perturbations to the time evolution of the flavor eigenstates.  
The longer the experimental
baseline, the higher the sensitivity to a new long-distance potential
acting on neutrinos. For example, some of the models for such
long-range interactions (LRI) as described in~\cite{Davoudiasl:2011sz} 
(Figure~\ref{fig:lri}) could contain discrete symmetries that
stabilize the proton and give rise to a dark-matter candidate particle,
thus providing new
connections between neutrino, proton decay and dark matter
experiments. The longer baseline of LBNE improves the sensitivity to
LRI beyond that possible with the current generation of long-baseline
neutrino experiments. The sensitivity will be determined by the amount
of $\nu_\mu/\overline{\nu}_\mu$-CC statistics accumulated and the accuracy
with which the unoscillated and oscillated $\nu_\mu$ spectra can be
determined.

\begin{figure}[!htb]
\centering\includegraphics[width=\textwidth]{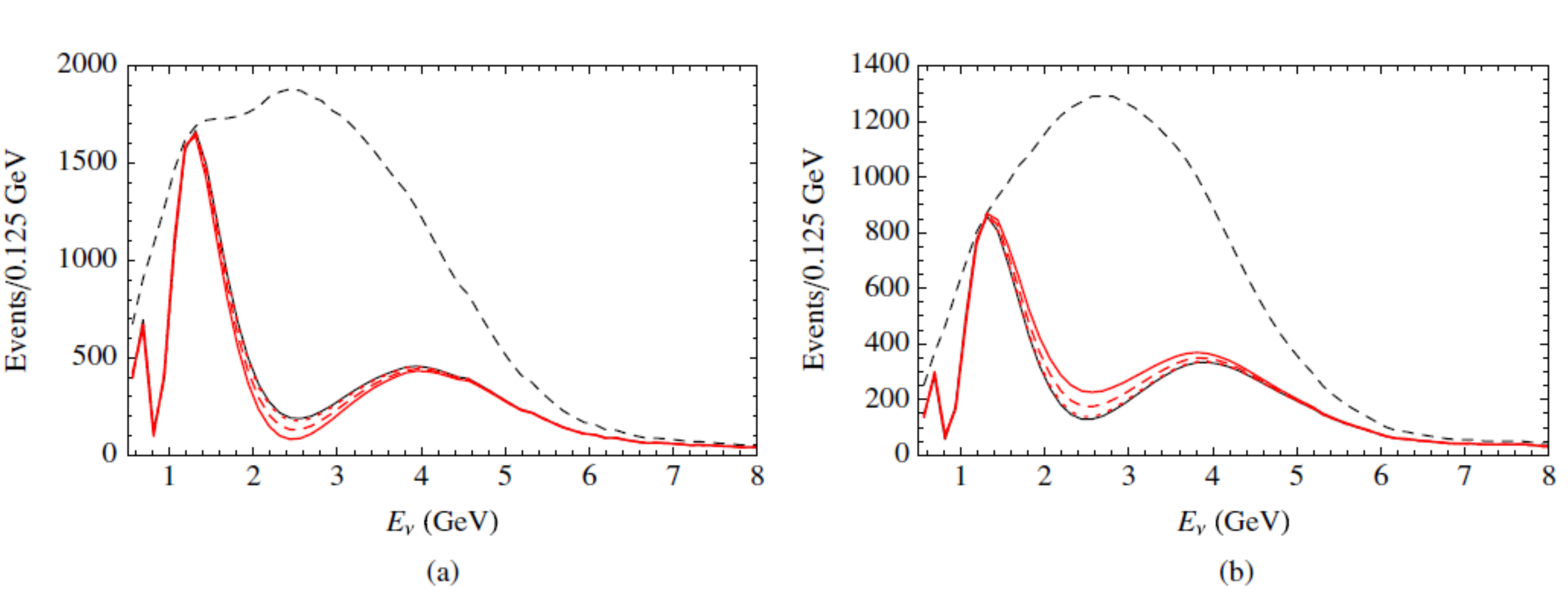}
\caption[Long-range interactions in LBNE]{Long-range interactions in LBNE. The number of (a) neutrino
  and (b) antineutrino events versus
  $E_{\nu}$, in a long-baseline
experiment with a \kmadj{1300} baseline. The unoscillated case (top black
dashed curves) and the case of no new physics
(thin black solid curves) are displayed, as well as the cases with
$\alpha'= (1.0, 0.5, \textrm{and }0.1)  \times 10^{-52}$, corresponding to 
red solid, dashed, and
dotted curves, respectively. $\alpha'$ is the \emph{fine structure
constant} of such interactions, which is constrained 
to be $\alpha' \leq 10^{-47}$~\cite{Davoudiasl:2011sz}.}
\label{fig:lri}
\end{figure}

\subsection{Search for Mixing between Active and Sterile Neutrinos}

Searches for evidence of active-sterile neutrino mixing at LBNE can be
conducted by examining the NC event rate at the far detector and
comparing it to a precise estimate of the expected rate extrapolated
from $\nu_\mu$ flux measurements from the  
near detector and from 
beam and detector simulations. Observed deficits in the NC rate could be evidence for mixing between the
active neutrino states and unknown sterile neutrino states. The most recent such search
in a long-baseline experiment was conducted by the MINOS
experiment~\cite{Adamson:2010wi}.

LBNE will provide a unique
opportunity to revisit this search with higher precision over a large
range of neutrino energies and a longer baseline. 
The expected rate of 
NC interactions with visible energy $>$ 0.5 GeV in a \ktadj{10} detector 
over three years is approximately 2,000 events
 (Table~\ref{tab:lbl_event_rates}) in the low-energy beam
tune and 3,000 events 
in the medium-energy beam tune. The NC
identification efficiency is high, with a low rate of $\nu_\mu$-CC
background misidentification as shown in
Table~\ref{tab:lar-nuosc-totaltable}.
The high-resolution
LArTPC far detector
will enable a coarse measurement of the incoming neutrino
energy in a NC interaction by using the event topology and correcting
for the missing energy of the invisible neutrino. This will greatly
improve the sensitivity of LBNE to active-sterile mixing as compared
to current long-baseline experiments such as MINOS+ since both the
energy spectrum and the rate of NC interactions can be measured
at both near and far detectors. Studies are currently underway to
quantify LBNE's sensitivity to active-sterile mixing.

\subsection{Search for Large Extra Dimensions}

Several theoretical models propose that right-handed neutrinos
propagate in large compactified extra dimensions, whereas the standard
left-handed neutrinos are confined to the four-dimensional 
brane~\cite{Machado:2011wx}. Mixing between the right-handed \emph{Kaluza-Klein} 
modes and the standard
neutrinos would change the mixing patterns predicted by
the three-flavor model. The effects could manifest, for example, as distortions in the
disappearance spectrum of $\nu_\mu$.  The rich oscillation
structure visible in LBNE, measured with 
its high-resolution detector
using both beam and atmospheric oscillations, 
could provide further opportunities to probe for
this type of new physics. 
Studies are underway to understand the
limits that LBNE
could impose
relative to current limits and those
expected 
from other experiments.

\section{Comparison of LBNE Sensitivities to other Proposed Experiments}

\begin{introbox}
  With tight control of systematics, LBNE will reach $5\sigma$
  sensitivity to CP violation for a large fraction of \deltacp
  values. LBNE delivers the best resolution of the value of \deltacp with
  the lowest combination of power-on-target and far detector mass when
  compared to other future proposed neutrino oscillation experiments
  (Figure~\ref{fig:cpvcomp}).
\end{introbox}

\begin{figure}[!htb]
  \centering\includegraphics[width=0.7\textwidth]{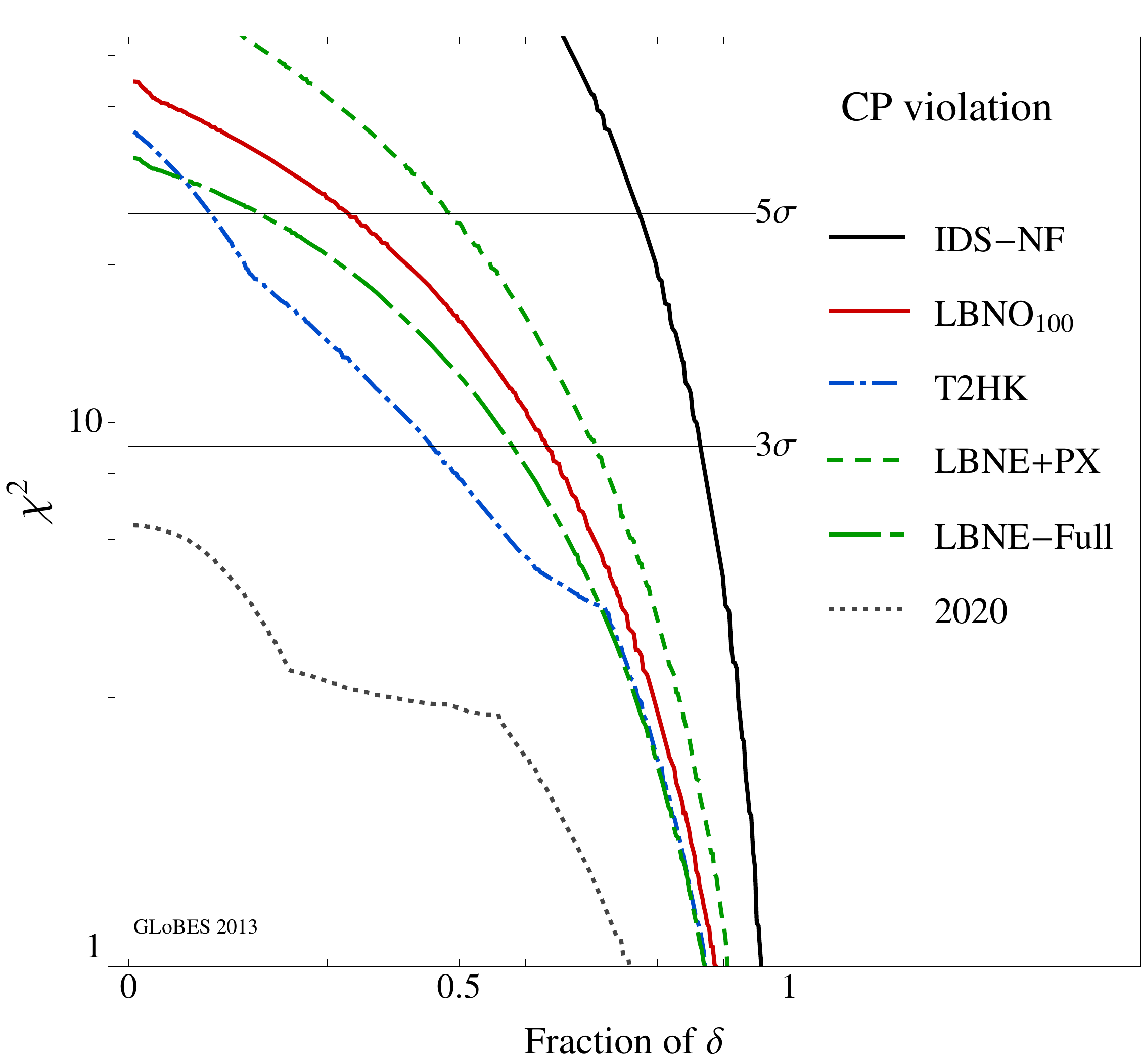}
  \caption[CPV in LBNE and other proposed experiments]
{The minimal CP-violation sensitivity for a given
    fraction of \deltacp values for different proposed neutrino
    oscillation experiments. The exposure and baseline of each experiment
    is described in the text. Figure is based on the studies detailed in~\cite{Coloma:2012ji}.}
\label{fig:cpvcomp}
\end{figure}
In Figure~\ref{fig:cpvcomp}, the CP-violation sensitivity of LBNE is
compared to that of other proposed neutrino oscillation experiments from an
\emph{independent study} with updated LBNE input based on~\cite{Coloma:2012ji}.  The
dashed black curve labeled ``2020'' is the expected sensitivity from
the current generation of experiments that could be achieved by 2020.
``LBNE-Full'' represents a \SIadj{34}{\kt} LArTPC running in a \MWadj{1.2} beam
for 3 ($\nu$) +3 ($\overline{\nu}$)  years. ``LBNE-PX''
is LBNE staged with PIP-II and further upgraded beams with power up to \SI{2.}{\MW} as shown in
Figure~\ref{fig:lar-cp-frac2}. ``T2HK'' is a \SIadj{560}{\kt} (fiducial mass) water Cherenkov
detector running in a \MWadj{1.66} beam for 1.5 ($\nu$) + 3.5 ($\overline{\nu}$)
years~\cite{Abe:2011ts}.  ``LBNO$_{\rm 100}$'' is a \SIadj{100}{\kt} LArTPC at a
baseline of 2,300$\,$km running in a \MWadj{0.8} beam from CERN for 5 ($\nu$)
+ 5 ($\overline{\nu}$)  years~\cite{Stahl:2012exa}. ``IDS-NF'' is the Neutrino
Factory with a neutrino beam generated from muon decays in a \GeVadj{10}
muon storage ring produced from a \MWadj{4}, \GeVadj{8} Project X proton beam
coupled with \SIadj{100}{\kt} magnetized iron
detectors at a baseline of \SI{2000}{\km} for 10 ($\nu + \overline{\nu}$)
simultaneously)~\cite{Apollonio:2012hga}.  LBNE can reach $5\sigma$
sensitivity to CP violation for a large fraction of \deltacp
values with the lowest combination of power-on-target and far
detector mass when compared to current and future proposed neutrino
oscillation experiments.

\begin{figure}[!htb]
  \includegraphics[width=\textwidth]{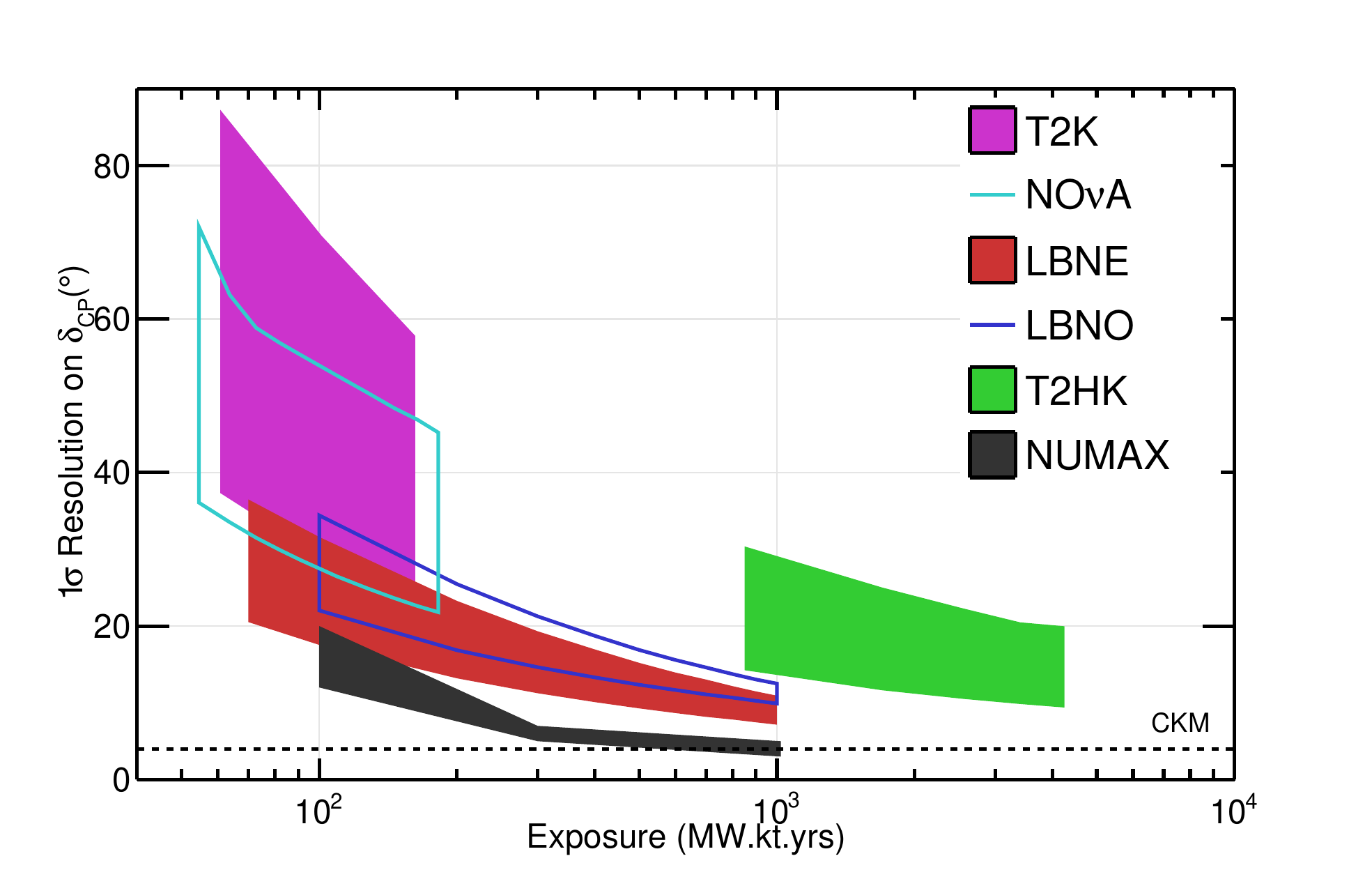}
  \caption[Resolution on \deltacp in LBNE and other
  experiments]{The 1$\sigma$ resolution on \deltacp that can be
    achieved by existing and proposed beamline neutrino oscillation
    experiments as a function of exposure in terms of mass $\times$
    beam power $\times$ years of running. The band represents the
    variation in the resolution as a function of \deltacp with
    the lower edge representing the best resolution and the upper edge
    the worst. The bands start and stop at particular milestones. For
    example, the LBNE band starts with the resolutions achieved by
    the \SIadj{10}{\kt} LBNE and ends with the full-scope LBNE running with the \MWadj{2.3}
    upgrades beyond PIP-II. With the exception of the NuMAX
    sensitivity, which is taken from \cite{Christensen:2013va}, the 
    resolutions in the colored bands are calculated independently by LBNE 
    using GLoBES. The dashed line
    denotes the $4^\circ$ resolution point which is the resolution of
    $\mdeltacp^{\rm CKM}$ from the 2011 global fits.}
\label{fig:snowmassdcpres}
\end{figure}
Alone, LBNE can potentially reach a precision on
\deltacp 
between roughly $6^\circ$ and $10^\circ$, i.e., close to the $4^\circ$
CKM precision on $\mdeltacp^{\rm CKM}$ --- but an exposure of 
$\sim$\SI[inter-unit-product=\ensuremath{{}\cdot{}}]{700}{\kt\MW\year}s
is needed. Nevertheless, as shown in
Figure~\ref{fig:snowmassdcpres}, wide-band, long-baseline experiments
such as LBNE (and LBNO) can achieve nearly CKM precision on
\deltacp with much less exposure than is required for existing
experiments such as NO$\nu$A, T2K and proposed 
short-baseline, off-axis experiments such as T2HK. With the exception
of the NuMAX sensitivity, which is taken from~\cite{Christensen:2013va}, the
resolutions in the colored bands in Figure~\ref{fig:snowmassdcpres}
are calculated independently by LBNE using GLoBES and found to be in
good agreement with the values reported by the experiments themselves 
(T2HK~\cite{Kearns:2013lea}, NO$\nu$A~\cite{Messier:2013sfa}, LBNO~\cite{Agarwalla:2013kaa}).

It is important to note that 
the precision on \deltacp in the off-axis experiments
shown in Figure~\ref{fig:snowmassdcpres} assumes the mass hierarchy (MH) 
is resolved. If the MH is unknown, the resolution of T2K,
NO$\nu$A and T2HK will be much poorer than indicated. LBNE does not
require external information on the MH to reach the
precisions described in this section.  Only a neutrino factory can
possibly out-perform a wide-band, long-baseline experiment --- but not by
much --- for equivalent power, target mass and years of running. 
To achieve this precision, however, LBNE will need to tightly
control the systematic uncertainties on the $\nu_e$ appearance
signal. Its high-resolution near detector will enable it to reach this
level of precision, as described in Section~\ref{sec:ndproj}.

An independent study comparing LBNE's sensitivity to the mass ordering
to that of current and future proposed experiments highlights its
potential~\cite{Blennow:2013oma}.  The study uses frequentist
methods of hypothesis testing to define sensitivities. The validity of
the approach is tested using toy MC simulations of the various
experiments. The comparison of expected MH sensitivities
for a variety of current and proposed experiments using different
approaches with reasonable estimates as to the start time of the
different experiments is summarized in Figure~\ref{fig:mhcomp}.
\begin{figure}[!htb]
   \centering\includegraphics[height=0.4\textheight]{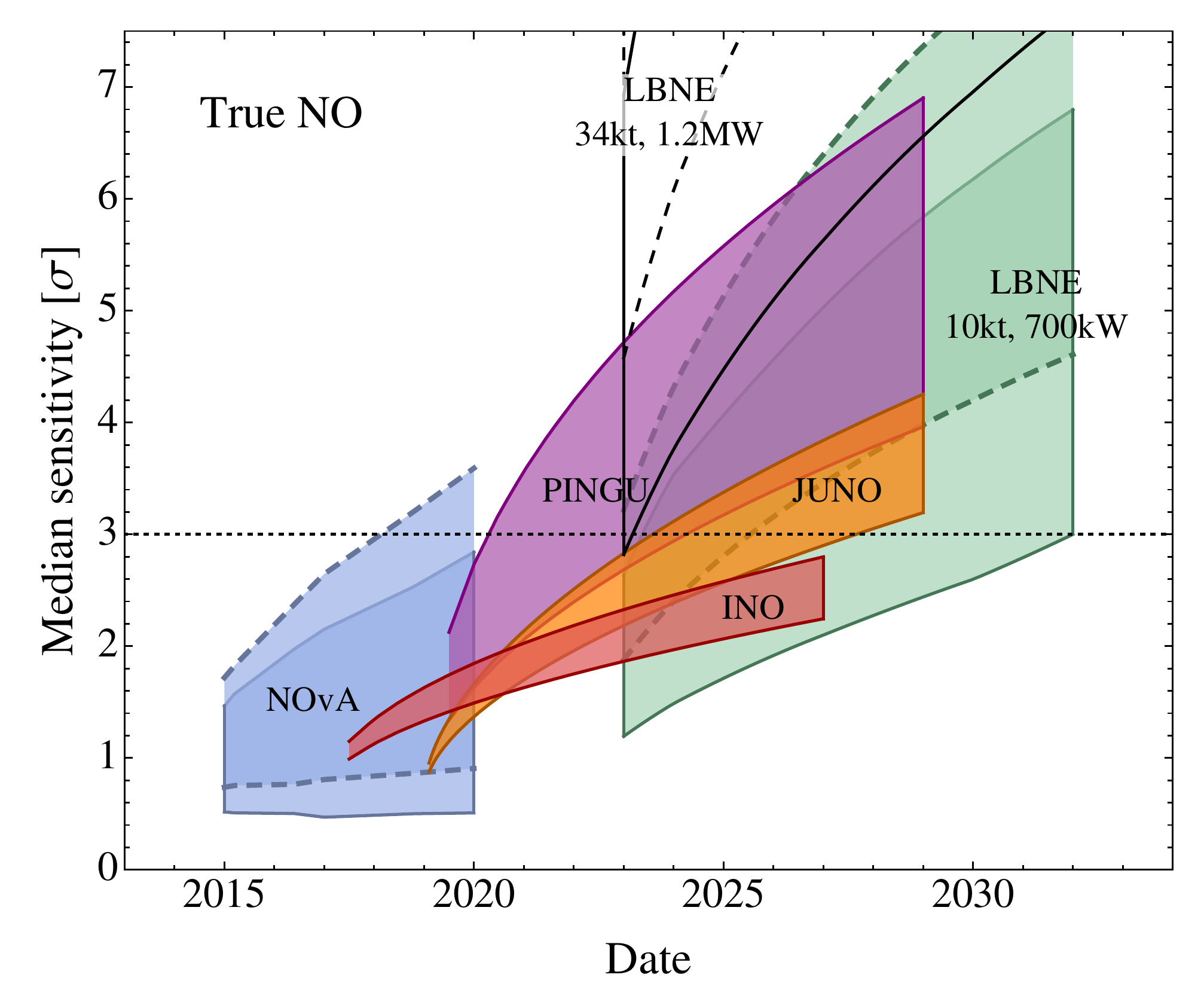}
   \centering\includegraphics[height=0.4\textheight]{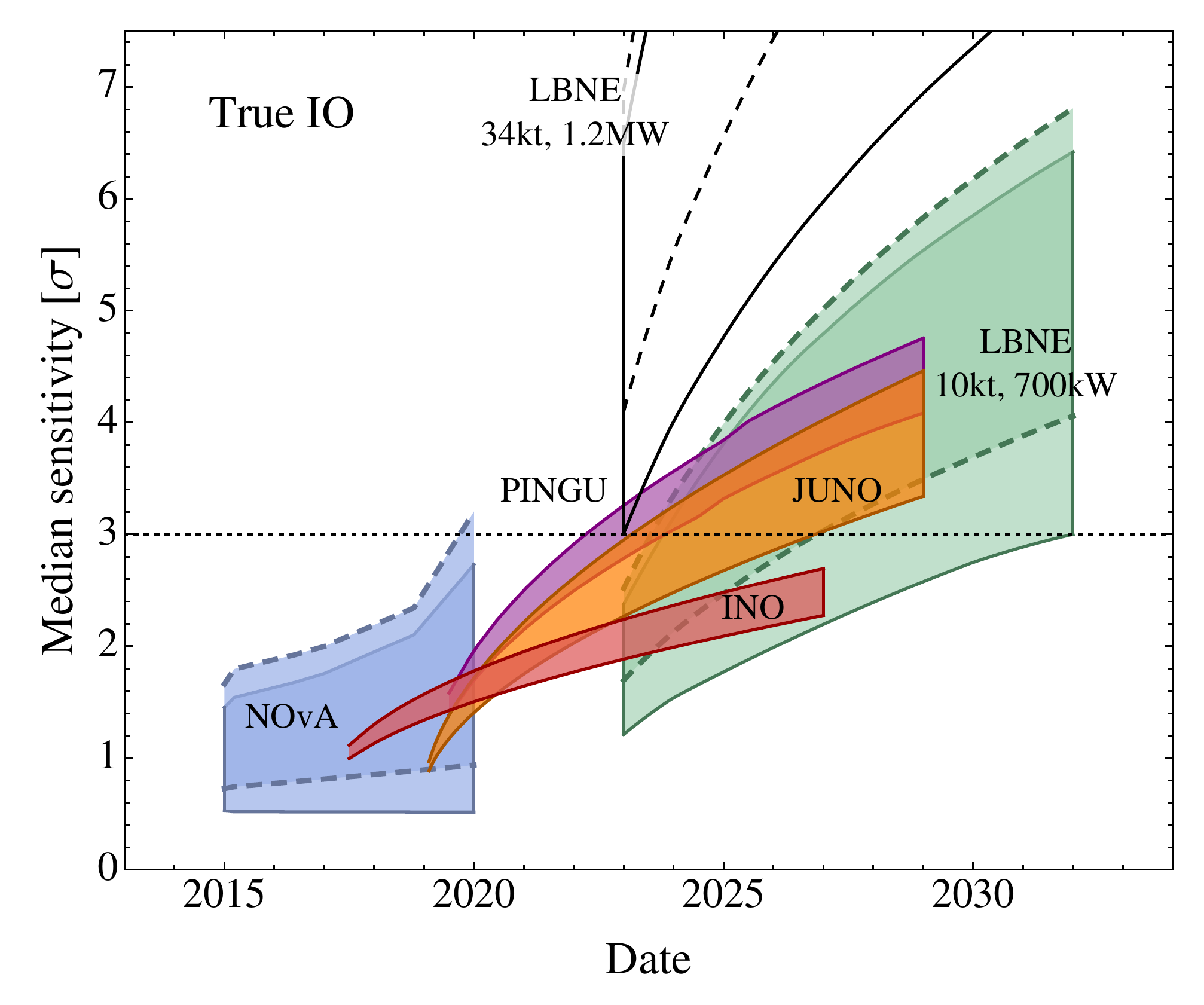}
  \caption[MH sensitivity in LBNE with time compared to
  other proposed experiments] 
    {The top (bottom) figure shows the 
    median sensitivity in number of sigmas 
    for rejecting the inverted (normal) hierarchy if the
    normal (inverted) hierarchy is true for different facilities as a 
    function of the
    date. The width of the bands corresponds to different true values
    of the CP phase \deltacp for NO$\nu$A and LBNE, different true values of
    $\theta_{23}$ between $40^\circ$ and $50^\circ$ for INO and PINGU, and 
    energy resolution between $3\%/\sqrt{E \ {\rm (MeV})}$  and 
    $3.5\%/\sqrt{E \ {\rm (MeV})}$ for JUNO. 
    For the long-baseline experiments,
    the bands with solid (dashed) contours correspond to a true value for
    $\theta_{23}$ of $40^\circ$ ($50^\circ$). 
    In all cases, octant degeneracies
    are fully considered.
    This figure is from the analysis presented in~\cite{Blennow:2013oma},
    however, for
    the plots shown here, the beam power for the full-scope, \ktadj{34} LBNE
    has been changed to
    \SI{1.2}{\MW} to reflect the Fermilab PIP-II upgrade plan.
}
\label{fig:mhcomp}
\end{figure}

Future upgrades to the Fermilab accelerator complex --- in particular
the prospect of high-power, low-energy proton beams such as the \MWadj{3}, \GeVadj{8} beam originally proposed as Stage 4 of Project X --- could open up
further unique opportunities for LBNE to probe CP violation using on-axis, low-energy
beams specifically directed at the second oscillation maximum where CP
effects dominate the asymmetries~\cite{Bishai:2013yqo}. Such high-power, low-energy beams could even enable studies in $\nu_1$-$\nu_2$ mixing in very
long-baseline experiments.

%% file: chapter-pdk.tex

\begin{introbox}
  Baryon number conservation is an unexplained symmetry in the
  Universe with deep connections to both cosmology and particle
  physics.  As one of the conditions underlying the observed
  matter-antimatter asymmetry of the Universe, baryon number
  \emph{should} be violated. Nucleon decay, which is a manifestation of 
  baryon number violation, is a hallmark of many Grand Unified
  Theories (GUTs), theories that connect quarks and leptons in ways
  not envisioned by the Standard Model. Observation of proton or
  bound-neutron decay would provide a clear experimental signature of
  baryon number violation.

  Predicted rates for nucleon decay based on GUTs are uncertain but
  cover a range directly accessible with the next generation of large
  underground detectors. LBNE, configured with its massive,
  deep-underground LArTPC far detector, offers unique opportunities
  for the discovery of nucleon decay, with sensitivity to key decay
  channels an order of magnitude beyond that of the current generation
  of experiments.
\end{introbox}

\section{LBNE and the Current Experimental Context}

Current limits on nucleon decay via numerous channels are dominated by
\superk\ (SK)~\cite{Raaf:2012pva}, for which the most recently
reported preliminary results are based on an overall exposure of
\SI{260}{\ktyr}.
Although the SK search has so far not observed nucleon decay, it
has established strict limits ($90\%$ CL) on the partial lifetimes for
decay modes of particular interest to GUT models such as $\tau/B(p\to
e^+\pi^0) > 1.3\times 10^{\mathrm{34}}\,$year and $\tau/B(p\to
K^+\overline{\nu}) > 0.59\times
10^{\mathrm{34}}\,$year~\cite{kearns_isoups}.  These are significant
limits on theoretical models that constrain model builders and set a
high threshold for the next-generation detectors such as LBNE and
Hyper-Kamiokande (Hyper-K). After more than ten years of exposure, the SK limits
will improve only slowly. A much more massive detector such as
Hyper-K --- which will have a \ktadj{560} fiducial mass 
--- is required to make a significant (order-of-magnitude) improvement 
using the water Cherenkov technique.

The uniqueness of proton decay signatures in a
LArTPC and the potential for reconstructing them with redundant
information has long been recognized as a key strength of this
technology. A LArTPC can reconstruct all final-state charged
particles and make an accurate assessment of particle type,
distinguishing between muons, pions, kaons and
protons. Electromagnetic showers are readily measured, and those that
originate from photons generated by $\pi^0$ decay can be distinguished
to a significant degree from those that originate from $\nu_e$ charged-current (CC)
 interactions.  Kiloton-per-kiloton, LArTPC technology
is expected to outperform water Cherenkov in both detection efficiency
and atmospheric-neutrino background rejection for most nucleon decay
modes, although intranuclear effects, which can smear out some of the
proton decay signal, are smaller for oxygen and nonexistent for
hydrogen.

When mass and cost are taken into account, water Cherenkov technology
is optimum for the $p\to e^+\pi^0$ final-state topology, where the
signal efficiency is roughly 40\% and the background rate is two events
per \SI{}{\Mtyr}.
The efficiency estimate
for this mode~\cite{Bueno:2007um} for a 
LArTPC is 45\% with one event per  \SI{}{\Mtyr} --- not a significant
enough improvement in efficiency to overcome the penalty of the higher
cost per kiloton for liquid argon.

For the $p \rightarrow K^+ \overline{\nu}$ channel, on the other hand, the
LArTPC technology is superior based on the same criteria.  In a
LArTPC, the $K^+$ track is reconstructed and identified as a charged
kaon. The efficiency for the $K^+ \overline{\nu}$ mode in a LArTPC is estimated
to be as high as 97.5\% with a background rate of one event per 
 \SI{}{\Mtyr}. In water Cherenkov detectors the efficiency for this mode is roughly
19\% for a low-background search, with a background rate of four events
per  \SI{}{\Mtyr}.
Based on these numbers and a ten-year exposure, LBNE's 
\ktadj{34} LArTPC and the \ktadj{560} Hyper-K WCD have comparable
sensitivity (at 90\% CL), but the estimated LArTPC background of 0.3
events is dramatically better than the 22 estimated for Hyper-K 
(assuming no further improvement in analysis technique past that
currently executed for SK~\cite{kearns_isoups}).
 
\section{Signatures for Nucleon Decay in Liquid Argon}

\begin{introbox}
  The LBNE LArTPC's superior detection efficiencies for decay modes
  that produce kaons will outweigh its relatively low mass compared
  with multi-hundred-kiloton water Cherenkov detectors.  Because the
  LArTPC can reconstruct protons that are below
  Cherenkov threshold, it can reject many atmospheric-neutrino
  background topologies by vetoing on the presence of a recoil proton.
  Due to its excellent spatial resolution, it also performs better for event
  topologies with displaced vertices, such as $p \rightarrow K^+
  \overline{\nu}$ (for multi-particle $K^+$ decay topologies) 
  and $p \rightarrow K^0 \mu^+$.  The latter mode is
  preferred in some SUSY GUTs.
\end{introbox}
For modes with no
electron in the final state, the same displaced vertex performance
that underpins long-baseline neutrino oscillation measurements allows
the rejection of CC interactions of atmospheric $\nu_e$'s.
As will be stressed for the key mode of $p \rightarrow K^+ \overline{\nu}$
described in detail below, the capability to reconstruct the charged
kaon with the proper range and $dE/dx$ profile allows for a high-efficiency,
background-free analysis.  In general, these criteria favor all
modes with a kaon, charged or neutral, in the final state. Conversely,
the efficiency for decay modes to a lepton plus light meson will be
limited by intranuclear reactions that plague liquid argon to a greater extent
than they do $^{\mathrm{16}}$O in a water Cherenkov detector.

An extensive survey~\cite{Bueno:2007um} of nucleon decay efficiency 
and background rates for large LArTPCs with various depth/overburden 
conditions, published in 2007, provides the starting point for the 
assessment of LBNE's capabilities.  Table~\ref{tab:pdecay} lists selected
modes where LArTPC technology exhibits a significant performance 
advantage (per kiloton) over the water Cherenkov technology.
The remainder of this chapter focuses on the capabilities 
of LBNE for the $p\to K^+\overline{\nu}$ channel, as the most 
promising from theoretical and experimental 
considerations.  Much of the discussion that follows can be 
applied to cover the other channels with kaons listed in 
the table.
\begin{table}[!htbp]
\caption[Efficiencies and background rates for nucleon decay modes]
        {Efficiencies and background rates (events per \SI{}{\Mtyr}) for nucleon decay 
         channels of interest for a large underground LArTPC~\cite{Bueno:2007um}, and 
         comparison with water Cherenkov detector capabilities.  
         The entries for the water Cherenkov capabilities are based 
         on experience with the \superk{} detector~\cite{kearns_isoups}.  
        }
\begin{center}
\begin{tabular}{$L^c^c^c^c} 
\toprule
\rowtitlestyle
Decay Mode   & \multicolumn{2}{^>{\columncolor{\ChapterBubbleColor}}c}{Water Cherenkov} & 
\multicolumn{2}{^>{\columncolor{\ChapterBubbleColor}}c}{Liquid Argon TPC} \\
\rowtitlestyle
   & Efficiency &   Background & Efficiency &   Background \\ \toprowrule
$p \rightarrow K^+ \overline{\nu}$       & 19\%  &  4   &  97\%   &     1  \\ \colhline
$p \rightarrow K^0 \mu^+$      & 10\%  &  8   &  47\%   &  $<2 $ \\ \colhline
$p \rightarrow K^+ \mu^- \pi^+$ &       &      &  97\%   &     1  \\ \colhline
$n \rightarrow K^+ e^- $        & 10\%  &  3   &  96\%   &  $<2$  \\ \colhline
$n \rightarrow e^+\pi^-$      & 19\%  &  2   &  44\%   &  0.8   \\
\bottomrule
\end{tabular}
\end{center}
\label{tab:pdecay}
\end{table}

The key signature for $p\to K^+\overline{\nu}$ is the presence of an
isolated charged kaon (which would also be monochromatic 
for the case of free protons, with $p=$\SI{340}{\MeV}).  
Unlike the case of $p\to e^+\pi^0$, where the maximum
detection efficiency is limited to 40--45\% because of inelastic
intranuclear scattering of the $\pi^0$, the kaon in $p\to
K^+\overline{\nu}$ emerges intact (because the kaon momentum is 
below threshold for inelastic reactions)
from the nuclear environment of the decaying proton $\sim 97\%$ of the
time.  Nuclear effects come into play in other ways, however: the kaon
momentum is smeared by the proton's Fermi motion and shifted downward
by re-scattering~\cite{Stefan:2008zi}. 
The kaon emerging from this process is below Cherenkov threshold,
therefore a water detector would need to detect it after it stops, via its
decay products.  Not all $K$ decay modes are reconstructable, however,
and even for those that are, insufficient information exists to
determine the initial $K$ momentum.  Still, water detectors can
reconstruct significant hadronic channels such as $K^+\to\pi^+\pi^0$
decay, and the \MeVadj{6} gamma from de-excitation of $O^{16}$ provides an
added signature to help with the $K^+\to\mu^+\nu$ channel. The overall
detection efficiency in SK~\cite{kearns_isoups} thus approaches
$20\%$.

\begin{figure}[!htb]
\centering
\includegraphics[width=0.72\textwidth]{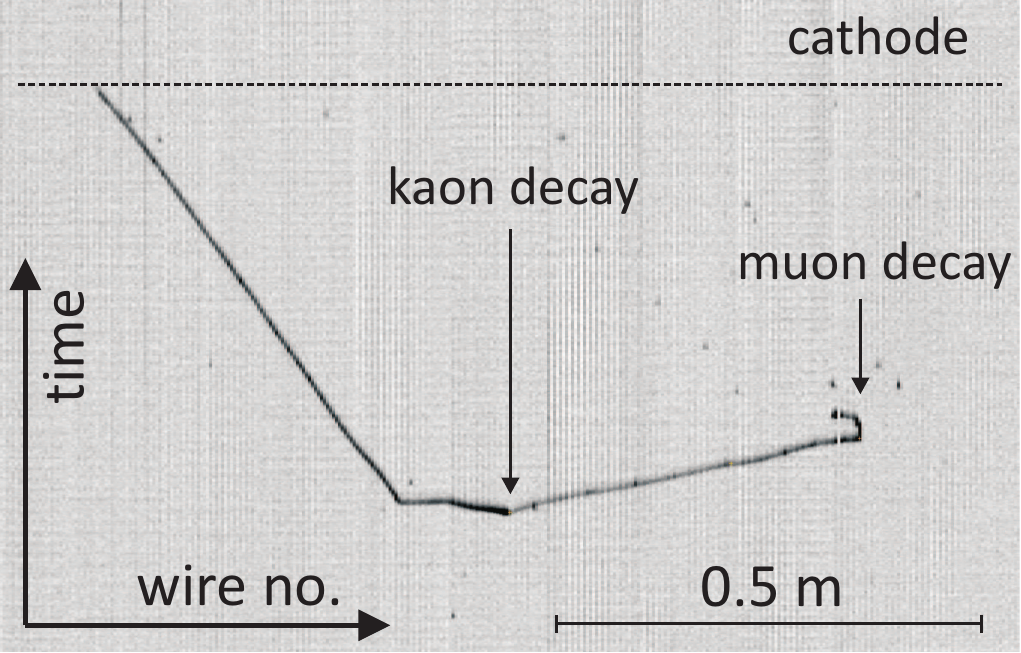}
\includegraphics[width=0.72\textwidth]{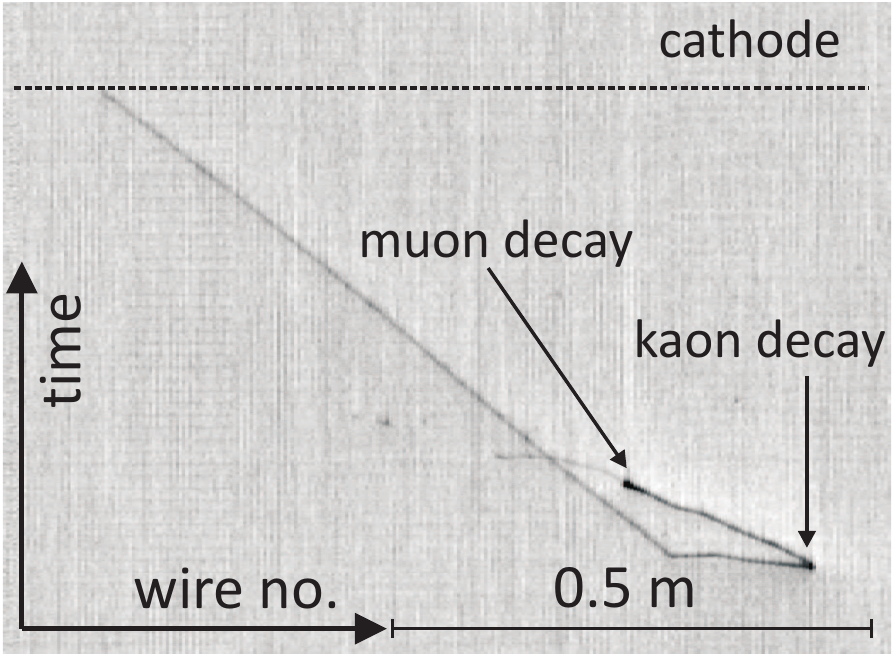}
\caption[Decaying kaon observed during the ICARUS run at CNGS]
{Event display for a decaying kaon candidate $K \rightarrow \mu \nu_\mu \ \mu \rightarrow e \nu_e \nu_\mu$ 
in the ICARUS T600 detector observed
in the CNGS data ($K$: \SI{90}{\cm}, \SI{325}{\MeV}; $\mu$ : \SI{54}{\cm}, \SI{147}{\MeV}; 
$e$ : \SI{13}{\cm}, \SI{27}{\MeV}). The top figure shows the signal on the collection plane,
  and the bottom figure shows the signal on the second induction plane~\cite{Antonello:2012hu}.}
\label{fig:icaruskaon}
\end{figure}

\begin{figure}[!htb]
\includegraphics[width=0.7\textwidth]{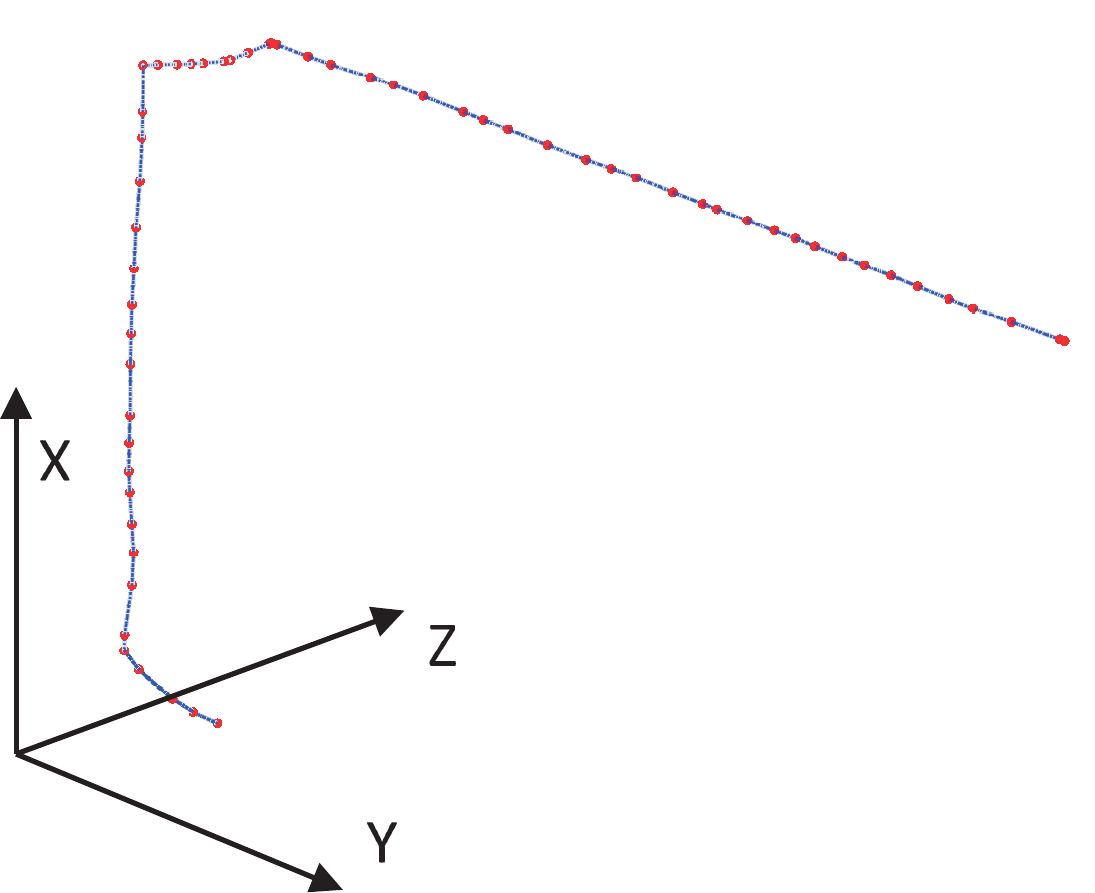}
\caption[3D construction of decaying kaon in the ICARUS detector]{3D reconstruction of the decaying kaon event observed in the ICARUS T600 detector and shown in Figure~\ref{fig:icaruskaon}.}
\label{fig:icarusk3d}
\end{figure}

In LArTPC detectors, the $K^+$ can be tracked, its momentum measured
by range, and its identity positively resolved via detailed analysis
of its energy-loss profile.  Additionally, all decay modes can be
cleanly reconstructed and identified, including those with neutrinos,
since the decaying proton is essentially at rest.  With this level of
detail, it is possible for a single event to provide overwhelming
evidence for the appearance of an isolated kaon of the right momentum
originating from a point within the fiducial volume.  The strength of
this signature is clear from cosmogenic-induced kaons observed by the
ICARUS Collaboration in the cosmic-ray (CR) test run of half of the T600
detector, performed at a surface installation in Pavia~\cite{Amerio:2004ze} 
and in high-energy neutrino interactions with the full T600 in the recent 
CNGS (CERN Neutrinos to Gran Sasso) run~\cite{Antonello:2012hu}.
Figure~\ref{fig:icaruskaon} shows a sample event from the CNGS run in
which the kaon is observed as a progressively heavily-ionizing track 
that crosses into the active liquid argon volume, stops, and
decays to $\mu\nu$, producing a muon track that also stops and decays
such that the Michel-electron track is also visible. The 3D
reconstruction of the event is shown in Figure~\ref{fig:icarusk3d}.

If it can be demonstrated that background processes mimicking this
signature can be rejected at the appropriate level, 
a single $p\to K^+\overline{\nu}$ candidate could constitute 
evidence for proton decay. 

\section{Background Levels and Rejection Capabilities}
\label{sec:pdk:background-rej}

This section discusses the key background processes and 
their signatures, focusing on the $p\to K^+\overline{\nu}$ 
channel as 
the benchmark mode\footnote{Much of this discussion applies 
equally well to other nucleon decay modes involving charged 
or neutral kaons.}.  The two potential sources 
of background are cosmic-ray muons and atmospheric 
neutrinos, described separately below.  

\subsection{Cosmic-Ray Muon Backgrounds}

Cosmic-ray (CR) muons contribute background signals when they
penetrate the detector.  Hence, the self-shielding feature of 
the LArTPC and the depth of the site are important assets for 
controlling the rate of signals that can mimic a proton decay 
event.  Additionally, the energy deposition associated 
with spallation products is well below the hundreds-of-MeV 
range for depositions from proton decay final-state particles.

The most pernicious CR-muon background in liquid argon for proton decay with kaon 
final states thus comes from particular pathological processes.  
Specifically, CR muons that
produce kaons via photonuclear interactions in the rock near
the detector or in the liquid argon itself but outside the active volume are 
capable of producing signatures that mimic $p\to K^+\overline{\nu}$ 
and other modes with kaons.  CR-induced kaon backgrounds as 
a function of depth have been studied for
liquid argon~\cite{Bueno:2007um,Bernstein:2009ms,DOCDB5904}.

In particular, at the 4,850-ft level, the vertical rock overburden will be
approximately 4-km water equivalent, at which depth the muon rate
through a \ktadj{34} LArTPC will be approximately \SI{0.1}{s}$^{-1}$. This is
low enough that a veto on the detection of a muon in the liquid argon volume can
be applied with negligible loss of live-time.  Specifically, assuming
a maximum drift time of \SI{2}{\ms}, the probability of a muon passing
through the detector in time with any candidate event (i.e., a
candidate for proton decay or other signal of interest) will be $2
\times 10^{-4}$.  Thus, any candidate event that coincides in time
with a large energy deposition from a muon or muon-induced cascade can
be rejected with a negligible signal efficiency loss of 0.02\%.
Only background from events associated with CR muons in which
the muon itself does not cross the active region of the detector 
remain to be considered.

One class of such backgrounds involves production of a charged kaon 
outside the active volume, which then enters the active region.  
Assuming unambiguous determination of the drift time (via the 
scintillation-photon detection system and other cues such as 
detailed analysis of the $dE/dx$ profile of the kaon candidate), 
it will be possible to identify and reject such entering kaons 
with high efficiency.  It should be noted that, through studies 
of CR muons that interact within the active volume of the 
detector, backgrounds of this type can be well characterized 
with data from the detector itself.

A potentially less tractable background 
for the decay mode $p^+ \rightarrow K^+\overline{\nu}$
occurs when a neutral particle (e.g., a $K^0_L$) originating in a
muon-induced cascade outside the detector propagates into the detector
volume and undergoes a charge-exchange reaction in the fiducial
volume.  To further understand the possible rate for this background
at LBNE, simulations of CR muons and their secondaries at
depth have been run.  The rate of positive kaons produced inside the
\ktadj{34} detector by a neutral particle entering from outside (and with
no muon inside) has been found to be 0.9 events per year before any other selection 
criteria are applied. Further studies included the following additional
selection criteria:
\begin{enumerate}
\item No muon is in the detector active volume.
\item The $K^+$ candidate is produced 
      inside the liquid argon active volume 
      at a distance from the wall greater than \SI{10}{\cm}. 
\item The energy deposition from the $K^+$ and its descendants 
        (excluding decay products) is less than \SI{150}{\MeV}. 
\item The total energy deposition from the $K^+$, its descendants
      and decay products is less than \SI{1}{\GeV}.
\item Energy deposition from other particles in the muon-induced 
      cascade (i.e., excluding the energy deposition
      from the positive kaon, its descendants and decay products) 
      is less than \SI{100}{\MeV}. 
\end{enumerate}

No event survived the additional selection criteria, resulting in an
upper bound on the rate of this type of background event of 0.07
events per year in a \ktadj{34} LArTPC, equivalent to two events per 
 \SI{}{\Mtyr}.  A key factor contributing to the rejection of 
CR backgrounds to this level is that although a large number 
of $K^+$'s generated by cosmic rays deposit an 
energy similar to that expected from proton decay, the
energy depositions from $K^+$'s are not the only ones recorded for
these events.  Other particles from the CR-muon interaction 
tend also to enter the detector and deposit additional visible
energy, making the rejection of background events simpler than
would be expected assuming only the appearance of a kaon in the detector.

In addition to the impact of an active veto system for detectors at 
various depths, the studies of~\cite{Bueno:2007um} also consider 
impacts of progressively restrictive fiducial volume cuts. 
Together, these and the above studies demonstrate that proton decay 
searches in the LBNE LArTPC at the 4,850-ft level can be made immune 
to CR-muon backgrounds, without the requirement of an external 
active veto system.  To the extent that there are uncertainties on 
the rate of kaon production in CR-muon interactions, one has flexibility 
to suppress background from this source further by application of modest 
fiducial volume cuts.

\subsection{Background from Atmospheric-Neutrino Interactions}

Unlike the case of CR-muon backgrounds, the contamination of 
a nucleon decay candidate set due to interactions of atmospheric 
neutrinos cannot be directly controlled by changing the depth 
or fiducial volume definition of the LBNE detector.  Furthermore 
the atmospheric-neutrino flux is naturally concentrated around 
the energy range relevant for proton decay.  In the analysis 
of~\cite{Bueno:2007um}, a single simulated neutral-current (NC) event 
survived the requirement of having an isolated single kaon 
with no additional tracks or $\pi^0$'s, and total deposited 
energy below \SI{800}{\MeV}.  This event is responsible 
for the estimated background rate of 1.0 per  \SI{}{\Mtyr}.

While this rate is acceptable for LBNE, it is natural to ask 
to what extent simulations are capable of providing reliable 
estimates for such rare processes.  What if the actual rate 
for single-kaon atmospheric-neutrino events is higher by a 
factor of ten or more?  Is that even conceivable?  To set the 
scale, it is useful to recall that the atmospheric-neutrino 
sample size in LBNE is expected to be of order $10^5$ per 
 \SI{}{\Mtyr} of exposure (Table~\ref{tab:atmos_event_rates}).  
Hence, ``rare-but-not-negligible'' in this context denotes
a process that occurs at a level of no less than $10^{-6}$.

\superk\ has given considerable attention to atmospheric-neutrino backgrounds in its nucleon decay searches (e.g.,~\cite{Kobayashi:2005pe}). 
In the SK analyses, data obtained with 
relaxed cuts have been studied to validate 
the atmospheric-neutrino flux and interaction models 
employed.  Consequently, the atmospheric-neutrino 
backgrounds for nucleon decay searches are well established 
at the level required for the water Cherenkov detector 
approach to this physics. 

For the case of LBNE, however, with a different detector 
technology, and with a goal of being sufficiently background-free 
to enable a discovery based on observation of a single candidate  
event, one would like to go further to understand at a detailed 
level what the rates for the specific background processes are. 
The first question to ask is 
what are the physical processes that could 
produce the exact signature of a $p\to K^+\overline{\nu}$ 
event?  Some possibilities are discussed below.  

\textbf{{\boldmath Strange particle production in $\Delta S = 0$ processes:}}
An identified source of background events 
for SK~\cite{Kobayashi:2005pe}
involves associated production of 
a pair of strange hadrons, nominally in the strong decay of a
nucleonic resonance excited 
via an inelastic NC
neutrino-nucleon interaction.  This could be in the form of a kaon 
accompanying a $\Lambda$ baryon.  Again, conservation of 
strangeness holds that the baryon cannot be absorbed, and thus a 
weak decay of the strange quark is guaranteed.  For water Cherenkov 
detectors the strange baryon is produced with a small enough 
momentum that its decay products are typically below Cherenkov 
threshold.  For a liquid argon detector, these final state particles 
should be detectable, leaving distinctive signatures that can be 
reconstructed.  Thus in principle, this source of background 
can be suppressed with appropriate event reconstruction and analysis 
tools.   To understand this prospect in quantitative terms, the range of kinematic 
distributions are currently under investigation.

It is possible to imagine yet more contrived scenarios, for example where 
the meson produced is a $K^0_L$ that escapes detection, while a
charged kaon ($K^-$ in this case) results from the decay of an 
excited $\Lambda$ or $\Sigma$ baryon produced in association.
However, one would expect such processes to be even more rare than 
those described above.  Thus if the rates for (say) the $K^+\Lambda$ 
production channel described above can be constrained as being sufficiently 
small, it can be argued that the more contrived scenarios can be ignored.

\textbf{{\boldmath Strange particle production 
               in $\Delta S = 1$ processes:}}
A potentially challenging source of background is production 
of a single charged kaon (in this case a $K^-$) 
in a $\Delta S = 1$ process.  
In the simplest case, one could think of it as the Cabibbo-suppressed 
version of single $\pi$ production in a CC
antineutrino interaction.  In contrast to the $\Delta S = 0$ processes described above,
no strange baryon is
produced in association, and so there are no other hadrons to detect.  
(Similarly, one could imagine the kaon originating in 
the decay of a strange baryon resonance produced in a Cabibbo-suppressed 
neutrino interaction, accompanied by a neutron that goes undetected.)
On the other hand, such processes can only occur in CC 
interactions, and thus a charged lepton will accompany the 
kaon.  This therefore constitutes a background only for cases where 
the charged lepton is missed, which should be rare.  The combination 
of probabilities associated with (1) Cabibbo-suppression, (2) single 
hadron production, and (3) circumstances causing the charged lepton 
to be missed, lead to an overall suppression of this source of 
background.  Thus it should be possible to rule it out 
as a source of concern for LBNE on the basis of these features alone.

\textbf{\boldmath Misidentification of pions in 
               atmospheric neutrino events:}
While misidentification of 
leading pions as kaons in atmospheric-neutrino scattering events is a potential problem,
 it can be argued that the rate for such 
misidentification events can be controlled.  Key signatures 
for the kaon are found in the distinctive residual-range dependence 
of its energy deposition near the end of its 
trajectory (nominally \SI{14}{\cm}) as well as in the explicit reconstruction of its decay products. 
Similarly, tails in the measurement of $dE/dx$ would be a concern if they
led a pion track to mimic a kaon, however the momentum (\SI{30}{\MeV})
and hence range of the muon produced in the decay of a stopping pion 
would not match that of the corresponding muon (\SI{236}{\MeV}) 
in a $K^+\to \mu^+\nu$ decay.  Thus, it should be possible to 
control this background experimentally.

\begin{figure}[!htb]
\centering
\centering\includegraphics[width=0.8\textwidth]{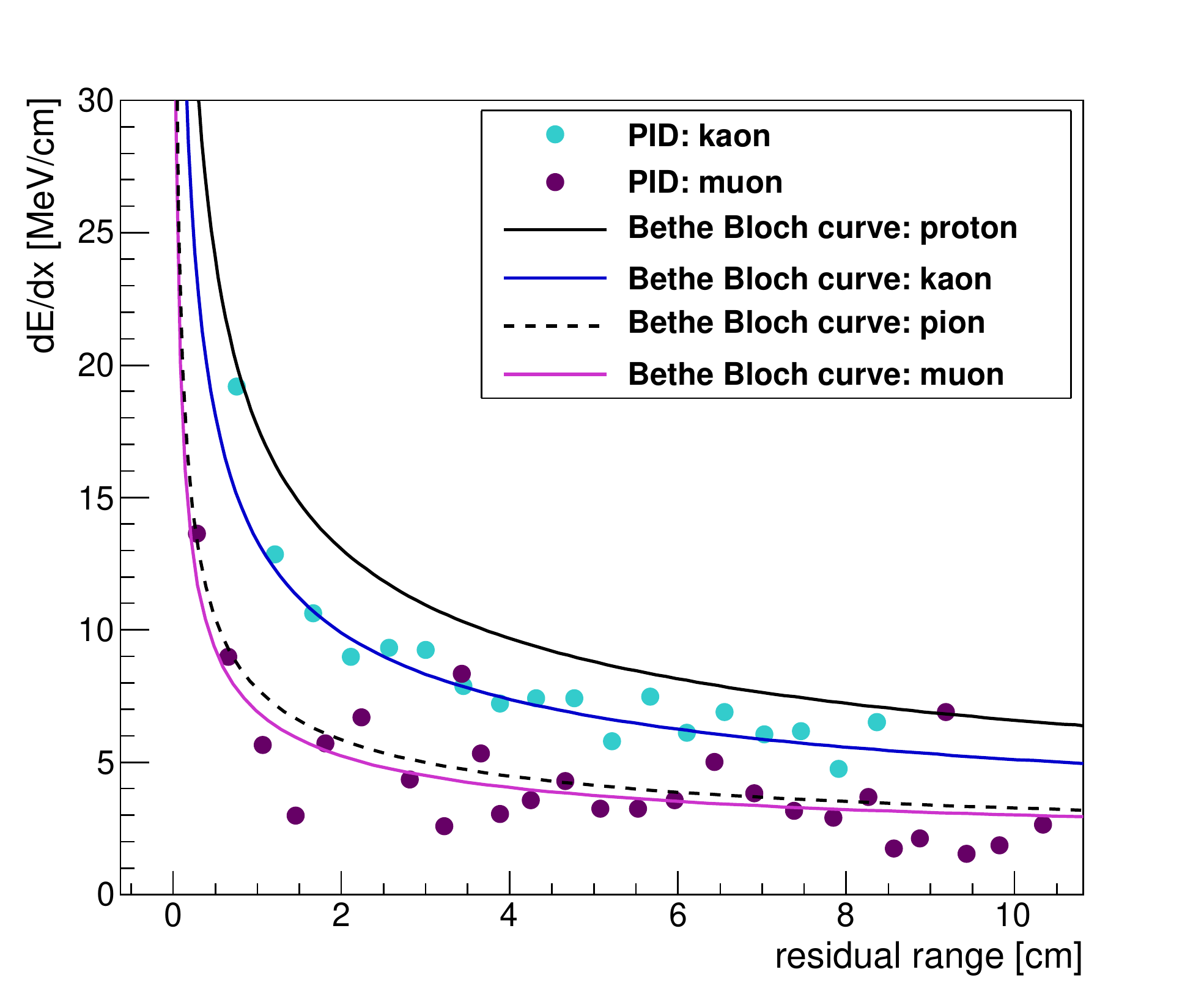}
\caption[$dE/dx$ profile of decaying kaon in the ICARUS CNGS run]
        {Measurements of $dE/dx$ versus residual range for signals
          associated with the kaon track in Figure~\ref{fig:icaruskaon} (cyan points) and the decay muon
          (magenta points).  Overlaid are the expected $dE/dx$ profiles for the
          two particle identities~\cite{Antonello:2012hu}. }
\label{fig:pdkdedx}
\end{figure}
One variant of this background source occurs for the case where the
pion decays in flight.  Two experimental handles on this background
can be immediately identified.  First is the deviation from the
expected $dE/dx$ profile for a kaon, which will be more dramatic than
in the case of the stopping pion.  Second is the correlation of the
direction of the decay muon with that of the pion, which is absent in
the decay of a particle at rest.  Assessment of the cumulative impact
of event rejection based on these features is under study.  However,
the decaying kaon observed in the ICARUS CNGS run displayed in 
Figure~\ref{fig:icaruskaon} can be used to give a sense of the 
$\pi/K$ discrimination possible in a LArTPC via $dE/dx$.   
In Figure~\ref{fig:pdkdedx}, the measurements
of $dE/dx$ versus residual range for the anode wires registering
signals from the kaon and muon tracks in this event are plotted
against the expected $dE/dx$ profiles~\cite{Antonello:2012hu}. The
data from the kaon track (cyan points) agree very well with the
expected $dE/dx$ profile (blue curve) and are quite distinguishable
from the expected pion profile (dashed curve).

\textbf{\boldmath Event reconstruction pathologies:}
While consideration of rare event topologies in atmospheric-neutrino 
interactions is important, it will be equally important to understand 
ways in which more typical events might be misreconstructed so as 
to mimic nucleon decay processes.  For example, a quasi-elastic 
$\nu_\mu$-CC interaction will produce a muon and a recoil
proton from a common vertex.  However, it may be possible to interpret 
the vertex as the kink associated with the decay of a stopping kaon, 
where the proton track is confused with a kaon traveling in the opposite 
direction.  Tools are still under development to be able to understand 
the degree to which this possibility poses a potential background.  
Naively, the $dE/dx$ profile of the proton as a function of residual range 
will not match the time-reversed version of this for a kaon, 
and distributions of kinematic quantities will be 
distinct.  Additionally, such a background will only affect 
the portion of the $p\to K^+\overline{\nu}$ analysis focused 
on $K^+\to \mu^+\nu$; other $K^+$ decays will be immune to this 
pathology.  

The point of this example is to illustrate that although the 
exquisite performance characteristics of the LArTPC technique 
enables unambiguous identification of nucleon decay signatures, 
an extensive program of detailed analysis will be required 
to fully exploit these capabilities.

\textbf{Conclusions on atmospheric-neutrino backgrounds:}
The above examples suggest that it will be possible to demonstrate 
the desired level of suppression of atmospheric-neutrino background 
without undue reliance on simulations 
via a combination of arguments based on existing experimental data 
(from SK proton decay searches, as well as data from various 
sources on exclusive and inclusive neutrino-interaction processes 
that yield rare topologies), physics considerations, and detailed 
analysis of anticipated detector response.  For the latter, 
ongoing LBNE event-reconstruction efforts will play a role with 
simulated atmospheric-neutrino samples.  Additionally, useful 
input is expected to come in over the short/intermediate term 
from analyses of LArTPC data from ArgoNeuT, MicroBooNE and 
the proposed LArIAT.
Finally, while the state of neutrino flux and interaction models 
is already quite advanced, vigorous theoretical work is ongoing 
to improve these further, exploiting existing data from neutrino 
and electron-scattering experiments.  In particular, kaon production 
in neutrino interactions in relevant energy ranges is receiving 
renewed attention~\cite{gallagher-private}.
\clearpage
\section{\boldmath Summary of Expected Sensitivity to Key Nucleon Decay Modes}

\begin{figure}[!htb]
\centering
\includegraphics[width=0.8\textwidth]{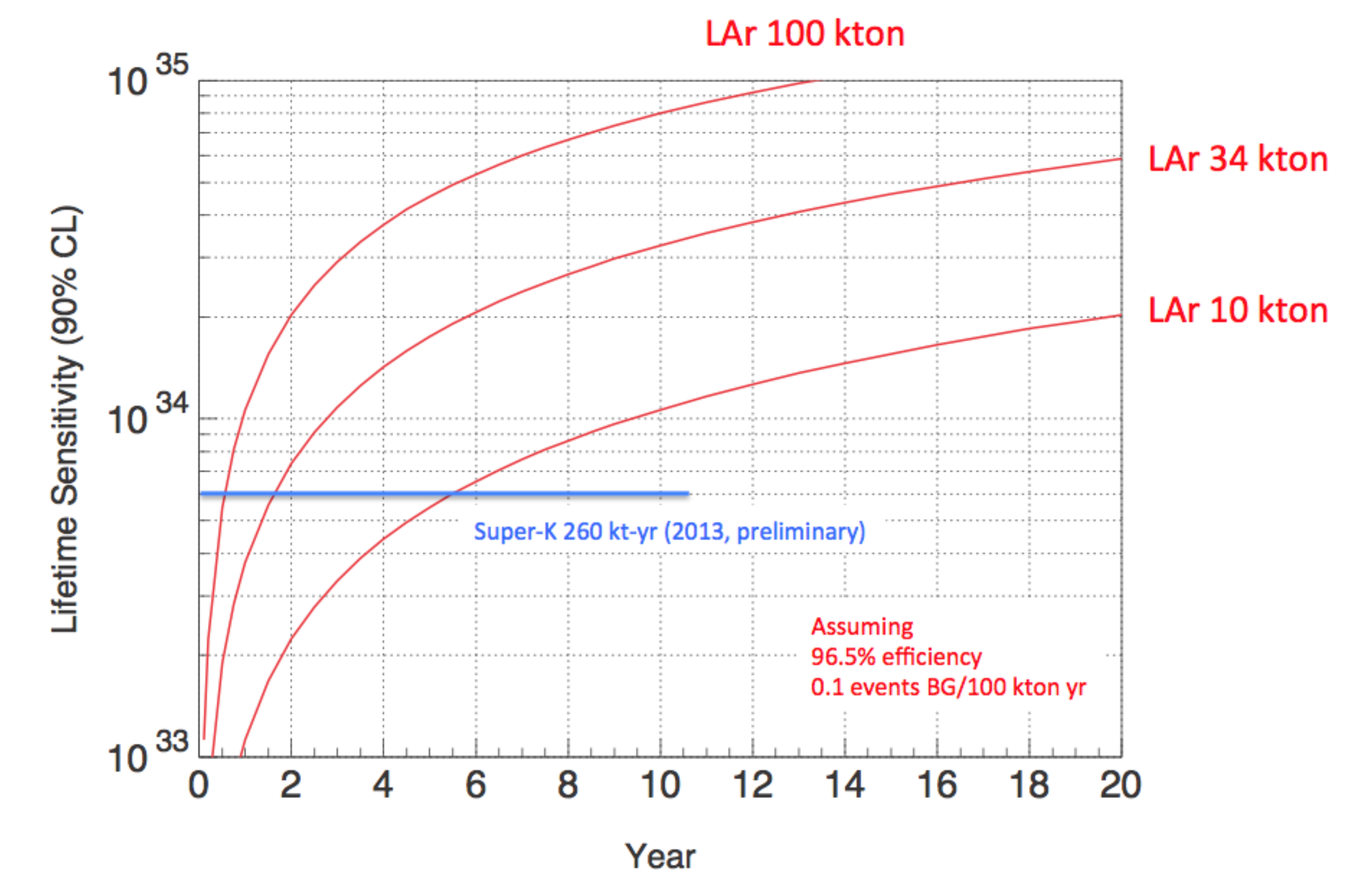}
\caption[Proton decay lifetime limit for $p \rightarrow K^+ \overline{\nu}$
  versus time]{Proton decay lifetime limit for $p
  \rightarrow K^+ \overline{\nu}$ as a function of time for
  underground LArTPCs of fiducial masses 10, 34 and 100~kt.
  For comparison, the current limit from SK is also shown.
  The limits are at 90\% C.L., calculated for
  a Poisson process including background, assuming that the detected events
  equal the expected background.}
\label{fig:kdklimit}
\end{figure}

Based on the expected signal efficiency and the upper limit on the
background rates estimated in Section~\ref{sec:pdk:background-rej}, 
the expected limit on the proton
lifetime as a function of running time in LBNE for $p \rightarrow K^+
\overline{\nu}$ is shown in 
Figure~\ref{fig:kdklimit}. 
\begin{introbox}
Figure~\ref{fig:kdklimit} demonstrates that 
to improve the current limits on
the $p \rightarrow \overline{\nu} K^+$, set by \superk, significantly
beyond that experiment's sensitivity, 
a LArTPC
detector of at least 10~kt, installed deep underground, is needed.
A \ktadj{34} detector will improve the current limits by an order of
magnitude after running for two decades.  Clearly a larger detector
mass would improve the limits even more in that span of time.
\end{introbox}

While the background rates are thought to be no higher than those assumed 
in generating the above sensitivity projections, it is possible to estimate 
the impact of higher rates.  For $p\to K^+\overline{\nu}$, 
Table~\ref{tab:pdecay-bgvariation} shows a comparison of the 
$90\%$ CL lower bounds on proton lifetime for an exposure  of \SI{340}{\ktyr} 
assuming the nominal 1.0 per  \SI{}{\Mtyr} background rate with the 
corresponding bounds for a rate that is ten times higher, as well as for 
a fully background-free experiment.
%
%
While a factor of ten increase in the background would hurt the 
sensitivity, useful limits can still be obtained.  As stated 
above, however, there is good reason to believe such a case 
is highly unlikely.
\begin{table}[!htb]
\caption[Sensitivity for $p\to K^+\overline{\nu}$ with different background rates]
        {The impact of different assumed background rates on the expected 
         $90\%$ CL lower bound for the partial proton lifetime for 
         the $p\to K^+\overline{\nu}$ channel, for a \ktadj{34} detector 
         operating for ten years.  The expected background rate is 
         one event per  \SI{}{\Mtyr}.  Systematic uncertainties are not included 
         in these evaluations.
        }
\begin{center}
\begin{tabular}{$L^c} 
\toprule
\rowtitlestyle
Background Rate & Expected Partial Lifetime Limit\\ \toprowrule
0 events/\SI{}{\Mtyr}    & $3.8 \times 10^{34}$ years  \\ \colhline
1 events/\SI{}{\Mtyr}    & $3.3 \times 10^{34}$ years  \\ \colhline
10 events/\SI{}{\Mtyr}    & $2.0 \times 10^{34}$ years  \\
\bottomrule
\end{tabular}
\end{center}
\label{tab:pdecay-bgvariation}
\end{table}
%

\begin{figure}[!htb]
\centering
\includegraphics[width=\textwidth]{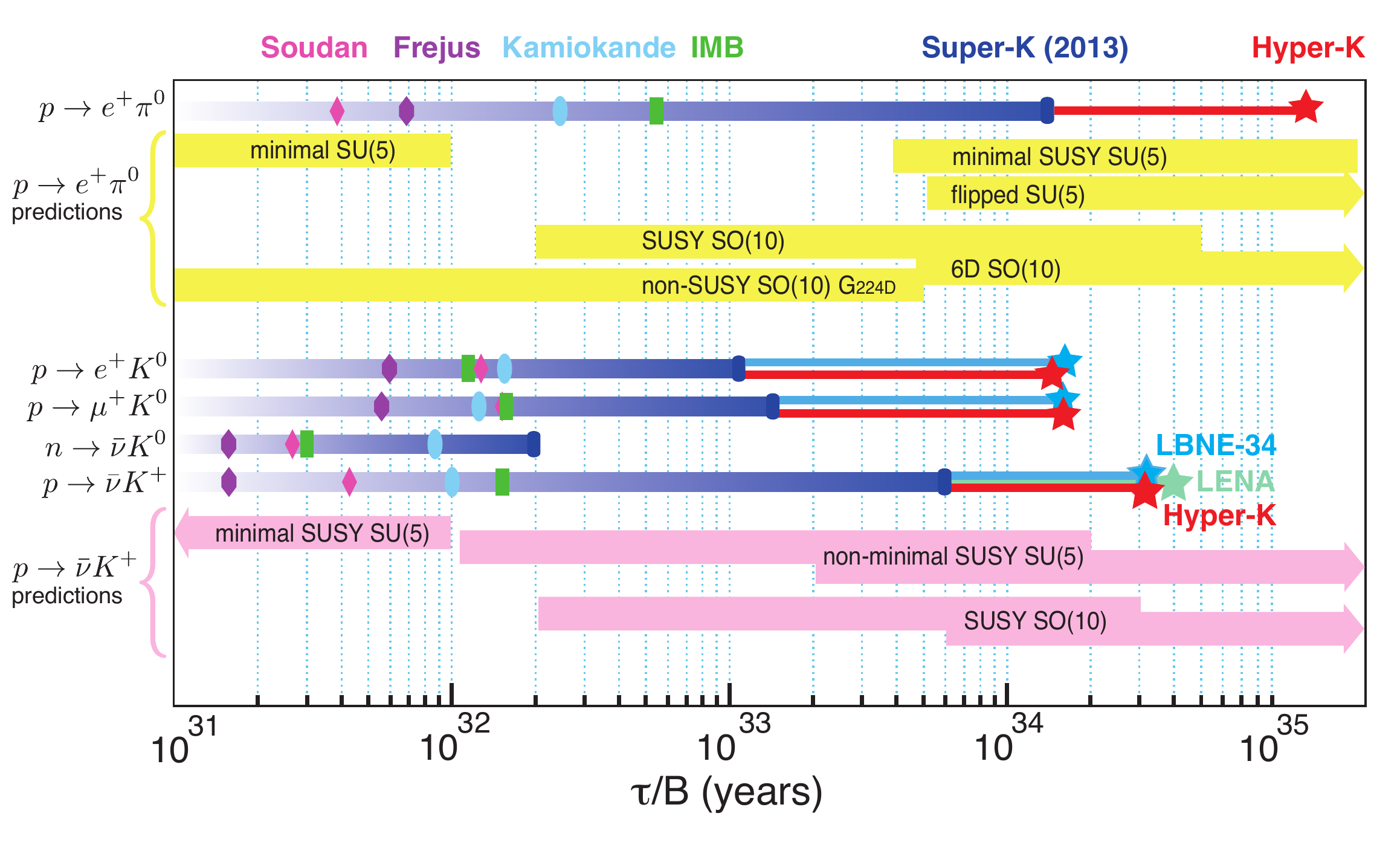}
\caption[Proton decay lifetime limits achievable by \ktadj{34} LBNE;
comparison to others]{Proton decay
  lifetime limits that can be achieved by the LBNE \ktadj{34} detector compared
  to other proposed future experiments.  The limits are at 90\% C.L.,
  calculated for a Poisson process including background, assuming that
  the detected events equal the expected background.}
\label{fig:nnn13}
\end{figure}

%
Sensitivities have been computed for some of the other
decay channels listed in Table~\ref{tab:pdecay}. The limits that could
be obtained from an LBNE \ktadj{34} detector in ten years of running as
compared to other proposed future experiments and theoretical
expectations are shown in Figure~\ref{fig:nnn13}.

%% file: chapter-sn.tex

\begin{introbox}
Neutrinos emitted in the first few seconds of a core-collapse
  supernova carry with them the potential for great insight into the mechanisms 
behind some of the most spectacular events that have played key roles in the
  evolution of the Universe.  Collection and analysis of this
  high-statistics neutrino signal from a supernova within our galaxy
  would provide a rare opportunity to witness the energy and flavor
  development of the burst as a function of time. This would in turn
  shed light on the astrophysics of the collapse as well as on
  neutrino properties.  
\end{introbox}

\section{The Neutrino Signal and Astrophysical Phenomena}

A core-collapse supernova\footnote{\emph{Supernova} always
  refers to a \emph{core-collapse supernova} in this chapter unless
  stated otherwise.} occurs when a massive star reaches the end of its
life, and stellar burning can no longer support the star's weight.
This catastrophic collapse results in a compact remnant such as a
neutron star, or possibly a black hole, depending on the mass of the
progenitor.  The infall is followed by a \emph{bounce} when sufficiently
high core density is reached, and in some unknown (but nonzero)
fraction of cases, the shock wave formed after the bounce results in a
bright explosion~\cite{Janka:2012wk}.  The explosion energy represents
only a small fraction of the enormous total gravitational binding
energy of the resulting compact remnant, however --- thanks to the
neutrinos' weak coupling, which allows them to escape --- within a few
tens of seconds almost all of the energy is emitted in the form of
neutrinos in the tens-of-MeV range.  In spite of their weak coupling,
the neutrinos are copious enough to (very likely) play a significant
role in the explosion.

Neutrinos from the celebrated SN1987A core
collapse~\cite{Bionta:1987qt,Hirata:1987hu} in the Large Magellanic
Cloud outside the Milky Way were observed; however, the
statistics 
were sparse 
and a great many questions remain.  A high-statistics observation of a
neutrino burst from a nearby supernova would be possible with the current
generation of detectors. Such an observation would shed light
on 
the nature of the astrophysical event, as well as on the nature of
neutrinos themselves.  Sensitivity to the different flavor components
of the flux is highly desirable.

\begin{figure}[!htb]
\centering
\includegraphics[width=\textwidth]{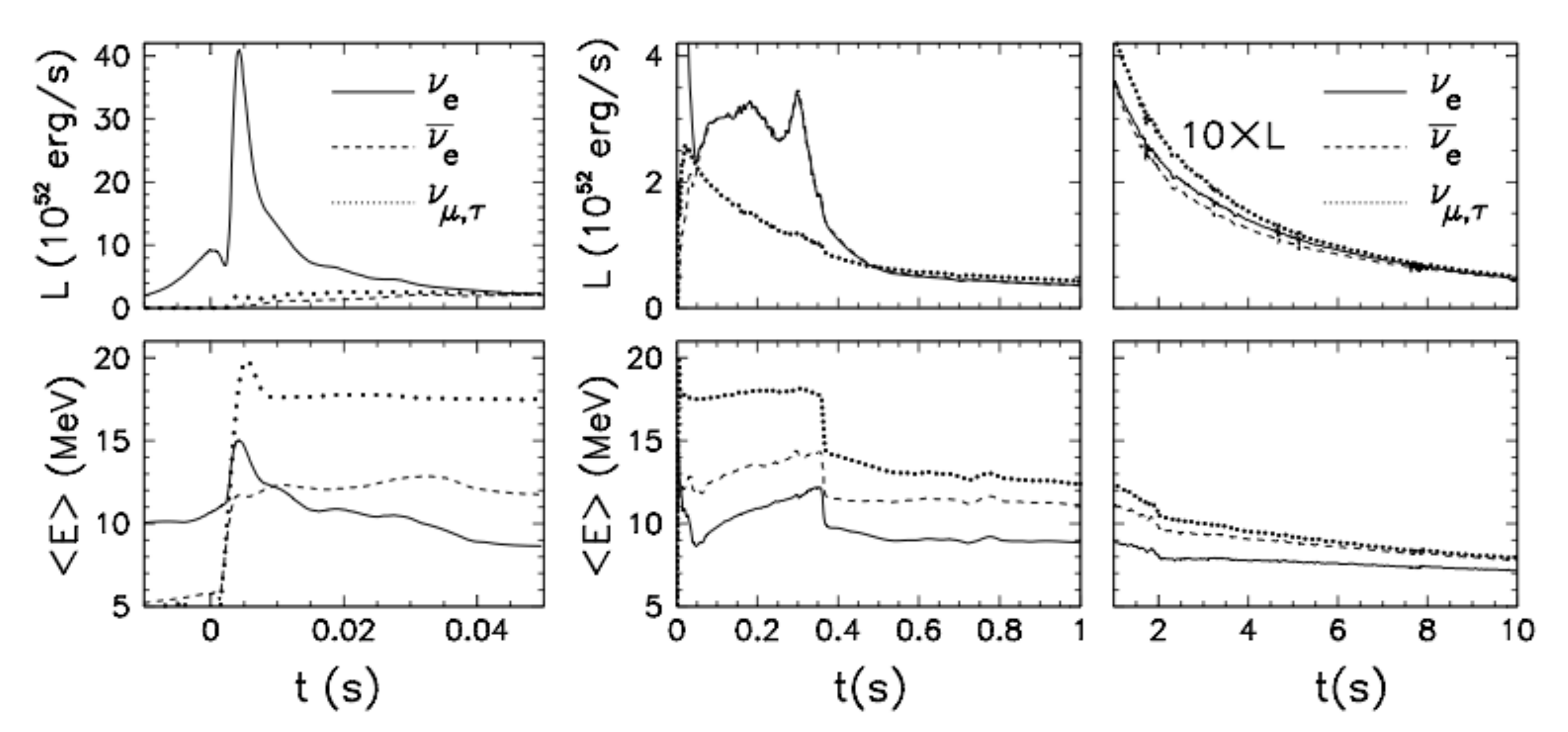}
\caption[Expected core-collapse neutrino signal]{Expected
  core-collapse neutrino signal from the \emph{Basel}
  model~\cite{Fischer:2009af}, for a
  10.8 $M_{\odot}$ progenitor.  The left plots show the very early
  signal, including the neutronization burst; the middle plots show
  the accretion phase, and the right plots show the cooling
  phase. Luminosities as a function of time are shown across the top plots. 
  The bottom plots show average energy as a function of time for the
  $\nu_e$, $\overline{\nu}_e$ and $\nu_{\mu,\tau}$ flavor components of the
  flux (fluxes for $\nu_\mu$, $\overline{\nu}_\mu$, $\nu_\tau$,
  and $\overline{\nu}_\tau$ should be identical).  Figure courtesy of~\cite{Wurm:2011zn}.}
\label{fig:spectrum}
\end{figure}

The core-collapse neutrino signal starts with a short, sharp
\emph{neutronization} burst primarily composed of
$\nu_e$ (originating from $p+e^- \rightarrow n + \nu_e$, as protons
and electrons get squeezed together), and is followed by an
\emph{accretion} phase lasting some hundreds of milliseconds, as matter falls onto the collapsed core.  The later
\emph{cooling} phase over $\sim$10~seconds represents the main part of
the signal, over which the proto-neutron star sheds its gravitational
binding energy.  The neutrino flavor content and spectra change
throughout these phases, and the supernova's temperature evolution can
be followed with the neutrino signal. Some fairly generic supernova
signal features are illustrated in Figure~\ref{fig:spectrum}, based on~\cite{Fischer:2009af} and reproduced from~\cite{Wurm:2011zn}.
%

The supernova-neutrino spectrum at a given moment in time is expected
to be well described by a
parameterization~\cite{Minakata:2008nc,Tamborra:2012ac} given by:
\begin{equation}
        \label{eq:pinched}
        \phi(E_{\nu}) = \mathcal{N} 
        \left(\frac{E_{\nu}}{\langle E_{\nu} \rangle}\right)^{\alpha} \exp\left[-\left(\alpha + 1\right)\frac{E_{\nu}}{\langle E_{\nu} \rangle}\right] \ ,
\end{equation}
where $E_{\nu}$ is the neutrino energy, $\langle E_\nu \rangle$ is the
mean neutrino energy, $\alpha$ is a \emph{pinching parameter}, and
$\mathcal{N}$ is a normalization constant.
Large $\alpha$ corresponds to a more pinched spectrum (suppressed
high-energy tail). This parameterization is referred to as a
\emph{pinched-thermal} form. The different $\nu_e$, $\overline{\nu}_e$ and
$\nu_x, \, x = \mu, \tau$ flavors are expected to have different
average energy and $\alpha$ parameters and to evolve differently in
time.

A wide variety of astrophysical phenomena affect the
flavor-energy-time evolution of the spectrum, including neutrino
oscillation effects that are determined by the mass hierarchy (MH) and
\emph{collective} effects due to neutrino-neutrino interactions.  A
voluminous literature exists exploring these collective phenomena,
e.g.,~\cite{Duan:2005cp,Fogli:2007bk,Raffelt:2007cb,Raffelt:2007xt,EstebanPretel:2008ni,Duan:2009cd,Dasgupta:2009mg,Duan:2010bg,Duan:2010bf}.
%
\begin{introbox}
A number of astrophysical phenomena associated with supernovae are expected to be observable
in the supernova-neutrino signal, providing a remarkable window into the event, for example: 
\begin{itemize}
\item The initial burst, primarily composed of $\nu_e$ and called the
  \emph{neutronization} or \emph{breakout}
  burst, 
  represents only a small component of the total signal.  However,
  oscillation effects can manifest in an observable manner
  in this burst, and flavor transformations can be modified by the
  \emph{halo} of neutrinos generated in the supernova envelope by
  scattering~\cite{Cherry:2013mv}.
\item The formation of a black hole would cause a sharp signal cutoff
  (e.g.,~\cite{Beacom:2000qy,Fischer:2008rh}).
\item Shock wave effects (e.g.,~\cite{Schirato:2002tg}) would cause a
  time-dependent change in flavor and spectral composition as the
  shock wave propagates.
\item The standing accretion shock instability
  (SASI)~\cite{Hanke:2011jf,Hanke:2013ena}, a \emph{sloshing} mode
  predicted by 3D neutrino-hydrodynamics simulations of
  supernova cores, would give an oscillatory flavor-dependent
  modulation of the flux.
\item Turbulence effects~\cite{Friedland:2006ta,Lund:2013uta} would
  also cause flavor-dependent spectral modification as a function of
  time.
\end{itemize}
\end{introbox}

This list is far from comprehensive.  Furthermore, signatures of
\emph{collective} effects and signatures that depend on the MH will make an impact on many of the above signals (examples
will be presented in Section~\ref{ch:sn-sig-in-lar}).
Certain phenomena are even postulated to indicate
beyond-the-Standard-Model physics~\cite{Raffelt:1999tx} such as
axions, extra dimensions and an anomalous neutrino magnetic moment;
non-observation of these effects, conversely, would enable constraints
on these phenomena.

The supernova-neutrino burst signal is prompt with respect to the
electromagnetic signal and therefore can be exploited to provide an
early warning to astronomers~\cite{Antonioli:2004zb,Scholberg:2008fa}.
Additionally, a LArTPC signal~\cite{Bueno:2003ei} is expected to
provide some pointing information, primarily from elastic scattering
on electrons.

Even non-observation of a burst, or non-observation of
a $\nu_e$ component of a burst in the presence of supernovae (or other
astrophysical events) observed in electromagnetic or gravitational
wave channels, would still provide valuable information about the
nature of the sources.  Moreover, a long-timescale, sensitive search
yielding no bursts will also provide limits on the rate of
core-collapse supernovae.

\section{Expected Signal and Detection in Liquid Argon}
\label{ch:sn-sig-in-lar}

As discussed in
Section~\ref{ss:snphysics}, liquid argon is known to exhibit a singular
sensitivity to the $\nu_e$ component of a supernova-neutrino burst.
This feature is especially important, as it will make LBNE a unique
source in the global effort to combine data from a variety of detectors
with different flavor sensitivities to obtain a complete picture of the
physics of the burst.

\begin{figure}[!htb]
\centering
\includegraphics[width=0.6\textwidth]{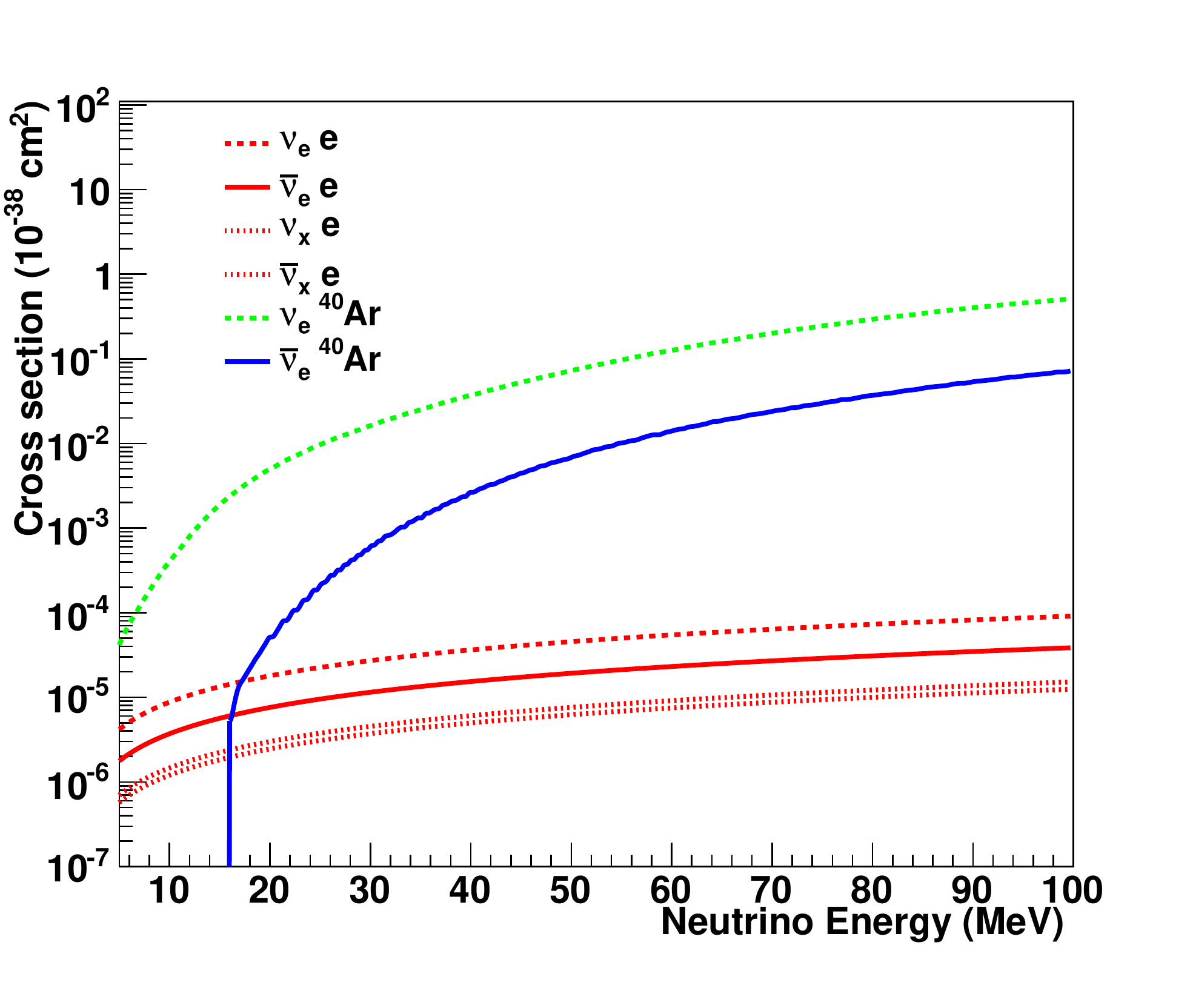}
\caption[]{Cross sections for supernova-relevant interactions in argon.}
\label{fig:xscns}
\end{figure}
The predicted event rate from a supernova-neutrino burst may be calculated by
folding expected neutrino differential energy spectra in with cross
sections for the relevant channels, and with detector response.
For event rate estimates in liquid argon, a detection threshold of
\SI{5}{\MeV} is assumed. The photon-detection system of the LBNE
far detector, coupled with charge collection and simple pattern recognition,
is expected to provide a highly efficient trigger.  Most LBNE
supernova physics sensitivity studies so far have been done using
parameterized detector responses from~\cite{Amoruso:2003sw}
implemented in the SNOwGLoBES software
package~\cite{snowglobes}. SNOwGLoBES takes as input fluxes, cross
sections (Figure~\ref{fig:xscns}), \emph{smearing matrices} (that
incorporate both interaction product spectra and detector response)
and post-smearing efficiencies. The energy resolution used is
\begin{equation}
\frac{\sigma}{E \ {\rm (MeV)}} = \frac{11\%}{\sqrt{E \ {\rm MeV}}} + 2\%
\end{equation}
Work is currently underway using the full Geant4 simulation~\cite{Agostinelli:2002hh} framework and the LArSoft software package~\cite{Church:2013hea} to characterize low-energy response for
realistic LBNE detector configurations. Preliminary studies of the
detector response with the full simulation are summarized in Section~\ref{sec:snlarsoft} and are found to be consistent with the
parameterized response implemented in SNOwGLoBES.

\begin{figure}[!htb]
\centerline{
\includegraphics[width=0.5\textwidth]{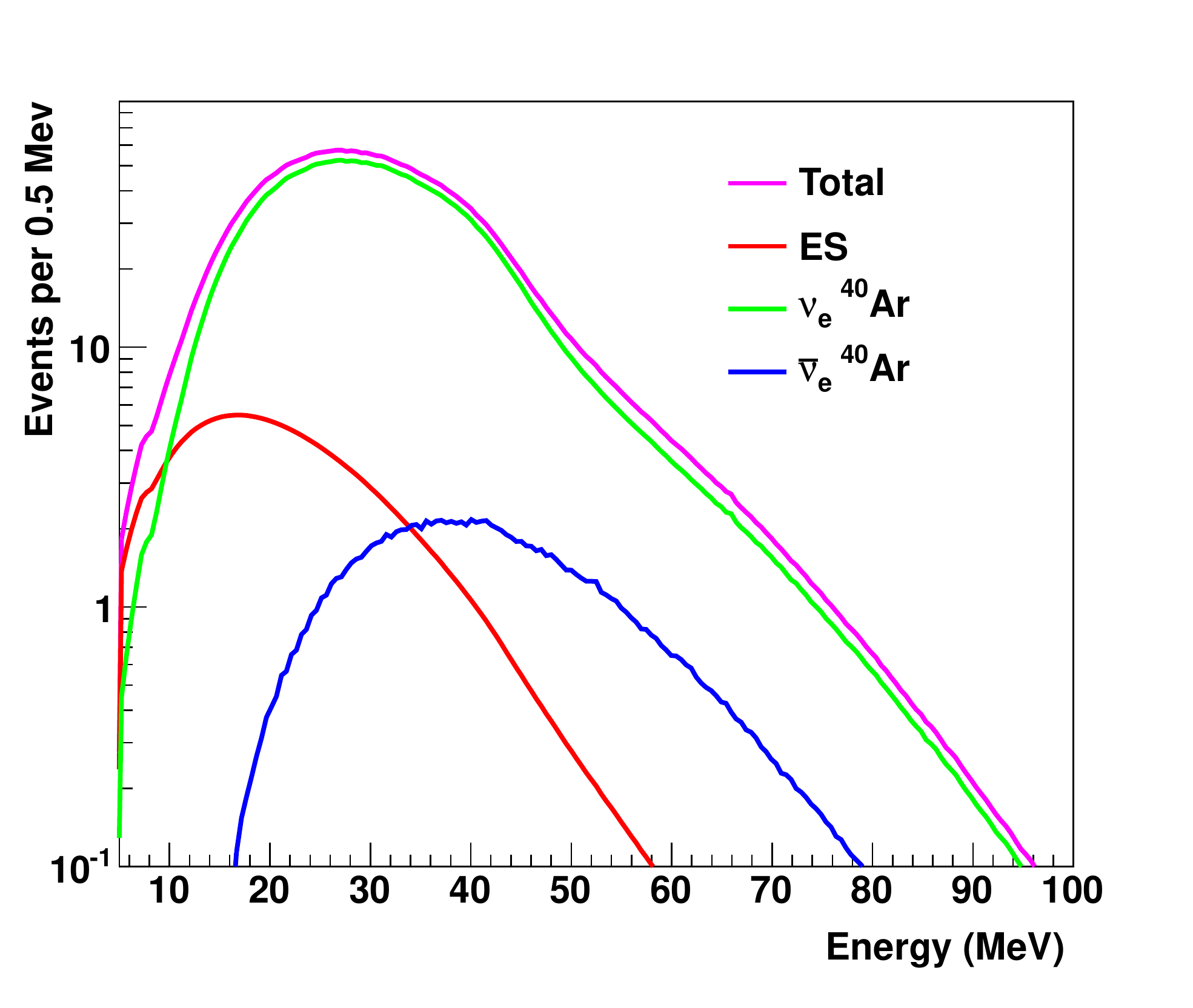}
\includegraphics[width=0.5\textwidth]{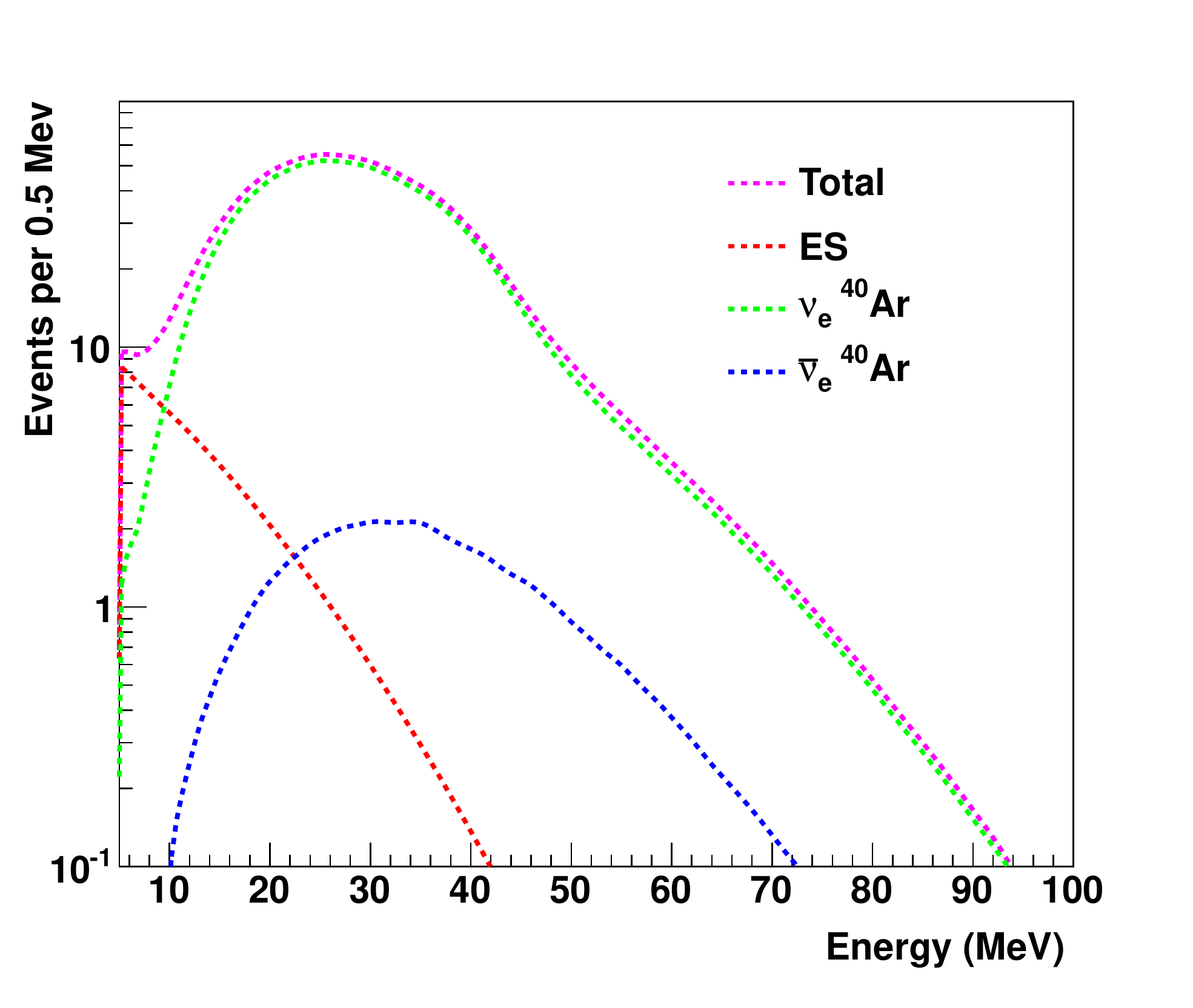}
}
\caption[SN $\nu$ event rates in \SI{34}{kt} of LAr for a core
  collapse at 10~kpc, GKVM]{Supernova-neutrino event rates in 34~kt of argon for a core
  collapse at 10~kpc, for the GKVM model~\cite{Gava:2009pj} (events
  per 0.5~MeV), showing three relevant interaction channels. Left:
  interaction rates as a function of true neutrino energy.  Right:
  \emph{smeared} rates as a function of detected energy, assuming
  resolution from~\cite{Amoruso:2003sw}.}
  \label{fig:eventrates}
\end{figure}

Table~\ref{tab:argon_events} shows rates calculated with SNOwGLoBES for the dominant interactions in argon for
the \emph{Livermore} model~\cite{Totani:1997vj}, and the \emph{GKVM}
model~\cite{Gava:2009pj}.  Figure~\ref{fig:eventrates} shows the
expected observed differential event spectra for these fluxes.  Clearly, the $\nu_e$
flavor dominates.
\begin{table}[!htb]
  \caption[Event rates for different models in \SI{34}{\kt} of LAr for
    a core collapse at 10~kpc]{Event rates for different
    supernova models in \SI{34}{\kt} of liquid argon for a core collapse at 10~kpc, for $\nu_e$ and \ane\ charged-current channels and elastic scattering (ES) on electrons.
    Event rates will simply scale by active detector mass and inverse square of supernova distance.}
\label{tab:argon_events}\centering
\begin{tabular}{$L^c^c}
\toprule
\rowtitlestyle
Channel & Events & Events \\
\rowtitlestyle
& \emph{Livermore} model & \emph{GKVM} model  \\ 
\toprowrule

$\nu_e + ^{40}{\rm Ar} \rightarrow e^- + ^{40}{\rm K^*}$ & 2308  & 2848 \\ \colhline

$\overline{\nu}_e + ^{40}{\rm Ar} \rightarrow e^+ + ^{40}{\rm Cl^*}$ & 194 & 134\\ \colhline

$\nu_x + e^- \rightarrow \nu_x + e^-$                           & 296 &  178\\

\toprule
\rowtitlestyle
Total &  2794& 3160 \\ 
\bottomrule
\end{tabular}
\end{table}

\begin{figure}[!htb]
\centering
\includegraphics[width=0.9\textwidth]{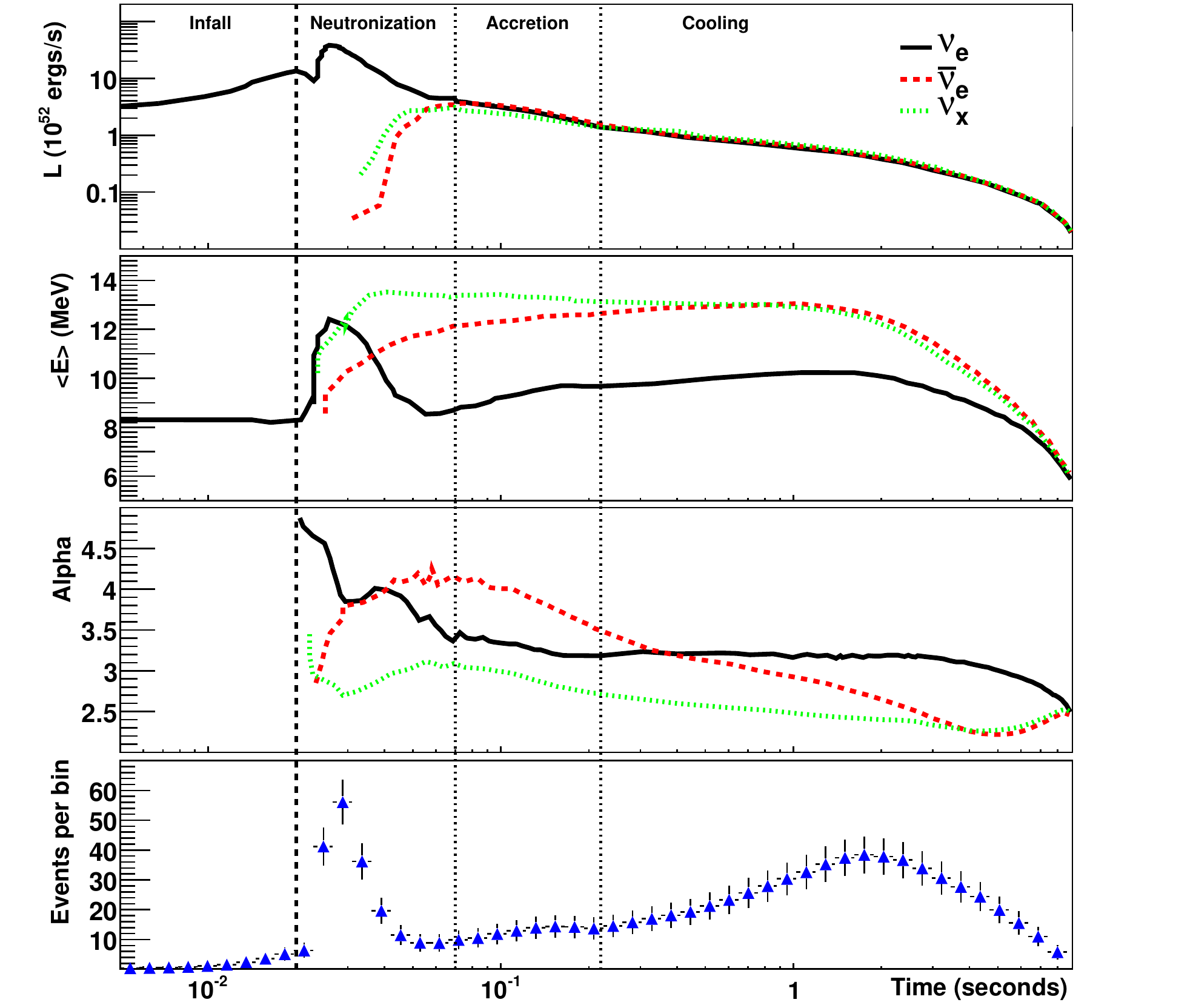}
\caption[Garching flux signal with neutronization burst]{ Expected
  time-dependent signal for a specific flux model for an
  electron-capture supernova~\cite{Huedepohl:2009wh} at 10~kpc.  The
  top plot shows the luminosity, the second plot
  shows average neutrino energy, and the third plot shows the $\alpha$
  (pinching) parameter.  The fourth (bottom) plot shows the total number of
  events (mostly $\nu_e$) expected in 34 kt of liquid argon, calculated using
  SNoWGLoBES.  Note the logarithmic binning in time; the plot shows
  the number of events expected in the given bin and the error bars
  are statistical. The vertical dashed line at 0.02 seconds indicates
  the time of core bounce, and the vertical lines indicate different
  eras in the supernova evolution.  The leftmost time interval
  indicates the infall period.  The next interval, from core bounce to
  50~ms, is the neutronization burst era, in which the flux is
  composed primarily of $\nu_e$.  The next period, from 50 to 200~ms,
  is the accretion period. The final era, from 0.2 to 9~seconds, is
  the proto-neutron-star cooling period.  }
\label{fig:garching}
\end{figure}

Figure~\ref{fig:garching} gives another example of an expected burst
signal, for which a calculation with detailed time dependence of the
spectra is available~\cite{Huedepohl:2009wh} out to nine seconds
post-bounce.  This model has relatively low luminosity but a robust
neutronization burst.  Note that the relative fraction of
neutronization-burst events is quite high.

\begin{figure}[!htb]
\centerline{
\includegraphics[width=.5\textwidth]{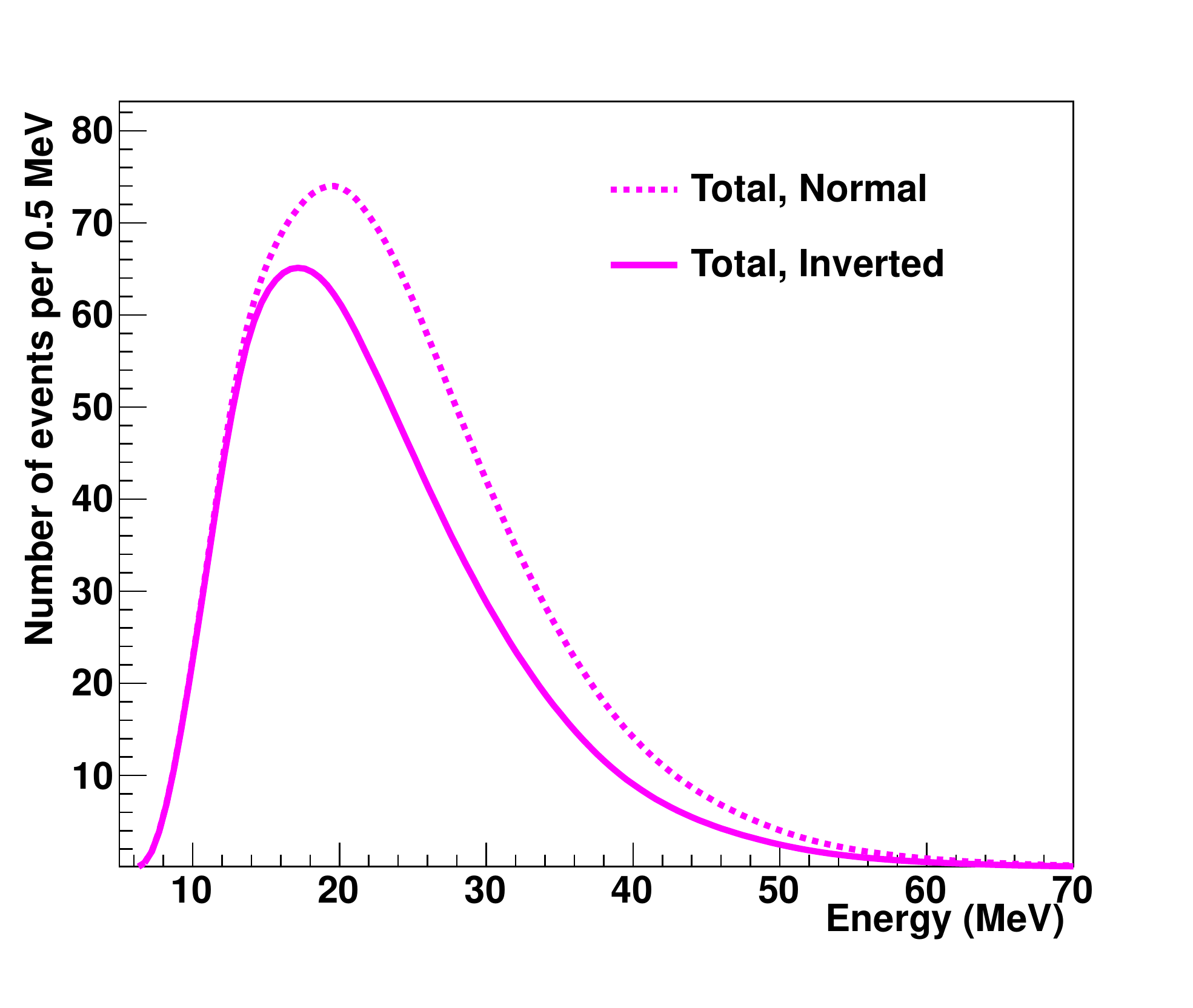}
\includegraphics[width=.5\textwidth]{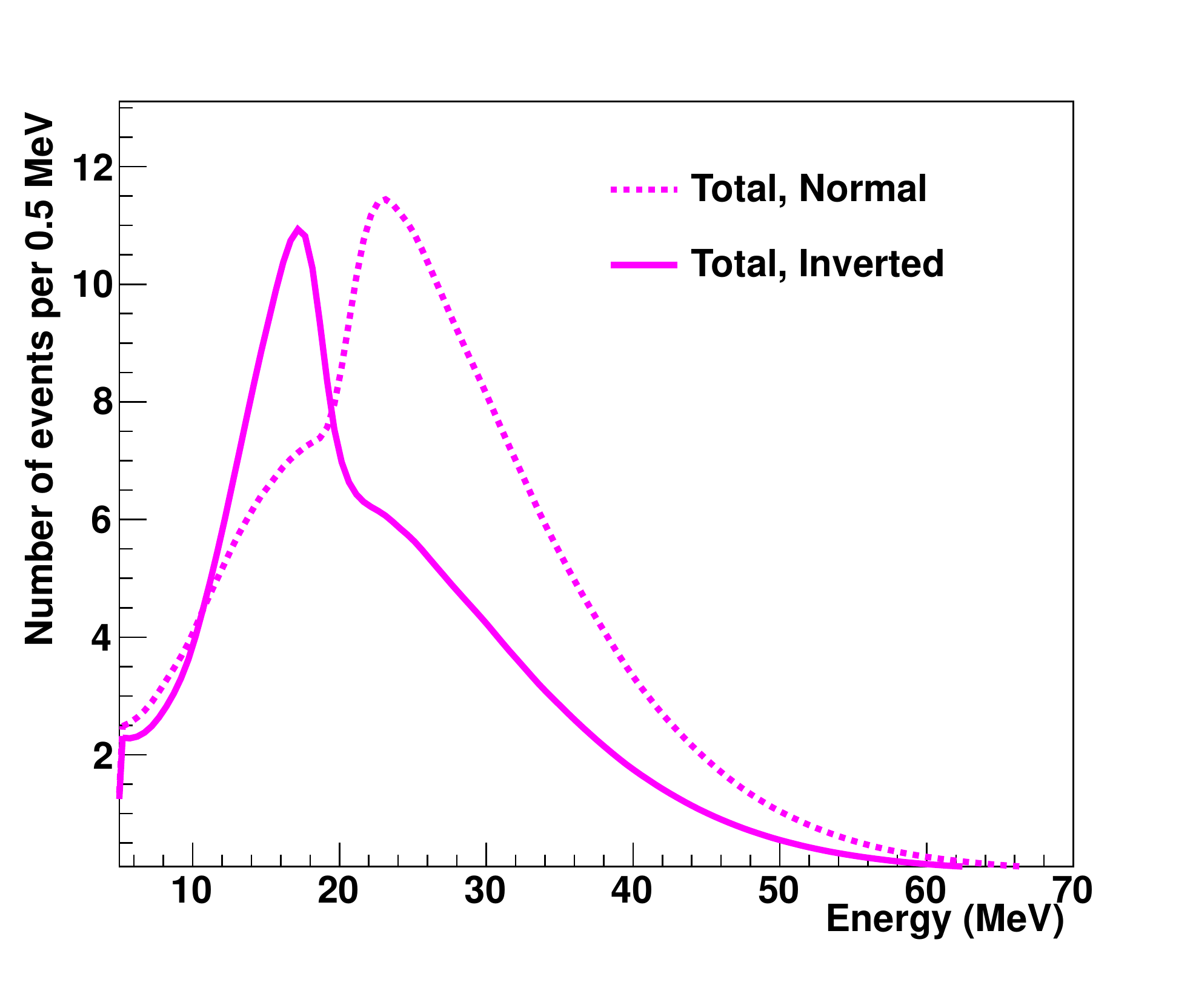}
}
\caption[Comparison of total event rates, normal and inverted
 MH, for WCD and LAr]{Comparison of total event rates for normal and inverted
  MH, for a specific flux example, for a \ktadj{100} water Cherenkov
  detector (left) and for a \ktadj{34} LArTPC (right)
  configuration, in events per \SI{0.5}{\MeV}.  There are distinctive
  features in liquid argon for different neutrino mass hierarchies for this supernova 
model~\cite{cherrypriv}.}
\label{fig:hierarchy_comparison}
\end{figure}

 In Figure~\ref{fig:hierarchy_comparison}, different oscillation
 hypotheses have been applied to \emph{Duan}
 fluxes~\cite{Duan:2010bf}. The Duan flux represents only a single
 late time slice of the supernova-neutrino burst and not the full flux;
 MH information will be encoded in the time evolution of the
 signal, as well. The figure illustrates, if only anecdotally,
 potential MH signatures.

\begin{figure}[!htb]
\centerline{
\includegraphics[width=.5\textwidth]{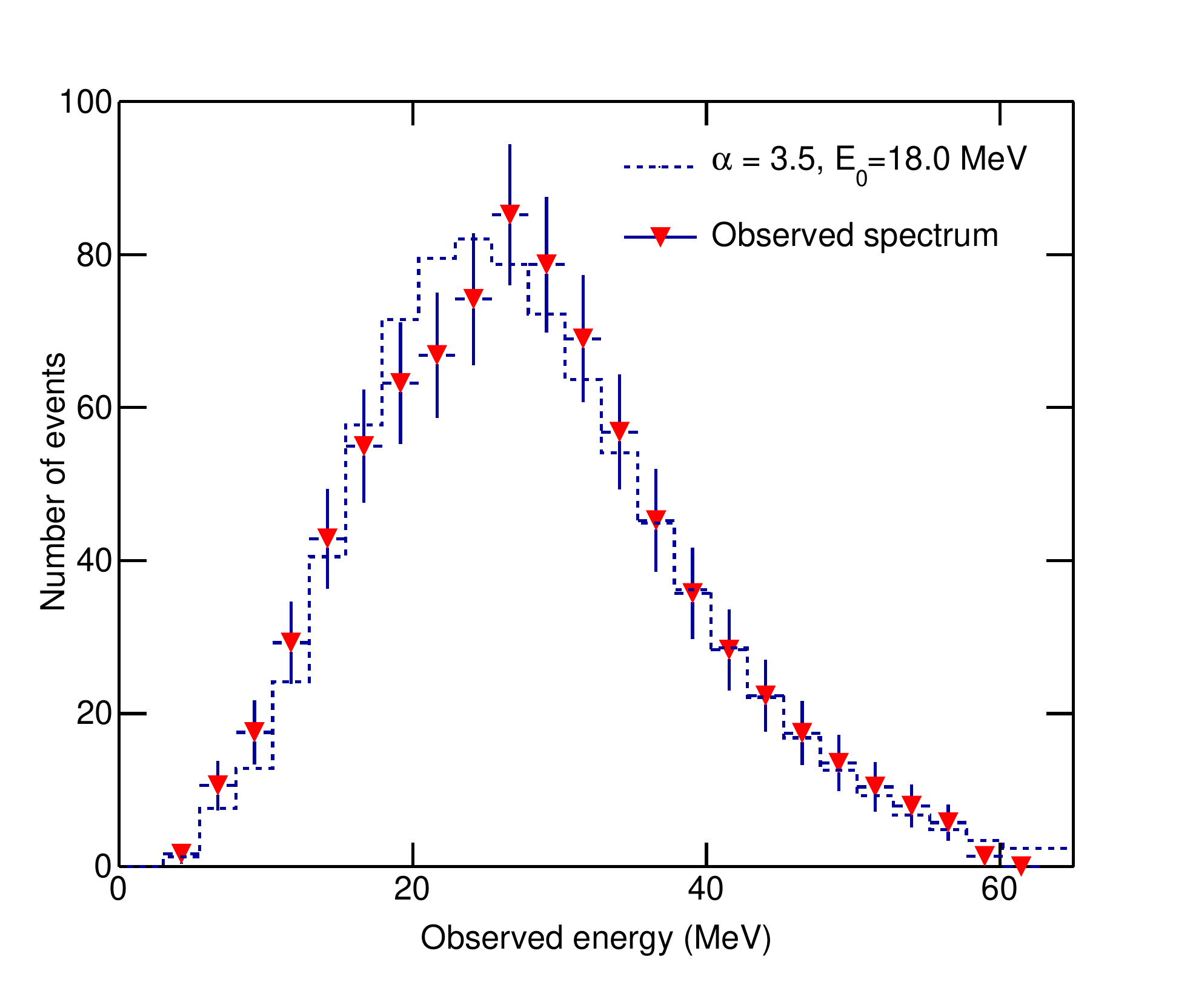}
\includegraphics[width=.5\textwidth]{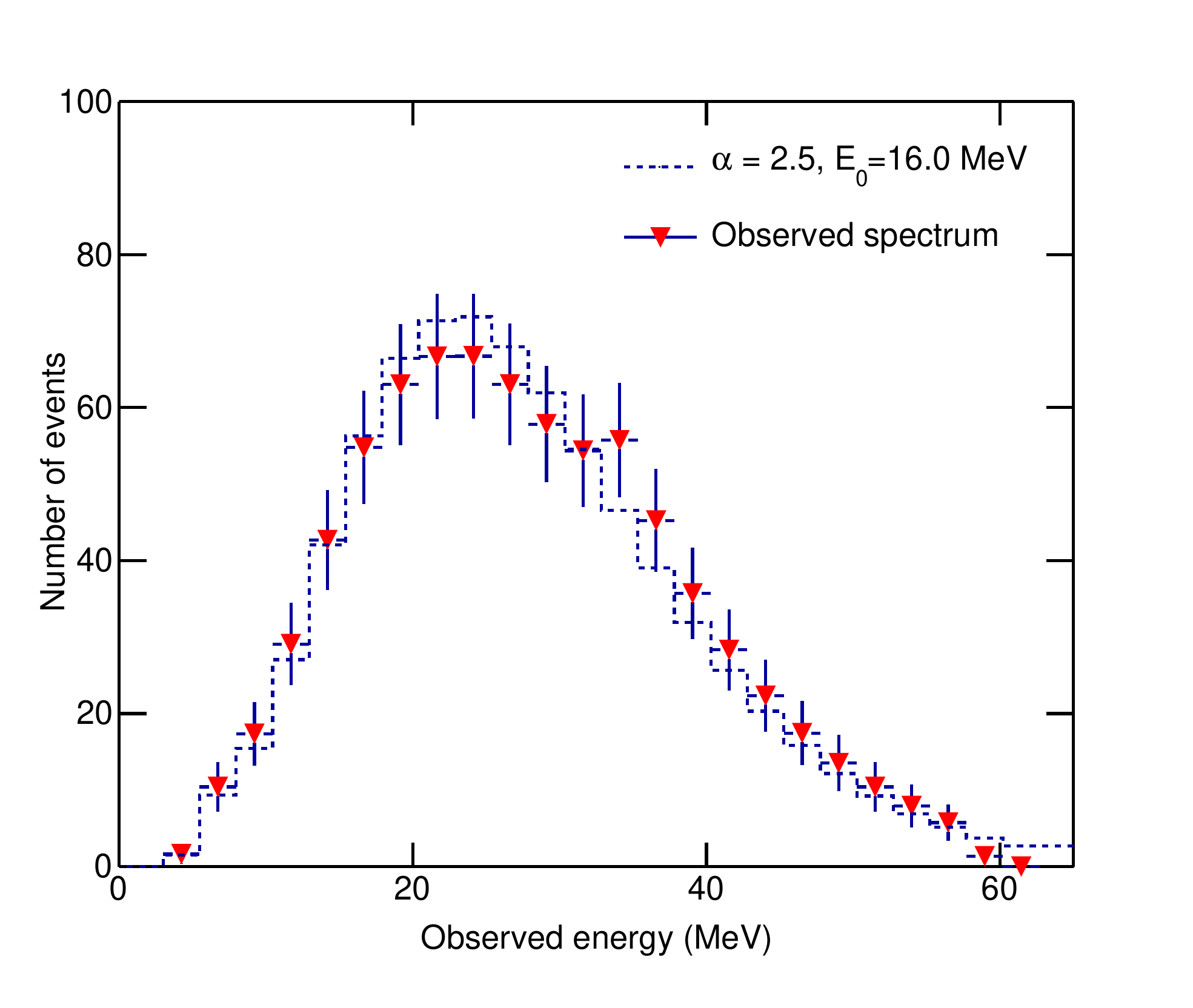}
}
 \caption[Observed $\nu_e$ spectra in 34~kt of LAr for a 10-kpc core
   collapse]{Observed $\nu_e$ spectra in 34~kt of liquid argon for a 10-kpc core
   collapse, representing about one second of integration time each at
   one-second intervals during the supernova cooling phase.  The dashed
   line represents the best fit to a parameterized pinched-thermal
   spectrum.  Clear \emph{non-thermal} features in the spectrum that
   change with time are visible, on the left at around 20~MeV and on
   the right at around 35~MeV.  Error bars are statistical.  These
   features are present \emph{only} for the normal MH.
}
\label{fig:shock_comparison}
\end{figure}

 Another potential MH signature is shown in
 Figure~\ref{fig:shock_comparison}, for which a clear time-dependent shock-wave-related feature is
 visible for the normal MH case. 

\begin{figure}[htb]
\centering\includegraphics[width=.7\textwidth]{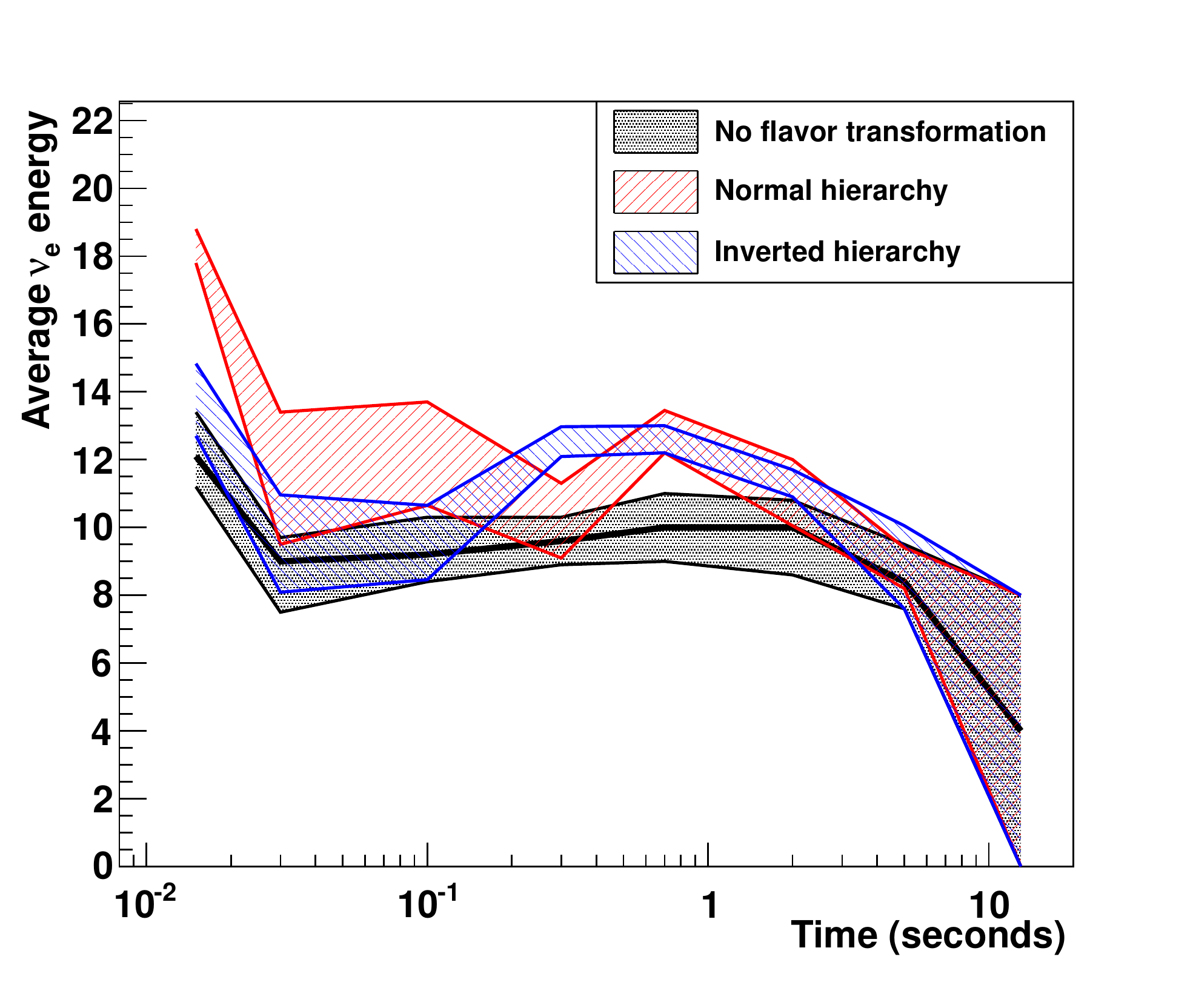}
\caption[Avg $\nu_e$ energy, SNOwGLoBES-smeared,
  pinched-thermal spectrum, 34~kt, 10~kpc]{Average $\nu_e$ energy from fit to SNOwGLoBES-smeared,
  pinched-thermal spectrum as a function of time (34~kt at 10~kpc), for a flux model
  based on~\cite{Keil:2002in} and including collective oscillations,
  for two different MH assumptions.  The
  bands represent 1$\sigma$ error bars from the fit. The solid black
  line is the truth $\langle E_\nu\rangle $ for the unoscillated
  spectrum.  
Clearly, meaningful information can
  be gleaned by tracking $\nu_e$ spectra as a function of time. 
}
\label{fig:temp_comparison}
\end{figure}

Figure~\ref{fig:temp_comparison} shows yet another example of a
preliminary study showing how one might track supernova temperature as
a function of time with the $\nu_e$ signal in liquid argon.  Here, a
fit is made to the pinched-thermal form of Equation~\ref{eq:pinched}.
Not only can the internal temperature of the supernova be effectively
measured, but the time evolution is observably different for the
different hierarchies.

\section{Low-Energy Backgrounds}
\label{sec:sn-le-backgrounds}

\subsection{Cosmic Rays}

Due to their low energy, supernova-neutrino events are subject to background
from cosmic rays, although the nature of the signal --- a
short-timescale burst --- is such that the background from these muons
and their associated Michel electrons can in principle be well known, 
easily distinguished and subtracted. Preliminary studies~\cite{COSMICBKGD} suggest that the shielding provided 
by the \ftadj{4850} depth available at the \SURF is acceptable.

\subsection{Local Radiation Sources}

It is possible that radioactive decays will directly
overlap with the energy spectrum created by supernova-neutrino events
in LBNE.  It is also possible for an ensemble of radioactive-decay
events in and around higher-energy particle interactions (e.g., from
beam neutrinos) to obscure the edges of electromagnetic showers from
highly scattering particles such as electrons and pions; this would
appear as the radiological equivalent of dark noise in a digital
image, and could potentially introduce a systematic uncertainty in the
energy calculated for events, even at much higher energy than the
decays themselves.
It is therefore very important to calculate the radioactive-decay
backgrounds in the LBNE far detector with sufficient accuracy to
properly account for their presence, either as direct backgrounds or
as systematic effects in energy calculations.
To this end, LBNE collaborators are in the process of creating a
physics-driven, radioactive-background budget and associated event
generator for low-energy background events in the far detector.

\begin{figure}[!hbt]
\centering
\includegraphics[width=0.78\textwidth]{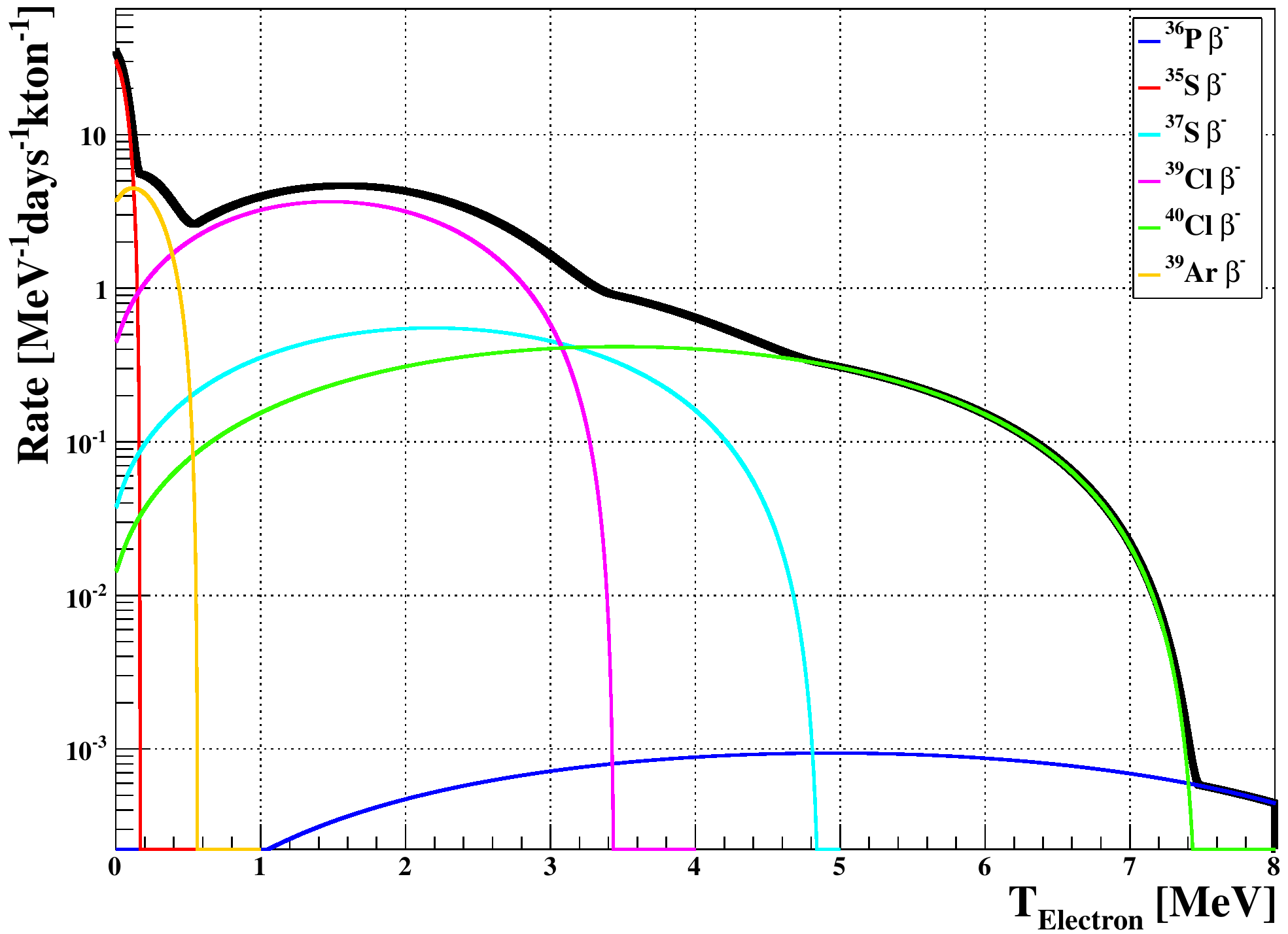} 
\caption[Cosmogenic backgrounds in the LArTPC at the 4850-ft
  level]{Cosmogenic background rates in the LBNE LArTPC as a function
  of the decay beta kinetic energy calculated at the
  \SIadj{4850}{\ft} level of the \SURF.}
\label{fig:cosmicbkg}
\end{figure}

The radioactive-background budget will have many components, each of which 
 will fall into one of two categories: 
\begin{enumerate}
\item intrinsic radioactive
contamination in the argon or support materials, or 
\item cosmogenic
radioactivity produced in situ from cosmic-ray showers
interacting with the argon or the support materials. 
\end{enumerate}
 The former is
dependent on the detector materials, and is
therefore independent of far detector depth.  
The latter is
strongly coupled to the cosmic-ray flux and spectrum. A preliminary
estimate~\cite{DOCDB8419} of the cosmogenic radioactivity from beta
emitters produced from cosmic-ray
interactions with argon in the LBNE far detector at the \SI{4850}{\ft} level of the
Sanford Underground Research Facility is shown in Figure
~\ref{fig:cosmicbkg}.
Both of these
background categories add to the direct energy
depositions from cosmic rays themselves and associated showers.
%

\subsection{Intrinsic Radioactive Background Mitigation}
\label{sec:Intrinsic}

Intrinsic backgrounds in the far detector come from the radioactive
material that is prevalent in the detector materials (both active and
instrumentation/support materials and the cryostat itself), in the
cavern walls and in the dust~\cite{Concrete}.  The isotopes of
primary interest are ``the usual suspects'' in experiments where
radioactive backgrounds must be controlled: $^{232}${Th} and
$^{238}${U} (and their associated decay chains), $^{40}${K}, and
$^{60}${Co}.  In addition, $^{39}${Ar} will contribute a significant
component, since it is present in natural argon harvested from the
atmosphere at the level of approximately 1 Bq/kg.  In consequence, a
\ktadj{10} far detector filled with $^{\rm nat.}${Ar} will experience
a rate from $^{39}${Ar} of approximately 10~MHz across the whole
detector.  The beta decay spectrum from $^{39}${Ar} is thankfully
quite low in energy (Q$_{\beta}$ = 0.565 MeV), so it will not
interfere directly with the supernova signal, but it may contribute to
the \emph{dark noise} effect.
Furthermore, the 
product of the average beta energy with this rate indicates the level at which 
the background due to introduction of power into the detector becomes a problem.
This radioactive power from $^{39}${Ar} is approximately:
\begin{equation}
P_{Rad} \sim 0.25 \mbox{ MeV} \times 10 \mbox{ MHz} = 2.5 \times 10^{6} \mbox{ MeV/s}.
\end{equation}
Because this category of background can come from the cavern walls,
the concrete cavern lining, the cryostat materials or the materials
that compose the submersed instrumentation,
it is important to know which type of radioactive decay is produced by
each isotope as well as the total energy it releases.  
For instance,
an alpha decay from an isotope in the U or Th decay chain will deposit
its full energy into the detector if it occurs in the active region of
the detector, but will deposit no energy if it occurs inside of some
macroscopically thick piece of support material because of its very
short range ($\lesssim$\SI{1}{\micro\meter}) in most solids.  This requires
different accounting for energy depositions from intrinsic radioactive
contamination measured in different locations (or groups of
locations).  This is clearly a tractable problem, but one which must
be handled with care and forethought.

Since a large body of work has been compiled on the control of
radiological background in previous experiments that have
encountered similar conditions, much of the work in this area will be
cited from these experiments (e.g., DARKSIDE~\cite{DARKSIDE},
EXO~\cite{Leonard:2007uv}, ICARUS, BOREXINO, KamLAND and \superk{}).
Work remains, however, on understanding the background particular to
the LBNE far detector location/depth (e.g., radon levels and dust
activity, for instance), and on integrating existing and new work into
the LBNE simulation, reconstruction and analysis framework.

\clearpage
\section{Summary of Core-Collapse Supernova Sensitivities}

\begin{introbox}
  LBNE, with its high-resolution LArTPC far detector, is uniquely
  sensitive to the $\nu_e$ component of the neutrino flux from a
  core-collapse supernova within our galaxy. The $\nu_e$ component of
  the neutrino flux dominates the initial neutronization burst of the
  supernova. Preliminary studies indicate that such a supernova at a
  distance of \SI{10}{\kilo \parsec} would produce $\sim$\num{3000} events
  in a \SIadj{34}{\kt} LArTPC. The time dependence of the signal will
  allow differentiation between different neutrino-driven
  core-collapse dynamical models, and will exhibit a discernible
  dependence on the neutrino mass hierarchy.

  A low energy threshold of $\sim$ \SI{5}{\MeV} will enable the detector
  to extract the rich information available from the $\nu_e$ supernova
  flux. LBNE's photon detection system is being designed to provide a
  high-efficiency trigger for supernova events. Careful design and quality control of the detector
  materials will minimize low-energy background from radiological
  contaminants.

\end{introbox} 

%% file: chapter-nd-physics.tex

\begin{introbox}
The LBNE near neutrino detector provides scientific value beyond
      its essential role of calibrating beam and neutrino interaction
      properties for the long-baseline physics program described in
      Chapter~\ref{nu-oscil-chap}.
      By virtue of the theoretically clean, purely weak leptonic
      processes involved,
      neutrino beams have historically served as unique probes for
      new physics in their interactions with matter.
      The high intensity and broad energy range of the LBNE beam
      will open the door for a highly capable near detector
      to perform its own diverse
program of incisive investigations.
\end{introbox}

The reduction of systematic uncertainties for the neutrino oscillation
program 
requires excellent resolution in the
reconstruction of neutrino events. Combined with the unprecedented
neutrino fluxes available 
--- which will
allow the collection of ${\cal{O}}$(\num{e8}) inclusive neutrino charged
current (CC) interactions for 
\num{e22} protons-on-target (POT) just downstream of the
beamline --- the 
near detector (ND)  
will significantly enhance the LBNE long-baseline 
oscillation program and produce a range of short-baseline neutrino
scattering physics measurements.  The combined statistics and
resolution expected in the ND will allow precise tests of fundamental
interactions resulting in a better understanding of the structure of
matter. 

Table~\ref{tab:rates} lists the expected number of 
beam-neutrino interactions per ton of detector at the LBNE ND site,
located \SI{459}{\meter} downstream from
the 
target.  

This chapter presents a short description of some of the studies that
can be performed with LBNE's fine-grained near neutrino detector
and gives a flavor of the outstanding physics potential. A more
detailed and complete discussion of the ND physics
potential can be found in~\cite{docdb-6704}.

Appendix~\ref{app-dis} describes neutrino scattering 
kinematics and includes
definitions of the kinematic variables used in this chapter.

\begin{table}[!htb]
\centering
\caption[Interaction rates, $\nu$ mode, per ton
for \SI{1e20}{\POT}, \SI{459}{\meter}, \SI{120}{\GeV}]{Estimated interaction rates in the neutrino (second column) and antineutrino (third column) beams per ton of detector (water) 
  for \SI{1e20}{\POT} at \SI{459}{\meter} assuming neutrino
  cross-section predictions from NUANCE~\cite{Casper:2002sd} and a \GeVadj{120}
  proton beam using the CDR reference design.  Processes are defined at the initial neutrino
  interaction vertex and thus do not include final-state effects. These estimates do not
  include detector efficiencies or acceptance~\cite{DOCDB740,DOCDB783}. 
}
\label{tab:rates}
\begin{tabular}[!htbp]{$L^r^r}
\toprule
\rowtitlestyle
Production mode & $\nu_\mu$ Events & $\overline\nu_\mu$ Events\\
\toprowrule
CC QE ($\nu_\mu n \rightarrow \mu^- p$)                             & 50,100 & 26,300 \\ \colhline
NC elastic ($\nu_\mu N \rightarrow \nu_\mu N$)                      & 18,800 & 8,980 \\ \colhline
CC resonant $\pi^+$ ($\nu_\mu N \rightarrow \mu^- N \pi^+$)         & 67,800 & 0 \\ \colhline
CC resonant $\pi^-$ ($\overline{\nu}_\mu N \rightarrow \mu^+ N \pi^-$)   & 0      & 20,760 \\ \colhline
CC resonant $\pi^0$ ($\nu_\mu n \rightarrow \mu^- \ p \, \pi^0$)    & 16,200 & 6,700 \\ \colhline
NC resonant $\pi^0$ ($\nu_\mu N \rightarrow \nu_\mu \, N \, \pi^0$) & 16,300 & 7,130 \\ \colhline
NC resonant $\pi^+$ ($\nu_\mu p \rightarrow \nu_\mu \, n \, \pi^+$) & 6,930  & 3,200 \\ \colhline
NC resonant $\pi^-$ ($\nu_\mu n \rightarrow \nu_\mu \, p \, \pi^-$) & 5,980  & 2,570 \\ \colhline
CC DIS ($\nu_\mu N \rightarrow \mu^- X$ or 
$\overline{\nu}_\mu N \rightarrow \mu^+ X$, $W>2$)                     & 66,800 & 13,470 \\ \colhline
NC DIS ($\nu_\mu N \rightarrow \nu_\mu X$ or 
$\overline{\nu}_\mu N \rightarrow \overline{\nu}_\mu X$, $W>2$)                   & 24,100 & 5,560 \\ \colhline
NC coherent $\pi^0$ ($\nu_\mu A \rightarrow \nu_\mu A \pi^0$ or 
$\overline{\nu}_\mu A \rightarrow \overline{\nu}_\mu A \pi^0$
)       & 2,040  & 1,530 \\
CC coherent $\pi^+$ ($\nu_\mu A \rightarrow \mu^- A \pi^+$)         & 3,920  &  0 \\ \colhline
CC coherent $\pi^-$ ($\overline{\nu}_\mu A \rightarrow \mu^+ A \pi^-$)   & 0      & 2,900 \\ \colhline
NC resonant radiative decay ($N^* \rightarrow N \gamma $)          & 110    & 50 \\ \colhline
NC elastic electron ($\nu_\mu e^- \rightarrow \nu_\mu e^-$  
or  $\overline{\nu}_\mu e^- \rightarrow \overline{\nu_\mu} e^-$)              & 30  & 17 \\ \colhline
Inverse Muon Decay ($\nu_\mu e \rightarrow \mu^- \nu_e$)            & 12  & 0\\ \colhline
Other                                                              & 42,600 & 15,800 \\ 
\toprule
\rowtitlestyle
Total CC  (rounded)                                                       & 236,000 & 81,000 \\ 
\rowtitlestyle
Total NC+CC  (rounded)                                                      & 322,000 & 115,000 \\
\bottomrule
\end{tabular}
\end{table}

\section{Precision Measurements with Long-Baseline Oscillations}
\label{sec-fluxosc}

From the studies of uncertainties and the impact of the spectral shape
presented in Section~\ref{sec:systs}, it is evident that to fully
realize the goals of the full LBNE scientific program --- in
particular, sensitivity to CP violation and the precision measurement
of the three-flavor oscillation parameters --- it is necessary to
characterize the expected unoscillated neutrino flux with high
precision. In addition to the precise determination of the neutrino
flux, shape and flavor composition, the characterization of different
neutrino interactions and interaction cross sections on a liquid argon target
is necessary to estimate physics backgrounds to the oscillation
measurements.  The high-resolution near tracking detector 
described in Section~\ref{sec:ndproj} can measure the unoscillated flux
normalization, shape and flavor to a few percent using systematically
independent techniques that are 
discussed in the following sections.

\subsection{Determination of the Relative Neutrino and Antineutrino Flux} 
\label{sec-lownu0}

The most promising method of determining the shape of the \nm\ and
\anm\ flux is by measuring CC events with low 
hadronic-energy deposition (low-$\nu$) where $\nu$ is the total energy of the
hadrons that are produced after a neutrino interaction, $E_\nu -
E_\mu$. It is important to note that not all the hadrons escape the
remnant nucleus, and intranuclear effects will smear the visible energy
of the hadronic system.  A method of relative flux determination known
as low-$\nu_0$ --- where $\nu_0$ is a given value of visible hadronic
energy in the interaction that is selected to minimize the fraction of
the total interaction energy carried by the hadronic system
--- is well developed~\cite{srmishra-reviewtalk}.  The method follows
from the general expression of the $\nu$-nucleon differential cross
section:
\begin{equation}
{\cal N} (\nu < \nu_0) \simeq C \Phi(E_\nu) \nu_0 \left[ {\cal A} +
\left( \frac{\nu_0}{E_\nu} \right) {\cal B} + \left( \frac{\nu_0}{E_\nu} \right)^2 {\cal C} +
{\cal O} \left( \frac{\nu_0}{E_\nu} \right)^3 \right],
\end{equation}
\noindent
where the coefficients are ${\cal A} = {\cal F}_2$, ${\cal B} = ({\cal
  F}_2 \pm {\cal F}_3)/2$, ${\cal C} = ({\cal F}_2 \mp {\cal F}_3)/6$, 
and ${\cal F}_i =\int^1_0 \int^{\nu_0}_0 F_i(x) dx d\nu$ is the
integral of structure function $F_i(x)$.  
The dynamics of
neutrino-nucleon scattering  implies that the number of events in a
given energy bin with hadronic energy $E_{\rm had} < \nu_0$ is
proportional to the (anti)neutrino flux in that energy bin up
to corrections ${\cal O}(\nu_0/E_\nu)$ and ${\cal O}(\nu_0/E_\nu)^2$.
The number ${\cal
  N}(\nu<\nu_0)$ is therefore proportional to the flux up to correction factors
of the order ${\cal O} (\nu_0/E_\nu)$ or smaller, which are not
significant for small values of $\nu_0$ at energies $\geq \nu_0$. 
 The coefficients ${\cal A}$, ${\cal B}$ and ${\cal C}$ are
determined for each energy bin and neutrino flavor within the ND data.

LBNE's primary interest is the relative flux
determination, i.e., the neutrino flux in one energy bin relative to that in
another; variations in the coefficients do not affect the
relative flux. The prescription for the relative flux determination is
simple: count the number of 
neutrino CC events below a certain small
value of hadronic energy ($\nu_0$).  The observed number of events, up
to the correction of the order ${\cal O} (\nu_0/E_\nu)$ due to the
finite $\nu_0$ in each total visible energy bin, is proportional to
the relative flux. The smaller the factor $\nu_0/E_\nu$ is, the smaller
is the correction.  Furthermore, the energy of events passing the
low-$\nu_0$ cut is dominated by the corresponding lepton energy. 

It is
apparent from the above discussion that this method of relative flux
determination is not very sensitive to nucleon structure, QCD
corrections or types of neutrino interactions such as scaling or
nonscaling. With the excellent granularity and resolution foreseen in
the low-density magnetized tracker, it will be possible to use a value
of $\nu_0\sim$\SI{0.5}{\GeV} or lower, thus allowing flux predictions down to
$E_\nu\sim$\SI{0.5}{\GeV}. A preliminary analysis with the high-resolution
tracker achieved a precision $\leq 2\%$ on the relative $\nu_\mu$
flux with the low-$\nu_0$ method in the energy region $1 \leq E_\nu
\leq 30$ \si{GeV} in the fit with $\nu_0 < 0.5$ \si{\GeV}. Similar uncertainties
are expected for the $\overline{\nu}_\mu$ component (the dominant one) in
the antineutrino beam mode (negative focusing).

\subsection{Determination of the Flavor Content of the Beam: $\boldsymbol{\nu_\mu, \overline{\nu}_\mu, \nu_e, \overline{\nu}_e}$}

The empirical parameterization 
of the pion and kaon neutrino parents produced from the proton target,
determined from the low-$\nu_0$ flux at the ND, allows prediction of
the $\nu_\mu$ and $\overline{\nu}_\mu$ flux at the far detector
location.  This parameterization provides a measure of the
$\pi^+/K^+/\mu^+(\pi^-/K^-/\mu^-)$ distributions of neutrino parents
of the beam observed in the ND.  Additionally, with the capability to
identify $\overline{\nu}_e$ CC interactions, it is possible to
directly extract the elusive $K^0_L$ content of the beam.  Therefore,
an accurate measurement of the $\nu_\mu, \overline{\nu}_\mu$ and
$\overline{\nu}_e$ CC interactions provides a prediction of the
$\nu_e$ content of the beam, which is an irreducible background for
the $\nu_e$ appearance search in the far detector:

\begin{eqnarray} \label{eqn:nueparents}
\nu_e & \equiv & \mu^+(\pi^+\to \nu_\mu) \oplus K^+(K^+\to \nu_\mu) \oplus K^0_L\\
\overline{\nu}_e & \equiv & \mu^-(\pi^-\to \overline{\nu}_\mu) \oplus K^-(K^-\to \overline{\nu}_\mu) \oplus K^0_L
\end{eqnarray}

The $\mu$ component is well constrained from $\nu_\mu
(\overline{\nu}_\mu)$ CC data at low energy, while the $K^\pm$
component is only partially constrained by the $\nu_\mu
(\overline{\nu}_\mu)$ CC data at high energy and requires external
hadro-production measurements of $K^\pm/\pi^\pm$ ratios at low energy
from hadro-production experiments such as MIPP~\cite{Raja:2005sh} and
NA61~\cite{Korzenev:2013gia}.  Finally, the $K_L^0$ component can be
constrained by the $\overline{\nu}_e$ CC data and by external
dedicated measurements at hadron-production experiments.  In the
energy range $1 (5) \leq E_\nu \leq 5 (15)$ \si{GeV}, the approximate
relative contributions to the $\nu_e$ spectrum are 85\% (55\%) from
$\mu^+$, 10\% (30\%) from $K^+$ and 3\% (15\%) from $K_L^0$.

Based on the NOMAD experience, 
a precision of $\leq 0.1\%$ on the flux ratio $\nu_e/\nu_\mu$ is
expected at high energies. Taking into account the projected precision
of the $\nu_\mu$ flux discussed in Section~\ref{sec-lownu0}, this
translates into an absolute prediction for the $\nu_e$ flux at the
level of $2\%$.

Finally, the fine-grained ND can directly identify $\nu_e$ CC
interactions from the LBNE beam. The relevance of this measurement is
twofold:
\begin{enumerate}
\item It provides an independent
validation for the flux predictions obtained from the low-$\nu_0$ method.
\item It can
further constrain the uncertainty on the knowledge of the absolute $\nu_e$ flux.
\end{enumerate}

\subsection{Constraining the Unoscillated $\boldsymbol{\nu}$ Spectral Shape with the QE Interaction}

In any long-baseline neutrino oscillation program, including LBNE, the
quasi-elastic (QE) interactions are special.  First, the QE cross
section is substantial at lower energies~\cite{Formaggio:2013kya}.
Second, because of the simple topology (a $\mu^-$ and a proton), the
visible interaction energy provides, to first order, a close
approximation to the neutrino energy ($E_\nu$).  
In the context of a fine-grained tracker, a precise measurement of QE
will impose direct constraints on nuclear effects related to both the
primary and final-state interaction (FSI) dynamics 
(Section~\ref{sec-nuclear}), which can affect the overall neutrino
energy scale and, thus, the entire oscillation program.  To this end,
the key to reconstructing a high-quality sample of $\nu_\mu$ QE
interactions is the two-track topology where both final-state
particles are visible: $\mu^-$ and $p$. A high-resolution ND can
efficiently identify the recoil proton and measure its momentum vector
as well as $dE/dx$. Preliminary studies indicate that in a
fine-grained tracking detector the efficiency (purity) for the proton
reconstruction in QE events is $52\%$ ($82\%$). A comparison between
the neutrino energy reconstructed from the muon momentum through the
QE kinematics (assuming a free target nucleon) with the visible
neutrino energy measured as the sum of $\mu$ and $p$ energies is
sensitive to both nuclear effects and FSI. Furthermore, comparing the
two-track sample ($\mu$ and $p$) with the single-track sample (in which only $\mu$
is reconstructed) empirically constrains the rate of FSI.

\subsection{Low-Energy Absolute Flux: Neutrino-Electron NC Scattering}
\label{ssec:ncscatter}

Neutrino neutral current (NC) interaction with the atomic electron in the
target, $\nu_\mu e^- \rightarrow \nu_\mu e^-$, provides an elegant
measure of the absolute flux.  The total cross section for NC elastic
scattering off electrons is given by~\cite{Marciano:2003eq}:
\begin{eqnarray}
\sigma (\nu_l e \to \nu_l e) & = & \frac{G_\mu^2 m_e E_\nu}{2\pi} \left[ 1 -4 \sin^2 \theta_W + \frac{16}{3} \sin^4 \theta_W \right], \\
\sigma (\overline{\nu}_l e \to \overline{\nu}_l e) & = & \frac{G_\mu^2 m_e E_\nu}{2\pi} \left[ \frac{1}{3} -\frac{4}{3} \sin^2 \theta_W + \frac{16}{3} \sin^4 \theta_W \right], 
\end{eqnarray}

\noindent
where $\theta_W$ is the weak mixing angle (WMA).  For the currently
known value of $\sin^2 \theta_W\simeq0.23$, the above cross sections
are very small: $\sim 10^{-42} (E_\nu/{\rm GeV})$~cm$^2$. The NC
elastic scattering off electrons can be used to determine the absolute
flux normalization since the cross section only depends on the
knowledge of $\sin^2 \theta_W$. Within the Standard Model, the value
of $\sin^2 \theta_W$ at the average momentum transfer expected at
LBNE, $Q\sim0.07$~\si{\GeV}, can be extrapolated down from the
LEP/SLC\footnote{LEP was the Large Electron-Positron Collider at CERN
  that operated from 1989 to 2000 and provided a detailed study of the
  electroweak interaction.}  measurements with a precision of $\leq
1\%$. The $\nu_\mu e^- \rightarrow \nu_\mu e^-$ will produce a single
$e^-$ collinear with the $\nu$-beam ($\leq 40$~mrad).  The background,
dominated by the asymmetric conversion of a photon in an ordinary
$\nu$-nucleon NC event, will produce $e^-$ and $e^+$ in equal measure
with much broader angular distribution.  A preliminary analysis of the
expected elastic scattering signal in the high-resolution tracking ND
shows that the scattering signal can be selected with an efficiency of
about 60\% with a small background contaminant. The measurement will
be dominated by the statistical error. 
The determination of the absolute flux of the LBNE neutrinos is
estimated to reach a precision of $\simeq 2.5\%$ for $E_\nu \leq
10$~\si{\GeV}.  The measurement of NC elastic scattering off electrons
can only provide the integral of all neutrino flavors.

\subsection{High-Energy Absolute Flux: Neutrino-Electron CC Scattering}

The \nm-$e^-$ CC interaction, $\nu_\mu + e^- \rightarrow \mu^- +
\nu_e$ (\emph{inverse muon decay} or \emph{IMD}), offers an elegant
way to determine the absolute flux. Given the energy threshold needed
for this process, IMD requires 
$E_\nu \geq 10.8$~\si{\GeV}.  The high-resolution ND in the
LBNE neutrino beam will observe $\geq$ \num{2000} IMD events in three
years. The reconstruction efficiency of the single, energetic 
forward $\mu^-$ will be $\geq$ 98\%; the angular resolution of the
IMD $\mu$ is $\leq$ \SI{1}{\mrad}. The background, primarily from the
$\nu_\mu$-QE interactions, can be precisely constrained using control
samples.  In particular, the systematic limitations of the CCFR
(\cite{Mishra:1989jn,Mishra:1990yf}) and 
the CHARM-II~\cite{Vilain:1996yf} IMD measurements can be
substantially alleviated in LBNE with the proposed ND design. A
preliminary analysis indicates that the absolute flux can be
determined with an accuracy of $\approx 3\%$ for $E_\nu \geq$
\SI{11}{\GeV} (average $E_\nu \approx$\SI{25}{\GeV}).

\subsection{Low-Energy Absolute Flux: QE in Water and Heavy-Water Targets}

Another  
independent method to extract the absolute flux is through the
QE-CC scattering ($\nu_\mu n(p) \to \mu^- p(n)$) on
deuterium at low $Q^2$. Neglecting terms in $(m_\mu/M_n)^2$ at $Q^2=0$,
the QE cross section is independent of neutrino energy for $(2E_\nu
M_n)^{1/2} > m_\mu$:
\begin{equation}
\frac{d \sigma}{d Q^2}  \mid {Q^2=0}\mid = \frac{G_\mu^2 \cos^2 \theta_c}{2\pi}
\left[ F_1^2(0) + G_A^2(0) \right] = 2.08 \times 10^{-38}~\rm cm^2{\rm GeV}^{-2},
\end{equation}
\noindent 
which is determined by neutron $\beta$ decay and has a theoretical
uncertainty $<1\%$.  The flux can be extracted experimentally by
measuring low $Q^2$ QE interactions ($ \leq 0.05$ GeV) and extrapolating
the result to the limit of $Q^2=0$. The measurement requires a
deuterium (or hydrogen for antineutrino) target to minimize the
smearing due to Fermi motion and other nuclear effects. This
requirement can only be achieved by using both H$_2$O and D$_2$O
targets embedded in the fine-grained tracker and extracting the events
produced in deuterium by statistical subtraction of the larger oxygen
component.  The experimental resolution on the muon and proton
momentum and angle is crucial.  Dominant uncertainties of the method
are related to the extrapolation to $Q^2=0$, to the theoretical cross
section on deuterium, to the experimental resolution and to the
statistical subtraction.  Sensitivity studies and the experimental
requirements are under study.

\subsection{Neutral Pions, Photons and $\boldsymbol{\pi^{\pm}}$ in NC and CC Events}
\label{sec-bkgnds}

The principal background to the $\nu_e$ and $\overline{\nu}_e$
appearance comes from the NC events where a photon from the $\pi^0$
decay produces a signature similar to that produced by $\nu_e$-induced
electron; the second source of background is due to $\pi^0$'s from
$\nu_\mu$ CC where the $\mu^-$ evades identification --- typically at
high $y_{Bj}$.  Since the energy spectra of NC and CC interactions are
different, it is critical for the ND to measure $\pi^0$'s in NC and CC
interactions in the full kinematic phase space.
 
The proposed ND is designed to measure $\pi^0$'s with 
high accuracy in three topologies: 
\begin{enumerate}
\item Both photons convert 
in the tracker ($\simeq$25\%).
\item One photon converts  
in the tracker and the other in the calorimeter ($\simeq$50\%). 
\item Both photons convert in the calorimeter;  
the first two topologies afford the best resolution 
because the tracker provides precise $\gamma$-direction measurement. 
\end{enumerate}

The $\pi^0$ reconstruction efficiency in the proposed fine-grained tracker is
expected to be $\geq$75\% if photons that reach the ECAL are
included.   By contrasting the $\pi^0$ mass  in the tracker
versus in the calorimeter, the relative efficiencies 
of photon reconstruction will be well constrained. 

Finally, the $\pi^{\pm}$ track momentum and $dE/dx$ information will
be measured by the tracker.  An in situ determination of the charged
pions in the $\nu_{\mu}/\overline{\nu}_\mu$ CC events --- with $\mu$ID and
without $\mu$ID --- and in the $\nu$ NC events is crucial to constrain
the systematic error associated with the \nm(\anm) disappearance,
especially at low $E_\nu$.

\subsection{Signal and Background Predictions for the Far Detector} 
\label{sec-extfd} 

In order to achieve reliable predictions for signal and backgrounds in the far detector, near detector measurements --- including (anti)neutrino fluxes, nuclear cross sections and detector 
smearing --- must be unfolded and extrapolated to the far detector location. 
The geometry of the beam and detectors (point source versus extended source) 
as well as the expected neutrino oscillations imply differences in the (anti)neutrino fluxes 
 in the near and far detectors. 
These differences, in turn, will result in increased sensitivity of the long-baseline analysis to cross-section uncertainties, in particular between neutrinos and antineutrinos and for exclusive background topologies. 
Furthermore, the much higher event rates at the near site and the 
smaller detector size (i.e., reduced containment) make it virtually impossible to achieve identical measurement 
conditions in both the near and far detectors. However, as discussed in 
Sections~\ref{sec-lownu0} to~\ref{sec-bkgnds}, the energy, angular and 
space resolution of the low-density 
ND are key factors in reducing the systematic uncertainties achievable 
on the event predictions for the far detector; the ND can offer a precise \emph{in situ} 
measurement of the absolute flux of all flavor components of the beam, 
$\nu_\mu, \nu_e, \bar\nu_\mu, \bar \nu_e$, resulting in constraints on the parent 
$\pi^\pm/K^\pm/\mu^\pm$ distributions. 
In addition, measurements of momenta and energies of final-state particles produced 
in (anti)neutrino interactions will allow a detailed study of exclusive topologies affecting the 
signal and background rates in the far detector. 
All of these measurements will be used to cross-check and fine-tune the simulation programs  
needed for the actual extrapolation from the near to the far detector. 

It is important to note that several of these techniques have already been used and \emph{proven to work} 
in neutrino experiments such as MINOS~\cite{Adamson:2009ju} and 
NOMAD~\cite{Wu:2007ab,Lyubushkin:2008pe,Samoylov:2013xoa}. 
The higher segmentation and resolution in the LBNE ND with respect to past experiments 
will increase the available information about the (anti)neutrino event topologies, allowing further 
reduction of systematic uncertainties both in the ND measurements and in the Monte Carlo extrapolation.  

For a more detailed discussion of the impact of ND measurements on the long-baseline oscillation analysis see 
Section~\ref{sec:systs}.

\clearpage
\section{Electroweak Precision Measurements} 
\label{sec-ew-wma}

\begin{introbox}
  Neutrinos and antineutrinos are the most effective probes for
  investigating electroweak physics.  Interest in a precise
  determination of the weak mixing angle ($\sin^2 \theta_W$) at LBNE
  energies via neutrino scattering is twofold: (1) it provides a
  direct measurement of neutrino couplings to the $Z$ boson and (2) it
  probes a different scale of momentum transfer than LEP 
did by virtue of not being at the $Z$ boson mass peak. 
\end{introbox}

\begin{figure}[!htb]
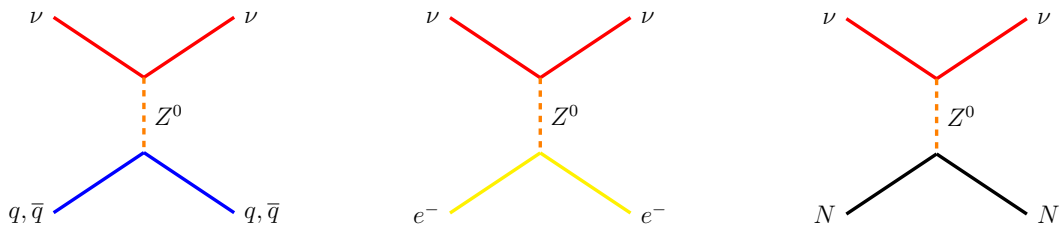

\centering

  \feynmanNC{$\nu$}{neutrino}{$q,\overline{q}$}{quark}{0.3\linewidth}
  \feynmanNC{$\nu$}{neutrino}{$e^-$}{lepton}{0.3\linewidth}
  \feynmanNC{$\nu$}{neutrino}{$N$}{hadron}{0.3\linewidth}
  
  \caption[Feynman diagrams for the three main NC
  processes]{Feynman diagrams for the three main neutral current
    processes that can be used to extract $\sin^2 \theta_W$ with the
    LBNE near detector.  From left, deep inelastic scattering off
    quarks, elastic scattering off electrons and elastic scattering
    off nucleons.  }
\label{fig:graphs}
\end{figure}

The weak mixing angle can be extracted
experimentally from three main NC physics processes:
\begin{enumerate}
\item deep inelastic scattering off quarks inside nucleons: $\nu N \to \nu X$
\item elastic scattering off electrons: $\nu e^- \to \nu e^-$
\item elastic scattering off protons: $\nu p \to \nu p$
\end{enumerate}
Figure~\ref{fig:graphs} shows the Feynman diagrams corresponding to the three processes.

\subsection{Deep Inelastic Scattering} 
\label{ssec:nd:dis}
The most precise measurement of $\sin^2 \theta_W$ in
neutrino deep inelastic scattering (DIS) comes from the NuTeV experiment, which reported
a value that is $3\sigma$ from the Standard Model~\cite{Zeller:2001hh}. 
The LBNE ND can perform a similar
analysis in the DIS channel by measuring the ratio of NC and CC interactions induced by
neutrinos:
\begin{equation}
{\cal R}^\nu \equiv \frac{\sigma^\nu_{\rm NC}}{\sigma^\nu_{\rm CC}}
 \simeq \rho^2 \left( \frac{1}{2} - \sin^2 \theta_W +\frac{5}{9} \left(1 + r \right) \sin^4 \theta_W  \right).
\end{equation}

\noindent
Here $\rho$ is the relative coupling strength of the
neutral-to-charged current interactions ($\rho =1$ at tree-level in
the Standard Model) and $r$ is the ratio of antineutrino to neutrino
cross section ($r\sim0.5$).  The absolute sensitivity of ${\cal
  R}^\nu$ to $\sin^2 \theta_W$ is 0.7, which implies that a
measurement of ${\cal R}^\nu$ to 1\% precision would in turn provide a
1.4\% precision on $\sin^2 \theta_W$.  This technique was used by the
CDHS~\cite{Abramowicz:1986vi}, CHARM~\cite{Allaby:1987vr} and CCFR~\cite{Reutens:1985hv} 
experiments. In contrast to the NuTeV experiment, the antineutrino
interactions cannot be used for this analysis at LBNE due to the large
number of $\nu_\mu$ DIS interactions in the $\overline{\nu}_\mu$ beam
compared to the $\overline{\nu}_\mu$ DIS interactions.

The measurement of $\sin^2 \theta_W$ from DIS interactions can only be
performed with a low-density magnetized tracker since an accurate
reconstruction of the NC event kinematics and of the $\nu$ CC
interactions are crucial for keeping the systematic uncertainties on
the event selection under control. The analysis selects events in the
ND after imposing a cut on the visible hadronic energy of $E_{\rm had}
>$~\SI{5}{\GeV} (the CHARM analysis had $E_{\rm had} >$~\SI{4}{\GeV}).
With an exposure of $5\times 10^{21}$ POT in the \SIadj{120}{\GeV}
beam using the CDR reference design, about $7.7 \times 10^6$ CC events
and $2.4 \times 10^6$ NC events are expected, giving a statistical
precision of 0.074\% on ${\cal R}^\nu$ and 0.1\% on $\sin^2 \theta_W$
(Table~\ref{tab:NuTeV-sin2tw}).

\begin{table}[!htb]
\centering
\caption[Uncertainties on the ${\cal R}^\nu$ measurement, NuTeV versus LBNE]{Comparison of uncertainties on the ${\cal R}^\nu$ measurement between NuTeV and LBNE with a 5 t fiducial mass 
  after an exposure of $5\times 10^{21}$ POT (5 year) with the CDR reference \SIadj{120}{\GeV} beam. The corresponding relative uncertainties on $\sin^2 \theta_W$ must
  be multiplied by a factor of 1.4, giving for LBNE a projected overall precision of 0.35\%.}
\label{tab:NuTeV-sin2tw}
\begin{tabular}{$L^l^l^l}
\toprule
\rowtitlestyle
Source of uncertainty & \multicolumn{2}{^>{\columncolor{\ChapterBubbleColor}}c}{~~~~~~~ $\delta R^{\nu}/R^{\nu}$~~~~~~~ } & 
Comments \\
 \rowtitlestyle
& NuTeV & LBNE & \\ 
\toprowrule
 Data statistics & 0.00176 & 0.00074 & \\ \colhline

 Monte Carlo statistics & 0.00015   &  & \\ \colhline

 \textit{Total Statistics} &  \textit{0.00176} &  \textit{0.00074} & \\
 \midrule

$\nu_{e}, \overline{\nu}_{e}$ flux ($\sim1.7\%$) & 0.00064 &  0.00010 & 
$e^-/e^+$ identification \\ \colhline

 Energy measurement &  0.00038 &  0.00040 & \\ \colhline
 Shower length model &  0.00054 &  n.a. & \\ \colhline
 Counter efficiency, noise &  0.00036 &  n.a. & \\ \colhline

 Interaction vertex & 0.00056 &  n.a. & \\ \colhline
 $\overline{\nu}_\mu$ flux    &  n.a. &  0.00070 & Large $\bar \nu$ contamination \\ \colhline
 Kinematic selection    &  n.a. &  0.00060 & Kinematic identification of NC \\  \colhline

  \textit{Experimental systematics} &  \textit{0.00112} &   \textit{0.00102} & \\ 
\midrule 
 d,s$\rightarrow$c, s-sea &  0.00227 &  0.00140 & Based on existing knowledge \\ \colhline

 Charm sea &  0.00013  &   n.a. & \\
 $r = \sigma^{\overline{\nu}}/\sigma^{\nu}$ &  0.00018 &  n.a. & \\ \colhline
 Radiative corrections & 0.00013 &  0.00013 & \\ \colhline

 Non-isoscalar target &  0.00010 &  N.A. &  \\ \colhline
 Higher twists &  0.00031 &   0.00070 & Lower $Q^2$ values \\ \colhline

 $R_{L}$ ($F_2,F_T,xF_3$) &  0.00115 &   0.00140 & Lower $Q^2$ values \\ \colhline
 Nuclear correction    &        &  0.00020 &  \\  \colhline
  \textit{Model systematics} &   \textit{0.00258} &    \textit{0.00212} & \\ 
\toprule
\rowtitlestyle
 Total  &  0.00332 &     0.00247 &  \\
\bottomrule
\end{tabular}
\end{table}

The use of a low-density magnetized tracker can substantially reduce
systematic uncertainties compared to a massive
calorimeter. Table~\ref{tab:NuTeV-sin2tw} shows a comparison of the
different uncertainties on the measured ${\cal R}^\nu$ between NuTeV
and LBNE.  While NuTeV measured both ${\cal R}^\nu$ and ${\cal
  R}^{\overline{\nu}}$, the largest experimental uncertainty in the
measurement of ${\cal R}^\nu$ is related to the subtraction of the
$\nu_e$ CC contamination from the NC sample. Since the low-density
tracker at LBNE can efficiently reconstruct the electron tracks, the
$\nu_e$ CC interactions can be identified on an event-by-event basis,
reducing the corresponding uncertainty to a negligible
level. Similarly, uncertainties related to the location of the
interaction vertex, noise, counter efficiency and so on are removed by
the higher resolution and by changing the analysis selection. The
experimental selection at LBNE will be dominated by two uncertainties:
the knowledge of the $\overline{\nu}_\mu$ flux and the kinematic
selection of NC interactions. The former is relevant due to the larger
NC/CC ratio for antineutrinos. The total experimental systematic
uncertainty on $\sin^2 \theta_W$ is expected to be about 0.14\%.

The measurement of ${\cal R}^\nu$ will be dominated by theoretical
systematic uncertainties on the structure functions of the
target nucleons.  The estimate of these uncertainties for LBNE is
based upon the extensive work performed for the NOMAD analysis and
includes a Next-to-Next-Leading-Order (NNLO) QCD calculation of
structure functions (NLO for charm
production)~\cite{Alekhin:2007fh,Alekhin:2008ua,Alekhin:2008mb},
parton distribution functions (PDFs) extracted from dedicated low-$Q$
global fits, high-twist contributions~\cite{Alekhin:2007fh},
electroweak corrections~\cite{Arbuzov:2004zr} and nuclear
corrections~\cite{Kulagin:2004ie,Kulagin:2007ju,Kulagin:2010gd}. The
charm quark production in CC, which has been the dominant source of
uncertainty in all past determinations of $\sin^2 \theta_W$ from
$\nu$N DIS, is reduced to about 4\% of the total $\nu_\mu$ CC DIS for
$E_{\rm had}>5$~GeV with the low-energy beam spectrum at LBNE.  This
number translates into a systematic uncertainty of 0.14\% on ${\cal
  R}^\nu$ (Table~\ref{tab:NuTeV-sin2tw}), assuming the current
knowledge of the charm production cross section.  It is worth noting
that the recent measurement of charm dimuon production by the NOMAD
experiment allowed a reduction of the uncertainty on the strange sea
distribution to $\sim3\%$ and on the charm quark mass $m_c$ to
$\sim75$~MeV~\cite{Samoylov:2013xoa}. The
lower neutrino energies available at LBNE reduce the accessible $Q^2$
values with respect to NuTeV, increasing in turn the effect of
non-perturbative contributions (high twists) and $R_L$. The
corresponding uncertainties are reduced by the recent studies of
low-$Q$ structure functions and by improved modeling with respect to
the NuTeV analysis (NNLO vs. LO).  The total model systematic
uncertainty on $\sin^2 \theta_W$ is expected to be about 0.21\% with
the reference beam configuration. The corresponding total uncertainty
on the value of $\sin^2 \theta_W$ extracted from $\nu$N DIS is 0.35\%.

Most of the model uncertainties will be constrained by dedicated in
situ measurements using the large CC samples and employing
improvements in theory that will have evolved over the course of the
experiment. The low-density tracker will collect about \num{350000}
neutrino-induced inclusive charm events in a five-year run with the
\SIadj{120}{\GeV} \MWadj{1.2} beam.  The precise
reconstruction of charged tracks will allow measurement of exclusive
decay modes of charmed hadrons (e.g., $D^{*+}$) and measurement of
charm fragmentation and production parameters. The average
semileptonic branching ratio $B_\mu$ is of order $5\%$ with the
low-energy LBNE beam, and the low-density ND will be able to
reconstruct both the $\mu \mu$ and $\mu e$ decay channels. Currently,
the most precise sample of \num{15400} dimuon events has been
collected by the NOMAD experiment.  Finally, precision measurements of
CC structure functions in the LBNE ND would further reduce the
uncertainties on PDFs and on high-twist contributions.

The precision that can be achieved from $\nu$N DIS interactions is
limited by both the event rates and the energy spectrum of the
standard beam configuration.  The high-statistics beam
exposure with the low-energy default beam-running configuration
(described in Chapter~\ref{project-chap}) combined with a dedicated
run with the high-energy beam option would increase the statistics by
more than a factor of ten. This major step forward would not only
reduce the statistical uncertainty to a negligible level, but would
provide large control samples and precision auxiliary measurements to
reduce the systematic uncertainties on structure functions. The two
dominant systematic uncertainties, charm production in CC interactions
and low $Q^2$ structure functions, are essentially defined by the
available data at present.  Overall, the use of a high-energy beam
with upgraded intensity can potentially improve the precision
achievable on $\sin^2 \theta_W$ from $\nu$N DIS to better than 0.2\%.

\subsection{Elastic Scattering} 

A second independent measurement of $\sin^2 \theta_W$ can be obtained
from NC $\nu_\mu e$ elastic scattering. This channel has lower
systematic uncertainties since it does not depend on knowledge of
the structure of nuclei, but it has limited statistics due to its very
low cross section. The value of $\sin^2 \theta_W$ can be extracted
from the ratio of interactions~\cite{Marciano:2003eq} as follows:
\begin{equation} \label{eqn:NCel}
{\cal R}_{\nu e} (Q^2) \equiv \frac{\sigma(\overline{\nu}_\mu e \to \overline{\nu}_\mu e)}{\sigma(\nu_\mu e \to \nu_\mu e)} (Q^2)
\simeq \frac{1 - 4 \sin^2 \theta_W + 16 \sin^4 \theta_W}{3 -12 \sin^2 \theta_W + 16 \sin^4 \theta_W},
\end{equation}
\noindent 
in which systematic uncertainties related to the selection and the
electron identification cancel out.  The absolute sensitivity of this
ratio to $\sin^2 \theta_W$ is 1.79, which implies that a measurement of
${\cal R}_{\nu e}$ to 1\% precision would provide a 
measurement of $\sin^2 \theta_W$ to 0.65\% precision.

The best measurement of NC elastic scattering off electrons was
performed by CHARM II, which observed 2677$\pm82$ $\nu$ and 2752$\pm$88
$\overline{\nu}$ events~\cite{Vilain:1994qy}. 
The CHARM II analysis was characterized by a
sizable uncertainty related to the extrapolation of the background
into the signal region.  

The event selection for NC elastic scattering is described in
Section~\ref{ssec:ncscatter}.  Since the NC elastic scattering off
electrons is also used for the absolute flux normalization, the WMA
analysis can be performed only with the low-density, magnetized tracker
in conjunction with a large liquid argon detector. In the case of the flux
normalization measurement, the total reconstructed statistics is
limited to about 4,500 (2,800) $\nu(\bar \nu)$ events.  These numbers
do not allow a competitive determination of $\sin^2 \theta_W$ by using
the magnetized tracker alone.  However, a \tonneadj{100} liquid argon detector
in the ND 
would be expected to collect about 90,000 (60,000) reconstructed $\nu
(\overline{\nu})$ events with the standard beam, and an additional factor of two with 
an upgraded \MWadj{2.3} beam. 

A combined analysis of both detectors can achieve the optimal
sensitivity: the fine-grained tracker is used to reduce systematic
uncertainties (measurement of backgrounds and calibration), while the
liquid argon 
detector provides the statistics required for a competitive measurement.
Overall, the use of the complementary liquid argon detector can provide a statistical
accuracy on $\sin^2 \theta_W$ of about 0.3\%.  However, the extraction
of the WMA is dominated by the systematic uncertainty on the
$\overline{\nu}_\mu / \nu_\mu$ flux ratio in
Equation~(\ref{eqn:NCel}).  This uncertainty has been evaluated with
the low-$\nu_0$ method for the flux extraction and a systematic
uncertainty of about 1\% was obtained on the ratio of the
$\overline{\nu}_\mu / \nu_\mu$ flux integrals.  An improved precision
on this quantity could be achieved from a measurement of the
ratios $\pi^-/\pi^+$ and $\rho^-/\rho^+$ from coherent production in
the fine-grained tracker.  Due to the excellent angular and momentum
resolution and to large cancellations of systematic uncertainties,
preliminary studies indicate that an overall precision of about 0.3\% can
be achieved on the $\overline{\nu}_\mu / \nu_\mu$ flux ratio using
coherent production.

\begin{figure}[!htb]
\centering\includegraphics[width=.8\textwidth]{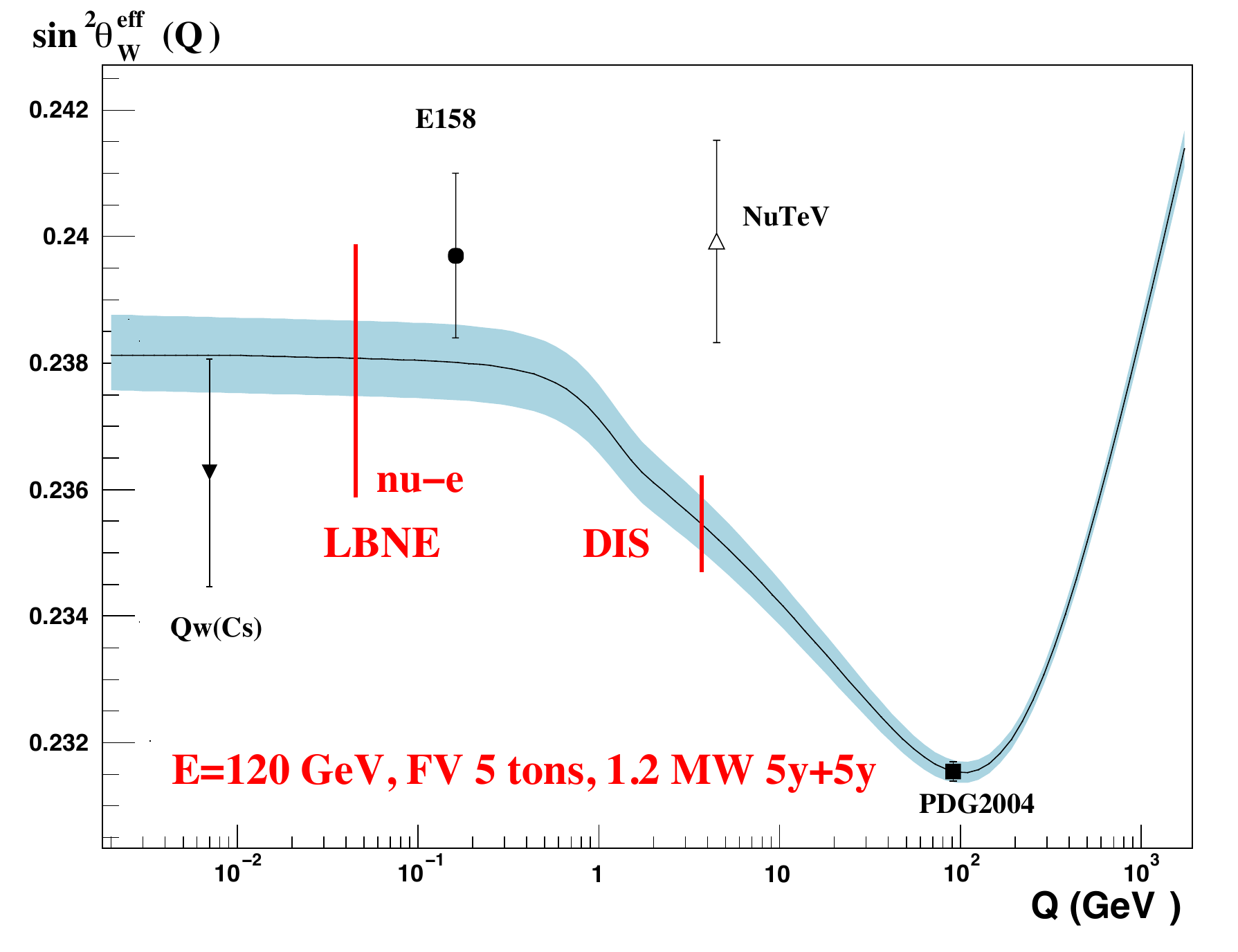}
\caption[Expected near detector sensitivity to $\sin^2 \theta_W$ 
for a \MWadj{1.2} beam]{Expected sensitivity to the measurement of $\sin^2 \theta_W$ from the LBNE ND
with the reference \MWadj{1.2} beam and an exposure of $5\times 10^{21}$ POT with a neutrino beam (five years) and 
$5\times 10^{21}$ POT with an antineutrino beam (five years). 
The curve shows the Standard Model prediction as a function of the 
momentum scale~\cite{Czarnecki:2000ic}.
Previous measurements from Atomic Parity Violation~\cite{Bennett:1999zza,Yao:2006px}, Moeller
scattering (E158~\cite{Anthony:2005pm}), $\nu$ DIS (NuTeV~\cite{Zeller:2001hh}) 
and the combined $Z$ pole  measurements (LEP/SLC)~\cite{Yao:2006px}  are also shown for comparison.
The use of a high-energy beam tune
can reduce the LBNE uncertainties by almost a factor of two.
}
\label{fig:sin2thetaw}
\end{figure}

Together, the DIS and the NC elastic scattering channels involve
substantially different scales of momentum transfer, providing a tool
to test the running of $\sin^2 \theta_W$ in a single experiment. To
this end, the study of NC elastic scattering off protons can provide
additional information since it occurs at a momentum scale that is
intermediate between the two other processes.
Figure~\ref{fig:sin2thetaw} summarizes the target sensitivity from the
LBNE ND, compared with existing measurements as a function of the
momentum scale.

In the near future, another precision measurement of $\sin^2 \theta_W$
is expected from the $Q_{\rm weak}$ experiment~\cite{Lee:2013kya}
at Jefferson Laboratory. From the
measurement of parity-violating asymmetry in elastic electron-proton
scattering, the $Q_{\rm weak}$ experiment should achieve a precision
of 0.3\% on $\sin^2 \theta_W$ at $Q^2=0.026$ GeV$^2$.  It should be
noted that the $Q_{\rm weak}$ measurement is complementary to those
from neutrino scattering given the different scale of momentum
transfer and the fact that neutrino measurements are the only direct
probe of the $Z$ coupling to neutrinos. With the \GeVadj{12} upgrade
of Jefferson Laboratory, the $Q_{weak}$ experiment~\cite{Nuruzzaman:2013bwa} could
potentially reach precisions on the order of 0.2-0.1 \%.

\section{Observation of the Nucleon's Strangeness Content}
\label{sec-deltas} 
\begin{introbox}
  The strange-quark content of the proton and its contribution to the
  proton spin remain enigmatic~\cite{Jaffe:1989jz}.  The question is whether the strange
  quarks contribute substantially to the vector and axial-vector
  currents of the nucleon.  A large observed value of the
  strange-quark contribution to the nucleon spin (axial current),
  $\Delta s$, would enhance our understanding of the proton structure.

The spin structure of the nucleon also affects the couplings of axions and
supersymmetric particles to dark matter. 

\end{introbox}

\subsection{Strange Form Factors of Nucleons}

The strange quark \emph{vector} elastic form factors\footnote{Nucleon form factors describe the scattering amplitudes off
different partons in a nucleon. They are usually given as a function of
$Q^2$ the momentum transfer to the nucleon from the scattering lepton
(since the structure of the nucleon looks different depending on the
energy of the probe).}
 of the nucleon have been
measured to high precision in parity-violating electron scattering
(PVES) at Jefferson Lab, Mainz and elsewhere.
A recent global analysis~\cite{Young:2006jc} 
of PVES data finds a strange 
magnetic moment $\mu_s = 0.37 \pm 0.79$ (in units of the nucleon
magneton), so that the strange quark contribution to proton magnetic
moment is less than 10\%.
For the strange electric charge radius parameter, $\rho_s$, one finds a very
small value, $\rho_s\ = -0.03 \pm 0.63$~GeV$^{-2}$, consistent with zero. 
Both  results are consistent with theoretical expectations
based on lattice QCD and phenomenology~\cite{Leinweber:2004tc}.

In contrast, the strange \emph{axial vector} form factors are poorly 
determined.  A global study of PVES data~\cite{Young:2006jc} 
finds
$\widetilde{G}_A^N(Q^2)
= \widetilde{g}_A^N \left( 1 + {Q^2 / M_A^2} \right)^2$,
where $M_A = 1.026 $ GeV is the axial dipole mass, with the
effective proton and neutron axial charges 
$\widetilde{g}_A^p = -0.80 \pm 1.68$ and 
$\widetilde{g}_A^n = 1.65 \pm 2.62$.

The strange quark axial form factor at $Q^2=0$ is related to the
\emph{spin} carried by strange quarks, $\Delta s$.
Currently the world data on the spin-dependent $g_1$ structure function
constrain $\Delta s$ to be $\approx -0.055$ at a scale $Q^2=1$~GeV$^2$,
with a significant fraction coming from the region $x < 0.001$. 

An independent extraction of $\Delta s$, which does not rely on the difficult
measurements of the $g_1$ structure function at very small values of the Bjorken variable $x$, can be obtained from (anti)neutrino NC elastic scattering off protons 
 (Figure~\ref{fig-delta-s}). Indeed, 
this process provides the most direct measurement of $\Delta s$.
The differential cross section for NC-elastic and CC-QE scattering of
(anti)neutrinos from protons can be written as:
\begin{equation} \label{eqn:QE}
\frac{d \sigma}{d Q^2} = \frac{G_\mu^2}{2\pi} \frac{Q^2}{E_\nu^2} \left( A \pm BW + C W^2 \right); \;\;\;\;  W=4E_\nu/M_p - Q^2/M_p^2,
\end{equation}
where the positive (negative) sign is for neutrino (antineutrino) scattering and the coefficients
$A, B,$ and $C$ contain the vector and axial form factors as follows:
\begin{eqnarray*}
A & = &  \frac{1}{4} \left[ G_1^2 \left( 1 +\tau \right) - \left( F_1^2 - \tau F_2^2 \right)
\left( 1 - \tau \right) + 4 \tau F_1 F_2 \right]\\
B & = &- \frac{1}{4} G_1 \left( F_1 + F_2 \right)\\
C & = &  \frac{1}{16} \frac{M_p^2}{Q^2} \left( G_1^2 + F_1^2 + \tau F_2^2 \right) \\
\end{eqnarray*}

The axial-vector form factor, $G_1$, for NC scattering can be written as the sum of the known axial
form factor $G_A$ plus a strange form factor $G_A^s$:
\begin{equation}
G_1 = \left[ - \frac{G_A}{2} + \frac{G_A^s}{2} \right],
\end{equation}
while the NC vector form factors can be written as:
\begin{equation}
F_{1,2} = \left[ \left(\frac{1}{2} - \sin^2 \theta_W \right) \left( F_{1,2}^p - F_{1,2}^n \right)
- \sin^2 \theta_W \left( F_{1,2}^p + F_{1,2}^n \right) - \frac{1}{2} F_{1,2}^s \right],
\end{equation}
where $F_1^{p(n)}$ is the Dirac form factor of the proton (neutron),
$F_2^{p(n)}$ is the corresponding Pauli form factor, and $F_{1,2}^s$
are the strange-vector form factors.  These latter form factors are
expected to be small from the PVES measurements summarized above.  
In the limit $Q^2 \to 0$, the differential cross section is proportional
to the square of the axial-vector form factor $d \sigma / d Q^2
\propto G_1^2$ and $G_A^s \to \Delta s$.  The value of $\Delta s$ can
therefore be extracted experimentally by extrapolating the NC
differential cross section to $Q^2=0$.


\subsection{Extraction of the Strange Form Factors}

Previous neutrino scattering experiments have been limited by the
statistics and by the systematic uncertainties on background
subtraction.  One of the earliest measurements available comes from
the analysis of 951 NC $\nu p$ and 776 NC $\overline{\nu}p$ collected
by the experiment BNL
E734~\cite{Ahrens:1986xe,Garvey:1992cg,Alberico:1998qw}. There are
also more recent results with high statistics from MiniBooNE where a
measurement of $\Delta s$ was carried out using neutrino NC elastic
scattering with 94,531 $\nu N$ events~\cite{AguilarArevalo:2010cx}.
The MiniBooNE measurement was limited by the inability to distinguish
the proton and neutron from $\nu N$ scattering. The LBNE neutrino beam
will be sufficiently intense that a measurement of NC elastic
scattering on protons  
in the fine-grained ND can provide a definitive
statement on the contribution of the strange sea to either the axial
or vector form factor.

\begin{figure}[!htb]
\centering\includegraphics[width=.7\textwidth]{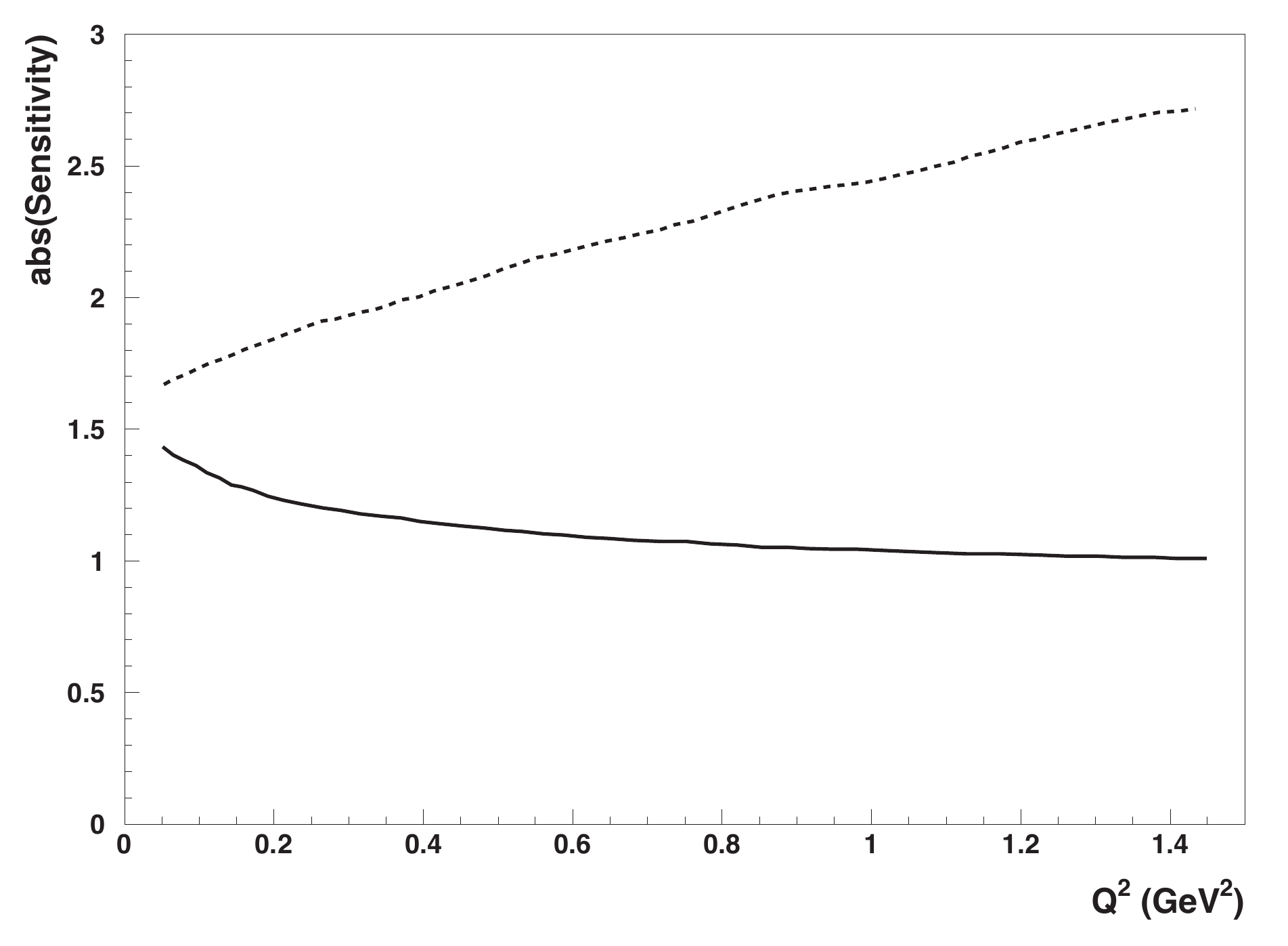}
\caption[Sensitivity of NC/CC to the strange contribution to nucleon spin]
{Sensitivity (magnitude) of the ratios ${\cal R}_{\nu p}$ (solid) and
${\cal R}_{\overline{\nu} p}$ (dashed) to a variation of the strange contribution to the
spin of the nucleon, $\Delta s$,
as a function of $Q^2$. Values greater than one imply that the relative uncertainty on $\Delta s$ is smaller than that of the corresponding ratio (see text).}
\label{fig-delta-s}
\end{figure}

Systematic uncertainties can be reduced by measuring the NC/CC ratios
for both neutrinos and antineutrinos as a function of $Q^2$:
\begin{equation}
{\cal R}_{\nu p} (Q^2) \equiv \frac{\sigma(\nu_\mu p \to \nu_\mu p)}{\sigma(\nu_\mu n \to \mu^- p)}(Q^2); \;\;\;\;\;
{\cal R}_{\overline{\nu} p} (Q^2) \equiv \frac{\sigma(\overline{\nu}_\mu p \to \overline{\nu}_\mu p)}{\sigma(\overline{\nu}_\mu p \to \mu^+ n)}(Q^2),
\end{equation}
Figure~\ref{fig-delta-s} shows the absolute sensitivity of both ratios
to $\Delta s$ for different values of $Q^2$. The sensitivity for
$Q^2\sim0.25$~GeV$^2$ is about 1.2 for neutrinos and 1.9 for
antineutrinos, which implies that a measurement of ${\cal R}_{\nu p}$
and ${\cal R}_{\overline{\nu} p}$ of 1\% precision would enable the
extraction of $\Delta s$ with an uncertainty of 0.8\% and 0.5\%,
respectively.

%
%

The design of the 
tracker includes several
different nuclear targets.  Therefore, most of the neutrino scattering
is from nucleons embedded in a nucleus, requiring nuclear effects to
be taken into account. Fortunately, in the ratio of NC/CC, the nuclear
corrections are expected to largely cancel out.  The $\Delta s$
analysis requires a good proton reconstruction efficiency as well as
high resolution on both the proton angle and energy. To this end, the
low-density 
tracker can increase the range of the
protons inside the ND, allowing the reconstruction of proton tracks
down to $Q^2\sim0.07$~GeV$^2$. This capability will reduce the
uncertainties in the extrapolation of the form factors to the limit
$Q^2 \to 0$.

Table~\ref{tab:prange} summarizes the expected proton range for the
low-density ($\rho\sim$~\SI{0.1}{\gram\per\cubic\centi\meter}) straw-tube 
tracker (STT) in the ND tracking detector design described in
Section~\ref{sec:ndproj}.  About $2.0 (1.2) \times 10^6$ $\nu p
(\overline{\nu} p)$ events are expected after the selection cuts in
the low-density 
tracker, yielding a statistical precision on the order
of 0.1\%.

\begin{table}[htb]
\centering
\caption[Expected proton range for the  low-density tracker]{Expected proton range for the  low-density
($\rho\sim$\SI{0.1}{\gram\per\cubic\centi\meter}) tracker. The first column gives the proton kinetic energy
and the last column the proton momentum. The $Q^2$ value producing $T_p$ is calculated
assuming the struck nucleon is initially at rest.}
\label{tab:prange}
\begin{tabular}{$L^c^r^c}  
\toprule
\rowtitlestyle
$ T_p$  &  $Q^2$          &  Range STT &  $P_p$  \\
\rowtitlestyle
MeV  &  GeV$^2/c^2$ &  $cm$        &  GeV$/c$ \\ 
\toprowrule
         20 &             0.038  &     4.2          & 0.195  \\ \colhline
        40 &              0.075  &   14.5          & 0.277  \\ \colhline
        60 &              0.113  &   30.3          & 0.341   \\ \colhline
        80 &              0.150  &  50.8          & 0.395  \\ \colhline
      100 &               0.188 &  75.7          & 0.445  \\ 
\bottomrule
\end{tabular}

\end{table}

The determination of $\Delta s$ in the STT 
utilizes
analysis techniques performed by the FINeSSE
Collaboration~\cite{Bugel:2004yk} and used by the SciBooNE experiment.  In
particular, based on the latter, LBNE
expects a purity of about 50\%, with background contributions of 20\%
from neutrons produced outside of the detector, 10\% $\nu n$ events
and 10\% NC pion backgrounds.  The dominant systematic uncertainty
will be related to the background subtraction.  The low-energy beam
spectrum at LBNE provides the best sensitivity for this measurement
since the external background from neutron-induced proton recoils will
be reduced by the strongly suppressed high-energy tail.  The
low-density magnetized tracker is expected to increase the purity by
reducing the neutron background and the NC pion background.  The
outside neutron background, it should be noted, can be determined
using the $n \rightarrow p + \pi^-$ process in the STT.  The
sensitivity analysis is still in progress, however LBNE is confident
of achieving a precision on $\Delta s$ of about \numrange[range-phrase
= --]{0.02}{0.03}.

\clearpage
\section{Nucleon Structure and QCD Studies}
\label{sec-nucleon}
\begin{introbox}
  Precision measurements of (anti)neutrino differential cross sections
  in the LBNE near detector will provide additional constraints on
  several key nucleon structure functions that are complementary to
  results from electron scattering experiments.

  In addition, these measurements would directly improve LBNE's
  oscillation measurements by providing accurate simulation of
  neutrino interactions in the far detector and offer an estimate of
  all background processes that are dependent upon the angular
  distribution of the outgoing particles in the far detector.
  Furthermore, certain QCD analyses --- i.e., global fits used for extraction of
  parton distribution functions (PDFs) via 
  the differential cross sections measured in ND data ---
  would constrain the systematic error in 
  precision electroweak measurements. This would apply  
  not only in neutrino physics but also in hadron collider measurements.  
\end{introbox}

\subsection{\boldmath Determination of the $F_3$
Structure Function and GLS Sum Rule}

For quantitative studies of inclusive deep-inelastic lepton-nucleon
scattering, it is vital to have precise measurements of the $F_3$
structure functions as input into global PDF fits.  Because it depends
on weak axial quark charges, the $F_3$ structure function can only be
measured with neutrino and antineutrino beams and is unique in its
ability to differentiate between the quark and antiquark content of
the nucleon.  On a proton target, for instance, the neutrino and
antineutrino $F_3$ structure functions (at leading order in
$\alpha_s$) are given by
\begin{eqnarray}
xF_3^{\nu p}(x) 
&=& 2 x \left( d(x) - \overline u(x) + s(x) + \cdots \right)\, , \\
xF_3^{\overline\nu p}(x) 
&=& 2 x \left( u(x) - \overline d(x) - \overline s(x) + \cdots \right)\, , \\ 
xF_3^{\nu n}(x) 
&=& 2 x \left( u(x) - \overline d(x) + s(x) + \cdots \right)\, , \\
xF_3^{\overline\nu n}(x) 
&=& 2 x \left( d(x) - \overline u(x) - \overline s(x) + \cdots \right)\, .
\end{eqnarray}

where $u_v=u-\bar u$ and $d_v=d-\bar d$ are the valence sea quark
distributions. Under the assumption of a symmetric strange sea,
i.e., $s(x)=\bar s(x)$, the above expressions show that a measurement
of the average $xF_3=(xF_3^{\nu N}+xF_3^{\bar\nu N})/2$ for neutrino
and antineutrino interactions on isoscalar targets provides a direct
determination of the valence quark distributions in the proton. This
measurement is complementary to the measurement of Drell-Yan
production at colliders, which is essentially proportional to the sea
quark distributions.

\clearpage
The first step in the structure function analysis is the measurement
of the differential cross section:
\begin{equation} 
\frac{1}{E_\nu} \frac{d \sigma^2}{dx dQ^2} = \frac{N(x,Q^2,E_\nu)}{N(E_\nu)} \frac{\sigma_{\rm tot}/E_\nu}{dx dQ^2} 
\end{equation} 
where $N(x,Q^2,E_\nu)$ is the number of events in each $(x,Q^2,E_\nu)$ bin and $N(E_\nu)$ is the number of events in each $E_\nu$ 
bin integrated over $x$ and $Q^2$. The average $xF_3$ structure function can be extracted by taking the difference between neutrino and 
antineutrino differential cross sections: 
\begin{equation} 
\frac{1}{E_\nu} \frac{d^2 \sigma^\nu}{dx dQ^2} - \frac{1}{E_\nu} \frac{d^2 \sigma^{\bar \nu}}{dx dQ^2} = 2 \left[ y \left( 1 - \frac{y}{2} \right) \frac{y}{Q^2} \right] xF_3  
\end{equation} where $xF_3$ denotes the sum for neutrino and antineutrino interactions. 

The determination of the $xF_3$ structure functions will, in turn,
allow a precision measurement of the Gross-Llewellyn-Smith (GLS) QCD
sum rule:
\begin{eqnarray}
\label{eq:GLS}
S_{\rm GLS} (Q^2) & = & 
\frac{1}{2} \int^1_0 \frac{1}{x} \left[ xF_3^{\nu N} + xF_3^{\bar \nu N} \right] dx \nonumber \\ 
& = & 3 \left[ 1 - \frac{\alpha_s(Q^2)}{\pi} - a(n_f) \left( \frac{\alpha_s(Q^2)}{\pi} \right)^2
-b(n_f) \left( \frac{\alpha_s(Q^2)}{\pi} \right)^3 \right] + \Delta {\rm HT}
\end{eqnarray}
where $\alpha_s$ is the strong coupling constant, $n_f$ is the number
of quark flavors, $a$ and $b$ are known functions of $n_f$, and the
quantity $\Delta {\rm HT}$ represents higher-twist contributions.  The
equation above can be inverted to determine $\alpha_s(Q^2)$ from the
GLS sum rule. The most precise determination of the GLS sum rule was
obtained by the CCFR experiment on an iron target~\cite{Leung:1992yx}
$S_{\rm GLS} (Q^2=3~GeV^2) = 2.50 \pm 0.018 \pm 0.078$. 
The high-resolution ND combined with the unprecedented statistics
would substantially reduce the systematic uncertainty on the low-$x$
extrapolation of the $xF_3$ structure functions entering the GLS
integral.  In addition, the presence of different nuclear targets, as
well as the availability of a target with free protons
will allow investigation of isovector and nuclear corrections, and
adding a tool to test isospin (charge) symmetry (Section~\ref{sec-isospin}).

\subsection{\boldmath Determination of the Longitudinal Structure Function $F_L(x,Q^2)$}
The structure
function $F_L$ is directly related to the gluon distribution
$G(x,Q^2)$ of the nucleon, as can be seen from the
Altarelli-Martinelli relation:
\begin{equation} 
F_L(x,Q^2) = \frac{\alpha_s(Q^2)}{\pi} \left[ \frac{4}{3}\int^1_x \frac{dy}{y} \left(\frac{x}{y} \right)^2 F_2(x,Q^2) + 
n_f \int^1_x \frac{dy}{y}\left(\frac{x}{y} \right)^2 \left( 1-\frac{x}{y} \right) G(y,Q^2) \right]
\end{equation}  
where $n_f$ is the number of parton flavors. In the leading order 
approximation the longitudinal structure function $F_L$ is zero, while
at higher orders a nonzero $F_L(x,Q^2)$ is originated as a consequence of the violation
of the Callan-Gross relation:
\begin{equation} 
F_L(x,Q^2) = \left(  1+\frac{4M^2x^2}{Q^2} \right) F_2(x,Q^2) - 2x F_1(x,Q^2) 
\end{equation}  
where $2x F_1=F_T$ is the transverse structure function.  A
measurement of $R=F_L/F_T$ is therefore both a test of perturbative QCD at
large $x$ and a clean probe of the gluon density at small $x$ where the
quark contribution is small. A poor knowledge of $R$, especially at
small $x$, results in uncertainties in the structure functions extracted
from deep inelastic scattering cross sections, and in turn, in
electroweak measurements.  It is instructive to compare the low-$Q^2$
behavior of $R$ for charged-lepton versus  neutrino scattering. In both
cases CVC implies that $F_T \propto Q^2$ as $Q^2 \to 0$. However,
while $F_L \propto Q^4$ for the electromagnetic current, for the weak
current $F_L$ is dominated by the finite PCAC (partial conservation of
the axial current) contribution \cite{Kulagin:2007ju}.
The behavior of
$R$ at $Q^2\ll 1$ GeV$^2$ is therefore very different for
charged-lepton and neutrino scattering.  A new precision measurement
of the $Q^2$ dependence of $R$ with (anti)neutrino data would also
clarify the size of the high-twist contributions to $F_L$ and $R$,
which reflect the strength of multi-parton correlations (qq and
qg). 

The ratio of longitudinal to transverse structure functions can be
measured from the $y$ dependence of the deep inelastic scattering
data. Fits to the following function:
\begin{equation} 
F(x,Q^2, \epsilon) = \frac{\pi (1- \epsilon)}{y^2 G_F^2 M E_\nu} \left[ \frac{d^2 \sigma^\nu}{dx dy} + \frac{d^2 \sigma^{\bar \nu}}{dx dy} \right] = 2 x F_1(x,Q^2) \left[ 1 + \epsilon R(x,Q^2) \right] 
\end{equation} 
have been used by CCFR and NuTeV to determine
$R=\sigma_L/\sigma_T$. In this equation $\epsilon \simeq 2
(1-y)/(1+(1-y)^2)$ is the polarization of the virtual $W$ boson. This
equation assumes $xF_3^\nu = xF_3^{\bar \nu}$, and a correction must be
applied if this is not the case. The values of $R$ are extracted from
linear fits to $F$ versus $\epsilon$ at fixed $x$ and $Q^2$ bins.

\subsection{\boldmath Determination of $F_2^n$ and the $d/u$ Ratio of Quark Distribution Functions}

Because of the larger electric charge on the $u$ quark than on the
$d$, the electromagnetic proton $F_2$ structure function data provide
strong constraints on the $u$-quark distribution, but are relatively
insensitive to the $d$-quark distribution.  To constrain the $d$-quark
distribution a precise knowledge of the corresponding $F_2^n$
structure functions of free neutrons is required, which in current practice is
extracted from inclusive deuterium $F_2$ data.  At large values of $x$
($x>0.5$) the nuclear corrections in deuterium become large and, more
importantly, strongly model-dependent, leading to large uncertainties
on the resulting $d$-quark distribution.  Using the isospin relation
$F_2^{\bar \nu p} = F_2^{\nu n}$ and $F_2^{\nu p} = F_2^{\bar \nu n}$
it is possible to obtain a direct determination of $F_2^{\nu n}$ and
$F_2^{\bar \nu n}$ with neutrino and antineutrino scattering off a target with free
protons. This determination is free from model uncertainties
related to nuclear targets. The extraction of $F_2^{\nu n}$ and
$F_2^{\bar \nu n}$ will allow a precise extraction on the $d$-quark
distribution at large $x$.  Existing neutrino data on hydrogen  
have relatively large errors and do not extend beyond
$x\sim0.5$~\cite{Bodek:1985tv,Jones:1987gk}.

The $F_2^{\bar \nu p}$ and $F_2^{\nu p}$ structure functions can be
obtained from interactions on a target with free protons  after subtracting
the contributions from $xF_3$ and $R$. These latter can either be
modeled within global PDF fits or taken from the other two
measurements described above. As discussed in Section~\ref{sec-isospin} the LBNE 
ND can achieve competitive measurements of $F_2^{\bar \nu p}$ and $F_2^{\nu p}$ 
with an increase of statistics of three orders of magnitude with respect to the 
existing hydrogen data~\cite{Bodek:1985tv,Jones:1987gk}. 

\subsection{Measurement of Nucleon Structure Functions}

At present neutrino scattering measurements of cross sections have
considerably larger uncertainties than those of the electromagnetic
inclusive cross sections.  The measurement of the differential cross
sections~\cite{Formaggio:2013kya} is dominated by three
uncertainties: (1) muon energy scale, (2) hadron energy scale, and (3)
knowledge of the input (anti)neutrino flux.  Table~\ref{tab:expcomp}
shows a comparison of past and present experiments and the
corresponding uncertainties on the energy scales.  The most precise
measurements are from the CCFR, NuTeV and NOMAD experiments, which are
limited to a statistics of about \num{e6} neutrino events.
\begin{table}[!htb] 
  \caption[Past experiments' structure function measurements]{Summary of past experiments performing structure function measurements. The expected
    numbers in the LBNE near detector for a five-year run with the \SIadj{1.2}{\MW} \SIadj{120}{\GeV} reference beam  ($5 \times 10^{21}$ POT) are also given for comparison.
  }  
\label{tab:expcomp} 
\begin{tabular}{$L^c^c^c^c^r^r}
\toprule
\rowtitlestyle
Experiment & Mass & \multicolumn{1}{^>{\columncolor{\ChapterBubbleColor}}c}{$\nu_{\mu}$ CC Stat.} & Target & $E_\nu$ (GeV)
& $\Delta E_\mu$  & $\Delta E_{\rm H}$ \\ \toprowrule
            CDHS~\cite{Berge:1989hr} &  750 t &  { $10^{7}$}   &  p,Fe & 20-200 & 2.0\% & 2.5\% \\ \colhline
            BEBC~\cite{Allasia:1983vw,Allasia:1985hw} & various &   5.7$\times$$10^{4}$   & p,D,Ne & 10-200 &  & \\ \colhline
            CCFR~\cite{Yang:2000ju,Yang:2001xc} & 690 t & { 1.0$\times$$10^{6}$}   & Fe & 30-360 & 1.0\% & 1.0\% \\  \colhline
            NuTeV~\cite{Tzanov:2005kr} &  690 t & { 1.3$\times$$10^{6}$}  &  Fe &  30-360 &  0.7\% &  0.43\% \\ \colhline
            CHORUS~\cite{Onengut:2005kv} & 100 t & { 3.6$\times$$10^{6}$}   &  Pb &  10-200 &  2.5\% &  5.0\% \\ \colhline
            NOMAD~\cite{Wu:2007ab} & 2.7 t & { 1.3$\times$$10^{6}$}   &  C &  5-200 &  0.2\% &  0.5\% \\ \colhline
            ~~~~~~~~~~~~~~~~~\cite{Samoylov:2013xoa}     & 18 t & { 1.2$\times$$10^{7}$}   &  Fe  &  5-200 &  0.2\% &  0.6\% \\ \colhline
            MINOS ND~\cite{Adamson:2009ju} & 980 t &  3.6$\times$$10^{6}$   &  Fe & 3-50 & 2-4\% & 5.6\% \\  \colhline
            LBNE ND  & 5 t &  5.9$\times$$10^{7}$   & (C$_3$H$_6$)$_n$  & 0.5-30 & $<0.2$\% & $<0.5$\% \\  \bottomrule
\end{tabular} 
\end{table}

The MINER$\nu$A~\cite{Osmanov:2011ig} experiment is expected to provide new structure
function measurements on a number of nuclear targets including He, C,
Fe and Pb in the near future.  Since the structure function
measurement mainly involves DIS events, the MINER$\nu$A measurement
will achieve a competitive statistics after the completion of the new
run with the medium-energy beam. 
MINER$\nu$A will focus on a measurement of the ratio of different nuclear
targets to measure nuclear corrections in (anti)neutrino
interactions. It must be noted that the MINER$\nu$A experiment relies
on the MINOS ND for muon identification.  The corresponding
uncertainty on the muon-energy scale (Table~\ref{tab:expcomp}) is
substantially larger than that in other modern experiments, e.g.,
NuTeV and NOMAD, thus limiting 
the potential of absolute
structure function measurements. Furthermore, the muon-energy scale is
also the dominant source of uncertainty in the determination of the
(anti)neutrino fluxes with the low-$\nu$ method.  Therefore, the flux
uncertainties in MINER$\nu$A are 
expected to be larger than in
NOMAD and NuTeV. 
 
Given its reference beam design and \MWadj{1.2} proton-beam power, LBNE
expects to collect about \num{2.3e7} neutrino DIS events and
about \num{4.4e6}  antineutrino DIS events in the ND. 
These numbers correspond to an improvement
by more than one order of magnitude with respect to the most precise
past experiments, e.g., NuTeV~\cite{Tzanov:2005kr} and 
NOMAD~\cite{Wu:2007ab,Samoylov:2013xoa}. 
With these high-statistics
samples, LBNE will be able to significantly reduce the gap between the
uncertainties on the weak and electromagnetic structure functions.
A possible high-energy run with the upgraded \MWadj{2.3} beam would offer a 
further increase by more than a factor of ten in statistics.  

In addition to the large data samples, the use of a high-resolution,
low-density spectrometer allows LBNE to reduce systematic
uncertainties with respect to previous measurements. The LBNE ND is
expected to achieve precisions better than 0.2\% and 0.5\% on the muon-
and hadron-energy scales, respectively. 
These numbers are based on the results achieved by the NOMAD experiment
(Table~\ref{tab:expcomp}), which had 
much lower statistics and
poorer resolution than is expected in the LBNE ND. The calibration of the momentum and energy scales
will be performed with the large sample of reconstructed $K^0_S \to \pi \pi$,
$\Lambda \to p \pi$, and $\pi^0 \to \gamma \gamma$ decays.
In addition, the overall hadronic energy scale can be calibrated by exploiting the
well-known structure of the Bjorken $y$ distribution in (anti)neutrino DIS
interactions~\cite{Wu:2007ab,Petti:2011zz}.
The relative fluxes as a function of energy can be extracted to a precision of 
about 2\% with the low-$\nu$ method, due to the small uncertainty on the muon-energy
scale. The world average absolute normalization of the differential
cross sections $\sigma_{\rm tot}/E$, is known to 2.1\%
precision~\cite{Beringer:1900zz}. 
However, with the \MWadj{1.2} beam available from the PIP-II
upgrades, it will be possible to improve the absolute normalization
using $\nu$-e NC elastic scattering events, coherent meson production, etc. 
An overall precision of 1-2\% would make (anti)neutrino
measurements comparable to or better than the complementary measurements from
charged-lepton DIS.

On the time scale of 
LBNE, comparable measurements from
(anti)neutrino experiments are not expected, primarily due to the low
energy of competing beamlines (J-PARC neutrino beamline in Japan~\cite{Sekiguchi:2012xma}) or to the poorer resolution of the detectors
used (MINER$\nu$A~\cite{Osmanov:2011ig} , T2K~\cite{Abe:2011ks},
NO$\nu$A~\cite{Ayres:2007tu}). The 
experimental program 
most likely to compete with the LBNE ND measurements is the \GeVadj{12} upgrade at
Jefferson Laboratory (JLab)~\cite{Dudek:2012vr}.  However, it must be
emphasized that the use of electron beams at JLab makes this program
\emph{complementary} to LBNE's.  In particular, the three topics
discussed above are specific to the (anti)neutrino interactions.

Several planned experiments at JLab with the energy-upgraded \GeVadj{12}
beam will measure the $d/u$ ratio from D targets up to $x\sim0.85$, 
using different methods to minimize the nuclear corrections.  
The LBNE measurement 
will be competitive with the
proposed JLab \GeVadj{12} experiments, since the large statistics expected will allow
a precise determination of $F_2^{\nu
  n}$ and $F_2^{\bar \nu n}$ up to $x\sim0.85$. Furthermore,
the use of a weak probe coupled with a wide-band beam will provide
a broader $Q^2$ range than in JLab experiments, thus allowing a separation of
higher twist and other sub-leading effects in $1/Q^2$.

\section{Tests of Isospin Physics and Sum-Rules}
\label{sec-isospin}
\begin{introbox}
  One of the most compelling physics topics accessible to 
  LBNE's high-resolution near detector 
is the isospin physics using
  neutrino and antineutrino interactions. This physics involves the
  Adler sum rule and tests isospin (charge) symmetry in nucleons and
  nuclei.
\end{introbox}

The Adler sum rule relates the integrated difference of the
antineutrino and neutrino $F_2$ structure functions to the isospin of the target:
\begin{equation}
\label{ASR}
{\cal S}_A (Q^2) =\int_0^1 \; dx \; \left[ F_2^{\overline\nu} (x,Q^2) - F_2^{\nu}(x,Q^2) \right]/(2x)
= 2\,I_z,
\end{equation}
where the integration is performed over the entire kinematic range of the
Bjorken variable $x$ and $I_z$ is the projection of
the target isospin vector on the quantization axis ($z$ axis).
For the proton ${\cal S}_A^{p}=1$ and for the neutron ${\cal S}_A^{n}=-1$.

In the quark-parton model the Adler sum is the difference between the
number of valence $u$ and $d$ quarks of the target. The Adler sum rule
survives the strong-interaction effects because of the conserved vector 
current (CVC) and provides an
exact relation to test the local current commutator algebra of the weak
hadronic current. 
In the derivation of the Adler sum rule the effects of both
non-conservation of the axial current and heavy-quark production are
neglected. 

Experimental tests of the Adler sum rule require the use of a hydrogen target
to avoid nuclear corrections to the bound nucleons inside the nuclei.
The structure functions $F_2^{\overline{\nu}}$ and $F_2^\nu$ have to be determined
from the corresponding differential cross sections and must be extrapolated
to small $x$ values in order to evaluate the integral. 
The test performed in bubble chambers by the BEBC
Collaboration --- the only test available ---  is limited by the modest statistics;
it used about 9,000 $\overline{\nu}$ and 5,000 $\nu$ events
collected on hydrogen~\cite{Allasia:1985hw}.

The LBNE program can provide the first high-precision test of the
Adler sum rule.  To this end, the use of the high-energy beam tune
shown in Figure~\ref{fig:beamtunes}, although not essential, would
increase the sensitivity, allowing attainment of higher $Q^2$
values. Since the use of a liquid H$_2$ bubble chamber is excluded in
the ND hall due to safety concerns, the (anti)neutrino interactions
off a hydrogen target can only be extracted with a subtraction method
from the composite materials of the ND targets.  Using this technique
to determine the position resolution in the location of the primary
vertex is crucial to reducing systematic uncertainties.  For this
reason, a precision test of the Adler sum rule is best performed with
the low-density magnetized ND.

A combination of two different targets --- the polypropylene
$(C_3H_6)_n$ foils placed in front of the STT modules and pure carbon
foils --- are used in the low-density, magnetized 
ND to provide a
fiducial hydrogen mass of about 1 t.  With the LBNE fluxes from
the standard exposure, $5.0(1.5) \times 10^6 \pm 13(6.6)\times 10^3
(sub.)$ $\nu(\overline{\nu})$ CC events (where the quoted uncertainty is dominated by the
statistical subtraction procedure) 
would be collected on the hydrogen
target.  The level of precision that can be achieved is sufficient to
open up the possibility of making new discoveries in the quark and
hadron structure of the proton. No other comparable measurement is
expected on the timescale of LBNE.

\section{Studies of (Anti)Neutrino-Nucleus Interactions} 
\label{sec-nuclear} 

An integral part of the physics program envisioned for the LBNE ND
involves detailed measurements of (anti)neutrino interactions in a
variety of nuclear targets.  The LBNE ND offers substantially
larger statistics coupled with a much higher resolution and, in turn,
lower systematic uncertainties with respect to past experiments
(Table~\ref{tab:expcomp}) or ongoing and future ones
(MINER$\nu$A~\cite{Osmanov:2011ig}, T2K~\cite{Abe:2011ks},
NO$\nu$A~\cite{Ayres:2007tu}).  The most important nuclear target is
of course the argon target, which matches the LBNE far detector.
The ND standard target is polypropylene
(C$_3$H$_6$)$_n$, largely provided by the mass of the STT radiators.
An additional proposed ND target is argon gas in pressurized aluminum
tubes with sufficient mass to provide $\simeq$10 times the \nm CC and
NC statistics as expected in the LBNE far detector.  Equally important
nuclear targets are carbon (graphite), which is essential in order to
get (anti)neutrino interactions on free protons through a statistical
subtraction procedure from the main polypropylene target 
(Section~\ref{sec-isospin}), and calcium.  In particular, this latter
target has the same atomic weight ($A=40$) as argon but is
isoscalar. One additional nuclear target is iron, which is used in the
proposed India-based Neutrino Observatory
(INO)~\cite{Mondal:2012fn}. 
The modularity of the STT provides for successive measurements using
thin nuclear targets (thickness $< 0.1 X_0$), while the excellent
angular and space resolution allows a clean separation of events
originating in different target materials.  Placing an arrangement of
different nuclear targets upstream of the detector
provides the desired nuclear samples in (anti)neutrino interactions.
For example, a single \SIadj{7}{\milli\meter}-thick calcium layer at the upstream 
end of the detector will provide  
about \num{3.1e5} \nm CC interactions in one year. 

Potential ND studies in nuclear effects include the following: 
\begin{itemize}
\item nuclear modifications of form factors
\item nuclear modifications of structure functions
\item mechanisms for nuclear effects in coherent and incoherent regimes
\item a dependence of exclusive and semi-exclusive processes
\item effect of final-state interactions
\item effect of short-range correlations
\item two-body currents
\end{itemize}

The study of nuclear effects in (anti)neutrino interactions off nuclei
is directly relevant for the long-baseline oscillation studies.  The
use of heavy nuclei like argon in the LBNE far detector requires a measurement
of nuclear cross sections on the same targets in the ND in order to reduce signal 
and background uncertainties in the oscillation analyses.  Cross-section
measurements obtained from other experiments using different nuclei
are not optimal; in addition to the different $p/n$ ratio in argon
compared to iron or carbon where measurements from other experiments
exist, nuclear modifications of cross sections can differ from 5\% to
15\% between carbon and argon for example, while the difference in the
final-state interactions could be larger.
Additionally, nuclear modifications can introduce a substantial
smearing of the kinematic variables reconstructed from the observed
final-state particles.  Detailed measurements of the dependence on the
atomic number $A$ of different exclusive processes are then required
in order to understand the absolute energy scale of neutrino event
interactions and to reduce the corresponding systematic uncertainties
on the oscillation parameters.

It is worth noting that the availability of a free-proton target
through statistical subtraction of the (C$_3$H$_6$)$_n$ and carbon
targets (Section~\ref{sec-isospin}) will allow for the first time a
direct model-independent measurement of nuclear effects --- including
both the primary and final-state interactions --- on the argon target
relevant for the far detector oscillation analysis.

Furthermore, an important question in nuclear physics is how the
structure of a nucleon is modified when said nucleon is inside the
medium of a heavy nucleus as compared to a free nucleon like the
proton in a hydrogen nucleus.  Studies of the ratio of structure
functions of nuclei to those of free nucleons (or in practice, the
deuteron) reveal nontrivial deviations from unity as a function of $x$
and $Q^2$.  These have been well explored in charged-lepton scattering
experiments, but little empirical information exists from neutrino
scattering. Measurements of structure using neutrino scattering are
complementary to those in charged-lepton scattering.

Another reason to investigate the nuclear-medium modifications 
of neutrino structure functions is that most neutrino scattering
experiments are performed on nuclear targets, from which information
on the free nucleon is inferred by performing a correction for the
nuclear effects.  
In practice this often means applying the same nuclear correction as
for the electromagnetic structure functions, which introduces an
inherent model-dependence in the result.  In particular, significant
differences between photon-induced and weak-boson-induced nuclear
structure functions are predicted, especially at low $Q^2$ and low
$x$, which have not been tested.  A striking example is offered by the
ratio $R$ of the longitudinal-to-transverse structure
functions~\cite{Kulagin:2007ju}.  While the electromagnetic ratio tends
to zero in the photoproduction limit, $Q^2 \to 0$, by current
conservation, the ratio for neutrino structure functions is predicted
to be \emph{finite} in this limit.  Thus, significant discovery
potential exists in the study of neutrino scattering from nuclei.

The comparison of argon and calcium targets (${}^{40}_{18}$Ar and
${}^{40}_{20}$Ca) in the LBNE ND would be particularly
interesting. Since most nuclear effects depend on the atomic weight
$A$, inclusive properties of (anti)neutrino interactions are expected
to be the same for these two
targets~\cite{Kulagin:2007ju,Butkevich:2012zr,Butkevich:2007gm,Ankowski:2007uy}.
This fact would allow the use of both targets to model signal and
backgrounds in the LBNE far detector (argon target), as well as to
compare LBNE results for nuclear effects on argon with the extensive
data on calcium from charged lepton DIS. In addition, a high-precision
measurement of (anti)neutrino interactions in both argon and calcium
opens the possibility for studying a potential flavor and isovector
dependence of nuclear effects and to further test the isospin (charge
symmetry) in nuclei (Section~\ref{sec-isospin}).  Evidence for any
of these effects would constitute important discoveries.

Finally, the extraction of (anti)neutrino interactions on deuterium
from the statistical subtraction of H$_2$O from D$_2$O, which is
required to measure the fluxes (Section~\ref{sec-fluxosc}), would
allow the first direct measurement of nuclear effects in deuterium.
This measurement can be achieved since the structure function of a
free isoscalar nucleon is given by the average of neutrino and
antineutrino structure functions on hydrogen ($F_2^{\nu
  n}=F_2^{\overline{\nu} p}$).  A precise determination of nuclear
modifications of structure functions in deuterium would play a crucial
role in reducing systematic uncertainties from the global PDF fits.

\section{Search for Heavy Neutrinos} 

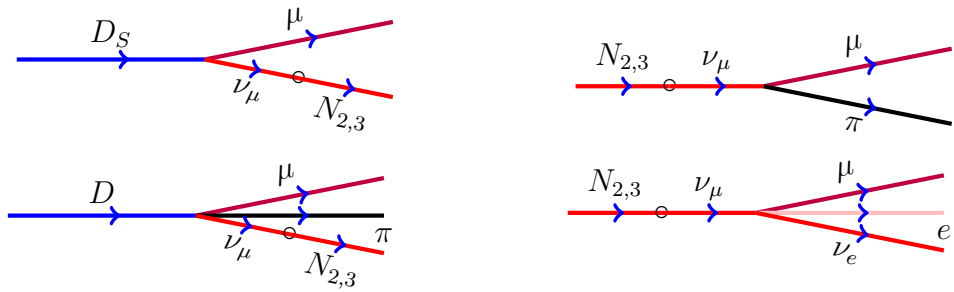
\begin{figure}[!htb]
\centering

\tikzset{
  quark/.style={draw=blue, postaction={decorate},
    decoration={markings,mark=at position .6 with {\arrow[draw=blue]{>}}}},
  electron/.style={draw=pink, postaction={decorate},
    decoration={markings,mark=at position .6 with {\arrow[draw=blue]{>}}}},
  neutrino/.style={draw=red, postaction={decorate},
    decoration={markings,mark=at position .6 with {\arrow[draw=blue]{>}}}},
  heavy/.style={draw=red, postaction={decorate},
    decoration={markings,mark=at position .6 with {\arrow[draw=blue]{>}}}},
  pion/.style={draw=black,postaction={decorate},
    decoration={markings,mark=at position .6 with {\arrow[draw=blue]{>}}}},
  muon/.style={draw=purple, postaction={decorate},
    decoration={markings,mark=at position .6 with {\arrow[draw=blue]{>}}}},
  gamma/.style={decorate, decoration={snake,amplitude=4pt, segment length=5pt}, draw=red},
}

  \begin{minipage}[h]{0.45\textwidth}
    \begin{center}
  \begin{tikzpicture}[ultra thick,scale=0.25]
    \draw[quark] (-10,0) -- node[black,above,sloped] {$D_S$} (0,0);
    \draw[muon] (0,0) -- node[black,above,sloped] {$\mu$} (10,2);
    \draw[neutrino] (0,0) -- node[black,below,sloped] {$\nu_\mu$} (5,-1) node{$\circ$};
    \draw[heavy] (5,-1) -- node[black,below,sloped] {$N_{2,3}$} (10,-2);
  \end{tikzpicture}
  \begin{tikzpicture}[ultra thick,scale=0.25]
    \draw[quark] (-10,0) -- node[black,above,sloped] {$D$} (0,0);
    \draw[muon] (0,0) -- node[black,above,sloped] {$\mu$} (10,2);
    \draw[pion] (0,0) --  (10,0) node[black,below,sloped] {$\pi$};
    \draw[neutrino] (0,0) -- node[black,below,sloped] {$\nu_\mu$} (5,-1) node{$\circ$};
    \draw[heavy] (5,-1) -- node[black,below,sloped] {$N_{2,3}$} (10,-2);
  \end{tikzpicture}
    \end{center}
  \end{minipage}%
  \begin{minipage}[h]{0.45\textwidth}
    \begin{center}
  \begin{tikzpicture}[ultra thick,scale=0.25]
    \draw[heavy] (-10,0) -- node[black,above,sloped] {$N_{2,3}$} (-5,0) node{$\circ$};
    \draw[neutrino] (-5,0) -- node[black,above,sloped] {$\nu_\mu$} (0,0);
    \draw[muon] (0,0) -- node[black,above,sloped] {$\mu$} (10,2);
    \draw[pion] (0,0) -- node[black,below,sloped] {$\pi$} (10,-2);
  \end{tikzpicture}
  \begin{tikzpicture}[ultra thick,scale=0.25]
    \draw[heavy] (-10,0) -- node[black,above,sloped] {$N_{2,3}$} (-5,0) node{$\circ$};
    \draw[neutrino] (-5,0) -- node[black,above,sloped] {$\nu_\mu$} (0,0);
    \draw[muon] (0,0) -- node[black,above,sloped] {$\mu$} (10,2);
    \draw[electron] (0,0) -- (10,0) node[black,below,sloped] {$e$};
    \draw[neutrino] (0,0) -- node[black,below,sloped] {$\nu_e$} (10,-2);
  \end{tikzpicture}
    \end{center}
  \end{minipage}

\caption[Feynman diagrams pertaining to sterile neutrinos]{Left: Feynman  diagrams of meson decays producing
heavy sterile neutrinos. Right: Feynman diagrams of sterile-neutrino decays.}
\label{fig:production-and-decays}
\end{figure}

\begin{introbox}
  The most economical way to handle the problems of neutrino masses,
  dark matter and the Baryon Asymmetry of the Universe in a unified
  way may be to add to the Standard Model (SM) three Majorana singlet
  fermions with masses roughly on the order of the masses of known
  quarks and leptons using the seesaw
  mechanism~\cite{Yanagida:1980xy}. The appealing feature of this
  theory (called the $\nu$MSM for \emph{Neutrino Minimal
    SM})~\cite{Asaka:2005pn} is that every left-handed fermion has a
  right-handed counterpart, leading to a consistent way of treating
  quarks and leptons.

  The most efficient mechanism proposed for producing these heavy
  sterile singlet states experimentally is through weak decays of
  heavy mesons and baryons, as can be seen from the left-hand diagram
  in Figure~\ref{fig:production-and-decays}, showing some examples of
  relevant two- and three-body decays~\cite{Gorbunov:2007ak}. These
  heavy mesons can be produced by energetic protons scattering off the
  LBNE neutrino production target and the heavy singlet neutrinos from their
  decays detected in the near detector.
\end{introbox}

The lightest of the three new singlet fermions in the $\nu$MSM, is
expected to have a mass from \SIrange{1}{50}{\keV}~\cite{Boyarsky:2009ix} 
and could play the role of the dark matter particle~\cite{Dodelson:1993je}. 
The two other neutral fermions are
responsible for giving mass to ordinary neutrinos via the seesaw
mechanism at the {\em electroweak scale} and for creation of the
Baryon Asymmetry of the Universe (BAU; for a review
see~\cite{Boyarsky:2009ix}). The masses of these particles and their
coupling to ordinary leptons are constrained by particle physics
experiments and cosmology~\cite{Gorbunov:2007ak,Atre:2009rg}. 
They should be almost degenerate, thus nearly forming Dirac fermions (this is
dictated by the requirement of successful baryogenesis). Different
considerations indicate that their mass should be in the region of
${\cal O}(1)$~GeV~\cite{Shaposhnikov:2008pf}.  The mixing angle,
$U^2$, between the singlet fermions and the three active-neutrino
states must be small~\cite{Asaka:2005pn,Akhmedov:1998qx} 
--- otherwise the large mixing
would have led to equilibration of these particles in the early
Universe above the electroweak temperatures, and, therefore, to
erasing of the BAU --- explaining why these new particles
have not been seen previously.

Several experiments have conducted searches for heavy neutrinos, for
example BEBC~\cite{CooperSarkar:1985nh}, \linebreak
CHARM~\cite{Bergsma:1985is},
NuTeV~\cite{Vaitaitis:1999wq} and the CERN PS191
experiment~\cite{Bernardi:1985ny,Bernardi:1987ek} (see also a discussion
of different experiments in~\cite{Atre:2009rg}). 
In the search for heavy
neutrinos, the strength of the LBNE 
ND, compared
to earlier experiments, lies in reconstructing the exclusive decay
modes, including electronic, hadronic and muonic.  Furthermore, the
detector offers a means to constrain and measure the backgrounds using
control samples.

In case of the LBNE experiment the relevant heavy mesons are charmed. 
With a typical lifetime (in the rest frame) of about
\SI{e-10}{s}, 
these mesons mostly decay before further interaction,
yielding the sterile-neutrino flux.  Since these sterile neutrinos are
very weakly interacting they can cover quite a large distance before
decay, significantly exceeding the distance of roughly \SI{500}{\meter} from the target
to the ND.  The ND can search for decays of neutrinos into SM particles due
to mixing with active neutrinos,
provided a sufficiently long instrumented decay region is available.
Two examples of the interesting decay modes are presented on the right
panel of Figure~\ref{fig:production-and-decays}. More examples can be found
in~\cite{Gorbunov:2007ak}. 

An estimate of sterile-neutrino events that can be observed in the
LBNE ND, $N^{LBNE}_{signal}$, is obtained by comparing the
relevant parameters of the LBNE and CHARM experiments.  The number of
events grows linearly with the number of protons on target, the number
of produced charmed mesons, the detector length (decay region) and the
detector area.  In particular, this latter linear increase   is valid if the
angular spread of the neutrino flux, which is on the order of
$N_mM_D/E_{beam}$, is larger than the angle at which the ND is seen
from the target.  Here $N_m$ is the multiplicity of the produced
hadrons, and the above condition is valid for both LBNE and CHARM. The
number of events 
decreases linearly when the energy increases,
since this increases the lifetime, reducing the decay probability within
the detector.  Finally, the number of mesons decreases quadratically
with the distance between the target and the detector.

\begin{figure}[!htb]
\centerline{
\includegraphics[width=0.5\textwidth]{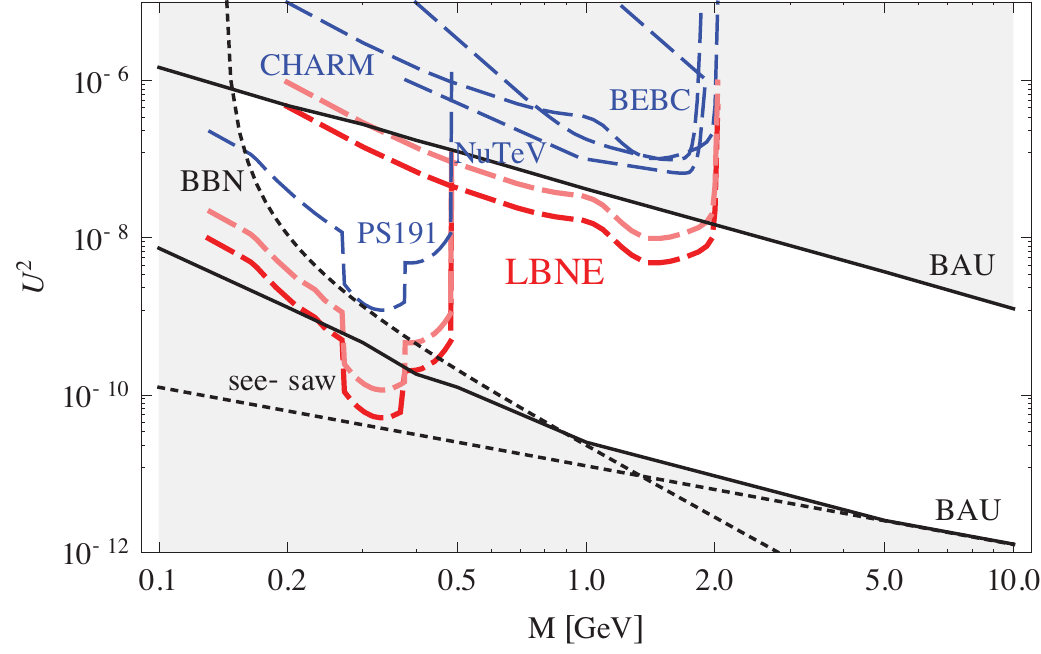}
\includegraphics[width=0.5\textwidth]{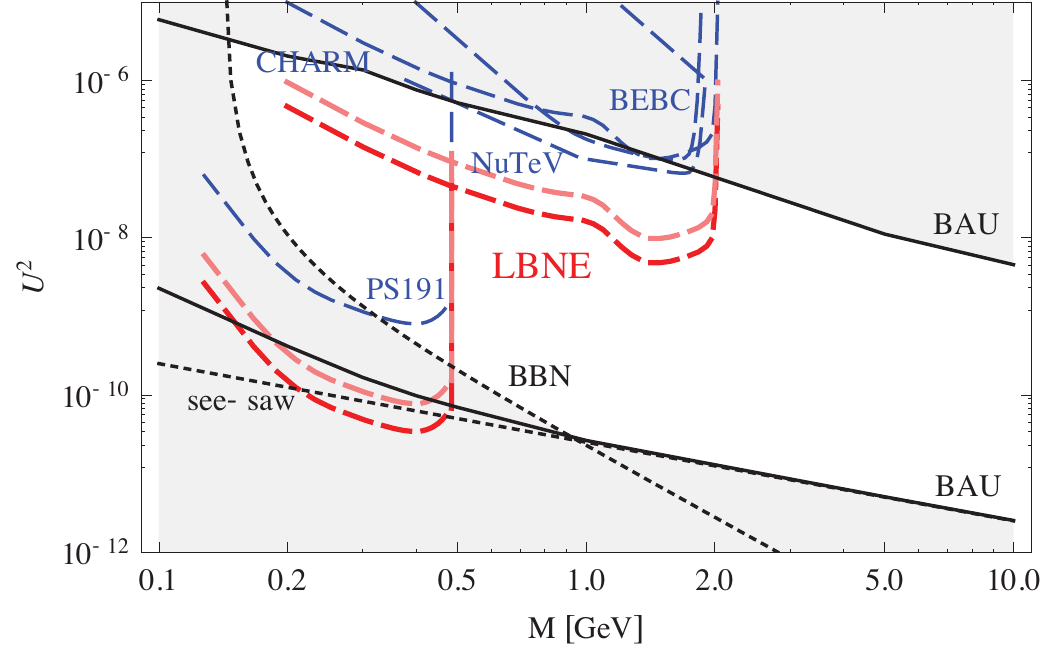}}
\caption[Near detector constraints on heavy Majorana neutrinos]
{Upper limits on $U^2$, the mixing angle between heavy sterile
  neutrinos and the light active states, coming from the Baryon
  Asymmetry of the Universe (solid lines), from the seesaw mechanism
  (dotted line) and from the Big Bang nucleosynthesis (dotted
  line). The regions corresponding to different experimental searches
  are outlined by blue dashed lines. Left panel: normal hierarchy;
  right panel: inverted hierarchy (adopted
  from~\cite{Canetti:2010aw}).  Pink and red curves indicate the
  expected sensitivity of the LBNE near detector with an exposure of
  $5\times 10^{21}$ POT ($\sim 5$ years) with the \MWadj{1.2}
  reference beam at 120 GeV for detector lengths of \SI{7}{m} and
  \SI{30}{m} , respectively (see text for details).}
\label{fig:heavynu}
\end{figure}

The considerations above imply that a search for $\nu$MSM sterile
neutrinos in the LBNE ND can be 
competitive after only five years of running with the reference beam,
corresponding to an overall integrated exposure of about \num{5e21}
POT with a proton energy of \SI{120}{GeV}.  The use of a low-density,
high-resolution spectrometer in the ND substantially reduces
backgrounds and allows the detection of both leptonic and hadronic
decay modes.  Assuming a fiducial length of the magnetized tracker of
\SI{7}{m} as decay region, the ratio between the signal event to be
observed in the LBNE ND and those in the CHARM experiment can be
estimated to be more than a factor of 50 after only four years of
running.  Since both production and decay rates are proportional to
the square of the neutrino mixing angles, the corresponding
improvement in the square of the neutrino mixing angle $U^2$ will be
about a factor of seven with respect to the CHARM
experiment. Figure~\ref{fig:heavynu} shows the projected LBNE
sensitivity in the $(U^2,M)$ plane.  At lower values of the mass of
the heavy neutrinos, additional constraints can be obtained for kaons
by comparing the LBNE and PS191 experiments, as shown in
Figure~\ref{fig:heavynu}.

It must be noted that exploitation of the complete 5 + 5 years ($\nu$ + $\overline\nu$) years 
of data taking would further improve the number of expected
events by a factor of  two, since it 
scales linearly with the number of protons on target.  With the beam
upgrade to \SIadj{2.3}{MW}, this improvement would become a factor of
four with respect to the initial five year run and the 
\SI{1.2}{MW} beam.

A better sensitivity to $\nu$MSM can be achieved by instrumenting the
upstream region of the ND hall (e.g., with the liquid argon detector and some
minimal tracking device upstream). The fiducial volume of the new
detector will need to be empty (material-free) or fully sensitive in order
to suppress background events. The geometry of the ND hall would allow
a maximal decay length of about \SI{30}{m}. The sensitivity of this
configuration can be estimated by rescaling the expected limits on
the neutrino mixing angle $U^2$. The expected number of signal events with a total decay
length of $\sim30$~m exceeds by about 200 (800) times the number of
events in CHARM after a five (5 +5) year run with the standard (upgraded)
beam. In turn, this implies an improvement by a factor of 15 (28) in
the sensitivity to $U^2$ with respect to the CHARM experiment.

If the magnetic moment of the sterile neutrinos
is sizeable, the dominant decay channel would be a radiative
electromagnetic decay into $\gamma \nu$, which has also been proposed
as a possible explanation for the observed MiniBooNE low-energy
excess~\cite{AguilarArevalo:2008rc}. This possibility,  in turn, requires a detector
capable of identifying and reconstructing single photon events.  The
low-density ND in LBNE can achieve an excellent 
sensitivity to this type of search as demonstrated by a similar analysis in
NOMAD~\cite{Kullenberg:2011rd}.

\section{Search for High $\boldsymbol{\Delta m^2}$ Neutrino Oscillations}
\label{sec-high-delmsq}

The evidence for neutrino oscillations obtained from atmospheric,
long-baseline accelerator, solar and long-baseline reactor data from
different experiments consistently indicates two different scales,
with $\Delta m_{32}^2\sim$\SI{2.4e-3}{\eV^2} defining the
atmospheric oscillations (also long-baseline accelerator and
short-baseline reactor scales) and $\Delta m_{21}^2\sim$\SI{7.9e-5}{\eV^2} defining the solar oscillations (and long-baseline
reactor oscillations).  The only way to accommodate oscillations with
relatively high $\Delta m^2$ at the \si{\eV^2} scale as suggested by the
results from the LSND experiment~\cite{Volpe:2001qe} is therefore to
add one or more sterile 
neutrinos to the conventional three light
neutrinos.

Recently, the MiniBooNE experiment reported that its antineutrino
data might be consistent with the LSND $\overline{\nu}_\mu \to \overline{\nu}_e$
oscillation with $\Delta m^2\sim$ \si{\eV^2}~\cite{Maltoni:2007zf}.
Contrary to the antineutrino data, the 
neutrino data seem to
exclude high $\Delta m^2$ oscillations, possibly indicating a 
different behavior between neutrinos and antineutrinos.

Models with five (3+2) or six (3+3) neutrinos can potentially explain
the MiniBooNE results. In addition to the cluster of the three neutrino
mass states (accounting for \emph{solar} and \emph{atmospheric} mass splitting), two
(or three) states at the eV scale are added, with a small
admixture of $\nu_e$ and $\nu_\mu$ to account for the LSND signal. 
One distinct prediction from such models is a significant probability
for $\overline{\nu}_\mu$ disappearance into sterile neutrinos, on the order
of 10\%, in addition to the small probability for $\overline{\nu}_e$ appearance.

\begin{introbox}
  Given a roughly \SIadj{500}{m} baseline and a low-energy beam, the LBNE ND
  can reach the same value $L/E_\nu\sim1$ as MiniBooNE and LSND. The
  large fluxes and the availability of fine-grained detectors make the
  LBNE program well suited to search for active-sterile neutrino
  oscillations beyond the three-flavor model with $\Delta m^2$ at the
  eV$^2$
  scale. 
\end{introbox}
Due to the potential differences between neutrinos and antineutrinos,
four possibilities have to be considered in the analysis: $\nu_\mu$
disappearance, $\overline{\nu}_\mu$ disappearance, $\nu_e$ appearance and
$\overline{\nu}_e$ appearance. As discussed in Section~\ref{sec-fluxosc},
the search for high $\Delta m^2$ oscillations has to be performed
simultaneously with the in situ determination of the fluxes.

To this end, an independent prediction of the $\nu_e$ and
$\overline{\nu}_e$ fluxes starting from the measured $\nu_\mu$ and $\overline{\nu}_\mu$ CC distributions are required since the $\nu_e$ and $\overline{\nu}_e$ CC distributions could
be distorted by the appearance signal. The low-$\nu_0$ method can provide
such predictions if external measurements for the $K_L^0$ component
are available from hadro-production experiments (Section~\ref{sec-fluxosc}). 

The study will implement an iterative procedure:
\begin{enumerate}
\item extraction of the fluxes from $\nu_\mu$ and $\overline{\nu}_\mu$ CC distributions assuming
no oscillations are present
\item comparison with data and determination of oscillation parameters (if any)
\item new flux extraction after subtraction of the oscillation effect
\item iteration until convergence
\end{enumerate}
The analysis has to be performed separately for neutrinos and antineutrinos due to
potential CP or CPT violation, according to MiniBooNE/LSND data.
The ratio of $\nu_e$ CC events to $\nu_\mu$ CC events will be measured: 
\begin{equation}
{\mathcal{R}}_{e \mu} (L/E)  \equiv  \frac{\#~of~\nu_e N \to e^- X}{\#~of~\nu_\mu N \to \mu^- X }(L/E); \;\;\;\;\;\;\;  \overline{\mathcal{R}}_{e \mu} (L/E) \equiv \frac{\#~of~\overline{\nu}_e N \to e^+ X}{\#~of~\overline{\nu}_\mu N \to \mu^+ X }(L/E)
\end{equation}
This is then compared with the predictions obtained from the low-$\nu_0$ method.
Deviations of ${\mathcal{R}}_{e \mu}$ or $\overline{\mathcal{R}}_{e \mu}$ from the expectations
as a function of $L/E$ would provide evidence for oscillations. 
This procedure only provides a relative measurement of $\nu_e (\overline{\nu}_e)$
versus $\nu_\mu (\overline{\nu}_\mu)$; since the fluxes
are extracted from the observed $\nu_\mu$ and $\overline{\nu}_\mu$ CC distributions, an analysis
of the ${\mathcal{R}}_{e \mu} (\overline{\mathcal{R}}_{e \mu})$ ratio cannot distinguish
between $\nu_\mu (\overline{\nu}_\mu)$ disappearance and $\nu_e (\overline{\nu}_e)$ appearance.

The process of NC elastic scattering off protons (Section~\ref{sec-deltas})
can provide the complementary measurement
needed to disentangle the two hypotheses of $\nu_\mu (\overline{\nu}_\mu)$ disappearance into
sterile neutrinos and $\nu_e (\overline{\nu}_e)$ appearance. In order to cancel systematic
uncertainties, the NC/CC ratio with respect to QE scattering will be measured:
\begin{equation}
{\mathcal{R}}_{NC} (L/E)  \equiv  \frac{\#~of~\nu p \to \nu p}{\#~of~\nu_\mu n \to \mu^- p }(L/E); \;\;\;\;\;\;\; \overline{\mathcal{R}}_{NC} (L/E) \equiv \frac{\#~of~\overline{\nu} p \to \overline{\nu} p}{\#~of~\overline{\nu}_\mu p \to \mu^+ n }(L/E)
\end{equation}

It is possible to reconstruct the neutrino energy from the proton
angle and momentum under the assumption that the nuclear smearing
effects are small enough to neglect (the same for the neutrino CC
sample). In the oscillation analysis, only the \emph{relative}
distortions of the ratio ${\mathcal{R}}_{NC}
(\overline{\mathcal{R}}_{NC})$ as a function of $L/E$ are of interest,
not their absolute values. For $Q^2>0.2$~GeV$^2$ the relative shape of
the total cross sections is not very sensitive to the details of the
form factors.  To improve the energy resolution, it is possible to use
neutrino interaction events originating from the deuterium inside the
D$_2$O target embedded into the fine-grained tracker. These events
have better energy resolution due to the smaller nuclear smearing
effects in D$_2$O.

An improved oscillation analysis is based on a simultaneous fit to
both ${\mathcal{R}}_{e \mu} (\overline{\mathcal{R}}_{e \mu})$ and
${\mathcal{R}}_{NC} (\overline{\mathcal{R}}_{NC})$. The first ratio
provides a measurement of the oscillation parameters while the latter
constrains the $\nu_e(\overline{\nu}_e)$ appearance versus the
$\nu_\mu(\overline{\nu}_\mu)$ disappearance. This analysis 
imposes two main requirements 
on the ND:
\begin{itemize}
\item $e^+/e^-$ separation to provide an unambiguous check of the different
behavior between neutrinos and antineutrinos suggested by MiniBooNE
\item accurate reconstruction of proton momentum and angle
\end{itemize}

Validation of the unfolding of the high $\Delta m^2$ oscillations from
the in situ extraction of the $\nu(\overline{\nu})$ flux would
also require changes to the beam conditions, since the ND cannot be
easily moved. This would require a short run with a high-energy beam
and the capability to change or switch off the beam focusing system.



\section{Light (sub-GeV) Dark Matter Searches}

According to the latest cosmological and astrophysical measurements,
nearly eighty percent of the matter in the Universe is in the form of
cold, non-baryonic dark matter (DM)~\cite{Ade:2013zuv,Bennett:2012zja}. 
The search to find evidence of the particle (or particles) that make
up DM, however, has so far turned up empty.  Direct detection
experiments and indirect measurements at the LHC, however, are
starting to severely constrain the parameter space of
Weakly-Interacting Massive Particles (WIMPs), one of the leading
candidates for DM.  The lack of evidence for WIMPs at these
experiments has forced many in the theory community to
reconsider.

Some theories consider an alternative possibility to the WIMP paradigm
in which the DM mass is much lighter than the electroweak scale (e.g.,
below the GeV level). In order to satisfy constraints on the relic
density of DM, these theories require that DM particles be accompanied
by light \emph{mediator} particles that would have allowed for efficient
DM annihilation in the early Universe. In the simplest form of these
theories an extra U(1) gauge field mixes with the SM
U(1) gauge field, but with an additional kinetic term.  This mixing
term provides a \emph{portal} from the dark sector to the charged
particles of the SM.  In this model, the mediators are called \emph{dark
photons} and are denoted by $V$.

\begin{introbox}
  Recently, a great deal of interest has been paid to the possibility
  of studying models of light (sub-GeV) Dark Matter at low-energy,
  fixed-target
  experiments~\cite{Batell:2009di,deNiverville:2011it,deNiverville:2012ij,Dharmapalan:2012xp}.
  High-flux neutrino beam experiments --- such as LBNE --- have been
  shown to potentially provide coverage of DM+mediator parameter space
  that cannot be covered by either direct detection or collider
  experiments.
\end{introbox}

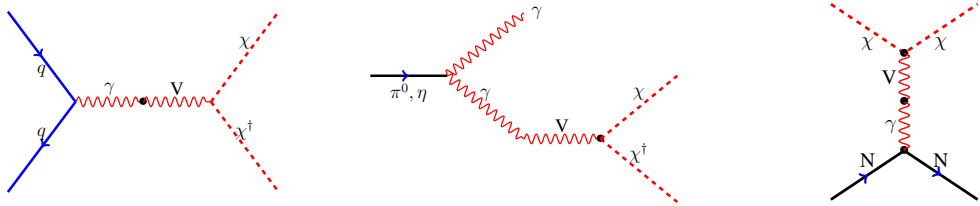
\begin{figure}[!htb]
\centering
\tikzset{
  quark/.style={draw=blue, postaction={decorate},
    decoration={markings,mark=at position .5 with {\arrow[draw=blue]{>}}}},
  electron/.style={draw=pink, postaction={decorate},
    decoration={markings,mark=at position .5 with {\arrow[draw=blue]{>}}}},
  neutrino/.style={draw=red, postaction={decorate},
    decoration={markings,mark=at position .5 with {\arrow[draw=blue]{>}}}},
  heavy/.style={draw=red, dashed},
  pion/.style={draw=black,postaction={decorate},
    decoration={markings,mark=at position .5 with {\arrow[draw=blue]{>}}}},
  nucleus/.style={ultra thick, draw=black,postaction={decorate},
    decoration={markings,mark=at position .5 with {\arrow[draw=blue]{>}}}},
  muon/.style={draw=purple, postaction={decorate},
    decoration={markings,mark=at position .5 with {\arrow[draw=blue]{>}}}},
  gamma/.style={thick, decorate, decoration={snake,amplitude=3pt, segment length=6pt}, draw=red},
}

  \begin{minipage}[c]{0.3\textwidth}
    \begin{center}
\scalebox{0.60}{
\begin{tikzpicture}[ultra thick, node distance=2cm and 1.5cm]
\coordinate[] (center);
\coordinate[left=of center] (gam);
\coordinate[below left=of gam] (qm);
\coordinate[above left=of gam] (qp);
\coordinate[right=of center] (vee);
\coordinate[below right=of vee] (chid);
\coordinate[above right=of vee] (chi);

\draw[quark] (qp) -- node[below]{$q$} (gam);
\draw[quark] (gam) -- node[above]{$q$} (qm);
\draw[gamma] (gam) -- node[above]{$\gamma$} (center) node {$\bullet$};
\draw[gamma] (center) -- node[above]{V} (vee);
\draw[heavy] (vee) -- node[above]{$\chi$} (chi);
\draw[heavy] (vee) -- node[above]{$\chi^\dagger$} (chid);
\end{tikzpicture}
}
    \end{center}
  \end{minipage}
  \begin{minipage}[c]{0.3\textwidth}
    \begin{center}
\scalebox{0.60}{
\begin{tikzpicture}[ultra thick, node distance=1.4cm and 1.7cm]
\coordinate[] (center);
\coordinate[left=of center] (gamgam);
\coordinate[left=of gamgam] (mes);
\coordinate[above right=of gamgam,label=right:{$\gamma$}] (gam1);
\coordinate[below right=of gamgam] (gam2);
\coordinate[right=of gam2] (veedk);
\coordinate[above right=of veedk] (chi);
\coordinate[below right=of veedk] (chid);

\draw[pion] (mes) -- node[below]{$\pi^0,\eta$} (gamgam);
\draw[gamma] (gamgam) -- (gam1);
\draw[gamma] (gamgam) -- node[above]{$\gamma$}(gam2);
\draw[gamma] (gam2) -- node[above]{V} (veedk) node{$\bullet$};
\draw[heavy] (veedk) -- node[above]{$\chi$}(chi);
\draw[heavy] (veedk) -- node[above]{$\chi^\dagger$}(chid);
\end{tikzpicture}
}
    \end{center}
  \end{minipage}
  \begin{minipage}[c]{0.3\textwidth}
    \begin{center}
\scalebox{0.65}{
\begin{tikzpicture}[ultra thick, node distance=1cm and 1.5cm]
\coordinate[] (center);
\coordinate[above=of center] (chichi);
\coordinate[above left=of chichi] (chi1);
\coordinate[above right=of chichi] (chi2);
\coordinate[below=of center] (enen);
\coordinate[below left=of enen] (en1);
\coordinate[below right=of enen] (en2);

\draw[heavy] (chi1) -- node[below]{$\chi$} (chichi)  node{$\bullet$} -- node[below]{$\chi$} (chi2);
\draw[nucleus] (en1) -- node[above]{N} (enen);
\draw[nucleus] (enen)  node{$\bullet$} -- node[above]{N} (en2);
\draw[gamma] (chichi) -- node[left]{V} (center) node{$\bullet$};
\draw[gamma] (center) -- node[left]{$\gamma$} (enen);
\end{tikzpicture}
}
    \end{center}
  \end{minipage}

\caption[Production mechanisms for dark matter at neutrino-beam
experiments] {On the left is shown the direct production of a dark
  photon, while, in the center, the dark photon is produced via the
  decay of a neutral pion or eta meson. In both cases, the dark photon
  promptly decays into a pair of DM particles. Right: Tree-level
  scattering of a DM particle off of nuclei. Analogous interactions
  with electrons in the detector are also possible.}
\label{fig:dm}
\end{figure}

Upon striking the target, the proton beam can produce the dark photons
either directly through $pp(pn)\rightarrow {V}$ 
as in the left-hand diagram of Figure~\ref{fig:dm} or indirectly
through the production of a $\pi^{0}$ or a $\eta$ meson which then
promptly decays into a SM photon and a dark photon as in the center
diagram in the figure. 
For the case where $m_{V} > 2m_{DM}$, the dark photons will quickly
decay into a pair of DM particles.

The LBNE ND together with the  high-intensity beam will provide an excellent setup for making this
measurement. The relativistic DM particles from the beam will travel along with
the neutrinos to the 
detector where they 
can  be detected through NC-like interactions either with
electrons or nucleons, 
as shown in the right-hand diagram of 
Figure~\ref{fig:dm}.  Since the signature of a DM event looks similar to that of
a neutrino event, the neutrino beam provides the major source of
background for the DM signal. 

Several ways have been proposed to suppress neutrino backgrounds using
the unique characteristics of the DM beam. Since DM will travel much
more slowly than the much lighter neutrinos, 
DM events in the ND will arrive out of time with the 
beam pulse.
In addition, since the electrons struck by DM will be in a much more
forward direction compared to neutrino interactions, the angle of
these electrons may be used to reduce backgrounds, taking advantage of
the ND's fine angular resolution. 

\begin{figure}[!tb]
\centerline{
\includegraphics[width=\textwidth]{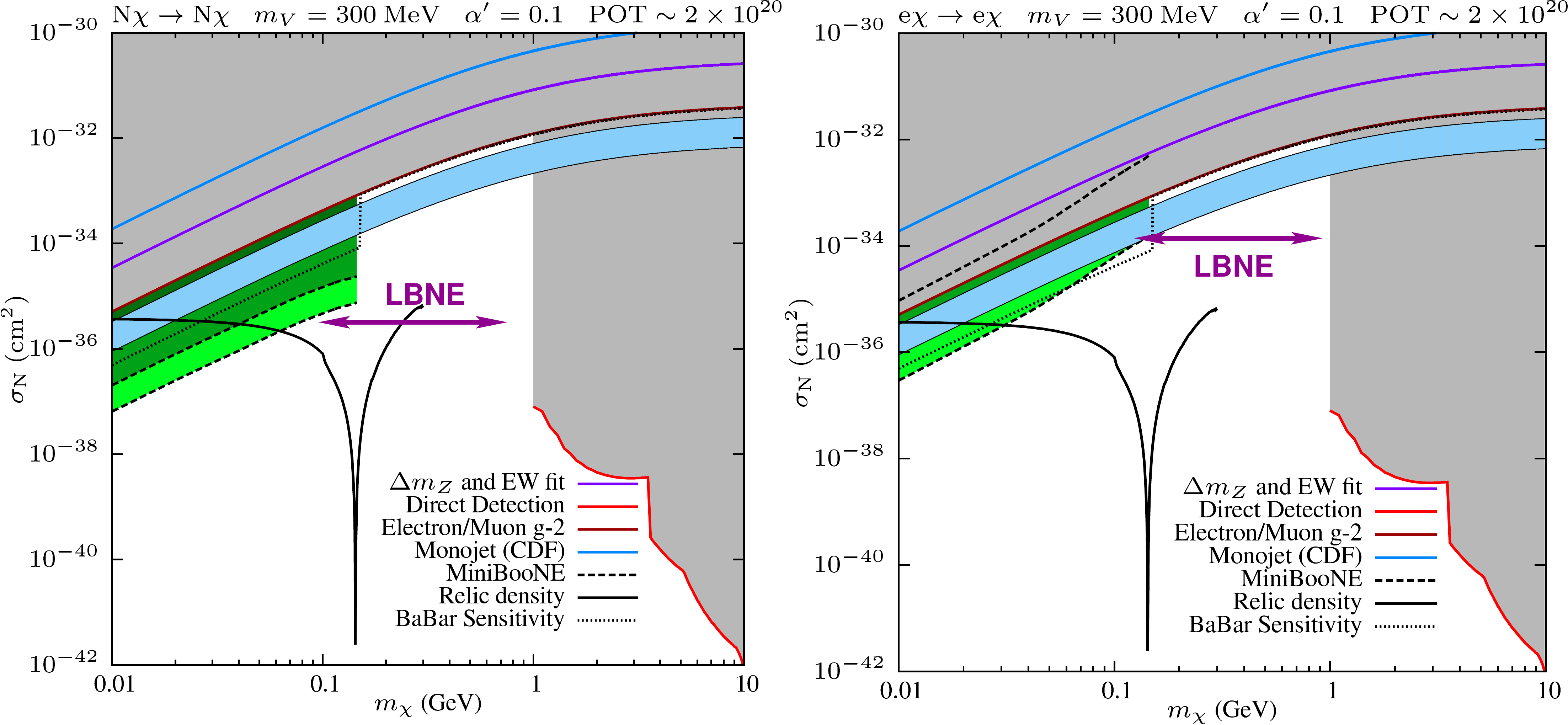}
}
\caption[Regions of nucleon-WIMP cross section versus WIMP
mass]{Regions of nucleon-WIMP scattering cross section (corresponding
  to dark matter in the lab moving with $v = 10^{-3}c$). The plot uses
  $m_V = 300$~MeV and $\alpha'=0.1$. Constraints are shown from
  different experiments.  The left plot shows the exclusion regions
  expected from MiniBooNE given 1-10 (light green), 10-1000 (green),
  and more than 1000 (dark green) elastic scattering events off
  nucleons. The right panel shows the same for elastic scattering off
  electrons. The magenta arrows indicate the region where LBNE can
  extend the MiniBooNE sensitivity. Figure is based on studies in ~\cite{Dharmapalan:2012xp}.}
\label{fig:wimp}
\end{figure}

Finally, a special run can be devised to turn off the focusing horn to
significantly reduce the charged particle flux that will produce
neutrinos.  
Figure ~\ref{fig:wimp} shows the expected sensitivity of the MiniBooNE
DM search using this technique~\cite{Dharmapalan:2012xp}. With a
wider-band, higher-energy, more intense beam, LBNE is expected to not
only cover the MiniBooNE sensitivity region with higher statistics,
but will also extend the sensitivity to cover the region between 
MiniBooNE and the direct DM searches.
If the LBNE ND were a LArTPC and the entire
detector volume active, the effective number of DM events detected
would be much higher when compared to a MINOS-like detector of the
same mass. Much more thorough studies must be conducted to obtain
reliable sensitivities. This requires an integration of theoretical
predictions into a simulation package for the detector.








%% file: chapter-other-goals.tex

\begin{introbox}
  The deep underground location of LBNE's LArTPC far
  detector will expand the range of science opportunities it can
  pursue to potentially include observation of solar and other
  low-energy neutrinos, dark matter, magnetic monopoles and
  nucleon-antinucleon transitions.
\end{introbox}

\section{Solar Neutrinos}
\label{sec:SolNu}

\begin{figure}[!htb]
\includegraphics[width=\textwidth]{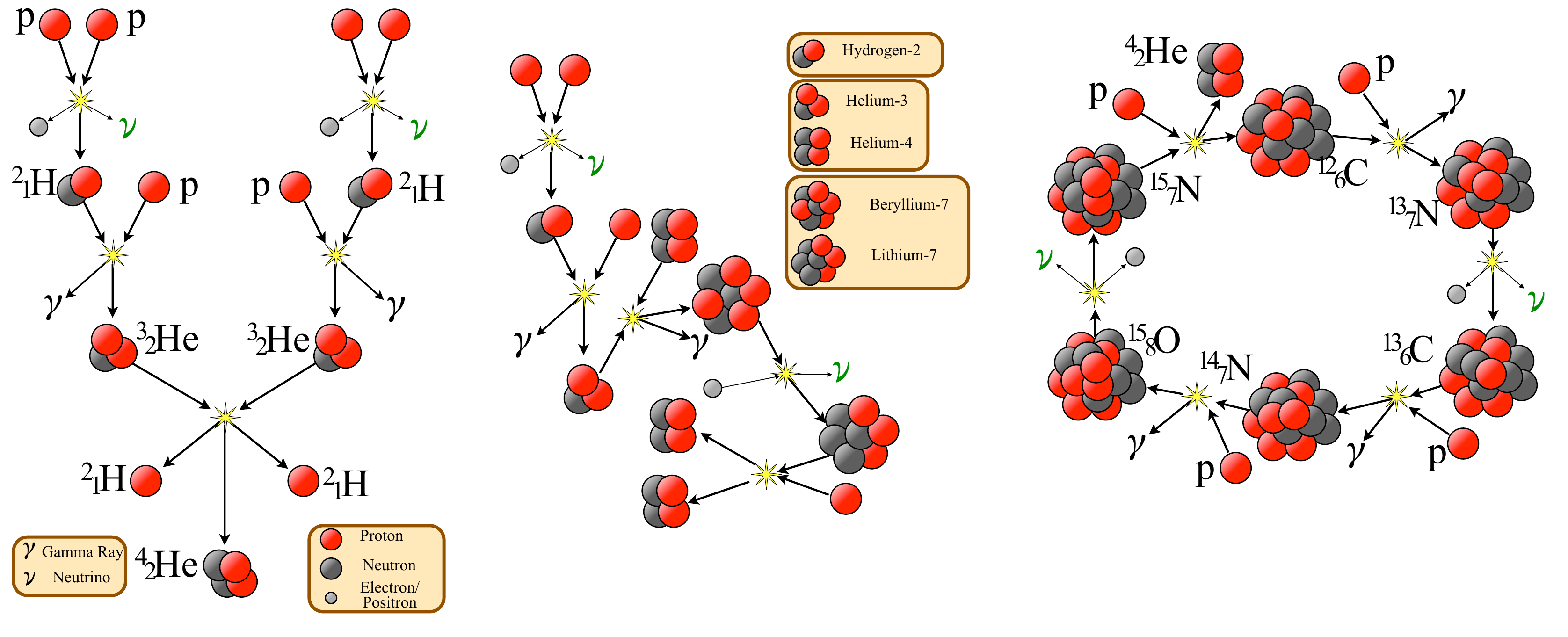}
\includegraphics[height=0.8\textwidth,angle=-90]{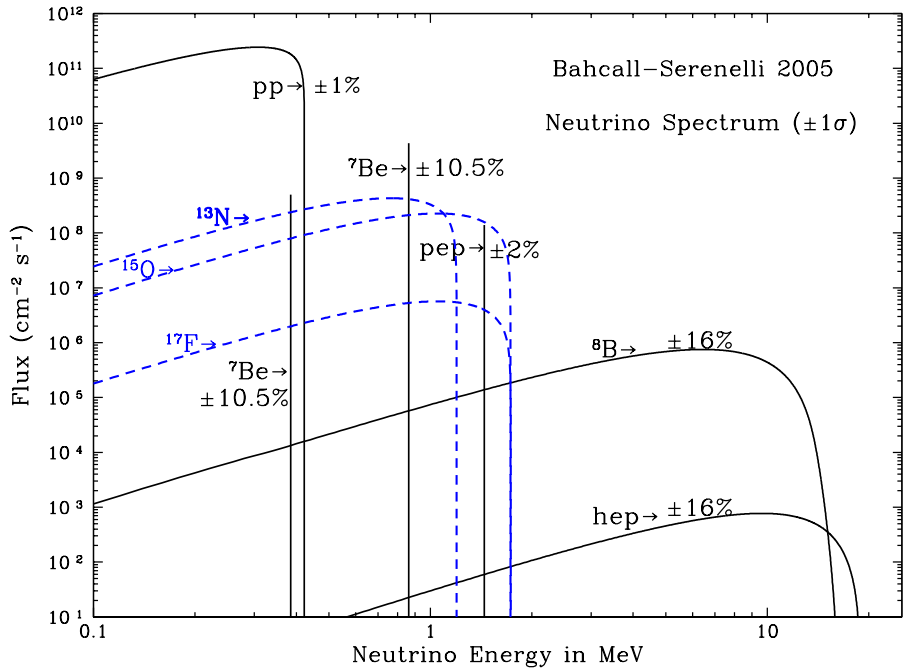}
\caption[The solar proton-proton cycle and solar $\nu$ spectrum]
{Top: the proton-proton and CNO reaction chain in the Sun. 
Bottom: solar-neutrino fluxes from~\cite{Bahcall:2004pz}.
}
\label{fig:solnuflux}
\end{figure}

In the early $20^{\mbox{th}}$ century, Arthur Stanley Eddington
suggested that nuclear reactions of protons fuel energy production in
the Sun.  After the discovery of the neutron, Hans
Bethe~\cite{Bethe:1939bt} proposed that the first stage of these
nuclear reactions involves the weak interaction: a $\beta$ decay of a
proton into a neutron, a positron and a neutrino accompanied by the
fusion of that neutron with another proton to form deuterium. This proton-proton 
($pp$) reaction $p+p\rightarrow ^1_1$H$+e^++\nu_e$ is the origin of most
solar neutrinos (called $pp$ neutrinos). In 0.2\% of the cases
deuterium is produced by the corresponding three-body reaction
$p+e^-+p\rightarrow ^1_1$H$+\nu_e$ (called $pep$) which produces
monoenergetic solar neutrinos at 1.4 MeV.  The $pp$ reaction is the
starting point of a chain of nuclear reactions which converts four
protons into a $^4_2$He nucleus, two positrons and two
neutrinos. This reaction chain, shown in Figure~\ref{fig:solnuflux},
produces 98\% of the energy from the Sun.  In addition to $pp$ and
$pep$, neutrinos are produced by the reactions
$^7_4$Be$+e^-\rightarrow ^7_3$Li$+\nu_e$ ($^7$Be neutrinos) and
$^3_2$He$+p\rightarrow ^4_2$He$+e^++\nu_e$ (\emph{hep} neutrinos) as well as
the $\beta$ decay $^8_5$B$\rightarrow ^8_4$Be$+e^++\nu_e\rightarrow
^4_2$He$+ ^4_2$He$+e^++\nu_e$ ($^8$B neutrinos).  Carl-Friedrich von
Weizs{\"a}cker~\cite{Weizsaecker:1938} complemented the pp-chain with
a cyclical reaction chain dubbed \emph{CNO cycle} after the principal
elements involved (shown in the top right illustration of
Figure~\ref{fig:solnuflux}). 
Although theorized to be responsible for only 2\% of energy
production in the Sun, the CNO cycle plays the dominant role in the energy production of
stars heavier than 1.3 solar masses.

The expected spectra of neutrinos from the $pp$ reaction
chain~\cite{Bahcall:2004pz} are shown as solid curves in the bottom
diagram of Figure~\ref{fig:solnuflux}. Neutrinos from the
CNO cycle are shown as dashed blue curves.

The chief motivation of Raymond Davis to build his pioneering solar-neutrino 
detector in the Homestake mine was the experimental
verification of stellar energy production by the observation of the
neutrinos from these nuclear processes. While he 
succeeded in
carrying out
the first measurements of solar neutrinos --- and shared the 2002 Nobel
Prize in physics for the results --- the measured
flux~\cite{Cleveland:1998nv} fell short of solar model calculations:
the \emph{solar-neutrino problem}.  Data from the \superk{} (SK) and
SNO~\cite{Fukuda:2001nj,Ahmad:2001an} experiments eventually explained
this mystery 30 years later as due to flavor transformation. However,
intriguing questions in solar-neutrino physics remain.
Some unknowns, such as the fraction of energy production via the CNO
cycle in the Sun, flux variation due to helio-seismological modes that
reach the solar core, or long-term stability of the solar core
temperature, are astrophysical in nature. Others directly impact
particle physics. Can the MSW model explain the amount of flavor
transformation as a function of energy, or are nonstandard neutrino
interactions required? Do solar neutrinos and reactor antineutrinos
oscillate with the same parameters?  Experimental data expected in the
immediate future (e.g., further data from
Borexino~\cite{Borexino7be:2011} and SK as well as 
SNO+~\cite{Kraus:2006qp}) will address some questions, but the
high-statistics measurements necessary to further constrain
alternatives to the standard oscillation scenario may need to wait for
a more capable experiment such as LBNE.

\begin{introbox}
  Detection of solar and other low-energy neutrinos is challenging in
  a LArTPC because of high intrinsic detection energy thresholds for
  the charged-current (CC) interaction on argon ($>$\SI{5}{\MeV}). To be
  competitive, this physics requires either a very low visible-energy
  threshold ($\sim$\SI{1}{\MeV}) or a very large mass (\SI{50}{\kt}).
  However, compared with other technologies, a LArTPC offers a large
  cross section and unique signatures from de-excitation
  photons. Aggressive R\&D efforts in low-energy triggering and
  control of background from radioactive elements may make detection
  in LBNE possible, and a large detector mass would make the pursuit
  of these measurements worthwhile.
\end{introbox}

\begin{figure}[!htb]
\centering\includegraphics[width=0.7\textwidth]{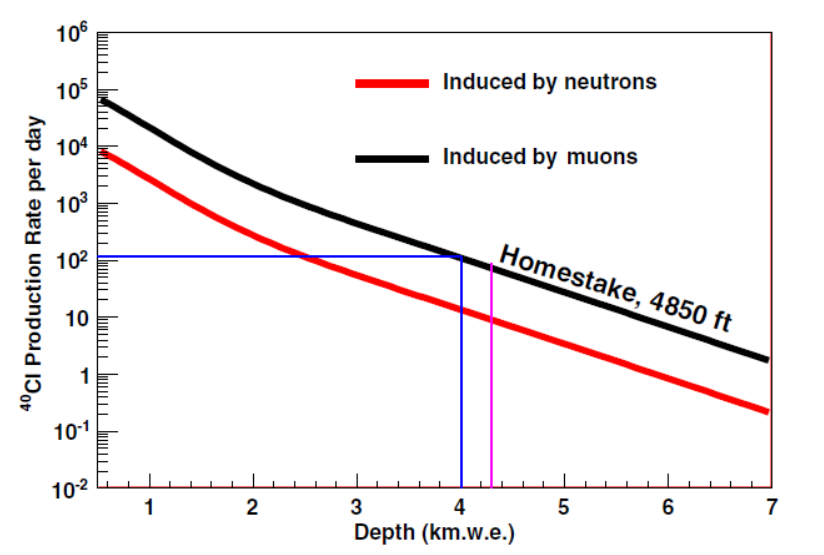}
\caption[$^{40}$Cl production rates produced by (n,p) reaction per depth, \SI{10}{\kt}]{$^{40}$Cl production rates in a \ktadj{10} detector produced by (n,p) reaction as a function of depth.}
\label{fig:cl40bkgd}
\end{figure}

The solar-neutrino physics potential of a large LArTPC depends
primarily on its energy threshold and depth. The energy threshold is
not only determined by the ability to pick up a low-energy electron,
but also by the light collection of the photon-triggering system as
well as background suppression. Only at a deep underground location
will it have a reasonable chance of detecting solar neutrinos. In any
detector of this kind, the decay of the naturally occurring $^{39}$Ar
produces $\beta$'s with a \keVadj{567} endpoint and an expected rate
of \SI{10}{\MHz} per \SI{10}{\kt} of liquid argon. This limits the
fundamental reach of LBNE to neutrino interactions with visible
energies above \SI{1}{\MeV}. Possible signatures of solar neutrinos in LBNE
are:
\begin{description}
\item [Elastic scattering of $^8$B neutrinos with electrons:] This
  signature would only reproduce the SK data; SK has
  already accumulated large %
  statistics (>60,000 solar-neutrino events). An energy threshold of
  about \SI{1}{\MeV} (lower than the SK threshold which
  is currently \SI{3.5}{\MeV}~\cite{Sekiya:2013hda}) would be
  required for a more interesting measurement of {\em pep} (defined in Figure~\ref{fig:solnuflux}) and CNO
  fluxes. Such solar-neutrino interactions are difficult to detect, as
  only low-energy single electrons (and neutrinos) are produced.
\item[Charged-current interactions with argon:] The signature for this interaction is:
\begin{equation}
 {\nu}_{e} + \ce{^{40}Ar} \rightarrow \ce{^{40}K^*} + e^{-}
\label{eqn:srninteract}
\end{equation}
  This signal is more
  interesting experimentally, as there is a signature of de-excitation
  photons and the visible energy is directly correlated with the
  neutrino energy; however, the reaction has an energy threshold of
  \SI{5}{\MeV}. 
\end{description}
 Cosmic-muon and fast-neutrino
  interactions with the $^{40}$Ar nucleus (which are rather complex
  compared to interactions on $^{16}$O or $^{12}$C) are likely to
  generate many long-lived spallation products that could limit the
  detection threshold for low-energy neutrinos.
Studies of the spallation background in the LBNE LArTPC are
underway. The production rate of $^{40}$Cl, a beta emitter with an
endpoint of \SI{7.48}{\MeV} that is a dominant source of background at
energies above \SI{5}{\MeV}, is shown in Figure~\ref{fig:cl40bkgd} as
a function of depth.
The cosmogenic background rates as a function of
beta kinetic energy from several other beta emitters at the \SIadj{4850}{\ft} 
level of Sanford Underground Research Facility are shown in
Figure~\ref{fig:cosmicbkg}.

In Table~\ref{tab:solarnu} the solar-neutrino event rate in a
\ktadj{34} LArTPC is shown, assuming a \MeVadj{4.5} neutrino energy
threshold and 31\% $\nu_e$.
\begin{table}[!htb]
\caption[Solar-neutrino rates in a LArTPC]{Solar-neutrino event rates in a \ktadj{34} LArTPC assuming 
a \MeVadj{4.5} neutrino energy threshold and an electron-flavor survival
probability $P_{ee}=31\%$.}
\label{tab:solarnu}
\begin{tabular}{$L^c}
\toprule
\rowtitlestyle
Transition & Rate (evts/day) \\ \toprowrule
Fermi  & 31 \\ \colhline
Gamow-Teller & 88 \\ \bottomrule
\end{tabular}
\end{table}

\begin{figure}[!htb]
\centering
\includegraphics[width=0.7\textwidth]{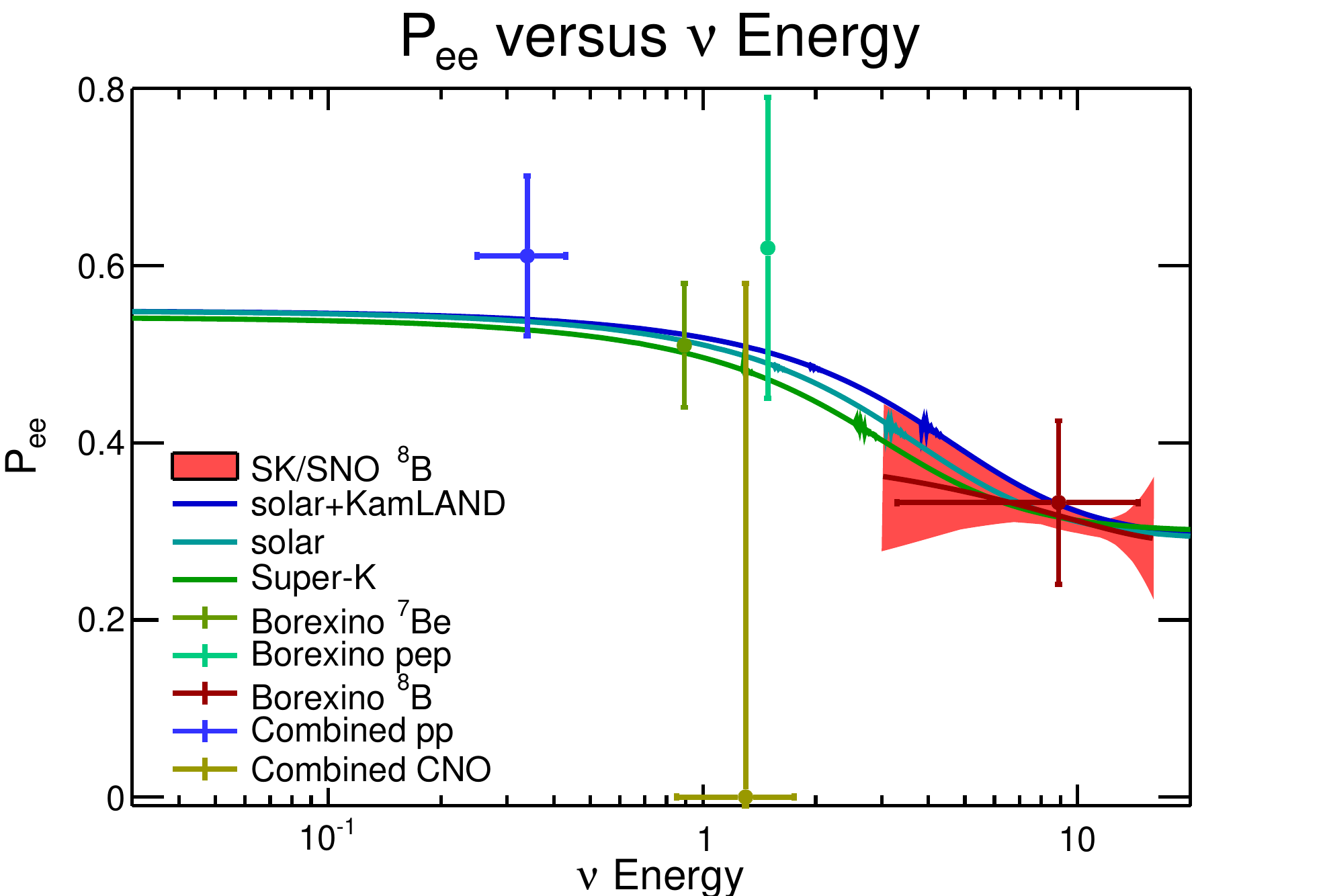}
\caption[Measurements of the solar MSW transition]{Measurements of the solar MSW transition. The red band combines
SK and SNO $^8$B data~\cite{Aharmim:2011vm}, the green measurements
of $^7$Be and pep are from Borexino~\cite{Borexino7be:2011,Borexinopep:2011} and the
red error bar is Borexino's $^8$B measurement~\cite{Borexino8b:2008}. The 
blue $pp$ point and the yellow error bar (CNO) combine all solar data. MSW resonance
curves for three different parameters are overlaid.}
\label{fig:solmsw}
\end{figure}

The ICARUS Collaboration has reported a \MeVadj{10}
threshold~\cite{Guglielmi:2012}. Assuming the 
detector itself
has low enough radioactivity levels, this threshold level would enable
a large enough detector to measure the electron flavor component of
the solar $^8$B neutrino flux with high statistical accuracy. It could 
thereby further test the MSW flavor transformation curve 
(Figure~\ref{fig:solmsw}) with higher statistical precision and
potentially better energy resolution. 

\begin{figure}[!htb]
\centering\includegraphics[width=0.7\textwidth]{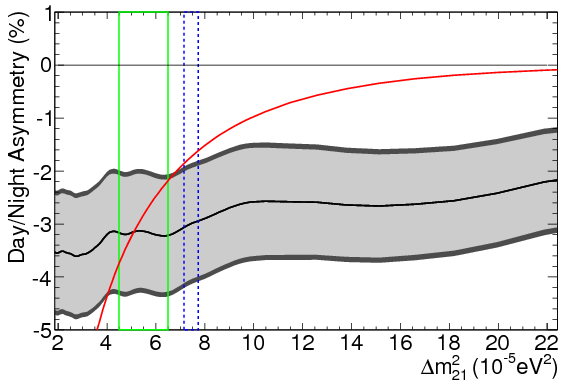}
\caption[Solar-neutrino day/night effect and dependence on $\Delta
m^2_{21}$]{Dependence of the measured day/night asymmetry (fitted
  day/night amplitude times the expected day/night asymmetry in red) on
  $\Delta m^2_{21}$, for $\sin^2 \theta_{12}=0.314$ and $\sin^2
  \theta_{13}=0.025$.  The 1$\sigma$ statistical uncertainties from
  the recent measurements by SK are given by the light
  grey band. The additional dark grey width to the band shows the
  inclusion of the systematic uncertainties. Overlaid are the 1$\sigma$ 
allowed ranges from the solar global fit (solid green) and
  the KamLAND experiment (dashed blue). Figure is from ~\cite{Renshaw:2013}.}
\label{fig:daynight}
\end{figure}

In addition to these solar
matter effects, solar neutrinos also probe terrestrial matter effects
with the variation of the $\nu_e$ flavor observed with solar zenith
angle while the Sun is below the horizon --- the \emph{day/night effect}. A
sizable effect is predicted only for the highest solar-neutrino
energies, so while the comparatively high energy threshold is a
handicap for testing the solar MSW resonance curve, it has a smaller
impact on the high-statistics test of terrestrial matter
effects. Recently, indication of the existence of the terrestrial
matter effects were reported~\cite{Renshaw:2013}. Measurements of this
effect currently give the best constraints on the solar mass ($\Delta
m_{21}^2$) splitting (Figure~\ref{fig:daynight}) using neutrinos
rather than antineutrinos~\cite{Gando:2013}. 

The comparison
of $\nu$ disappearance to $\overline{\nu}$ disappearance tests CPT
invariance. For good sensitivity to either solar-neutrino measurement,
a liquid argon far detector of at least \SI{34}{\kt} is required.

%

\section{Indirect Searches for WIMP Dark Matter}
\label{sec:wimp}

If the true nature of Dark Matter (DM) involves a
weakly-interacting massive particle (WIMP) with a mass on the order of \SI{1}{\GeV},
an experiment could look for anomalous signals in
astrophysical data from the annihilation (or decay) of DM into Standard Model particles,
e.g., neutrinos~\cite{Silk:1985ax}.  
Neutrinos produced by DM decay are expected to come from such distant objects as
the galactic center, the center of the Sun or even from the Earth.

As our solar system moves through the DM halo, WIMPs interact with the
nuclei of celestial bodies and become trapped in a body's
gravitational well.  Over time, the WIMPs accumulate near the core of
the body, enhancing the possibility of annihilation. The high-energy
neutrinos ($E\sim m_{\rm WIMP}$) from these annihilations can
free-stream through the astrophysical body and emerge roughly
unaffected, although oscillation and matter effects can slightly alter
the energy spectrum.
Neutrinos produced via the nuclear-fusion processes in the Sun have energies close
to \SI{1}{\MeV}, 
much lower than likely DM-decay neutrino energies. 

\begin{introbox}
  The LBNE far detector's large mass and directional tracking
  capabilities will enable it to act as a \emph{neutrino telescope}
  and search for neutrino signals produced by annihilations of dark
  matter particles in the Sun and/or the core of the Earth.  Detection
  of high-energy neutrinos coming exclusively from the Sun's
  direction, for example, would provide clear evidence of dark matter  
annihilation~\cite{Cirelli:2005gh}.
\end{introbox}

IMB~\cite{LoSecco:1986fu}, IceCube~\cite{Aartsen:2012kia} and
SK, all water Cherenkov-based detectors, have searched for
signals of DM annihilations coming from these sources, so far with
negative results.  A LArTPC can provide much better angular resolution
than can water Cherenkov detectors, therefore providing better
separation of the directional solar WIMP signal from the atmospheric-neutrino background.
More thorough studies~\cite{Blennow:2013pya} are needed to determine
whether LBNE could provide a competitive detection of dark matter.

\section{Supernova Relic Neutrinos}
\label{sec:srn}

Galactic supernovae are relatively rare, occurring somewhere between
one and four times a century (Chapter \ref{sn-chap}). In the Universe
at large, however, thousands of neutrino-producing explosions occur
every hour.  The resulting neutrinos --- in fact most of the neutrinos
emitted by all the supernovae since the onset of stellar formation ---
suffuse the Universe.  Known both as \emph{supernova relic neutrinos
  (SRN)} and as the \emph{diffuse supernova-neutrino background
  (DSNB)}, their energies are in the few-to-\MeVadj{30} range.  SRN
have not yet been observed, but an observation would greatly enhance
our understanding of supernova-neutrino emission and the overall
core-collapse rate.

\begin{figure}[!htb]
     \centering\includegraphics[width=.8\textwidth]{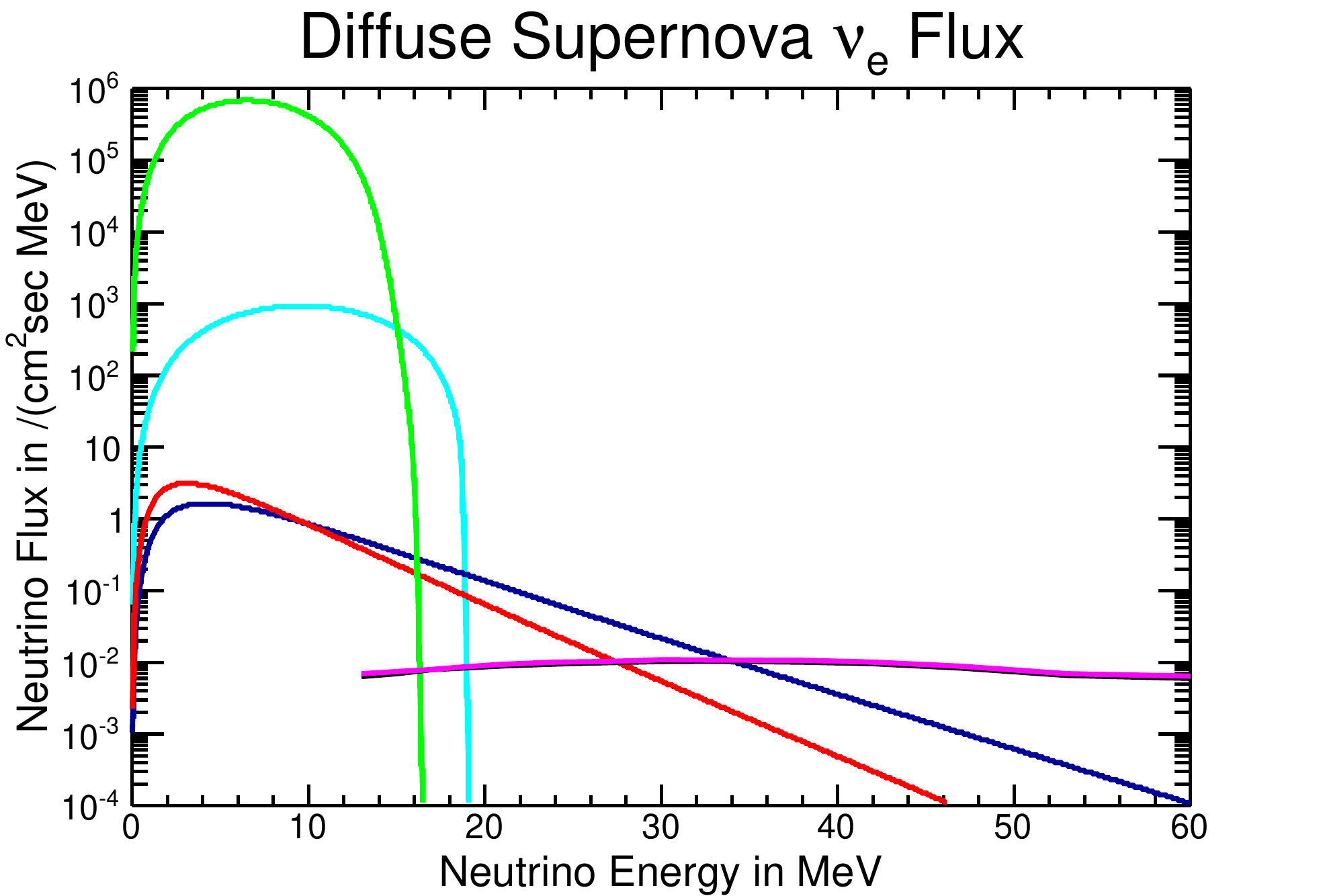}
     \caption[Relic supernova spectra and key neutrino
     backgrounds]{Predicted relic supernova $\nu_e$ spectra from two
       different models (red and blue) and some key neutrino
       backgrounds: $^8$B solar $\nu_e$ (green), hep solar $\nu_e$
       (cyan) and atmospheric $\nu_e$ (magenta).}
     \label{fig:pred_srnspec}
\end{figure}

\begin{introbox}
  A liquid argon detector such as LBNE's far detector is sensitive to
  the $\nu_e$ component of the diffuse relic supernova-neutrino flux,
  whereas water Cherenkov and scintillator detectors are sensitive to
  the \ane\ component.  However, backgrounds in liquid argon are as
  yet unknown, and a huge exposure ($>$\SI{500}{\ktyr}s)
  would likely be required for observation.  Given a detector of the
  scale required to achieve these exposures (\SIrange{50}{100}{\kt})
  together with tight control of
  backgrounds, 
  LBNE --- in the long term --- could 
play a unique and
  complementary role in the physics of relic neutrinos.
\end{introbox}
In the current LBNE design, the irreducible background from solar
neutrinos will limit the search for these relic neutrinos to an energy
threshold greater than \SI{18}{MeV}.  Similarly, a search for relic
antineutrinos is limited by the reactor-antineutrino background to a
threshold greater than $\sim$\SI{10}{MeV}. The lower threshold and the
smaller average $\nu_e$ energy relative to that for \ane\ 
(Figure~\ref{fig:pred_srnspec}) leads to the need for a larger detector
mass.

A small but dedicated industry devotes itself to trying to predict the
flux of these relic supernova neutrinos here on
Earth~\cite{Totani:1995dw,Sato:1997sc,Hartmann:1997qe,Malaney:1996ar,Kaplinghat:1999xi,Ando:2005ka,Lunardini:2006pd,Fukugita:2002qw}. Examples
of two different predicted SRN spectra are shown in
Figure~\ref{fig:pred_srnspec}, along with some of the key physics
backgrounds from other neutrino sources.

In the LBNE LArTPC, relic supernova electron neutrinos would be
detected primarily via the CC process as described by
Equation~\ref{eqn:srninteract}. The electron track should be
accompanied by evidence of ionization from the de-excitation of the
potassium, e.g., shorter tracks sharing a common vertex; this is
expected to help reduce backgrounds, but a detailed study has not yet
been undertaken.  In water Cherenkov and scintillator detectors, it is
the $\overline{\nu}_e$ SRN flux that is detected through the
process of inverse-beta decay.  Unlike inverse-beta decay, for which
the cross section is known to the several-percent level in the energy
range of interest~\cite{Vogel:1999zy,Strumia:2003zx}, the cross
section for neutrino interactions on argon is uncertain at the 20\%
level~\cite{Ormand:1994js,Kolbe:2003ys,SajjadAthar:2004yf}. Another
limitation is that the solar {\em hep} neutrinos (defined in Figure~\ref{fig:solnuflux}), which have an                
endpoint at \SI{18.8}{\MeV}, will determine the lower bound of the SRN
search window ($\sim$ \SI{16}{\MeV}).  The upper bound is determined
by the atmospheric ${\nu}_{e}$ flux as shown in
Figure~\ref{fig:pred_srnspec} and is around \SI{40}{MeV}.
Although the LArTPC provides a unique sensitivity to the
$\nu_e$ component of the SRN flux, early studies indicate
that due to this lower bound of $\sim$\SI{16}{\MeV}, LBNE would need a huge
mass of liquid argon --- of order \SI{100}{\kt} --- to get more than 4$\sigma$
evidence for the diffuse supernova flux in five
years~\cite{Cocco:2004ac}.
The expected number of relic
supernova neutrinos, $N_{\rm SRN}$, that could be observed in a
\SIadj{100}{\kt} LArTPC detector in five years~\cite{Cocco:2004ac}
assuming normal hierarchy is:
\begin{equation}
N_{\rm SRN} = 57 \pm 12,  \ \ \ 16 \, {\rm MeV} \leq E_e \leq 40 \, {\rm MeV},
\label{eqn:srnrate}
\end{equation}
where $E_e$ is the energy of the electron from the CC interaction as
shown in Equation~\ref{eqn:srninteract}. The estimate of the SRN rate
 in Equation~\ref{eqn:srnrate} has a weak dependence on the value
of $\sin ^2 \theta_{13}$. The above calculation is valid for values of
$\sin ^2 \theta_{13} > 10^{-3}$.  The main challenge for detection of such
a low rate of relic neutrinos in a LArTPC is understanding how much of
the large spallation background from cosmic-ray interactions with the
heavy argon nucleus (some of which are shown in Figure~\ref{fig:cosmicbkg}) leaks into the SRN search window. 

\section{GUT Monopoles}\label{sec:monop}

Searches for massive, slow-moving magnetic monopoles produced in the early 
Universe continue to be of pressing interest.  Magnetic monopoles left over 
from the Big Bang are predicted by Grand Unified Theories, but to date have not 
been observed.  Because of the very large masses set by the GUT scale, these 
monopoles are normally non-relativistic, however searches for relativistic 
and ultra-relativistic monopoles are also of interest.  

Relativistic monopoles are expected to be heavily ionizing, and hence best 
suited for detection in the large-area, neutrino-telescope 
Cherenkov detectors deployed 
in natural bodies of water or ice (e.g., \cite{Abbasi:2012eda,Aartsen:2013mla}).
With its much smaller active area, LBNE will most likely 
not be competitive in searches for fast monopoles.

\begin{figure}
\includegraphics[width=0.5\textwidth]{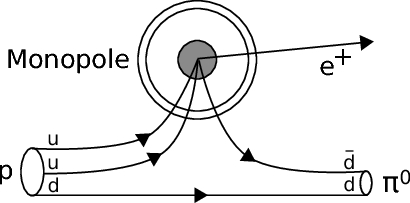}
\caption[Proton decay catalyzed by a GUT monopole]{Illustration of a proton decay into
  a positron and a neutral pion catalyzed by a GUT monopole from ~\cite{Aartsen:2014tna}.}
\label{fig:monopole}
\end{figure}

Massive GUT monopoles are postulated 
to catalyze nucleon decay (Figure~\ref{fig:monopole}).  
It is possible that large
underground detectors could detect this type of signal from transiting
monopoles~\cite{Ueno:2012md,Aartsen:2014tna} via a signature
consisting of multiple proton decays 
concurrent with the monopole's passage through the detector. 
Proton decay catalyzed by magnetic monopoles may be easier to observe in a 
LArTPC due to its superior imaging capability as compared to 
Cherenkov detectors, namely its high detection efficiency for a wider variety of
proton decay modes, and its low energy thresholds.  
Whether these features 
are sufficient to overcome the limitation of smaller detector area relative 
to the very large neutrino telescopes has yet to be studied.

It should also be possible for LBNE to detect slow-moving monopoles 
via time-of-flight measurements, thereby eliminating reliance  
on the assumption of a proton-decay catalysis signature.  
The most stringent limits from direct searches for GUT monopoles with 
velocities in the range $4\times 10^{-5} < \beta < 1$ have been obtained
by the MACRO experiment~\cite{Ambrosio:2002qq}, 
which has excluded fluxes at 
 the level of 
\SI{1.4e-16}{\per\cm\squared\per\second\per\steradian}.
These limits probe the flux region just beyond that excluded by the 
existence of the galactic magnetic field (as characterized in variants
of the Parker Bound).

The LBNE LArTPC far detector provides an opportunity to extend the 
reach of direct searches for slow monopoles, thanks to excellent timing 
and ionization measurement capabilities.  Quantitative
studies of sensitivity have yet to be carried out, but it is likely that 
the full-scope LBNE far detector will exceed the 
\SI{10000}{\msr}
isotropic-flux acceptance of MACRO.

\section{Neutron-Antineutron Oscillations ($\boldsymbol{\Delta B = 2}$)}
\label{sec:n-nbar}

Some Grand Unified Theories suggest the existence of double
baryon-number-violating transitions that change nucleons into
antinucleons~\cite{Mohapatra:2009wp}.  The nucleon-antinucleon
annihilation resulting from such a transition would provide an
unmistakable signal in the LBNE LArTPC.  

The imaging properties of the
detector --- superior to those of water detectors --- would enable
observation of nucleon annihilation final states in which the signal
is broadened by the mix of charged and neutral hadrons.  This signal
could, however, be suppressed in a LArTPC if the neutron-to-antineutron 
transition rate is suppressed for bound neutrons due to
interactions with the other nucleons.

\section{Geo and Reactor $\boldsymbol{\overline{\nu}_e}$'s }
\label{sec:GeoRea}  

Electron antineutrinos ($\overline{\nu}_e$'s) produced by radioactive
decays of the uranium, thorium and potassium present in the Earth are
referred to as \emph{geo-antineutrinos}. Decays of these three
elements are currently understood to be the dominant source of the
heat that causes mantle convection, the fundamental geological process
that regulates the thermal evolution of the planet and shapes its
surface. Detection of these geo-antineutrinos near the Earth's surface
can provide direct information about the deep-Earth uranium and
thorium content. 

Geo-antineutrino energies are typically below \SI{3.5}{MeV}. Reactor
antineutrinos are somewhat more energetic, up to \SI{8}{MeV}.

In a LArTPC, electron antineutrinos can in principle be detected by
argon inverse-beta decay, represented by
\begin{equation}
\overline{\nu}_e + \ce{^{40}Ar} \rightarrow \ce{^{40}Cl^*} + e^+.
\end{equation}

However, the threshold for this reaction is about \SI{8.5}{MeV},
leading to the conclusion that an $^{40}$Ar detector cannot use this
method to detect either geo-antineutrinos or reactor antineutrinos.

Interaction via elastic scattering with electrons, another potential
avenue, presents other obstacles. Not only are the recoil electrons
from this interaction produced at very low energies, but solar
neutrinos scatter off electrons and form an irreducible background
roughly a thousand times larger than the geo-antineutrino signal.
Although LBNE's location far away from any nuclear reactors leaves
only a small reactor-antineutrino background and is thus favorable for
geo-antineutrino detection, another detector technology (e.g., liquid
scintillator) would be required to do so.









%% file: chapter-conclusion.tex
The preceding chapters of this document describe the design of
the Long-Baseline Neutrino Experiment, its technical
capabilities, and the breadth of physics topics at the forefront of
particle and astrophysics the experiment can address. 
This chapter 
concludes the document with several discussions that look forward in time,
specifically:
\begin{itemize}  
\item a consideration of how the design and construction of
  the LBNE experiment might unfold from this point on
  for a general class of staging scenarios,
\item a summary of the grand vision for the science of LBNE
  and its potential for transformative discovery,
\item a summary of the compelling reasons --- such as LBNE's
  current advanced state of technical development and planning, and
  its alignment with the national High Energy Physics (HEP) program
  --- for which \emph{LBNE represents the world's best chance for
    addressing this science on a reasonable timescale},
\item comments on the broader impacts of LBNE, including the
  overarching benefits to the field of HEP, both within and
  beyond the U.S. program. 
\end{itemize}

\section{LBNE Staging Scenarios and Timeline}
\label{sec:staging-timeline}

\begin{introbox}
  With DOE CD-1 (``Alternate Selection and Cost Range'') approval in
  hand, the LBNE Project is working toward its technical design
  specifications, including detailed costs and schedule, in
  preparation for CD-2 (``Performance Baseline'').  It should be noted
  that the Project already has fully developed schedules for both the
  CD-1 scope (\ktadj{10} far detector on the surface at the \SURF, 
  no near neutrino detector), and for the full-scope (\ktadj{34} far
  detector located deep underground and near neutrino detector)
  for the scenario of funding solely from DOE.  
Partnerships with non-DOE groups are
being sought to enable the construction of LBNE 
with a near neutrino detector and an underground far detector mass greater
than \SI{10}{\kt} in the first phase.
\end{introbox}

Section~\ref{sec:global-partner} described the substantial
progress that has been achieved so far toward making LBNE a fully
international project. While the specific form and timing of
contributions from new partners are not yet known, there are several
plausible scenarios in which the Project can be implemented to
accommodate non-DOE contributions.  A review of the DOE project
milestones, indicating where flexibility and potential for
incorporating non-DOE contributions exist, provides a starting point.

\begin{figure}[!htb]
\includegraphics[width=\textwidth]{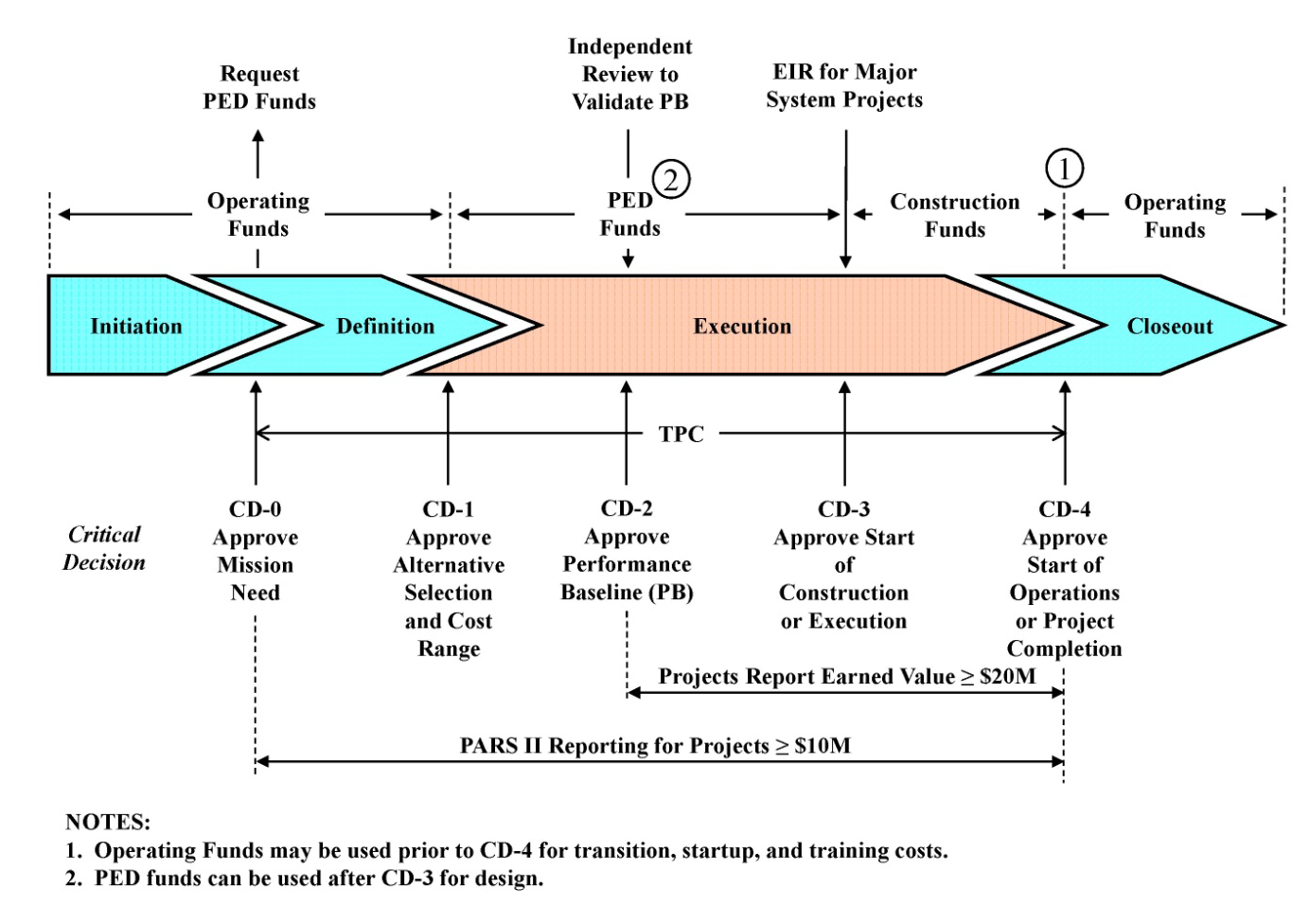}
\caption[Typical DOE Acquisition Management System for line item capital asset projects]{Typical DOE Acquisition Management System for line item capital asset projects~\cite{doe_o413_2010}.} 
\label{fig:cdprocess}
\end{figure}

DOE-funded projects are subject to several \emph{critical decision
  (CD)} milestones as shown in Figure~\ref{fig:cdprocess} and explained
in DOE Order O 413.3B~\cite{doe_o413_2010}.
At CD-2 the first-phase LBNE Project will be baselined.  Currently, the
timescale for CD-2 is projected to be toward the end of FY 2016,
although the DOE has indicated flexibility in the project approval
process specifically to allow for incorporation of scope changes
enabled by additional partners.  For example, it has been suggested
that the design and construction approval for different portions of
the Project can be approved at different times to facilitate proper
integration of international partners. It is also expected that CD-3a
approval (start of construction/execution) may take place for some
parts of the Project before CD-2, thereby authorizing expenditures for
long-leadtime components and construction activities, such as the
advanced site preparation at Fermilab for the new beamline.  The CD-4
milestone (completion of the construction project and transition to
experiment operations) is currently projected for 2025.  However, it
is expected that commissioning and operations for LBNE will have
started approximately a year before CD-4, which is considered the
formal termination of the construction project.

The actual timeframe for achieving LBNE science goals will depend on
the manner in which a complex sequence of developments takes place,
including the actions of partners as well as implementation of the
milestones above for the DOE-funded elements of the Project.  Various
scenarios for incorporating contributions from new partners/sources of
funding have been identified~\cite{DOCDB8694}.
%
\begin{introbox}
  Using the 
current understanding of DOE funding profiles, we
  outline one plausible long-term timeline that integrates evolution
  of LBNE detector mass with development of the Fermilab accelerator
  complex (i.e., PIP-II) and contributions from non-DOE partners.
  Implicit in this timeline is an assumption that agreements with new
  partners be put in place on a timescale of three years (by 2017).  In this
  scenario, the milestones that bear on the physics are as follows:
\begin{enumerate}
  \item LBNE begins operation in 2025 with a \MWadj{1.2} beam 
        and a \ktadj{15} far detector.  (In such a scenario, 
        a significant fraction of the far detector mass might be 
        provided in the form of a standalone LArTPC module 
        developed, funded, and constructed by international 
        partners.)
  \item Data are recorded for five years, for a net exposure of 
        \SI[inter-unit-product=\ensuremath{{}\cdot{}}]{90}{\kt\MW\year}.
  \item In 2030, the LBNE far detector mass is increased to \SI{34}{\kt}, 
        and proton beam power is increased to \SI{2.3}{\MW}.
  \item By 2035, after five years of additional running, a net exposure of 
        \SI[inter-unit-product=\ensuremath{{}\cdot{}}]{490}{\kt\MW\year} is attained.  
\end{enumerate}
\end{introbox}
Physics considerations will dictate the 
desired extent of operation of LBNE beyond 2035.

This very coarse timeline is indicative of the degree of flexibility available for the
staging of various elements of LBNE. For example, near detector construction (and the
 corresponding funding) could be undertaken by partners outside  the
U.S., on a timescale driven by the constraints they face, and could
be completed somewhat earlier or later than the far detector or
beamline.

With this timeline as a guide, the discussion of LBNE physics
milestones can 
be anchored by plausible construction scenarios.  
\clearpage
\section{Science Impact}

While considering the practical challenges implicit in the  
discussion in Section~\ref{sec:staging-timeline} for the realization of LBNE, 
it is important to reiterate the compelling science motivation in 
broad terms.  

The discovery that neutrinos have mass constitutes the only palpable
evidence \emph{within the body of particle physics data} that the
Standard Model of electroweak and strong interactions 
does not
describe all observed phenomena. In the Standard Model, the simple
Higgs mechanism --- now confirmed with the observation of the Higgs
boson --- is responsible for quark as well as lepton masses, mixing
and CP violation. Puzzling features such as the extremely small masses
of neutrinos compared to other fermions and the large extent of mixing
in the lepton sector relative to the quark sector, suggest that new
physics not included in the current Standard Model is needed
to connect the two sectors. These discoveries have moved the study of
neutrino properties to the forefront of experimental and theoretical
particle physics as a crucial tool for understanding the fundamental
nature and underlying symmetries of the physical world.

\begin{introbox}
 The measurement of 
the neutrino mass hierarchy and search for CP violation in LBNE will 
further clarify the pattern of mixing and mass ordering in the lepton sector 
and its relation to the patterns in the quark sector. The impact of exposures of
  \SI[inter-unit-product=\ensuremath{{}\cdot{}}]{90}{\kt\MW\year}
  (2030) and
  \SI[inter-unit-product=\ensuremath{{}\cdot{}}]{490}{\kt\MW\year}
  (2035) for Mass Hierarchy and CP-violation signatures is easily
  extracted from Figure~\ref{fig:lar-cp-frac}.  Should CP be violated
  through neutrino mixing effects, the typical signal in LBNE
  establishing this would have a significance of at least three (2030)
  and five standard deviations (2035), respectively for $50\%$ of
  \deltacp values (and greater than three standard deviations for
  nearly $75\%$ of \deltacp by 2035). In such a scenario, the
  mass hierarchy can be resolved with a sensitivity for a typical
  experiment of $\sqrt{\overline{\Delta \chi^2}} \geq 6$ for $50\%$ (100\%)
  of \deltacp by 2030 (2035).
\end{introbox}
If CP is violated maximally with a CP phase of $\mdeltacp \sim
-\pi/2$ as hinted at by global analyses of recent data~\cite{Capozzi:2013csa}, the
significance would be in excess of \num{7}$\sigma$.  
This opportunity to
establish the paradigm of leptonic CP violation is highly compelling,
particularly in light of the implications for leptogenesis as an
explanation for the Baryon Asymmetry of the Universe (BAU).  With tight
control of systematic uncertainties, additional data taking beyond 2035 would
provide an opportunity to strengthen a marginally significant signal
should \deltacp take a less favorable value.

Similarly, the typical LBNE data set will provide evidence for a
particular mass ordering by 2030 in the scenario described in
Section~\ref{sec:staging-timeline}, and will exclude the incorrect
hypothesis at a high degree of confidence by 2035, over the full range
of possible values for \deltacp, $\theta_{23}$ and the mass
ordering itself.  In addition to the implications for models of
neutrino mass and mixing directly following from this measurement,
such a result could take on even greater importance.  Should LBNE
exclude the normal hierarchy hypothesis, the predicted rate for 
neutrinoless double-beta decay would then be high enough so as to be 
accessible to the next generation of experiments~\cite{Bilenky:2012qi}.  
A positive result from these experiments would provide 
unambiguous --- and exciting --- evidence that neutrinos are Majorana 
particles\footnote{A Majorana particle is an elementary particle that is also
  its own antiparticle}, and that the empirical law of lepton number 
conservation --- a law lacking deeper theoretical explanation --- is not exact. Such a discovery would indicate that there
may be heavier sterile right-handed neutrinos that mix with ordinary
neutrinos, giving rise to the tiny observed neutrino masses as
proposed by the seesaw mechanism~\cite{Yanagida:1980xy}.  
On the other hand, a rejection of the normal neutrino mass hierarchy by
LBNE coupled with a null result from the next generation of
neutrinoless double-beta decay experiments would lead to the conclusion
that neutrinos are purely Dirac particles.  This would be a profound
and astonishing realization, since it is extremely difficult
theoretically to explain the tiny masses of Dirac neutrinos.
High-precision neutrino oscillation measurements carried out by LBNE beyond 2035
may provide evidence for Majorana neutrino mass effects that are outside of the
ordinary Higgs mechanism or for new interactions that differentiate the
various neutrino species. 

Within the program of underground physics, LBNE's most exciting 
milestones would correspond to observations of rare events.  
By 2035, LBNE will have been live for galactic supernova neutrino 
bursts for ten years in the above scenario.  Such an event would 
provide a spectacular data set that would likely be studied for 
years and even decades to follow.  

For proton decay, the net exposure obtained by 2035 in the above scenario 
also provides a compelling opportunity.  A partial lifetime 
for $p\to K^+\overline{\nu}$ of $1\times 10^{34}$ years, 
beyond the current limit from Super-Kamiokande by roughly a factor of 
two, would correspond to six candidate events in LBNE 
by 2035, with 0.25 background events expected.  Running for seven more 
years would double this sample.  (Similarly, one should not ignore 
the corresponding value of an LBNE construction scenario that 
has a larger detector mass operating from the start, in 2025). With
careful study of backgrounds, it may also be possible to 
suppress them further and/or relax fiducial cuts to gain 
further in sensitivity.  

Finally, the proposed high-resolution near detector, operating in the
high-intensity LBNE neutrino beam, will not only constrain the systematic
errors that affect the oscillation physics but will also conduct precise and 
comprehensive measurements of neutrino interactions --- from cross sections
to electroweak constants. 

\section{Uniqueness of Opportunity}

Considering the time and overall effort taken to reach the current
state of development of LBNE, it will be challenging for alternative
programs of similarly ambitious scope to 
begin operation
before 2025, particularly in light of the current constrained budget
conditions in HEP.  It should be noted that similar-cost alternatives
for the first phase of LBNE utilizing the existing NuMI beam were
considered during the reconfiguration exercise in 2012~\cite{LBNEreconfig}. 
 The panel concluded that none of
these alternatives presented a path toward an experiment capable of a
 CP-violation signal of \num{5}$\sigma$.  Furthermore, 
a large water Cherenkov far detector option for LBNE was carefully considered prior to
selection of the LArTPC technology~\cite{CDRwcd}.  
While both detector options are capable of satisfying the
scientific requirements, the LArTPC was judged to have a better
potential for scientific performance while also presenting the attraction of 
an advanced technological approach.

In the broader context of planned experimental programs 
with overlapping aims for portions of the LBNE science 
scope, it must be recognized that progress will be made 
toward some of these during the period before LBNE operations 
commence.  For example, indications for a preferred neutrino 
mass ordering may emerge from currently running experiments 
and/or from dedicated initiatives that can be realized on 
a shorter timescale.  Global fits will continue to be done to 
capitalize, to the extent possible, on the rich phenomenology 
of neutrino oscillation physics where disparate effects are 
intertwined.  At the same time, each experimental arena will 
be subject to its own set of systematic uncertainties and limitations.

It is in this sense that the power of LBNE is especially compelling.
LBNE will on its own be able to measure the full suite of neutrino
mixing parameters, and with redundancy in some cases.  To use the MH
example just given, it is notable that LBNE will have sensitivity both
with beam and atmospheric neutrinos.  Control of the relative
$\nu_\mu$/$\overline{\nu}_\mu$ content of the beam as well as the
neutrino energy spectrum itself, provides additional handles and
cross-checks absent in other approaches.

\section{Broader Impacts}

\subsection{Intensity Frontier Leadership}

The U.S. HEP community faces serious challenges to maintain its
vibrancy in the coming decades.  As is currently the case with the
LHC, the next-generation energy frontier facility is likely to be
sited outside the U.S.  
It is critical that the U.S. host facilities aimed at pursuing science
at the HEP scientific frontiers (Figure~\ref{fig:frontiers}), the lack
of which could result in erosion of expertise in key technical and
scientific sectors (such as accelerator and beam physics).

\begin{introbox}
LBNE represents a world-class U.S.-based effort to address the science 
of neutrinos with technologically advanced experimental techniques. 
By anchoring the U.S. Intensity Frontier
program~\cite{Hewett:2014qja}, LBNE provides a platform around which
to grow and sustain core infrastructure for the community.
Development of the Fermilab accelerator systems, in particular, will not only
advance  progress toward achieving the science goals of LBNE, it will also 
greatly expand the capability of Fermilab to host other key
experimental programs at the Intensity Frontier. 
\end{introbox}

\subsection{Inspirational Project for a New Generation}

Attracting young scientists to the field demands a future that is rich
with ground-breaking scientific opportunities.  LBNE provides such a
future, both in the technical development efforts required and its
physics reach.  The unparalleled potential of LBNE to address
fundamental questions about the nature of our Universe by making
high-precision, unambiguous measurements with
the ambitious
technologies it incorporates 
will attract the best and brightest scientists of the next generation
to the U.S. HEP effort.

A young scientist excited by these prospects can already participate
in current experiments --- some of which use medium-scale LArTPCs ---
and make contributions to leading-edge R\&D activities that provide
important preparation for LBNE, both scientifically and technically.

\section{Concluding Remarks}

\begin{introbox}
Understanding the fundamental nature of fermion flavor, the existence
     of CP violation in the lepton sector and how this relates to the Baryon
     Asymmetry of the Universe; knowing whether proton decay occurs and how;
     and elucidating the dynamics of supernova explosions all stand among
     the grand scientific questions of our times. The bold approach
   adopted for LBNE provides the most rapid and cost-effective means of
   addressing these questions.  With the support of the global HEP community,
   the vision articulated in this document can be realized in a way
   that maintains the level of excitement for particle physics and the
   inspirational impact it has in the U.S. and worldwide.
\end{introbox}

%% file: appendix-fd-simulation.tex
A \ktadj{10} or larger LArTPC far detector 
fulfills the high-mass requirement for LBNE and provides excellent
particle identification with high signal-selection efficiency ($\geq$
80\%) over a wide range of energies. The far detector is described in detail in the LBNE
Conceptual Design Report Volume~1~\cite{CDRv1} and briefly in Section~\ref{sec:fdproj} of
this document.  This appendix summarizes the status of the LBNE LArTPC simulation and
reconstruction efforts and their expected performance.

\section{Far Detector Simulation}

\subsection{Tools and Methods}  

In the full simulation of the far detector,
neutrino interactions 
are simulated with
Geant4~\cite{Agostinelli:2002hh} using the LArSoft~\cite{Church:2013hea} package.
LArSoft is being developed to provide an integrated,
experiment-agnostic set of software tools to perform simulation, data
reconstruction and analysis for LArTPC neutrino experiments. 
Individual experiments provide experiment-specific components
including a detector geometry description and analysis code, and they
contribute to  the LArSoft software development itself.

\begin{figure}[!htp]
\centering\includegraphics[width=1.0\textwidth]{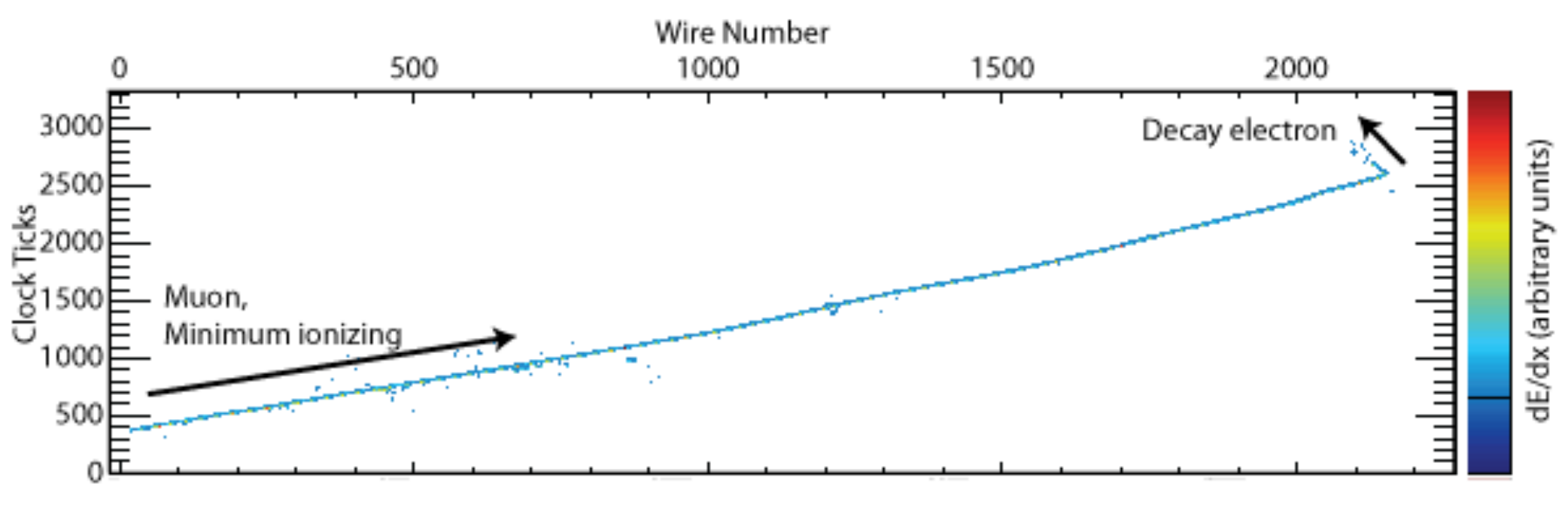}
\centering\includegraphics[width=1.0\textwidth]{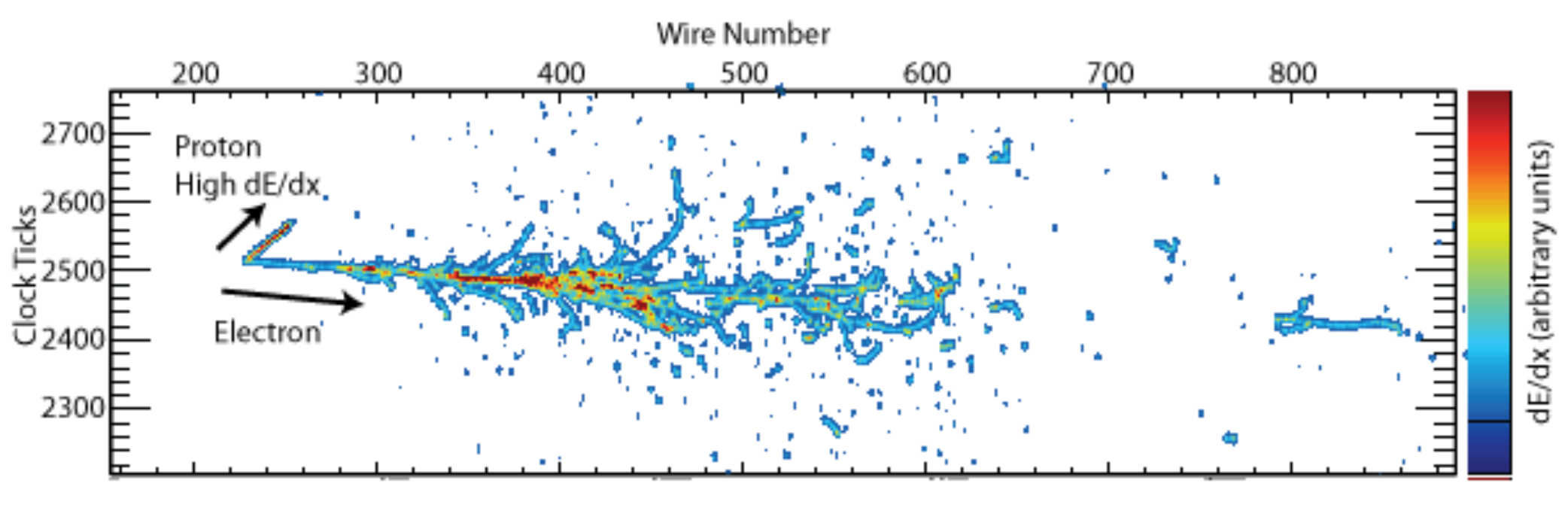}
\centering\includegraphics[width=1.0\textwidth]{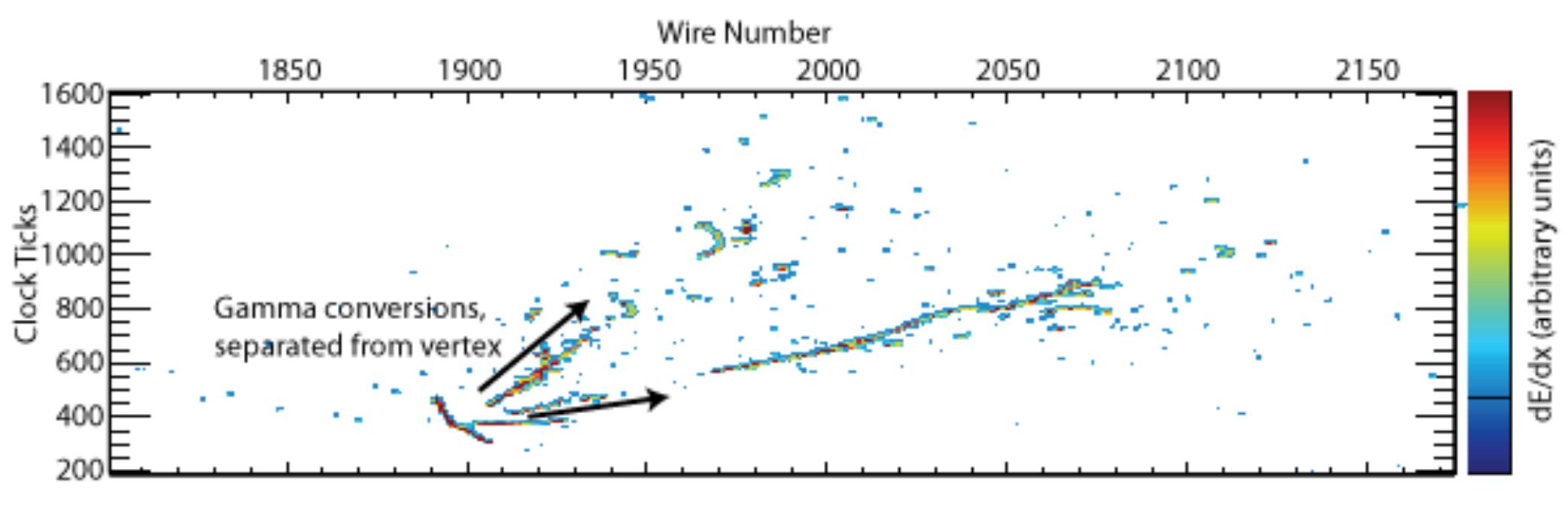}
\caption[Event displays of beam interactions in a LArTPC]{Examples of
  neutrino beam interactions in a LArTPC obtained from a Geant4
  simulation~\cite{Church:2013hea}. A $\nu_\mu$-CC interaction with a stopped $\mu$ followed
  by a decay Michel electron (top), a $\nu_e$-CCQE interaction with a
  single electron and a proton (middle), and an NC interaction which
  produced a $\pi^0$ that then decayed into two $\gamma$'s with
  separate conversion vertices (bottom).}
\label{fig:lar-scan-events}
\end{figure}

LArSoft is based on \emph{art}~\cite{Green:2012gv},   
an event-processing framework
developed and supported by the Fermilab Scientific Computing Division.
\emph{Art} is designed to be shared by multiple experiments and is currently
used by several intensity frontier experiments, including NO$\nu$A, Mu2e,
MicroBooNE~\cite{Katori:2011uq} and ArgoNeuT~\cite{Soderberg:2009qt}.
The last two have liquid argon TPC-based detectors and thus share many simulation 
and reconstruction requirements with LBNE.
Reconstruction algorithms developed in LArSoft for the ArgoNeuT and
MicroBooNE experiments can readily benefit
LBNE. 
Examples of neutrino beam interactions in a LArTPC obtained from the
LArSoft package using the MicroBooNE detector geometry are shown in
Figure~\ref{fig:lar-scan-events}.

The LBNE far detector geometries currently available in LArSoft
are the LBNE \ktadj{10} surface detector and the \ktadj{34} underground
detector.  Also included is geometry for a \SIadj{35}{\tonne}  prototype that LBNE has 
constructed at Fermilab\footnote{One of the goals of the
  \SIadj{35}{t} prototype is to test key elements of the TPC module
  design for the \SIadj{10}{\kt} and \SIadj{34}{\kt} detectors
  including the wrapped wire planes and drift distances.}.  
The LBNE far detector geometry description is generated in a flexible
way that allows the simulation of various detector design parameters
such as the wire spacing and angles, drift distances, and materials.
The photon-detector models are based on the design that uses acrylic
bars coated with wavelength-shifting tetraphenyl butadiene (TPB), 
read out with silicon photomultiplier tubes (SiPMs).

Geant4 is used to simulate particles traveling through the active and
inactive detector volumes and the surrounding materials such as the
cryostat and rock.  
The tens of thousands of photons and electrons produced 
(by the ionization of the argon) per MeV 
deposited are
simulated using a parameterization rather than a full Geant4 Monte
Carlo, as tracking them individually would be prohibitive.
The drifting electrons are modeled as many small clouds of charge that
diffuse as they travel toward the collection wires.  The response of
the channels to the drifting electrons is parameterized as a function
of drift time, with a separate response function for collection and
induction wires.  The signals on the induction-plane wires result from
induced currents and are thus bipolar as a function of time as charge
drifts past the wires, while the signals on the collection-plane wires
are unipolar. The response functions include the expected response of
the electronics.  Noise is simulated using a spectrum measured in the
ArgoNeuT detector.  The decays of $^{39}$Ar are included, but some
work is required to make them more realistic.

For the
\ktadj{10} far detector, a 1.5-ms readout of the TPC signals at 2~MHz gives a 
simulated data volume of just under 2~GB per event.
If the readout is extended 
to include the beam window, then in order to collect charge deposited by
cosmic rays (which would otherwise be partially contained), a greater data volume will be required.
To reduce the data
volume and speed up the calculation, long strings of consecutive ADC counts
below a settable threshold are suppressed in the readout.  Huffman
coding of the remaining data is  included in the digitization~\cite{huffman1952}. 

The photon-detection system likewise requires a full Monte Carlo
simulation. Photons propagating from the TPC to the acrylic bars have been
fully simulated using Geant4, and their probabilities of striking each
bar (as a function of the emission location and the position along the
bar at which the photon strikes) have been computed.  Smooth
parameterizations of these functions are currently used in the
simulation to compute the average number of photons expected to 
 strike a bar (as a function of position along it).   Given the
current design of the optical detectors, approximately 2-3\% of VUV (vacuum ultraviolet)
photons produced uniformly in the fiducial detector volume strike the
bars.  This low number is largely due to the small fraction of the
total area in contact with the argon that is represented by the bars,
and the low reflectivity of the stainless steel cathode planes, the field
cage and the CuBe wires.  

A second function is used to parameterize the
attenuation of light within the bar as a function of position along
the bar. 
 The total response of a SiPM to light produced in the detector is the product of the number
of photons produced, the probability of the
photons to survive propagation, the interaction with the wavelength shifter (commonly called
{\it downconversion}), the attenuation in the bar, and the detection efficiency of the SiPM.  This
product is used as the mean of a Poisson distribution from which the number of
photoelectrons is randomly drawn to simulate the measurement of the SiPM.
 Measured waveforms for cold SiPMs
are used in simulating the digitized response.  Measurements in
prototype dewars will be used to normalize the yield for signals in
the SiPMs as a function of the incident location of the VUV photon on
the bar.  The NEST~\cite{Szydagis:2011tk} model, which describes the conversion of
ionization energy into both electrons and photons in an anticorrelated
manner, and which has been shown to model a large
range of data from noble liquid detectors, is currently being incorporated 
into the LBNE detector
simulation. 

A variety of event generators are available for use in
the simulation.  Neutrino hard-scattering interactions and
subsequent nuclear breakup are simulated using GENIE~\cite{Andreopoulos:2009zz},
though the use of other generators is possible.  Cosmic rays are simulated with
CRY~\cite{cry}.  Single particles can be generated one at a time, and
general text-file interfaces are available allowing arbitrary
generators to be used without linking them with LArSoft.

Currently, samples of single electrons, muons, charged and neutral
pions, protons and tau leptons have been generated and simulated
using the \ktadj{10} surface geometry and the 35-ton geometry, though
without photon-detector simulation.  These samples are being used to
develop reconstruction algorithms.

Planned improvements to the simulation
 include creating an interface to a calibration
database, updating the response functions with measured responses from
MicroBooNE, which uses an electronics design very similar to that of LBNE,
simulating the effects of space-charge buildup in the drift
volume, and creating more detailed maps of the drift in the gaps between the
APAs and the charge that is deposited between the wire planes.

\subsection{Low-Energy Neutrino-Response Studies with LArSoft}
\label{sec:snlarsoft}

\begin{figure}[!htb]
\begin{centering}
\includegraphics[width=0.9\textwidth]{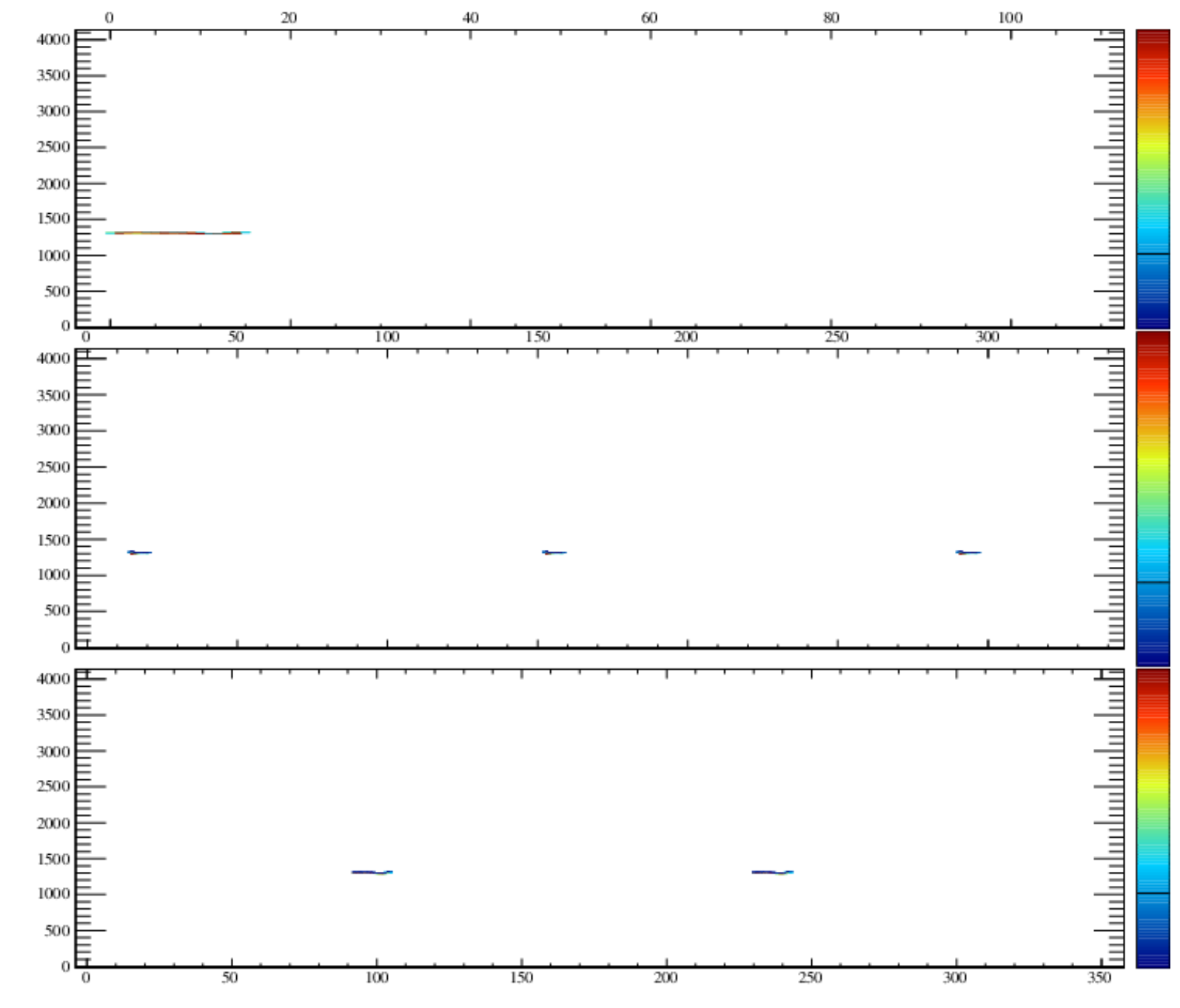}
\includegraphics[width=0.84\textwidth]{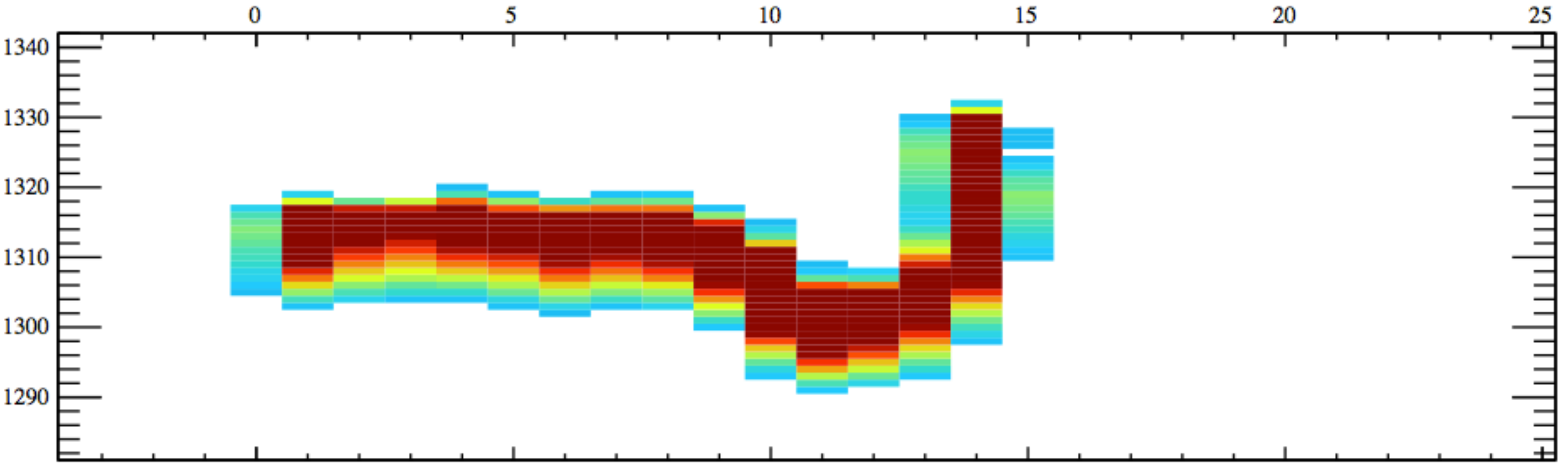}
\caption[Typical \MeVadj{20} event in the LBNE
  \SIadj{35}{t} prototype geometry]{Raw event display of a simulated \MeVadj{20} event in the LBNE
  \SIadj{35}{t} prototype; the top panel shows the collection plane, and the
  lower two panels show the induction planes (with multiple images due
  to wire wrapping). The bottom panel shows a zoom of the collection plane image.
 }
\label{fig:evdisplays}
\end{centering}
\end{figure}

\begin{figure}[!htb] 
 \centering
 \includegraphics[width=0.45\textwidth]{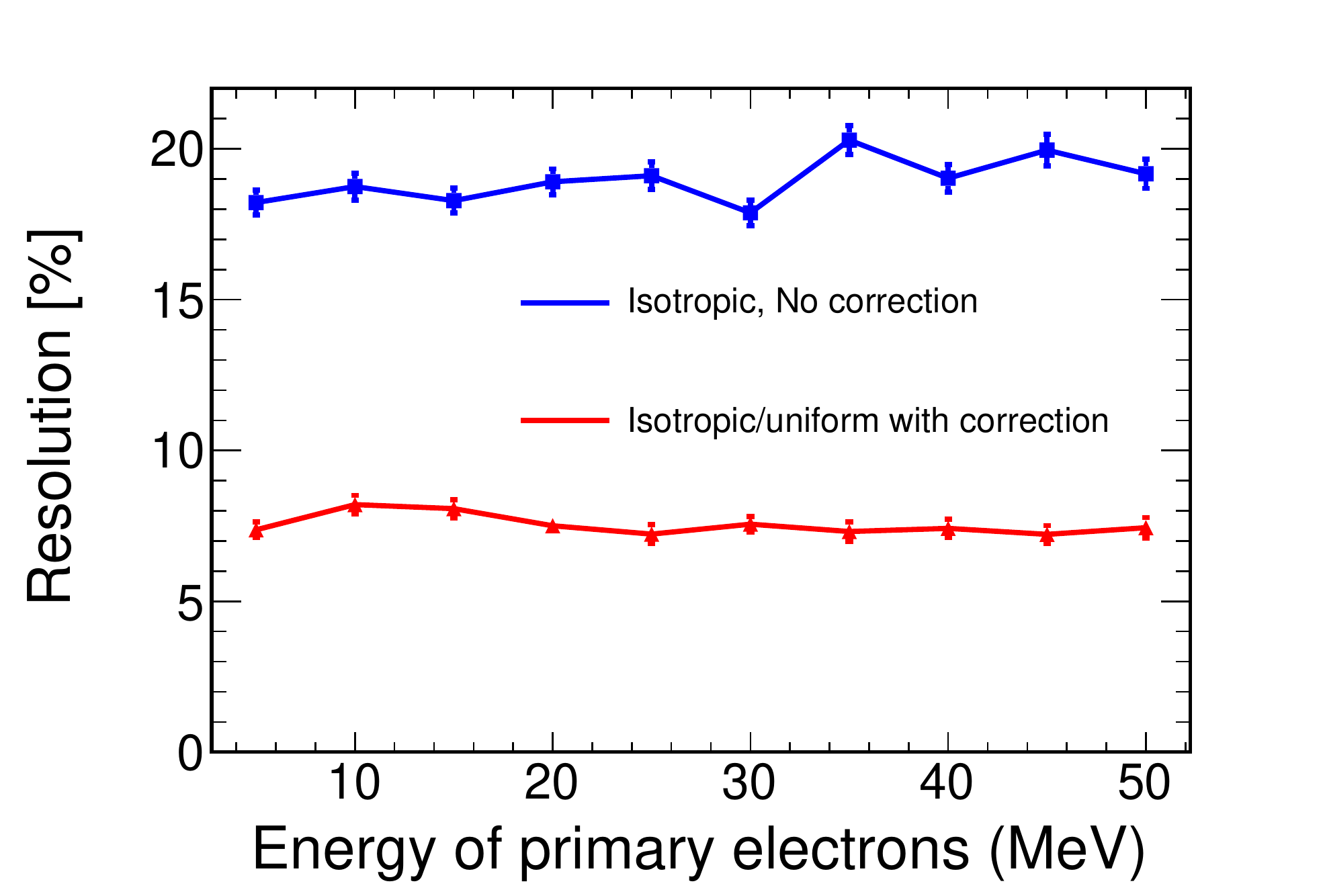} 
 \includegraphics[width=0.45\textwidth]{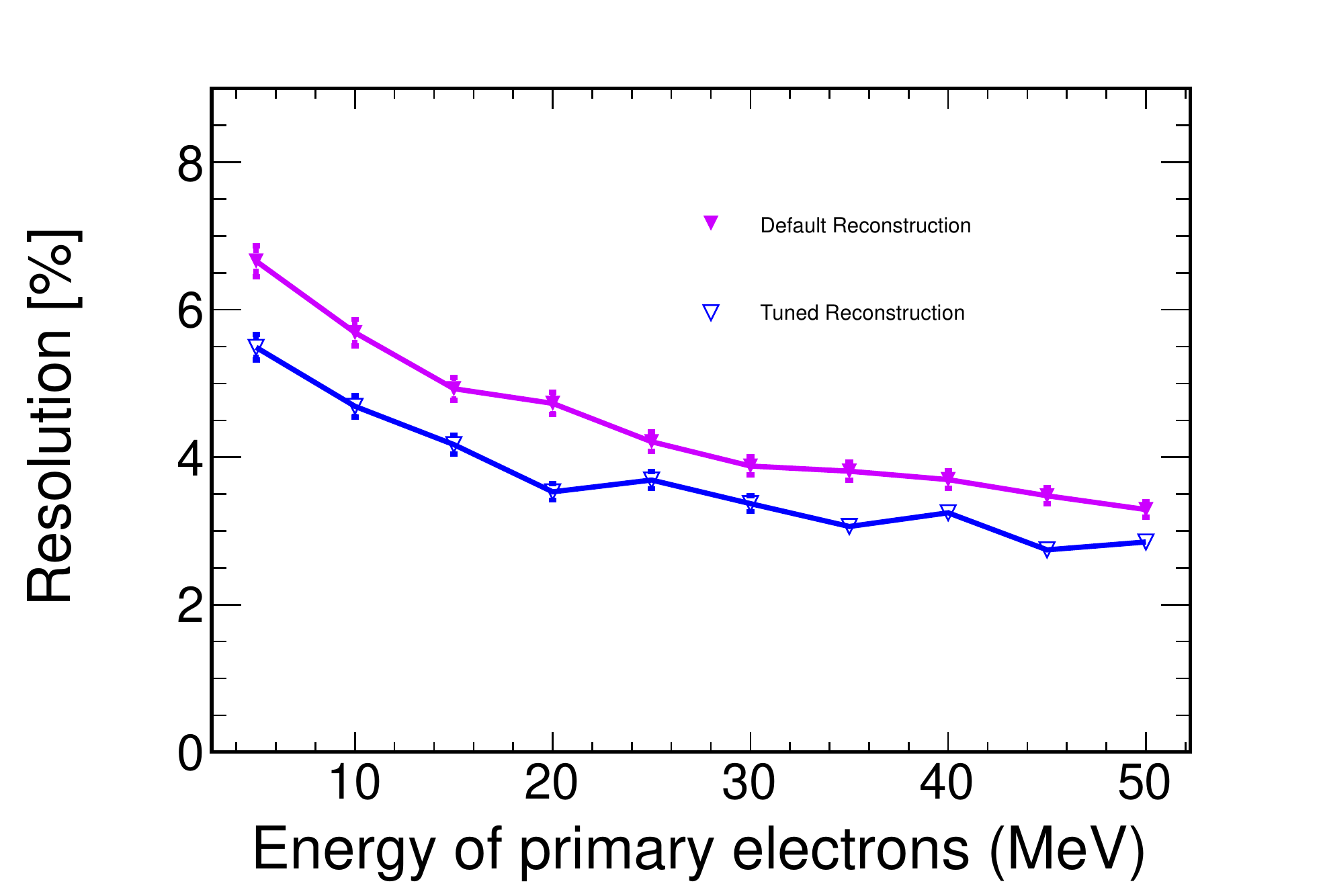} 
 \caption[Comparisons of energy resolution]{Left: Comparison of energy
   resolution (defined as $\sigma/E$, where $\sigma$ is the spread of
   the collection-plane-charge-based event energy $E$ for a
   monoenergetic electron), with and without electron-lifetime
   correction, as a function of electron energy. The blue curve is the
   energy resolution of isotropic and uniform electrons without
   electron-lifetime correction. The red curve is the energy
   resolution with electron-lifetime correction based on MC truth.
   Right: Comparison of energy resolution before and after tuning the
   reconstruction algorithm (for fixed position/direction electron
   events).}\label{fig:lowe_res}
\end{figure}

Work is currently underway using the LArSoft simulation package
to characterize low-energy
response for realistic LBNE detector configurations.
Figure~\ref{fig:evdisplays} shows a sample \MeVadj{20} event in the LBNE
\SIadj{35}{t} prototype
geometry simulated with LArSoft.  So far,
most studies have been done with the MicroBooNE geometry, with the
results expected to be 
generally applicable to the larger LBNE detector.  For a preliminary
understanding of achievable energy resolution, isotropic and uniform
monoenergetic electrons with energies of 5-50~MeV (which should
approximate the $\nu_e$-CC electron products) were simulated and
reconstructed with the LArSoft
package. 
The charge of reconstructed hits on the collection plane was used to
reconstruct the energy of the primary electrons. (Induction-plane charge
as well as track-length-based reconstruction were also considered, but
with inferior results). Figure~\ref{fig:lowe_res} shows the results.
A correction to compensate for loss of electrons during drift,
$Q_{collection}=Q_{production}\times e^{-T_{\rm drift} / T_{\rm
    electron}}$ (where $T_{drift}$ is the drift time of the ionization
electrons, and $T_{electron}$ is the electron lifetime), using Monte
Carlo truth to evaluate $T_{\rm drift}$, improved resolution
significantly.  This study indicated that photon time information will
be valuable for low-energy event reconstruction.  Some of the
resolution was determined to be due to imperfect hit-finding by the
nominal reconstruction software.  A tuned hit-finding algorithm did
somewhat better (Figure~\ref{fig:lowe_res}), and further
improvements for reconstruction algorithms optimized for low-energy
events are expected.

Also under study is the potential for tagging $\nu_e$-CC absorption
events ($\nu_e +{}^{40}{\rm Ar} \rightarrow e^- +{}^{40}{\rm
  K^*}$) using the cascade of de-excitation $\gamma$ rays, which should
serve the dual purposes of rejecting background and isolating the CC
component of the signal.  

\section{Far Detector Reconstruction}

The first stage of reconstruction of TPC data is unpacking and
deconvoluting the electronics and field response of the wire planes.
The deconvolution function includes a noise filter that currently
is parameterized with ArgoNeuT's noise, but will be tuned
for the eventual noise observed in the LBNE detector.  
The deconvolution makes sharp,
unipolar pulses from the bipolar induction-plane signals and also
sharpens the response to collection-plane signals.  Hits are then
identified in the deconvoluted signals by fitting Gaussian functions,
allowing for sums of several overlapping hits in each cluster.  
In LBNE, because of the large quantity of channels in the far detector, any
inefficiency in CPU and memory is magnified.  Improvements
in the memory-usage efficiency relative to the ArgoNeuT and MicroBooNE
implementations have been realized by 
rearrangement of the processing order and limiting the storage of the intermediate
uncompressed raw data and the deconvoluted waveforms. 

After signal deconvolution, line-finding and clustering based on a
Hough transform in two dimensions is done using an algorithm called
\emph{fuzzy clustering}~\cite{Sandhir:2012cr}.  This clustering is
performed separately on data from each induction plane.
Since the hit data on LArTPCs are inherently 2D --- wire number
and arrival time of the charge --- the location of the initial ionization
point has a 2D ambiguity if the deposition time is
unknown.  For beam events, the $t_0$ is known, and thus only a
1D ambiguity remains;  this 1D ambiguity is broken by
angling the induction-plane wires relative to the collection-plane
wires, in order to measure the $y$ location of the hits for which $t$
(thus $x$) and $z$ are known.  For (non-beam) 
cosmic-ray signals which arrive uniformly in time, the photon system 
provides $t_0$.
After clustering, 3D track-fitting 
is performed using a
Kalman filter~\cite{kalman}. Dedicated algorithms have been
developed to optimize electromagnetic shower reconstruction and energy
resolution. 

LBNE poses a unique challenge for reconstruction because
the induction-plane wires wrap around the edges of the APA frames.  
This introduces discrete
ambiguities that are not present in other LArTPC designs.  Whereas a hit on a
collection-plane wire identifies uniquely the side of the APA from
which it came, this is not known for a hit on an induction-plane
wire.  The angles between the $U$ and $V$ plane wires  are slightly
different from 45$^\circ$ and from each other in order to 
break the ambiguities.  
A combinatoric issue arises, however, if
many hits arrive on different wires at nearly the same time, for instance 
when a track, or even a track segment, propagates in a plane
parallel to the wire planes (i.e., at constant drift distance).  Showers will
also contain many hits on different wires that arrive at similar
times.  
Hits that arrive at different times can be 
clustered separately 
in the $Z$, $U$, and $V$ views without ambiguity, 
while hits that arrive at similar
times must be associated using a topological pattern-recognition
technique.  LBNE is developing a version of the fuzzy clustering tool
for use as a pattern-recognition step to allow association of
$Z$, $U$ and $V$ hits, a step that is needed to
assign the correct $y$ position to a track
segment or portion of a cluster.  
This process is called \emph{disambiguation} of the induction
hits.  Misassignment can affect particle-ID performance and
reconstructed-energy resolution because fully contained tracks may appear
partially contained and vice versa.   After disambiguation has been
performed, standard track, vertex and cluster reconstruction
algorithms are applied.

\begin{figure}[!htb]
\begin{center}
\includegraphics[width=0.8\textwidth]{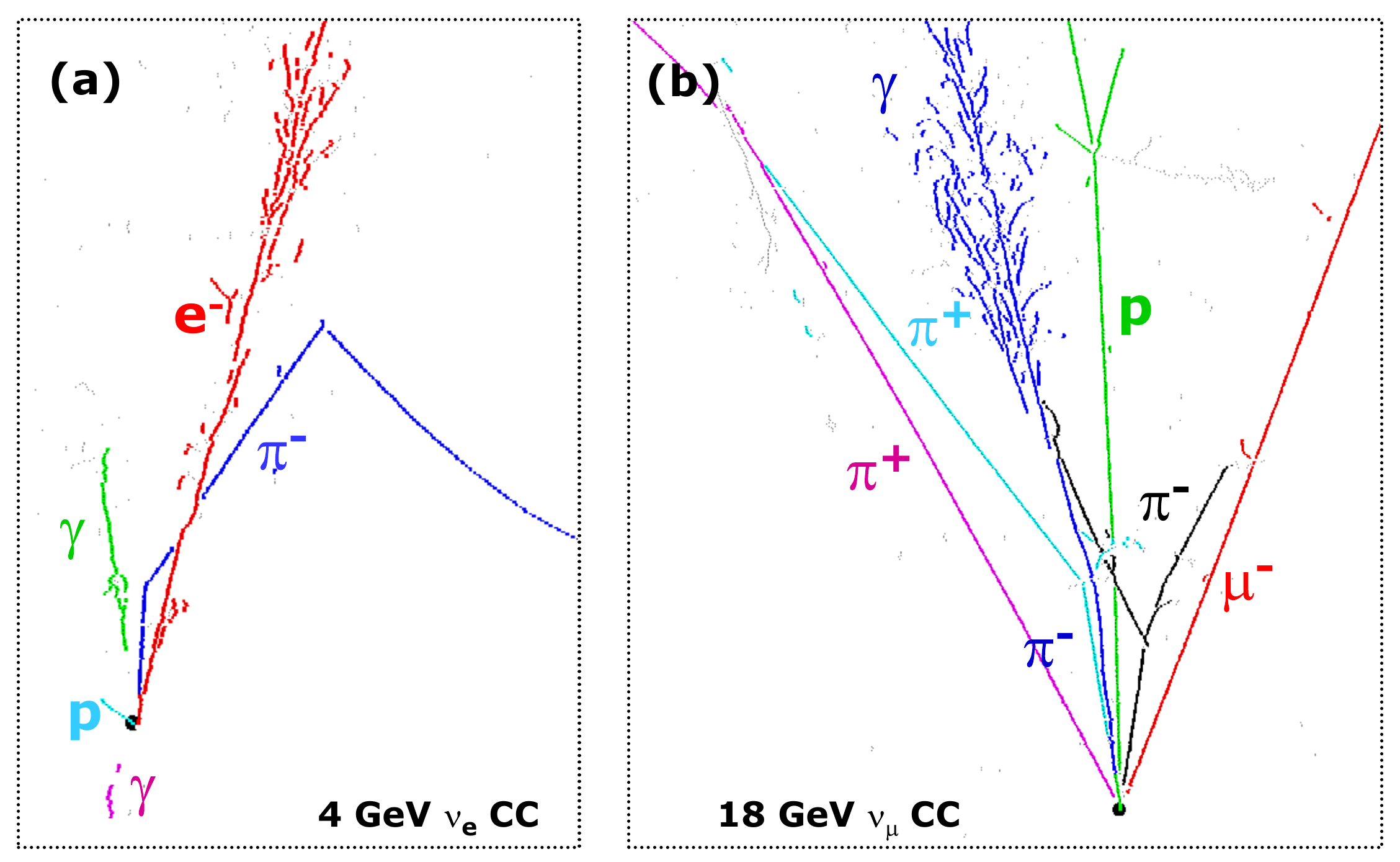}
\end{center}
\caption[2D clusterings of hits created by particles in
 two CC neutrino interactions in LAr]{ \label{fig:pandoraevent} PANDORA's 2D clusterings of hits created by the particles in
 two CC neutrino interactions in liquid argon.
Panel (a) shows a \SIadj{4}{\GeV} $\nu_e$ interaction, and panel (b) shows an \SIadj{18}{\GeV} $\nu_\mu$ interaction.  The colors indicate
the clusters into which PANDORA has divided the hits, and the particle labels are from the MC truth.
}
\end{figure}
\begin{figure}[!htb]
\begin{center}
\includegraphics[width=0.9\textwidth]{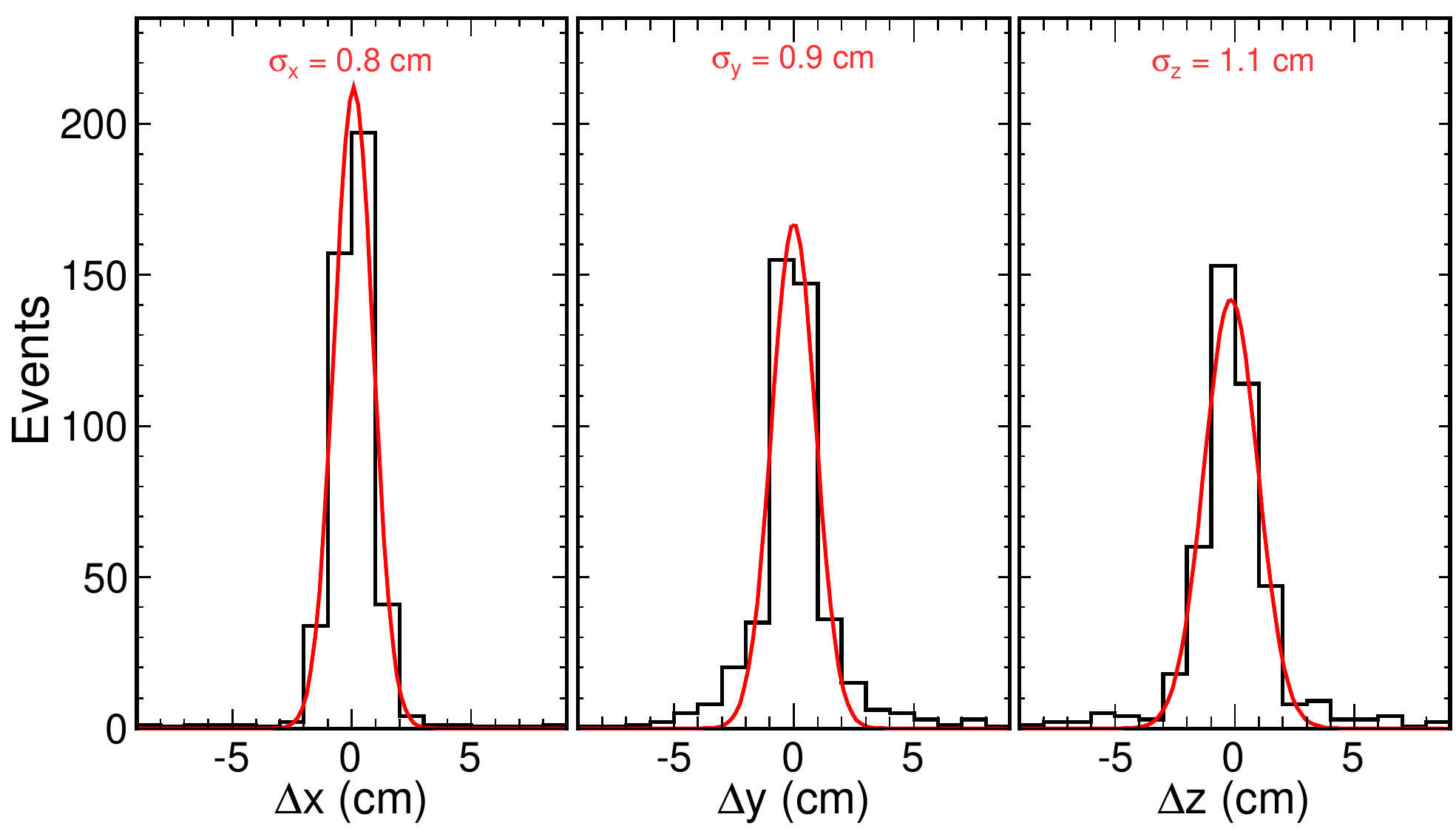}
\end{center}
\caption[Distributions of
residuals between reconstructed and MC primary vertices]{ \label{fig:pandoravertexresolution} Distributions of the
residuals between the reconstructed and the Monte Carlo true locations of primary vertices in neutrino interactions
in the MicroBooNE geometry using the LBNE beam spectrum.  The $x$ axis is oriented along the drift field, the $y$ axis
is parallel to the collection-plane wires,  and the $z$ axis points along the beam direction.
}
\end{figure}

A promising suite of algorithms for event reconstruction is provided
by the PANDORA toolkit~\cite{Marshall:2012hh}, which provides a framework for
reconstruction algorithms and visualization tools.  Currently it is
being used to develop pattern-recognition algorithms and to
reconstruct primary vertices.  PANDORA's pattern-recognition algorithm merges
hits based on proximity and pointing to form 2D clusters.  Vertices
are then identified from the clusters that best connect to the same event. 
Clusters that best correspond to particles emitted from the primary vertex 
are identified in 2D.  These particle candidates are then used to seed
3D reconstructed particles, and a 3D primary vertex is identified.
Examples of PANDORA's 2D clustering are shown in
Figure~\ref{fig:pandoraevent} for two simulated CC neutrino-scattering events.  Figure~\ref{fig:pandoravertexresolution} shows the
primary vertex spatial resolution in 3D with well-contained simulated
beam-neutrino events, using the nominal LBNE spectrum and MicroBooNE
geometry. 

\section{Fast Monte Carlo}
\label{appxsec:fastmc}

The LBNE full Monte Carlo (MC) simulation will use a Geant4 simulation
of the beamline to estimate the neutrino flux, a neutrino interaction
generator (e.g., GENIE), and detailed detector simulation that mimics
the real detector output for data events. Both data and MC will have
the same reconstruction algorithms applied to produce quantities that
will be used to analyze the data.  The full MC detector simulation and
reconstruction algorithms are still under development.  Due to their
detailed nature, these algorithms are CPU-intensive and time-consuming
to run.

In parallel, a Fast Monte Carlo simulation has been developed and is
available for use in place of the full MC to explore long-baseline
physics analysis topics.  A preliminary version of the Fast MC is
currently available.
Results from the latest detector simulations and advancements in
reconstruction algorithms are actively being incorporated to improve
the physics models and detector parameterization.
Because the Fast MC replaces CPU-intensive portions of the full MC
simulation with a fast parameterized model, it offers a quick, dynamic
alternative which is useful for trying out new ideas before
implementing them in the full MC.  This usefulness is expected to
remain even after the full MC simulation is mature.

To accurately approximate a full MC simulation, the Fast MC combines
the Geant4 LBNE beamline flux predictions, the GENIE event interaction
generator, and a parameterized detector response that is used to
simulate the measured (reconstructed) energy and momentum of each
final-state particle. The simulated energy deposition of the particles
in each interaction is then used to calculate reconstructed kinematic
quantities (e.g., the neutrino energy), and classify the type of
neutrino interaction, including backgrounds and misidentified
interactions.

The Fast MC is designed primarily to perform detailed sensitivity
studies that allow for the propagation of realistic systematic
uncertainties. 
It incorporates effects due to choices of models and their
uncertainties and design decisions and tolerances.  The neutrino
flux predictions, the neutrino-interaction cross-section models, and
the uncertainties related to these are also incorporated.
The parameterized detector response is informed by Geant4 simulations
of particle trajectories in liquid argon, by studies of detector response
simulation in MicroBooNE~\cite{Katori:2011uq}, results reported by the
ICARUS Collaboration, and by the expected LBNE detector geometry. The
realistic parameterization of reconstructed energy and angle
resolution, missing energy, and detector and particle identification
acceptances provide a simulation that respects the physics and
kinematics of the interaction and allows for propagation of model
changes to final-state reconstructed quantities.

Future efforts will allow for propagation of uncertainties in detector
effects and of detector design choices.  It should be noted that the
same 
GENIE files 
generated for the Fast MC can be used as
inputs for the full detector simulation and the results of the two
simulations can be compared both on an event-by-event basis and
in aggregate. Studies of this nature can be used to tune
the Fast MC and to cross-check the full simulations.

In the current configuration of the Fast MC, GENIE generates
interactions on $^{40}$Ar nuclei with neutrinos selected from the
energy spectra predicted by the Collaboration's Geant4 flux
simulations (described 
in Section~\ref{beamline-chap}).  For
each interaction simulated in GENIE, a record of the interaction
process, its initial kinematics, and 
the identity and four-momenta of the final-state particles 
is 
produced.
The parameterized detector response applies spatial and
energy/momentum smearing to each of the final-state particles based on
the particle properties and encoded detector-response parameters.
Detection thresholds are applied to determine if a final-state
particle will deposit energy in the detector and if that energy
deposition will allow for particle identification.  The detector
responses for neutrons and charged pions 
account for 
a variety of possible
outcomes that describe the way these particles deposit energy in the
detector.
Neutral pions are decayed into two photons.  Their conversion distance
from the point of decay determines the starting position of the
resulting electromagnetic showers. This distance is chosen from 
an exponential distribution with a characteristic length based on the
radiation length of photons in liquid argon. 
Tau leptons are also decayed by the Fast MC and their decay products
are dealt with appropriately.  The spatial extent of tracks and
showers in liquid argon is simulated in Geant4 and encoded as a
probability distribution function (PDF) 
or parameterization. Combined with vertex placement in a fiducial volume,
the fraction of particle energy and/or track length visible in the
detector is determined.

Once the Fast MC reconstructs the kinematics of the event ($E_{\nu}$,
$E_{had}$, $Q^{2}$, $x$, $y$, and so on), based on the smeared four-vectors
of particles that are above detection threshold, it searches
interaction final-state particle lists for lepton candidates to be
used in event classification algorithms.  The resulting
classifications are used to isolate samples for the $\nu_{e}$ appearance and
the $\nu_{\mu}$ disappearance analyses which are in turn used to build energy
spectra on an event-by-event basis.

Currently the classification algorithm categorizes each event as either $\nu_{e}$-CC,
$\nu_{\mu}$-CC, or NC. Events with a candidate muon are classified as $\nu_{\mu}$-CC.
Events without a candidate muon, but with a candidate electron/positron are classified as $\nu_{e}$-CC.
Events without a candidate muon or a candidate electron/positron are classified as NC.
A $\nu_{\tau}$-CC classification, which would identify $\nu_{\tau}$ candidates is under development.

A muon candidate is defined as a MIP-like track
that is greater than \SI{2.0}{\meter} long, and is not consistent with 
the behavior of a charged pion. Charged pions will often \emph{shower},
depositing a relatively large amount of energy in the detector at the 
end of its track, as compared to a muon. There are several situations
in which a pion topology will be indistinguishable from a muon:
(1) the pion stops at the end of its range without interacting,
(2) the kinetic energy of the pion is sufficiently small when is showers,
(3) the pion is absorbed cleanly by a nucleus with no hadronic debris, 
(4) the pion decays in flight, and (5) the track exists the detector. 
The \SIadj{2.0}{\meter} cut was chosen because the probability of (1)
or (2) is very small for pion tracks above this threshold.

An additional selection probability is enforced for
low-energy tracks to simulate acceptance losses due to
increased difficulty in particle identification for short
tracks, especially in high-multiplicity events. (The
falling edge of the selection probability is well below
the energy required to generate a \SIadj{2.0}{\meter}
track, minimizing the effect of this criterion.)

An electron candidate is defined as the highest-momentum 
electromagnetic (EM) shower in an event that is not consistent 
with a photon. An EM shower is identified as a photon 
(1) if it converts \SI{2.0}{\centi\meter} or more from the event vertex,
(2) if it can be matched with another EM shower in the events to reconstruct
the $\pi^0$ mass~($135\pm40$ MeV), or (3) if $dE/dx$ information from the
first several planes of the track is more photon-like than e$\pm$-like.
The latter is determined on a probabilistic basis as a function of 
EM-shower energy and hadronic-shower multiplicity. Signal and background
efficiencies from the $dE/dx$ e/$\gamma$ discriminant are based on MicroBooNE 
simulations. Cut values are tuned to preserve 95\% of the signal across all 
neutrino energies. As with muon candidates a low-energy selection probability 
is enforced to account for acceptance losses at low EM-shower energies,
especially in high-multiplicity events. For the electron candidates this
selection probability is tuned to agree with hand scan studies.

An event with no muon candidate and no electron
candidate is assumed to be an NC interaction. Preliminary
studies evaluating the use of transverse-momentum imbalance to
identify $\nu_{\tau}$-CC interaction candidates have shown promising results
for identifying NC candidates as well, and are likely to be included in the near future.

Currently no attempt is made to identify
tau lepton candidates in order to isolate a $\nu_\tau$-CC sample. A
preliminary algorithm to remove $\tau \rightarrow \mu+\nu+\nu$
and $\tau \rightarrow e+\nu+\nu$ backgrounds has recently been
incorporated in the Fast MC. This algorithm may also prove useful for isolating a
sample of $\nu_{\tau}$-CC interactions, in which the tau decays to a lepton.  Development
of an algorithm to identify taus that decay to hadrons is under discussion.

All of the selection criteria can easily be updated to reflect
improved simulations or new understanding of particle-identification
capabilities and analysis sample acceptances. Changes can also be made
to investigate alternate analysis techniques, or more conservative or
optimistic assumptions on signal acceptance and/or background-rejection 
rates. Furthermore, the information required to simulate
effects related to particle identification is available in the Fast MC
files and users are encouraged to construct and evaluate their own
selection criteria.

\begin{figure}[!htb]
\centerline{
\includegraphics[width=0.5\textwidth]{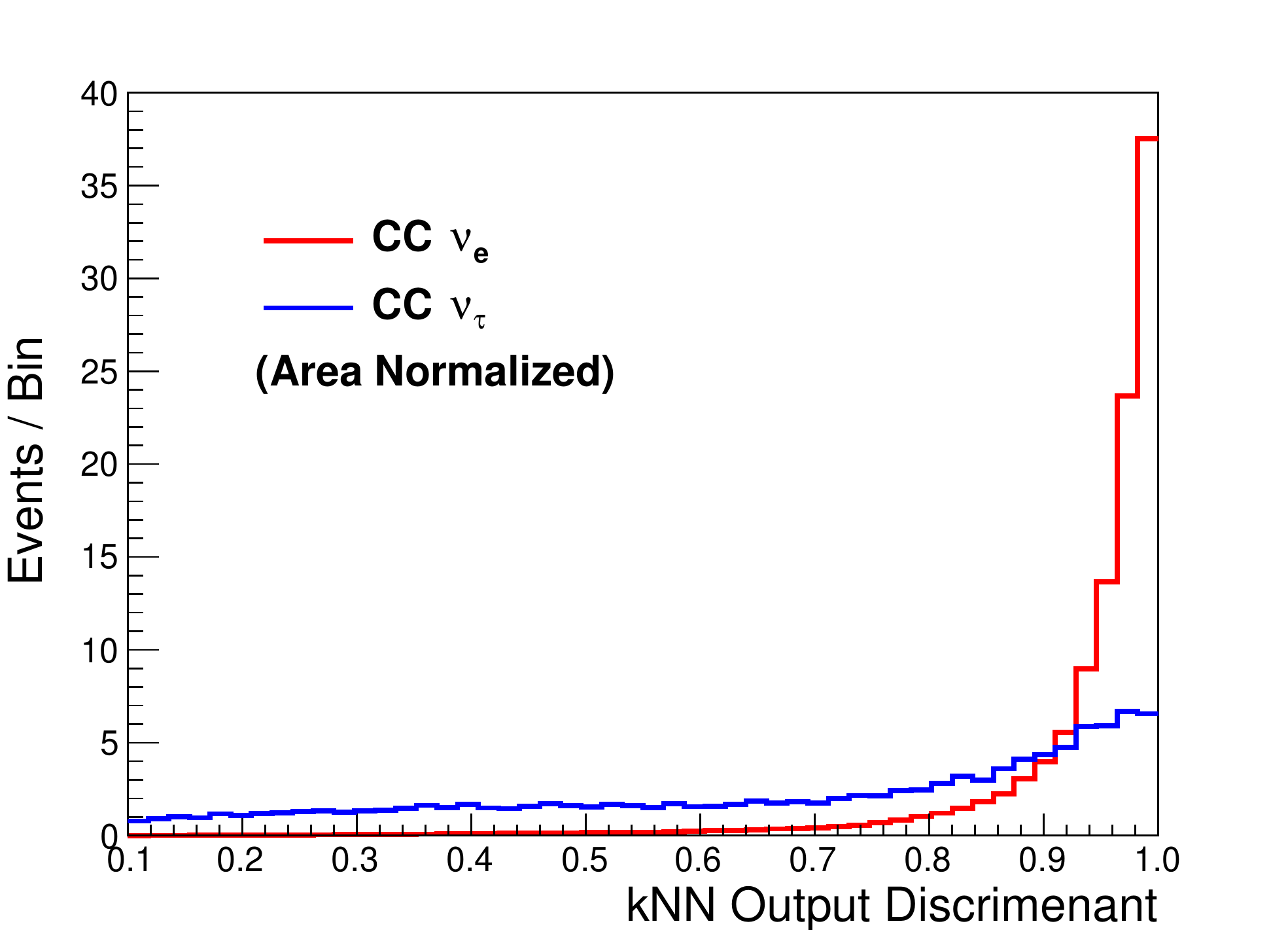}
\includegraphics[width=0.5\textwidth]{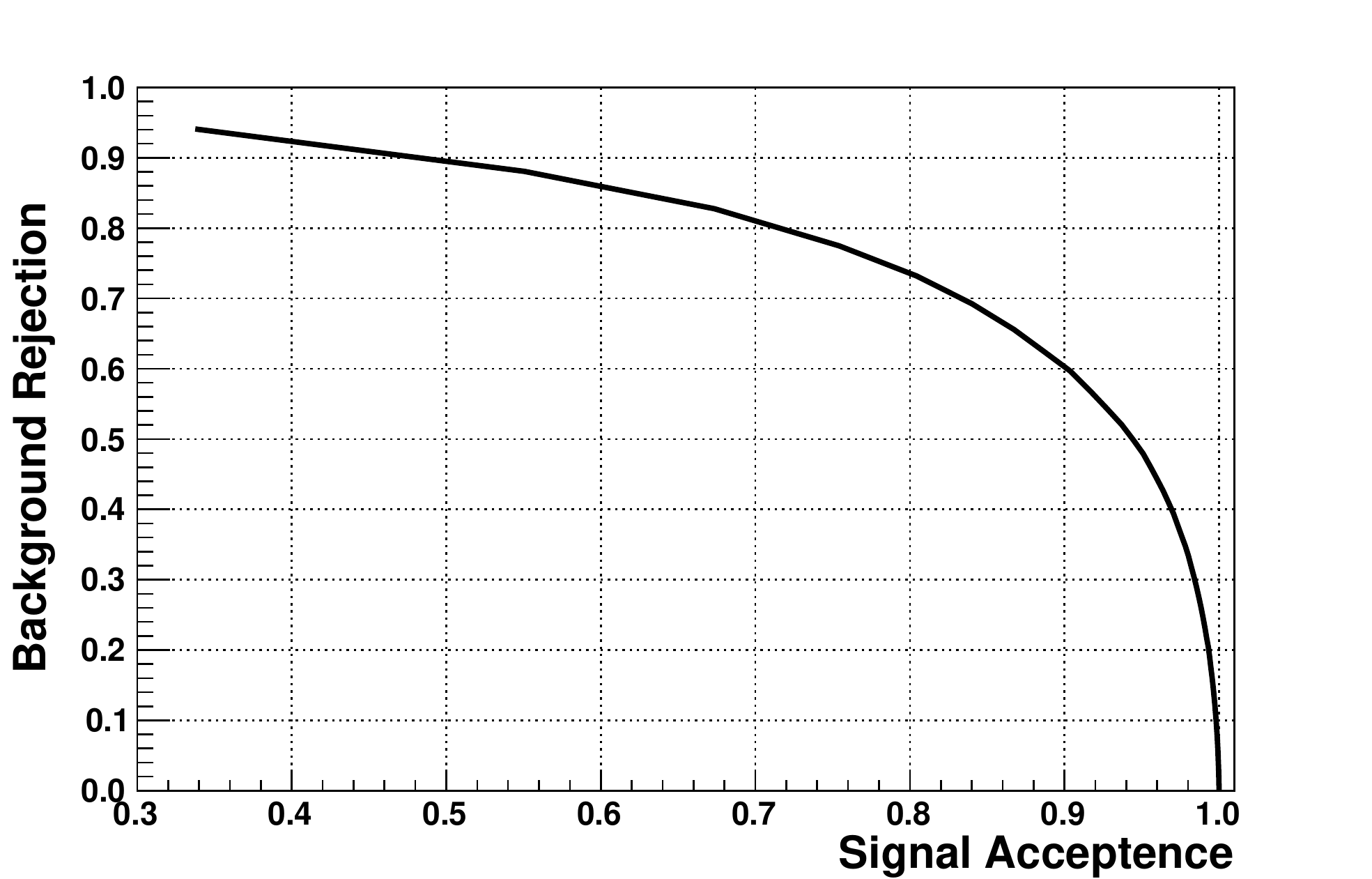}
}
\caption[kNN discriminant for removal of $\nu_\tau$-CC-induced backgrounds]
        {The output discriminant of a kNN (left) created to remove  $\nu_\tau$-CC-induced
         backgrounds from the $\nu_{\mu} \rightarrow \nu_{e}$ oscillation analysis sample.
         Signal events (red) tend toward high values, while the  $\nu_\tau$-CC-induced
         background events (blue) are more evenly distributed. 
         The fraction of $\nu_\tau$-CC-induced backgrounds removed from the
         $\nu_{\mu} \rightarrow \nu_{e}$ appearance candidate sample
         as a function of the corresponding signal efficiency (right). The curve is
         generated by varying the cut value on the kNN discriminant. }
\label{fig:kNN_resp}
\end{figure}
A preliminary algorithm for removing $\nu_\tau$-CC-induced backgrounds
from from the $\nu_\mu$-CC and the $\nu_{e}$-CC samples has been
developed.  It employs a k-Nearest Neighbor (kNN) machine-learning
technique as implemented in the ROOT TMVA package. The inputs to the
kNN are (1) the sum of the transverse momentum with respect to the
incoming neutrino direction, (2) the reconstructed energy of the
incoming neutrino, and (3) the reconstructed energy of the resulting
hadronic shower. Figure~\ref{fig:kNN_resp} (right) shows the
distribution of the output discriminant for true $\nu_{e}$-CC signal
events, and for true $\nu_\tau$-CC-induced backgrounds. The algorithm
is still being optimized but initial results are promising. 

As can be
seen in Figure~\ref{fig:kNN_resp} (left), cuts on the discriminant
that preserve 90\% of the signal remove roughly 60\% of the
$\nu_\tau$-CC-induced background in the $\nu_{e}$-CC sample. Similar
results are expected for the $\nu_\tau$-CC-induced background in the
$\nu_\mu$-CC sample.

A similar approach is being studied to isolate the $\nu_\tau$-CC
sample for the $\nu_\tau$-CC appearance analysis. Current efforts
are focused on identifying a set of reconstructed quantities that 
separate $\nu_\tau$-CC interactions from potential backgrounds.
For leptonic decay channels the quantities used in the above kNN 
are prime candidates. Attempts to reconstruct a $\rho$ mass from 
tracks originating at the vertex are expected to help to isolate 
hadronic $\tau$ decays. The parameterized pion response will allow
for selection of high-energy charged pions produced in hadronic 
$\tau$ decays.

\begin{figure}[!htb]
\includegraphics[width=0.48\textwidth]{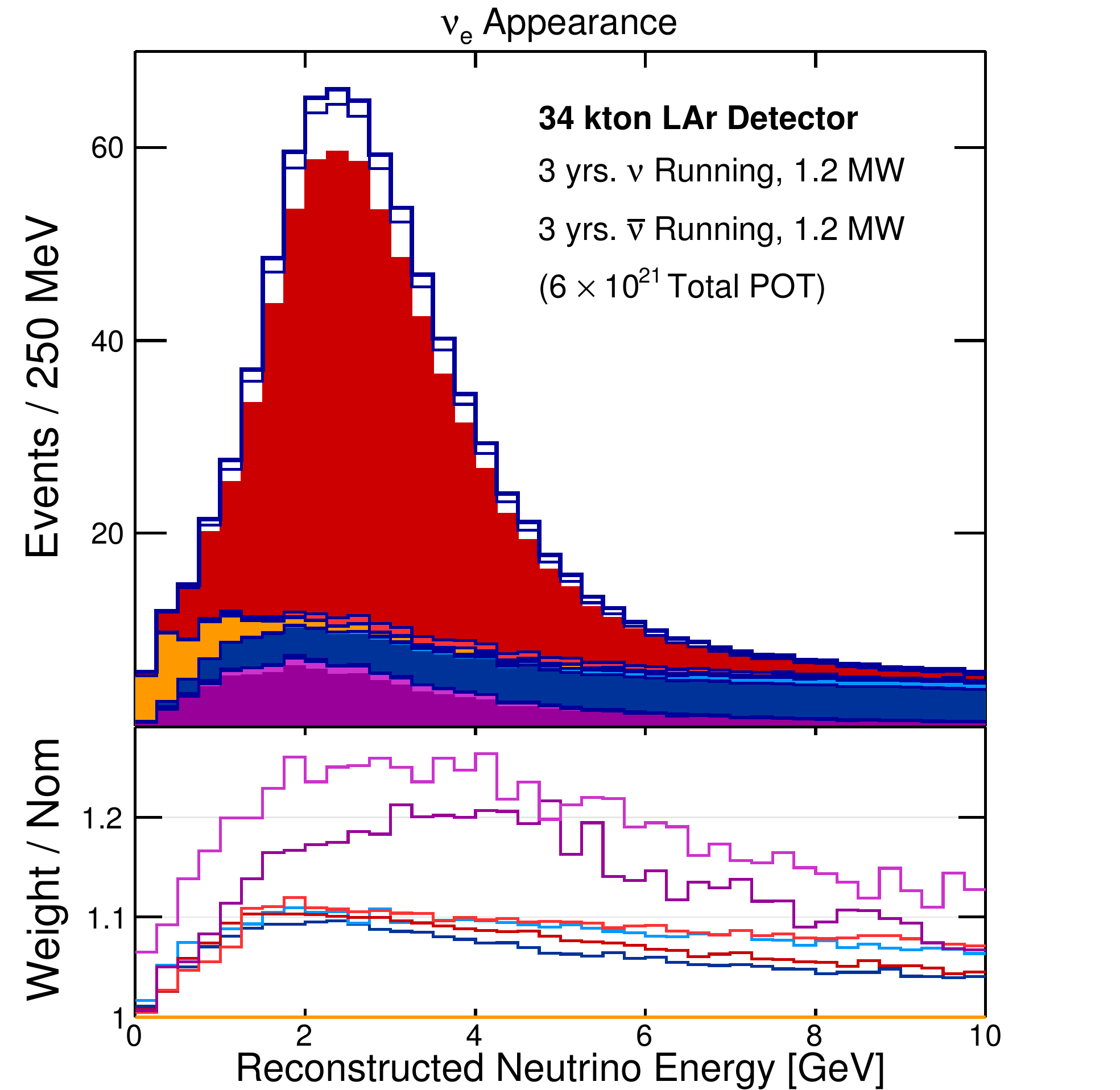}
\includegraphics[width=0.48\textwidth]{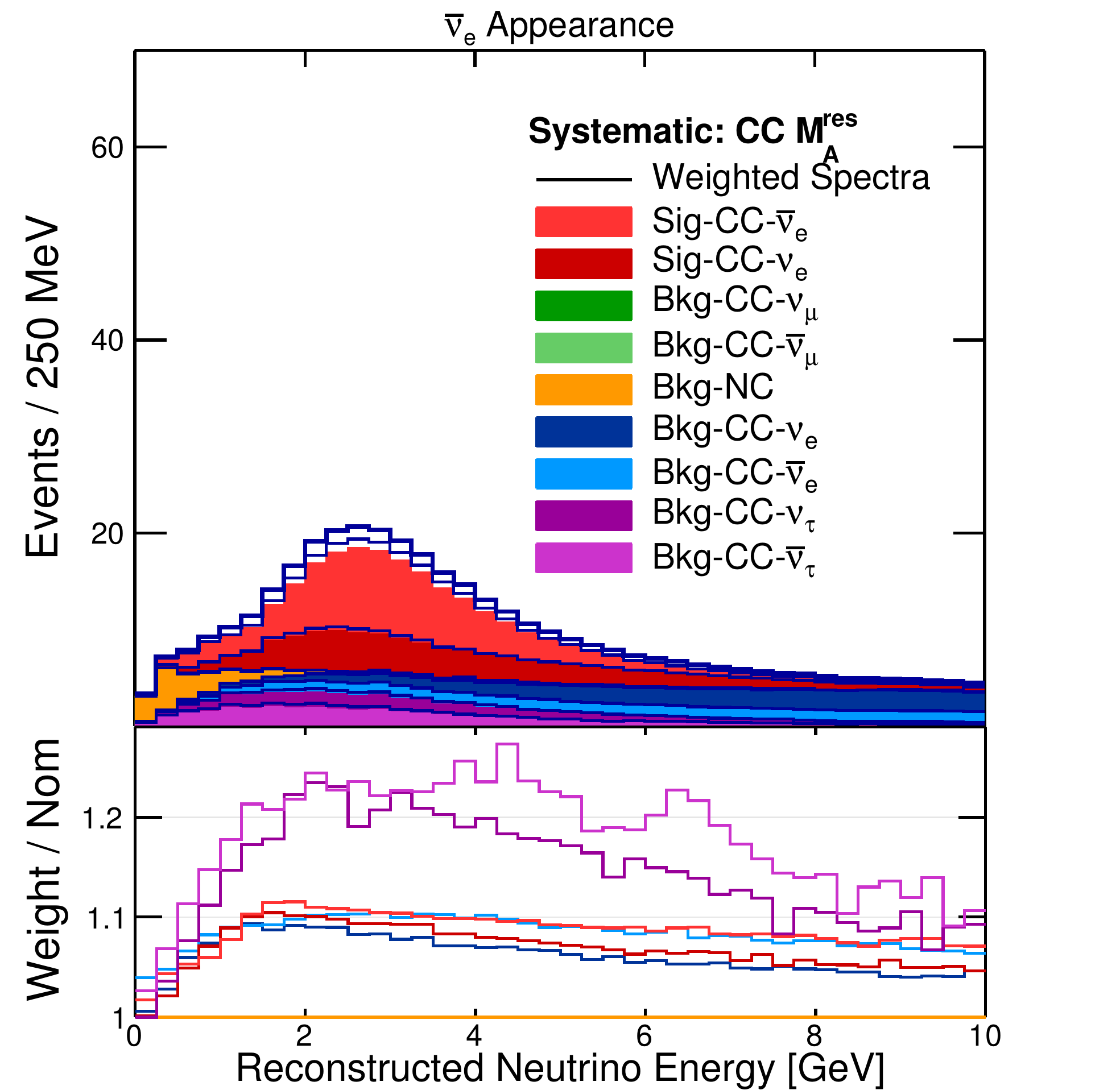}
\caption[Energy distributions for $\nu_e$ and  $\overline{\nu}_e$ appearance signals]
{The reconstructed energy distributions for the signals and backgrounds in
the $\nu_e$- (left) and  $\overline{\nu}_e$ appearance (right) samples, as predicted
by the Fast MC. The black histograms and bottom insert in each plot shows,
for each event type, the variation in the spectrum that is
induced by changing the value of CC~$M^{res}_{A}$ by +15\%.}
\label{fig:detspecapp}
\end{figure}

\begin{figure}[!htb]
\includegraphics[width=0.48\textwidth]{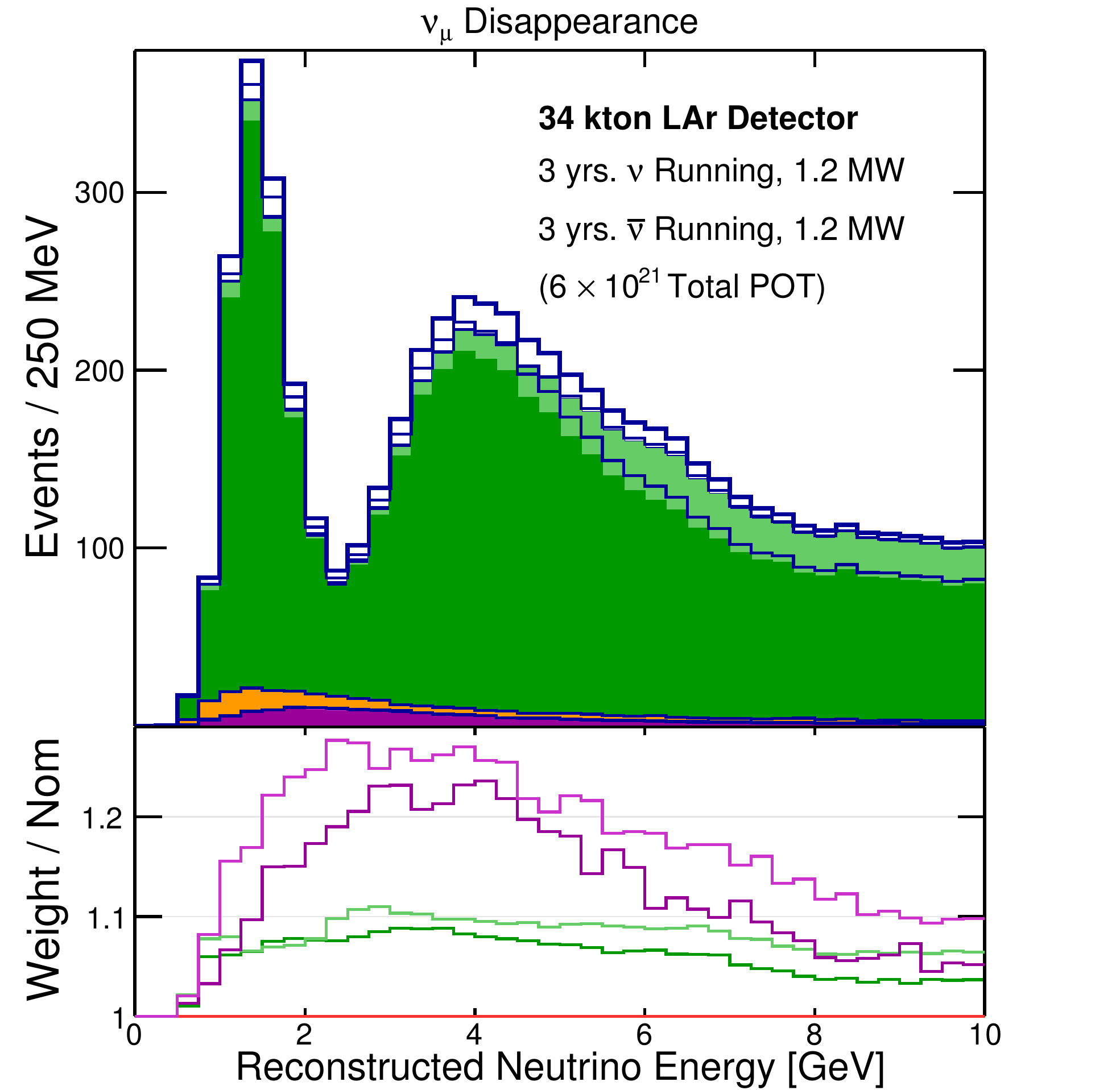}
\includegraphics[width=0.48\textwidth]{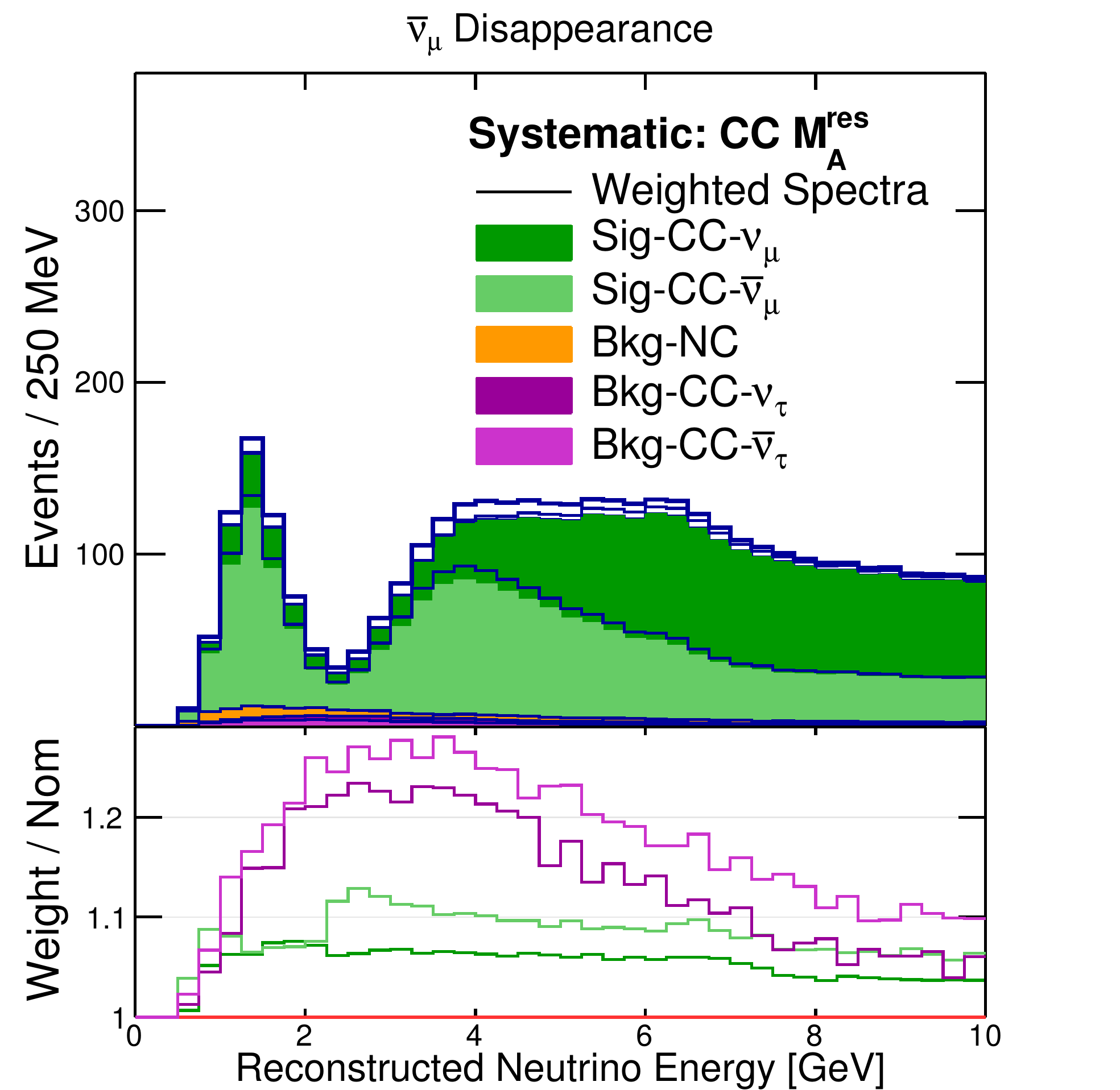}
\caption[Energy distributions for $\nu_\mu$ and $\overline{\nu}_\mu$ disappearance signals]{The reconstructed
energy distributions for the signals and backgrounds in
the $\nu_\mu$ (left) and $\overline{\nu}_\mu$ disappearance (right) samples, as predicted
by the Fast MC. The black histograms and bottom insert in each plot shows, for each event type, the
variation in the spectrum that is induced by changing the value of CC~$M^{res}_{A}$ by +15\%.}
\label{fig:detspecdisap}
\end{figure}

Figures~\ref{fig:detspecapp} and~\ref{fig:detspecdisap} show the
Fast MC reconstructed energy spectra of the signal and background
for the $\nu_e$ appearance and the $\nu_\mu$ disappearance samples, respectively. 
As an example of the cross-section and nuclear-effect systematics that can be studied,
the black histograms and the bottom insert in each plot show
the variation of the spectrum for each event type induced by changing 
the value of CC~$M^{res}_{A}$ by +1$\sigma$ (+15\%, 2014 GENIE official uncertainty). 
CC~$M^{res}_{A}$ is the axial mass parameter appearing in the axial form factor 
describing resonance production interactions in GENIE. 
This particular example demonstrates a spectral distortion
that is not a simple normalization and is different for signal and for background. 
The effect of varying CC~$M^{res}_{A}$ 
on the $\nu_\mu \rightarrow \nu_e$ analysis sample exhibits a strong correlation
with the changes induced in the $\nu_\mu \rightarrow \nu_\mu$ analysis sample. 


\begin{figure}[!htb]
\includegraphics[width=0.49\textwidth]{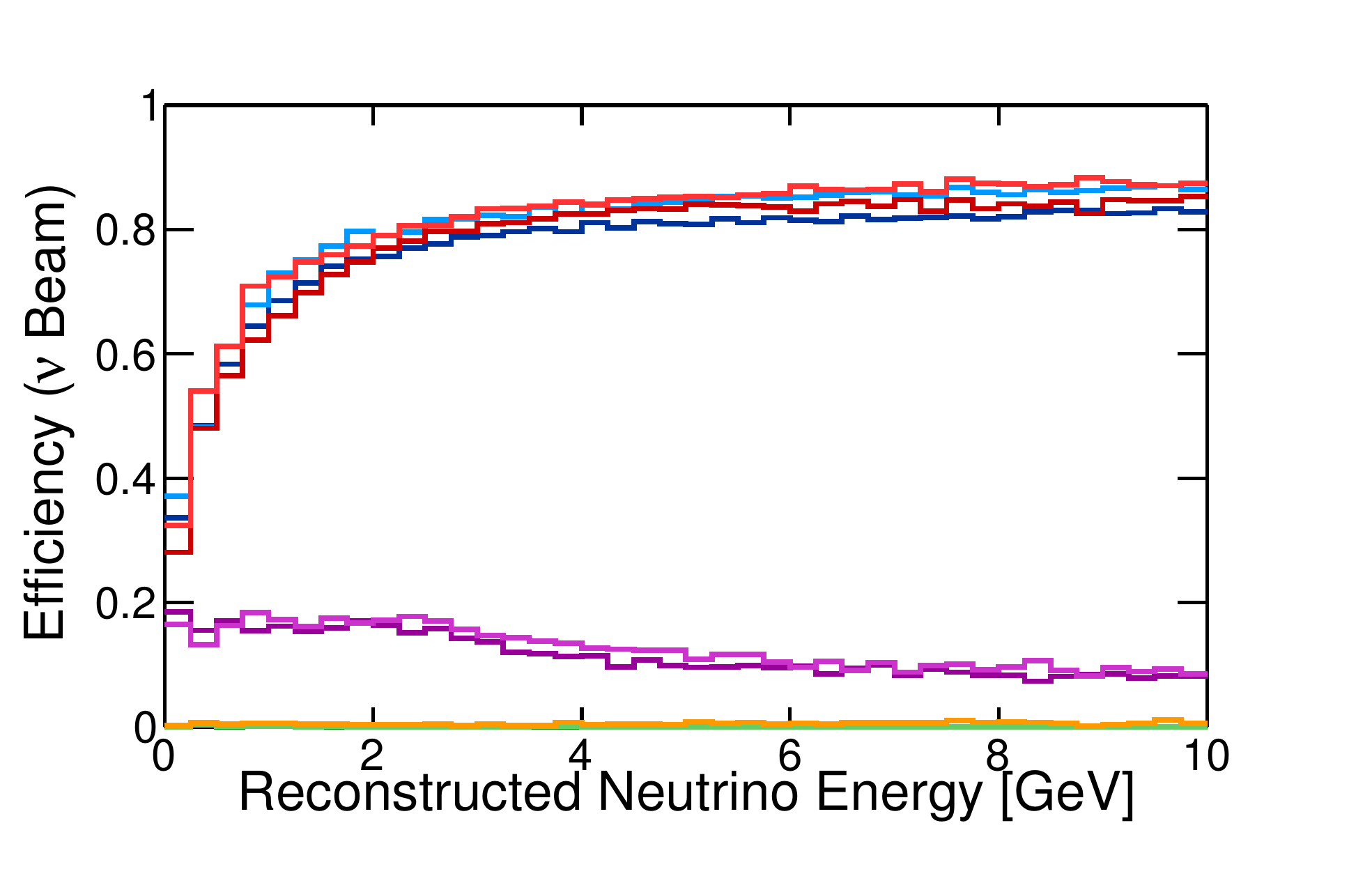}
\includegraphics[width=0.49\textwidth]{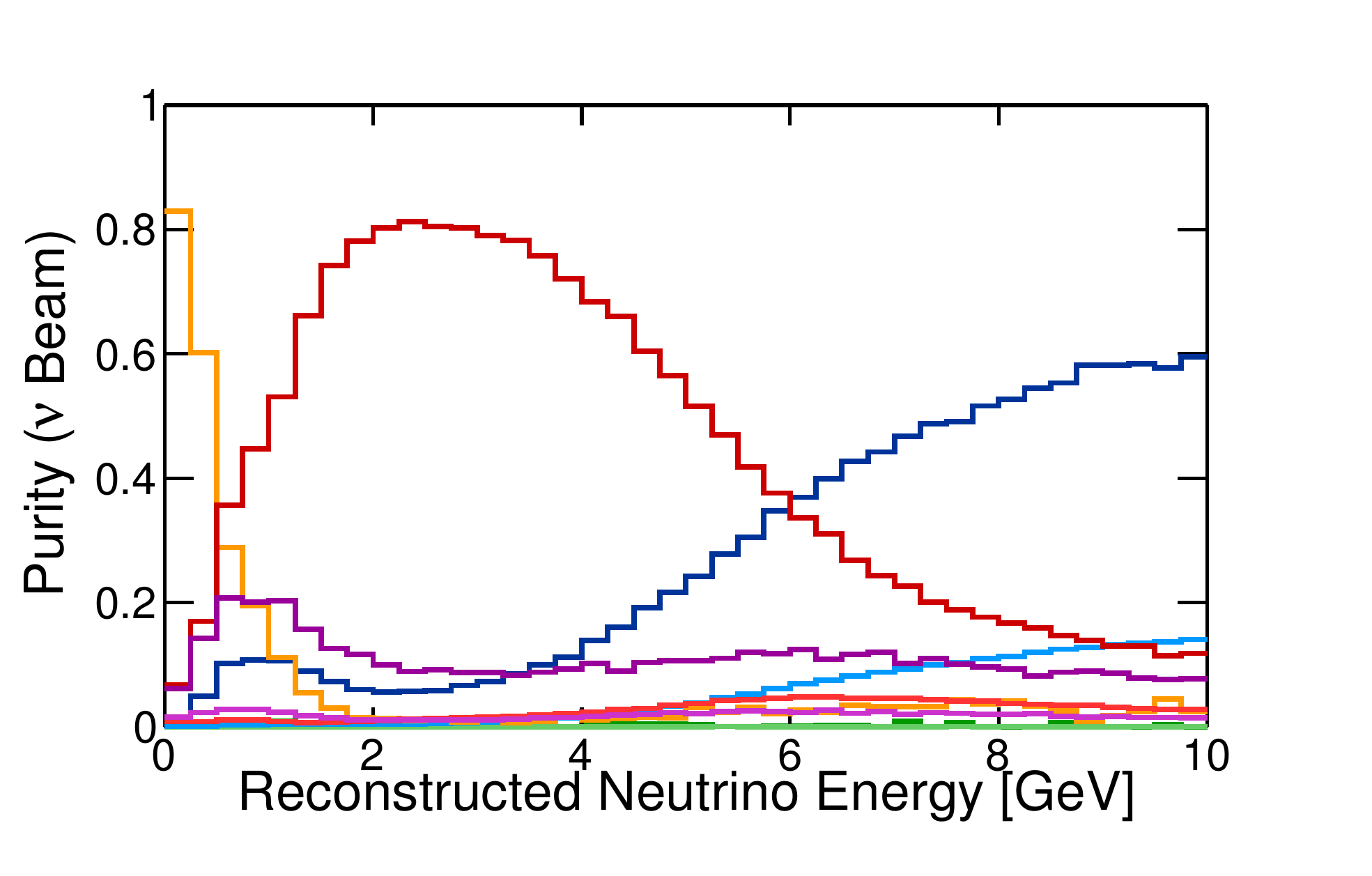}
\includegraphics[width=0.49\textwidth]{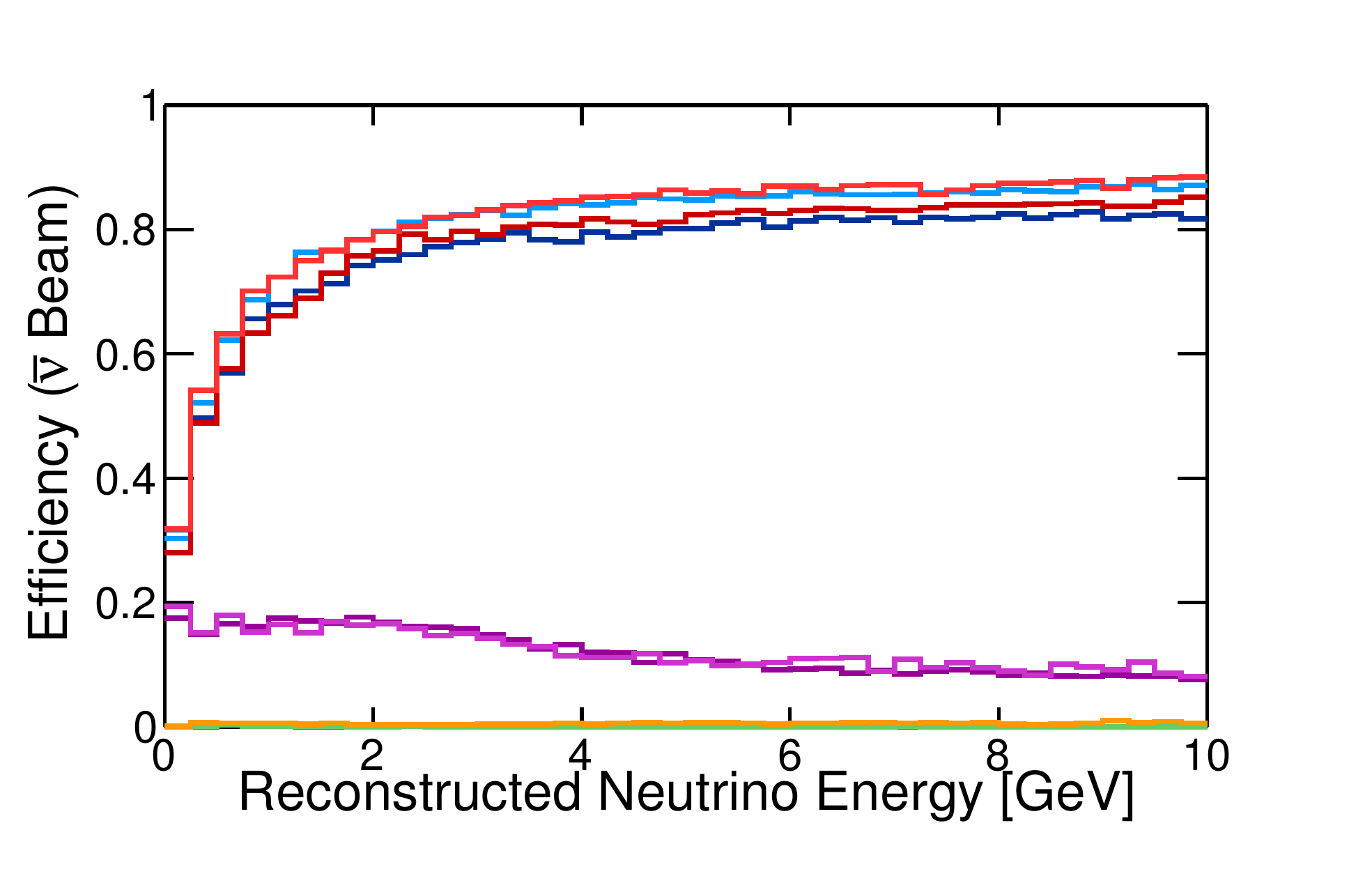}
\includegraphics[width=0.49\textwidth]{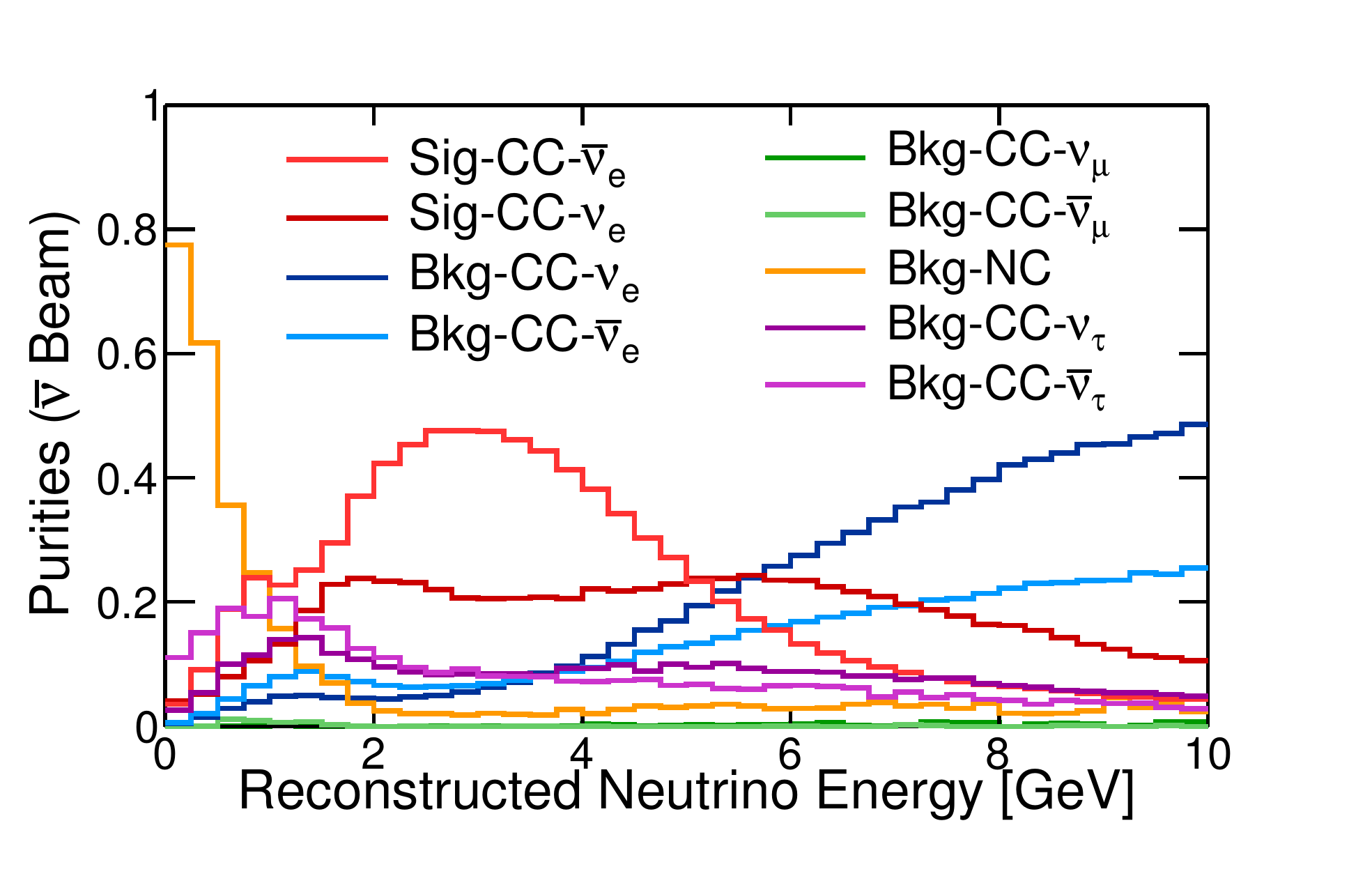}
\caption[Selection efficiency and sample purities for $\nu_e$ appearance in a LArTPC]{The
expected efficiencies and purities of selecting $\nu_e$ appearance
events in a LArTPC, obtained from the Fast MC.}
\label{fig:deteffapp}
\end{figure}

\begin{figure}[!htb]
\includegraphics[width=0.49\textwidth]{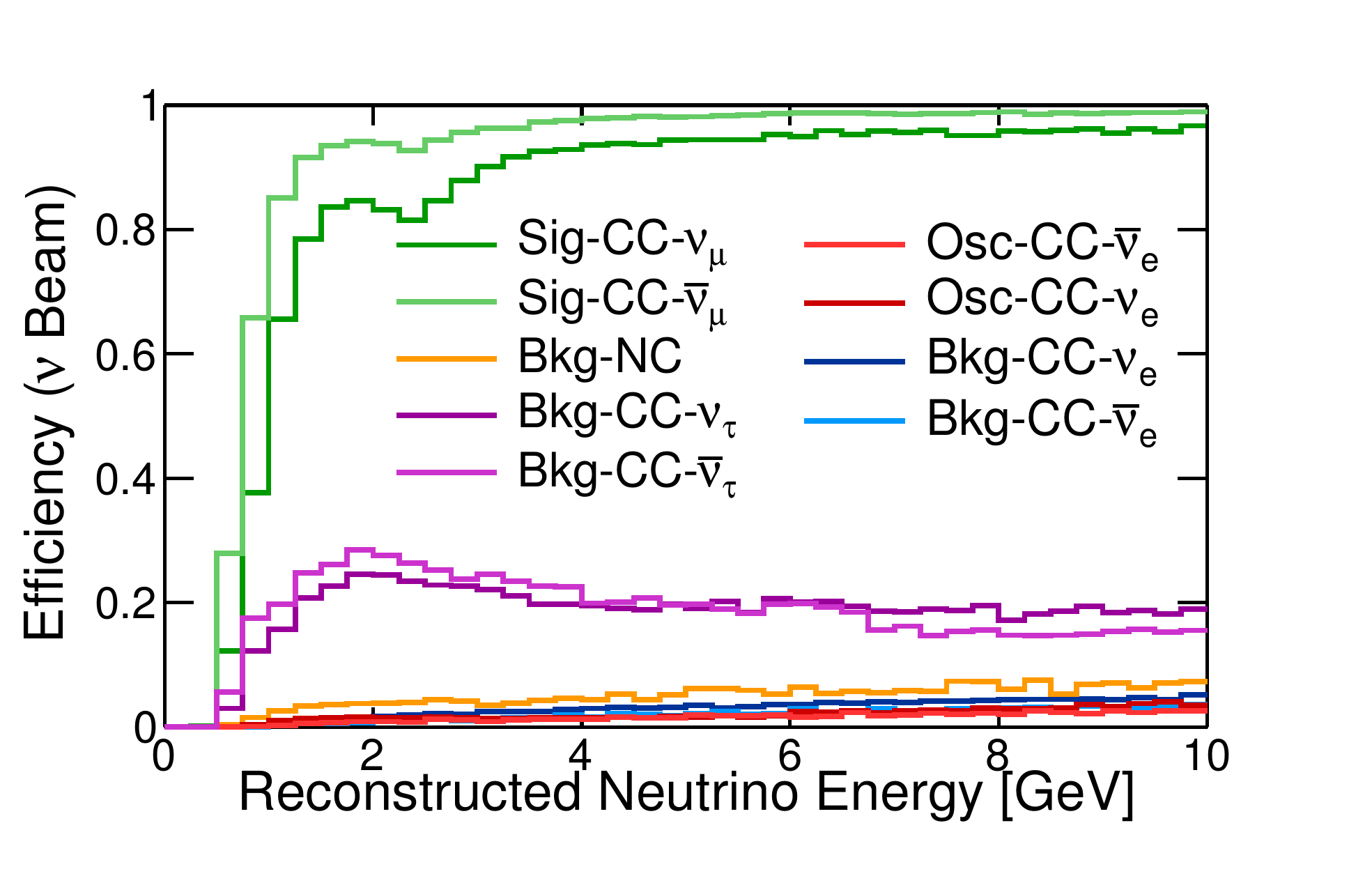}
\includegraphics[width=0.49\textwidth]{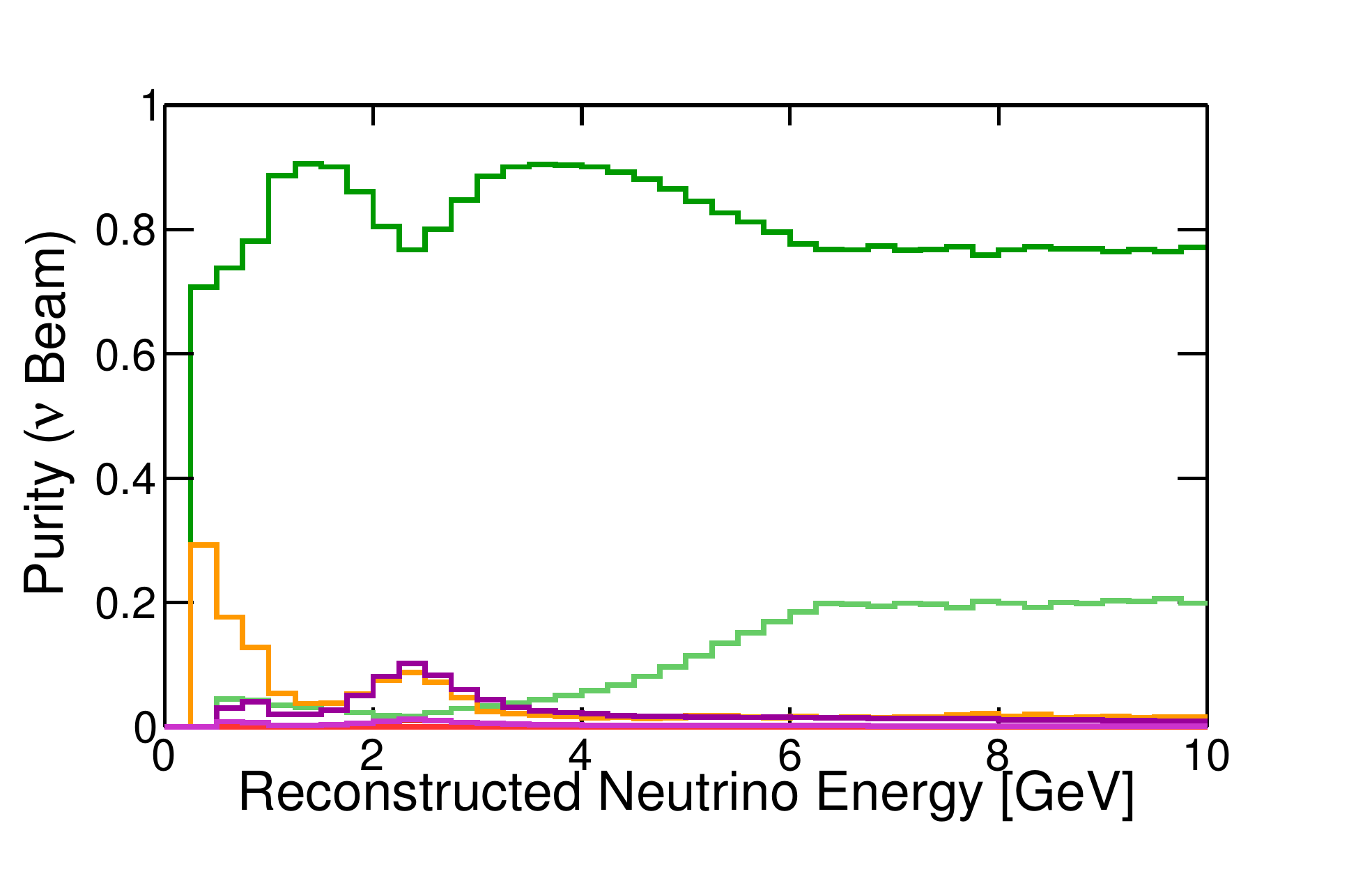}
\includegraphics[width=0.49\textwidth]{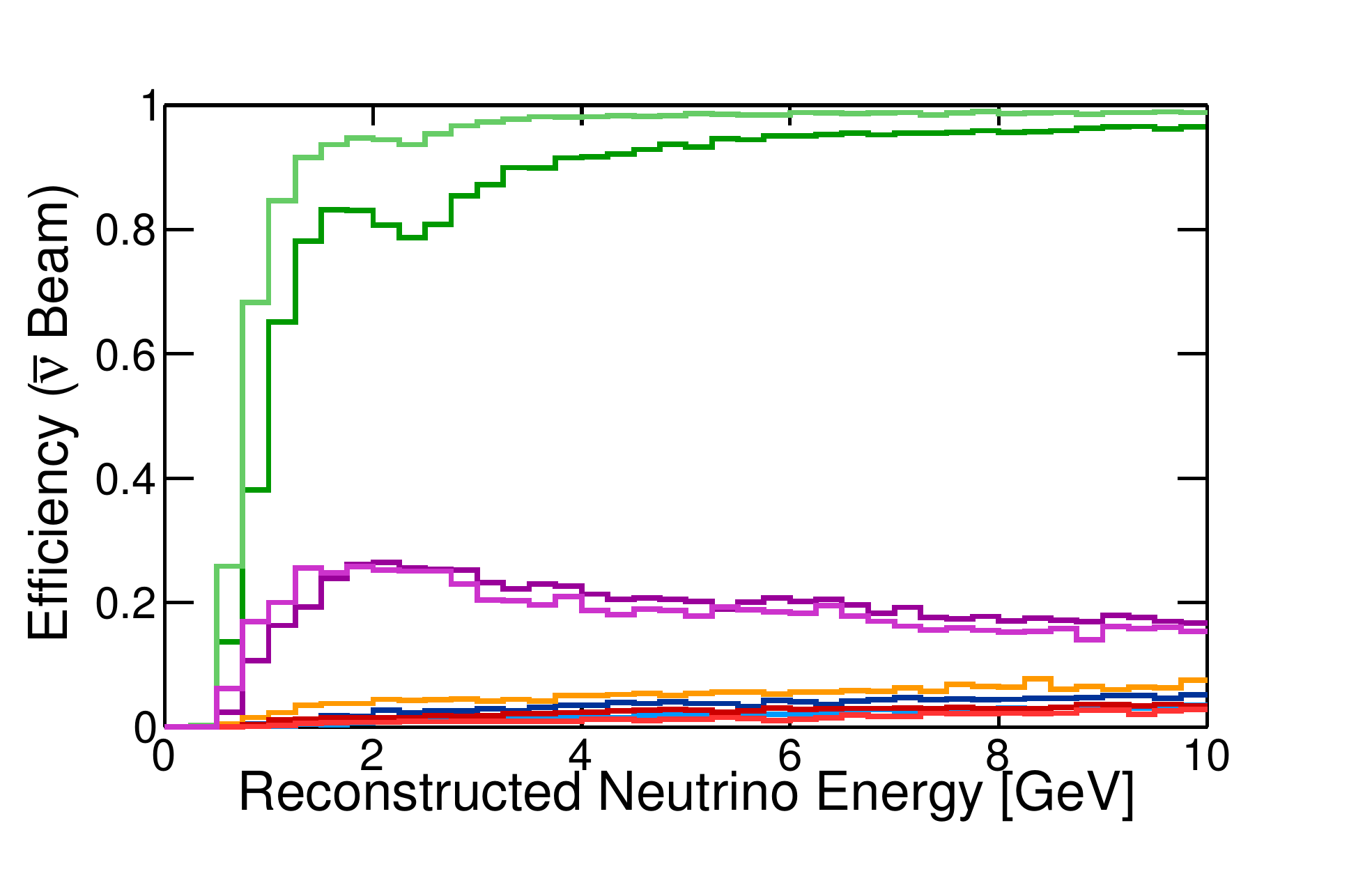}
\includegraphics[width=0.49\textwidth]{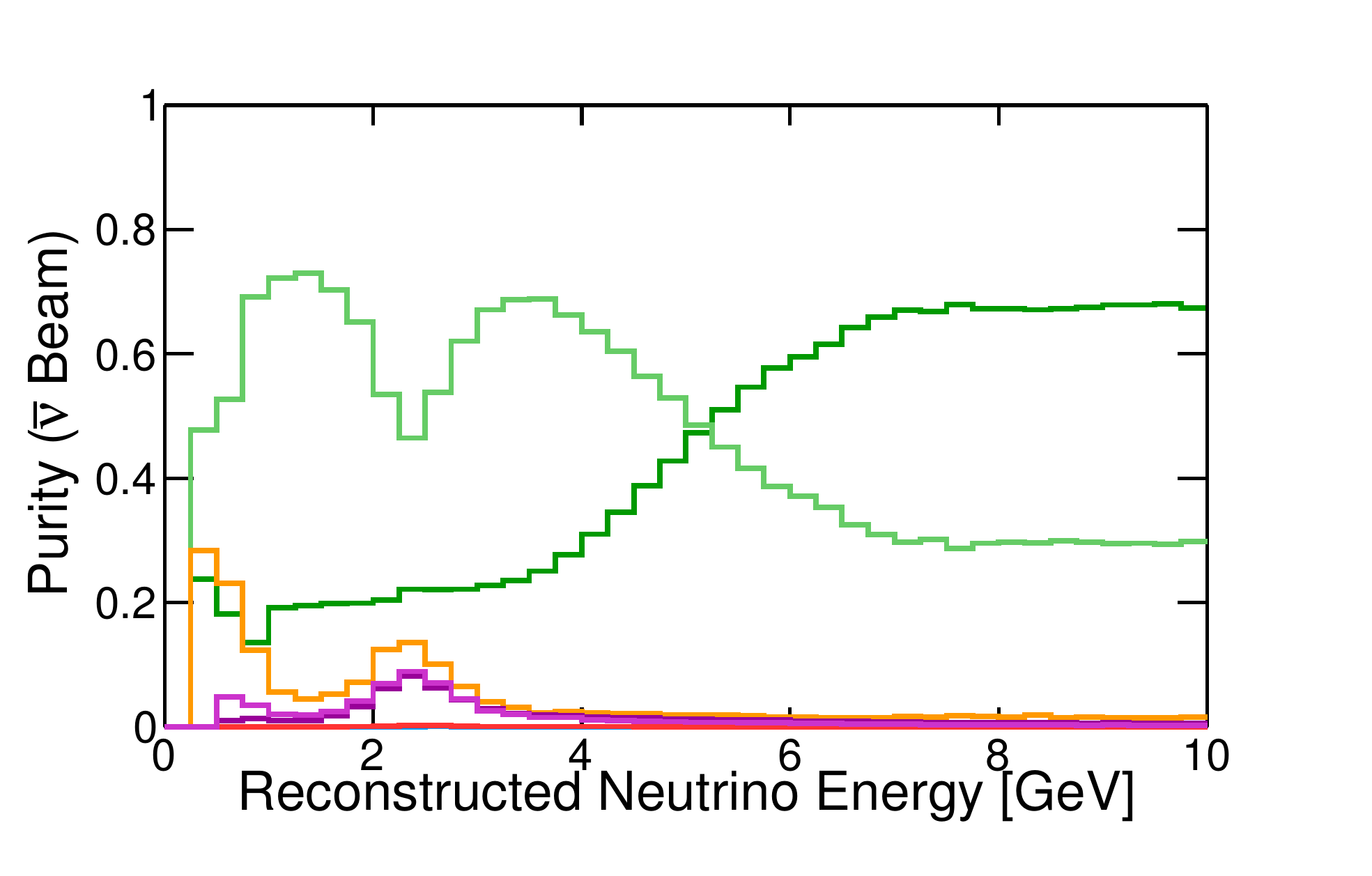}
\caption[Selection efficiency and sample purities for $\nu_\mu$ disappearance in a LArTPC]{The
expected efficiencies and purities of selecting $\nu_\mu$ disappearance 
events in a LArTPC obtained from the Fast MC.}
\label{fig:deteffdisap}
\end{figure}

The left-hand plots of Figures~\ref{fig:deteffapp}
and~\ref{fig:deteffdisap} show the acceptance (efficiency) of the
signal and the background for the Fast MC $\nu_e$ appearance
and $\nu_\mu$ disappearance selections, respectively. The effects of
the low-energy selection probabilities induce the observed low-energy 
fall off in the $\nu_e$ appearance sample. On the other hand,
the \SIadj{2.0}{\meter} track length requirement is mainly responsible 
for the low-energy behavior in the $\nu_\mu$ disappearance sample. The
corresponding plots on the right-hand side show the relative fraction
(purity) of the signal and each background sample for the Fast MC 
$\nu_e$ appearance and $\nu_\mu$ disappearance selections. The increased
wrong-sign contamination is evident in the $\overline{\nu}$ beam samples as
compared to the $\nu$ beam samples. No attempt has been made to reduce
the $\nu_\tau$ background in these plots.

The output of the Fast MC is a file containing the information one 
would expect from a full MC simulation. There are truth level quantities 
that describe the generated event, and reconstructed quantities that are 
calculated from simulated observables. The latter mimic the information that 
is expected to be available from reconstructing data or full simulation
and can be used in designing analyses 
aimed at measuring physics parameters. Analyses based on the simulated 
reconstruction produce event samples that can be used to estimate the 
sensitivity of LBNE to physics model parameters, specifically the parameters 
of the PMNS matrix, as a function of a variety of input parameters. Currently 
these studies are done using the GLoBES~\cite{Huber:2004ka} software package. 
However, instead of constructing the event-rate spectra as a function of true 
neutrino energy from predictions of the flux and neutrino-interaction cross sections, 
they are built event-by-event from the Fast MC. Similarly, smearing functions that
give the distribution of measured (reconstructed) neutrino energies as a 
function of the true neutrino energy are built event-by-event from the Fast MC, 
rather than estimated from external sources. 

In addition to the usual GLoBES inputs the Fast MC can provide
systematic uncertainty response functions, which encode the expected
changes to the energy spectra when input model parameters are varied
within their uncertainties. These response functions, along with an
augmented version of GLoBES, can be used to propagate realistic
systematic uncertainties in sensitivity studies.

The systematic uncertainty response functions are calculated from
weights stored in the Fast MC output files.
Each weight corresponds to the probability of producing the event with an alternate
physics model relative to the model used. 
Currently the Fast MC generates weights for parameters in interaction
models that can be reweighted in GENIE as well as a variety of
parameters related to the neutrino flux.
The flux parameters come in three varieties related to: 
changes to the beamline design, tolerances in the beamline design, and uncertainties 
in the physics models used in the simulations. The latter two contribute to systematic 
uncertainties while the first can be used to evaluate the impact of design optimizations. 

Propagation of systematic uncertainties through LBNE sensitivity studies using the Fast MC 
will require inclusion of new algorithms and improvements to existing reweighting algorithms. 
This includes (1) the introduction of new models into GENIE, (2) adding to and improving the 
reweighting functions currently in GENIE, (3) constructing flux files that correspond to 
the changes in the three aforementioned categories, (4) implementing a system for reweighting 
individual events based on changes to the models of hadronization from proton-target interactions,  
and (5) introducing detector parameterizations representing alternate detector designs, 
detector design tolerances, and model choices used in detector simulations.

The current focus of Fast MC studies is estimation of the effect of model uncertainties 
on sensitivity projections. This includes several steps, the first of which is to look 
at the changes in the analysis sample spectra induced by propagating individual systematic 
uncertainties. These studies are benchmarked by calculating the $\chi^{2}$ between the nominal 
and altered spectra. In the second step, sensitivities are calculated for combined fits of the four 
main analysis samples ($\nu_{\mu}/\overline{\nu}_{\mu}$ disappearance, $\nu_{e}/\overline{\nu}_{e}$ appearance). 
These studies must be done carefully to allow for realistic constraints of systematic 
uncertainties across analysis samples within GLoBES. Input covariance matrices can also be used to 
enforce external constraints on the relations between sources of systematic uncertainty. 
The results of these studies will inform the investigators as to which model uncertainties 
cause significant degradation of the sensitivities and therefore must be constrained by 
other methods. Methods to constrain these parameters will be sought from currently running 
experiments, proposed intermediate experiments, and from the LBNE beam monitoring and 
the LBNE near detector. Estimates of these constraints can then be propagated to sensitivity 
calculations to estimate the degree to which they mitigate the decline in sensitivity.
 
\begin{figure}[!htb]
\includegraphics[width=0.48\textwidth]{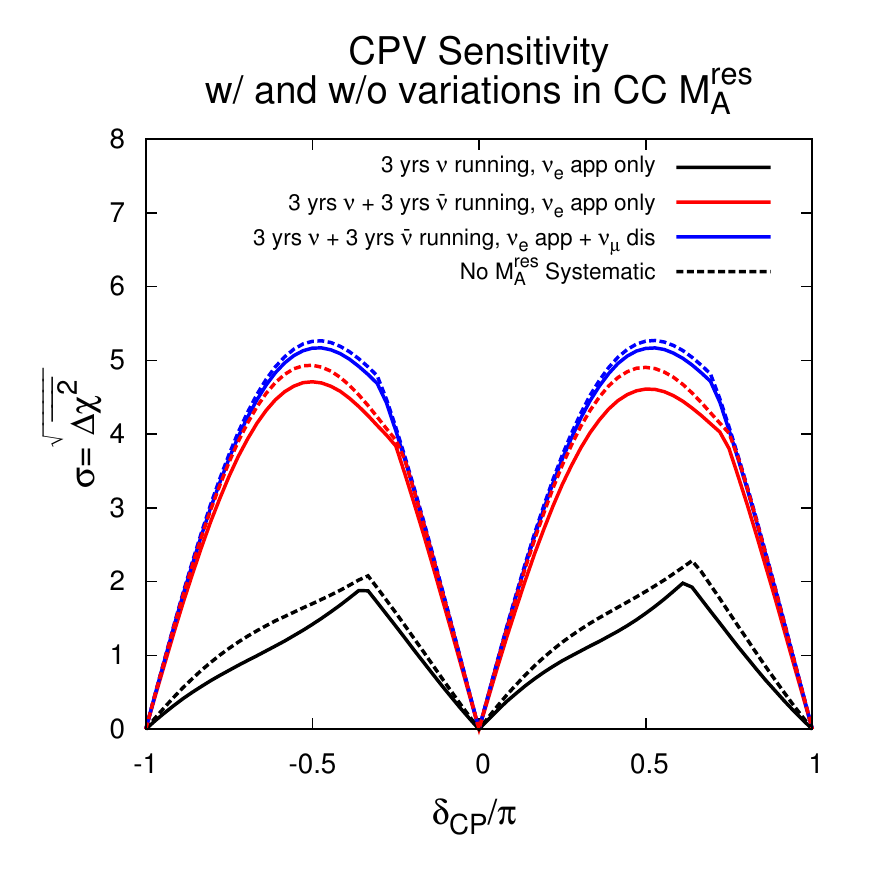}
\caption[CPV sensitivity with and without allowed variations in CC~$M_A^{res}$]
{The sensitivity to CP violation calculated using the
energy spectra generated by the Fast MC.  The sensitivities were
generated  with (solid) and without (dashed)  allowed variations in the CC~$M_A^{res}$
resonance production model parameter in GENIE. The
allowed variation degrades the sensitivity, however combined fits of
multiple analysis samples provide additional constraints and reduce
the impact.}  
\label{fig:CPVsensit_resCCma}
\end{figure}

Current studies focus on propagating uncertainties in flux and
GENIE model parameters via \linebreak reweighting techniques.  A example study shown in
Figure~\ref{fig:CPVsensit_resCCma} illustrates the effect of including the uncertainty on
CC~$M^{res}_{A}$ in the calculation of sensitivity to CP violation.
The sensitivity studies are performed for (1) a fit to the $\nu_{e}$ appearance
sample (three years of $\nu$-beam running), (2) a combined fit
of the $\nu_{e}$ appearance sample and the $\overline{\nu}_{e}$ appearance
sample (three years of $\nu$-beam plus three years of $\overline{\nu}$-beam running),
and (3) a combined fit of the $\nu_{e}/\overline{\nu}_{e}$ appearance
samples along with the corresponding $\nu_{\mu}/\overline{\nu}_{\mu}$ disappearance
samples. All three studies are done in two ways: with no allowance
for non-oscillation parameter systematic variation, and with allowed
15\% (width gaussian PDF) variations in CC~$M^{res}_{A}$.

As Figure~\ref{fig:CPVsensit_resCCma} shows, the inclusion of allowed variations in CC~$M_A^{res}$ degrades the sensitivity.
However, combined fits of multiple analysis samples provide additional constraints and reduce the impact.
The effect of these sample-to-sample constraints is dependent on the sample statistics, and the
curves in Figure~\ref{fig:CPVsensit_resCCma} include the statistical limitations on sample-to-sample constraints
from a six-year (three years $\nu$ + three years $\overline{\nu}$ running) exposure.
However, the software also allows for the inclusion of other possible limitations on sample-to-sample constraints related to the 
relative lack of experimental constraints on cross-section ratios (i.e., $\sigma_{\nu_{e}}/\sigma_{\nu_{\mu}}$, 
$\sigma_{\nu_{\tau}}/\sigma_{\nu_{\mu}}$, and $\sigma_{\overline{\nu}}/\sigma_{\nu}$), as well as theoretical considerations.

The preliminary Fast MC spectra shown in Figures~\ref{fig:detspecapp} and~\ref{fig:detspecdisap}
were generated with a different beam configuration than the ones shown in Figures~\ref{fig:lar-disapp-spectrum} 
and~\ref{fig:lar-event-spectrum}. 
Consequently, the sensitivities to CPV shown in Figure~\ref{fig:CPVsensit_resCCma} cannot be directly compared to
the corresponding figures in Section~\ref{sec:expt-sim}.
However, both the Fast MC and the methods discussed in Section~\ref{sec:expt-sim} have been used to generate comparable spectra
and to perform a series of sensitivity studies.
The two methods are consistent, except regarding known differences between the two simulations,
e.g., the inclusion of $\nu_{\tau}$-CC-induced
backgrounds. These differences are well understood, as are their impact on oscillation
parameter sensitivities.

Eventually the Fast MC seeks to incorporate near detector and atmospheric-neutrino analyses and directly 
perform combined fits with the long-baseline neutrino analysis samples. These studies will 
provide the most accurate estimate of the ultimate sensitivity of LBNE, and provide a template 
for future data analysis procedures.

\section{Simulation of Cosmic-Ray Background for a \SIadj{10}{\kt} Surface Detector}
\label{sec:cosmics}

A preliminary study of the  background events
expected from cosmic rays in the \SIadj{10}{\kt} far detector installed near the
surface at the \SURF is detailed in~\cite{COSMICBKGD}. The study simulated
cosmic-ray interactions in the far detector and focused on cosmic-ray
induced events from neutrons and muons that mimic electron-neutrino
interactions in the detector.  
These include electromagnetic cascades from knock-on electrons,
muon bremsstrahlung, and 
hadronic cascades with electromagnetic
components from photons and $\pi^0$'s. The background from decays of 
neutral hadrons 
into electrons such as $K^0_L \rightarrow \pi e \nu$ were also
studied. The energy of the cascades was required to be 
$>0.1$ GeV.  

These initial studies indicate that a combination of simple
kinematic and beam timing cuts will help to significantly reduce the
cosmic-ray background event rate in this far detector configuration.
In particular:
\begin{enumerate}
\item Only electromagnetic cascades with energies
greater than 0.25~GeV are considered background. For the neutrino
oscillation sensitivity calculations, only neutrino energies $\geq0.5$~GeV are considered.
\item $e^{\pm}$ background candidates are
tracked back to the parent muon; the distance between the muon
track and the point-of-closest-approach (PoCA) to the muon track is
required to be $>10$ cm.
\item The vertex of the $e^{\pm}$ shower is
required to be within the fiducial volume of the detector
(defined as 30~cm from the edge of the active detector volume). 
\item The $e^{\pm}$
cascade is required to be within a cone around the beam direction
(determined from the angular distribution of the beam signal
$e^{\pm}$ and the incoming neutrino beam).
\item It is assumed that EM
showers initiated by $\gamma$'s and $\pi^0 \rightarrow \gamma \gamma$
can be effectively distinguished from primary electron interactions
using particle ID techniques such as $dE/dX$.
\item Events are
timed with a precision of $\leq$\SI{1}{\micro\second} using the 
photon-detection system, which limits background to events occurring within
the \SI{10}{\micro\second} of the beam spill. 
\end{enumerate}

\begin{figure}[!htb]
\centerline{
\includegraphics[width=0.65\textwidth]{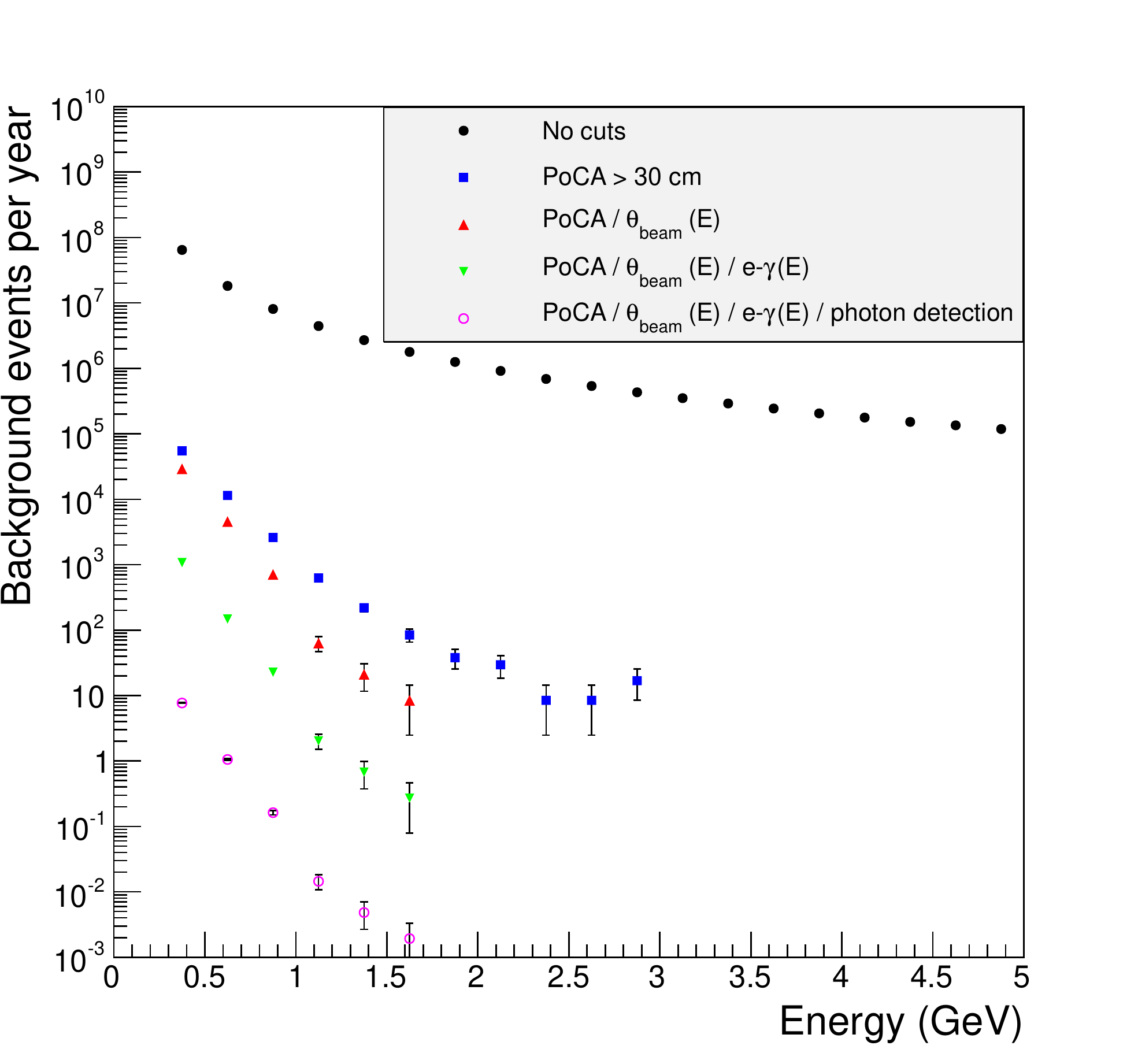}
}
\caption[Cosmic-ray background event distribution in the \SIadj{10}{\kt}
surface detector]{Energy spectra of muon-induced background events
for successively applied background rejection cuts.
 Simulations have
  been done for a muon spectrum at sea level. Correction for an
  altitude of 1,500~m above sea level has not been applied to the data.}
\label{fig:cosmic}
\end{figure}

The result of applying these selection criteria to the electromagnetic
showers initiated by cosmic rays is summarized in Table~\ref{tab:cosmic}
and Figure~\ref{fig:cosmic}. The background rates given in 
Table~\ref{tab:cosmic} include the recalculation for the cosmic flux at
1,500~m above sea level, which was not included in the previous
study~\cite{COSMICBKGD} (and is not included in
Figure~\ref{fig:cosmic}).  
 In the table, the initial background 
    event rate is calculated for one calendar year 
    assuming a  \SIadj{1.4}{\milli\second} drift time per beam pulse, a beam pulse every 1.33 seconds
    and \SI{2e7}{\second}/year of running. The expected event
    rate/yr after various selection criteria is applied from left to
    right in the table. The rates in all columns except the last are given for a time window of 
    1.4 ms, corresponding to the maximum electron drift time. The last column shows 
    the rate reduction assuming an efficient photon-detection system. 
    The first three rows show events with a muon in the detector where a PoCA cut (column 3) can be applied. 
    The row labeled `Missing $\mu$' shows events without a muon in the 
    detector; as there is no muon track, a PoCA cut can not be applied.
    The detector is assumed to be
    on the surface with three meters of rock overburden.
\begin{table}[!htb]
  \caption[Cosmic-ray-induced background in the surface \SIadj{10}{\kt} detector]{Cosmic-ray-induced background (at 1,500~m above sea level) 
    to the beam $\nu_e$-CC signal in the \SIadj{10}{\kt} detector. 
 }
\label{tab:cosmic}
\begin{tabular}{$l^c^c^c^c^c}
\toprule
\rowtitlestyle
Processes
& $E_e >$ \SI{0.25}{GeV} 
& PoCA $>$ \SI{10}{cm}
& Beam angle
& $\mathbf{e}\mathbf{/}\mathbf{\gamma}$ PID 
& Beam  timing \\  
\rowtitlestyle

& 
& and $\mathbf{D}$ $\mathbf{>}$ \SI{30}{cm}
& 
& 
& \\  \toprowrule

$\pi^0\rightarrow \gamma \rightarrow e^\pm$ & \num{2.2e6} &  \num{9.7e4} & \num{4.8e4} & \num{1.7e3}  & \num{12} \\ \colhline
$\mu \rightarrow \gamma \rightarrow e^\pm$& \num{7.1e6} & \num{12} & \num{0}  & \num{0}  &  $<$ \num{0.003} \\ \colhline
Ext $\gamma \rightarrow e^\pm$& \num{1.9e6} & \num{660}  & \num{340}   & \num{13}  & \num{0.1} \\ \colhline
$\pi^0, K^0 \rightarrow e^\pm$ & \num{1.4e6} & \num{810}    & \num{240}  & \num{240}  & \num{1.7} \\ \colhline
Missing $\mu$ & \num{1.3e6} & \num{1.8e3} & \num{580}   & \num{20}   & \num{0.1} \\ \colhline
Atm $n$ & \num{2.9e6} & \num{1.6e4} & \num{6.5e2} & \num{240}  & \num{1.7} \\ 
\toprule
\rowtitlestyle
Total & \num{1.1e7} & \num{1.2e5} & \num{5.6e4} & \num{2.2e3}  & \num{16} \\ 
\bottomrule
\end{tabular}
\end{table}
%

The dominant background is from $\pi^0 \rightarrow \gamma \rightarrow e^{\pm}$,
which contributes
12 out of the 16 total events per year and comes 
from $\pi^0$'s originating in cosmic
showers.  The study does not yet include specific $\pi^0$
reconstruction, only individual $e/\gamma$ separation. 
More sophisticated reconstruction techniques should further reduce the $\pi^0$
background.  The studies indicate that
application of these selection criteria coupled with a more detailed
background event reconstruction can potentially reduce the background
from cosmic rays to a few events per year --- mostly in the energy
region $<1$~GeV.

\clearpage

In Figure~\ref{fig:cosmic}, black-filled circles
  show events before any cuts are applied. The other point icons represent 
successively applied
cuts in the order listed below and in the figure's legend: 
\begin{enumerate}
\item Blue squares: PoCA to the muon track greater than 30~cm
\item Red triangles: angle with respect to the beam such that 99\% 
of signal events are retained
\item Green triangles: application of energy-dependent $e/\gamma$ 
discrimination 
\item Magenta open circles: application of efficient photon detection,
this allows the reduction of the time window from a maximum drift time
  of 1.4~ms down to a beam spill of \SI{10}{\micro\second} 
\end{enumerate}

%% file: appendix-dis-kinematics.tex
The following explanation of neutrino-nucleon scattering kinematics is adapted from~\cite{Accardi:2012qut}:

\tikzset{
lepton/.style={draw=red, postaction={decorate},
    decoration={markings,mark=at position .6 with {\arrow[red]{triangle 45}}}},
hadron/.style={draw=black, postaction={decorate},
    decoration={markings,mark=at position .6 with {\arrow[black]{triangle 45}}}},
noarrow/.style={draw=blue},
boson/.style={draw=orange,dashed},
}
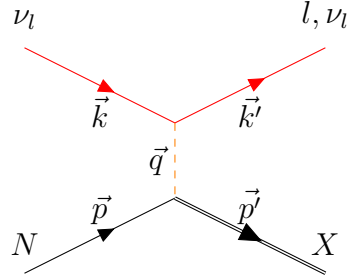
\begin{figure}[!htb]
\begin{center}

\begin{tikzpicture}[node distance=10mm and 20mm]
  \coordinate (qt);
  \coordinate[below=of qt] (qb);

  \coordinate[above left=of qt, label={[yshift=1mm]$\nu_l$}](nu);
  \coordinate[above right=of qt, label={[yshift=1mm]$l,\nu_l$}](lep);
  \coordinate[below left=of qb, label={[yshift=1mm]$N$}](N);
  \coordinate[below right=of qb, label={[yshift=1mm]$X$}](X);

  \draw[lepton] (nu) -- node[label=below:$\vec{k}$] {} (qt);
  \draw[lepton] (qt) -- node[label=below:$\vec{k'}$] {} (lep);
  \draw[boson] (qt) -- node[label=left:$\vec{q}$] {} (qb);
  \draw[hadron] (N) -- node[label=above:$\vec{p}$] {} (qb);
  \draw[hadron,double] (qb) -- node[label=above:$\vec{p'}$] {} (X);
\end{tikzpicture}%
\end{center}
\caption[Schematic of the neutrino-nucleon scattering]{
A schematic diagram of a neutrino-nucleon scattering process}
\label{fig:nuscatter}
\end{figure} 
The expression $\nu_l + N \longrightarrow
  l,\nu_l + X$ describes the scattering of a neutrino, $\nu_l$ off
a nucleon, $N$ as shown in Figure ~\ref{fig:nuscatter}. This interaction proceeds through the exchange  of a
$W^{\pm}$ or $Z^0$ boson, depending on whether it is a CC or NC
interaction, respectively. For the case of neutrino scattering, the
incoming lepton is a neutrino and the outgoing lepton is either a
neutrino (NC) or a charged lepton, $l$ (CC).  $X$ denotes the
resultant hadronic system.

The nucleon mass, $M$, is neglected where appropriate; the lepton
mass is neglected throughout. The following kinematic variables
describe the momenta and energies involved in the scattering process:
\begin{itemize}
\item $\vec{k}, \vec{k}^{\prime}$ are the four-momenta of the incoming and outgoing lepton. 
\item $\vec{p}$ is the initial four-momentum of the nucleon.
\item $E_\nu$ is the energy of the incoming neutrino.
\item  $E_N$ is the energy of the nucleon.
\end{itemize}

The Lorentz invariants are the following:
\begin{itemize}
\item The squared $\nu$+$N$ collision energy is $s= (|\vec{p} + \vec{k}|)^2
  = 4 E_N E_\nu$.
\item The squared momentum transfer to the lepton 
      $Q^2 = - q^2 = -(|\vec{k} - \vec{k}^{\prime}|)^2$ is
      equal to the virtuality of the exchanged boson. Large values of $Q^2$ 
      provide a hard scale to the process, which allows resolution of quarks 
      and gluons in the nucleon.
    \item The Bjorken variable $x_{Bj} = Q^2/(2 \vec{p} \cdot \vec{q})$ is often
      simply denoted by $x$. It determines the momentum fraction of
      the parton (quark or gluon) on which the boson scatters. Note
      that $0 < x < 1$ for $\nu$+$N$ collisions.
    \item The inelasticity $y = (\vec{q} \cdot \vec{p}) /(\vec{k}
      \cdot \vec{p})$ is limited to values $0 < y < 1$ and determines
      in particular the polarization of the virtual boson.  In the lab
      frame, the energy of the scattered lepton is $E_l = E_\nu (1 -
      y) + Q^2 /(4 E_\nu)$; detection of the scattered lepton thus
      typically requires a cut on $y < y_{max}$.
\end{itemize}

These invariants are related by $Q^2 = x y s$. The available phase
space is often represented in the plane of $x$ and $Q^2$. For a given
$\nu$+$N$ collision energy, lines of constant $y$ are then lines with a
slope of 45 degrees in a double logarithmic
$x-Q^2$ plot.

Two additional important variables are:
\begin{itemize}
\item The squared invariant mass of the produced hadronic system
  ($X$) is denoted by $W^2 = (|\vec{p} + \vec{q}|)^2 = Q^2 (1 -
  1/x)$. 
  Deep-inelastic scattering (DIS) is characterized by the Bjorken
  limit, where $Q^2$ and $W^2$ become large at a fixed value of $x$.
  Note: for a given $Q^2$, small $x$ corresponds to a high $W,Z$ - $N$
  collision energy.
\item The energy lost by the lepton (i.e., the energy carried away by
  the virtual boson) in the nucleon rest frame, is denoted $\nu = \vec{q} \cdot
  \vec{p} /M = y s /(2 M)$.
\end{itemize}

For scattering on a nucleus of atomic number $A$, the nucleon momentum
$\vec{p}$ would be replaced by $\vec{P}/A$ in the definitions, where
$\vec{P}$ is the momentum of the nucleus.  Note that the Bjorken
variable is then in the range $0 < x <A$.

%% file: acknowl.tex
This report is the result of an initial collaboration-wide effort to
prepare a whitepaper for the APS Division of Particles and Fields
Community Summer Study 2013~\cite{snowmass2013}. The paper has
evolved into LBNE's formal science document due to the hard work of
many LBNE Collaboration and Project members. We thank those colleagues
who made significant contributions and provided excellent feedback on
drafts of this document. The following is a nonexhaustive list of
LBNE collaborators who made major contributions to this document. A
major contribution is defined as $\geq$ a section and/or a study
reported in a figure prepared for this document.

\begin{center}
\begin{tabular}{$L^l}
\toprule
\rowtitlestyle
Contribution & LBNE Collaborator(s)  \\ \toprowrule
\bf{\emph{Overall editing}} & \bf{\emph{Anne Heavey, Mary Bishai, Jon Urheim, Brett Viren, Maury Goodman}} \\ \colhline
Technical/Figures & Brett Viren \\ \colhline
\emph{Original content} & \emph{Bob Wilson (Editor of the 2010 Interim Physics Report)} \\ \colhline 
General Guidance & Milind Diwan, Bob Wilson, Jim Strait, Sam Zeller, \\ 
& Bill Louis, Hank Sobel, Josh Klein, Nick Samios \\ 
\toprule
\rowcolor{\ChapterBubbleColor}
\multicolumn{2}{^>{\columncolor{\ChapterBubbleColor}}c}{\bf Major Contributors by Chapter. A $^{*}$ indicates the chapter editor(s)} \\ \toprowrule 
Chapter \ref{exec-sum-chap}& Jon Urheim$^{*}$, Jim Strait, Bob Wilson \\ \colhline
Chapter \ref{intro-chap} & Mary Bishai$^{*}$, Elizabeth Worcester$^{*}$, Jon Urheim, Kate Scholberg,\\ 
&  Ed Kearns, Bill Marciano, Xin Qian.  \\ \colhline
Chapter \ref{project-chap} & Mary Bishai$^{*}$, Jim Strait, Elaine McCluskey, Kevin Lesko, \\  
&  Jim Stewart, Vaia Papadimitriou,  Kevin Yarritu, Bill Louis \\ \colhline
Chapter~\ref{nu-oscil-chap} & Elizabeth Worcester$^{*}$, Mary Bishai$^{*}$, Matt Bass, Andy Blake, \\ 
& Xin Qian, Jon Urheim, Lisa Whitehead \\ \colhline
Chapter \ref{pdk-chap}& Ed Kearns$^{*}$, Jon Urheim$^{*}$,  Vitaly Kudryavtsev \\ \colhline
Chapter \ref{sn-chap} & Kate Scholberg$^{*}$, Vic Gehman, Alex Friedland \\ \colhline
Chapter \ref{nd-physics-chap} & Roberto Petti$^{*}$, Sanjib Mishra, Jaehoon Yu, Richard Van de Water \\ \colhline
Chapter \ref{chap-other-goals} & Michael Smy$^{*}$, Mary Bishai$^{*}$, Jaehoon Yu \\ \colhline
Chapter \ref{conclusion-chap} & Jon Urheim$^{*}$, Jim Strait \\ \colhline
Appendix \ref{app-sim} & Dan Cherdack$^{*}$, Tom Junk$^{*}$, Andy Blake,
Vitaly Kudryavtsev, \\ 
& Zepeng Li, Kate Scholberg\\ \colhline
Appendix \ref{app-dis} & Mary Bishai$^{*}$, Roberto Petti \\ \bottomrule
\end{tabular}
\end{center}

We would also like to express our gratitude to the following non-LBNE
collaborators who supplied us with invaluable information: {\bf Elke
  Aschenauer (Brookhaven Lab)}, the main editor of the Electron Ion
Collider (EIC) whitepaper~\cite{Accardi:2012qut} which inspired the
look and style of this document; {\bf Joachim Kopp (Max-Planck
  Institute, Heidelberg)} for his study of LBNE's sensitivity to
Non-Standard Interactions summarized in Section~\ref{sec:nsi}; {\bf
  Pilar Coloma (Virginia Tech)} for her studies comparing LBNE's
sensitivities to other proposed neutrino experiments shown in Figures~\ref{fig:cpvcomp} and \ref{fig:mhcomp}; {\bf Patrick Huber (Virginia
  Tech)} for a long and fruitful collaborative effort, and his
critical role in developing the case for a very long-baseline neutrino
oscillation experiment over the past decade; {\bf JJ Cherry (LANL)}
and {\bf Huaiyu Duan (U. of New Mexico)} for major input on the
supernova studies shown in Chapter~\ref{sn-chap}; {\bf Dmitry Gorbunov
  (Institute for Nuclear Research, Moscow)} for his studies on LBNE
sensitivities to $\nu$MSM heavy neutrinos shown in Figure~\ref{fig:heavynu}; {\bf Diana Brandonisio (Fermilab VMS)}, our
graphic designer for her gorgeous cover design and invaluable design
advice.

And last, but most importantly, our effusive thanks to {\bf Anne Heavey}
(AKA the \emph{\bf FIXME} monster) --- our devoted general editor
--- for her dogged insistence on clarity and quality, her hard work
well above and beyond the call of duty, and for being such an absolute
pleasure to work with.




This work was supported in part by the U.S. Department of Energy (DOE), the
National Science Foundation (NSF), the Sanford Underground Research Facility
and the South Dakota Science and Technology Authority (SDSTA); the Brazilian
Federal Agency for the Support and Evaluation of Graduate Education
(CAPES), the Sao Paulo Research Foundation (FAPESP) and the National
Council for Scientific and Technological Development (CNPq); the UK
Science and Technology Facilities Council (STFC); the Italian
government's Istituto Nazionale di Fisica Nucleare (INFN); the Indian
Department of Atomic Energy (DAE) and the Department of Science and
Technology (DST), Ministry of Science and Technology.

%% file: lbne-sci-opp-thebib.tex

\clearpage
\renewcommand{\headrulecolor}{CHAPACOL!70}
\renewcommand{\toccolor}{CHAPACOL!70}
\renewcommand{\ChapterTableColor}{CHAPACOL!100}
\renewcommand\ChapterTitleColor{CHAPACOL!30}
\renewcommand\ChapterBubbleColor{CHAPACOL!15}
\renewcommand\ChapterTabColor{CHAPACOL!30}
\renewcommand{\IntroBackgroundColor}{CHAPACOL!15}
\renewcommand{\IntroLineColor}{CHAPACOL!30}
\renewcommand\lbnechapterlabel{\ }
\pagestyle{\lbnerefstyle}
\chapter[\textcolor{\toccolor}{References}]{References}
\addtocounter{margintabsbump}{100}
\label{ref}
\begingroup
\renewcommand{\section}[2]{}
\renewcommand{\bibname}{}
\input{lbne-sci-opp-bib.bbl}
\endgroup

%% file: lbne-sci-opp-bib.bbl
\providecommand{\href}[2]{#2}\begingroup\raggedright\endgroup

%% file: lbne-sci-opp.bbl
\begin{thebibliography}{100}

\bibitem{snowmass2013}
``{APS Division of Particles and Fields Community Summer Study 2013},'' 2013.
\newblock \url{{http://www.snowmass2013.org}}.

\bibitem{Diwan:2000yb}
M.~Diwan and C.~Jung, ``{Next generation nucleon decay and neutrino detector.
  Proceedings, Workshop, NNN99, Stony Brook, USA, September 23-25, 1999},''
2000.

\bibitem{Marciano:2001tz}
W.~J. Marciano, ``{Extra long baseline neutrino oscillations and CP
  violation},'' BNL-HET-01-31,
  \href{http://arxiv.org/abs/hep-ph/0108181}{{\ttfamily arXiv:hep-ph/0108181
  [hep-ph]}},
2001.

\bibitem{Shrock:2003pb}
R.~Shrock, ``{Neutrinos and implications for physics beyond the standard model.
  Proceedings, Conference, Stony Brook, USA, October 11-13, 2002},''
2003.

\bibitem{Diwan:2002xc}
M.~Diwan, W.~Marciano, W.~Weng, D.~Beavis, M.~Brennan, {\em et~al.}, ``{Report
  of the BNL neutrino working group: Very long baseline neutrino oscillation
  experiment for precise determination of oscillation parameters and search for
  nu mu --> nu e appearance and CP violation},'' BNL-69395,
  \href{http://arxiv.org/abs/hep-ex/0211001}{{\ttfamily arXiv:hep-ex/0211001
  [hep-ex]}},
2002.

\bibitem{Diwan:2003bp}
M.~Diwan, D.~Beavis, M.-C. Chen, J.~Gallardo, S.~Kahn, {\em et~al.}, ``{Very
  long baseline neutrino oscillation experiments for precise measurements of
  mixing parameters and CP violating effects},''
  \href{http://dx.doi.org/10.1103/PhysRevD.68.012002}{{\em Phys.Rev.}
  {\bfseries D68} (2003) 012002},
\href{http://arxiv.org/abs/hep-ph/0303081}{{\ttfamily arXiv:hep-ph/0303081
  [hep-ph]}}.

\bibitem{Weng:2004zz}
W.~Weng, M.~Diwan, D.~Raparia, J.~Alessi, D.~Barton, {\em et~al.}, ``{The
  AGS-Based Super Neutrino Beam Facility Conceptual Design Report},''
  BNL-73210-2004-IR,
2004.

\bibitem{Diwan:2006qf}
M.~Diwan, S.~H. Kettell, L.~Littenberg, W.~Marciano, Z.~Parsa, {\em et~al.},
  ``{Proposal for an Experimental Program in Neutrino Physics and Proton Decay
  in the Homestake Laboratory},'' BNL-76798-2006-IR,
  \href{http://arxiv.org/abs/hep-ex/0608023}{{\ttfamily arXiv:hep-ex/0608023
  [hep-ex]}},
2006.

\bibitem{Barger:2007yw}
V.~Barger, M.~Bishai, D.~Bogert, C.~Bromberg, A.~Curioni, {\em et~al.},
  ``{Report of the US long baseline neutrino experiment study},''
  FERMILAB-0801-AD-E, BNL-77973-2007-IR,
  \href{http://arxiv.org/abs/0705.4396}{{\ttfamily arXiv:0705.4396 [hep-ph]}},
2007.

\bibitem{nusandbeyond}
N.~R.~C. Neutrino Facilities Assessment~Committee, {\em {Neutrinos and Beyond:
  New Windows on Nature}}.
\newblock The National Academies Press, 2003.
\newblock ISBN 0-309-08716-3.

\bibitem{potu}
{Interagency Working Group on the Physics of the Universe. National Science and
  Technology Council Committee on Science}, ``{A 21$^{st}$ Century Frontier of
  Discovery: The Physics of the Universe, a Strategic Plan for Federal Research
  at the Intersection of Physics and Astronomy}.''. February, 2004.
\newblock \url{{http://pcos.gsfc.nasa.gov/docs/Physics_of_the_Universe.pdf}}.

\bibitem{epp2010}
N.~R.~C. Committee on Elementary Particle Physics in the~21st Century, {\em
  {Revealing the Hidden Nature of Space and Time: Charting the Course for
  Elementary Particle Physics}}.
\newblock The National Academies Press, 2006.
\newblock ISBN 0-309-66039-4.

\bibitem{nusag}
{Neutrino Scientific Assessment Group}, ``{Recommendations to the Department of
  Energy and the National Science Foundation on a Future U.S. Program in
  Neutrino Oscillations. Report to the Nuclear Science Advisory Committee and
  the High Energy Physics Advisory Board}.''. July, 2007.
\newblock
  \url{{http://science.energy.gov/~/media/hep/pdf/files/pdfs/nusagfinalreportjuly13_2007.pdf}}.

\bibitem{p5report}
{Particle Physics Project Prioritization Panel}, ``{U.S. Particle Physics:
  Scientific opportunities, a plan for the next ten years}.''. May, 2008.
\newblock
  \url{{http://science.energy.gov/~/media/hep/pdf/files/pdfs/p5_report_06022008.pdf}}.

\bibitem{nasdusel}
{Ad Hoc Committee to Assess the Science Proposed for a Deep Underground Science
  and Engineering Laboratory (DUSEL); National Research Council}, {\em {An
  Assessment of the Deep Underground Science and Engineering Laboratory}}.
\newblock The National Academies Press, 2012.
\newblock ISBN 978-0-309-21723-1.

\bibitem{facilitiesreport}
{HEPAP Facilities Subpanel}, ``{Major High Energy Physics Facilities 2014-2024.
  Input to the prioritization of proposed scientific user facilities for the
  Office of Science}.''. March, 2013.
\newblock
  \url{{http://science.energy.gov/~/media/hep/hepap/pdf/Reports/HEPAP_facilities_letter_report.pdf}}.

\bibitem{europeanstrategy}
{CERN Council}, ``{The European Strategy for Particle Physics, Update 2013}.''.
  CERN-Council-S/106, May, 2013.
\newblock
  \url{http://council.web.cern.ch/council/en/EuropeanStrategy/esc-e-106.pdf}.

\bibitem{cdzero}
{DOE Office of Science, Office of High Energy Physics}, ``{Mission Need
  Statement for a Long-Baseline Neutrino Experiment (LBNE)},'' DOE,
\newblock
  \href{http://lbne2-docdb.fnal.gov/cgi-bin/ShowDocument?docid=6259}{{\ttfamily
  LBNE-doc-6259}}, 2009.

\bibitem{Kronfeld:2013uoa}
A.~S. Kronfeld, R.~S. Tschirhart, U.~Al-Binni, W.~Altmannshofer,
  C.~Ankenbrandt, {\em et~al.}, ``{Project X: Physics Opportunities},''
  FERMILAB-TM-2557, BNL-101116-2013-BC-81834, JLAB-ACP-13-1725,
  UASLP-IF-13-001, SLAC-R-1029, ANL-PHY-13-2, PNNL-22523, LBNL-6334E,
  \href{http://arxiv.org/abs/1306.5009}{{\ttfamily arXiv:1306.5009 [hep-ex]}},
2013.

\bibitem{deGouvea:2013onf}
A.~de~Gouvea {\em et~al.}, {\bfseries Intensity Frontier Neutrino Working
  Group} , ``{Neutrinos},'' FERMILAB-CONF-13-479-E,
  \href{http://arxiv.org/abs/1310.4340}{{\ttfamily arXiv:1310.4340 [hep-ex]}},
2013.

\bibitem{Babu:2013jba}
K.~Babu, E.~Kearns, U.~Al-Binni, S.~Banerjee, D.~Baxter, {\em et~al.},
  ``{Baryon Number Violation},''
  \href{http://arxiv.org/abs/1311.5285}{{\ttfamily arXiv:1311.5285 [hep-ph]}},
2013.

\bibitem{PIPII}
{Derwent, P. and others}, ``{Proton Improvement Plan II},''
\newblock
  {\href{http://projectx-docdb.fnal.gov/cgi-bin/ShowDocument?docid=1232}{Project
  X-doc-1232}}, November, 2013.

\bibitem{Holmes:2013vpa}
S.~Holmes, R.~Alber, B.~Chase, K.~Gollwitzer, D.~Johnson, {\em et~al.},
  ``{Project X: Accelerator Reference Design},'' FERMILAB-TM-2557,
  BNL-101116-2013-BC-81834, JLAB-ACP-13-1725, PNNL-22523, SLAC-R-1020,
  UASLP-IF-13-001, \href{http://arxiv.org/abs/1306.5022}{{\ttfamily
  arXiv:1306.5022 [physics.acc-ph]}},
2013.

\bibitem{mar2012review}
``{Final Report, Director's Independent Conceptual Design and CD-1 Readiness
  Review of the LBNE Project},''
\newblock
  \href{http://lbne2-docdb.fnal.gov/cgi-bin/ShowDocument?docid=5788}{{\ttfamily
  LBNE-doc-5788}}, March, 2012.

\bibitem{LBNEreconfig}
Y.~K. Kim {\em et~al.}, ``{LBNE Reconfiguration: Steering Committee Report},''
  2012.
\newblock
  \url{http://www.fnal.gov/directorate/lbne_reconfiguration/index.shtml}.

\bibitem{oct2012review}
``{Department of Energy Review Committee Report on the Technical, Cost,
  Schedule, and Management Review of the Long Baseline Neutrino Experiment
  (LBNE)},''
\newblock October, 2012.
\newblock
  \url{http://www.fnal.gov/directorate/OPMO/Projects/LBNE/DOERev/2012/10_30/1210_LBNE_rpt.pdf}.

\bibitem{nov2012review}
``{Independent Cost Review Closeout for the Long Baseline Neutrino Experiment
  (LBNE) Project},''
\newblock
  \href{http://lbne2-docdb.fnal.gov/cgi-bin/ShowDocument?docid=6522}{{\ttfamily
  LBNE-doc-6522}}, November, 2012.

\bibitem{dec2012approv}
``{Critical Decision 1 Approve Alternative Selection and Cost Range of the Long
  Baseline Neutrino Experiment (LBNE) Project},''
\newblock
  \href{http://lbne2-docdb.fnal.gov/cgi-bin/ShowDocument?docid=6681}{{\ttfamily
  LBNE-doc-6681}}, December, 2012.

\bibitem{CDRv1}
{\bfseries LBNE Project Management Team} , ``{LBNE Conceptual Design Report,
  Volume 1: The LBNE Project},''
\newblock
  \href{http://lbne2-docdb.fnal.gov/cgi-bin/ShowDocument?docid=5235}{{\ttfamily
  LBNE-doc-5235}}, 2012.

\bibitem{CDRv2}
{\bfseries LBNE Project Management Team} , ``{LBNE Conceptual Design Report,
  Volume 2: The Beamline at the Near Site},''
\newblock
  \href{http://lbne2-docdb.fnal.gov/cgi-bin/ShowDocument?docid=4317}{{\ttfamily
  LBNE-doc-4317}}, 2012.

\bibitem{CDRv3}
{\bfseries LBNE Project Management Team} , ``{LBNE Conceptual Design Report,
  Volume 3: Detectors at the Near Site},''
\newblock
  \href{http://lbne2-docdb.fnal.gov/cgi-bin/ShowDocument?docid=4724}{{\ttfamily
  LBNE-doc-4724}}, 2012.

\bibitem{CDRv4}
{\bfseries LBNE Project Management Team} , ``{LBNE Conceptual Design Report,
  Volume 4: The Liquid Argon Detector at the Far Site},''
\newblock
  \href{http://lbne2-docdb.fnal.gov/cgi-bin/ShowDocument?docid=4892}{{\ttfamily
  LBNE-doc-4892}}, 2012.

\bibitem{CDRv5}
{\bfseries LBNE Project Management Team} , ``{LBNE Conceptual Design Report,
  Volume 5: Conventional Facilities at the Near Site (MI-10 Shallow)},''
\newblock
  \href{http://lbne2-docdb.fnal.gov/cgi-bin/ShowDocument?docid=4623}{{\ttfamily
  LBNE-doc-4623}}, 2012.

\bibitem{CDRv6}
{\bfseries LBNE Project Management Team} , ``{LBNE Conceptual Design Report,
  Volume 6: Conventional Facilities at the Far Site},''
\newblock
  \href{http://lbne2-docdb.fnal.gov/cgi-bin/ShowDocument?docid=5017}{{\ttfamily
  LBNE-doc-5017}}, 2012.

\bibitem{wilson-nov2013}
R.~J. Wilson, ``{Long-Baseline Neutrino Experiment, presentation, November
  2013},'' 2013.
\newblock
  \url{{https://indico.fnal.gov/getFile.py/access?contribId=25&sessionId=7&resId=0&materialId=slides&confId=7485}}.

\bibitem{marxcommittee}
{Marx-Reichanadter Committee}, ``{Department of Energy Office of Science Review
  of Options for Underground Science}.''. June, 2011.
\newblock
  \url{{http://science.energy.gov/~/media/np/pdf/review_of_underground_science_report_final.pdf}}.

\bibitem{scicapreview}
{Grannis, P. and Green, D. and Nishikawa, K. and Robertson, H. and Sadoulet, B.
  and Wark, D.}, ``{The LBNE Science Capability Review},''
\newblock
  \href{http://lbne2-docdb.fnal.gov/cgi-bin/ShowDocument?docid=5333}{{\ttfamily
  LBNE-doc-5333}}, December, 2011.

\bibitem{Messier:2013sfa}
M.~Messier, {\bfseries NOvA Collaboration} , ``{Extending the NOvA Physics
  Program},'' FERMILAB-CONF-13-308-E,
  \href{http://arxiv.org/abs/1308.0106}{{\ttfamily arXiv:1308.0106 [hep-ex]}},
2013.

\bibitem{Huber:2010dx}
P.~Huber and J.~Kopp, ``{Two experiments for the price of one? -- The role of
  the second oscillation maximum in long baseline neutrino experiments},''
  \href{http://dx.doi.org/10.1007/JHEP03(2011)013,
  10.1007/JHEP05(2011)024}{{\em JHEP} {\bfseries 1103} (2011) 013},
\href{http://arxiv.org/abs/1010.3706}{{\ttfamily arXiv:1010.3706 [hep-ph]}}.

\bibitem{kearns_isoups}
E.~Kearns, ``{Future Experiments for Proton Decay. Presentation at ISOUPS
  (International Symposium: Opportunities in Underground Physics for Snowmass),
  Asilomar, May 2013},'' 2013.

\bibitem{DOCDB3056}
J.~Strait, ``{Physics Research Goals After Reconfiguration},''
\newblock
  \href{http://lbne2-docdb.fnal.gov/cgi-bin/ShowDocument?docid=3056}{{\ttfamily
  LBNE-doc-3056}}, 2011.

\bibitem{Mohapatra:2005wg}
R.~Mohapatra, S.~Antusch, K.~Babu, G.~Barenboim, M.-C. Chen, {\em et~al.},
  ``{Theory of neutrinos: A White paper},''
  \href{http://dx.doi.org/10.1088/0034-4885/70/11/R02}{{\em Rept.Prog.Phys.}
  {\bfseries 70} (2007) 1757--1867},
\href{http://arxiv.org/abs/hep-ph/0510213}{{\ttfamily arXiv:hep-ph/0510213
  [hep-ph]}}.

\bibitem{Aad:2012tfa}
G.~Aad {\em et~al.}, {\bfseries ATLAS Collaboration} , ``{Observation of a new
  particle in the search for the Standard Model Higgs boson with the ATLAS
  detector at the LHC},''
  \href{http://dx.doi.org/10.1016/j.physletb.2012.08.020}{{\em Phys.Lett.}
  {\bfseries B716} (2012) 1--29},
\href{http://arxiv.org/abs/1207.7214}{{\ttfamily arXiv:1207.7214 [hep-ex]}}.

\bibitem{Chatrchyan:2012ufa}
S.~Chatrchyan {\em et~al.}, {\bfseries CMS Collaboration} , ``{Observation of a
  new boson at a mass of 125 GeV with the CMS experiment at the LHC},''
  \href{http://dx.doi.org/10.1016/j.physletb.2012.08.021}{{\em Phys.Lett.}
  {\bfseries B716} (2012) 30--61},
\href{http://arxiv.org/abs/1207.7235}{{\ttfamily arXiv:1207.7235 [hep-ex]}}.

\bibitem{neutrinomatrix}
{DNP/DPF/DAP/DPB Joint Study on the Future of Neutrino Physics}, ``{The
  Neutrino Matrix}.''. November, 2004.
\newblock
  \url{{http://www.aps.org/policy/reports/multidivisional/neutrino/upload/main.pdf}}.

\bibitem{An:2012bu}
F.~An {\em et~al.}, {\bfseries Daya Bay Collaboration} , ``{Improved
  Measurement of Electron Antineutrino Disappearance at Daya Bay},''
  \href{http://dx.doi.org/10.1088/1674-1137/37/1/011001}{{\em Chin. Phys.}
  {\bfseries C37} (2013) 011001},
\href{http://arxiv.org/abs/1210.6327}{{\ttfamily arXiv:1210.6327 [hep-ex]}}.

\bibitem{Aguilar:2001ty}
A.~Aguilar-Arevalo {\em et~al.}, {\bfseries LSND Collaboration} , ``{Evidence
  for neutrino oscillations from the observation of anti-neutrino(electron)
  appearance in a anti-neutrino(muon) beam},''
  \href{http://dx.doi.org/10.1103/PhysRevD.64.112007}{{\em Phys.Rev.}
  {\bfseries D64} (2001) 112007},
\href{http://arxiv.org/abs/hep-ex/0104049}{{\ttfamily arXiv:hep-ex/0104049
  [hep-ex]}}.

\bibitem{AguilarArevalo:2007it}
A.~Aguilar-Arevalo {\em et~al.}, {\bfseries MiniBooNE Collaboration} , ``{A
  Search for electron neutrino appearance at the $\Delta m^{2} \sim 1$eV$^{2}$
  scale},'' \href{http://dx.doi.org/10.1103/PhysRevLett.98.231801}{{\em
  Phys.Rev.Lett.} {\bfseries 98} (2007) 231801},
\href{http://arxiv.org/abs/0704.1500}{{\ttfamily arXiv:0704.1500 [hep-ex]}}.

\bibitem{Aguilar-Arevalo:2013pmq}
A.~Aguilar-Arevalo {\em et~al.}, {\bfseries MiniBooNE Collaboration} ,
  ``{Improved Search for $\overline{\nu_\mu} \rightarrow \overline{ \nu_e}$
  Oscillations in the MiniBooNE Experiment},''
  \href{http://dx.doi.org/10.1103/PhysRevLett.110.161801}{{\em Phys.Rev.Lett.}
  {\bfseries 110} no.~16, (2013) 161801},
\href{http://arxiv.org/abs/1207.4809}{{\ttfamily arXiv:1207.4809 [hep-ex]}}.

\bibitem{Mention:2011rk}
G.~Mention, M.~Fechner, T.~Lasserre, T.~Mueller, D.~Lhuillier, {\em et~al.},
  ``{The Reactor Antineutrino Anomaly},''
  \href{http://dx.doi.org/10.1103/PhysRevD.83.073006}{{\em Phys.Rev.}
  {\bfseries D83} (2011) 073006},
\href{http://arxiv.org/abs/1101.2755}{{\ttfamily arXiv:1101.2755 [hep-ex]}}.

\bibitem{King:2014nza}
S.~F. King, A.~Merle, S.~Morisi, Y.~Shimizu, and M.~Tanimoto, ``{Neutrino Mass
  and Mixing: from Theory to Experiment},''
  \href{http://arxiv.org/abs/1402.4271}{{\ttfamily arXiv:1402.4271 [hep-ph]}},
2014.

\bibitem{Harrison:2002er}
P.~Harrison, D.~Perkins, and W.~Scott, ``{Tri-bimaximal mixing and the neutrino
  oscillation data},''
  \href{http://dx.doi.org/10.1016/S0370-2693(02)01336-9}{{\em Phys.Lett.}
  {\bfseries B530} (2002) 167},
\href{http://arxiv.org/abs/hep-ph/0202074}{{\ttfamily arXiv:hep-ph/0202074
  [hep-ph]}}.

\bibitem{Albright:2006cw}
C.~H. Albright and M.-C. Chen, ``{Model Predictions for Neutrino Oscillation
  Parameters},'' \href{http://dx.doi.org/10.1103/PhysRevD.74.113006}{{\em
  Phys.Rev.} {\bfseries D74} (2006) 113006},
\href{http://arxiv.org/abs/hep-ph/0608137}{{\ttfamily arXiv:hep-ph/0608137
  [hep-ph]}}.

\bibitem{Fogli:2012ua}
G.~Fogli, E.~Lisi, A.~Marrone, D.~Montanino, A.~Palazzo, {\em et~al.},
  ``{Global analysis of neutrino masses, mixings and phases: entering the era
  of leptonic CP violation searches},''
  \href{http://dx.doi.org/10.1103/PhysRevD.86.013012}{{\em Phys.Rev.}
  {\bfseries D86} (2012) 013012},
\href{http://arxiv.org/abs/1205.5254}{{\ttfamily arXiv:1205.5254 [hep-ph]}}.

\bibitem{Beringer:1900zz}
J.~Beringer {\em et~al.}, {\bfseries Particle Data Group} , ``{Review of
  Particle Physics (RPP)},''
\href{http://dx.doi.org/10.1103/PhysRevD.86.010001}{{\em Phys.Rev.} {\bfseries
  D86} (2012) 010001}.

\bibitem{Jarlskog:1985cw}
C.~Jarlskog, ``{A Basis Independent Formulation of the Connection Between Quark
  Mass Matrices, CP Violation and Experiment},''
\href{http://dx.doi.org/10.1007/BF01565198}{{\em Z.Phys.} {\bfseries C29}
  (1985) 491--497}.

\bibitem{Meroni:2012ty}
A.~Meroni, S.~Petcov, and M.~Spinrath, ``{A SUSY SU(5)xT' Unified Model of
  Flavour with large $\theta_{13}$},''
  \href{http://dx.doi.org/10.1103/PhysRevD.86.113003}{{\em Phys.Rev.}
  {\bfseries D86} (2012) 113003},
\href{http://arxiv.org/abs/1205.5241}{{\ttfamily arXiv:1205.5241 [hep-ph]}}.

\bibitem{Ding:2013bpa}
G.-J. Ding, S.~F. King, and A.~J. Stuart, ``{Generalised CP and $A_4$ Family
  Symmetry},'' \href{http://dx.doi.org/10.1007/JHEP12(2013)006}{{\em JHEP}
  {\bfseries 1312} (2013) 006},
\href{http://arxiv.org/abs/1307.4212}{{\ttfamily arXiv:1307.4212}}.

\bibitem{Luhn:2013lkn}
C.~Luhn, ``{Trimaximal TM$_{1}$ neutrino mixing in S$_{4}$ with spontaneous CP
  violation},'' \href{http://dx.doi.org/10.1016/j.nuclphysb.2013.07.003}{{\em
  Nucl.Phys.} {\bfseries B875} (2013) 80--100},
\href{http://arxiv.org/abs/1306.2358}{{\ttfamily arXiv:1306.2358 [hep-ph]}}.

\bibitem{Ding:2013nsa}
G.-J. Ding and Y.-L. Zhou, ``{Predicting Lepton Flavor Mixing from $\Delta(48)$
  and Generalized CP Symmetries},''
  \href{http://arxiv.org/abs/1312.5222}{{\ttfamily arXiv:1312.5222 [hep-ph]}},
2013.

\bibitem{Antusch:2013wn}
S.~Antusch, S.~F. King, and M.~Spinrath, ``{Spontaneous CP violation in $A_4
  \times SU(5)$ with Constrained Sequential Dominance 2},''
  \href{http://dx.doi.org/10.1103/PhysRevD.87.096018}{{\em Phys.Rev.}
  {\bfseries D87} no.~9, (2013) 096018},
\href{http://arxiv.org/abs/1301.6764}{{\ttfamily arXiv:1301.6764 [hep-ph]}}.

\bibitem{King:2013hoa}
S.~F. King, ``{A model of quark and lepton mixing},''
  \href{http://dx.doi.org/10.1007/JHEP01(2014)119}{{\em JHEP} {\bfseries 1401}
  (2014) 119},
\href{http://arxiv.org/abs/1311.3295}{{\ttfamily arXiv:1311.3295 [hep-ph]}}.

\bibitem{kolb94}
E.~Kolb and M.~Turner, {\em {The Early Universe}}.
\newblock Westview Press, 1994.
\newblock ISBN 978-0201626742.

\bibitem{weinberg08}
S.~Weinberg, {\em {Cosmology}}.
\newblock Oxford University Press, USA, first~ed., April, 2008.
\newblock ISBN 978-0198526827.

\bibitem{Steigman:2007xt}
G.~Steigman, ``{Primordial Nucleosynthesis in the Precision Cosmology Era},''
  \href{http://dx.doi.org/10.1146/annurev.nucl.56.080805.140437}{{\em
  Ann.Rev.Nucl.Part.Sci.} {\bfseries 57} (2007) 463--491},
\href{http://arxiv.org/abs/0712.1100}{{\ttfamily arXiv:0712.1100 [astro-ph]}}.

\bibitem{Fukugita:1986hr}
M.~Fukugita and T.~Yanagida, ``{Baryogenesis Without Grand Unification},''
\href{http://dx.doi.org/10.1016/0370-2693(86)91126-3}{{\em Phys.Lett.}
  {\bfseries B174} (1986) 45}.

\bibitem{Yanagida:1980xy}
T.~Yanagida, ``{Horizontal Symmetry and Masses of Neutrinos},''
\href{http://dx.doi.org/10.1143/PTP.64.1103}{{\em Prog.Theor.Phys.} {\bfseries
  64} (1980) 1103}.

\bibitem{Pascoli:2006ci}
S.~Pascoli, S.~Petcov, and A.~Riotto, ``{Leptogenesis and Low Energy CP
  Violation in Neutrino Physics},''
  \href{http://dx.doi.org/10.1016/j.nuclphysb.2007.02.019}{{\em Nucl.Phys.}
  {\bfseries B774} (2007) 1--52},
\href{http://arxiv.org/abs/hep-ph/0611338}{{\ttfamily arXiv:hep-ph/0611338
  [hep-ph]}}.

\bibitem{Capozzi:2013csa}
F.~Capozzi, G.~Fogli, E.~Lisi, A.~Marrone, D.~Montanino, {\em et~al.},
  ``{Status of three-neutrino oscillation parameters, circa 2013},''
  \href{http://arxiv.org/abs/1312.2878}{{\ttfamily arXiv:1312.2878 [hep-ph]}},
2013.

\bibitem{Adamson:2011ch}
P.~Adamson {\em et~al.}, {\bfseries MINOS Collaboration} , ``{Search for the
  disappearance of muon antineutrinos in the NuMI neutrino beam},''
  \href{http://dx.doi.org/10.1103/PhysRevD.84.071103}{{\em Phys.Rev.}
  {\bfseries D84} (2011) 071103},
\href{http://arxiv.org/abs/1108.1509}{{\ttfamily arXiv:1108.1509 [hep-ex]}}.

\bibitem{Mikheev:1986gs}
S.~Mikheev and A.~Y. Smirnov, ``{Resonance Amplification of Oscillations in
  Matter and Spectroscopy of Solar Neutrinos},''
{\em Sov.J.Nucl.Phys.} {\bfseries 42} (1985) 913--917.

\bibitem{Wolfenstein:1977ue}
L.~Wolfenstein, ``{Neutrino Oscillations in Matter},''
\href{http://dx.doi.org/10.1103/PhysRevD.17.2369}{{\em Phys.Rev.} {\bfseries
  D17} (1978) 2369--2374}.

\bibitem{Bellini:2008mr}
G.~Bellini {\em et~al.}, {\bfseries Borexino Collaboration} , ``{Measurement of
  the solar 8B neutrino rate with a liquid scintillator target and 3 MeV energy
  threshold in the Borexino detector},''
  \href{http://dx.doi.org/10.1103/PhysRevD.82.033006}{{\em Phys.Rev.}
  {\bfseries D82} (2010) 033006},
\href{http://arxiv.org/abs/0808.2868}{{\ttfamily arXiv:0808.2868 [astro-ph]}}.

\bibitem{Bellini:2011rx}
G.~Bellini, J.~Benziger, D.~Bick, S.~Bonetti, G.~Bonfini, {\em et~al.},
  ``{Precision measurement of the 7Be solar neutrino interaction rate in
  Borexino},'' \href{http://dx.doi.org/10.1103/PhysRevLett.107.141302}{{\em
  Phys.Rev.Lett.} {\bfseries 107} (2011) 141302},
\href{http://arxiv.org/abs/1104.1816}{{\ttfamily arXiv:1104.1816 [hep-ex]}}.

\bibitem{Aharmim:2011vm}
B.~Aharmim {\em et~al.}, {\bfseries SNO Collaboration} , ``{Combined Analysis
  of all Three Phases of Solar Neutrino Data from the Sudbury Neutrino
  Observatory},'' \href{http://dx.doi.org/10.1103/PhysRevC.88.025501}{{\em
  Phys.Rev.} {\bfseries C88} (2013) 025501},
\href{http://arxiv.org/abs/1109.0763}{{\ttfamily arXiv:1109.0763 [nucl-ex]}}.

\bibitem{Renshaw:2013}
A.~Renshaw {\em et~al.}, {\bfseries Super-Kamiokande Collaboration} , ``{First
  Indication of Terrestrial Matter Effects on Solar Neutrino Oscillation},''
  \href{http://arxiv.org/abs/1312.5176}{{\ttfamily arXiv:1312.5176 [hep-ex]}},
2013.

\bibitem{Freund:2001pn}
M.~Freund, ``{Analytic approximations for three neutrino oscillation parameters
  and probabilities in matter},''
  \href{http://dx.doi.org/10.1103/PhysRevD.64.053003}{{\em Phys.Rev.}
  {\bfseries D64} (2001) 053003},
\href{http://arxiv.org/abs/hep-ph/0103300}{{\ttfamily arXiv:hep-ph/0103300
  [hep-ph]}}.

\bibitem{Marciano:2006uc}
W.~Marciano and Z.~Parsa, ``{Intense neutrino beams and leptonic CP
  violation},'' \href{http://dx.doi.org/10.1016/j.nuclphysbps.2011.03.114}{{\em
  Nucl.Phys.Proc.Suppl.} {\bfseries 221} (2011) 166--172},
\href{http://arxiv.org/abs/hep-ph/0610258}{{\ttfamily arXiv:hep-ph/0610258
  [hep-ph]}}.

\bibitem{nuosc}
B.~Viren, ``{libnuosc++ - A library for calculating 3 neutrino oscillation
  probabilities.}.''
\newblock \url{{https://github.com/brettviren/nuosc}}.

\bibitem{PREM}
A.~M. Dziewonski and D.~L. Anderson, ``{Preliminary reference Earth model},''
  \href{http://dx.doi.org/10.1016/0031-9201(81)90046-7}{{\em Phys. Earth Plan.
  Int.} {\bfseries 25} (1981) 297}.

\bibitem{LBNEreconfigPWG}
J.~Appel {\em et~al.}, ``{Physics Working Group Report to the LBNE
  Reconfiguration Steering Committee},'' 2012.
\newblock
  \url{http://www.fnal.gov/directorate/lbne_reconfiguration/files/LBNE-Reconfiguration-PhysicsWG-Report-August2012.pdf}.

\bibitem{Brun:1987ma}
R.~Brun, F.~Bruyant, M.~Maire, A.~McPherson, and P.~Zanarini, ``{GEANT3},''
  CERN-DD-EE-84-1,
1987.

\bibitem{Bass:2013vcg}
M.~Bass {\em et~al.}, {\bfseries LBNE Collaboration} , ``{Baseline optimization
  for the measurement of CP violation and mass hierarchy in a long-baseline
  neutrino oscillation experiment},'' FERMILAB-PUB-13-506-E,
  \href{http://arxiv.org/abs/1311.0212}{{\ttfamily arXiv:1311.0212 [hep-ex]}},
2013.

\bibitem{Raidal:2004iw}
M.~Raidal, ``{Relation between the neutrino and quark mixing angles and grand
  unification},'' \href{http://dx.doi.org/10.1103/PhysRevLett.93.161801}{{\em
  Phys.Rev.Lett.} {\bfseries 93} (2004) 161801},
\href{http://arxiv.org/abs/hep-ph/0404046}{{\ttfamily arXiv:hep-ph/0404046
  [hep-ph]}}.

\bibitem{Minakata:2004xt}
H.~Minakata and A.~Y. Smirnov, ``{Neutrino mixing and quark-lepton
  complementarity},'' \href{http://dx.doi.org/10.1103/PhysRevD.70.073009}{{\em
  Phys.Rev.} {\bfseries D70} (2004) 073009},
\href{http://arxiv.org/abs/hep-ph/0405088}{{\ttfamily arXiv:hep-ph/0405088
  [hep-ph]}}.

\bibitem{Smirnov:2013uba}
A.~Y. Smirnov, ``{Neutrino mass, mixing and discrete symmetries},''
  \href{http://dx.doi.org/10.1088/1742-6596/447/1/012004}{{\em
  J.Phys.Conf.Ser.} {\bfseries 447} (2013) 012004},
\href{http://arxiv.org/abs/1305.4827}{{\ttfamily arXiv:1305.4827 [hep-ph]}}.

\bibitem{Harada:2013aja}
J.~Harada, ``{Non-maximal $\theta_{23}$, large $\theta_{13}$ and tri-bimaximal
  $\theta_{12}$ via quark-lepton complementarity at next-to-leading order},''
  \href{http://dx.doi.org/10.1209/0295-5075/103/21001}{{\em Europhys.Lett.}
  {\bfseries 103} (2013) 21001},
\href{http://arxiv.org/abs/1304.4526}{{\ttfamily arXiv:1304.4526 [hep-ph]}}.

\bibitem{Hu:2012eb}
B.~Hu, ``{Trimaximal-Cabibbo neutrino mixing: A parametrization in terms of
  deviations from tribimaximal mixing},''
  \href{http://dx.doi.org/10.1103/PhysRevD.87.053011}{{\em Phys.Rev.}
  {\bfseries D87} no.~5, (2013) 053011},
\href{http://arxiv.org/abs/1212.4079}{{\ttfamily arXiv:1212.4079 [hep-ph]}}.

\bibitem{ramond_isoups}
P.~Ramond, ``{Fundamental Physics Underground. Presentation at ISOUPS
  (International Symposium: Opportunities in Underground Physics for Snowmass),
  Asilomar, May 2013},'' 2013.

\bibitem{Antusch:2006vwa}
S.~Antusch, C.~Biggio, E.~Fernandez-Martinez, M.~Gavela, and J.~Lopez-Pavon,
  ``{Unitarity of the Leptonic Mixing Matrix},''
  \href{http://dx.doi.org/10.1088/1126-6708/2006/10/084}{{\em JHEP} {\bfseries
  0610} (2006) 084},
\href{http://arxiv.org/abs/hep-ph/0607020}{{\ttfamily arXiv:hep-ph/0607020
  [hep-ph]}}.

\bibitem{Qian:2013ora}
X.~Qian, C.~Zhang, M.~Diwan, and P.~Vogel, ``{Unitarity Tests of the Neutrino
  Mixing Matrix},'' \href{http://arxiv.org/abs/1308.5700}{{\ttfamily
  arXiv:1308.5700 [hep-ex]}},
2013.

\bibitem{Pati:1973rp}
J.~C. Pati and A.~Salam, ``{Is Baryon Number Conserved?},''
\href{http://dx.doi.org/10.1103/PhysRevLett.31.661}{{\em Phys.Rev.Lett.}
  {\bfseries 31} (1973) 661--664}.

\bibitem{Georgi:1974sy}
H.~Georgi and S.~Glashow, ``{Unity of All Elementary Particle Forces},''
\href{http://dx.doi.org/10.1103/PhysRevLett.32.438}{{\em Phys.Rev.Lett.}
  {\bfseries 32} (1974) 438--441}.

\bibitem{Dimopoulos:1981dw}
S.~Dimopoulos, S.~Raby, and F.~Wilczek, ``{Proton Decay in Supersymmetric
  Models},''
\href{http://dx.doi.org/10.1016/0370-2693(82)90313-6}{{\em Phys.Lett.}
  {\bfseries B112} (1982) 133}.

\bibitem{Langacker:1980js}
P.~Langacker, ``{Grand Unified Theories and Proton Decay},''
\href{http://dx.doi.org/10.1016/0370-1573(81)90059-4}{{\em Phys.Rept.}
  {\bfseries 72} (1981) 185}.

\bibitem{deBoer:1994dg}
W.~de~Boer, ``{Grand unified theories and supersymmetry in particle physics and
  cosmology},'' \href{http://dx.doi.org/10.1016/0146-6410(94)90045-0}{{\em
  Prog.Part.Nucl.Phys.} {\bfseries 33} (1994) 201--302},
\href{http://arxiv.org/abs/hep-ph/9402266}{{\ttfamily arXiv:hep-ph/9402266
  [hep-ph]}}.

\bibitem{Nath:2006ut}
P.~Nath and P.~Fileviez~Perez, ``{Proton stability in grand unified theories,
  in strings and in branes},''
  \href{http://dx.doi.org/10.1016/j.physrep.2007.02.010}{{\em Phys.Rept.}
  {\bfseries 441} (2007) 191--317},
\href{http://arxiv.org/abs/hep-ph/0601023}{{\ttfamily arXiv:hep-ph/0601023
  [hep-ph]}}.

\bibitem{Raby:2008pd}
S.~Raby, T.~Walker, K.~Babu, H.~Baer, A.~Balantekin, {\em et~al.}, ``{DUSEL
  Theory White Paper},'' SLAC-PUB-14734, FERMILAB-PUB-08-680-T,
  \href{http://arxiv.org/abs/0810.4551}{{\ttfamily arXiv:0810.4551 [hep-ph]}},
2008.

\bibitem{Senjanovic:2009kr}
G.~Senjanovic, ``{Proton decay and grand unification},''
  \href{http://dx.doi.org/10.1063/1.3327552}{{\em AIP Conf.Proc.} {\bfseries
  1200} (2010) 131--141},
\href{http://arxiv.org/abs/0912.5375}{{\ttfamily arXiv:0912.5375 [hep-ph]}}.

\bibitem{Li:2010dp}
T.~Li, D.~V. Nanopoulos, and J.~W. Walker, ``{Elements of F-ast Proton
  Decay},'' \href{http://dx.doi.org/10.1016/j.nuclphysb.2010.12.014}{{\em
  Nucl.Phys.} {\bfseries B846} (2011) 43--99},
\href{http://arxiv.org/abs/1003.2570}{{\ttfamily arXiv:1003.2570 [hep-ph]}}.

\bibitem{Noether:1918zz}
E.~Noether, ``{Invariant Variation Problems},''
  \href{http://dx.doi.org/10.1080/00411457108231446}{{\em Gott.Nachr.}
  {\bfseries 1918} (1918) 235--257},
\href{http://arxiv.org/abs/physics/0503066}{{\ttfamily arXiv:physics/0503066
  [physics]}}.

\bibitem{Nishino:2012ipa}
H.~Nishino {\em et~al.}, {\bfseries Super-Kamiokande Collaboration} , ``{Search
  for Nucleon Decay into Charged Anti-lepton plus Meson in Super-Kamiokande I
  and II},'' \href{http://dx.doi.org/10.1103/PhysRevD.85.112001}{{\em
  Phys.Rev.} {\bfseries D85} (2012) 112001},
\href{http://arxiv.org/abs/1203.4030}{{\ttfamily arXiv:1203.4030 [hep-ex]}}.

\bibitem{Bionta:1987qt}
R.~Bionta, G.~Blewitt, C.~Bratton, D.~Casper, A.~Ciocio, {\em et~al.},
  ``{Observation of a Neutrino Burst in Coincidence with Supernova SN 1987a in
  the Large Magellanic Cloud},''
\href{http://dx.doi.org/10.1103/PhysRevLett.58.1494}{{\em Phys.Rev.Lett.}
  {\bfseries 58} (1987) 1494}.

\bibitem{Hirata:1987hu}
K.~Hirata {\em et~al.}, {\bfseries KAMIOKANDE-II Collaboration} ,
  ``{Observation of a Neutrino Burst from the Supernova SN 1987a},''
\href{http://dx.doi.org/10.1103/PhysRevLett.58.1490}{{\em Phys.Rev.Lett.}
  {\bfseries 58} (1987) 1490--1493}.

\bibitem{Alekseev:1987ej}
E.~Alekseev, L.~Alekseeva, V.~Volchenko, and I.~Krivosheina, ``{Possible
  Detection of a Neutrino Signal on 23 February 1987 at the Baksan Underground
  Scintillation Telescope of the Institute of Nuclear Research},''
{\em JETP Lett.} {\bfseries 45} (1987) 589--592.

\bibitem{Scholberg:2007nu}
K.~Scholberg, ``{Supernova neutrino detection},''
  \href{http://dx.doi.org/10.1016/j.nuclphysbps.2011.09.012}{{\em
  Nucl.Phys.Proc.Suppl.} {\bfseries 221} (2011) 248--253},
\href{http://arxiv.org/abs/astro-ph/0701081}{{\ttfamily arXiv:astro-ph/0701081
  [astro-ph]}}.

\bibitem{Dighe:2008dq}
A.~Dighe, ``{Physics potential of future supernova neutrino observations},''
  \href{http://dx.doi.org/10.1088/1742-6596/136/2/022041}{{\em
  J.Phys.Conf.Ser.} {\bfseries 136} (2008) 022041},
\href{http://arxiv.org/abs/0809.2977}{{\ttfamily arXiv:0809.2977 [hep-ph]}}.

\bibitem{Tammann:1994ev}
G.~A. Tammann, W.~Loeffler, and A.~Schroder, ``{The Galactic supernova rate},''
\href{http://dx.doi.org/10.1086/192002}{{\em Astrophys. J. Suppl.} {\bfseries
  92} (1994) 487--493}.

\bibitem{Cappellaro:1999qy}
E.~Cappellaro, R.~Evans, and M.~Turatto, ``{A new determination of supernova
  rates and a comparison with indicators for galactic star formation},'' {\em
  Astron.Astrophys.} {\bfseries 351} (1999) 459,
\href{http://arxiv.org/abs/astro-ph/9904225}{{\ttfamily arXiv:astro-ph/9904225
  [astro-ph]}}.

\bibitem{Pagliaroli:2009qy}
G.~Pagliaroli, F.~Vissani, E.~Coccia, and W.~Fulgione, ``{Neutrinos from
  Supernovae as a Trigger for Gravitational Wave Search},''
  \href{http://dx.doi.org/10.1103/PhysRevLett.103.031102}{{\em Phys. Rev.
  Lett.} {\bfseries 103} (2009) 031102},
\href{http://arxiv.org/abs/0903.1191}{{\ttfamily arXiv:0903.1191 [hep-ph]}}.

\bibitem{Ott:2012jq}
C.~Ott, E.~O'Connor, S.~Gossan, E.~Abdikamalov, U.~Gamma, {\em et~al.},
  ``{Core-Collapse Supernovae, Neutrinos, and Gravitational Waves},''
  \href{http://dx.doi.org/10.1016/j.nuclphysbps.2013.04.036}{{\em
  Nucl.Phys.Proc.Suppl.} {\bfseries 235-236} (2013) 381--387},
\href{http://arxiv.org/abs/1212.4250}{{\ttfamily arXiv:1212.4250
  [astro-ph.HE]}}.

\bibitem{Mirizzi:2006xx}
A.~Mirizzi, G.~Raffelt, and P.~Serpico, ``{Earth matter effects in supernova
  neutrinos: Optimal detector locations},''
  \href{http://dx.doi.org/10.1088/1475-7516/2006/05/012}{{\em JCAP} {\bfseries
  0605} (2006) 012},
\href{http://arxiv.org/abs/astro-ph/0604300}{{\ttfamily arXiv:astro-ph/0604300
  [astro-ph]}}.

\bibitem{Choubey:2010up}
S.~Choubey, B.~Dasgupta, A.~Dighe, and A.~Mirizzi, ``{Signatures of collective
  and matter effects on supernova neutrinos at large detectors},''
  \href{http://arxiv.org/abs/1008.0308}{{\ttfamily arXiv:1008.0308 [hep-ph]}},
  2010.

\bibitem{Raffelt:1997ac}
G.~G. Raffelt, ``{Astrophysical axion bounds: An Update},''
  \href{http://arxiv.org/abs/astro-ph/9707268}{{\ttfamily
  arXiv:astro-ph/9707268 [astro-ph]}},
1997.

\bibitem{Hannestad:2001jv}
S.~Hannestad and G.~Raffelt, ``{New supernova limit on large extra
  dimensions},'' \href{http://dx.doi.org/10.1103/PhysRevLett.87.051301}{{\em
  Phys.Rev.Lett.} {\bfseries 87} (2001) 051301},
\href{http://arxiv.org/abs/hep-ph/0103201}{{\ttfamily arXiv:hep-ph/0103201
  [hep-ph]}}.

\bibitem{Antonioli:2004zb}
P.~Antonioli {\em et~al.}, ``Snews: The supernova early warning system,'' {\em
  New J. Phys.} {\bfseries 6} (2004) 114,
\href{http://arxiv.org/abs/astro-ph/0406214}{{\ttfamily astro-ph/0406214}}.

\bibitem{Scholberg:2008fa}
K.~Scholberg, ``{The SuperNova Early Warning System},'' {\em Astron. Nachr.}
  {\bfseries 329} (2008) 337--339,
\href{http://arxiv.org/abs/0803.0531}{{\ttfamily arXiv:0803.0531 [astro-ph]}}.

\bibitem{Scholberg:2010zz}
K.~Scholberg, ``{Future supernova neutrino detectors},''
\href{http://dx.doi.org/10.1088/1742-6596/203/1/012079}{{\em J.Phys.Conf.Ser.}
  {\bfseries 203} (2010) 012079}.

\bibitem{SURF}
``{Sanford Underground Research Facility}.''
\newblock \url{{http://www.sanfordlab.org}}.

\bibitem{Cleveland:1998nv}
B.~Cleveland, T.~Daily, R.~Davis~Jr., J.~R. Distel, K.~Lande, {\em et~al.},
  ``{Measurement of the solar electron neutrino flux with the Homestake
  chlorine detector},''
\href{http://dx.doi.org/10.1086/305343}{{\em Astrophys.J.} {\bfseries 496}
  (1998) 505--526}.

\bibitem{Gray:2010nc}
F.~Gray, C.~Ruybal, J.~Totushek, D.-M. Mei, K.~Thomas, {\em et~al.}, ``{Cosmic
  Ray Muon Flux at the Sanford Underground Laboratory at Homestake},''
  \href{http://dx.doi.org/10.1016/j.nima.2011.02.032}{{\em Nucl.Instrum.Meth.}
  {\bfseries A638} (2011) 63--66},
\href{http://arxiv.org/abs/1007.1921}{{\ttfamily arXiv:1007.1921 [nucl-ex]}}.

\bibitem{surfrock}
W.~Roggenthen and A.~Smith, ``{U, Th, K contents of materials associated with
  the Homestake DUSEL site, Lead, South Dakota},'' {\em {Private
  Communication}} .

\bibitem{Akerib:2013owa}
D.~Akerib {\em et~al.}, {\bfseries LUX Collaboration} , ``{First results from
  the LUX dark matter experiment at the Sanford Underground Research
  Facility},'' \href{http://arxiv.org/abs/1310.8214}{{\ttfamily arXiv:1310.8214
  [astro-ph.CO]}},
2013.

\bibitem{DOCDB3151}
M.~Bishai and Y.~Lu, ``{Conceptual Designs for a Wide-Band Low-Energy Neutrino
  Beam Target},''
\newblock
  \href{http://lbne2-docdb.fnal.gov/cgi-bin/ShowDocument?docid=3151}{{\ttfamily
  LBNE-doc-3151}}, November, 2010.

\bibitem{DOCDB8398}
B.~Lundberg, ``{A beginner guide to horn design and history of LBNE horn
  design},''
\newblock
  \href{http://lbne2-docdb.fnal.gov/cgi-bin/ShowDocument?docid=8398}{{\ttfamily
  LBNE-doc-8398}}, November, 2014.

\bibitem{Ayres:2007tu}
D.~Ayres {\em et~al.}, {\bfseries NOvA Collaboration} , ``{The NOvA Technical
  Design Report},'' FERMILAB-DESIGN-2007-01, 2007.
\url{http://lss.fnal.gov/archive/design/fermilab-design-2007-01.pdf}.

\bibitem{docdb-6599}
E.~Worcester, ``{Potential Sensitivity Improvements with 10 kT LBNE},''
\newblock
  \href{http://lbne2-docdb.fnal.gov/cgi-bin/ShowDocument?docid=6599}{{\ttfamily
  LBNE-doc-6599}}, 2012.

\bibitem{Mishra:2008nx}
S.~Mishra, R.~Petti, and C.~Rosenfeld, ``{A High Resolution Neutrino Experiment
  in a Magnetic Field for Project-X at Fermilab},'' {\em PoS} {\bfseries
  NUFACT08} (2008) 069,
\href{http://arxiv.org/abs/0812.4527}{{\ttfamily arXiv:0812.4527 [hep-ex]}}.

\bibitem{docdb-6704}
B.~Choudhary {\em et~al.}, {\bfseries Indian Institutions and Fermilab
  Collaboration} , ``{LBNE-India Detailed Project Report (DPR) submitted to
  DAE, India},''
\newblock
  \href{http://lbne2-docdb.fnal.gov/cgi-bin/ShowDocument?docid=6704}{{\ttfamily
  LBNE-doc-6704}}, 2012.

\bibitem{Huber:2004ka}
P.~Huber, M.~Lindner, and W.~Winter, ``{Simulation of long-baseline neutrino
  oscillation experiments with GLoBES (General Long Baseline Experiment
  Simulator)},'' \href{http://dx.doi.org/10.1016/j.cpc.2005.01.003}{{\em
  Comput.Phys.Commun.} {\bfseries 167} (2005) 195},
\href{http://arxiv.org/abs/hep-ph/0407333}{{\ttfamily arXiv:hep-ph/0407333
  [hep-ph]}}.

\bibitem{Huber:2007ji}
P.~Huber, J.~Kopp, M.~Lindner, M.~Rolinec, and W.~Winter, ``{New features in
  the simulation of neutrino oscillation experiments with GLoBES 3.0: General
  Long Baseline Experiment Simulator},''
  \href{http://dx.doi.org/10.1016/j.cpc.2007.05.004}{{\em Comput.Phys.Commun.}
  {\bfseries 177} (2007) 432--438},
\href{http://arxiv.org/abs/hep-ph/0701187}{{\ttfamily arXiv:hep-ph/0701187
  [hep-ph]}}.

\bibitem{Agostinelli:2002hh}
S.~Agostinelli {\em et~al.}, {\bfseries GEANT4} , ``{GEANT4: A simulation
  toolkit},''
\href{http://dx.doi.org/10.1016/S0168-9002(03)01368-8}{{\em Nucl. Instrum.
  Meth.} {\bfseries A506} (2003) 250--303}.

\bibitem{Andreopoulos:2009zz}
C.~Andreopoulos, {\bfseries GENIE Collaboration} , ``{The GENIE neutrino Monte
  Carlo generator},''
{\em Acta Phys.Polon.} {\bfseries B40} (2009) 2461--2475.

\bibitem{Abe:2011ks}
K.~Abe {\em et~al.}, {\bfseries T2K Collaboration} , ``{The T2K Experiment},''
  \href{http://dx.doi.org/10.1016/j.nima.2011.06.067}{{\em Nucl.Instrum.Meth.}
  {\bfseries A659} (2011) 106--135},
\href{http://arxiv.org/abs/1106.1238}{{\ttfamily arXiv:1106.1238
  [physics.ins-det]}}.

\bibitem{minos-numi-url}
{\bfseries {NuMI-MINOS}} .
\newblock \url{http://www-numi.fnal.gov/}.

\bibitem{Rubbia:2013zqa}
A.~Rubbia, ``{LAGUNA-LBNO: Design of an underground neutrino observatory
  coupled to long baseline neutrino beams from CERN},''
\href{http://dx.doi.org/10.1088/1742-6596/408/1/012006}{{\em J.Phys.Conf.Ser.}
  {\bfseries 408} (2013) 012006}.

\bibitem{Delahaye:2013jla}
J.-P. Delahaye, C.~Ankenbrandt, A.~Bogacz, S.~Brice, A.~Bross, {\em et~al.},
  ``{Enabling Intensity and Energy Frontier Science with a Muon Accelerator
  Facility in the U.S.: A White Paper Submitted to the 2013 U.S. Community
  Summer Study of the Division of Particles and Fields of the American Physical
  Society},'' FERMILAB-CONF-13-307-APC,
  \href{http://arxiv.org/abs/1308.0494}{{\ttfamily arXiv:1308.0494
  [physics.acc-ph]}},
2013.

\bibitem{Longhin:2012ae}
A.~Longhin, ``{Optimization of neutrino beams for underground sites in
  Europe},'' \href{http://arxiv.org/abs/1206.4294}{{\ttfamily arXiv:1206.4294
  [physics.ins-det]}},
2012.

\bibitem{Amoruso:2003sw}
S.~Amoruso {\em et~al.}, {\bfseries ICARUS Collaboration} , ``{Measurement of
  the mu decay spectrum with the ICARUS liquid argon TPC},''
  \href{http://dx.doi.org/10.1140/epjc/s2004-01597-7}{{\em Eur.Phys.J.}
  {\bfseries C33} (2004) 233--241},
\href{http://arxiv.org/abs/hep-ex/0311040}{{\ttfamily arXiv:hep-ex/0311040
  [hep-ex]}}.

\bibitem{T2K2kmprop}
{T2K Collaboration}, ``{A Proposal for a Detector 2km Away from the T2K
  Neutrino Source}.''. 2005.
\newblock
  \url{{http://www.phy.duke.edu/~cwalter/nusag-members/2km-proposal-05-05-30.pdf}}.

\bibitem{Ankowski:2006ts}
A.~Ankowski {\em et~al.}, {\bfseries ICARUS Collaboration} , ``{Measurement of
  through-going particle momentum by means of multiple scattering with the
  ICARUS T600 TPC},''
  \href{http://dx.doi.org/10.1140/epjc/s10052-006-0051-3}{{\em Eur.Phys.J.}
  {\bfseries C48} (2006) 667--676},
\href{http://arxiv.org/abs/hep-ex/0606006}{{\ttfamily arXiv:hep-ex/0606006
  [hep-ex]}}.

\bibitem{An:2013zwz}
F.~An {\em et~al.}, {\bfseries Daya Bay Collaboration} , ``{Spectral
  measurement of electron antineutrino oscillation amplitude and frequency at
  Daya Bay},'' \href{http://arxiv.org/abs/1310.6732}{{\ttfamily arXiv:1310.6732
  [hep-ex]}},
2013.

\bibitem{Adamson:2013ue}
P.~Adamson {\em et~al.}, {\bfseries MINOS Collaboration} , ``{Electron neutrino
  and antineutrino appearance in the full MINOS data sample},''
  \href{http://dx.doi.org/10.1103/PhysRevLett.110.171801}{{\em Phys.Rev.Lett.}
  {\bfseries 110} no.~17, (2013) 171801},
\href{http://arxiv.org/abs/1301.4581}{{\ttfamily arXiv:1301.4581 [hep-ex]}}.

\bibitem{Murtagh:1987xu}
M.~J. Murtagh, {\bfseries E734 Collaboration} , ``{A Search for muon-neutrino
  to electron-neutrino oscillations using the E734 detector},'' BNL-39667,
1987.

\bibitem{Seto:1988vg}
R.~Seto, ``{BNL E776: A Search for neutrino oscillations},''
\href{http://dx.doi.org/10.1063/1.37645}{{\em AIP Conf.Proc.} {\bfseries 176}
  (1988) 957--963}.

\bibitem{Borodovsky:1992pn}
L.~Borodovsky, C.~Chi, Y.~Ho, N.~Kondakis, W.-Y. Lee, {\em et~al.}, ``{Search
  for muon-neutrino oscillations muon-neutrino to electron-neutrino
  (anti-muon-neutrino to anti-electron-neutrino in a wide band neutrino
  beam},''
\href{http://dx.doi.org/10.1103/PhysRevLett.68.274}{{\em Phys.Rev.Lett.}
  {\bfseries 68} (1992) 274--277}.

\bibitem{Astier:2003gs}
P.~Astier {\em et~al.}, {\bfseries NOMAD Collaboration} , ``{Search for nu(mu)
  --> nu(e) oscillations in the NOMAD experiment},''
  \href{http://dx.doi.org/10.1016/j.physletb.2003.07.029}{{\em Phys.Lett.}
  {\bfseries B570} (2003) 19--31},
\href{http://arxiv.org/abs/hep-ex/0306037}{{\ttfamily arXiv:hep-ex/0306037
  [hep-ex]}}.

\bibitem{AguilarArevalo:2008rc}
A.~Aguilar-Arevalo {\em et~al.}, {\bfseries MiniBooNE Collaboration} ,
  ``{Unexplained Excess of Electron-Like Events From a 1-GeV Neutrino Beam},''
  \href{http://dx.doi.org/10.1103/PhysRevLett.102.101802}{{\em Phys.Rev.Lett.}
  {\bfseries 102} (2009) 101802},
\href{http://arxiv.org/abs/0812.2243}{{\ttfamily arXiv:0812.2243 [hep-ex]}}.

\bibitem{Abe:2013hdq}
K.~Abe {\em et~al.}, {\bfseries T2K Collaboration} , ``{Observation of Electron
  Neutrino Appearance in a Muon Neutrino Beam},''
  \href{http://arxiv.org/abs/1311.4750}{{\ttfamily arXiv:1311.4750 [hep-ex]}},
2013.

\bibitem{Qian:2012zn}
X.~Qian, A.~Tan, W.~Wang, J.~Ling, R.~McKeown, {\em et~al.}, ``{Statistical
  Evaluation of Experimental Determinations of Neutrino Mass Hierarchy},''
  \href{http://dx.doi.org/10.1103/PhysRevD.86.113011}{{\em Phys.Rev.}
  {\bfseries D86} (2012) 113011},
\href{http://arxiv.org/abs/1210.3651}{{\ttfamily arXiv:1210.3651 [hep-ph]}}.

\bibitem{Blennow:2013oma}
M.~Blennow, P.~Coloma, P.~Huber, and T.~Schwetz, ``{Quantifying the sensitivity
  of oscillation experiments to the neutrino mass ordering},''
  \href{http://arxiv.org/abs/1311.1822}{{\ttfamily arXiv:1311.1822 [hep-ph]}},
2013.

\bibitem{Cousins:priv2013}
R.~Cousins, ``Private communication,'' 2013.

\bibitem{Cousins:2005pq}
R.~Cousins, J.~Mumford, J.~Tucker, and V.~Valuev, ``{Spin discrimination of new
  heavy resonances at the LHC},''
\href{http://dx.doi.org/10.1088/1126-6708/2005/11/046}{{\em JHEP} {\bfseries
  0511} (2005) 046}.

\bibitem{Adamson:2011qu}
P.~Adamson {\em et~al.}, {\bfseries MINOS Collaboration} , ``{Improved search
  for muon-neutrino to electron-neutrino oscillations in MINOS},''
  \href{http://dx.doi.org/10.1103/PhysRevLett.107.181802}{{\em Phys.Rev.Lett.}
  {\bfseries 107} (2011) 181802},
\href{http://arxiv.org/abs/1108.0015}{{\ttfamily arXiv:1108.0015 [hep-ex]}}.

\bibitem{Adamson:2009ju}
P.~Adamson {\em et~al.}, {\bfseries MINOS Collaboration} , ``{Neutrino and
  Antineutrino Inclusive Charged-current Cross Section Measurements with the
  MINOS Near Detector},''
  \href{http://dx.doi.org/10.1103/PhysRevD.81.072002}{{\em Phys.Rev.}
  {\bfseries D81} (2010) 072002},
\href{http://arxiv.org/abs/0910.2201}{{\ttfamily arXiv:0910.2201 [hep-ex]}}.

\bibitem{Wu:2007ab}
Q.~Wu {\em et~al.}, {\bfseries NOMAD Collaboration} , ``{A Precise measurement
  of the muon neutrino-nucleon inclusive charged current cross-section off an
  isoscalar target in the energy range 2.5 < E(nu) < 40-GeV by NOMAD},''
  \href{http://dx.doi.org/10.1016/j.physletb.2007.12.027}{{\em Phys.Lett.}
  {\bfseries B660} (2008) 19--25},
\href{http://arxiv.org/abs/0711.1183}{{\ttfamily arXiv:0711.1183 [hep-ex]}}.

\bibitem{Lyubushkin:2008pe}
V.~Lyubushkin {\em et~al.}, {\bfseries NOMAD Collaboration} , ``{A Study of
  quasi-elastic muon neutrino and antineutrino scattering in the NOMAD
  experiment},'' \href{http://dx.doi.org/10.1140/epjc/s10052-009-1113-0}{{\em
  Eur.Phys.J.} {\bfseries C63} (2009) 355--381},
\href{http://arxiv.org/abs/0812.4543}{{\ttfamily arXiv:0812.4543 [hep-ex]}}.

\bibitem{Bodek:2012cm}
A.~Bodek, U.~Sarica, K.~Kuzmin, and V.~Naumov, ``{Extraction of Neutrino Flux
  with the Low $\nu$ Method at MiniBooNE Energies},''
  \href{http://dx.doi.org/10.1063/1.4826751}{{\em AIP Conf.Proc.} {\bfseries
  1560} (2013) 193--197},
\href{http://arxiv.org/abs/1207.1247}{{\ttfamily arXiv:1207.1247 [hep-ex]}}.

\bibitem{Adamson:2007gu}
P.~Adamson {\em et~al.}, {\bfseries MINOS Collaboration} , ``{A Study of Muon
  Neutrino Disappearance Using the Fermilab Main Injector Neutrino Beam},''
  \href{http://dx.doi.org/10.1103/PhysRevD.77.072002}{{\em Phys.Rev.}
  {\bfseries D77} (2008) 072002},
\href{http://arxiv.org/abs/0711.0769}{{\ttfamily arXiv:0711.0769 [hep-ex]}}.

\bibitem{Bishai:2012kta}
M.~Bishai, ``{Determining the Neutrino Flux from Accelerator Neutrino Beams},''
\href{http://dx.doi.org/10.1016/j.nuclphysbps.2012.09.034}{{\em
  Nucl.Phys.Proc.Suppl.} {\bfseries 229-232} (2012) 210--214}.

\bibitem{Osmanov:2011ig}
B.~Osmanov, {\bfseries MINERvA Collaboration} , ``{MINERvA Detector:
  Description and Performance},''
  \href{http://arxiv.org/abs/1109.2855}{{\ttfamily arXiv:1109.2855
  [physics.ins-det]}},
2011.

\bibitem{Korzenev:2013gia}
A.~Korzenev, {\bfseries NA61/SHINE} , ``{Hadron production measurement from
  NA61/SHINE},'' \href{http://arxiv.org/abs/1311.5719}{{\ttfamily
  arXiv:1311.5719 [nucl-ex]}},
2013.

\bibitem{Adamson:2011ig}
P.~Adamson {\em et~al.}, {\bfseries MINOS Collaboration} , ``{Measurement of
  the neutrino mass splitting and flavor mixing by MINOS},''
  \href{http://dx.doi.org/10.1103/PhysRevLett.106.181801}{{\em Phys.Rev.Lett.}
  {\bfseries 106} (2011) 181801},
\href{http://arxiv.org/abs/1103.0340}{{\ttfamily arXiv:1103.0340 [hep-ex]}}.

\bibitem{TingjunYang:2013vva}
T.~Yang, {\bfseries ArgoNeuT Collaboration} , ``{New Results from ArgoNeuT},''
  FERMILAB-CONF-13-510-E, \href{http://arxiv.org/abs/1311.2096}{{\ttfamily
  arXiv:1311.2096 [hep-ex]}},
2013.

\bibitem{Day:2012gb}
M.~Day and K.~S. McFarland, ``{Differences in Quasi-Elastic Cross-Sections of
  Muon and Electron Neutrinos},''
  \href{http://dx.doi.org/10.1103/PhysRevD.86.053003}{{\em Phys.Rev.}
  {\bfseries D86} (2012) 053003},
\href{http://arxiv.org/abs/1206.6745}{{\ttfamily arXiv:1206.6745 [hep-ph]}}.

\bibitem{Abe:2011ph}
K.~Abe {\em et~al.}, {\bfseries Super-Kamiokande Collaboration} , ``{Search for
  Differences in Oscillation Parameters for Atmospheric Neutrinos and
  Antineutrinos at Super-Kamiokande},''
  \href{http://dx.doi.org/10.1103/PhysRevLett.107.241801}{{\em Phys.Rev.Lett.}
  {\bfseries 107} (2011) 241801},
\href{http://arxiv.org/abs/1109.1621}{{\ttfamily arXiv:1109.1621 [hep-ex]}}.

\bibitem{Adamson:2013whj}
P.~Adamson {\em et~al.}, {\bfseries MINOS Collaboration} , ``{Measurement of
  Neutrino and Antineutrino Oscillations Using Beam and Atmospheric Data in
  MINOS},'' \href{http://dx.doi.org/10.1103/PhysRevLett.110.251801}{{\em
  Phys.Rev.Lett.} {\bfseries 110} (2013) 251801},
\href{http://arxiv.org/abs/1304.6335}{{\ttfamily arXiv:1304.6335 [hep-ex]}}.

\bibitem{Agrawal:1995gk}
V.~Agrawal, T.~Gaisser, P.~Lipari, and T.~Stanev, ``{Atmospheric neutrino flux
  above 1-GeV},'' \href{http://dx.doi.org/10.1103/PhysRevD.53.1314}{{\em
  Phys.Rev.} {\bfseries D53} (1996) 1314--1323},
\href{http://arxiv.org/abs/hep-ph/9509423}{{\ttfamily arXiv:hep-ph/9509423
  [hep-ph]}}.

\bibitem{ankowski2010energy}
A.~Ankowski {\em et~al.}, ``Energy reconstruction of electromagnetic showers
  from pi0 decays with the icarus t600 liquid argon tpc,'' {\em Acta Physica
  Polonica B} {\bfseries 41} no.~1, (2010) 103,
  \href{http://arxiv.org/abs/0812.2373}{{\ttfamily arXiv:0812.2373 [hep-ex]}}.

\bibitem{Arneodo:112001}
F.~Arneodo {\em et~al.}, {\bfseries The ICARUS-Milano Collaboration} ,
  ``Performance of a liquid argon time projection chamber exposed to the cern
  west area neutrino facility neutrino beam,''
  \href{http://dx.doi.org/10.1103/PhysRevD.74.112001}{{\em Phys. Rev. D}
  {\bfseries 74} (Dec, 2006) 112001}.
  \url{http://link.aps.org/doi/10.1103/PhysRevD.74.112001}.

\bibitem{Rubbia:2011ft}
C.~Rubbia {\em et~al.}, ``{Underground operation of the ICARUS T600 LAr-TPC:
  first results},'' \href{http://dx.doi.org/10.1088/1748-0221/6/07/P07011}{{\em
  JINST} {\bfseries 6} (2011) P07011},
\href{http://arxiv.org/abs/1106.0975}{{\ttfamily arXiv:1106.0975 [hep-ex]}}.

\bibitem{Davidson:2003ha}
S.~Davidson, C.~Pena-Garay, N.~Rius, and A.~Santamaria, ``{Present and future
  bounds on nonstandard neutrino interactions},''
  \href{http://dx.doi.org/10.1088/1126-6708/2003/03/011}{{\em JHEP} {\bfseries
  0303} (2003) 011},
\href{http://arxiv.org/abs/hep-ph/0302093}{{\ttfamily arXiv:hep-ph/0302093
  [hep-ph]}}.

\bibitem{GonzalezGarcia:2007ib}
M.~Gonzalez-Garcia and M.~Maltoni, ``{Phenomenology with Massive Neutrinos},''
  \href{http://dx.doi.org/10.1016/j.physrep.2007.12.004}{{\em Phys.Rept.}
  {\bfseries 460} (2008) 1--129},
\href{http://arxiv.org/abs/0704.1800}{{\ttfamily arXiv:0704.1800 [hep-ph]}}.

\bibitem{Biggio:2009nt}
C.~Biggio, M.~Blennow, and E.~Fernandez-Martinez, ``{General bounds on
  non-standard neutrino interactions},''
  \href{http://dx.doi.org/10.1088/1126-6708/2009/08/090}{{\em JHEP} {\bfseries
  0908} (2009) 090},
\href{http://arxiv.org/abs/0907.0097}{{\ttfamily arXiv:0907.0097 [hep-ph]}}.

\bibitem{Davoudiasl:2011sz}
H.~Davoudiasl, H.-S. Lee, and W.~J. Marciano, ``{Long-Range Lepton Flavor
  Interactions and Neutrino Oscillations},''
  \href{http://dx.doi.org/10.1103/PhysRevD.84.013009}{{\em Phys.Rev.}
  {\bfseries D84} (2011) 013009},
\href{http://arxiv.org/abs/1102.5352}{{\ttfamily arXiv:1102.5352 [hep-ph]}}.

\bibitem{Adamson:2010wi}
P.~Adamson {\em et~al.}, {\bfseries MINOS Collaboration} , ``{Search for
  sterile neutrino mixing in the MINOS long baseline experiment},''
  \href{http://dx.doi.org/10.1103/PhysRevD.81.052004}{{\em Phys.Rev.}
  {\bfseries D81} (2010) 052004},
\href{http://arxiv.org/abs/1001.0336}{{\ttfamily arXiv:1001.0336 [hep-ex]}}.

\bibitem{Machado:2011wx}
P.~Machado, H.~Nunokawa, F.~P.~d. Santos, and R.~Z. Funchal, ``{Large Extra
  Dimensions and Neutrino Oscillations},''
  \href{http://arxiv.org/abs/1110.1465}{{\ttfamily arXiv:1110.1465 [hep-ph]}},
2011.

\bibitem{Coloma:2012ji}
P.~Coloma, P.~Huber, J.~Kopp, and W.~Winter, ``{Systematic uncertainties in
  long-baseline neutrino oscillations for large $\theta_{13}$},''
  \href{http://dx.doi.org/10.1103/PhysRevD.87.033004}{{\em Phys.Rev.}
  {\bfseries D87} no.~3, (2013) 033004},
\href{http://arxiv.org/abs/1209.5973}{{\ttfamily arXiv:1209.5973 [hep-ph]}}.

\bibitem{Abe:2011ts}
K.~Abe, T.~Abe, H.~Aihara, Y.~Fukuda, Y.~Hayato, {\em et~al.}, ``{Letter of
  Intent: The Hyper-Kamiokande Experiment --- Detector Design and Physics
  Potential ---},'' \href{http://arxiv.org/abs/1109.3262}{{\ttfamily
  arXiv:1109.3262 [hep-ex]}},
2011.

\bibitem{Stahl:2012exa}
A.~Stahl, C.~Wiebusch, A.~Guler, M.~Kamiscioglu, R.~Sever, {\em et~al.},
  ``{Expression of Interest for a very long baseline neutrino oscillation
  experiment (LBNO)},'' CERN-SPSC-2012-021, SPSC-EOI-007,
2012.

\bibitem{Apollonio:2012hga}
M.~Apollonio, A.~Bross, J.~Kopp, and K.~Long, {\bfseries IDS-NF Collaboration}
  , ``{The International Design Study for the Neutrino Factory},''
\href{http://dx.doi.org/10.1016/j.nuclphysbps.2012.09.152}{{\em
  Nucl.Phys.Proc.Suppl.} {\bfseries 229-232} (2012) 515}.

\bibitem{Christensen:2013va}
E.~Christensen, P.~Coloma, and P.~Huber, ``{Physics Performance of a
  Low-Luminosity Low Energy Neutrino Factory},''
  \href{http://arxiv.org/abs/1301.7727}{{\ttfamily arXiv:1301.7727 [hep-ph]}},
2013.

\bibitem{Kearns:2013lea}
E.~Kearns {\em et~al.}, {\bfseries Hyper-Kamiokande Working Group} ,
  ``{Hyper-Kamiokande Physics Opportunities},''
  \href{http://arxiv.org/abs/1309.0184}{{\ttfamily arXiv:1309.0184 [hep-ex]}},
2013.

\bibitem{Agarwalla:2013kaa}
S.~Agarwalla {\em et~al.}, {\bfseries LAGUNA-LBNO Collaboration} , ``{The
  mass-hierarchy and CP-violation discovery reach of the LBNO long-baseline
  neutrino experiment},'' \href{http://arxiv.org/abs/1312.6520}{{\ttfamily
  arXiv:1312.6520 [hep-ph]}},
2013.

\bibitem{Bishai:2013yqo}
M.~Bishai, M.~Diwan, S.~Kettell, J.~Stewart, B.~Viren, {\em et~al.},
  ``{Precision Neutrino Oscillation Measurements using Simultaneous High-Power,
  Low-Energy Project-X Beams},'' BNL-101234-2013-CP, FERMILAB-FN-0962,
  \href{http://arxiv.org/abs/1307.0807}{{\ttfamily arXiv:1307.0807 [hep-ex]}},
2013.

\bibitem{Raaf:2012pva}
J.~L. Raaf, {\bfseries Super-Kamiokande Collaboration} , ``{Recent Nucleon
  Decay Results from Super-Kamiokande},''
\href{http://dx.doi.org/10.1016/j.nuclphysbps.2012.09.196}{{\em
  Nucl.Phys.Proc.Suppl.} {\bfseries 229-232} (2012) 559}.

\bibitem{Bueno:2007um}
A.~Bueno, Z.~Dai, Y.~Ge, M.~Laffranchi, A.~Melgarejo, {\em et~al.}, ``{Nucleon
  decay searches with large liquid argon TPC detectors at shallow depths:
  Atmospheric neutrinos and cosmogenic backgrounds},''
  \href{http://dx.doi.org/10.1088/1126-6708/2007/04/041}{{\em JHEP} {\bfseries
  0704} (2007) 041},
\href{http://arxiv.org/abs/hep-ph/0701101}{{\ttfamily arXiv:hep-ph/0701101
  [hep-ph]}}.

\bibitem{Stefan:2008zi}
D.~Stefan and A.~M. Ankowski, ``{Nuclear effects in proton decay},'' {\em Acta
  Phys.Polon.} {\bfseries B40} (2009) 671--674,
\href{http://arxiv.org/abs/0811.1892}{{\ttfamily arXiv:0811.1892 [nucl-th]}}.

\bibitem{Amerio:2004ze}
S.~Amerio {\em et~al.}, {\bfseries ICARUS Collaboration} , ``{Design,
  construction and tests of the ICARUS T600 detector},''
\href{http://dx.doi.org/10.1016/j.nima.2004.02.044}{{\em Nucl.Instrum.Meth.}
  {\bfseries A527} (2004) 329--410}.

\bibitem{Antonello:2012hu}
M.~Antonello, B.~Baibussinov, P.~Benetti, E.~Calligarich, N.~Canci, {\em
  et~al.}, ``{Precise 3D track reconstruction algorithm for the ICARUS T600
  liquid argon time projection chamber detector},''
  \href{http://dx.doi.org/10.1155/2013/260820}{{\em Adv.High Energy Phys.}
  {\bfseries 2013} (2013) 260820},
\href{http://arxiv.org/abs/1210.5089}{{\ttfamily arXiv:1210.5089
  [physics.ins-det]}}.

\bibitem{Bernstein:2009ms}
A.~Bernstein, M.~Bishai, E.~Blucher, D.~B. Cline, M.~V. Diwan, {\em et~al.},
  ``{Report on the Depth Requirements for a Massive Detector at Homestake},''
  FERMILAB-TM-2424-E, BNL-81896-2008-IR, LBNL-1348E,
  \href{http://arxiv.org/abs/0907.4183}{{\ttfamily arXiv:0907.4183 [hep-ex]}},
2009.

\bibitem{DOCDB5904}
V.~Kudryavtsev {\em et~al.}, ``Cosmic rays and cosmogenics. report to the lbne
  collaboration.,''
\newblock
  \href{http://lbne2-docdb.fnal.gov/cgi-bin/ShowDocument?docid=5904}{{\ttfamily
  LBNE-doc-5904}}, 2012.

\bibitem{Kobayashi:2005pe}
K.~Kobayashi {\em et~al.}, {\bfseries Super-Kamiokande Collaboration} ,
  ``{Search for nucleon decay via modes favored by supersymmetric grand
  unification models in Super-Kamiokande-I},''
  \href{http://dx.doi.org/10.1103/PhysRevD.72.052007}{{\em Phys.Rev.}
  {\bfseries D72} (2005) 052007},
\href{http://arxiv.org/abs/hep-ex/0502026}{{\ttfamily arXiv:hep-ex/0502026
  [hep-ex]}}.

\bibitem{gallagher-private}
H.~Gallagher, ``Private communication.''.

\bibitem{Janka:2012wk}
H.-T. Janka, ``{Explosion Mechanisms of Core-Collapse Supernovae},''
  \href{http://dx.doi.org/10.1146/annurev-nucl-102711-094901}{{\em
  Ann.Rev.Nucl.Part.Sci.} {\bfseries 62} (2012) 407--451},
\href{http://arxiv.org/abs/1206.2503}{{\ttfamily arXiv:1206.2503
  [astro-ph.SR]}}.

\bibitem{Fischer:2009af}
T.~Fischer, S.~Whitehouse, A.~Mezzacappa, F.-K. Thielemann, and
  M.~Liebendorfer, ``{Protoneutron star evolution and the neutrino driven wind
  in general relativistic neutrino radiation hydrodynamics simulations},''
  \href{http://dx.doi.org/10.1051/0004-6361/200913106}{{\em Astron.Astrophys.}
  {\bfseries 517} (2010) A80},
\href{http://arxiv.org/abs/0908.1871}{{\ttfamily arXiv:0908.1871
  [astro-ph.HE]}}.

\bibitem{Wurm:2011zn}
M.~Wurm {\em et~al.}, {\bfseries LENA Collaboration} , ``{The next-generation
  liquid-scintillator neutrino observatory LENA},''
  \href{http://dx.doi.org/10.1016/j.astropartphys.2012.02.011}{{\em
  Astropart.Phys.} {\bfseries 35} (2012) 685--732},
\href{http://arxiv.org/abs/1104.5620}{{\ttfamily arXiv:1104.5620
  [astro-ph.IM]}}.

\bibitem{Minakata:2008nc}
H.~Minakata, H.~Nunokawa, R.~Tomas, and J.~W. Valle, ``{Parameter Degeneracy in
  Flavor-Dependent Reconstruction of Supernova Neutrino Fluxes},''
  \href{http://dx.doi.org/10.1088/1475-7516/2008/12/006}{{\em JCAP} {\bfseries
  0812} (2008) 006},
\href{http://arxiv.org/abs/0802.1489}{{\ttfamily arXiv:0802.1489 [hep-ph]}}.

\bibitem{Tamborra:2012ac}
I.~Tamborra, B.~Muller, L.~Hudepohl, H.-T. Janka, and G.~Raffelt,
  ``{High-resolution supernova neutrino spectra represented by a simple fit},''
  \href{http://dx.doi.org/10.1103/PhysRevD.86.125031}{{\em Phys.Rev.}
  {\bfseries D86} (2012) 125031},
\href{http://arxiv.org/abs/1211.3920}{{\ttfamily arXiv:1211.3920
  [astro-ph.SR]}}.

\bibitem{Duan:2005cp}
H.~Duan, G.~M. Fuller, and Y.-Z. Qian, ``{Collective neutrino flavor
  transformation in supernovae},''
  \href{http://dx.doi.org/10.1103/PhysRevD.74.123004}{{\em Phys.Rev.}
  {\bfseries D74} (2006) 123004},
  \href{http://arxiv.org/abs/astro-ph/0511275}{{\ttfamily
  arXiv:astro-ph/0511275 [astro-ph]}}.

\bibitem{Fogli:2007bk}
G.~L. Fogli, E.~Lisi, A.~Marrone, and A.~Mirizzi, ``{Collective neutrino flavor
  transitions in supernovae and the role of trajectory averaging},''
  \href{http://dx.doi.org/10.1088/1475-7516/2007/12/010}{{\em JCAP} {\bfseries
  0712} (2007) 010}, \href{http://arxiv.org/abs/0707.1998}{{\ttfamily
  arXiv:0707.1998 [hep-ph]}}.

\bibitem{Raffelt:2007cb}
G.~G. Raffelt and A.~Y. Smirnov, ``{Self-induced spectral splits in supernova
  neutrino fluxes},'' \href{http://dx.doi.org/10.1103/PhysRevD.76.081301,
  10.1103/PhysRevD.77.029903, 10.1103/PhysRevD.76.081301,
  10.1103/PhysRevD.77.029903}{{\em Phys.Rev.} {\bfseries D76} (2007) 081301},
  \href{http://arxiv.org/abs/0705.1830}{{\ttfamily arXiv:0705.1830 [hep-ph]}}.

\bibitem{Raffelt:2007xt}
G.~G. Raffelt and A.~Y. Smirnov, ``{Adiabaticity and spectral splits in
  collective neutrino transformations},''
  \href{http://dx.doi.org/10.1103/PhysRevD.76.125008}{{\em Phys.Rev.}
  {\bfseries D76} (2007) 125008},
  \href{http://arxiv.org/abs/0709.4641}{{\ttfamily arXiv:0709.4641 [hep-ph]}}.

\bibitem{EstebanPretel:2008ni}
A.~Esteban-Pretel, A.~Mirizzi, S.~Pastor, R.~Tomas, G.~Raffelt, {\em et~al.},
  ``{Role of dense matter in collective supernova neutrino transformations},''
  \href{http://dx.doi.org/10.1103/PhysRevD.78.085012}{{\em Phys.Rev.}
  {\bfseries D78} (2008) 085012},
  \href{http://arxiv.org/abs/0807.0659}{{\ttfamily arXiv:0807.0659
  [astro-ph]}}.

\bibitem{Duan:2009cd}
H.~Duan and J.~P. Kneller, ``{Neutrino flavour transformation in supernovae},''
  \href{http://dx.doi.org/10.1088/0954-3899/36/11/113201}{{\em J.Phys.G}
  {\bfseries G36} (2009) 113201},
  \href{http://arxiv.org/abs/0904.0974}{{\ttfamily arXiv:0904.0974
  [astro-ph.HE]}}.

\bibitem{Dasgupta:2009mg}
B.~Dasgupta, A.~Dighe, G.~G. Raffelt, and A.~Y. Smirnov, ``{Multiple Spectral
  Splits of Supernova Neutrinos},''
  \href{http://dx.doi.org/10.1103/PhysRevLett.103.051105}{{\em Phys.Rev.Lett.}
  {\bfseries 103} (2009) 051105},
  \href{http://arxiv.org/abs/0904.3542}{{\ttfamily arXiv:0904.3542 [hep-ph]}}.

\bibitem{Duan:2010bg}
H.~Duan, G.~M. Fuller, and Y.-Z. Qian, ``{Collective Neutrino Oscillations},''
  \href{http://dx.doi.org/10.1146/annurev.nucl.012809.104524}{{\em
  Ann.Rev.Nucl.Part.Sci.} {\bfseries 60} (2010) 569--594},
  \href{http://arxiv.org/abs/1001.2799}{{\ttfamily arXiv:1001.2799 [hep-ph]}}.

\bibitem{Duan:2010bf}
H.~Duan and A.~Friedland, ``{Self-induced suppression of collective neutrino
  oscillations in a supernova},''
  \href{http://dx.doi.org/10.1103/PhysRevLett.106.091101}{{\em Phys.Rev.Lett.}
  {\bfseries 106} (2011) 091101},
  \href{http://arxiv.org/abs/1006.2359}{{\ttfamily arXiv:1006.2359 [hep-ph]}}.

\bibitem{Cherry:2013mv}
J.~F. Cherry, J.~Carlson, A.~Friedland, G.~M. Fuller, and A.~Vlasenko, ``{Halo
  Modification of a Supernova Neutronization Neutrino Burst},''
  \href{http://dx.doi.org/10.1103/PhysRevD.87.085037}{{\em Phys.Rev.}
  {\bfseries D87} (2013) 085037},
\href{http://arxiv.org/abs/1302.1159}{{\ttfamily arXiv:1302.1159
  [astro-ph.HE]}}.

\bibitem{Beacom:2000qy}
J.~F. Beacom, R.~Boyd, and A.~Mezzacappa, ``{Black hole formation in core
  collapse supernovae and time-of-flight measurements of the neutrino
  masses},'' \href{http://dx.doi.org/10.1103/PhysRevD.63.073011}{{\em
  Phys.Rev.} {\bfseries D63} (2001) 073011},
\href{http://arxiv.org/abs/astro-ph/0010398}{{\ttfamily arXiv:astro-ph/0010398
  [astro-ph]}}.

\bibitem{Fischer:2008rh}
T.~Fischer, S.~C. Whitehouse, A.~Mezzacappa, F.~K. Thielemann, and
  M.~Liebendorfer, ``{The neutrino signal from protoneutron star accretion and
  black hole formation},'' \href{http://arxiv.org/abs/0809.5129}{{\ttfamily
  arXiv:0809.5129 [astro-ph]}},
2008.

\bibitem{Schirato:2002tg}
R.~C. Schirato and G.~M. Fuller, ``{Connection between supernova shocks, flavor
  transformation, and the neutrino signal},'' LA-UR-02-3068,
  \href{http://arxiv.org/abs/astro-ph/0205390}{{\ttfamily
  arXiv:astro-ph/0205390 [astro-ph]}},
2002.

\bibitem{Hanke:2011jf}
F.~Hanke, A.~Marek, B.~Muller, and H.-T. Janka, ``{Is Strong SASI Activity the
  Key to Successful Neutrino-Driven Supernova Explosions?},''
  \href{http://dx.doi.org/10.1088/0004-637X/755/2/138}{{\em Astrophys.J.}
  {\bfseries 755} (2012) 138},
\href{http://arxiv.org/abs/1108.4355}{{\ttfamily arXiv:1108.4355
  [astro-ph.SR]}}.

\bibitem{Hanke:2013ena}
F.~Hanke, B.~Mueller, A.~Wongwathanarat, A.~Marek, and H.-T. Janka, ``{SASI
  Activity in Three-Dimensional Neutrino-Hydrodynamics Simulations of Supernova
  Cores},'' \href{http://dx.doi.org/10.1088/0004-637X/770/1/66}{{\em
  Astrophys.J.} {\bfseries 770} (2013) 66},
\href{http://arxiv.org/abs/1303.6269}{{\ttfamily arXiv:1303.6269
  [astro-ph.SR]}}.

\bibitem{Friedland:2006ta}
A.~Friedland and A.~Gruzinov, ``{Neutrino signatures of supernova
  turbulence},'' LA-UR-06-2202,
  \href{http://arxiv.org/abs/astro-ph/0607244}{{\ttfamily
  arXiv:astro-ph/0607244 [astro-ph]}},
2006.

\bibitem{Lund:2013uta}
T.~Lund and J.~P. Kneller, ``{Combining collective, MSW, and turbulence effects
  in supernova neutrino flavor evolution},''
  \href{http://arxiv.org/abs/1304.6372}{{\ttfamily arXiv:1304.6372
  [astro-ph.HE]}},
2013.

\bibitem{Raffelt:1999tx}
G.~G. Raffelt, ``{Particle Physics from Stars},''
  \href{http://dx.doi.org/10.1146/annurev.nucl.49.1.163}{{\em Ann. Rev. Nucl.
  Part. Sci.} {\bfseries 49} (1999) 163--216},
\href{http://arxiv.org/abs/hep-ph/9903472}{{\ttfamily arXiv:hep-ph/9903472}}.

\bibitem{Bueno:2003ei}
A.~Bueno, I.~Gil~Botella, and A.~Rubbia, ``{Supernova neutrino detection in a
  liquid argon TPC},'' ICARUS-TM-03-02,
  \href{http://arxiv.org/abs/hep-ph/0307222}{{\ttfamily arXiv:hep-ph/0307222
  [hep-ph]}},
2003.

\bibitem{snowglobes}
K.~Scholberg {\em et~al.}, ``{SNOwGLoBES: SuperNova Observatories with
  GLoBES}.''
\newblock \url{{http://www.phy.duke.edu/~schol/snowglobes}}.

\bibitem{Church:2013hea}
E.~D. Church, ``{LArSoft: A Software Package for Liquid Argon Time Projection
  Drift Chambers},'' \href{http://arxiv.org/abs/1311.6774}{{\ttfamily
  arXiv:1311.6774 [physics.ins-det]}},
2013.

\bibitem{Totani:1997vj}
T.~Totani, K.~Sato, H.~E. Dalhed, and J.~R. Wilson, ``{Future detection of
  supernova neutrino burst and explosion mechanism},''
  \href{http://dx.doi.org/10.1086/305364}{{\em Astrophys. J.} {\bfseries 496}
  (1998) 216--225},
\href{http://arxiv.org/abs/astro-ph/9710203}{{\ttfamily
  arXiv:astro-ph/9710203}}.

\bibitem{Gava:2009pj}
J.~Gava, J.~Kneller, C.~Volpe, and G.~C. McLaughlin, ``{A dynamical collective
  calculation of supernova neutrino signals},''
  \href{http://dx.doi.org/10.1103/PhysRevLett.103.071101}{{\em Phys. Rev.
  Lett.} {\bfseries 103} (2009) 071101},
\href{http://arxiv.org/abs/0902.0317}{{\ttfamily arXiv:0902.0317 [hep-ph]}}.

\bibitem{Huedepohl:2009wh}
L.~Hudepohl, B.~Muller, H.-T. Janka, A.~Marek, and G.~Raffelt, ``{Neutrino
  Signal of Electron-Capture Supernovae from Core Collapse to Cooling},''
  \href{http://dx.doi.org/10.1103/PhysRevLett.104.251101,
  10.1103/PhysRevLett.105.249901}{{\em Phys.Rev.Lett.} {\bfseries 104} (2010)
  251101},
\href{http://arxiv.org/abs/0912.0260}{{\ttfamily arXiv:0912.0260
  [astro-ph.SR]}}.

\bibitem{cherrypriv}
A.~Cherry, A.~Friedland, and H.~Duan, ``Private communication.''.

\bibitem{Keil:2002in}
M.~T. Keil, G.~G. Raffelt, and H.-T. Janka, ``{Monte Carlo study of supernova
  neutrino spectra formation},'' \href{http://dx.doi.org/10.1086/375130}{{\em
  Astrophys.J.} {\bfseries 590} (2003) 971--991},
\href{http://arxiv.org/abs/astro-ph/0208035}{{\ttfamily arXiv:astro-ph/0208035
  [astro-ph]}}.

\bibitem{COSMICBKGD}
E.~Church {\em et~al.}, ``{Muon-induced background for beam neutrinos at the
  surface},''
\newblock
  \href{http://lbne2-docdb.fnal.gov/cgi-bin/ShowDocument?docid=6232}{{\ttfamily
  LBNE-doc-6232}}, October, 2012.

\bibitem{DOCDB8419}
{Gehman, V. and Kadel, R}, ``{Calculation of intrinsic and cosmogenic
  backgrounds in the LBNE far detector for use in detection of supernova
  neutrinos},''
\newblock
  \href{http://lbne2-docdb.fnal.gov/cgi-bin/ShowDocument?docid=8419}{{\ttfamily
  LBNE-doc-8419}}, January, 2014.

\bibitem{Concrete}
J.~H. Harley {\em et~al.}, ``{Report No. 094 - Exposure of the Population in
  the United States and Canada from Natural Background Radiation},'' {\em
  National Council on Radiation Protection and Measurements} (2014) .
  \url{{http://www.ncrppublications.org/Reports/094}}.

\bibitem{DARKSIDE}
L.~Grandi, ``Darkside-50: performance and results from the first atmospheric
  argon run,'' February, 2014.
\newblock \href{https://hepconf.physics.ucla.edu/dm14/agenda.html}{UCLA's 11th
  Symposium on Sources and Detection of Dark Matter and Dark Energy in the
  Universe}.

\bibitem{Leonard:2007uv}
D.~Leonard, P.~Grinberg, P.~Weber, E.~Baussan, Z.~Djurcic, {\em et~al.},
  ``{Systematic study of trace radioactive impurities in candidate construction
  materials for EXO-200},''
  \href{http://dx.doi.org/10.1016/j.nima.2008.03.001}{{\em Nucl.Instrum.Meth.}
  {\bfseries A591} (2008) 490--509},
\href{http://arxiv.org/abs/0709.4524}{{\ttfamily arXiv:0709.4524
  [physics.ins-det]}}.

\bibitem{Casper:2002sd}
D.~Casper, ``{The Nuance neutrino physics simulation, and the future},''
  \href{http://dx.doi.org/10.1016/S0920-5632(02)01756-5}{{\em
  Nucl.Phys.Proc.Suppl.} {\bfseries 112} (2002) 161--170},
\href{http://arxiv.org/abs/hep-ph/0208030}{{\ttfamily arXiv:hep-ph/0208030
  [hep-ph]}}.

\bibitem{DOCDB740}
G.~Zeller, ``{Nuclear Effects in Water vs. Argon},''
\newblock
  \href{http://lbne2-docdb.fnal.gov/cgi-bin/ShowDocument?docid=740}{{\ttfamily
  LBNE-doc-740}}, 2010.

\bibitem{DOCDB783}
G.~Zeller, ``{Expected Event Rates in the LBNE Near Detector},''
\newblock
  \href{http://lbne2-docdb.fnal.gov/cgi-bin/ShowDocument?docid=783}{{\ttfamily
  LBNE-doc-783}}, 2010.

\bibitem{srmishra-reviewtalk}
S.R.Mishra, Apr, 1990.
\newblock Review talk presented at Workshop on Hadron Structure Functions and
  Parton Distributions, Fermilab.

\bibitem{Raja:2005sh}
R.~Raja, ``{The Main injector particle production experiment (MIPP) at
  Fermilab},'' \href{http://dx.doi.org/10.1088/1742-6596/9/1/058}{{\em
  Nucl.Instrum.Meth.} {\bfseries A553} (2005) 225--230},
\href{http://arxiv.org/abs/hep-ex/0501005}{{\ttfamily arXiv:hep-ex/0501005
  [hep-ex]}}.

\bibitem{Formaggio:2013kya}
J.~Formaggio and G.~Zeller, ``{From eV to EeV: Neutrino Cross Sections Across
  Energy Scales},'' \href{http://dx.doi.org/10.1103/RevModPhys.84.1307}{{\em
  Rev.Mod.Phys.} {\bfseries 84} (2012) 1307},
\href{http://arxiv.org/abs/1305.7513}{{\ttfamily arXiv:1305.7513 [hep-ex]}}.

\bibitem{Marciano:2003eq}
W.~J. Marciano and Z.~Parsa, ``{Neutrino-Electron Scattering Theory},''
  \href{http://dx.doi.org/10.1088/0954-3899/29/11/013}{{\em J. Phys.}
  {\bfseries G29} (2003) 2629--2645},
\href{http://arxiv.org/abs/hep-ph/0403168}{{\ttfamily arXiv:hep-ph/0403168}}.

\bibitem{Mishra:1989jn}
S.~Mishra, K.~Bachmann, R.~Bernstein, R.~Blair, C.~Foudas, {\em et~al.},
  ``{Measurement of Inverse Muon Decay $\nu_\mu + e \to \mu^- + \nu_e$ at
  {Fermilab} Tevatron Energies 15-{GeV} - 600-{GeV}},''
\href{http://dx.doi.org/10.1103/PhysRevLett.63.132}{{\em Phys.Rev.Lett.}
  {\bfseries 63} (1989) 132--135}.

\bibitem{Mishra:1990yf}
S.~Mishra, K.~Bachmann, R.~Blair, C.~Foudas, B.~King, {\em et~al.}, ``{Inverse
  Muon Decay, $\nu_\mu e \to \mu^- \nu_e$, at the {Fermilab} Tevatron},''
\href{http://dx.doi.org/10.1016/0370-2693(90)91099-W}{{\em Phys.Lett.}
  {\bfseries B252} (1990) 170--176}.

\bibitem{Vilain:1996yf}
P.~Vilain {\em et~al.}, {\bfseries CHARM-II Collaboration} , ``{A Precise
  measurement of the cross-section of the inverse muon decay muon-neutrino + e-
  --> mu- + electron-neutrino},''
\href{http://dx.doi.org/10.1016/0370-2693(95)01298-6}{{\em Phys.Lett.}
  {\bfseries B364} (1995) 121--126}.

\bibitem{Samoylov:2013xoa}
O.~Samoylov {\em et~al.}, {\bfseries NOMAD} , ``{A Precision Measurement of
  Charm Dimuon Production in Neutrino Interactions from the NOMAD
  Experiment},'' \href{http://dx.doi.org/10.1016/j.nuclphysb.2013.08.021}{{\em
  Nucl.Phys.} {\bfseries B876} (2013) 339--375},
\href{http://arxiv.org/abs/1308.4750}{{\ttfamily arXiv:1308.4750 [hep-ex]}}.

\bibitem{Zeller:2001hh}
G.~Zeller {\em et~al.}, {\bfseries NuTeV Collaboration} , ``{A Precise
  determination of electroweak parameters in neutrino nucleon scattering},''
  \href{http://dx.doi.org/10.1103/PhysRevLett.88.091802}{{\em Phys.Rev.Lett.}
  {\bfseries 88} (2002) 091802},
\href{http://arxiv.org/abs/hep-ex/0110059}{{\ttfamily arXiv:hep-ex/0110059
  [hep-ex]}}.

\bibitem{Abramowicz:1986vi}
H.~Abramowicz, R.~Belusevic, A.~Blondel, H.~Blumer, P.~Bockmann, {\em et~al.},
  {\bfseries CDHS Collaboration} , ``{A Precision Measurement of sin**2theta(W)
  from Semileptonic Neutrino Scattering},''
\href{http://dx.doi.org/10.1103/PhysRevLett.57.298}{{\em Phys.Rev.Lett.}
  {\bfseries 57} (1986) 298}.

\bibitem{Allaby:1987vr}
J.~Allaby {\em et~al.}, {\bfseries CHARM Collaboration} , ``{A Precise
  Determination of the Electroweak Mixing Angle from Semileptonic Neutrino
  Scattering},''
\href{http://dx.doi.org/10.1007/BF01630598}{{\em Z.Phys.} {\bfseries C36}
  (1987) 611}.

\bibitem{Reutens:1985hv}
P.~Reutens, F.~Merritt, D.~MacFarlane, R.~Messner, D.~Novikoff, {\em et~al.},
  {\bfseries CCFR Collaboration} , ``{Measurement of $\sin^2\theta_W$ and
  $\rho$ in Deep Inelastic Neutrino - Nucleon Scattering},''
\href{http://dx.doi.org/10.1016/0370-2693(85)90519-2}{{\em Phys.Lett.}
  {\bfseries B152} (1985) 404--410}.

\bibitem{Alekhin:2007fh}
S.~Alekhin, S.~A. Kulagin, and R.~Petti, ``{Modeling lepton-nucleon inelastic
  scattering from high to low momentum transfer},''
  \href{http://dx.doi.org/10.1063/1.2834481}{{\em AIP Conf.Proc.} {\bfseries
  967} (2007) 215--224},
\href{http://arxiv.org/abs/0710.0124}{{\ttfamily arXiv:0710.0124 [hep-ph]}}.

\bibitem{Alekhin:2008ua}
S.~Alekhin, S.~A. Kulagin, and R.~Petti, ``{Update of the global fit of PDFs
  including the low-Q DIS data},''
  \href{http://arxiv.org/abs/0810.4893}{{\ttfamily arXiv:0810.4893 [hep-ph]}},
2008.

\bibitem{Alekhin:2008mb}
S.~Alekhin, S.~A. Kulagin, and R.~Petti, ``{Determination of Strange Sea
  Distributions from Neutrino-Nucleon Deep Inelastic Scattering},''
  \href{http://dx.doi.org/10.1016/j.physletb.2009.04.033}{{\em Phys.Lett.}
  {\bfseries B675} (2009) 433--440},
\href{http://arxiv.org/abs/0812.4448}{{\ttfamily arXiv:0812.4448 [hep-ph]}}.

\bibitem{Arbuzov:2004zr}
A.~Arbuzov, D.~Y. Bardin, and L.~Kalinovskaya, ``{Radiative corrections to
  neutrino deep inelastic scattering revisited},''
  \href{http://dx.doi.org/10.1088/1126-6708/2005/06/078}{{\em JHEP} {\bfseries
  0506} (2005) 078},
\href{http://arxiv.org/abs/hep-ph/0407203}{{\ttfamily arXiv:hep-ph/0407203
  [hep-ph]}}.

\bibitem{Kulagin:2004ie}
S.~A. Kulagin and R.~Petti, ``{Global study of nuclear structure functions},''
  \href{http://dx.doi.org/10.1016/j.nuclphysa.2005.10.011}{{\em Nucl.Phys.}
  {\bfseries A765} (2006) 126--187},
\href{http://arxiv.org/abs/hep-ph/0412425}{{\ttfamily arXiv:hep-ph/0412425
  [hep-ph]}}.

\bibitem{Kulagin:2007ju}
S.~A. Kulagin and R.~Petti, ``{Neutrino inelastic scattering off nuclei},''
  \href{http://dx.doi.org/10.1103/PhysRevD.76.094023}{{\em Phys.Rev.}
  {\bfseries D76} (2007) 094023},
\href{http://arxiv.org/abs/hep-ph/0703033}{{\ttfamily arXiv:hep-ph/0703033
  [HEP-PH]}}.

\bibitem{Kulagin:2010gd}
S.~Kulagin and R.~Petti, ``{Structure functions for light nuclei},''
  \href{http://dx.doi.org/10.1103/PhysRevC.82.054614}{{\em Phys.Rev.}
  {\bfseries C82} (2010) 054614},
\href{http://arxiv.org/abs/1004.3062}{{\ttfamily arXiv:1004.3062 [hep-ph]}}.

\bibitem{Vilain:1994qy}
P.~Vilain {\em et~al.}, {\bfseries CHARM-II Collaboration} , ``{Precision
  measurement of electroweak parameters from the scattering of muon-neutrinos
  on electrons},''
\href{http://dx.doi.org/10.1016/0370-2693(94)91421-4}{{\em Phys.Lett.}
  {\bfseries B335} (1994) 246--252}.

\bibitem{Czarnecki:2000ic}
A.~Czarnecki and W.~J. Marciano, ``{Polarized Moller scattering asymmetries},''
  \href{http://dx.doi.org/10.1016/S0217-751X(00)00243-0}{{\em Int.J.Mod.Phys.}
  {\bfseries A15} (2000) 2365--2376},
\href{http://arxiv.org/abs/hep-ph/0003049}{{\ttfamily arXiv:hep-ph/0003049
  [hep-ph]}}.

\bibitem{Bennett:1999zza}
S.~Bennett and C.~Wieman, ``{Erratum: Measurement of the 6s --> 7s Transition
  Polarizability in Atomic Cesium and an Improved Test of the Standard Model
  [Phys. Rev. Lett. 82, 2484 (1999) ]},''
\href{http://dx.doi.org/10.1103/PhysRevLett.82.4153}{{\em Phys.Rev.Lett.}
  {\bfseries 82} (1999) 4153--4153}.

\bibitem{Yao:2006px}
W.~Yao {\em et~al.}, {\bfseries Particle Data Group} , ``{Review of Particle
  Physics},''
\href{http://dx.doi.org/10.1088/0954-3899/33/1/001}{{\em J.Phys.} {\bfseries
  G33} (2006) 1--1232}.

\bibitem{Anthony:2005pm}
P.~Anthony {\em et~al.}, {\bfseries SLAC E158 Collaboration} , ``{Precision
  measurement of the weak mixing angle in Moller scattering},''
  \href{http://dx.doi.org/10.1103/PhysRevLett.95.081601}{{\em Phys.Rev.Lett.}
  {\bfseries 95} (2005) 081601},
\href{http://arxiv.org/abs/hep-ex/0504049}{{\ttfamily arXiv:hep-ex/0504049
  [hep-ex]}}.

\bibitem{Lee:2013kya}
J.~H. Lee, ``{The Qweak: Precision measurement of the proton's weak charge by
  parity violating experiment},''
\href{http://dx.doi.org/10.1007/s00601-012-0339-9}{{\em Few Body Syst.}
  {\bfseries 54} (2013) 129--134}.

\bibitem{Nuruzzaman:2013bwa}
Nuruzzaman, ``{Q-weak: First Direct Measurement of the Weak Charge of the
  Proton},'' \href{http://arxiv.org/abs/1312.6009}{{\ttfamily arXiv:1312.6009
  [nucl-ex]}},
2013.

\bibitem{Jaffe:1989jz}
R.~Jaffe and A.~Manohar, ``{The G(1) Problem: Fact and Fantasy on the Spin of
  the Proton},''
\href{http://dx.doi.org/10.1016/0550-3213(90)90506-9}{{\em Nucl.Phys.}
  {\bfseries B337} (1990) 509--546}.

\bibitem{Young:2006jc}
R.~D. Young, J.~Roche, R.~D. Carlini, and A.~W. Thomas, ``{Extracting nucleon
  strange and anapole form factors from world data},''
  \href{http://dx.doi.org/10.1103/PhysRevLett.97.102002}{{\em Phys.Rev.Lett.}
  {\bfseries 97} (2006) 102002},
\href{http://arxiv.org/abs/nucl-ex/0604010}{{\ttfamily arXiv:nucl-ex/0604010
  [nucl-ex]}}.

\bibitem{Leinweber:2004tc}
D.~B. Leinweber, S.~Boinepalli, I.~Cloet, A.~W. Thomas, A.~G. Williams, {\em
  et~al.}, ``{Precise determination of the strangeness magnetic moment of the
  nucleon},'' \href{http://dx.doi.org/10.1103/PhysRevLett.94.212001}{{\em
  Phys.Rev.Lett.} {\bfseries 94} (2005) 212001},
\href{http://arxiv.org/abs/hep-lat/0406002}{{\ttfamily arXiv:hep-lat/0406002
  [hep-lat]}}.

\bibitem{Ahrens:1986xe}
L.~Ahrens, S.~Aronson, P.~Connolly, B.~Gibbard, M.~Murtagh, {\em et~al.},
  ``{Measurement of Neutrino - Proton and anti-neutrino - Proton Elastic
  Scattering},''
\href{http://dx.doi.org/10.1103/PhysRevD.35.785}{{\em Phys.Rev.} {\bfseries
  D35} (1987) 785}.

\bibitem{Garvey:1992cg}
G.~Garvey, W.~Louis, and D.~White, ``{Determination of proton strange
  form-factors from neutrino p elastic scattering},''
\href{http://dx.doi.org/10.1103/PhysRevC.48.761}{{\em Phys.Rev.} {\bfseries
  C48} (1993) 761--765}.

\bibitem{Alberico:1998qw}
W.~Alberico, M.~Barbaro, S.~M. Bilenky, J.~Caballero, C.~Giunti, {\em et~al.},
  ``{Strange form-factors of the proton: A New analysis of the neutrino
  (anti-neutrino) data of the BNL-734 experiment},''
  \href{http://dx.doi.org/10.1016/S0375-9474(99)00142-6}{{\em Nucl.Phys.}
  {\bfseries A651} (1999) 277--286},
\href{http://arxiv.org/abs/hep-ph/9812388}{{\ttfamily arXiv:hep-ph/9812388
  [hep-ph]}}.

\bibitem{AguilarArevalo:2010cx}
A.~Aguilar-Arevalo {\em et~al.}, {\bfseries MiniBooNE Collaboration} ,
  ``{Measurement of the Neutrino Neutral-Current Elastic Differential Cross
  Section on Mineral Oil at $E_\nu \sim 1$ GeV},''
  \href{http://dx.doi.org/10.1103/PhysRevD.82.092005}{{\em Phys.Rev.}
  {\bfseries D82} (2010) 092005},
\href{http://arxiv.org/abs/1007.4730}{{\ttfamily arXiv:1007.4730 [hep-ex]}}.

\bibitem{Bugel:2004yk}
L.~Bugel {\em et~al.}, {\bfseries FINeSSE Collaboration} , ``{A Proposal for a
  near detector experiment on the booster neutrino beamline: FINeSSE: Fermilab
  intense neutrino scattering scintillator experiment},''
  FERMILAB-PROPOSAL-0937, \href{http://arxiv.org/abs/hep-ex/0402007}{{\ttfamily
  arXiv:hep-ex/0402007 [hep-ex]}},
2004.

\bibitem{Leung:1992yx}
W.~Leung, P.~Quintas, S.~Mishra, F.~Sciulli, C.~Arroyo, {\em et~al.}, ``{A
  Measurement of the Gross-Llewellyn-Smith sum rule from the CCFR x(F3)
  structure function},''
\href{http://dx.doi.org/10.1016/0370-2693(93)91386-2}{{\em Phys.Lett.}
  {\bfseries B317} (1993) 655--659}.

\bibitem{Bodek:1985tv}
A.~Bodek and A.~Simon, ``{What Do Electron and Neutrino Experiments Tell Us
  About Nuclear Effects in the Deuteron},''
\href{http://dx.doi.org/10.1007/BF01550821}{{\em Z.Phys.} {\bfseries C29}
  (1985) 231}.

\bibitem{Jones:1987gk}
G.~Jones {\em et~al.}, {\bfseries Birmingham-CERN-Imperial
  Coll.-MPI(Munich)-Oxford-University Coll. Collaboration} , ``{A Measurement
  of the Proton Structure Functions From Neutrino Hydrogen and Anti-neutrino
  Hydrogen Charged Current Interactions},''
\href{http://dx.doi.org/10.1007/BF01415551}{{\em Z.Phys.} {\bfseries C44}
  (1989) 379--384}.

\bibitem{Berge:1989hr}
J.~Berge, H.~Burkhardt, F.~Dydak, R.~Hagelberg, M.~Krasny, {\em et~al.}, ``{A
  Measurement of Differential Cross-Sections and Nucleon Structure Functions in
  Charged Current Neutrino Interactions on Iron},''
\href{http://dx.doi.org/10.1007/BF01555493}{{\em Z.Phys.} {\bfseries C49}
  (1991) 187--224}.

\bibitem{Allasia:1983vw}
D.~Allasia {\em et~al.}, {\bfseries WA25 Collaboration} , ``{Measurement of the
  Neutron and Proton Structure Functions From Neutrino and Anti-neutrinos
  Scattering in Deuterium},''
\href{http://dx.doi.org/10.1016/0370-2693(84)90488-X}{{\em Phys.Lett.}
  {\bfseries B135} (1984) 231}.

\bibitem{Allasia:1985hw}
D.~Allasia, C.~Angelini, A.~Baldini, L.~Bertanza, A.~Bigi, {\em et~al.},
  ``{Q**2 Dependence of the Proton and Neutron Structure Functions from
  Neutrino and anti-neutrinos Scattering in Deuterium},''
\href{http://dx.doi.org/10.1007/BF01413595}{{\em Z.Phys.} {\bfseries C28}
  (1985) 321}.

\bibitem{Yang:2000ju}
U.-K. Yang {\em et~al.}, {\bfseries CCFR/NuTeV Collaboration} , ``{Measurements
  of $F_2$ and $xF^{\nu}_3 - x F^{\bar{\nu}}_3$ from CCFR $\nu_\mu-$Fe and
  $\bar{\nu}_\mu-$Fe data in a physics model independent way},''
  \href{http://dx.doi.org/10.1103/PhysRevLett.86.2742}{{\em Phys.Rev.Lett.}
  {\bfseries 86} (2001) 2742--2745},
\href{http://arxiv.org/abs/hep-ex/0009041}{{\ttfamily arXiv:hep-ex/0009041
  [hep-ex]}}.

\bibitem{Yang:2001xc}
U.-K. Yang {\em et~al.}, {\bfseries CCFR/NuTeV Collaboration} , ``{Extraction
  of R = sigma(L) / sigma(T) from CCFR Fe-neutrino(muon) and
  Fe-anti-neutrino(muon) differential cross-sections},''
  \href{http://dx.doi.org/10.1103/PhysRevLett.87.251802}{{\em Phys.Rev.Lett.}
  {\bfseries 87} (2001) 251802},
\href{http://arxiv.org/abs/hep-ex/0104040}{{\ttfamily arXiv:hep-ex/0104040
  [hep-ex]}}.

\bibitem{Tzanov:2005kr}
M.~Tzanov {\em et~al.}, {\bfseries NuTeV Collaboration} , ``{Precise
  measurement of neutrino and anti-neutrino differential cross sections},''
  \href{http://dx.doi.org/10.1103/PhysRevD.74.012008}{{\em Phys.Rev.}
  {\bfseries D74} (2006) 012008},
\href{http://arxiv.org/abs/hep-ex/0509010}{{\ttfamily arXiv:hep-ex/0509010
  [hep-ex]}}.

\bibitem{Onengut:2005kv}
G.~Onengut {\em et~al.}, {\bfseries CHORUS Collaboration} , ``{Measurement of
  nucleon structure functions in neutrino scattering},''
\href{http://dx.doi.org/10.1016/j.physletb.2005.10.062}{{\em Phys.Lett.}
  {\bfseries B632} (2006) 65--75}.

\bibitem{Petti:2011zz}
R.~Petti and O.~Samoylov, ``{Charm dimuon production in neutrino-nucleon
  interactions in the NOMAD experiment},''
\href{http://dx.doi.org/10.1134/S1547477111070168}{{\em Phys.Part.Nucl.Lett.}
  {\bfseries 8} (2011) 755--761}.

\bibitem{Sekiguchi:2012xma}
T.~Sekiguchi, ``{Neutrino facility and neutrino physics in J-PARC},''
\href{http://dx.doi.org/10.1093/ptep/pts020}{{\em PTEP} {\bfseries 2012} (2012)
  02B005}.

\bibitem{Dudek:2012vr}
J.~Dudek, R.~Ent, R.~Essig, K.~Kumar, C.~Meyer, {\em et~al.}, ``{Physics
  Opportunities with the 12 GeV Upgrade at Jefferson Lab},''
  \href{http://dx.doi.org/10.1140/epja/i2012-12187-1}{{\em Eur.Phys.J.}
  {\bfseries A48} (2012) 187},
\href{http://arxiv.org/abs/1208.1244}{{\ttfamily arXiv:1208.1244 [hep-ex]}}.

\bibitem{Mondal:2012fn}
N.~Mondal, ``{India-Based Neutrino Observatory (INO)},''
\href{http://dx.doi.org/10.1140/epjp/i2012-12106-y}{{\em Eur.Phys.J.Plus}
  {\bfseries 127} (2012) 106}.

\bibitem{Butkevich:2012zr}
A.~Butkevich, ``{Quasi-elastic neutrino charged-current scattering off
  medium-heavy nuclei: 40Ca and 40Ar},''
  \href{http://dx.doi.org/10.1103/PhysRevC.85.065501}{{\em Phys.Rev.}
  {\bfseries C85} (2012) 065501},
\href{http://arxiv.org/abs/1204.3160}{{\ttfamily arXiv:1204.3160 [nucl-th]}}.

\bibitem{Butkevich:2007gm}
A.~Butkevich and S.~A. Kulagin, ``{Quasi-elastic neutrino charged-current
  scattering cross sections on oxygen},''
  \href{http://dx.doi.org/10.1103/PhysRevC.76.045502}{{\em Phys.Rev.}
  {\bfseries C76} (2007) 045502},
\href{http://arxiv.org/abs/0705.1051}{{\ttfamily arXiv:0705.1051 [nucl-th]}}.

\bibitem{Ankowski:2007uy}
A.~M. Ankowski and J.~T. Sobczyk, ``{Construction of spectral functions for
  medium-mass nuclei},''
  \href{http://dx.doi.org/10.1103/PhysRevC.77.044311}{{\em Phys.Rev.}
  {\bfseries C77} (2008) 044311},
\href{http://arxiv.org/abs/0711.2031}{{\ttfamily arXiv:0711.2031 [nucl-th]}}.

\bibitem{Asaka:2005pn}
T.~Asaka and M.~Shaposhnikov, ``{The nuMSM, dark matter and baryon asymmetry of
  the universe},'' \href{http://dx.doi.org/10.1016/j.physletb.2005.06.020}{{\em
  Phys.Lett.} {\bfseries B620} (2005) 17--26},
\href{http://arxiv.org/abs/hep-ph/0505013}{{\ttfamily arXiv:hep-ph/0505013
  [hep-ph]}}.

\bibitem{Gorbunov:2007ak}
D.~Gorbunov and M.~Shaposhnikov, ``{How to find neutral leptons of the
  $\nu$MSM?},'' \href{http://dx.doi.org/10.1007/JHEP11(2013)101,
  10.1088/1126-6708/2007/10/015}{{\em JHEP} {\bfseries 0710} (2007) 015},
\href{http://arxiv.org/abs/0705.1729}{{\ttfamily arXiv:0705.1729 [hep-ph]}}.

\bibitem{Boyarsky:2009ix}
A.~Boyarsky, O.~Ruchayskiy, and M.~Shaposhnikov, ``{The Role of sterile
  neutrinos in cosmology and astrophysics},''
  \href{http://dx.doi.org/10.1146/annurev.nucl.010909.083654}{{\em
  Ann.Rev.Nucl.Part.Sci.} {\bfseries 59} (2009) 191--214},
\href{http://arxiv.org/abs/0901.0011}{{\ttfamily arXiv:0901.0011 [hep-ph]}}.

\bibitem{Dodelson:1993je}
S.~Dodelson and L.~M. Widrow, ``{Sterile-neutrinos as dark matter},''
  \href{http://dx.doi.org/10.1103/PhysRevLett.72.17}{{\em Phys.Rev.Lett.}
  {\bfseries 72} (1994) 17--20},
\href{http://arxiv.org/abs/hep-ph/9303287}{{\ttfamily arXiv:hep-ph/9303287
  [hep-ph]}}.

\bibitem{Atre:2009rg}
A.~Atre, T.~Han, S.~Pascoli, and B.~Zhang, ``{The Search for Heavy Majorana
  Neutrinos},'' \href{http://dx.doi.org/10.1088/1126-6708/2009/05/030}{{\em
  JHEP} {\bfseries 0905} (2009) 030},
\href{http://arxiv.org/abs/0901.3589}{{\ttfamily arXiv:0901.3589 [hep-ph]}}.

\bibitem{Shaposhnikov:2008pf}
M.~Shaposhnikov, ``{The nuMSM, leptonic asymmetries, and properties of singlet
  fermions},'' \href{http://dx.doi.org/10.1088/1126-6708/2008/08/008}{{\em
  JHEP} {\bfseries 0808} (2008) 008},
\href{http://arxiv.org/abs/0804.4542}{{\ttfamily arXiv:0804.4542 [hep-ph]}}.

\bibitem{Akhmedov:1998qx}
E.~K. Akhmedov, V.~Rubakov, and A.~Y. Smirnov, ``{Baryogenesis via neutrino
  oscillations},'' \href{http://dx.doi.org/10.1103/PhysRevLett.81.1359}{{\em
  Phys.Rev.Lett.} {\bfseries 81} (1998) 1359--1362},
\href{http://arxiv.org/abs/hep-ph/9803255}{{\ttfamily arXiv:hep-ph/9803255
  [hep-ph]}}.

\bibitem{CooperSarkar:1985nh}
A.~M. Cooper-Sarkar {\em et~al.}, {\bfseries WA66 Collaboration} , ``{Search
  for Heavy Neutrino Decays in the BEBC Beam Dump Experiment},''
\href{http://dx.doi.org/10.1016/0370-2693(85)91493-5}{{\em Phys.Lett.}
  {\bfseries B160} (1985) 207}.

\bibitem{Bergsma:1985is}
F.~Bergsma {\em et~al.}, {\bfseries CHARM Collaboration} , ``{A Search for
  Decays of Heavy Neutrinos in the Mass Range 0.5-GeV to 2.8-GeV},''
\href{http://dx.doi.org/10.1016/0370-2693(86)91601-1}{{\em Phys.Lett.}
  {\bfseries B166} (1986) 473}.

\bibitem{Vaitaitis:1999wq}
A.~Vaitaitis {\em et~al.}, {\bfseries NuTeV Collaboration, E815 Collaboration}
  , ``{Search for Neutral Heavy Leptons in a High-Energy Neutrino Beam},''
  \href{http://dx.doi.org/10.1103/PhysRevLett.83.4943}{{\em Phys.Rev.Lett.}
  {\bfseries 83} (1999) 4943--4946},
\href{http://arxiv.org/abs/hep-ex/9908011}{{\ttfamily arXiv:hep-ex/9908011
  [hep-ex]}}.

\bibitem{Bernardi:1985ny}
G.~Bernardi, G.~Carugno, J.~Chauveau, F.~Dicarlo, M.~Dris, {\em et~al.},
  ``{Search for Neutrino Decay},''
\href{http://dx.doi.org/10.1016/0370-2693(86)91602-3}{{\em Phys.Lett.}
  {\bfseries B166} (1986) 479}.

\bibitem{Bernardi:1987ek}
G.~Bernardi, G.~Carugno, J.~Chauveau, F.~Dicarlo, M.~Dris, {\em et~al.},
  ``{Further Limits on Heavy Neutrino Couplings},''
\href{http://dx.doi.org/10.1016/0370-2693(88)90563-1}{{\em Phys.Lett.}
  {\bfseries B203} (1988) 332}.

\bibitem{Canetti:2010aw}
L.~Canetti and M.~Shaposhnikov, ``{Baryon Asymmetry of the Universe in the
  NuMSM},'' \href{http://dx.doi.org/10.1088/1475-7516/2010/09/001}{{\em JCAP}
  {\bfseries 1009} (2010) 001},
\href{http://arxiv.org/abs/1006.0133}{{\ttfamily arXiv:1006.0133 [hep-ph]}}.

\bibitem{Kullenberg:2011rd}
C.~Kullenberg {\em et~al.}, {\bfseries NOMAD Collaboration} , ``{A Search for
  Single Photon Events in Neutrino Interactions in NOMAD},''
  \href{http://dx.doi.org/10.1016/j.physletb.2011.11.049}{{\em Phys.Lett.}
  {\bfseries B706} (2012) 268--275},
\href{http://arxiv.org/abs/1111.3713}{{\ttfamily arXiv:1111.3713 [hep-ex]}}.

\bibitem{Volpe:2001qe}
C.~Volpe, N.~Auerbach, G.~Colo, T.~Suzuki, and N.~Van~Giai, ``{Neutrino C-12
  reactions and the LSND and KARMEN experiments on neutrino oscillations},''
\href{http://dx.doi.org/10.1134/1.1389536}{{\em Phys. Atom. Nucl.} {\bfseries
  64} (2001) 1165--1168}.

\bibitem{Maltoni:2007zf}
M.~Maltoni and T.~Schwetz, ``{Sterile neutrino oscillations after first
  MiniBooNE results},''
  \href{http://dx.doi.org/10.1103/PhysRevD.76.093005}{{\em Phys.Rev.}
  {\bfseries D76} (2007) 093005},
\href{http://arxiv.org/abs/0705.0107}{{\ttfamily arXiv:0705.0107 [hep-ph]}}.

\bibitem{Ade:2013zuv}
P.~Ade {\em et~al.}, {\bfseries Planck Collaboration} , ``{Planck 2013 results.
  XVI. Cosmological parameters},''
  \href{http://arxiv.org/abs/1303.5076}{{\ttfamily arXiv:1303.5076
  [astro-ph.CO]}},
2013.

\bibitem{Bennett:2012zja}
C.~Bennett {\em et~al.}, {\bfseries WMAP} , ``{Nine-Year Wilkinson Microwave
  Anisotropy Probe (WMAP) Observations: Final Maps and Results},''
  \href{http://dx.doi.org/10.1088/0067-0049/208/2/20}{{\em Astrophys.J.Suppl.}
  {\bfseries 208} (2013) 20},
\href{http://arxiv.org/abs/1212.5225}{{\ttfamily arXiv:1212.5225
  [astro-ph.CO]}}.

\bibitem{Batell:2009di}
B.~Batell, M.~Pospelov, and A.~Ritz, ``{Exploring Portals to a Hidden Sector
  Through Fixed Targets},''
  \href{http://dx.doi.org/10.1103/PhysRevD.80.095024}{{\em Phys.Rev.}
  {\bfseries D80} (2009) 095024},
\href{http://arxiv.org/abs/0906.5614}{{\ttfamily arXiv:0906.5614 [hep-ph]}}.

\bibitem{deNiverville:2011it}
P.~deNiverville, M.~Pospelov, and A.~Ritz, ``{Observing a light dark matter
  beam with neutrino experiments},''
  \href{http://dx.doi.org/10.1103/PhysRevD.84.075020}{{\em Phys.Rev.}
  {\bfseries D84} (2011) 075020},
\href{http://arxiv.org/abs/1107.4580}{{\ttfamily arXiv:1107.4580 [hep-ph]}}.

\bibitem{deNiverville:2012ij}
P.~deNiverville, D.~McKeen, and A.~Ritz, ``{Signatures of sub-GeV dark matter
  beams at neutrino experiments},''
  \href{http://dx.doi.org/10.1103/PhysRevD.86.035022}{{\em Phys.Rev.}
  {\bfseries D86} (2012) 035022},
\href{http://arxiv.org/abs/1205.3499}{{\ttfamily arXiv:1205.3499 [hep-ph]}}.

\bibitem{Dharmapalan:2012xp}
R.~Dharmapalan {\em et~al.}, {\bfseries MiniBooNE Collaboration} , ``{Low Mass
  WIMP Searches with a Neutrino Experiment: A Proposal for Further MiniBooNE
  Running},'' FERMILAB-PROPOSAL-1032,
  \href{http://arxiv.org/abs/1211.2258}{{\ttfamily arXiv:1211.2258 [hep-ex]}},
2012.

\bibitem{Bethe:1939bt}
H.~Bethe, ``{Energy production in stars},''
\href{http://dx.doi.org/10.1103/PhysRev.55.434}{{\em Phys.Rev.} {\bfseries 55}
  (1939) 434--456}.

\bibitem{Weizsaecker:1938}
C.~Weizs{\"a}cker, ``{{\"U}ber Elementumwandlungen im Innern der Sterne II},''
\href{http://dx.doi.org/10.1103/Physik.Z.39.633}{{\em Physik.Z.} {\bfseries 39}
  (1938) 633--646}.

\bibitem{Bahcall:2004pz}
J.~N. Bahcall, A.~M. Serenelli, and S.~Basu, ``{New solar opacities,
  abundances, helioseismology, and neutrino fluxes},''
  \href{http://dx.doi.org/10.1086/428929}{{\em Astrophys.J.} {\bfseries 621}
  (2005) L85--L88},
\href{http://arxiv.org/abs/astro-ph/0412440}{{\ttfamily arXiv:astro-ph/0412440
  [astro-ph]}}.

\bibitem{Fukuda:2001nj}
S.~Fukuda {\em et~al.}, {\bfseries Super-Kamiokande Collaboration} , ``{Solar
  B-8 and hep neutrino measurements from 1258 days of Super-Kamiokande data},''
  \href{http://dx.doi.org/10.1103/PhysRevLett.86.5651}{{\em Phys.Rev.Lett.}
  {\bfseries 86} (2001) 5651--5655},
\href{http://arxiv.org/abs/hep-ex/0103032}{{\ttfamily arXiv:hep-ex/0103032
  [hep-ex]}}.

\bibitem{Ahmad:2001an}
Q.~Ahmad {\em et~al.}, {\bfseries SNO Collaboration} , ``{Measurement of the
  rate of nu/e + d -> p + p + e- interactions produced by B-8 solar neutrinos
  at the Sudbury Neutrino Observatory},''
  \href{http://dx.doi.org/10.1103/PhysRevLett.87.071301}{{\em Phys.Rev.Lett.}
  {\bfseries 87} (2001) 071301},
\href{http://arxiv.org/abs/nucl-ex/0106015}{{\ttfamily arXiv:nucl-ex/0106015
  [nucl-ex]}}.

\bibitem{Borexino7be:2011}
G.~Bellini, J.~Benziger, D.~Bick, S.~Bonetti, G.~Bonfini, {\em et~al.},
  ``{Precision measurement of the 7Be solar neutrino interaction rate in
  Borexino},'' \href{http://dx.doi.org/10.1103/PhysRevLett.107.141302}{{\em
  Phys.Rev.Lett.} {\bfseries 107} (2011) 141302},
\href{http://arxiv.org/abs/1104.1816}{{\ttfamily arXiv:1104.1816 [hep-ex]}}.

\bibitem{Kraus:2006qp}
C.~Kraus, {\bfseries SNO+ Collaboration} , ``{SNO with liquid scintillator:
  SNO+},''
\href{http://dx.doi.org/10.1016/j.ppnp.2005.12.001}{{\em Prog. Part. Nucl.
  Phys.} {\bfseries 57} (2006) 150--152}.

\bibitem{Sekiya:2013hda}
H.~Sekiya, {\bfseries Super-Kamiokande Collaboration} , ``{Solar neutrino
  analysis of Super-Kamiokande},''
  \href{http://arxiv.org/abs/1307.3686}{{\ttfamily arXiv:1307.3686}},
2013.

\bibitem{Guglielmi:2012}
A.~Guglielmi, {\bfseries ICARUS Collaboration} , ``{Status and early events
  from ICARUS T600},''
{\em Nucl.Phys} {\bfseries B (Proc. Suppl.) 229-232} (2012) 342--346.

\bibitem{Borexinopep:2011}
G.~Bellini {\em et~al.}, {\bfseries Borexino Collaboration} , ``{First evidence
  of pep solar neutrinos by direct detection in Borexino},''
  \href{http://dx.doi.org/10.1103/PhysRevLett.108.051302}{{\em Phys.Rev.Lett.}
  {\bfseries 108} (2012) 051302},
\href{http://arxiv.org/abs/1110.3230}{{\ttfamily arXiv:1110.3230 [hep-ex]}}.

\bibitem{Borexino8b:2008}
G.~Bellini {\em et~al.}, {\bfseries Borexino Collaboration} , ``{Measurement of
  the solar 8B neutrino rate with a liquid scintillator target and 3 MeV energy
  threshold in the Borexino detector},''
  \href{http://dx.doi.org/10.1103/PhysRevD.82.033006}{{\em Phys.Rev.}
  {\bfseries D82} (2010) 033006},
\href{http://arxiv.org/abs/0808.2868}{{\ttfamily arXiv:0808.2868 [astro-ph]}}.

\bibitem{Gando:2013}
A.~Gando {\em et~al.}, {\bfseries KamLAND Collaboration} , ``{Reactor On-Off
  Antineutrino Measurement with KamLAND},''
  \href{http://arxiv.org/abs/1303.4667}{{\ttfamily arXiv:1303.4667 [hep-ex]}},
2013.

\bibitem{Silk:1985ax}
J.~Silk, K.~A. Olive, and M.~Srednicki, ``{The Photino, the Sun and High-Energy
  Neutrinos},''
\href{http://dx.doi.org/10.1103/PhysRevLett.55.257}{{\em Phys.Rev.Lett.}
  {\bfseries 55} (1985) 257--259}.

\bibitem{Cirelli:2005gh}
M.~Cirelli, N.~Fornengo, T.~Montaruli, I.~A. Sokalski, A.~Strumia, {\em
  et~al.}, ``{Spectra of neutrinos from dark matter annihilations},''
  \href{http://dx.doi.org/10.1016/j.nuclphysb.2005.08.017,
  10.1016/j.nuclphysb.2007.10.001}{{\em Nucl.Phys.} {\bfseries B727} (2005)
  99--138},
\href{http://arxiv.org/abs/hep-ph/0506298}{{\ttfamily arXiv:hep-ph/0506298
  [hep-ph]}}.

\bibitem{LoSecco:1986fu}
J.~LoSecco, J.~Van~der Velde, R.~Bionta, G.~Blewitt, C.~Bratton, {\em et~al.},
  ``{Limits on the Flux of Energetic Neutrinos from the Sun},''
\href{http://dx.doi.org/10.1016/0370-2693(87)91403-1}{{\em Phys.Lett.}
  {\bfseries B188} (1987) 388}.

\bibitem{Aartsen:2012kia}
M.~Aartsen {\em et~al.}, {\bfseries IceCube Collaboration} , ``{Search for dark
  matter annihilations in the Sun with the 79-string IceCube detector},''
  \href{http://dx.doi.org/10.1103/PhysRevLett.110.131302}{{\em Phys.Rev.Lett.}
  {\bfseries 110} (2013) 131302},
\href{http://arxiv.org/abs/1212.4097}{{\ttfamily arXiv:1212.4097
  [astro-ph.HE]}}.

\bibitem{Blennow:2013pya}
M.~Blennow, M.~Carrigan, and E.~F. Martinez, ``{Probing the Dark Matter mass
  and nature with neutrinos},''
  \href{http://dx.doi.org/10.1088/1475-7516/2013/06/038}{{\em JCAP} {\bfseries
  1306} (2013) 038},
\href{http://arxiv.org/abs/1303.4530}{{\ttfamily arXiv:1303.4530 [hep-ph]}}.

\bibitem{Totani:1995dw}
T.~Totani, K.~Sato, and Y.~Yoshii, ``{Spectrum of the supernova relic neutrino
  background and evolution of galaxies},''
  \href{http://dx.doi.org/10.1086/176970}{{\em Astrophys.J.} {\bfseries 460}
  (1996) 303--312},
\href{http://arxiv.org/abs/astro-ph/9509130}{{\ttfamily arXiv:astro-ph/9509130
  [astro-ph]}}.

\bibitem{Sato:1997sc}
K.~Sato, T.~Totani, and Y.~Yoshii, ``{Spectrum of the supernova relic neutrino
  background and evolution of galaxies},''
1997.

\bibitem{Hartmann:1997qe}
D.~Hartmann and S.~Woosley, ``{The cosmic supernova neutrino background},''
\href{http://dx.doi.org/10.1016/S0927-6505(97)00018-2}{{\em Astropart.Phys.}
  {\bfseries 7} (1997) 137--146}.

\bibitem{Malaney:1996ar}
R.~Malaney, ``{Evolution of the cosmic gas and the relic supernova neutrino
  background},'' \href{http://dx.doi.org/10.1016/S0927-6505(97)00012-1}{{\em
  Astropart.Phys.} {\bfseries 7} (1997) 125--136},
\href{http://arxiv.org/abs/astro-ph/9612012}{{\ttfamily arXiv:astro-ph/9612012
  [astro-ph]}}.

\bibitem{Kaplinghat:1999xi}
M.~Kaplinghat, G.~Steigman, and T.~Walker, ``{The Supernova relic neutrino
  background},'' \href{http://dx.doi.org/10.1103/PhysRevD.62.043001}{{\em
  Phys.Rev.} {\bfseries D62} (2000) 043001},
\href{http://arxiv.org/abs/astro-ph/9912391}{{\ttfamily arXiv:astro-ph/9912391
  [astro-ph]}}.

\bibitem{Ando:2005ka}
S.~Ando, J.~F. Beacom, and H.~Yuksel, ``{Detection of neutrinos from supernovae
  in nearby galaxies},''
  \href{http://dx.doi.org/10.1103/PhysRevLett.95.171101}{{\em Phys.Rev.Lett.}
  {\bfseries 95} (2005) 171101},
  \href{http://arxiv.org/abs/astro-ph/0503321}{{\ttfamily
  arXiv:astro-ph/0503321 [astro-ph]}}.

\bibitem{Lunardini:2006pd}
C.~Lunardini, ``{Testing neutrino spectra formation in collapsing stars with
  the diffuse supernova neutrino flux},''
  \href{http://dx.doi.org/10.1103/PhysRevD.75.073022}{{\em Phys.Rev.}
  {\bfseries D75} (2007) 073022},
\href{http://arxiv.org/abs/astro-ph/0612701}{{\ttfamily arXiv:astro-ph/0612701
  [astro-ph]}}.

\bibitem{Fukugita:2002qw}
M.~Fukugita and M.~Kawasaki, ``{Constraints on the star formation rate from
  supernova relic neutrino observations},''
  \href{http://dx.doi.org/10.1046/j.1365-8711.2003.06507.x}{{\em
  Mon.Not.Roy.Astron.Soc.} {\bfseries 340} (2003) L7},
\href{http://arxiv.org/abs/astro-ph/0204376}{{\ttfamily arXiv:astro-ph/0204376
  [astro-ph]}}.

\bibitem{Vogel:1999zy}
P.~Vogel and J.~F. Beacom, ``{Angular distribution of neutron inverse beta
  decay, anti-neutrino(e) + p ---> e+ + n},''
  \href{http://dx.doi.org/10.1103/PhysRevD.60.053003}{{\em Phys.Rev.}
  {\bfseries D60} (1999) 053003},
\href{http://arxiv.org/abs/hep-ph/9903554}{{\ttfamily arXiv:hep-ph/9903554
  [hep-ph]}}.

\bibitem{Strumia:2003zx}
A.~Strumia and F.~Vissani, ``{Precise quasielastic neutrino nucleon cross
  section},'' \href{http://dx.doi.org/10.1016/S0370-2693(03)00616-6}{{\em Phys.
  Lett.} {\bfseries B564} (2003) 42--54},
\href{http://arxiv.org/abs/astro-ph/0302055}{{\ttfamily
  arXiv:astro-ph/0302055}}.

\bibitem{Ormand:1994js}
W.~E. Ormand, P.~M. Pizzochero, P.~F. Bortignon, and R.~A. Broglia, ``{Neutrino
  capture cross-sections for Ar-40 and Beta decay of Ti-40},''
  \href{http://dx.doi.org/10.1016/0370-2693(94)01605-C}{{\em Phys. Lett.}
  {\bfseries B345} (1995) 343--350},
\href{http://arxiv.org/abs/nucl-th/9405007}{{\ttfamily arXiv:nucl-th/9405007}}.

\bibitem{Kolbe:2003ys}
E.~Kolbe, K.~Langanke, G.~Martinez-Pinedo, and P.~Vogel, ``{Neutrino nucleus
  reactions and nuclear structure},''
  \href{http://dx.doi.org/10.1088/0954-3899/29/11/010}{{\em J. Phys.}
  {\bfseries G29} (2003) 2569--2596},
\href{http://arxiv.org/abs/nucl-th/0311022}{{\ttfamily arXiv:nucl-th/0311022}}.

\bibitem{SajjadAthar:2004yf}
M.~Sajjad~Athar and S.~K. Singh, ``{nu/e (anti-nu/e) - Ar-40 absorption cross
  sections for supernova neutrinos},''
\href{http://dx.doi.org/10.1016/j.physletb.2004.04.025}{{\em Phys. Lett.}
  {\bfseries B591} (2004) 69--75}.

\bibitem{Cocco:2004ac}
A.~Cocco, A.~Ereditato, G.~Fiorillo, G.~Mangano, and V.~Pettorino, ``{Supernova
  relic neutrinos in liquid argon detectors},''
  \href{http://dx.doi.org/10.1088/1475-7516/2004/12/002}{{\em JCAP} {\bfseries
  0412} (2004) 002},
\href{http://arxiv.org/abs/hep-ph/0408031}{{\ttfamily arXiv:hep-ph/0408031
  [hep-ph]}}.

\bibitem{Abbasi:2012eda}
R.~Abbasi {\em et~al.}, {\bfseries IceCube Collaboration} , ``{Search for
  Relativistic Magnetic Monopoles with IceCube},''
  \href{http://dx.doi.org/10.1103/PhysRevD.87.022001}{{\em Phys.Rev.}
  {\bfseries D87} (2013) 022001},
\href{http://arxiv.org/abs/1208.4861}{{\ttfamily arXiv:1208.4861
  [astro-ph.HE]}}.

\bibitem{Aartsen:2013mla}
M.~Aartsen {\em et~al.}, {\bfseries IceCube Collaboration} , ``{The IceCube
  Neutrino Observatory Part IV: Searches for Dark Matter and Exotic
  Particles},'' \href{http://arxiv.org/abs/1309.7007}{{\ttfamily
  arXiv:1309.7007 [astro-ph.HE]}},
2013.

\bibitem{Ueno:2012md}
K.~Ueno {\em et~al.}, {\bfseries Super-Kamiokande Collaboration} , ``{Search
  for GUT Monopoles at Super-Kamiokande},''
  \href{http://dx.doi.org/10.1016/j.astropartphys.2012.05.008}{{\em
  Astropart.Phys.} {\bfseries 36} (2012) 131--136},
\href{http://arxiv.org/abs/1203.0940}{{\ttfamily arXiv:1203.0940 [hep-ex]}}.

\bibitem{Aartsen:2014tna}
M.~Aartsen {\em et~al.}, {\bfseries IceCube Collaboration} , ``{Search for
  non-relativistic Magnetic Monopoles with IceCube},''
  \href{http://arxiv.org/abs/1402.3460}{{\ttfamily arXiv:1402.3460
  [astro-ph.CO]}},
2014.

\bibitem{Ambrosio:2002qq}
M.~Ambrosio {\em et~al.}, {\bfseries MACRO Collaboration} , ``{Final results of
  magnetic monopole searches with the MACRO experiment},''
  \href{http://dx.doi.org/10.1140/epjc/s2002-01046-9}{{\em Eur.Phys.J.}
  {\bfseries C25} (2002) 511--522},
\href{http://arxiv.org/abs/hep-ex/0207020}{{\ttfamily arXiv:hep-ex/0207020
  [hep-ex]}}.

\bibitem{Mohapatra:2009wp}
R.~Mohapatra, ``{Neutron-Anti-Neutron Oscillation: Theory and Phenomenology},''
  \href{http://dx.doi.org/10.1088/0954-3899/36/10/104006}{{\em J.Phys.}
  {\bfseries G36} (2009) 104006},
\href{http://arxiv.org/abs/0902.0834}{{\ttfamily arXiv:0902.0834 [hep-ph]}}.

\bibitem{doe_o413_2010}
{The United States Department of Energy}, ``{Program and Project Management for
  the Acquisition of Capital Assets},'' DOE,
\newblock
  {\href{https://www.directives.doe.gov/directives/0413.3-BOrder-b/view}{DOE O
  413.3B}}, November, 2010.

\bibitem{DOCDB8694}
J.~Strait, R.~Wilson, and V.~Papadimitriou, ``{LBNE Presentations to P5},''
\newblock
  \href{http://lbne2-docdb.fnal.gov/cgi-bin/ShowDocument?docid=8694}{{\ttfamily
  LBNE-doc-8694}}, November, 2013.

\bibitem{Bilenky:2012qi}
S.~Bilenky and C.~Giunti, ``{Neutrinoless double-beta decay: A brief review},''
  \href{http://dx.doi.org/10.1142/S0217732312300157}{{\em Mod.Phys.Lett.}
  {\bfseries A27} (2012) 1230015},
\href{http://arxiv.org/abs/1203.5250}{{\ttfamily arXiv:1203.5250 [hep-ph]}}.

\bibitem{CDRwcd}
{\bfseries LBNE Project Management Team} , ``{LBNE Conceptual Design Report:
  The LBNE Water Cherenkov Detector},''
\newblock
  \href{http://lbne2-docdb.fnal.gov/cgi-bin/ShowDocument?docid=5118}{{\ttfamily
  LBNE-doc-5118}}, 2012.

\bibitem{Hewett:2014qja}
J.~Hewett, H.~Weerts, K.~Babu, J.~Butler, B.~Casey, {\em et~al.}, ``{Planning
  the Future of U.S. Particle Physics (Snowmass 2013): Chapter 2: Intensity
  Frontier},'' FERMILAB-CONF-14-019-CH02,
  \href{http://arxiv.org/abs/1401.6077}{{\ttfamily arXiv:1401.6077 [hep-ex]}},
2014.

\bibitem{Green:2012gv}
C.~Green, J.~Kowalkowski, M.~Paterno, M.~Fischler, L.~Garren, {\em et~al.},
  ``{The art framework},''
\href{http://dx.doi.org/10.1088/1742-6596/396/2/022020}{{\em J.Phys.Conf.Ser.}
  {\bfseries 396} (2012) 022020}.

\bibitem{Katori:2011uq}
T.~Katori, {\bfseries MicroBooNE Collaboration} , ``{MicroBooNE, A Liquid Argon
  Time Projection Chamber (LArTPC) Neutrino Experiment},''
  \href{http://dx.doi.org/10.1063/1.3661595}{{\em AIP Conf.Proc.} {\bfseries
  1405} (2011) 250--255},
\href{http://arxiv.org/abs/1107.5112}{{\ttfamily arXiv:1107.5112 [hep-ex]}}.

\bibitem{Soderberg:2009qt}
M.~Soderberg, {\bfseries ArgoNeuT Collaboration} , ``{ArgoNeuT: A Liquid Argon
  Time Projection Chamber Test in the NuMI Beamline},'' FERMILAB-CONF-09-516-E,
  \href{http://arxiv.org/abs/0910.3433}{{\ttfamily arXiv:0910.3433
  [physics.ins-det]}},
2009.

\bibitem{huffman1952}
D.Huffman, ``{A Method for the Construction of Minimum-Redundancy Codes},'' in
  {\em Proceedings of the IRE}.
\newblock 1952.

\bibitem{Szydagis:2011tk}
M.~Szydagis, N.~Barry, K.~Kazkaz, J.~Mock, D.~Stolp, {\em et~al.}, ``{NEST: A
  Comprehensive Model for Scintillation Yield in Liquid Xenon},''
  \href{http://dx.doi.org/10.1088/1748-0221/6/10/P10002}{{\em JINST} {\bfseries
  6} (2011) P10002},
\href{http://arxiv.org/abs/1106.1613}{{\ttfamily arXiv:1106.1613
  [physics.ins-det]}}.

\bibitem{cry}
C.~Hagman, D.~Lange, J.~Verbeke, and D.~Wright, ``{Cosmic-ray Shower Library
  (CRY)},'' Lawrence Livermore National Laboratory,
\newblock UCRL-TM-229453, March, 2012.
\newblock \url{{http://nuclear.llnl.gov/simulation/doc_cry_v1.7/cry.pdf}}.

\bibitem{Sandhir:2012cr}
R.~P. Sandhir, S.~Muhuri, and T.~Nayak, ``{Dynamic Fuzzy c-Means (dFCM)
  Clustering and its Application to Calorimetric Data Reconstruction in High
  Energy Physics},'' \href{http://dx.doi.org/10.1016/j.nima.2012.04.023}{{\em
  Nucl.Instrum.Meth.} {\bfseries A681} (2012) 34--43},
\href{http://arxiv.org/abs/1204.3459}{{\ttfamily arXiv:1204.3459 [nucl-ex]}}.

\bibitem{kalman}
R.~E. Kalman, ``A new approach to linear filtering and prediction problems,''
  {\em Transactions of the ASME--Journal of Basic Engineering} {\bfseries 82}
  no.~Series D, (1960) 35--45.

\bibitem{Marshall:2012hh}
J.~Marshall and M.~Thomson, ``{The Pandora software development kit for
  particle flow calorimetry},''
\href{http://dx.doi.org/10.1088/1742-6596/396/2/022034}{{\em J.Phys.Conf.Ser.}
  {\bfseries 396} (2012) 022034}.

\bibitem{Accardi:2012qut}
A.~Accardi, J.~Albacete, M.~Anselmino, N.~Armesto, E.~Aschenauer, {\em et~al.},
  ``{Electron Ion Collider: The Next QCD Frontier - Understanding the glue that
  binds us all},'' BNL-98815-2012-JA, JLAB-PHY-12-1652,
  \href{http://arxiv.org/abs/1212.1701}{{\ttfamily arXiv:1212.1701 [nucl-ex]}},
2012.

\end{thebibliography}
